Universidade de São Paulo
Instituto de Física

# Esquemas de Medição em TQCA, Contextualidade e o Gedankenexperiment Amigo de Wigner

Felipe Dilho Alves

Orientador: Prof. Dr. João Carlos Alves Barata

Dissertação de mestrado apresentada ao Instituto de Física da Universidade de São Paulo, como requisito parcial para a obtenção do título de Mestre em Ciências.

São Paulo
2024

# University of São Paulo
# Physics Institute

# Measurement Schemes in AQFT, Contextuality and the Wigner's Friend Gedankenexperiment

Felipe Dilho Alves

Supervisor: Prof. Dr. João Carlos Alves Barata

# Agradecimentos

*Wittgenstein once greeted me with the question: "Tell me, why do people always say it was natural to assume that the sun went round the Earth, rather than that the Earth was rotating on its axis?" I replied: "I suppose, because it looked as if the sun went round the Earth."; "Well," he then asked: "what would it have looked like **if it had looked as if** the Earth was rotating?".*

— G. E. M. Anscombe, in
*An Introduction to Wittgenstein's Tractatus (1959, p. 151).*

# Abstract

Measurements have historically presented a problem for the consistent description of quantum theories, be it in non-relativistic quantum mechanics or in quantum field theory. Drawing on a recent surge of interest in the description of measurements in Algebraic Quantum Field theory, it was decided that this dissertation would be focused on trying to close the gap between the description of measurements proposed by K. Hepp in the 70's, considering decoherence of states in quasilocal algebras and the new framework of generally covariant measurement schemes proposed recently by C. Fewster and R. Verch. Another recent result that we shall also consider is the Frauchinger-Renner Gedankenexperiment, that has taken inspiration on Hepp's article about decoherence based measurements to arrive at a no-go result about the consistency of quantum descriptions of systems containing rational agents, we shall seek to provide a closure for the interpretation of this result. In doing so we naturally arrive at the study of the contextual properties of measurement setups.

# Resumo

Medições têm historicamente representado um problema para a descrição consistente de teorias quânticas, seja na mecânica quântica não relativística ou na teoria quântica de campos. Com base numa recente onda de interesse na descrição de medições na Teoria Quantica de Campos Algébrica, estipulou-se que esta dissertação se concentraria em tentar preencher a lacuna entre a descrição de medições proposta por K. Hepp na década de 70, considerando a decoerência de estados em álgebras quasilocais e a nova estrutura de esquemas de medição com covariância geral proposta recentemente por C. Fewster e R. Verch. Outro resultado recente que também consideraremos é o *Gedankenexperiment* de Frauchinger-Renner, que se inspirou no artigo do Hepp sobre medições baseadas em decoerência para chegar a uma obstrução na consistência das descrições de sistemas quânticos contendo agentes racionais, procuramos fornecer uma interpretação definitiva deste resultado. Ao fazer isso, chegamos naturalmente ao estudo das propriedades contextuais de configurações de medição.

**Palavras-chaves**: teoria quântica de campos algébrica, teoria de medição quântica, contextualidade, álgebras de von Neumann.

# Abbreviations

| | |
|---|---|
| AQFT | Algebraic Quantum Field Theory |
| EVM | Effect Valued Measure |
| FV *frm.* | Fewster-Verch framework |
| GNS | Gel'fand—Naĭmark—Segal |
| GR | General Relativity |
| KMS | Kubo—Martin—Schwinger |
| LHS | Left-hand side |
| pAQFT | Perturbative Algebraic Quantum Field Theory |
| POVM | Positive Operator-Valued-Measure |
| PVM | Projection-Valued-Measure |
| QFT | Quantum Field Theory |
| QM | Quantum Mechanics |
| QMT | Quantum Measurement Theory |
| RHS | Right-hand side |
| SOT | Strong Operator Topology |
| SR | Special Relativity |
| WAY | Wigner—Araki—Yanase |
| WOT | Weak Operator Topology |

# Contents

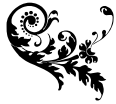 # CHAPTER 1 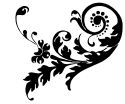

# Introduction

*A brief historical overview of the subject matter of this dissertation, which will serve not only to introduce the fundamental elements that shall constitute the body of the text, but also to motivate the topic.*

Quantum mechanics had it's first modern mathematical formulation in 1925 in the form of the **matrix mechanics** of Heisenberg, Born and Jordan (before that there existed a phenomenological understanding of quantum physics present in the work of Planck, Bohr, Sommerfeld, amongst others); just one year latter, in 1926, Schrödinger published a paper with his equation inaugurating **wave mechanics** as another formulation of quantum mechanics. The mathematical equivalence of both formulations eluded physicists until what was then called *Transformation theory* was developed by Dirac and Schrödinger, in a casuistic way by working with explicit calculations of known examples leading to the formal manipulation rules of the, until then, ill-defined mathematical objects of the theory such as the Dirac $\delta$-"function" and its derivatives [96], as mathematical rigour wasn't of first importance, citing Dirac: "problems were tackled top-down, by working on the great principles, with the details left to look after themselves" [81, 29].

Von Neumann in his seminal work *Mathematische Grundlagen der Quantenmechanik* [98] established the first complete mathematical formulation of quantum mechanics through the framework of a abstract, axiomatic theory of separable Hilbert spaces and their linear operators, in this formulation states are represented by the rays of Hilbert space vectors and the measurable quantities by hermitian operators of the underlying Hilbert space.

The separable condition means that the Hilbert space contains a subset that is *countable* and *dense*. Since a Hilbert space is a complete metric space there is a proposition[1] that says that for metric spaces being separable is equivalent to being second countable, *i.e.* having a countable base of open sets, that is, any open set in this space can be represented by the union of a subfamily of this countable base of open sets. But since every vector space has a complete orthonormal basis (a fact that is logically equivalent to Zorn's Lemma), we then get that every separable Hilbert space has a countable orthonormal basis by taking the intersection between each element of its countable base of open sets and its complete orthonormal basis.

Considering the space of states in *matrix mechanics*, the necessity of normalization means that only those elements for which the sum of the square of its components is finite can be included, that is the space:

$$\left\{ (z_n)_{n\in\mathbb{N}} \in \mathbb{C}^{\mathbb{N}} \ \middle| \ \left\|(z_n)_{n\in\mathbb{N}}\right\|_{\ell_2} = \left(\sum_{n=1}^{\infty} z_n \bar{z}_n\right)^{\frac{1}{2}} < \infty \right\} \equiv \ell_2(\mathbb{C})$$

---

[1] the proper statement of this proposition together with a proof and a elaboration on the following reasoning about Hilbert space theory is in Appendix B, proposition B.0.5.

On the other hand the space of states in *wave machanics* must also have its wave functions, at a time $t$, $\Psi_t$ normalizable:

$$\left\{\Psi_t : \mathbb{R}^n \to \mathbb{C} \;\middle|\; \Psi_t \text{ is } \mu\text{-measurable}, \; \|\Psi_t\|_{L^2} = \left(\int_{\mathbb{R}^n} \Psi_t(\mathbf{x})\overline{\Psi_t(\mathbf{x})}d\mu(\mathbf{x})\right)^{\frac{1}{2}} < \infty \right\} \Big/ \sim_\mu \equiv L^2(\mathbb{R}^n, d\mu)$$

The condition of the $\Psi_t$ being measurable is necessary for the integral to be defined, the equivalence relation $\sim_\mu$ is defined as $f \sim_\mu g \iff \mu(\{\mathbf{x} \in \mathbb{R}^n \mid f(\mathbf{x}) \neq g(\mathbf{x})\}) = 0$ and the technicality of the necessity to consider the quotient set comes from the fact that the functions $\Psi_t$ have not been required to be continuous and therefore we may have two different functions that are operationally the same, also because of the following proposition, we can argue that the measure $\mu$ should be **separable**:

**Proposition 1.0.1.** *A $\sigma$-additive positive measure $\mu$ on a measurable space $(X, \Sigma)$, and hence also for $L^p(X, d\mu)$, **is separable** if the following conditions hold:*

*(i) $\mu$ is $\sigma$-finite (that is $\exists \{M_n\}_{n\in\mathbb{N}}$ such that $\forall n \in \mathbb{N}$, $\mu(M_n) < \infty$ and we have that $\bigcup_{n\in\mathbb{N}} M_n = X$).*

*(ii) the $\sigma$-algebra $\Sigma$ is generated by a countable collection of measurable sets at most.*

While it is true that for an arbitrary $\sigma$-algebra and measure $\mu$ in $\mathbb{R}$ it can be that the measure $\mu$ is not separable, but when we are talking about wave functions on an euclidean space, the intuitive property of translation invariance of the measure would imply that we are speaking of a Lebesgue measure, furthermore the natural and simplest $\sigma$-algebra to consider in physical situations is the Borel $\sigma$-algebra of $\mathbb{R}^n$. With these physical assumptions both conditions hold since we can partition $\mathbb{R}^n$ by hyperrectangles $I_m = [a_{m1}, b_{m1}] \times ... \times [a_{mn}, b_{mn}]$, $m \in \mathbb{Z}^n$ with finite volume and this same enumerable set of hyperrectangles generate the Borel $\sigma$-algebra of $\mathbb{R}^n$.[2] Therefore the space of wave functions $L^2(\mathbb{R}^n, d\mu)$ must be a separable space.[3] We shall make a abuse of notation from here on where we will denote $f \in L^2(X, \mu)$ for a function $f : X \to \mathbb{C}$, when in fact it should be the equivalence class $[f]_{\sim_\mu} \in L^2(X, \mu)$ instead of the representative $f \in [f]_{\sim_\mu}$.

---

[2] The proof of 1.0.1 as well as elements of measure theory will be given at Appendix C, specifically prop. C.0.27.

[3] The general assumption in QM that the underlying Hilbert space of the theory must be separable has been originally imposed because it was sufficiently general to be able to make calculations with its countable basis on any know physical system, initially without any strong physical reason that imposed separability.

V. Moretti in [77] elaborates that one could make the case that the Hilbert space of a physical system is separable, by considering thermodynamical states of a system in equilibrium that are quantum mechanically represented by mixed states of the form $Ze^{-\beta H}$, where $Z := \text{tr}(e^{-\beta H})$, $H$ is the Hamiltonian of the system, $\beta := 1/(k_B T)$, $k_B$ is Boltzmann's constant and $T$ is the absolute temperature. Then obviously $e^{-\beta H}$ must be of trace-class which implies that $e^{-\beta H}$ and also $H$ itself must either have purely point spectrum or must have a spectrum composed of a point spectrum part and a $0 \in \sigma_c$ part in the continuous spectrum, for if the Hilbert space was not separable then $e^{-\beta H}$ would have an uncountable set of orthonormal eigenvectors with strictly positive eigenvalue, giving rise to a divergent trace. And a bounded below Hamiltonian, even if equipped with pure point spectrum only, could not produce thermal mixed states of the usually know form.

As it turns out it is possible in the algebraic approach to deal with systems with possibly non-separable Hilbert space representation, in the algebraic formalism it is possible to deal with extended thermodynamical systems defined in algebraic states using the KMS condition in order to characterize algebraic thermodynamical states. More about the algebraic formalism will be discussed in Chapter 3.

With these facts we can get, in a completely analogous way as von Neumann did [98], a rigorous demonstration that *matrix mechanics* is mathematically equivalent to *wave machanics* through the Riesz–Fischer theorem[4]:

**Theorem 1.0.2** (Riesz-Fischer).[5] *Let $\{\phi_i\}_{i \in I} \subseteq L^2(X, \mu)$ be a enumerable orthonormal basis, then there is an isometric isomorphism between $L^2(X, d\mu)$ and $\ell_2(\mathbb{C})$ given by:*

$$\Phi : L^2(X, d\mu) \to \ell_2(\mathbb{C}),\ L^2(X, d\mu) \ni \psi \overset{\Phi}{\mapsto} \left(\int_X \psi(x)\overline{\phi_i(x)}d\mu(x)\right)_{i \in I} \in \ell_2(\mathbb{C}).$$

This then establishes a isomorphism between the space of *wave functions* and the space of *discrete amplitudes of matrix mechanics*, as we can see the inverse of $\Phi$ being given by the map $\Phi^{-1} : \ell_2(\mathbb{C}) \to L^2(X, d\mu),\ \ell_2(\mathbb{C}) \ni (c_i)_{i \in I} \overset{\Phi^{-1}}{\longmapsto} \sum_{i \in I} c_i \phi_i$, and that this map is isometric follows because of Parseval's identity $\|f\|_{L^2}^2 = \sum_{i \in I} |\int_X f\overline{\phi_i} d\mu|^2 = \|\Phi(f)\|_{\ell_2}^2$, through this isomorphism follows that *matrix mechanics* and *wave mechanics* are in a 1-1 correspondence that preserves probabilities and therefore both have the same descriptive power via this mathematical equivalence. Therefore von Neumann established that both theories could be thought of as different representations of the same abstract Hilbert space. As is alluded in the Riesz-Fischer theorem, maps that realize isometric isomorphisms are of particular importance in Hilbert space theory.

**Definition 1.0.3** (Unitary transformation). *An isomorphism of inner product spaces between Hilbert spaces, is in particular a isomorphism of Hilbert spaces, and is called a* unitary transformation *or equivalently a* unitary operator.

In this definition the inner product isomorphism is equivalent to a isometry for the metrics given by $d_{\mathscr{H}}(a, b) := \|a - b\|_{\mathscr{H}} = \sqrt{\langle a - b, a - b\rangle_{\mathscr{H}}}$ for each Hilbert space, and if the isomorphism is realized though the action of a linear operator $L$, then, by definition, this operator is bounded since by the isometry $\|Lr\|_{\mathscr{H}_1} = C\|r\|_{\mathscr{H}_2}$, with the constant $C = 1$.

## 1.1 Measurement in quantum mechanics

The field often called Quantum Measurement Theory (QMT) explores certain questions relevant to the description of the measurement process, as first noted by von Neumann in [98] the usual description of quantum mechanics involves two different rules for updating a quantum system, one due to continuous, deterministic and unitary time evolution given by a dynamics governed by a strongly continuous one-parameter unitary group generated from the Hamiltonian operator, in the case of a vector state at time $t$, $\Psi_t$, this is:

$$\Psi_{t+t'} = e^{-it'\frac{H}{\hbar}}\Psi_t \ ,$$

---

[4] Sometimes, as in [8], the Riesz—Fischer theorem is enunciated as:
For $p \geq 1$ the spaces $L^p(X, d\mu)$ are complete metric spaces in the metric induced by the norm $\|\cdot\|_{L^p}$,
$d_p(f, g) := \|f - g\|_{L^p} = \left(\int_X |f - g|^p d\mu\right)^{\frac{1}{p}}$, $f, g \in L^p(X, d\mu)$
As it happens, a proof will be given at Appendix B, corollary B.0.11, this enunciation is equivalent to the enunciation given at the body of the text only when p=2, since it is the only case in which $L^p$ is a Hilbert space.
This happens because the Riesz–Fischer theorem was originally only proved for p=2 and since both enunciations were equivalent both received the same name in the literature, but as developments progressed one of the enunciations was generalized keeping the same name, nevertheless it is not possible to generalize the other enunciation in the same way.

[5] This result may be extended [43] for a non-countable $I$ and non-separable $\mu$ with a redefinition of $\ell_2(\mathbb{C})$ in the case of uncountable sequences as $\widetilde{\ell_2} := \left\{(z_\alpha)_{\alpha \in I} \in X^I \,\big|\, \|(z_\alpha)_{\alpha \in I}\|_{\widetilde{\ell_2}} = \left(\sum_{\alpha \in I} z_\alpha \overline{z_\alpha}\right)^{\frac{1}{2}} := \left(\sup_J \{\sum_{j \in J} z_j \overline{z_j} \mid J \subset I, J \text{ finite}\}\right)^{\frac{1}{2}} < \infty\right\}$,
but the complicated space resulting from this expression does not have a clear physical motivation or applicability.

And another update rule, present only during measurement processes, which is discontinuous, probabilistic, non-unitary and non-local sometimes called "collapse of the wave function" and by others of "reduction of the wave packet", first explicitly stated by von Neumann[98] that basically just considered that if a physical quantity in a quantum system was measured twice in succession then the same value would be obtained each time, this would only describe the post-measurement state in the case of a quantity associated with a operator with only a point spectrum and no degeneracy, this measurement *ansatz* would later be generalized by Lüders [74] adding the requirement that, in the case of a vector state $\Psi$, the measurement of a operator with a degenerate point spectrum, the final state $\Psi_\lambda$ after the measurement of the eigenvalue $\lambda$ will be the normalized projection of the initial state in the eigenspace of the measured eigenvalue $\lambda$:

$$\Psi_\lambda = \frac{P_\lambda \Psi}{\|P_\lambda \Psi\|},$$

This von Neumann-Lüders postulate refers to idealized non-destructive measurements, a type of measurement that was coined by Pauli [83] to be a measurement of the first kind, a prescription that allows one to specify uniquely and in a standardized way the state of the system after a measurement of a physical quantity with a given result. Pauli's measurements of the second kind involve generally either the absorption or destruction of the physical system being measured and these processes need not happen in a standardized way, they represent any way of capturing information from the system, almost like torturing the system in a controlled fashion until it spills the beans, so to speak. Although part of the argument for focusing on measurements of the first kind has to do with it being possible to treat them in a unified way, it also occurs that there is no definite example of a physical quantity not admitting at least one measuring instrument satisfying, even if only approximately, this postulate [10].

As it can be seen, after a von Neumann-Lüders type of measurement, the process through which this happens need not be unitary:

Consider two states represented by the vectors $\Psi$ and $\phi$ which are eigenstates of a given operator $L$ such that $\Psi$ has a $s$ eigenvalue and $\phi$ has a $p$ eigenvalue. For a measurement that results in the value $s$ we shall have that both $\Psi$ and the superposition $\alpha\Psi + \beta\phi$; $\alpha, \beta \in \mathbb{C}$, $\alpha^2 + \beta^2 = 1$ are taken to $\Psi$ by a Lüders-von Neumann measurement, which isn't bijective and therefore not unitary.

For pragmatic reasons many people object to saying that there is a "measurement problem", after all, the usual description of quantum mechanics does not reveal any inconsistency with the data of experiments carried out within a non-relativistic limit of validity to which it adheres. subject, however a more descriptive analysis reveals at least one logical incompatibility, if we think that quantum systems are necessarily measured by other quantum systems, that is, the measurement apparatus, then we must expect that the evolution of the system is unitary, on the other hand, it seems as if when one places a observer tag in a given apparatus, the theory then tells us that the evolution of the system must then be given by a reduction of the wave packet.

It seems logically reasonable that if there are physical differences between the types of evolution described by both processes, these differences must come from other physical characteristics that differentiate the measuring apparatus from the rest of the quantum systems, and this is partly what the field of quantum measurement theory (QMT) seeks to evidentiate. Although it seems only a matter of philosophical debate, QMT methods have come to have relevant physical applications in the field of quantum information.

## 1.2 From AQFT and back again

The beginning of Algebraic Quantum Field Theory (AQFT) came from the development of local operator algebras as a way to reconcile quantum mechanics with special relativity in order to maintain the algebraic spirit established by **matrix mechanics** of Heisenberg *et al*, instead of the usual formulation through Feynman path integrals, those, perhaps, more in the analytic spirit of Schrödinger's attempts to extend **wave mechanics**, whose full and mathematically rigorous description still eludes us.

AQFT sets out to formulate an axiomatic framework under which all quantum field theories that deserve the name can be treated. Later developments [46] seem to indicate that small adaptations in the AQFT formalism can be employed to lead to a description of quantum field theories in curved spacetimes.

As it happens, a singular result in QMT was deduced in 1972 by Klaus Hepp [56] using the formalism of AQFT, in his paper Hepp uses a version of the Haag-Kastler axioms for QFT's [52] to deduce several results about ways in which the reduction of the wave packet can be obtained through a unitary evolution of quantum systems solving, in a way, the problem of measurement, at least for those systems that may be described by that specific model, and considering that infinite limits of both time and physical extension of the measuring apparatus could be applied, this is one of the criticisms of J.S. Bell about the results of Hepp. Nevertheless even in finite cases Hepp had shown that the errors would be limited by a rapidly decreasing Gaussian fo the number of elements of the measuring apparatus.

Another interesting coincidence is that one of the main proponents of AQFT, Huzihiro Araki (sometimes the Haag-Kastler axioms are called Araki-Haag-Kastler axioms in the literature) had also worked on a theorem [6] that gave limitations on measurements of systems that have a additive conserved quantity, or at least the accuracy with which observables that fail to commute with the conserved quantity can be measured. This paper in specific was not written with the algebraic framework (it was published before [6] after all), but one lemma present in it that is a corollary of the Naĭmark extension theorem shows that the developments on that paper had at least been informed by the algebraic methods that were still only just inscribed on the domain of operator algebras in mathematics.

What can be seen from these descriptions exposed so far is that one could expect that AQFT, initially formulated with algebras of local observables, could serve as a language in which one could express QMT results and obtain a more mathematically precise foundation for them.

However, what can be observed is that in the most recent developments of a *framework* proposed by C. Fewster and R. Verch [41, 35] with the ability to describe measurement processes having general covariance, no mention is made of older results of measurement processes in AQFT [6, 104, 56], in the case of the subject of the article by K. Hepp [56] which deals with an explicit model of reduction of the package of wave, Fewster and Verch [41] choose not to include the measurement problem as something that could be dealt with by the proposed *framework*, raising the question about the possibility that a reformulation of the Coleman-Hepp model in the framework by Fewster and Verch may bring either new results or refinements of old results about the wave packet reduction process as an emergent rather than inherent process in quantum systems.

Furthermore, the article [56] also came to influence another article recently published by D. Frauchiger and R. Renner [45] on a generalization of the gedankenexperiment proposed by Wigner [102] known as the gedankenexperiment of the friend of Wigner, whose clear interpretation of its implications has eluded several researchers [18, 44, 80] and that due to the way the impossibility theorem ("no-go" theorem) was constructed from the situation stipulated by the gedankenexperiment strongly suggests a relationship with results arising from quantum contextuality, such as the one present in the Kochen-Specker theorem [67], so an explanation of how contextuality is involved in Frauchiger and Renner's result can not only help to interpret it, but also help to simplify and generalize the same, in order to also contribute to the understanding of the Coleman-Hepp model itself in order to understand how contextuality is presented in the situation of measuring a system by an apparatus.

As a reminder, in Wigner's Gedankenexperiment a quantum system $d$ that can be measured to be in any of the two eigenstates of a operator $\sigma \in \mathfrak{B}(\mathscr{H})$, for simplicity let's consider this a measurement of spin in the $z$ direction for a spin-$\frac{1}{2}$ particle, $d$ is located in a laboratory $L$ and is measured upon by experimenter $F$ getting an measurement outcome for the spin projection quantity $\sigma$, this laboratory is completely isolated from the rest of the world with the exception of $W$ that can measure the whole of $L$ getting the state of $d$ and the state of the experimenter $F$ himself.

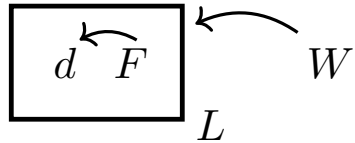

Figure 1 — Representation of Wigner's friend Gedankenexperiment, where the quantum system $d$ is measured by $F$ in the laboratory $L$ and $W$ measures the entirety of $L$.

The state of $d$ is prepared to be in the superposition $\alpha \begin{pmatrix} 1 \\ 0 \end{pmatrix} + \beta \begin{pmatrix} 0 \\ 1 \end{pmatrix}$, $\alpha^2 + \beta^2 = 1$; of the elements of the eigenbasis $\{\begin{pmatrix} 1 \\ 0 \end{pmatrix}, \begin{pmatrix} 0 \\ 1 \end{pmatrix}\}$ of $\sigma \in \mathfrak{B}(\mathbb{C}^2)$, upon measuring the system $d$ the experimenter $F$ finds the system in either the state $\begin{pmatrix} 1 \\ 0 \end{pmatrix}$ or the state $\begin{pmatrix} 0 \\ 1 \end{pmatrix}$.

The observer $W$ on the other hand considers the entire laboratory $L$ as a quantum system and therefore to be able to describe the state of $F$ with relation to its remembered result of the measurement of system $d$, models the memory of $F$ itself as a quantum system with two possible results by a measurement of a operator $R \in \mathfrak{B}(\mathscr{H}_F)$ that is considered in a basis $\{\uparrow, \downarrow\}$ of its eigenstates, this can be more precisely be realized if the experimenter is substituted by a mechanical system that realizes the measurement of $d$ and saves the result in a memory that is a quantum q-bit with relation to the operator $R$, and in a idealization all other components of the laboratory can be ignored in the description of the Gedankenexperiment.

Before the measurement $W$ then has to consider the laboratory to be in the state given by $\alpha \left( \begin{pmatrix} 1 \\ 0 \end{pmatrix} \otimes \left( \frac{1}{\sqrt{2}}(\uparrow + \downarrow) \right) \right) + \beta \left( \begin{pmatrix} 0 \\ 1 \end{pmatrix} \otimes \left( \frac{1}{\sqrt{2}}(\uparrow + \downarrow) \right) \right)$, this is so because by the imposition that the state of $F$ is possible to be observed with the operator $R$, then even before the measurement this state must exist and be amenable to be measured with $R$ with outcomes $\uparrow$ or $\downarrow$, even though this is not *a priori* clear, $W$ knows that $d$ is initially in the state $\alpha \begin{pmatrix} 1 \\ 0 \end{pmatrix} + \beta \begin{pmatrix} 0 \\ 1 \end{pmatrix}$ and incorporates this into the description of the state of the laboratory, also, as there is no way to know if any is more likely than the other a completely symmetric superposition of both eigenstates must be postulated. After the measurement realized by experimenter $F$ the state of the laboratory will be $\alpha \begin{pmatrix} 1 \\ 0 \end{pmatrix} \otimes \uparrow + \beta \begin{pmatrix} 0 \\ 1 \end{pmatrix} \otimes \downarrow$.

This makes the processes of measurement, and therefore of the so-called collapse of the wave function, be asymmetric between the agent that actually performs the measurement and a outside observer that tries to factor in the information that a measurement has been realized within the larger system that he is observing, originally this aspect has been associated with a difficulty in precisely delimiting the system from the observer, *i.e.* the joining of a observer with a quantum system can be seen as another quantum system by another outside observer, and so on. Nevertheless it is somewhat clear that this asymmetry happens because of a underlying asymmetry in the flow of information, as experimenter $F$ knows the *result* of his measurement of system $d$, whereupon observer $W$ is only aware that a measurement *has been realized*, without knowledge of its result, so the states are updated with different specificities depending on the amount of information available.

Several recently published papers on AQFT [35, 41, 38, 36, 103] and a book written by N.P. Landsman [71] bring the latest developments in the intersection between AQFT and QMT, *i.e.*, in the description of the measurement process through the formalism of algebras of local observables in AQFT, which reveals a new period of interest in this area , which despite having resulted in some articles during the birth period of the Haag-Kastler axiomatic approach, that is, AQFT, in the 60s and 70s [6, 104, 56, 57], after that first moment passed for a period of waning interest.

This dissertation seeks to examine the Coleman-Hepp model of wave packet reduction and to explain the connections and relationships between it and several other results in QMT that emerged later (and before!), and that, to the knowledge of the author, have not yet been used in an analysis of the Coleman-Hepp model.

## 1.3   Dissertation Structure

Before even beginning to write this dissertation, I came to the conclusion that, given the reduced niche of people that are actually specialists in AQFT, or even by considering the number of physicists that actively use elements of $C^*$-algebra theory, von Neumann algebra theory or category theory; I should make the most self-contained text that I could with the allotted time that was available to me for writing this dissertation. Although I have used the expression *the most self-contained*, it should be pointed out that I have chosen to enunciate and prove mainly those results that are in direct logical dependency with the principal line of discussion of the dissertation, meaning that there are several propositions, lemmas and theorems that were left out and that substantiate each of the subjects of the dissertation way beyond what was actually included, these results although not directly related to the exposition, could be necessary for a better holistic comprehension of the topics to a less accustomed reader.

The dissertation is organized so as to try to bridge several disperse results on Quantum Measurement Theory that can be either directly or indirectly generalized for QFT, and when possible to QFT on curved spacetimes though the AQFT approach to the subject. Of main interest are the consequences of decoherence based measurements and of the contextuality property of observables in quantum systems for the argued conceptual inconsistencies of quantum descriptions, be it in their more old and already sorted out instances, such as the Wigner's friend Gedankenexperiment, or in their more modern instantiations, namely the Frauchinger-Renner Gedankenexperiment, that although wrapped in much controversy, has not been given a completely definite closure, at least from the vast majority of the papers that the writer of this dissertation has, humbly, come across so far; with the notable exception of [80], that, by the conclusion of the writer of this dissertation, misses the mark of a more satisfactory closure by considering a approach to contextuality that is unecessarily general for physical situations, leaving behind some open considerations that would be much more diminished in a completely physical scenario.

The organization of the dissertation progresses as follows:

- In chapter 2 we take the, rather unusual path, of re-deriving classical logic from classical mechanics and in somewhat the same spirit we derive the Birkhoff-von Neumann quantum logic from quantum mechanics, this was necessary for completeness, as each of these logical systems determines the way assertions about physical systems are to be meaningfully handled.

  We also introduce a logical description of rational agents that can ascertain the validity of physical assertions and deduce conclusions from them, by the inclusion of Kripke semantics within logical systems. This will be useful when talking about Gedankenexperiment in which there are several experimenters making measurements and arriving at conclusions from those measurements, such as in the Wigner's friend Gedankenexperiment or the Frauchinger-Renner Gedankenexperiment.

- In chapter 3 we consider how the usual notion of a sharp measurement alludes to the existence of a unitary operator capable of realizing a Lüders-von Neumann type of evolution on a composed system. Also, how there are constraints upon sharp measurements that go beyond simple incompatibility of observables, in particular we analyze the Wigner-Araki-Yanase theorem and how it gives limitations on the accuracy of the measurements of a quantum system when there is a underlying conservation law, in that case for quantized discrete observables, with a approach that is reminiscent of spin systems. We also consider how in this situation, approximate measurements can be realized with arbitrary precision,so long as the apparatus is prepared carefully enough, and how macroscopic measurement apparatuses can play a role in greatly reducing measurement errors in unsharp measurements.

- In chapter 4 we describe AQFT from the Haag-Kastler axioms adapted for describing QFT in possibly curved spacetimes and how the measurement scheme proposed by Fewster and Verch allows several notions of QMT, that are usually defined for non-relativistic QM, to be generalized for QFT and even possibly QFT in curved spacetimes, by using the AQFT approach.

- In chapter 5 we further delve into AQFT as a way of obtaining results of how decoherence can be used to close the gap between the unitary evolution of isolated quantum systems, and the non-unitary evolution that is present when measurements are performed, this involves a deep excursion into the properties of sequences of states in AQFT that are weakly converging, results on quasilocal algebras and their representations, until we finally arrive at the main objective of the chapter, namely the Coleman-Hepp model, of which some effort was made to try to clarify its exposition.

Still on the Coleman-Hepp model we bring a argument made originally by W. Wreszinski and H. Narnhofer, that to the understanding of the author solves a criticism raised by J. Bell about that specific model, other than the fact that there are some approximations on the derivation of the solution for that scenario.

- In chapter 6 we consider the property of contextuality in quantum systems, that was originally considered as a way of showing the impossibility of hidden variable theories in quantum mechanics, as these hidden variables would have values that would be dependent on which type of measurement was to be applied to them. We also consider how this notion evolved from this to the deeper understanding that in QM it is not only that the values are undefined before measurements are realized, and that they become defined after a measurement, only just to be latter on perturbed by other interactions that can then make the value undefined again, but that they are undefinable without making reference to which context of observables the value measurement is realized in, this is the Kochen-Specker theorem.

  We also consider how contextuality can appear with a state dependency, with Hardy's paradox Gedankenexperiment, how this type of contextuality can have strange consequences on measurements, such as the possibility of, in principle, realizing interaction free measurements. We then go on to use what was covered on the previous sections of the chapter to elucidate the no-go result of Frauchinger-Renner, understood to be the conclusion of the Frauchinger-Renner Gedankenexperiment, here armed with the formal language of Kripke semantics, that was introduced in chapter 2, to simplify the exposition. We also propose a simple solution to the supposed inconsistency that this result should bring.

- In chapter 7 we abstract the notions of contextuality, so as to be able to arrive at the conclusion that although contexts are generally seem to convey a local information about a set of observables, where this locality is thought initially in the sense of concordant compatible information, these contexts actually reflect global limitations.

  This is made clear with a analysis of the presheaves present in quantum systems, namely that the presence of contextuality in quantum systems is a consequence of the nonexistence of global sections of the observable presheaf, this is done in a very general description using category theory, and as such the chapter begins by presenting the necessary basics of category theory that will be used thereupon.

- In chapter 8 lies the results of this project, in which we give the general conclusions that can be drawn from each of the considerations and resulting consequences of these considerations addressed in the previous chapters. In this chapter we try to argue that AQFT presented in light of the decoherence and contextuality results presets a much more flexible foundation for QFT, be it in usual Minkowsky or in curved spacetimes, since these results can very naturally be incorporated into the AQFT description, without the need to make any fundamental changes in the description of the theory.

Lastly, the author has tried to make the expositions of the topics as elementary as he could, writing explicitly, whenever possible, the definitions of the structures that are being referred in a proof, the reasons why intermediary facts, that would otherwise be considered trivial for specialists, hold, and mentioning the logical dependencies, be it with previous results, other results that are proven in the appendices or at least to where that result can be found proven in the literature.

In doing so, there is always the (almost certain) possibility that very elegant proofs have been rendered rather unappealing for the trained eye, as several logical redundancies are made for the sake of clarity of exposition for unaccustomed readers (statements like the notorious "a monad in $X$ is just a monoid in the category of endofunctors of $X$, with product $\times$ replaced by composition of endofunctors and unit set by the identity endofunctor."[6] are very elegant both synthetically and semantically, as they make very efficient use of the definitions and correlated meanings, but do very little in explaining any deeper meaning that isn't already present in the reader's mind).

No original theorem or proposition was proven in this text, the only claims of originality rest in the considerations about the consequences of these previously known results, that being said, any typos, mistakes or more serious errors present in this dissertation are of complete responsibility of its author.

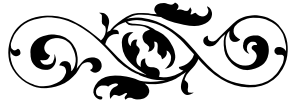

---

[6] A famous excerpt from [72], that has given many headaches for those studying category theory with practical applications in mind, *i.e.* mainly computer scientists and the like.

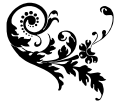 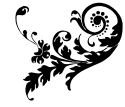

# CHAPTER 2

# The Propositional Logic of Quantum Mechanics

*This chapter begins by considering a quick exposition of classical mechanics through the techniques of symplectic geometry, which allows a more synthetic, coordinate free expression of the classical dynamics in phase space. To the non initiated reader, in this formalism vector fields are seen as maps that take functionals of the manifold to other functionals of the manifold, and covector fields, also called 1-forms, are maps that take vector fields to other vector fields, good references for differential geometry are [65, 66] and for classical mechanics in this formalism [7, 2]. We then infer classical logic from the classical dynamics in phase space, from this we set out to define a quantum logic from QM.*

We as humans experience our everyday world of middle experiences, that is, we don't perceive spatial extensions that are either cosmological or subatomic, or velocities that are either quasi-static or relativistic, or temperatures that are close to absolute zero or close to the nucleus of a blue giant star. Therefore we had a evolutionary pressure to develop a deductive reasoning that is compatible with classical mechanics. Classical mechanics can be seen as a geometric theory, a physical classical system with $n$ spatial degrees of freedom, has also $n$ kinetic degrees of freedom, and therefore $2n$ degrees of freedom overall.

The ambient space where all the physical quantities are immersed, for the system with $2n$ degrees of freedom, is the *phase spacetime* $\mathscr{P}_{2n+1}$, a smooth manifold of real dimension $2n+1$ that is also a *fibre bundle* with a base manifold given by $\mathbb{R}$ called the time axis, and fibres $\mathscr{F}_t$ called the *phase spaces at time* $t$ that are sub-manifolds of $\mathscr{P}_{2n+1}$, each being a 2n-dimensional symplectic manifold, all diffeomorphic and smoothly depending on $t \in \mathbb{R}$, so that $\mathscr{P}_{2n+1}$ can be written as the disjoint union of all $\mathscr{F}_t$:

$$\mathscr{P}_{2n+1} = \bigsqcup_{t \in \mathbb{R}} \mathscr{F}_t$$

Where as the base manifold $\mathbb{R}$ represents **time**, the points in $\mathscr{F}_t$ represent the **states** of a classical system at time $t$, and the functions $C^\infty(\mathscr{F}_t)$ are the **observables** of the system. Since each of the $\mathscr{F}_t$ is a symplectic manifold each has its own symplectic form $\omega^{\mathscr{F}_t}\big|_{s \in \mathscr{F}_t} : T_s\mathscr{F}_t \times T_s\mathscr{F}_t \to \mathbb{R}$, nevertheless since each phase space is diffeomorphic to every other phase space we can consider only one symplectic form $\omega$ where the diffeomorphism maps the vectors from each $\mathscr{F}_{t'}$ to the $\mathscr{F}_t$ where $\omega$ was originally defined.

So considering that for a $C^\infty(\mathscr{F}_t)$ function $H : \mathscr{F}_t \to \mathbb{R}$ we have the differential covector field $dH \in \mathfrak{D}^1(\mathscr{F}_t)$, the space of smooth covector fields (or smooth 1-forms), defined by $dH(V) := $
$:= VH, \forall V \in \mathfrak{X}(\mathscr{F}_t)$, the space of smooth vector fields, calling this function the Hamiltonian we are then able to define Hamiltonian vector fields $\xi_H \in \mathfrak{X}(\mathscr{F}_t)$ by the relation $\omega(V, \xi_H) := dH(V)$, where the symplectic form evaluated at vector fields is understood as a real function of the manifold, this can be made more clear by considering the musical isomorphisms[1] $\flat: T\mathscr{F}_t \to T^*\mathscr{F}_t$ and $\sharp: T^*\mathscr{F}_t \to T\mathscr{F}_t$ defined through real functions of $\mathscr{F}_t$ by $(\flat \circ X)(Y) \equiv X^\flat(Y) := \omega(Y, X), \forall X, Y \in \mathfrak{X}(\mathscr{F}_t)$ and $\omega(Y, (\sharp \circ \alpha)) \equiv \omega(Y, \alpha^\sharp) := \alpha(Y), \forall \alpha \in \mathfrak{D}^1(\mathscr{F}_t)$ and $Y \in \mathfrak{X}(\mathscr{F}_t)$; so that a Hamiltonian vector field $\xi_H$ can also be defined more explicitly as $\xi_H := dH^\sharp$.

With all of that done we can define **Poisson brackets** $\{\cdot, \cdot\}: C^\infty(\mathscr{F}_t) \times C^\infty(\mathscr{F}_t) \to C^\infty(\mathscr{F}_t)$ that is induced by the symplectic structure as $\{f, g\} := -\omega(\xi_f, \xi_g)$, and combining all these definitions we get that calculating the Poisson bracket of two smooth functions of the phase space $\mathscr{F}_t$, one of them being the Hamiltonian, a function of particular importance, we get $\{f, H\} = \omega(\xi_H, \xi_f) = df(\xi_H) = \xi_H f$ so that the Poisson bracket of any function together with the Hamiltonian is equal to the Hamiltonian vector field applied to the function.

Considering then the Hamiltonian vector flow $\varphi: \mathscr{P}_{2n+1} \to \mathscr{P}_{2n+1}$ that is a map of class $C^\infty(\mathscr{P}_{2n+1}, \mathscr{P}_{2n+1})$ and is defined by:
$$\left.\frac{d}{dt}\right|_{\varphi_t(X)} = \xi_H\big|_{\varphi_t(X)} \quad, \varphi((t, X)) \equiv \varphi_t(X) \text{ for } (t, X) \in \mathscr{P}_{2n+1}$$
that is an equality of vectors, and can be understood if considering that for every $t \in \mathbb{R}$, $X \in \mathscr{F}_t$ and function $f \in C^\infty(\mathscr{P}_{2n+1})$ we have that $\left.\frac{df}{dt}\right|_{\varphi_t(X)} = \xi_H f\big|_{\varphi_t(X)}$, which means that if this $f$ is a given coordinate function $x^i: \mathscr{F}_t \to \mathbb{R}$, and since this equation applies to every coordinate $(x^1, ..., x^{2n})$, then we see that the solution to this equation is exactly the *integral curves* of $\xi_H$.

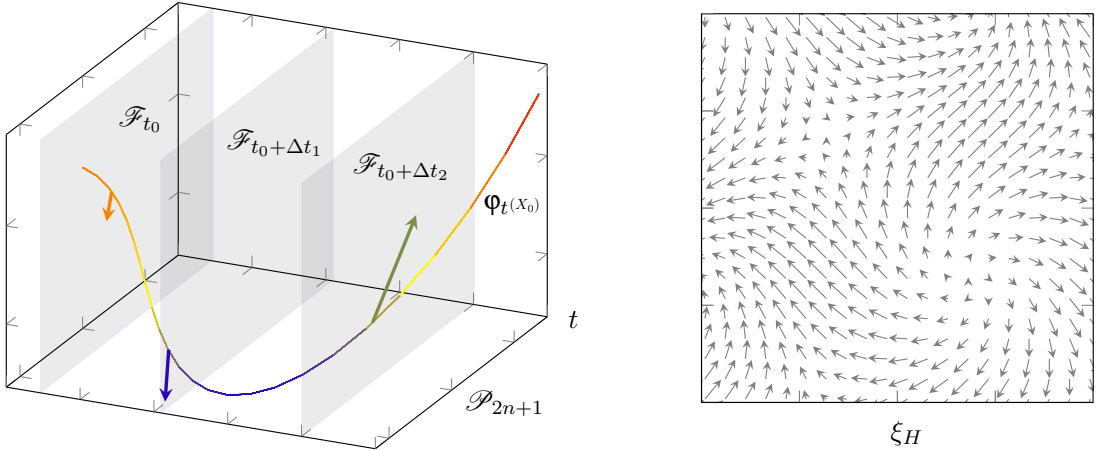

Figure 2 — Sketch of the Phase spacetime $\mathscr{P}_{2n+1}$ and Hamiltonian vector field $\xi_H$

---

[1] Isomorphisms thus guaranteed since $\mathscr{F}_t$ is a finite dimensional symplectic manifold, and considering the map for $T\mathscr{F}_t \to T^*\mathscr{F}_t$ fibre by fibre of the respective tangent and cotangent bundles given by $T_s\mathscr{F}_t \ni X \mapsto \omega(X, \cdot) \in T_s^*\mathscr{F}_t$ for each $s \in \mathscr{F}_t$, we have that this map is linear on $T_s^*\mathscr{F}_t$ since $\omega$ is bilinear in $T_s\mathscr{F}_t$, it is injective because $\omega(X, Y) = 0, \forall Y \in T_s\mathscr{F}_t \Rightarrow X = 0$ since $\omega$ is always non-degenerate for symplectic manifolds, which means that if $\omega(V, \cdot) = \omega(W, \cdot) \Rightarrow \omega(V - W, \cdot) = 0 \Rightarrow V - W = 0 \Rightarrow V = W$; and it is surjective since by a modified Gram—Schmidt process we can form a Darboux basis $\{G_k, \tilde{G}_k\}_{k=1}^n$ of a given tangent space $T_s\mathscr{F}_t$, such that $\omega(G_i, G_j) = \omega(\tilde{G}_i, \tilde{G}_j) = 0, \forall i, j$ and $\omega(G_i, \tilde{G}_j) = -\omega(\tilde{G}_j, G_i) = \delta_{ij}$ the Kronecker delta, from which an arbitrary vector $\Lambda \in T_s\mathscr{F}_t$ can be written as $\Lambda = \sum_{k=1}^n (c_k G_k + c_{k+n}\tilde{G}_k)$ and hence considering a given co-vector $\varphi \in T_s^*\mathscr{F}_t$ the calculation of this functional applied to $\Lambda$ yields $\varphi(\Lambda) = \varphi\left(\sum_{k=1}^n (c_k G_k + c_{n+k}\tilde{G}_k)\right) = \sum_{k=1}^n (c_k \varphi(G_k) + c_{n+k}\varphi(\tilde{G}_k)) = \sum_{k=1}^n (\omega(\tilde{G}_k, \Lambda)\varphi(G_k) + \omega(G_k, \Lambda)\varphi(\tilde{G}_k)) = \sum_{k=1}^n (\omega(\varphi(G_k)\tilde{G}_k, \Lambda) + \omega(\varphi(\tilde{G}_k)G_k, \Lambda) = \omega\left(\sum_{k=1}^n (\varphi(G_k)\tilde{G}_k + \varphi(\tilde{G}_k)G_k), \Lambda\right)$ so that to every $\varphi \in T_s^*\mathscr{F}_t$ we can write it as $\omega\left(\sum_{k=1}^n (\varphi(G_k)\tilde{G}_k + \varphi(\tilde{G}_k)G_k), \cdot\right)$, and so the map $X \mapsto \omega(X, \cdot)$ is surjective, establishing at last a isomorphism $T_s\mathscr{F}_t \simeq T_s^*\mathscr{F}_t, \forall s \in \mathscr{F}_t$ and therefore $T\mathscr{F}_t \simeq T^*\mathscr{F}_t$.

Requiring the Hamiltonian $H$ to be defined in the entire manifold $\mathscr{P}_{2n+1}$ and of class $C^\infty(\mathscr{P}_{2n+1})$ then the Hamiltonian vector field associated with this Hamiltonian function is also defined on the entire manifold $\mathscr{P}_{2n+1}$, by the Picard-Lindelöf theorem, since we supposed the field to be smooth then it is Lipschitz continuous, which means that the solutions exist and are unique within $\mathcal{B}_\varepsilon(\varphi_{t_0}(X_0))$ for a initial condition of $(t_0, X_0) \in \mathscr{P}_{2n+1}$ and a small real value $\varepsilon > 0$. Arnol'd in [7] mentions that $H$ having all its level sets compact is sufficient for the vector field $\xi_H$ to be complete, that is for all its integral curves to be unlimitedly extendable, this rests in the following theorem which we enunciate and whose proof can be found in [2]:

**Theorem 2.0.1.** *Let $V$ be a vector field of class $C^k(T\mathscr{P}_{2n+1})$, $1 \le k \le \infty$. Let $c(t)$ be be a maximal integral curve of $V$ such that for every open finite interval $(a, b)$ in the domain of $c$, $c((a, b))$ lies in a compact subset of $\mathscr{P}_{2n+1}$. Then $c$ is defined for all $t \in \mathbb{R}$.*

This theorem has as a corollary that every vector field with compact support on a manifold $M$ is complete. This is relevant since we can use smooth functions of compact support in $\mathscr{P}_{2n+1}$ to define a vector field with compact support between any two regions of $\mathscr{P}_{2n+1}$.

Considering then that $\mathscr{P}_{2n+1}$ admits an atlas with local coordinates $\varphi_t(X) = (t, \mathbf{q}, \mathbf{p}) \equiv (t, q^1, ..., q^n, p_1, ..., p_n)$ given in relation to a Darboux basis and are called symplectic coordinates, in these coordinates the symplectic form is given by $\omega = dp_i \wedge dq^i$ where the Einstein summation convention is being used, and since $\forall\, V \in T_s\mathscr{F}_t$, $\omega(V, \xi_{p_i}) = dp_i(V) \Rightarrow^2 \xi_{p_i} = \frac{\partial}{\partial q^i}$ and since $\omega(V, \xi_{q^i}) = dq^i(V) \Rightarrow \xi_{q^i} = -\frac{\partial}{\partial p_i}$, and since by symmetry $\xi_H f = df(\xi_H) = \omega(\xi_H, \xi_f) = -\omega(\xi_f, \xi_H) = -dH(\xi_f) = -\xi_f H$. From these calculations and the vector flow we arrive at the ubiquitous Hamilton's equations for a Hamiltonian $H: \mathscr{P}_{2n+1} \to \mathscr{P}_{2n+1}$ understood as $H(t, \mathbf{q}, \mathbf{p}) \equiv H_t(\mathbf{q}, \mathbf{p})$ where each $H_t: \mathscr{F}_t \to \mathbb{R}$:

$$\begin{cases} \dot{q}^i = \phantom{-}\partial_{p^i} H &= \{q^i, H\} \quad , i \in \{1, ..., n\} \\ \dot{p}_i = -\partial_{q_i} H &= \{p_i, H\} \quad , i \in \{1, ..., n\} \end{cases} \tag{2.1}$$

The system's evolution in time is described by the integral curves of Hamilton's differential equations. If $s(t) = (\mathbf{q}(t), \mathbf{p}(t)) \in \mathscr{F}_t$ is the state of the system at time $t$, each integral curve determines, at any given time $t \in \mathbb{R}$, a point $(t, s(t)) \in \mathscr{P}_{2n+1}$ where the curve meets $\mathscr{F}_t$. For any given coordinate frame the value of the Hamiltonian $H$ is under usual circumstances the total mechanical *energy* of the system, its numerical value may depend on the coordinate frame, but as we have constructed, Hamilton's equations of motion are independent of any frame. For ease of exposition, we considered only functions of class $C^\infty(\mathscr{P}_{2n+1})$, but in retrospect we see that for Hamilton's equations to be well defined it was enough to consider functions of class $C^1(\mathscr{P}_{2n+1})$.

From the onset we have considered that the phase spacetime $\mathscr{P}_{2n+1}$ has a inherent symplectic structure, and from this structure we derived a Poisson bracket induced by this symplectic structure, but one could also, more generally, consider manifolds that have this Poisson structure, but that may not be symplectic. The physical systems that are so modeled can be said to possess **classical superselection rules** [70], this means that not every $C^1(\mathscr{P}_{2n+1})$ function is an observable of the system, since not every point of the manifold has a Hamiltonian flow leading to it from any other point of the manifold, making it impossible for a system in one such superselection sector, that is a

---

[2] Since $\omega(V, \xi_{p_i}) = dp_j(V)dq^j(\xi_{p_i}) - dp_j(\xi_{p_i})dq^j(V) = dp_i(V)$ from which $dq^j(\xi_{p_i}) = \delta^j_i$ and $dp_j(\xi_{p_i}) = 0$ which is solved by $\xi_{p_i} = \frac{\partial}{\partial q_i}$ and analogously $\omega(V, \xi_{q^i}) = dp_j(V)dq^j(\xi_{q^i}) - dp_j(\xi_{q^i})dq^j(V) = dq^i(V)$ when $dp_j(\xi_{q^i}) = -\delta^i_j$ and $dq^j(\xi_{q^i}) = 0$ arriving at $\xi_{q^i} = -\frac{\partial}{\partial p_i}$.

*symplectic leaf* of the foliation of the manifold by Hamiltonian flows, to be able to possibly interact with, and therefore measure, any point of a distinct superselection sector.

In general it is never the case, even in classical mechanics, for one to have perfect information of the positions in configuration space and momenta in conjugate space, this would suggest that perfect measurements could be realized in our mathematical model of mechanics of such nature as to obtain real numbers with infinite precision for the coordinate values of position and momentum, this clearly represents a informational impossibility when considering the storage and transmission of such a infinite amount of data. Although it must be pointed out the obvious fact that, differently from quantum mechanics, classical mechanics has no intrinsically ontological uncertainty about any set of it's measurement values, which means that, in principle, one could measure any set of values to any arbitrary finite precision.

Nevertheless, there exists a essential need of modeling physical systems with a certain degree of ignorance of its exact state, which is the ubiquitous scenario in systems studied by Statistical Mechanics and Thermodynamics where the usual systems are composed of several moles of microsystems each with their own microstate, nonwithstandingly one can also consider a system whose influence of its internal structure is negligible, but that also has uncertainties in the measured value of the components of it's state and consider the time evolution of such a system as a way of considering the propagation of uncertainty between a initial and final time to consider all of the possible final conditions of the system.

This modeling of uncertainties is made through **statistical ensembles**, a set of independent copies of the system, whose states are distributed in the various $\mathscr{F}_t$ with a certain probability given locally by a $C^1(\mathscr{P}_{2n+1})$ probability density function $\rho : \mathscr{P}_{2n+1} \to \mathbb{R}$. This function when considered in a Hamiltonian flow $\varphi$ must evolve and then considering $\frac{d\rho}{dt} = \frac{\partial \rho}{\partial q^i}\dot{q}^i + \frac{\partial \rho}{\partial p_i}\dot{p}_i + \frac{\partial \rho}{\partial t} = \frac{\partial \rho}{\partial q^i}\frac{\partial H}{\partial p_i} - \frac{\partial \rho}{\partial p_i}\frac{\partial H}{\partial q^i} + \frac{\partial \rho}{\partial t} = \{\rho, H\} + \frac{\partial \rho}{\partial t}$ where again Einstein summation notation was used. Considering a volume element $d\mu_t = dq_1 \wedge ... \wedge dq_n \wedge dp^1 \wedge ... \wedge dp^n$ we have that since $\rho$ is a density $\rho \geq 0$ and $\int_{\mathscr{F}_t} \rho d\mu_t = 1$, more specifically if $\mathcal{P} = \mathrm{supp}\left(\rho\Big|_{\mathscr{F}_{t_0}}\right)$ for a given time $t_0$, then considering the pullback $\varphi_t^* f(s) := f(\varphi_t(s))$ and the Lie derivatives of functions $(\mathcal{L}_{\xi_H} f)(p) = \frac{d}{dt}\Big|_{t=0}(\varphi_t^* f)(p)$ and of differential forms $(\mathcal{L}_{\xi_H} d\alpha)\big|_s = \frac{d}{dt}\big|_{t=0}(\varphi_t^* d\alpha)\big|_s$:

$$0 = \frac{d}{dt}\int_{\varphi_t(\mathcal{P})} \rho d\mu_t = \frac{d}{dt}\int_{\mathcal{P}} \varphi_t^*(\rho d\mu_t) = \int_{\mathcal{P}} \frac{d}{dt}(\varphi_t^*(\rho)\varphi_t^*(d\mu_t)) = \int_{\mathcal{P}}\left(\mathcal{L}_{\xi_H}\rho + \frac{\partial \rho}{\partial t}\right)d\mu_t + \rho\mathcal{L}_{\xi_H}(d\mu_t)$$

Where in the last equality the lie derivative of the volume element through a Hamiltonian field is zero because of Liouville's volume theorem C.0.28[3], which implies that volume elements transformed by Hamiltonian flows are preserved. Therefore $\int_{\mathcal{P}}\left(\mathcal{L}_{\xi_H}\rho + \frac{\partial \rho}{\partial t}\right)d\mu_t = 0$ for any probability measure $\rho$ which implies that the integrated part is zero, $\mathcal{L}_{\xi_H}\rho + \frac{\partial \rho}{\partial t} = 0$ or in other notation:

$$\frac{\partial \rho}{\partial t} + \{\rho, H\} = 0$$

Which is Liouville's equation for the time evolution of a probability distribution in phase space, in symplectic coordinates we get the usual equation:

$$\frac{\partial \rho}{\partial t} + \sum_{i=1}^{n}\left(\frac{\partial \rho}{\partial q^i}\frac{\partial H}{\partial p_i} - \frac{\partial H}{\partial q^i}\frac{\partial \rho}{\partial p_i}\right) = 0$$

---

[3] Proven in Appendix C.

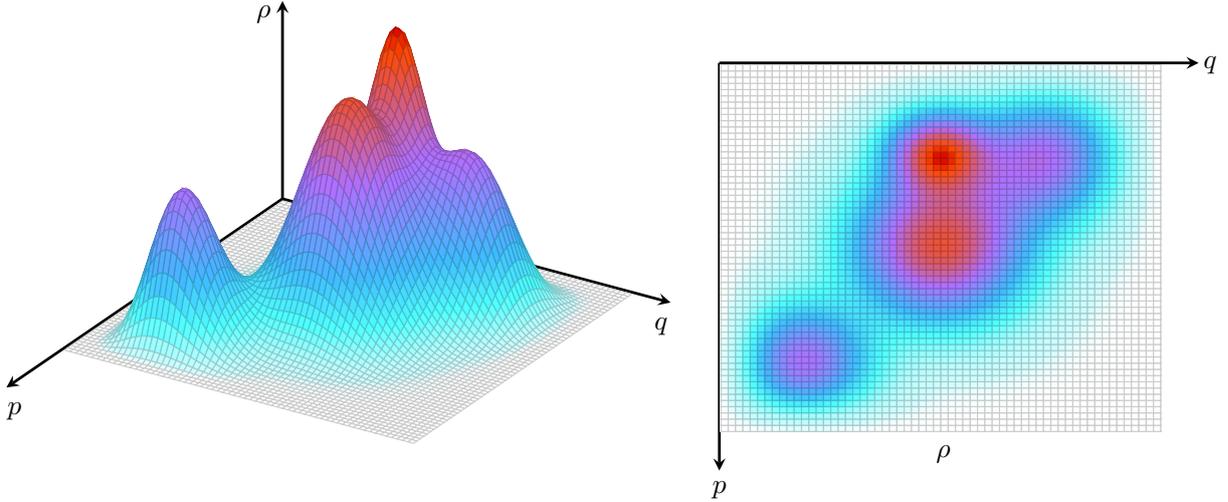

Figure 3 — Schematic representation of a probability density function $\rho$.

Since the best characterization of the system in those cases are sill given by $\rho(t,\mathbf{p},\mathbf{q}) \equiv \rho_t(\mathbf{p},\mathbf{q})$, then $\rho_t$ must also be seen as a state of the system, from these considerations we define:

**Definition 2.0.2** (Classical states). *A classical system state at time $t$ can be understood as a Borel probability measure $\nu_t : \mathscr{B}(\mathscr{F}_t) \to \mathbb{R}_+$, defined in $\mathscr{B}(\mathscr{F}_t)$, the Borel $\sigma$-algebra of $\mathscr{F}_t$; two cases are of particular interest:*

*(i) A **probabilistic state** defined as $\nu_t(E) := \int_E \rho_t(s) d\mu_t$, for $s \in \mathscr{F}_t$ and $E \in \mathscr{B}(\mathscr{F}_t)$.*

*(ii) A **sharp state** defined as $\nu_t(E) := \delta_{s(t)}(E) = \begin{cases} 1 & , \text{if } E \ni s(t); \\ 0 & , \text{if } E \not\ni s(t). \end{cases}$ for points $s(t) \in \mathscr{F}_t$.*

The function $\delta_s$ is a measure called a Dirac measure concentrated at $s$, it must not be confused with the Dirac delta distribution $\delta : C_c^\infty(\mathscr{F}_t) \to \mathbb{C}$ which is a continuous linear functional on the space of test functions $C_c^\infty(\mathscr{F}_t)$.

## 2.1 Classical Logic

The propositions about a classical mechanical system constitute the characteristics that this system may have, since we postulated that all mechanical properties are representable on the phase space, then all propositions about mechanical characteristics of the system are associated with the points of the phase space, that is to say, the propositions of the physical system are identified with subsets of the phase space.

**Definition 2.1.1.** *A **proposition** $P$ on a classical mechanical system is a subset of the phase space, $P \subset \mathscr{F}_t$, it is a proposition about the system at time $t$. As in classical propositional logic we may ascribe the truth values **false**, numerically represented by $0$, and **true**, numerically represented by $1$, but unlike in classical propositional calculus we can also consider probabilities of a given proposition being right by a numerical value between $0$ and $1$. The assignment of the truth value of a proposition on a classical mechanical system is given by it's evaluation by the state of the system at the same time $\nu_t(P)$.*

We may also generalise our notion of the observable of a classical physical system by considering not only $C^1(\mathscr{H}_{2n+1})$ class functions that would have their exact time evolution given by Hamilton's equation, but to consider possibly discontinuous quantities $w : \mathscr{H}_{2n+1} \to \mathbb{R}$ for the physical system, which would imply that their time evolution would also be discontinuous and could be given by the evaluation of $w$ on the Hamiltonian vector flow $\varphi_t$ of the initial conditions of the system. Nevertheless, if $w$ were to still need to represent a measurement value, then it would need to preserve the measurable structures of each space[4] and therefore $w$ must be a Borel measurable function.

Von Neumann and Birkhoff in the first paper dedicated to the logic of quantum mechanics [11], only consider in classical systems the probabilistic states, since for them points in the phase space and other non-Lebesgue measurable subsets of $\mathscr{F}_t$ don't make experimental sense, citing the article: "how absurd it would be to call an 'experimental proposition,' the assertion that the angular momentum (in radians per second) of the earth around the sun was at a particular instant a rational number!", nevertheless there are known problems in classical mechanics that depend on rationally commensurable values for solutions, such as the problem of periodic trajectories on billiards in a rectangle table.

As in classical propositional logic we may define all logical connectives through either the use of a non-conjunction $\overline{\wedge}$, also called a NAND logical operator, or by a non-disjunction $\overline{\vee}$, also called a NOR logical operator. Following [100] we opt for a non-conjunction, defined by:

$$P \overline{\wedge} Q := \mathscr{F}_t \setminus (P \cup Q).$$

that is, quite simply the negation of the conjunction of propositions $P$ and $Q$. From this functionally complete Boolean operator we get all the other logical operators by the standard definitions $\neg P := P \overline{\wedge} P$, $P \wedge Q := (P \overline{\wedge} Q) \overline{\wedge} (P \overline{\wedge} Q)$, $P \vee Q := (P \overline{\wedge} P) \overline{\wedge} (Q \overline{\wedge} Q)$, $P \to Q := P \overline{\wedge} (Q \overline{\wedge} Q)$, $P \leftrightarrow Q := (P \overline{\wedge} Q) \overline{\wedge} ((P \overline{\wedge} P) \overline{\wedge} (Q \overline{\wedge} Q))$; these simplify by considering the set theoretic expressions as $\neg P := \mathscr{F}_t \setminus P$, $P \wedge Q := P \cap Q$, $P \vee Q := P \cup Q$; Ignoring then the cases in which we start from a false premise, to avoid *ex falso sequitur quodlibet*, that is, the logical principle that from a false premise one can prove anything true or false, then considering only the cases in which a true premise implies a true conclusion we can simplify the expressions of implication and biconditional as relations within set theory, with $P \Rightarrow Q \equiv P \subset Q$, $P \Leftrightarrow Q \equiv (P = Q)$.

Since proposition are subsets of $\mathscr{F}_t$ for each time $t$, then we can construct in this logic expressions involving infinitely many propositions, this can be clear by considering the following proposition, for a physical quantity given by a function $f : \mathscr{F}_t \to \mathbb{R}$, the proposition $P_E^{(f)} \equiv$ "*the value that $f$ assumes on the system state belongs to the Borel set $E \subset \mathbb{R}$*", this proposition can be true or false and since propositions are sets then this one must be $P_E^{(f)} = f^{-1}(E)$, $E \in \mathscr{B}(\mathbb{R})$, considering a decomposition of $E$ into infinitely many sets, which can be more clearly seen if $E$ is

---

[4]   In this passage the different notions of physical measurement and the mathematical notion of measurability are being stressed, the requirement that measurements must preserve the measurable structures is clear if one were to consider that states are localized on Borel sets, for the states to be given by a associated measure, and sets of measurement values themselves must also be Borel sets, since a set of measurements must have it's size be quantifiable.

The author witnessed a occasion in which the measure problem in measure theory and the measurement problem in quantum mechanics were misunderstood as synonyms in a presentation given to less mathematically inclined physicists, to clarify this difference the problem of measurability is enunciated and proven in Appendix C, prop. C.0.3, suffice to say that it is about the impossibility of the existence of a non-zero translational invariant measure on all subsets of $\mathbb{R}$, or of any $\mathbb{R}^n$.

a interval $[a, b)$, then a decomposition such as $[a,b) = \bigcup_{k=1}^{\infty} [a_k, a_{k+1})$ where each $a_k = b - \frac{(b-a)}{k}$, then $a_1 = a$, $a_k < a_{k+1}$ and $a_k \xrightarrow{k \to \infty} b$; then clearly $P_{[a,b)}^{(f)} = \bigcup_{k=1}^{\infty} P_{[a_k, a_{k+1})}^{(f)}$ or in other words the proposition:

$$P_{[a,b)}^{(f)} = \bigvee_{k=1}^{\infty} P_{[a_k, a_{k+1})}^{(f)}.$$

One could also consider approximating this interval by intersections such as $[a,b) = \bigcap_{k=1}^{\infty} [a, b_k)$ in which $b_k = b + \frac{1}{k}$ such that $b_k > b_{k+1}$ and $b_k \xrightarrow{k \to \infty} b$; so that $P_{[a,b)}^{(f)} = \bigcap_{k=1}^{\infty} P_{[a, b_k)}^{(f)}$ or idem:

$$P_{[a,b)}^{(f)} = \bigwedge_{k=1}^{\infty} P_{[a, b_k)}^{(f)}$$

.

The usual quantifiers $\forall, \exists$ are also available, imported from set theory, since this logic is constructed within set theory, which takes this logic from a propositional logic to a first order logic, not only that, but again by the properties of set theory into a second order logic, where in this logic the usual classical semantics of Tarski are employed [94], so that we have the usual understanding that the negation of truth is false, and all other ways in which a logical expression can be evaluated as truth follows the usual relations present in truth tables for that expression. This way we recover classical logic from classical mechanics.

This construction in which we consider the elementary sentences of a, in this case physical, theory $\mathfrak{T}$ and then collect all such sentences in equivalence classes where two such sentences $A$ and $B$ are equivalent if it "is possible to prove $B$ from $A$ in theory $\mathfrak{T}$", this is the construction of the Lindenbaum—Tarski algebra of the given theory,[5] it is an algebra considering disjunction as the sum of the "sum" of the algebra, conjuction as the "multiplication". This process of constructing the Lindenbaum—Tarski algebra of a theory is the aspect of the construction of classical logic from classical mechanics that will be invariant in the subsequent construction of a quantum logic from quantum mechanics.

## 2.2   Quantum Logic

Since in order to consider a logic in quantum theory, we must abandon the phase space formalism, since quantum mechanics expressed in phase space be it through Wigner functions, or the Glauber—Sudarshan P representation or even in Husimi Q representation; these representations have a very idiosyncratic and convoluted way to represent states, to begin with they are all quasiprobability distributions, the Wigner function and the Glauber—Sudarshan P representation may ascribe negative values for certain points of phase space, and the Husimi Q representation although always positive can make different points of its "phase space" be correlated by the same point in a Wigner function phase space, since the Husimi Q representation can be defined as a gaussian convolution of the Wigner function for the same system [25]. Also, these representations all have the problem that they aren't isomorphic to the usual representations of QM, to begin with there are way more distributions in these phase spaces than there are states. Nevertheless, although unknown to the writer a derivation of the logic of quantum mechanics from these phase space formalisms would be quite interesting, even though incredibly convoluted.

---

[5]   For more on this topic of semantic approaches to quantum logic, we recomend the chapter 5 of the book [90].

Another novel approach could have been to begin with an approach to classical mechanics through the Koopman—von Neumann formalism, in which classical mechanics is described in a Hilbert space with wave functions, but that would have been very artificial with no intuitive motivation as to why one would devise such treatment.

We now ask our selves which type of generalizations of the underlying logic imbued in the phase space description of classical mechanics, as it turns out the algebra of classical logic is a Boolean algebra, and boolean algebras can be generalized by *lattices*, which just happen to be what we need in constructing a quantum logic.

To define a Lattice we first have to define a partially ordered set:

**Definition 2.2.1** (Poset). *A partially ordered set $(X, \preccurlyeq)$ is composed of a set $X$ and a partial relation $\preccurlyeq$, such that:*

(i) $x \preccurlyeq x$ , $\forall x \in X$.

(ii) *being $a, b, c \in X$, if $a \preccurlyeq b$ and $b \preccurlyeq c$ then $a \preccurlyeq c$.*

(iii) *being $a, b \in X$, if $a \preccurlyeq b$ and $b \preccurlyeq a$ then $a = b$.*

As we can see a partial order is partial since it doesn't require that either $x \preccurlyeq y$ or $y \preccurlyeq x$ have to be true.

**Definition 2.2.2** (Lattice). *A partially ordered set $(X, \preccurlyeq)$ is a **lattice** when for any $a, b \in X$:*

(i) $\sup\{a, b\}$ *exists, and we denote* $\sup\{a, b\} \equiv a \vee b$*, calling it the join of $a$ and $b$.*

(ii) $\inf\{a, b\}$ *exists, and we denote* $\inf\{a, b\} \equiv a \wedge b$*, calling it the meet of $a$ and $b$.*

*As a consequence of this definition we have the following properties $\forall a, b, c \in X$:*

- **Associativity:** $(a \wedge b) \wedge c = a \wedge (b \wedge c)$ and $(a \vee b) \vee c = a \vee (b \vee c)$.
- **Commutativity:** $a \wedge b = b \wedge a$ and $a \vee b = b \vee a$.
- **Absorption:** $a \wedge (a \vee b) = a$ and $a \vee (a \wedge b) = a$.
- **Idempotency:** $a \wedge a = a$ and $a \vee a = a$.

*We could have only asked these properties for the relations and we would have a partially ordered set through $a \preccurlyeq b \Leftrightarrow a \wedge b = a$.*

We may further characterize Lattices by the following names:

**Definition 2.2.3** (Types of Lattices). *A lattice $(X, \preccurlyeq)$ is called:*

(1) **distributive** *if $\forall a, b, c \in X$:*

$$a \wedge (b \vee c) = (a \wedge b) \vee (a \wedge c), \quad a \vee (b \wedge c) = (a \vee b) \wedge (a \vee c).$$

(2) **bounded** *if it admits a minimum* **0** *and a maximum* **1**.

(3) **orthocomplemented** *if it is bounded and admits a mapping* $X \ni a \mapsto \neg a \in X$, *where* $\neg a$ *is called the* orthogonal complement *of $a$, and is such that,* $\forall a, b \in X$:

     *a)* $a \wedge \neg a = \mathbf{0}$;

     *b)* $a \vee \neg a = \mathbf{1}$;

     *c)* $\neg(\neg a) = a$;

     *d)* $a \preccurlyeq b$ *implies* $\neg b \preccurlyeq \neg a$;

*two elements* $a, b \in X$ *are called:*

- **orthogonal** *and denoted* $a \perp b$, *if* $a \preccurlyeq \neg b$ *or equivalently* $b \preccurlyeq \neg a$;

- **commuting** *if, for* $c_1, c_2, c_3 \in X$, $a = c_1 \vee c_3$ *and* $b = c_2 \vee c_3$ *with* $c_i \perp c_j$ *if* $i \neq j$;

*the* **centre** *of the lattice is the set of elements commuting with every element of the lattice, the lattice is* **irreducible** *if its centre is* $\{\mathbf{0}, \mathbf{1}\}$;

(4) **modular** *if* $a \preccurlyeq b$ *implies that,* $\forall r \in X$, $(a \vee r) \wedge b = a \vee (r \wedge b)$.

(5) **orthomodular**, *if orthocomplemented and* $a \preccurlyeq b$ *implies* $b = a \vee ((\neg a) \wedge b)$.

(6) **$\sigma$-complete** *if every countable set* $\{a_n\}_{n \in \mathbb{N}} \subset X$ *admits a least upper bound, indicated by* $\bigvee_{n \in \mathbb{N}} a_n$ *and a greatest lower bound, indicated by* $\bigwedge_{n \in \mathbb{N}} a_n$.

(7) **complete** *if every set* $K \subset X$ *admits both least upper bound and greatest lower bound.*

(8) **Boolean algebra** *if has properties (1), (2) and (3) (which imply (4) and (5)).*

(9) **Boolean $\sigma$-algebra** *if is a Boolean algebra satisfying at least (6), since (7) implies (6).*

It is clear from these definitions that the logic that emerges from classical mechanics in phase space is a lattice, were the poset is the phase space itself and the order relation comes from the subset relation, the meet is the intersection and the join is the union of sets, therefore it is distributive, since unions and intersections are, it is bounded, since there is a minimum given by the empty set and a maximum given by the whole phase space, and it is orthocomplemented with relation to the set complement over the whole phase space acting as a orthogonal complement, and it is a $\sigma$-algebra since the propositions $P_E^{(f)}$ map the Borel $\sigma$-algebra of $\mathbb{R}$ into $\mathscr{F}_t$. Therefore the lattice of classical mechanical logic is a Boolean $\sigma$-algebra.

As in the classical mechanics case, we interpret propositions on a physical system to mean the value of a possible measurement of the properties of the system, so that the truth values of these propositions are **1** or **true** if the value of the proposition is the same as the measured value and **0** or **false** if it isn't. We remember that in quantum mechanics this association of propositions to experiments is not so simple as in classical mechanics, since in quantum mechanics we have incompatible observables, that is, measurements that are mutually exclusive in the sense that a state can never have well defined values for both observables.

Von Neumann's and Birkhoff's [11] solution to the problem of describing propositions in a quantum system is to identify the propositions with the orthogonal projectors of a complex Hilbert space, this will be the way to construct the Lindenbaum–Tarski algebra of quantum theory by considering equivalent two propositions of the type "the probability that the value of the observable $Q$ lies in $\mathcal{V}$ is equal to $p$", when the probability is exactly 1 we can think of the closed linear space formed from all the vectors of the Hilbert space belonging to the range of the spectral projection of $Q$ for the set $\mathcal{V}$, instead of the individual equivalent propositions, this suggests the following definition for our logical propositions:

**Definition 2.2.4.** *A proposition $P$ on a quantum system $S$ described in a non-zero, **separable** and complex Hilbert space $\mathcal{H}_S$, corresponds to a orthogonal projector in this Hilbert space, denoted by the same letter, in such a way that compatible propositions correspond to commuting orthogonal projectors, and if the proposition $P$ logically implies a compatible proposition $Q$, $P \Rightarrow Q$, then the associated projectors are such that $P(\mathcal{H}_S) \subset Q(\mathcal{H}_S)$, i.e. $P \leq Q$.*

*Therefore the set of propositions on $S$ is in a bijection with a subset of the set of projectors of $S$ denoted by $\mathcal{P}(\mathcal{H}_S)$, which is also called the **logic of elementary propositions** of the system.*

From this we can then define the logical connectives:

**Definition 2.2.5.** *If $\{P_\alpha\}_{\alpha \in I}$ is a family, of arbitrary cardinality, of pairwise commuting orthogonal projectors, then we define:*

- *A conjunction of these projectors by:*

$$\bigwedge_{\alpha \in I} P_\alpha \text{ is the projector onto } \bigcap_{\alpha \in I} P_\alpha(\mathcal{H});$$

- *A disjunction of these projectors by:*

$$\bigvee_{\alpha \in I} P_\alpha \text{ is the projector onto } \overline{\text{span}\left(\bigcup_{\alpha \in I} P_\alpha(\mathcal{H})\right)};$$

*where the closure is taken with respect to the norm induced by the inner product of the Hilbert space.*

- *The negation of a projector $P$ is $\neg P := \mathbb{1}_\mathcal{H} - P$;*

As we can see the set of propositions on a quantum system $\mathcal{P}(\mathcal{H}_S)$ is lattice that is bounded, above by the identity on $\mathcal{H}_S$, $\mathbb{1}_{\mathcal{H}_S}$, and bellow by the zero projector $\mathbf{0} : \mathcal{H}_S \to \{0\}$. Since $\forall P \in \mathcal{P}(\mathcal{H}_S)$, $P \wedge (\neg P) = P \wedge (\mathbb{1}_{\mathcal{H}_S} - P)$ which is the projector onto $P(\mathcal{H}_S) \cap (\mathcal{H}_S \setminus P(\mathcal{H}_S)) = \{0\}$ that is the zero projector $\mathbf{0}$, also $\forall P \in \mathcal{P}(\mathcal{H}_S)$, $P \vee (\neg P) = P \vee (\mathbb{1}_{\mathcal{H}_S} - P)$ which is the projector that is surjective on $\overline{\text{span}[P(\mathcal{H}_S) \cup (\mathcal{H}_S \setminus P(\mathcal{H}_S))]} = \mathcal{H}_S$ which is the identity $\mathbb{1}_{\mathcal{H}_S}$, also clearly $\forall P \in \mathcal{P}(\mathcal{H}_S)$, $\neg(\neg P) = P$. Also if for $P, Q \in \mathcal{P}(\mathcal{H}_S)$, $P \leq Q$ if and only if $P(\mathcal{H}_S) \subset Q(\mathcal{H}_S)$ which by the complement imply $Q(\mathcal{H}_S)^\complement \subset P(\mathcal{H}_S)^\complement$ and since the negations of $Q$ and $P$ are the projectors onto the respective complements then from that last inclusion $(\neg Q) \leq (\neg P)$; all of these properties make $\mathcal{P}(\mathcal{H}_S)$ a orthocomplemented lattice.

More than that, since considering that for $P \leq Q$ then considering part by part of $P \vee ((\neg P) \wedge Q)$, we see that $(\neg P) \wedge Q$ is the projector onto $(\mathcal{H}_S \setminus P(\mathcal{H}_S)) \cap Q(\mathcal{H}_S)$ so that the disjunction of this projector with $P$ gives the projector onto $\overline{\text{span}[P(\mathcal{H}_S) \cup ((\mathcal{H}_S \setminus P(\mathcal{H}_S)) \cap Q(\mathcal{H}_S))]} =$

$\overline{\operatorname{span}\left[\mathscr{H}_S \cap (P(\mathscr{H}_S) \cup Q(\mathscr{H}_S))\right]}$ and since $P \leq Q$ this is equal to $\overline{\operatorname{span}(\mathscr{H}_S \cap Q(\mathscr{H}_S))} = Q(\mathscr{H}_S)$ so that the resulting projector is the projector onto $Q(\mathscr{H}_S)$, that is $Q$, therefore $P \vee ((\neg P) \wedge Q) = Q$ for $P \leq Q$, which makes $\mathcal{P}(\mathscr{H}_S)$ a orthomodular lattice. For a infinite dimensional Hilbert space the lattice properties stop at orthomodularity, for finite dimensional Hilbert spaces we have:

**Theorem 2.2.6.** *If the Hilbert space $\mathscr{H}$ is finite dimensional then $\mathcal{P}(\mathscr{H})$ is a modular lattice.*

*Proof:* We must show that the modular property, that is, for $M, N \subset \mathcal{P}(\mathscr{H})$, with $M \leq N$ then $\forall L \in \mathcal{P}(\mathscr{H})$:

$$M \vee (L \wedge N) = (M \vee L) \wedge N$$

holds for finite dimensional Hilbert spaces, in such case we have that for a vector $\vartheta \in (M \vee (L \wedge N))(\mathscr{H})$, we can factor this vector into a $M(\mathscr{H})$ component $\vartheta_M$ and a $(L \wedge N)(\mathscr{H})$ component $\vartheta_{(L \wedge N)}$ such that $\vartheta = \vartheta_M + \vartheta_{(L \wedge N)}$. Also it is true that since $\vartheta_{(L \wedge N)} \in (L \wedge N)(\mathscr{H})$ then $\vartheta_{(L \wedge N)} \in L(\mathscr{H})$ and also $\vartheta_{(L \wedge N)} \in N(\mathscr{H})$, so that $\vartheta = \left(\vartheta_M + \vartheta_{(L \wedge N)}\right) \in (M \vee L)(\mathscr{H})$ since it is the sum of vectors in $M(\mathscr{H})$ and $L(\mathscr{H})$. Also the vector $\vartheta$ is in $N(\mathscr{H})$, as $\vartheta_M \in N(\mathscr{H})$ for $M \leq N$ and $\vartheta_{(L \wedge N)}$ is also in $N(\mathscr{H})$.

Conversely, if $\vartheta \in ((M \vee L) \wedge N)(\mathscr{H})$, then $\vartheta \in N(\mathscr{H})$ and $\vartheta \in (M \vee L)(\mathscr{H})$ this second condition holds if and only if $\vartheta$ can be written as $\vartheta = \vartheta_M + \vartheta_L$ which considering the first condition and the hypothesis that $M \leq N$ implies:

$$\vartheta_L = (\vartheta - \vartheta_M) \in N(\mathscr{H}).$$

and so $\vartheta_L \in (L \wedge N)(\mathscr{H})$, consequently $\vartheta$ is the sum of two vectors one lying on the subspace projected by $M$ and other on the subspace projected by $L \wedge N$ so that $\vartheta \in (M \vee (L \wedge N))(\mathscr{H})$. ∎

In the above proof the fact that $M(\mathscr{H}), N(\mathscr{H})$ and $L(\mathscr{H})$ are finite dimensional vector spaces played an instrumental role, for if that were not the case we couldn't use that whenever we had disjunctions of projectors we could write the elements of the subspace projected by this disjunction as the sum of elements of each member of the disjunction, this is not necessarily true for infinite dimensional vector spaces since in the disjunction we take the closure of the set, that is, $(A \wedge B)(\mathscr{H}) = \overline{\operatorname{span}(A(\mathscr{H}) \cup B(\mathscr{H}))}$. In fact, as we have already said, for a infinite dimensional Hilbert space the lattice properties stop at orthomodularity:

**Theorem 2.2.7.** *If $\mathscr{H}$ is not finite dimensional, then $\mathcal{P}(\mathscr{H})$ is not modular.*

*Proof:* Let $\{\vartheta_n\}_{n \in \mathbb{N}}$ be a orthonormal basis in $\mathscr{H}$, and let us define the elements $\xi_n$ by:

$$\xi_n = \vartheta_{2n} + a^{-n}\vartheta_1 + a^{-2n}\vartheta_{2n+1}, \quad \text{for a real } a > 1.$$

Consider the subspaces of projectors $M, N$ and $L$ defined as follows:

$$M(\mathscr{H}) \equiv \overline{\operatorname{span}\left(\bigcup_{n \in \mathbb{N}} \{\xi_n\}\right)};$$

$$N(\mathscr{H}) \equiv \overline{\operatorname{span}\left(\left(\bigcup_{n \in \mathbb{N}} \{\xi_n\}\right) \cup \{\vartheta_1\}\right)};$$

$$L(\mathscr{H}) \equiv \overline{\operatorname{span}\left(\bigcup_{n \in \mathbb{N}} \{\vartheta_{2n}\}\right)}.$$

Since $M(\mathscr{H})$ and $N(\mathscr{H})$ are composed of vectors displaced by vectors $a^{-2n}\vartheta_{2n+1}$ that are not in $L(\mathscr{H})$, and therefore $M \wedge L = 0$ and $N \wedge L = 0$, obviously. Also it is easily seen that $M \leq N$, since all the elements $\left(\sum_{n \in \mathbb{N}} \lambda_n \xi_n\right) \in M(\mathscr{H})$ are also in $N(\mathscr{H})$, in this particular case $M$ is strictly smaller $N$ since:

$$N(\mathscr{H}) \ni \xi'_n = \vartheta_{2n} + a^{-n}\vartheta_1 + a^{-2n}\vartheta_{2n+1} - a^{-n}\vartheta_1 = (\vartheta_{2n} + a^{-2n}\vartheta_{2n+1}) \notin M(\mathscr{H}).$$

for any n. Since $M < N$, it holds that:

$$(M \vee L) \leq (N \vee L)$$

The subspace $N(\mathscr{H})$ is the closure of the linear subspace of elements of the forms $\left(\sum_{n=1}^{N} \lambda_n \xi_n + \lambda \vartheta_1\right)$ and the subspace $L(\mathscr{H})$ is the closure of the linear subspace of elements $\sum_{n=1}^{N} \lambda_n \vartheta_{2n}$. Since $\left(\sum_{n=1}^{N} \lambda_n \xi_n\right)$ generates $M$, and considering that $(M \vee L)(\mathscr{H}) \ni (a^n \xi_n - a^n \vartheta_{2n}) = \vartheta_1 + a^{-n}\vartheta_{2n+1}$, and as disjunctions include the closures we can then take the limit $\lim_{n \to \infty}(a^n \xi_n - a^n \vartheta_{2n}) = \vartheta_1$, then $\vartheta_1 \in (M \vee L)(\mathscr{H})$, so that the elements $\left(\sum_{n=1}^{N} \lambda_n \xi_n + \lambda \vartheta_1\right)$ are also in $(M \vee L)(\mathscr{H})$, since they are a sum of elements from $L(\mathscr{H})$ and elements themselves from $(M \vee L)(\mathscr{H})$. Therefore this would imply that $N = (M \vee L)$ and therefore $(N \vee L) = (M \vee L)$. Assuming the modularity property holds $M \vee (L \wedge N) = (M \vee L) \wedge N$ and using that last equality of $(N \vee L) = (M \vee L)$, and those first observations that $L \wedge M = 0$ and $L \wedge N = 0$ we can see that:

$$M = M \vee 0 = M \vee (L \wedge N) = (M \vee L) \wedge N = N \wedge N = N.$$

But that is $M = N$, which contradicts the fact that $M$ is strictly smaller than $N$ and therefore $M \neq N$. ∎

Classical logic is not only modular, but it has the distributive property, which is also a property that is necessary for probability measures to be definable on these propositions, von Neumann explored the possibility of defining probability measures in modular lattices, but since the lattice of propositions for a general, possibly infinite, Hilbert space is not modular then it is impossible to define usual probability measures on the propositions of a quantum system, so as to make them serve as the event structure in the sense of probability theory.[6]

**Definition 2.2.8** (Upper and Lower sets). *A subset $\mathcal{S}$ of a lattice $(X, \preccurlyeq)$ is called **upwardly closed** or **upper** if for $x \in X$ and $s \in \mathcal{S}$, if $s \preccurlyeq x$ then $x \in \mathcal{S}$. Analogously, a **downwardly closed** or **lower** subset $\mathcal{U}$ of a lattice $(X, \preccurlyeq)$ is a set such that for $x \in X$ and $u \in \mathcal{U}$, if $x \preccurlyeq u$ then $x \in \mathcal{U}$.*

**Definition 2.2.9** (Filter). *A non-empty subset $\mathcal{F}$ of a lattice $\mathcal{L} = (X, \preccurlyeq)$ is called a **filter** if $\mathcal{F}$ is an upper set of $\mathcal{L}$ and if it is closed with respect to finite meets. A filter $\mathcal{F}$ is said to be proper if $\mathcal{F} \subsetneq X$.*

---

[6] Sometimes this is cited as the reason that von Neumann began to disfavour the Hilbert space formalism [1] for the foundations of QM in behalf of the study of what was then called rings of operators which gave rise to what is now called von Neumann algebras or $W^*$−algebras, which has properties that allow it to be called, also, as noncommutative measure theory.

A filter $\mathcal{F}$ is called a **prime filter** if for $A, B \in \mathcal{L}$ and $(A \vee B) \in \mathcal{F}$ we have that either $A \in \mathcal{F}$ or $B \in \mathcal{F}$.

**Definition 2.2.10** (Lattice Homomorphism). *Considering two lattices $\mathcal{L}_1, \mathcal{L}_2$, a map $h : \mathcal{L}_1 \to \mathcal{L}_2$ is called a **lattice homomorphism** if, $\forall A, B \in \mathcal{L}_1$:*

$$h(A \vee B) = h(A) \vee h(B);$$
$$h(A \wedge B) = h(A) \wedge h(B).$$

*A lattice homomorphism is called a* lattice imbedding *if $A \neq B$ implies $h(A) \neq h(B)$.*

**Proposition 2.2.11.** *Let $h : \mathcal{L}_1 \to \mathcal{L}_2$ be a lattice homomorphism and $\mathcal{F}$ be a filter in $\mathcal{L}_2$, then:*

(i) $h^{-1}(\mathcal{F})$ *is a filter in $\mathcal{L}_1$,*

(ii) *If $\mathcal{F}$ is a prime filter, then $h^{-1}(\mathcal{F})$ is also a prime filter.*

*Proof:* Item $(i)$ is clear since by the property of lattice homomorphism $h^{-1}(\mathscr{F})$ is closed with respect to finite meets operations, since for $A, B \in h^{-1}(\mathscr{F})$ we have that $h(A), h(B) \in \mathscr{F}$, which is a filter and therefore $h(A) \wedge h(B) \in \mathscr{F}$, by the homomorphism $h(A) \wedge h(B) = h(A \wedge B)$ which implies that $(A \wedge B) \in h^{-1}(\mathscr{F})$, it is also clear that the homomorphism preserves the ordering of elements since $A \preccurlyeq B \Leftrightarrow A \wedge B = A$ and therefore assuming $h(A) \preccurlyeq_{L_2} h(B)$ then $h(A) \wedge h(B) = h(A)$ and by the homomorphism $h(A) \wedge h(B) = h(A \wedge B) = h(A) \Rightarrow A \preccurlyeq_{L_1} B$, so that $h^{-1}(\mathscr{F})$ is a upper set, closed with respect to finite meets operations and therefore a filter.

Item $(ii)$ is direct since for $h(A) \vee h(B) = h(A \vee B) \in \mathscr{F}$, if $\mathscr{F}$ is a prime filter, then either $h(A) \in \mathscr{F}$ or $h(B) \in \mathscr{F}$ so that either $A \in h^{-1}(\mathscr{F})$ or $B \in h^{-1}(\mathscr{F})$, whenever $(A \vee B) \in h^{-1}(\mathscr{F})$. ∎

**Proposition 2.2.12.** *If $\mathcal{F}$ is a subset in a lattice $\mathcal{L}$, then the following are equivalent:*

(i) $\mathcal{F}$ *is a* prime filter,

(ii) $\mathcal{F} = h^{-1}(\mathbf{1})$ *for some lattice homomorphism $h : \mathcal{L} \to \{\mathbf{0}, \mathbf{1}\}$.*

*Proof:* The implication $(ii) \Rightarrow (i)$ comes from Proposition 2.2.11 since $\{\mathbf{1}\}$ is a prime filter and so is $\mathcal{F} = h^{-1}(\mathbf{1})$ by hypothesis. The converse $(i) \Rightarrow (ii)$ follows from defining $h : \mathcal{L} \to \{\mathbf{0}, \mathbf{1}\}$ by:

$$h(A) = \begin{cases} \mathbf{1}, & \text{if } A \in \mathcal{F} \\ \mathbf{0}, & \text{if } A \notin \mathcal{F} \end{cases} \quad \forall A \in \mathcal{L}$$

So $h^{-1}(\mathbf{1}) = \mathcal{F}$. There is still the need for proving the homomorphism properties of $h$, if $(A \vee B) \in \mathcal{F}$ then $h(A \vee B) = \mathbf{1}$, by the hypothesis of $\mathcal{F}$ being a prime ideal then either $A \in \mathcal{F}$ or $B \in \mathcal{F}$ so that either $h(A) = \mathbf{1}$ or $h(B) = \mathbf{1}$ and therefore $h(A) \vee h(B) = \mathbf{1}$. If $(A \vee B) \notin \mathcal{F}$ then $h(A \vee B) = \mathbf{0}$ and since $\mathcal{F}$ is upwardly closed and it holds that $A \preccurlyeq (A \vee B)$ and $B \preccurlyeq (A \vee B)$, for any $A, B \in \mathcal{L}$ then $A \notin \mathcal{F}$ and $B \notin \mathcal{F}$ for if any of them were so would $(A \vee B)$ since it is upper than either, therefore $h(A) = \mathbf{0}$ and $h(B) = \mathbf{0}$ for which $h(A) \vee h(B) = \mathbf{0}$ and the first equality is established.

For the second equality the thinking is analogous, for $(A \wedge B) \in \mathcal{F}$ then $h(A \wedge B) = \mathbf{1}$, assuming by contradiction that $h(A) \wedge h(B) = \mathbf{0}$ then either $h(A) = \mathbf{0}$ or $h(B) = \mathbf{0}$ which implies that either $A \notin \mathcal{F}$ or $B \notin \mathcal{F}$, since we made the assumption that $(A \wedge B) \in \mathcal{F}$ and remembering that it always holds that $(A \wedge B) \preccurlyeq A$ and $(A \wedge B) \preccurlyeq B$, by the hypothesis that $\mathcal{F}$ is upwardly closed, that is absurd, therefore $h(A) \wedge h(B) = \mathbf{1}$. If $(A \wedge B) \notin \mathcal{F}$ then $h(A \wedge B) = \mathbf{0}$, if it were the case then that $h(A) \wedge h(B) = \mathbf{1}$ then it would follow that $A, B \in \mathcal{F}$, but then the contradiction of $(A \wedge B) \in \mathcal{F}$ would occur since $\mathcal{F}$ is closed with respect to the meet operation. So the second homomorphism equality is established. ∎

This last proposition shows us that there is a one-to-one correspondence between prime filters in a lattice $\mathcal{L}$ and the two-valued homomorphisms on the lattice, since the second lattice of the homomorphism is two valued we can understand this assignment as a **valuation map** that assigns either the value $\mathbf{1}$ for "**true**" or the value $\mathbf{0}$ for "**false**" to the elements in $\mathcal{L}$. In other words, there exists a valuation map on a lattice if and only if the lattice contains a prime filter.

**Definition 2.2.13.** *A state $\mu$ at time t in a quantum system S is a $\sigma-$additive probability measure on $\mathcal{P}(\mathcal{H}_S)$, i.e., a map $\mu : \mathcal{P}(\mathcal{H}_S) \to [0, 1]$ such that:*

(i) $\mu(\mathbb{1}) = 1$.

(ii) *if $\{P_i\}_{i \in \mathbb{N}} \subset \mathcal{P}(\mathcal{H}_S)$ are two by two disjoint, $P_i P_j = 0, i \neq j$, then:*

$$\mu\left(s\text{-}\sum_{i=0}^{\infty} P_i\right) = \sum_{i=0}^{\infty} \mu(P_i).$$

It must be also noted that differently from classical logic, if $P$ is not true for a given state $\xi$, that is $P\xi \neq \xi$, then it doesn't follow that $\neg P$ is true, since for example $\xi$ could be the sum of $\kappa \in P(\mathcal{H}_S)$ and $\nu \in P(\mathcal{H}_S)^\perp$, so that $P\xi \neq \xi$ and $(\neg P)\xi \neq \xi$.

However, we shall see in chapter 6 that the lattice of projections on a Hilbert space of dimension 3 or greater does not have a prime filter, by the Kochen—Specker Theorem, hence those lattices do not allow a valuation map.

## 2.3   Kripke semantics

Given the need arising from Frauchinger-Renner's Gedankenexperiment to deal with systems that contain a logic of agents that contain information and that can act based on that information, the need to explain the logic present in these systems as a epistemic modal logic becomes evident. Epistemic modal logic is a generalization of modal logic, itself being a extension of propositional logic, the epistemic part comes from the intention of creating a logic pertinent to information and its receptacles. In modal logic two other types of unary operators are considered on the propositions $\square q$ ("it is necessary $q$") and $\lozenge p$ ("it is possible $p$"), in such a way that they are dual since $\square p = \neg \lozenge(\neg p)$, $p$ is necessary is equal to not possibly not $p$, i.e. the necessity of $p$ is the same as the negation of the impossibility of $p$; and $\lozenge q = \neg \square(\neg q)$, $q$ is possible is equal to not necessarily not $q$, i.e. the possibility of $q$ is the same as the negation of the unnecessarity of $q$; Epistemic modal logic is generalised from modal logic in such a way that $\square q$ are replaced by $\mathcal{K} q$ that can be read as "it is epistemically necessary that $q$" or "it is known that $q$", the $\Diamond$ is not generalized or given any importance, since it does not have a clear interpretation as to what it would represent in terms of knowledges.

In a situation where we have several agents, we now have many unary operators of the type $\mathscr{K}_A, \mathscr{K}_B, \mathscr{K}_C, \ldots$ each referring to a agent A, B, C, ... so that we can now deal with a Kripke semantic [80]:

**Definition 2.3.1** (Kripke structure). *A Kripke structure $\mathcal{M}$ for $n$ agents over a set of propositions $\Phi$ is a $(n+2)$-tuple $(\Sigma, \pi, \mathbf{K}_1, \ldots, \mathbf{K}_n)$, where $\Sigma$ is a nonempty set of states (or configurations), $\pi: \Sigma \times \Phi \to \{0,1\}$ is an interpretation which defines the truth value of a proposition $\phi \in \Phi$ in a state $s \in \Sigma$ and $\mathbf{K}_i$ is a binary equivalence relation on the set of states $\Sigma$ in which $(s,t) \in \mathbf{K}_i$ if agent $i$ can derive state $t$ from state $s$. Formally, we say that agent $i$ "knows" $\phi$ in a configuration $s$, that is, $(\mathcal{M}, s) \models \mathscr{K}_i \phi$, if and only if for all possible worlds $t$ such that $(s,t) \in \mathbf{K}_i$ (that is, all the equivalent configurations admitted by the agent given their knowledge), it holds that $(\mathcal{M}, t) \models \phi$.*

For this logical system we consider the following 5 axioms, that represent possible deductions that are able to be made in a system with agents, each with the capacity of reasoning about information. The notation $(\mathcal{M}, s) \models \phi$ [7] means that in $\mathcal{M}$, $\pi(s, \phi) = 1$, videlicet, within the Kripke structure $\mathcal{M}$ and for a state $s \in \Sigma$, the proposition $\phi$ is true, whereas the notation $\models \psi$ is used to indicate that $\phi$ is always true, independent of any configuration state of the formal system, i.e. $\psi$ is a **tautology**.

**Definition 2.3.2** (Distribution axiom). *This axiom establishes that if a agent $i$ knows $\phi$ and that $\phi \Rightarrow \psi$ then $i$ knows $\psi$, or in symbols:*

$$(\mathcal{M}, s) \models (\mathscr{K}_i \phi) \wedge (\mathscr{K}_i [\phi \Rightarrow \psi]) \iff (\mathcal{M}, s) \models \mathscr{K}_i [\phi \wedge (\phi \Rightarrow \psi)] \implies (\mathcal{M}, s) \models \mathscr{K}_i \psi.$$

*This is a realization of* modus ponens *as a rule of inference.*

**Definition 2.3.3** (Knowledge generalization axiom). *This axiom says that if a proposition $\phi$ is a tautology, then this proposition is available to all agents, and therefore any agent $i$ knows $\phi$, that is:*

$$\models \phi \implies \mathcal{M} \models \mathscr{K}_i \phi, \forall i.$$

**Definition 2.3.4** (Positive introspection axiom). *Agents must intuitively be able to introspect about their own knowledge, that is if a agent $i$ knows $\phi$ then he must also know that he knows $\phi$, so we add this property as a axiom:*

$$(\mathcal{M}, s) \models \mathscr{K}_i \phi \implies (\mathcal{M}, s) \models \mathscr{K}_i \mathscr{K}_i \phi.$$

*This is equivalent by contrapositive to the statement that if it is false that agent $i$ knows that he knows $\phi$ then $i$ doesn't know $\phi$:*

$$(\mathcal{M}, s) \models \neg \mathscr{K}_i \mathscr{K}_i \phi \implies (\mathcal{M}, s) \models \neg \mathscr{K}_i \phi.$$

---

[7]   This notation is usually used to indicate that a proposition is implied for a formal system in a given configuration, it is a variation on $\vdash$, sometimes called "Frege's stroke" because it is a vestigial notation from Gottlob Frege's *Begriffsschrift*, it stands for *syntactic consequence* or *derivability* and can be understood as meaning "is provable", where $\mathfrak{T} \vdash p$ would mean "$p$ is provable by system $\mathfrak{T}$", where as $\models$ stands for *semantic consequence* and can be understood as meaning "this configuration entails".

**Definition 2.3.5** (Negative introspection axiom)**.** *Agents must intuitively also be able to introspect about what they do not know, this makes sense in a logical system about knowledge where there is no confusion about what is known and about that which is unknown, that is knowledge is always known or not known, therefore if a agent i doesn't know $\phi$ then he knows that he doesn't know $\phi$, or in symbols:*

$$(\mathcal{M}, s) \models \neg \mathcal{K}_i \phi \implies (\mathcal{M}, s) \models \mathcal{K}_i \neg \mathcal{K}_i \phi.$$

*Again by contrapositive this statement is equivalent to saying that if it is false that agent i knows that he doesn't know $\phi$ then he knows $\phi$, otherwise he would know that he doesn't know $\phi$, in symbols:*

$$(\mathcal{M}, s) \models \neg \mathcal{K}_i \neg \mathcal{K}_i \phi \implies (\mathcal{M}, s) \models \mathcal{K}_i \phi.$$

**Definition 2.3.6** (Knowledge axiom)**.** *Since we are talking about knowledge, it is reasonable that if a agent i knows $\phi$ then $\phi$ must be necessarily true, otherwise i wouldn't **know** $\phi$ but would instead be believing $\phi$, or in other words, whereas one can believe something that is false, one cannot know as true something that is false.*

$$(\mathcal{M}, s) \models \mathcal{K}_i \phi \implies (\mathcal{M}, s) \models \phi.$$

*By considering the above formula for a negation of a proposition $\phi$, that is $(\mathcal{M}, s) \models \mathcal{K}_i(\neg \phi) \implies (\mathcal{M}, s) \models \neg \phi$, and using the contrapositive on this case, we get that if a proposition $\phi$ is true then it is impossible for any agent i to know the negation of $\phi$ to be true:*

$$(\mathcal{M}, s) \models \phi \implies (\mathcal{M}, s) \models \neg \mathcal{K}_i \neg \phi, \forall i.$$

*This axiom allows us to obtain knowledge from introspective knowledge since $(\mathcal{M}, s) \models \mathcal{K}_i \mathcal{K}_i \phi \implies (\mathcal{M}, s) \models \mathcal{K}_i \phi$, but not only that, it also allows one agent to obtain knowledge from knowing another agent's knowledge, that is for a agent i that knows that agent j knows $\phi$, we have: $(\mathcal{M}, s) \models \mathcal{K}_i \mathcal{K}_j \phi \implies (\mathcal{M}, s) \models \mathcal{K}_i \phi$.*

This last axiom is too strong for use in *QM* and will have to be weakened so as to allow contextual knowledge, as it is this axiom would imply that for one specific global configurational situation all knowledge is absolute and does not depend on the specific local situation of each observing agent. The detailed discussion of the reasoning of the necessity of weakening this axiom will be made in chapter 6, section 6.3.

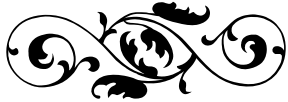

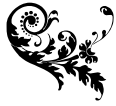 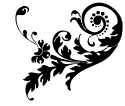

# CHAPTER 3

# Quantum Measurement Theory

*The procedure of measurement in QM deserves close scrutiny, specially considering that the core of all questions about the ambiguities (being them descriptive ambiguities or interpretative ambiguities) that emerge from the formal description of QM, lie in the experimentally accessible part of QM, that is, in measurements.*

The first idea about the origins of ambiguities in QM, comes from Bohr [12], he prescribes that measurement outcomes must always be treated as *classical concepts*, this vague prescription has given rise to several different understandings, especially in the philosophy of physics, of how these classical concepts would exactly factor in the interpretation of results in QM. On this note, Josef Jauch's article [62] proposed a particular realization of this idea that at the level of measurement outcomes the quantities we have can only be faced as their respective classical concepts, this proposed operationality of measurement outcomes will be necessary for Hepp's model in chapter 5.

The basic description of a process of measurement of a physical system with spin $\frac{1}{2}$, such as that of a Stern-Gerlach experiment, where the state of the physical system being measured is preserved after measurement, i.e. Pauli's measurements of the *first kind*, can be made as follows:

Let the Hilbert space of the states of the two-dimensional system $\mathscr{H}_s$ be represented by $\mathscr{H}_s = \mathbb{C}^2$. The states $\psi_+ = \binom{1}{0}$ and $\psi_- = \binom{0}{1}$ are eigenstates of $\sigma_z$ with eigenvalues $\pm 1$ and projectors $P_\pm$. $\sigma_z$ has values $\pm 1$ with probabilities $p_\pm$ in state $\rho = p_+ P_+ + p_- P_-$, in coherent superposition $\psi = c_+ \psi_+ + c_- \psi_-$ with amplitudes $c_\pm$, $|c_+|^2 + |c_-|^2 = 1$, $\sigma_z$ with no definitive value.

In the measurement the system is coupled to a measuring apparatus, which is itself a quantum system, in the state space $\mathscr{H}_A$, initially in the state $\varphi_0$.

The combined system is in state $\psi_\pm \otimes \varphi_0 \in \mathscr{H}_s \otimes \mathscr{H}_A$. After the experiment we have:

$$\psi_\pm \otimes \varphi_0 \longrightarrow \psi_\pm \otimes \varphi_\pm$$

Where $\varphi_\pm \in \mathscr{H}_A$ corresponds to some macroscopic pointer state of the measuring apparatus. However, if we consider the unitary evolution of the system by $U(t)$, which depends on the specific Hamiltonian of the system, we will have linearity:

$$U(t)\left((c_+ \psi_+ + c_- \psi_-) \otimes \varphi_0\right) = c_+ \psi_+ \otimes \varphi_+ + c_- \psi_- \otimes \varphi_-$$

Then, following [62]: Being $Q\pm$ the projectors of $\varphi_\pm$, we identify the previous coherent overlap with the mixture:

$$|c_+|^2 P_+ \otimes Q_+ + |c_-|^2 P_- \otimes Q_-,$$

this replacement is then called the "wave packet reduction".

If $\mathscr{A} \subseteq \mathfrak{B}(\mathscr{H}_s \otimes \mathscr{H}_A)$ corresponds to all possible observations, then we understand the measurement process of $\sigma_z$ in Stern-Gerlach as:

$$\langle \psi_+ \otimes \varphi_+, A \psi_- \otimes \varphi_- \rangle = 0, \quad \forall A \in \mathscr{A}, \tag{3.1}$$

so the two identifiers are equivalent with respect to $\mathscr{A}$.

More precisely we have that if $\mathfrak{K}$ is the set of observables of a physical system, its elements being self-adjoint, then we may say that two states $W_1$ and $W_2$ are *equivalent* with respect to our system of observables $\mathfrak{K}$ if:

$$\text{tr}(W_1 A) = \text{tr}(W_2 A), \quad \forall A \in \mathfrak{K}.$$

We shall denote that two states $W_1$ and $W_2$ are *equivalent* with respect to the set of observables $\mathfrak{K}$ by $W_1 \sim_\mathfrak{K} W_2$, this relation is trivially a equivalence relation since it obeys reflexivity $\text{tr}(WA) = \text{tr}(WA), \forall A \in \mathfrak{K}$, symmetry since if $\text{tr}(W_1 A) = \text{tr}(W_2 A), \forall A \in \mathfrak{K} \Rightarrow \text{tr}(W_2 A) = \text{tr}(W_1 A), \forall A \in \mathfrak{K}$ and transitivity, for if $\text{tr}(W_1 A) = \text{tr}(W_2 A) = \text{tr}(W_3 A), \forall A \in \mathfrak{K}$ then clearly $\text{tr}(W_1 A) = \text{tr}(W_3 A)$. Physically, the relation $W_1 \sim_\mathfrak{K} W_2$ means that the two states $W_1$ and $W_2$ cannot be distinguished by any measurement, whatsoever, with observables from the system $\mathfrak{K}$. We then consider that for the previously discussed case, for a pure state $(c_+ \psi_+ + c_- \psi_-) \in \mathscr{H}_S$ to be equivalent with respect to a set of observables $\mathfrak{K}$ to a mixed state $(|c_+|^2 P_+ + |c_-|^2 P_-) \in \mathfrak{B}(\mathscr{H})$ then:

$$\text{tr}\big(P_{(c_+ \psi_+ + c_- \psi_-)} A\big) = \text{tr}\big((|c_+|^2 P_+ + |c_-|^2 P_-) A\big), \quad \forall A \in \mathfrak{K}$$

we note that calculating the LHS:

$$\text{tr}\big(P_{(c_+ \psi_+ + c_- \psi_-)} A\big) = \langle (c_+ \psi_+ + c_- \psi_-), A(c_+ \psi_+ + c_- \psi_-) \rangle =$$

$$= |c_+|^2 \langle \psi_+, A\psi_+ \rangle + \overline{c_+} c_- \langle \psi_+, A\psi_- \rangle + \overline{c_-} c_+ \langle \psi_-, A\psi_+ \rangle + |c_-|^2 \langle \psi_-, A\psi_- \rangle,$$

whereas from the RHS:

$$\text{tr}\big((|c_+|^2 P_+ + |c_-|^2 P_-) A\big) = |c_+|^2 \langle \psi_+, A\psi_+ \rangle + |c_-|^2 \langle \psi_-, A\psi_- \rangle.$$

For both these expressions to be equal we must have for the cross terms:

$$\langle \psi_+, A\psi_- \rangle = \overline{\langle \psi_-, A\psi_+ \rangle} = 0, \quad \forall A \in \mathfrak{K},$$

where the self-adjointness of observable $A$ was used, from this equation we obtain condition (3.1) in the scenario of measurement schemes.

This ability to aggregate states that are indiscernible to a given collection of observables $\mathfrak{K}$ is relevant for the study of the process of wave packet reduction, since by this equivalence condition we may create classes of equivalent states $[W]_{\sim_\mathfrak{K}}$ and as we have seen, these classes allows one to relate pure states to mixed states, in fact the class itself may be seen as a type of state called a *macrostate*, which explicitly represents the possibility of perceptual ambiguities of states, especially when these are dealt macroscopically.

## 3.1　Wigner-Araki-Yanase Theorem

Considering measurements in systems that have conservation of a given additive quantity one can ask if this conservation law in any way disrupts the measurement process, the following theorem proved by H. Araki and M. Yanase in [6] and based on investigations proposed by Wigner in [101] gives conditions for the measurement process to not be affected by this conservation law, but also serves to elucidate the possible constraints on approximate measurements:

**Theorem 3.1.1** (Wigner-Araki-Yanase). *Let $M$ be a discrete and self-adjoint operator on $\mathcal{H}_1$, and let $L_1$ and $L_2$ bounded self-adjoint operators on $\mathcal{H}_1$ and $\mathcal{H}_2$, respectively, let $U(t) : \mathcal{H}_1 \otimes \mathcal{H}_2 \to \mathcal{H}_1 \otimes \mathcal{H}_2$ be a unitary operator describing the time development of the combined system after a measurement realized, so that by the Lüders-von Neumann measurement rule:*

$$U(t)(\phi_{\mu k} \otimes \xi) = \sum_{k'} \phi_{\mu k'} \otimes X_{\mu k k'},$$

*where $\phi_{\mu k}$ is a eigenstate of $M$ with a eigenvalue $\mu$ and possible degeneracy $k$, and $\xi$ is a prepared state of the apparatus. If there is a universal conservation law for a self-adjoint operator $L : \mathcal{H}_1 \otimes \mathcal{H}_2 \to \mathcal{H}_1 \otimes \mathcal{H}_2$ which is additive by:*

$$L = L_1 \otimes \mathbb{1} + \mathbb{1} \otimes L_2,$$

*where $L_1 : \mathcal{H}_1 \to \mathcal{H}_1$ and $L_2 : \mathcal{H}_2 \to \mathcal{H}_2$ are self-adjoint operators in $\mathcal{H}_1$ and $\mathcal{H}_2$, respectively. The universal conservation law is understood as:*

$$[U(t), L] = 0, \qquad \forall\, t \in \mathbb{R}.$$

*Then such a unitary operator $U(t)$ doesn't exist unless:*

$$[L_1, M] = 0.$$

*This relation is called the Yanase condition.*

*Proof:* Since $M$ has discrete eigenvalues, each of which will be represented with the variable $\mu$, each with associated eigenvectors $\phi_{\mu k}$, with a degeneracy index $k$, which are orthonormal and complete in $\mathcal{H}_1$:

$$M\phi_{\phi_{\mu k}} = \mu \phi_{\mu k},$$

$$\langle \phi_{\mu k},\, \phi_{\mu' k'} \rangle = \delta_{\mu \mu'} \delta_{k k'}. \tag{3.2}$$

With the apparatus prepared in a state $\xi \in \mathcal{H}_2$, that is a fixed normalized state independent of the soon to be measured system state, such that a Lüders-von Neumann measurement of the combination in the state $\phi_{\mu k} \otimes \xi$ evolves after a finite time $t$ into:

$$U(t)(\phi_{\mu k} \otimes \xi) = \sum_{k'} \phi_{\mu k'} \otimes X_{\mu k k'}, \tag{3.3}$$

Where $U(t)$ is a unitary operator describing the time development of the combined system. The distinguishability of different measured values of the operator $M$ in the measured apparatus is realized by the imposition of the orthogonality of apparatus states associated with different eigenvalues:

$$\langle X_{\mu k k'},\, X_{\mu' k' k''} \rangle = 0, \quad \text{for } \mu \neq \mu. \tag{3.4}$$

And considering the additive operator given by:

$$L = L_1 \otimes I + I \otimes L_2 \tag{3.5}$$

whose conservation law can be written as:

$$[U(t), L] = 0,\ \forall t \tag{3.6}$$

Calculating then:

$$\langle\phi_{\mu'k'}\otimes\xi,\,L(\phi_{\mu k}\otimes\xi)\rangle \stackrel{U \text{ is unitary}}{=} \langle U(t)(\phi_{\mu'k'}\otimes\xi),\,U(t)L(\phi_{\mu k}\otimes\xi)\rangle \stackrel{(3.6)}{=} \langle U(t)(\phi_{\mu'k'}\otimes\xi),\,LU(t)(\phi_{\mu k}\otimes\xi)\rangle \stackrel{(3.3)}{=}$$

$$\stackrel{(3.3)}{=} \left\langle \sum_{k'''}\phi_{\mu'k'''}\otimes X_{\mu'k'k'''},\,L\sum_{k''}\phi_{\mu k''}\otimes X_{\mu kk''}\right\rangle. \tag{3.7}$$

On the other hand, by considering the additivity of $L$ the sesquilinearity of the inner product and the definition of inner produts of tensor products, we get:

$$\langle\phi_{\mu'k'}\otimes\xi,\,L(\phi_{\mu k}\otimes\xi)\rangle \stackrel{(3.7)}{=} \left\langle \sum_{k'''}\phi_{\mu'k'''}\otimes X_{\mu'k'k'''},\,L\sum_{k''}\phi_{\mu k''}\otimes X_{\mu kk''}\right\rangle \stackrel{(3.5)}{=}$$

$$= \left\langle \sum_{k'''}\phi_{\mu'k'''}\otimes X_{\mu'k'k'''},\,\sum_{k''}L_1\phi_{\mu k''}\otimes X_{\mu kk''}\right\rangle + \left\langle \sum_{k'''}\phi_{\mu'k'''}\otimes X_{\mu'k'k'''},\,\sum_{k''}\phi_{\mu k''}\otimes L_2 X_{\mu kk''}\right\rangle =$$

$$= \sum_{k'',k'''}\left(\langle\phi_{\mu'k'''},\,L_1\phi_{\mu k''}\rangle\langle X_{\mu'k'k'''},\,X_{\mu kk''}\rangle + \langle\phi_{\mu'k'''},\,\phi_{\mu k''}\rangle\langle X_{\mu'k'k'''},\,L_2 X_{\mu kk''}\rangle\right),$$

because of the orthogonalities 3.2 and 3.4 we see that the last equality is equal to 0 if $\mu \neq \mu'$, as it must also be true for:

$$\langle\phi_{\mu'k'}\otimes\xi,\,L(\phi_{\mu k}\otimes\xi)\rangle \stackrel{(3.5)}{=} \langle\phi_{\mu'k'},\,L_1\phi_{\mu k}\rangle\underbrace{\langle\xi,\,\xi\rangle}_{=1} + \underbrace{\langle\phi_{\mu'k'},\,\phi_{\mu k}\rangle}_{\mu\neq\mu'\atop =0}\langle\xi,\,L_2\xi\rangle \stackrel{\mu\neq\mu'}{=} 0,$$

as a consequence of this orthogonality we get:

$$\langle\phi_{\mu'k'},\,L_1\phi_{\mu k}\rangle = \delta_{\mu\mu'}\langle\phi_{\mu'k'},\,L_1\phi_{\mu k}\rangle \tag{3.8}$$

Then we can go on to prove that $L_1$ must commute with $M$. Decomposing $M$ into projection operators:

$$M = \sum_\mu \mu P_\mu;$$

$$P_\mu \phi_{\mu'k} = \delta_{\mu\mu'}\phi_{\mu'k}. \tag{3.9}$$

We will now prove the commutativity of $L_1$ and $P_\mu$. Considering the self-adjoint nature of $P_\mu$ and calculating the following inner product:

$$\langle\phi_{\mu'k'},\,P_\mu L_1\phi_{\mu''k''}\rangle \stackrel{P_\mu \text{ is self-adjoint}}{=} \langle P_\mu\phi_{\mu'k'},\,L_1\phi_{\mu''k''}\rangle \stackrel{(3.9)}{=} \delta_{\mu\mu'}\langle\phi_{\mu'k'},\,L_1\phi_{\mu''k''}\rangle \stackrel{(3.8)}{=}$$

$$= \delta_{\mu\mu'}\delta_{\mu'\mu''}\langle\phi_{\mu'k'},\,L_1\phi_{\mu''k''}\rangle, \tag{3.10}$$

and for the inverse order:

$$\langle\phi_{\mu'k'},\,L_1 P_\mu\phi_{\mu''k''}\rangle \stackrel{(3.9)}{=} \delta_{\mu\mu''}\langle\phi_{\mu'k'},\,L_1\phi_{\mu''k''}\rangle \stackrel{(3.8)}{=} \delta_{\mu\mu''}\delta_{\mu'\mu''}\langle\phi_{\mu'k'},\,L_1\phi_{\mu''k''}\rangle. \tag{3.11}$$

By linearity:

$$\langle\phi_{\mu'k'},\,P_\mu L_1\phi_{\mu''k''}\rangle - \langle\phi_{\mu'k'},\,L_1 P_\mu\phi_{\mu''k''}\rangle = \langle\phi_{\mu'k'},\,[L_1,\,P_\mu]\phi_{\mu''k''}\rangle$$

$$= (\delta_{\mu\mu'}\delta_{\mu'\mu''} - \delta_{\mu\mu''}\delta_{\mu'\mu''})\langle\phi_{\mu'k'},\,L_1\phi_{\mu''k''}\rangle$$

The only possibility of this last line being different from zero would necessitate that $\mu = \mu'$ and $\mu' = \mu''$ but this implies $\mu = \mu' = \mu''$ but for these the line is equal to zero, alternatively $\mu = \mu''$ and $\mu'' = \mu'$ but then again $\mu = \mu' = \mu''$, which manifestly demonstrates that $L_1$ commutes with $P_\mu$. Then we see the Yanase condition as a consequence:

$$[L_1,\, P_\mu] = 0 \implies \left[L_1,\, \sum_\mu P_\mu\right] = [L_1,\, M] = 0.$$

∎

Although sharp measurements have this limitation, we can try to relax the requirement of a perfectly discernible sharp measurement to a unsharp one that doesn't perfectly discern the results and see what implications this conservation law has in the uncertainties in the measurement.

## 3.2 Araki-Yanase fiduciary apparatus for approximate measurements

It is know that approximate measurements with a arbitrary precision exist for configurations of the probed system and apparatus in which the additive conserved quantity $L$ has a discrete spectrum with the system to be measured being limited by a minimum and a maximum value, that is, $L_1$ has only a finite number of eigenvalues. We may assume, for simplicity of exposition, that the eigenvalues of $L_1$ are $0, \pm 1, \pm 2, \ldots, \pm l$, otherwise we could use the same following construction with a more complicated indexation of eigenvalues, a more cumbersome delimitation of intervals of eigenvalues and a less homogeneous set of normalization coefficients. We can decompose the $L$'s into projection operators:

$$L = \sum_\lambda \lambda P(\lambda),$$
$$L_i = \sum_\lambda \lambda P_i(\lambda), \quad i = 1, 2.$$

The additivity (3.5) and our particular assumption that $\pm l$ are the extremal values of $L_1$ on our system, imply that the values may be distributed between system and apparatus in such a way that:

$$P(\lambda) = \sum_{|\lambda'| \leq l} P_1(\lambda') P_2(\lambda - \lambda'). \tag{3.12}$$

The following lemma shall also be useful in our construction of a apparatus for approximate measurements, guaranteeing that the measurement process, with our carefully prepared apparatus and the considered resulting states, can in fact be realized by a unitary transformation (even though it does not guarantee the existence of a Hamiltonian that generates this transformation).

**Lemma 3.2.1.** *Given two sets of vectors $\{\Psi_\alpha^i\}_\alpha$ and $\{\Psi_\alpha^f\}_\alpha$ in a Hilbert space $\mathcal{H} = \mathcal{H}_1 \otimes \mathcal{H}_2$ satisfying:*

$$\langle \Psi_\alpha^i,\, P(\lambda) \Psi_\beta^i \rangle = \langle \Psi_\alpha^f,\, P(\lambda) \Psi_\beta^f \rangle \tag{3.13}$$

*for every possible measurement value $\lambda$ of an additive quantity $L : \mathcal{H} \to \mathcal{H}$ defined as $L := L_1 \otimes \mathbb{1} + \mathbb{1} \otimes L_2$ and where each $L_i$ is a self-adjoint operator in $\mathcal{H}_i$, $i = 1, 2$; then there exists a Hilbert space $\widetilde{\mathcal{H}_2}$ containing $\mathcal{H}_2$ and a unitary operator $U$ on $\widetilde{\mathcal{H}} = \mathcal{H}_1 \otimes \widetilde{\mathcal{H}_2}$ such that:*

1) a self-adjoint operator $\widetilde{L}_2$ (representing the conserved quantity in $\widetilde{\mathscr{H}_2}$) is defined on $\widetilde{\mathscr{H}_2}$ coinciding with $L_2$ on $\mathscr{H}_2$,

2) $U$ commutes with the conserved quantity $\widetilde{L}$ on $\widetilde{\mathscr{H}}$, $\widetilde{L} = L_1 \otimes \mathbb{1} + \mathbb{1} \otimes \widetilde{L}_2$, and

$$U\Psi_\alpha^i = \Psi_\alpha^f. \tag{3.14}$$

*If the set of the indices $\alpha$ is finite, $\widetilde{\mathscr{H}_2}$ can be taken to be $\mathscr{H}_2$.*

*Proof:* To prove this lemma we first introduce the following notations for: the subspace with eigenvalue $\lambda$ written as $\mathscr{H}_\lambda \equiv \overline{\text{span}\{v \in \mathscr{H} \mid Lv = \lambda v\}}$, calling by $I$ the set of indexes $\alpha$ we can consider the subspaces spanned by the components of $\Psi_\alpha^i$ with eigenvalue $\lambda$ written as $\mathscr{H}_\lambda^i \equiv \overline{\text{span}\{P(\lambda)\Psi_\alpha^i \in \mathscr{H}_\lambda \mid \alpha \in I, \lambda \in \sigma(L)\}}$, the subspaces spanned by the components of $\Psi_\alpha^f$ with eigenvalue $\lambda$ written as $\mathscr{H}_\lambda^f \equiv \overline{\text{span}\{P(\lambda)\Psi_\alpha^f \in \mathscr{H}_\lambda \mid \alpha \in I, \lambda \in \sigma(L)\}}$, and the orthogonal complements of $\mathscr{H}_\lambda^i$ within $\mathscr{H}_\lambda$ given by $\mathscr{H}_{\lambda\perp}^i \equiv \left(\mathscr{H}_\lambda^i\right)^\perp \cap \mathscr{H}_\lambda$ and the orthogonal complement of $\mathscr{H}_\lambda^f$ within $\mathscr{H}_\lambda$ given by $\mathscr{H}_{\lambda\perp}^f \equiv \left(\mathscr{H}_\lambda^f\right)^\perp \cap \mathscr{H}_\lambda$.

From these definitions we obviously get that:

$$\mathscr{H} = \bigoplus_\lambda \left(\mathscr{H}_\lambda^i \oplus \mathscr{H}_{\lambda\perp}^i\right) = \bigoplus_\lambda \left(\mathscr{H}_\lambda^f \oplus \mathscr{H}_{\lambda\perp}^f\right), \tag{3.15}$$

where the $\oplus$ operation is the usual direct sum of Hilbert spaces.

Defining a linear mapping $U_\lambda : \mathscr{H}_\lambda^i \to \mathscr{H}_\lambda^f$, through linearity by the action:

$$U_\lambda \left(\sum_{\alpha \in I} C_\alpha P(\lambda)\Psi_\alpha^i\right) = \sum_{\alpha \in I} C_\alpha P(\lambda)\Psi_\alpha^f, \quad C_\alpha \in \mathbb{C}, \forall \alpha \in I.$$

As a consequence of the hypothesis (3.13) and the sesquilinearity of the inner product, then either $\sum_{\alpha \in I} C_\alpha P(\lambda)\Psi_\alpha^i$ and $\sum_{\alpha \in I} C_\alpha P(\lambda)\Psi_\alpha^f$ both diverge or both converge, in them converging, if one of them converges to zero so does the other; as $U_\lambda$ is linear, then $U_\lambda$ is injective since if for $a, b \in \mathscr{H}_\lambda^i$, $a \neq b$ we were to have $U_\lambda a = U_\lambda b$ then $U_\lambda(a-b) = 0 \stackrel{(3.13)}{\Longrightarrow} \langle U_\lambda(a-b), P(\lambda)U_\lambda(a-b)\rangle = \langle (a-b), P(\lambda)(a-b)\rangle = \langle (a-b), (a-b)\rangle = \|a-b\|^2 = 0 \Rightarrow a - b = 0 \Rightarrow a = b$ which is a contradiction, therefore $U_\lambda$ is injective.

Due to (3.13) the map $U_\lambda$ is a isometry between $\mathscr{H}_\lambda^i$ and $\mathscr{H}_\lambda^f$, which makes $U_\lambda$ a unitary transformation between these two spaces, and therefore they have the same dimension.

- If this dimension is finite, then because $\mathscr{H}_\lambda = \mathscr{H}_\lambda^i \oplus \mathscr{H}_{\lambda\perp}^i = \mathscr{H}_\lambda^f \oplus \mathscr{H}_{\lambda\perp}^f$ then if $\dim(\mathscr{H}_\lambda^i) = \dim(\mathscr{H}_\lambda^f) = d$ this forces $\dim(\mathscr{H}_{\lambda\perp}^i) = \dim(\mathscr{H}_{\lambda\perp}^f) = \dim(\mathscr{H}_\lambda) - d$, where if $\dim(\mathscr{H}_\lambda)$ is infinite then $\dim(\mathscr{H}_\lambda) - d$ is also obviously infinite. Therefore there always exists a unitary mapping $U_{\lambda\perp}$ of $\mathscr{H}_{\lambda\perp}^i$ onto $\mathscr{H}_{\lambda\perp}^f$.

Now we define the operator $U : \mathscr{H} \to \mathscr{H}$ by:

$$U = \bigoplus_{\lambda \in \sigma(L)} (U_\lambda \oplus U_{\lambda\perp}) \tag{3.16}$$

Because of the unitarity of $U_\lambda$ and $U_{\lambda\perp}$ and the decomposition (3.16) of $U$, the operator $U$ is obviously unitary. Whence, for each $\Psi \in \mathscr{H}$ we have:

$$U\Psi = \sum_{\lambda \in \sigma(L)} \left(U_\lambda \Psi_\lambda^i + U_{\lambda\perp} \Psi_{\lambda\perp}^i\right), \tag{3.17}$$

from where $\Psi$ must be writable as:

$$\Psi = \sum_{\lambda \in \sigma(L)} \left( \Psi^i_\lambda + \Psi^i_{\lambda\perp} \right), \quad \Psi^i_\lambda \in \mathscr{H}^i_\lambda, \; \Psi^i_{\lambda\perp} \in \mathscr{H}^i_{\lambda\perp}, \tag{3.18}$$

being a unique decomposition of $\Psi$ according to the first equation of (3.15). Since the subspace $\mathscr{H}_\lambda$ of $\mathscr{H}$ is spanned by eigenvectors of $L = L_1 \otimes \mathbb{1} + \mathbb{1} \otimes L_2$ with the eigenvalue $\lambda$ is mapped onto itself by $U$, then $L$ commutes with $U$; this can be seen from the fact that $\mathscr{H}_\lambda$ is only composed of eigenvectors of $L$ with eigenvalue $\lambda$ and therefore $L$ does not change the vectors of $\mathscr{H}_\lambda$ apart from a multiplication by a scalar $\lambda$, where as $U$ does change a vector $v \in \mathscr{H}_\lambda$ to a vector $Uv$ that is still a eigenvectors of $L$ with eigenvalue $\lambda$, therefore changing the order of application of $L$ and $U$ only changes in which step the factor of $\lambda$ that will multiply $Uv$ will appear, and therefore they commute. In this case we take the $\widetilde{\mathscr{H}_2}$ of the statement of the lemma as being $\mathscr{H}_2$, and therefore $\widetilde{L}_2 = L_2$. This completes the proof for the case where the dimension of $\mathscr{H}^i_\lambda$ or equally $\mathscr{H}^f_\lambda$ is finite.

- If this dimension is infinite, then the dimensions of $\mathscr{H}^i_{\lambda\perp}$ and $\mathscr{H}^f_{\lambda\perp}$ can be different. In such a case we introduce a new residual Hilbert space $\mathscr{H}^r_2$ which has its own $L$ valued operator given by $L^r_2$ in such a way that the dimension of $\mathscr{H}^r_2$ is at least the cardinality of the set of indexes $I$, with this we can define the space $\mathscr{H}^r \equiv \mathscr{H}_1 \otimes \mathscr{H}^r_2$, the additive operator $L^r := L_1 \otimes \mathbb{1} + \mathbb{1} \otimes L^r_2$ and the subspaces $\mathscr{H}^r_\lambda \equiv \overline{\text{span}\{v \in \mathscr{H}^r \mid L^r v = \lambda v\}}$, spanned by the eigenstates of $L^r$ with eigenvalues $\lambda$.

Since the dimension of $\mathscr{H}^i_\lambda$ or equally $\mathscr{H}^f_\lambda$ does not exceed the cardinality of the set of indexes $I$ (which makes the cardinality of $I$ infinite when the dimension of $\mathscr{H}^i_\lambda$ or equally $\mathscr{H}^f_\lambda$ is infinite, as is the case now), then the spaces $\mathscr{H}^{ir}_\lambda \equiv \mathscr{H}^i_{\lambda\perp} \oplus \mathscr{H}^r_\lambda$ and $\mathscr{H}^{fr}_\lambda \equiv \mathscr{H}^f_{\lambda\perp} \oplus \mathscr{H}^r_\lambda$ have the same dimension (the enumerable dimension of $\mathscr{H}^r_\lambda$). Hence, there always exists a unitary mapping $U_{\lambda\perp}$ of $\mathscr{H}^{ir}_\lambda$ onto $\mathscr{H}^{fr}_\lambda$.

Constructing now the extension Hilbert space as $\widetilde{\mathscr{H}_2} := \mathscr{H}_2 \oplus \mathscr{H}^r_2$, the extended self-adjoint operator $\widetilde{L}_2 : \widetilde{\mathscr{H}_2} \to \widetilde{\mathscr{H}_2}$ is defined as $\widetilde{L}_2 := L_2 \oplus L^r_2$.

We can then define $\widetilde{\mathscr{H}}$ to be the composition:

$$\widetilde{\mathscr{H}} := \bigoplus_{\lambda \in \sigma(L)} \left( \mathscr{H}^i_\lambda \oplus \mathscr{H}^{ir}_\lambda \right) = \bigoplus_{\lambda \in \sigma(L)} \left( \mathscr{H}^f_\lambda \oplus \mathscr{H}^{fr}_\lambda \right)$$

The unitary operator $U$ in this case will then be such that $U : \widetilde{\mathscr{H}} \to \widetilde{\mathscr{H}}$ given by:

$$U := \bigoplus_{\lambda \in \sigma(L)} \left( \widetilde{U}_\lambda \oplus \widetilde{U}_{\lambda\perp} \right)$$

for which, instead of (3.17) and (3.18), we have that $\forall \Psi \in \widetilde{\mathscr{H}}$:

$$U\Psi = \sum_{\lambda \in \sigma(L)} \left( \widetilde{U}_\lambda \Psi^i_\lambda + \widetilde{U}_{\lambda\perp} \Psi^{ir}_\lambda \right),$$

$$\Psi = \sum_{\lambda \in \sigma(L)} \left( \Psi^i_\lambda + \Psi^{ir}_\lambda \right), \quad \Psi^i_\lambda \in \mathscr{H}^i_\lambda, \; \Psi^{ir}_\lambda \in \mathscr{H}^{ir}_\lambda,$$

Then by the same argument as in the previous case, we can show the unitarity of $U$, and commutativity with the respective $\widetilde{L} : \widetilde{\mathscr{H}} \to \widetilde{\mathscr{H}} (= \mathscr{H}_1 \otimes \mathscr{H}_2 \otimes \mathscr{H}^r_2)$ that is defined as $\widetilde{L} := L_1 \otimes \mathbb{1} + \mathbb{1} \otimes \widetilde{L}_2$. ■

We are now ready for the specification of the fiduciary approximate measurement arrangement of the apparatus and resulting states, if we consider particular states for the measurement apparatus before the measurement $\xi$ and after the measurement $\{X_{\mu k}\}_{\mu k}$, where just like before the analogous of (3.4) is valid, and reminiscent states after the measurement in both the mensurated system $\psi$ and apparatus $\eta_{\mu k}$, these represent noise which doesn't allow for the perfect discernability on a measurement of a carefully prepared system in a state $\phi_{\mu k}$, which differently from (3.3) evolves after a Lüders-von Neumann measurement to:

$$U(t)(\phi_{\mu k} \otimes \xi) = \phi_{\mu k} \otimes X_{\mu k} + \psi \otimes \eta_{\mu k} \tag{3.19}$$

Requiring that, beyond this, the measuring arrangement has the following properties:

$$\langle X_{\mu k}, X_{\mu' k'} \rangle = 0, \quad \text{if } \mu \neq \mu'. \tag{3.20}$$

$$\langle X_{\mu k}, \eta_{\mu' k'} \rangle = 0, \quad \text{for any } \mu, \mu', k, k'. \tag{3.21}$$

$$\langle \eta_{\mu k}, \eta_{\mu' k'} \rangle = 0, \quad \text{for } (\mu, k) \neq (\mu', k'). \tag{3.22}$$

$$\|\psi \otimes \eta_{\mu k}\|^2 \leq \varepsilon \tag{3.23}$$

And such that the vectors given by $\Psi_\gamma^i = \phi_\gamma \otimes \xi$, $\Psi_\gamma^f = \phi_\gamma \otimes X_\gamma + \psi \otimes \eta_\gamma$ satisfy lemma 3.2.1 hypothesis (3.13) for the substitutions $\gamma \rightleftarrows \alpha = (\mu, k)$ and $\gamma \rightleftarrows \beta = (\mu', k')$.

Then we can see this arrangement as a fiduciary measurement experiment, such that for any system state $\phi \in \mathscr{H}_1$ that has norm 1, one can conceive the measurement of $\phi$ by making reference to this arrangement, since:

$$U(t)(\phi \otimes \xi) = U(t)\left(\sum_{\mu,k} \langle \phi_{\mu k}, \phi \rangle (\phi_{\mu k} \otimes \xi)\right) = \sum_{\mu,k} \langle \phi_{\mu k}, \phi \rangle (\phi_{\mu k} \otimes X_{\mu k}) + \sum_{\mu,k} \langle \phi_{\mu k}, \phi \rangle (\psi \otimes \eta_{\mu k})$$

where we see, from (3.23), that the inaccuracy from the noise is such that:

$$\left\| \sum_{\mu,k} \langle \phi_{\mu k}, \phi \rangle (\psi \otimes \eta_{\mu k}) \right\|^2 \leq \sum_{\mu,k} \left( |\langle \pi_{\mu k}, \phi \rangle|^2 \|\psi \otimes \eta_{\mu k}\|^2 \right) < \varepsilon$$

since by the orthonormality (3.2) of the $\phi_{\mu k}$ and the fact that $\phi$ is normalized, $1 = \|\phi\|^2 = \sum_{\mu,k} |\langle \phi_{\mu k}, \phi \rangle|^2 \|\phi_{\mu k}\|^2 = \sum_{\mu,k} |\langle \phi_{\mu k}, \phi \rangle|^2$.

To specify such an arrangement we construct the required $\xi$, $X_{\mu k}$, $\psi$ and $\eta_{\mu k}$ in the following way:

$\psi$ is taken to be a normalized eigenstate of $L_1$ with the eigenvalue 0,

$$L_1 \psi = 0, \quad \langle \psi, \psi \rangle = 1. \tag{3.24}$$

Considering now the vectors decomposed in the eigenbasis, where we denote by $\mathscr{H}_{2,\lambda}$ the subspace of $\mathscr{H}_2$ which is spanned by eigenvectors of $L_2$ with eigenvalue $\lambda$, these vector will then be written as:

$$\xi = \sum_\lambda \xi_\lambda, \quad X_{\mu k} = \sum_\lambda X_{\mu k \lambda}, \quad \eta_{\mu k} = \sum_\lambda \eta_{\mu k \lambda}. \tag{3.25}$$

We choose $\xi_\lambda \in \mathscr{H}_{2\lambda}$ arbitrarily such that the norm is determined by:

$$\langle \xi_\lambda, \xi_\lambda \rangle = \begin{cases} 0 & \text{, for } |\lambda| > N, \\ (2N+1)^{-1} & \text{, for } |\lambda| \leq N. \end{cases} \quad \begin{array}{r}(3.26\text{a})\\(3.26\text{b})\end{array}$$

which makes $\xi = \sum_\lambda \xi_\lambda$ normalized to 1, since eigenvectors with different eigenvalues of a self-adjoint operator are orthogonal, the number $N \in \mathbb{N}$ is a cutoff value for the constituent possible values that the prepared measuring apparatus state $\xi$ initially has, it is presciently chosen to be such that:

$$N > \frac{2l}{\varepsilon} - \frac{1}{2} \tag{3.27}$$

The $X_{\mu k\lambda} \in \mathscr{H}_{2,\lambda}$ are orthogonal to each other and with a norm, obeying this orthogonality, determined by:

$$\langle X_{\mu k\lambda}, X_{\mu' k'\lambda} \rangle = \begin{cases} 0 & \text{, for } |\lambda| > N - 2l, \\ \delta_{\mu\mu'}\delta_{kk'}(2N+1)^{-1} & \text{, for } |\lambda| \leq N - 2l. \end{cases} \quad \begin{array}{r}(3.28\text{a})\\(3.28\text{b})\end{array}$$

which makes $X_{\mu k} = \sum_\lambda X_{\mu k\lambda}$ normalized to $1 - \frac{4l}{2N+1}$. The complement of the set $\{X_{\mu k\lambda} \,|\, \mu, k \text{ varying }\}$ will be denoted by $\mathscr{H}_{2,\lambda}^\eta \equiv \mathscr{H}_{2,\lambda} \setminus \overline{\text{span}\{X_{\mu k\lambda} \,|\, \mu, k \text{ varying }\}}$.

The $\eta_{\mu k\lambda}$ are taken from $\mathscr{H}_{2,\lambda}^\eta$ and chosen such that:

$$\langle \eta_{\mu k\lambda}, \eta_{\mu' k'\lambda} \rangle = \begin{cases} 0 & \text{, for } |\lambda| > N+l \text{ or } |\lambda| \leq N - 3l, \\ \dfrac{1}{2N+1}\displaystyle\sum_{\substack{|\lambda'|\leq l \\ |\lambda-\lambda'|>N-2l}} \langle \phi_{\mu k}, P_1(\lambda')\phi_{\mu' k'}\rangle & \text{, for } N+l \geq |\lambda| > N-l, \\ \delta_{\mu\mu'}\delta_{kk'}\dfrac{1}{2N+1}\displaystyle\sum_{\substack{|\lambda'|\leq l \\ |\lambda-\lambda'|>N-2l}} \langle \phi_{\mu k}, P_1(\lambda')\phi_{\mu k}\rangle & \text{, for } N-l \geq |\lambda| > N-3l. \end{cases} \quad \begin{array}{r}(3.29\text{a})\\ \\(3.29\text{b})\\ \\(3.29\text{c})\end{array}$$

We now show that $\xi, X_{\mu k}, \psi$, and $\eta_{\mu k}$ thus constructed have the desired properties. As we have seen $\xi$ is normalized, the orthogonality condition 3.20 for $X_{\mu k}$ is held by the $\delta_{\mu\mu'}$ imposed in the construction of each $X_{\mu k\lambda}$. Whereas the orthogonality condition 3.21 between the $\{X_{\mu k}\}_{\mu k}$ and the $\{\eta_{\mu k}\}_{\mu k}$ is trivially satisfied by the choice of taking $\{X_{\mu k}\}_{\mu k} \subset \mathscr{H}_{2,\lambda}$ and $\{\eta_{\mu k}\}_{\mu k} \subset \left((\{X_{\mu k\lambda}\}_{\mu k})^\perp \cap \mathscr{H}_{2,\lambda}\right) \equiv \mathscr{H}_{2,\lambda}^\eta$.

We then tacitly decide to prove first that the property of inner product preservation (3.13) holds for $\Psi_{\mu k}^i = \phi_{\mu k} \otimes \xi$ and $\Psi_{\mu k}^f = \phi_{\mu k} \otimes X_{\mu k} + \psi \otimes \eta_{\mu k}$, to do that we first manipulate equation (3.13) using (3.12) to get:

$$\left\langle \phi_{\mu k} \otimes \xi, \sum_{|\lambda'|\leq l} P_1(\lambda')P_2(\lambda-\lambda')\phi_{\mu' k'} \otimes \xi \right\rangle = \sum_{|\lambda'|\leq l} \langle \phi_{\mu k}, P(\lambda')\phi_{\mu' k'}\rangle \langle \xi, P_2(\lambda-\lambda')\xi\rangle =$$

$$\overset{(3.13)}{=} \left\langle \phi_{\mu k} \otimes X_{\mu k} + \psi \otimes \eta_{\mu k}, \sum_{|\lambda'|\leq l} P_1(\lambda')P_2(\lambda-\lambda')\left(\phi_{\mu' k'} \otimes X_{\mu' k'} + \psi \otimes \eta_{\mu' k'}\right) \right\rangle =$$

$$= \sum_{|\lambda'|\leq l} \left(\langle \phi_{\mu k}, P_1(\lambda')\phi_{\mu' k'}\rangle\langle X_{\mu k}, P_2(\lambda-\lambda')X_{\mu' k'}\rangle + \langle \psi, P_1(\lambda')\psi\rangle\langle \eta_{\mu k}, P_2(\lambda-\lambda')\eta_{\mu' k'}\rangle\right),$$

because of (3.24) $\psi$ is a eigenvector of $L_1$ with eigenvalue 0 and therefore only for $\lambda' = 0$ is $P(\lambda')\psi \neq 0$ in fact $P(\lambda')\psi = \psi$, which implies:

$$\sum_{|\lambda'|\leq l} \langle\phi_{\mu k}, P_1(\lambda')\phi_{\mu'k'}\rangle\langle X_{\mu k}, P_2(\lambda-\lambda')X_{\mu'k'}\rangle + \cancel{\langle\psi,\psi\rangle_1}\langle\eta_{\mu k}, P_2(\lambda)\eta_{\mu'k'}\rangle.$$

Then:

$$\sum_{|\lambda'|\leq l}\langle\phi_{\mu k},P(\lambda')\phi_{\mu'k'}\rangle\langle\xi,P_2(\lambda-\lambda')\xi\rangle = \sum_{|\lambda'|\leq l}\langle\phi_{\mu k},P_1(\lambda')\phi_{\mu'k'}\rangle\langle X_{\mu k},P_2(\lambda-\lambda')X_{\mu'k'}\rangle + \langle\eta_{\mu k},P_2(\lambda)\eta_{\mu'k'}\rangle. \quad (3.30)$$

Considering then the eigendecompositions (3.25) of these vectors we see that $P(\lambda-\lambda')\xi = \xi_{\lambda-\lambda'}$, $P(\lambda-\lambda')X_{\mu k} = X_{\mu k\,(\lambda-\lambda')}$ and $P(\lambda)\eta_{\mu k} = \eta_{\mu k\,\lambda}$, so that:

$$\stackrel{(3.30)}{\Longrightarrow} \sum_{|\lambda'|\leq l} \langle\phi_{\mu k}, P(\lambda')\phi_{\mu'k'}\rangle \left(\langle\xi, P_2(\lambda-\lambda')P_2(\lambda-\lambda')\xi\rangle - \langle X_{\mu k}, P_2(\lambda-\lambda')P_2(\lambda-\lambda')X_{\mu'k'}\rangle\right) =$$

$$= \langle\eta_{\mu k}, P_2(\lambda)P_2(\lambda)\eta_{\mu'k'}\rangle$$

$$\Longrightarrow \sum_{|\lambda'|\leq l} \langle\phi_{\mu k}, P(\lambda')\phi_{\mu k}\rangle \left(\|\xi_{\lambda-\lambda'}\|^2 - \langle X_{\mu k\,(\lambda-\lambda')}, X_{\mu'k'\,(\lambda-\lambda')}\rangle\right) = \langle\eta_{\mu k}, \eta_{\mu'k'}\rangle \quad (3.31)$$

Having manipulated (3.13) into (3.31) we can now go on to prove (3.31) instead of (3.13), to do this we will divide the range of $\lambda$ into 4 parts and prove the equation separately for $\lambda$ in each of these 4 regions.

1) If $|\lambda| > N + l$, then $|\lambda - \lambda'| > N$ and (3.31) is trivially satisfied because all terms vanish.

2) If $N + l \geq |\lambda| > N - l$, then $|\lambda - \lambda'| > N - 2l$, and hence the term containing $X$ still vanishes. Due to (3.26b), the left-hand side of (3.31) becomes:

$$\sum_{|\lambda'|\leq l} \langle\phi_{\mu k}, P(\lambda')\phi_{\mu k}\rangle\|\xi_{\lambda-\lambda'}\|^2 = \frac{1}{2N+1} \sum_{\substack{|\lambda'|\leq l \\ |\lambda-\lambda'|\leq N}} \langle\phi_{\mu k}, P(\lambda')\phi_{\mu k}\rangle,$$

which is exactly the right hand-side of (3.31) when (3.29b) is used to calculate the inner product with the $\eta$'s.

3) If $N - l \geq |\lambda| > N - 3l$, then $|\lambda - \lambda'| \leq N$ and hence by (3.26b) the $\|\xi_{\lambda-\lambda'}\|^2$ is always $(2n+1)^{-1}$. By the orthogonality (3.2), the definition (3.28b) and the equation:

$$\sum_{|\lambda'|\leq l} P_1(\lambda') = \mathbb{1}, \quad (3.32)$$

the left-hand side of (3.31) becomes

$$\sum_{|\lambda'|\leq l} \langle\phi_{\mu k}, P(\lambda')\phi_{\mu k}\rangle\left((2N+1)^{-1} - \langle X_{\mu k\,(\lambda-\lambda')}, X_{\mu'k'\,(\lambda-\lambda')}\rangle\right) =$$

$$(2N+1)^{-1}\delta_{\mu\mu'}\delta_{kk'} \sum_{|\lambda'|\leq l} \langle\phi_{\mu k}, P(\lambda')\phi_{\mu k}\rangle\left[1 - (2N+1)\langle X_{\mu k\,(\lambda-\lambda')}, X_{\mu'k'\,(\lambda-\lambda')}\rangle\right]. \quad (3.33)$$

Because of (3.28b), the inside of the square bracket of (3.33) vanishes for $|\lambda - \lambda'| \leq N - 2l$ and, by (3.28a) is unity for $|\lambda - \lambda'| > N - 2l$. Thus, due to (3.29c) when the square bracket in (3.33) vanishes so does $\langle \eta_{\mu k \lambda}, \eta_{\mu' k' \lambda} \rangle$ since the sum which defines this inner product has no terms for $|\lambda - \lambda'| \leq N - 2l$, and when the square bracket goes to 1 the sum has to be done for $|\lambda - \lambda'| > N - 2l$ and therefore by (3.29c) both will also be equal, this means that for all $N - l \geq |\lambda| > N - 3l$ we have that (3.33) is equal to the right-hand side of (3.31).

4) If $N - 3l \geq |\lambda|$, then $|\lambda - \lambda'| \leq N - 2l$ and the left-hand side of (3.31) becomes by (3.28b):

$$(2N+1)^{-1} \sum_{|\lambda'| \leq l} \langle \phi_{\mu k}, P_1(\lambda') \phi_{\mu' k'} \rangle (1 - \delta_{\mu\mu'} \delta_{kk'})$$

Because of (3.32) and the orthogonality (3.2), this expression vanishes and hence is equal to the right-hand side of (3.31) which also vanishes due to (3.29a). This completes the proof of (3.13), which by lemma 3.2.1 implies (3.14).

Finally, to prove (3.21) and (3.22) we consider that, since $\psi$ is normalized then $\|\psi \otimes \eta_{\mu k}\| = \sqrt{\langle \psi \otimes \eta_{\mu k}, \psi \otimes \eta_{\mu k} \rangle} = \sqrt{\langle \psi, \psi \rangle_1 \langle \eta_{\mu k}, \eta_{\mu k} \rangle} = \|\eta_{\mu k}\|$. By (3.25) we have:

$$\langle \eta_{\mu k}, \eta_{\mu' k'} \rangle = \sum_{\lambda} \langle \eta_{\mu k \lambda}, \eta_{\mu' k' \lambda} \rangle.$$

By (3.31), (3.28a) and (3.28b), we get:

$$\sum_{\lambda} \langle \eta_{\mu k \lambda}, \eta_{\mu' k' \lambda} \rangle = \sum_{\substack{k, k' \\ |\lambda'| \leq l}} \langle \phi_{\mu k}, P_1(\lambda') \phi_{\mu' k'} [\|\xi_{\lambda-\lambda'}\|^2 - \delta_{\mu\mu'} \delta_{kk'} \|X_{\mu k (\lambda-\lambda')}\|^2] \rangle =$$

$$= \sum_{\substack{k, k' \\ |\lambda'| \leq l}} \langle \phi_{\mu k}, P_1(\lambda') \phi_{\mu' k'} [\|\xi_{\lambda}\|^2 - \delta_{\mu\mu'} \delta_{kk'} \|X_{\mu k (\lambda)}\|^2] \rangle.$$

By considering this re-indexation and the completeness (3.32) together with the orthogonality (3.2):

$$\sum_{|\lambda'| \leq l} \langle \phi_{\mu k}, P_1(\lambda') \phi_{\mu' k'} \rangle = \delta_{\mu\mu'} \delta kk'. \tag{3.34}$$

By The already discussed norms of $\xi$ and $X_{\mu k}$ we get:

$$\sum_{\lambda} [\|\xi_{\lambda}\|^2 - \|X_{\mu k \lambda}\|^2] = \frac{4l}{2N+1},$$

combining this with (3.27) we obtain:

$$\|\psi \otimes \eta_{\mu k}\|^2 = \frac{4l}{2N+1} < \varepsilon$$

and because of (3.34) property (3.21) is also proved:

$$\langle \eta_{\mu k}, \eta_{\mu' k'} \rangle = 0, \quad \text{for } (\mu, k) \neq (\mu', k').$$

In the above fiduciary arrangement we must consider that for values of $\lambda$ such that $|\lambda| \leq N - 3l$, to be able to differentiate the states of the system, $\mathscr{H}_{2,\lambda}$ should at least have the dimension of $\mathscr{H}_1$. Which also implies that $\mathscr{H}_{2,\lambda}$ must have a higher dimension than $\mathscr{H}_1$ for $N - 3l < |\lambda| \leq N$.

We now follow Ozawa's expositon in [82] to consider the uncertainty relations present in this exposition.

Since $L_1$ and $L_2$ are supposed to be statistically independent, the variance of their sum is the sum of their variances, *i.e.*:

$$\text{Var}(L_1 \otimes \mathbb{1} + L_2 \otimes \mathbb{1}; \psi \otimes \xi) = \text{Var}(L_1; \psi \otimes \xi) + \text{Var}(L_2; \psi \otimes \xi), \tag{3.35}$$

If we consider $A : \mathcal{H}_2 \to \mathcal{H}_2$ to be a operator that measures the error vector $\eta_{\mu k} \in \mathcal{H}_2$, since $M$ and $L_1$ are operating on the system and $A, L_2$ are acting on the apparatus system, then:

$$[U^*(t)(M \otimes \mathbb{1})U(t), U^*(t)(\mathbb{1} \otimes L_2)U(t)] = [L_1 \otimes \mathbb{1}, \mathbb{1} \otimes A] = 0 \tag{3.36}$$

If we define the so-called noise operator $N : \mathcal{H}_1 \otimes \mathcal{H}_2 \to \mathcal{H}_1 \otimes \mathcal{H}_2$, by:

$$N := U^*(t)(M \otimes \mathbb{1})U(t) - \mathbb{1} \otimes A,$$

then, by equation (3.36):

$$[N, L] = [U^*(t)(M \otimes \mathbb{1})U(t) - \mathbb{1} \otimes A, L] =$$

$$= [U^*(t)(M \otimes \mathbb{1})U(t) - \mathbb{1} \otimes A, L] = [U^*(t)(M \otimes \mathbb{1})U(t), U^*(t)(L_1 \otimes \mathbb{1})U(t)] - [\mathbb{1} \otimes A, \mathbb{1} \otimes L_2]. \tag{3.37}$$

where it was used that as $[U(t), L] = 0 \Rightarrow L = U^*(t)LU(t)$ and equation (3.36).

Considering then, that the noise $\epsilon(\psi)$ for measuring $A$ on input state $\psi$, is defined by:

$$\epsilon(\psi)^2 = \langle \psi \otimes \xi, N^2 \psi \otimes \xi \rangle,$$

then by the formula for variance $\text{Var}(X; \psi) := \langle \psi, X^2 \psi \rangle - \langle \psi, X\psi \rangle^2$, we have:

$$\epsilon(\psi)^2 \geq \text{Var}(N; \psi \otimes \xi) \tag{3.38}$$

Then using the Robertson-Schrödinger uncertainty relation:

$$\text{Var}(N; \psi \otimes \xi)\text{Var}(L_1 \otimes \mathbb{1} + L_2 \otimes \mathbb{1}; \psi \otimes \xi) \geq \frac{1}{4}|\langle \psi \otimes \xi, [N, L_1 \otimes \mathbb{1} + L_2 \otimes \mathbb{1}]\psi \otimes \xi \rangle|^2 \tag{3.39}$$

Hence we can combine relations (3.38) and (3.39) to get:

$$\epsilon(\psi)^2 \geq \frac{1}{4}\frac{|\langle \psi \otimes \xi, [N, L_1 \otimes \mathbb{1} + L_2 \otimes \mathbb{1}]\psi \otimes \xi \rangle|^2}{\text{Var}(L_1 \otimes \mathbb{1} + L_2 \otimes \mathbb{1}; \psi \otimes \xi)} \tag{3.40}$$

further combine (3.35) with (3.40):

$$\epsilon(\psi)^2 \geq \frac{|\langle \psi \otimes \xi, [N, L_1 \otimes \mathbb{1} + L_2 \otimes \mathbb{1}]\psi \otimes \xi \rangle|^2}{4\text{Var}(L_1; \psi \otimes \xi) + 4\text{Var}(L_2; \psi \otimes \xi)} \tag{3.41}$$

and finally combining (3.41) with (3.37), to arrive at:

$$\epsilon(\psi)^2 \geq \frac{|\langle \psi \otimes \xi, ([U^*(t)(M \otimes \mathbb{1})U(t), U^*(t)(L_1 \otimes \mathbb{1})U(t)] - [\mathbb{1} \otimes A, \mathbb{1} \otimes L_2])\psi \otimes \xi \rangle|^2}{4\text{Var}(L_1; \psi \otimes \xi) + 4\text{Var}(L_2; \psi \otimes \xi)}.$$

Since we can, in principle prepare the apparatus to be in an arbitrary state $\xi$, before realizing the measurement, we can choose $\xi$ is such a way as to cancel the term $\langle \psi \otimes \xi, [\mathbb{1} \otimes A, \mathbb{1} \otimes L_2]\psi \otimes \xi \rangle$, hence we get the following noise lower bound estimate:

$$\epsilon(\psi)^2 \geq \frac{|\langle \psi \otimes \xi, \, [U^*(t)(M \otimes \mathbb{1})U(t), U^*(t)(L_1 \otimes \mathbb{1})U(t)]\psi \otimes \xi \rangle|^2}{4\mathrm{Var}(L_1; \psi \otimes \xi) + 4\mathrm{Var}(L_2; \psi \otimes \xi)}$$

This indicates that the only way to try to decrease the error in the measurement, here understood to be directly associated with the noise $\epsilon(\psi)$, is to chose a apparatus whose state $\xi$ allows for a large variance of the conserved quantity in the apparatus, namely a large value of $\mathrm{Var}(L_2; \psi \otimes \xi)$, this is a good indication that for more precise measurements, apparatuses that are extensive, such as macroscopic ones, since the additive property of the conserved quantity would allow for a greater range of values in these systems, and so one could also expect a larger variance of the conserved quantity in the apparatus. Although we did not get a upper bound for the error, so as to exactly determine the conditions that would certainly diminish the error in measurements, we did arrive at a necessary condition for the lower bound in the error to be made small, which although doesn't guarantee a smaller error, gives a required condition for such a outcome.

We also note that the WAY theorem 3.1.1, only applies to observables with discrete value spaces, and does not apply to the measurement of observables with continuous value spaces, such as position, for which repeatable measurements do not exist, as can be ascertained in theorem 10.4 of [20], which raises the question of whether conservation laws constrain measurements in such cases.

This question of compatibility suggests a new element relevant to the question of measurement, the systematic study of the compatibility properties of observables is carried out by the concepts of contextuality that emerged from [67] and which were later shown to be consequences of Gleason's theorem [48], as will be seen in chapter 6.

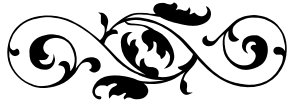

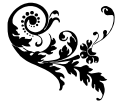 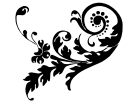

# CHAPTER 4

# Algebraic Quantum Field Theory

*In this chapter we establish several results that will form the groundwork for Algebraic Quantum Field Theory, we then postulate the fundamental general principles of the formalism through a modified version of the Haag-Kastler axioms, in a more general, using $*$-algebras instead of full on $C^*$-algebras as it is usually done, this gives more flexibility in the description, containing the usual Haag-Kastler operator-algebra description as a special case, this is the general idea behind what is called Local Quantum Field Theory [17, 40] and allows one to treat QFT's in curved spacetimes in the AQFT formalism.*

*After this simplified pedagogical introduction, we go on to prove some more useful facts that will receive applications latter, after which we make a review of the measurement scheme proposed by Fewster and Verch, that is applicable for QFT and for QFT in curved spacetimes, as it is locally covariant.*

**Definition 4.0.1** (Unital $*$-algebra)**.** *A $*$-algebra is a structure $(\mathscr{A}, \mathbb{K}, \cdot, +, \bullet, *)$ consisting of a set $\mathscr{A}$, a field $\mathbb{K} = \mathbb{R}$ or $\mathbb{C}$, three binary operations, a scalar product $\cdot : \mathbb{R} \times \mathscr{A} \to \mathscr{A}$, a vector sum $+ : \mathscr{A} \times \mathscr{A} \to \mathscr{A}$ and a associative algebra product $\bullet : \mathscr{A} \times \mathscr{A} \to \mathscr{A}$; the last element is an involution $* : \mathscr{A} \to \mathscr{A}$, these are such that $(\mathscr{A}, \mathbb{K}, \cdot, +, \bullet)$ is a associative algebra over the field $\mathbb{K}$, the $*$-algebra is unital if $\mathscr{A}$ has a unit element $\mathbb{1}_{\mathscr{A}}$ for the associative algebra, and the involution has the following properties, for $x, y \in \mathscr{A}$, $\alpha \in \mathbb{K}$:*

- $(x^*)^* = x$;
- $(\alpha \cdot x + y)^* = \overline{\alpha} \cdot x^* + y^*$;
- $(x \bullet y)^* = y^* \bullet x^*$;
- *for $\mathbb{1}_{\mathscr{A}} \in \mathscr{A}$, $\mathbb{1}_{\mathscr{A}}^* = \mathbb{1}_{\mathscr{A}}$;*

**Definition 4.0.2** (Positive elements)**.** *For a $*$-algebra $(\mathscr{A}, \mathbb{K}, \cdot, +, \bullet, *)$, an element $a \in \mathscr{A}$ is called **positive** if there are finitely many elements $\{a_k\}_{k=1}^n \subset \mathscr{A}$, such that $a = \sum_{k=1}^n a_k^* a_k$, the positivity of $a$ is denoted by $a \geq 0$.[1] The set of positive elements is denoted by $\mathscr{A}_+$.*

*A linear functional $f : \mathscr{A} \to \mathbb{C}$ is also called positive if it is true that $f(a) \geq 0$, $\forall a \in \mathscr{A}_+$.*

A $C^*$-algebra is then defined from a $*$-algebra together with a norm $\|\cdot\|$ that makes it into a Banach algebra (only this would make a Banach $*$-algebra) with a regularity property between the involution and the norm called the $C^*$ condition.

**Definition 4.0.3** ($C^*$-algebra)**.** *A (abstract) $C^*$-algebra is a structure $(\mathscr{A}, \mathbb{K}, \cdot, +, \bullet, *, \|\cdot\|)$, such that $(\mathscr{A}, \mathbb{K}, \cdot, +, \bullet, *)$ is a $*$-algebra, $(\mathscr{A}, \mathbb{K}, \cdot, +, \|\cdot\|)$ is a Banach space over the field $\mathbb{K}$, and the following relations between the norm and the product are valid, $\forall A, B \in \mathscr{A}$:*

---

[1] This notation is justified because this positive property defines a partial order on the set of self-adjoint elements of $\mathscr{A}$, by considering that if $a - b \in \mathscr{A}_+$, for $a, b \in \mathscr{A}_\dagger$ then $a \geq b$.

- *Submultiplicative property:* $\|A \cdot B\| \leq \|A\|\|B\|$;

- **C*condition:** $\|A^* \cdot A\| = \|A^*\|\|A\|$;

*We shall now denote the associative algebra product, as usual, by concatenation $A \cdot B \equiv AB$.*

*A concrete $C^*$-algebra is the name given for a norm closed $*$-invariant sub-algebra of $\mathfrak{B}(\mathscr{H})$, where $\mathscr{H}$ is a Hilbert space.*

As it turns out the $C^*$ condition is equivalent to the $B^*$ condition $\|A^*A\| = \|A\|^2$, that the $B^*$ condition implies the $C^*$ condition is clear since $\|A^*\|\|A\| \overset{submultiplicative}{\geq} \|A^*A\| = \|A\|^2$ and making the substitution $A \rightleftarrows A^*$ we get $\|A\|\|A^*\| \overset{submultiplicative}{\geq} \|AA^*\| = \|A^*\|^2$, follows then that $\|A\| = \|A^*\|$, and therefore $\|A^*A\| = \|A^*\|\|A\|$, a sketch of the proof of the converse implication is found in Appendix A.

**Definition 4.0.4** (von Neuman algebra). *A (abstract) von Neumann algebra, also called a $W^*$-algebra, $\mathfrak{M}$ is a $C^*$-algebra which has a **pre-dual** $\mathfrak{M}_*$, i.e., $\mathfrak{M}_*$ is a Banach space such that the vector space of all linear functionals to the field of complex numbers, also know as its algebraic dual space and represented with $(\mathfrak{M}_*)^*$, is such that $(\mathfrak{M}_*)^* = \mathfrak{M}$ as a Banach space.*

**Definition 4.0.5** (Commutant). *Let $\mathscr{A}$ be a concrete $C^*$-algebra. We define the commutant of $\mathscr{A}$ to be the set:*
$$\mathscr{A}' = \{B \in \mathfrak{B}(\mathscr{H}) \,|\, AB = BA, \, \forall A \in \mathscr{A}\}$$

Given a $C^*$-algebra we can also speak about the space of linear functionals of such $C^*$-algebra:

**Definition 4.0.6.** *Let $A: Dom(A) \subseteq \mathscr{H} \to \mathscr{H}$ be a linear operator with a dense domain $Dom(A)$, the **adjoint operator** $A^*: Dom(A^*) \subseteq \mathscr{H} \to \mathscr{H}$ is defined as acting on the subspace $Dom(A^*)$ consisting of those elements $y$ for which there is a $z \in \mathscr{H}$ such that $\langle y, Ax \rangle = \langle z, x \rangle, \forall x \in Dom(A)$, with $A^*y := z$, owing to the density of $Dom(A)$ and the Riesz representation theorem B.0.13 $z$ is uniquely defined.*

*A operator $A$ is called **self-adjoint** if $Dom(A) = Dom(A^*)$ and $A = A^*$.*

**Definition 4.0.7.** *For any set $S \subseteq \mathscr{A}$ of a $C^*$-algebra $\mathscr{A}$ we define the self-adjoint part of the set as $S_\dagger := \{A \in S \,|\, A = A^*\}$.*

**Corollary 4.0.8.** *If $B$ is a sub-$C^*$-algebra of a $C^*$-algebra $\mathscr{A}$, then $B_\dagger$ is convex.*

*Proof:* Since $B$ is a $C^*$-algebra it is closed by linear combinations of elements of $B$, since the involution $*$ is also anti-linear then for $\forall t \in [0,1]$ and any $r, s \in B$ then $B_\dagger \ni (tr + (1-t)s)^* = tr^* + (1-t)s^*$, therefore $B_\dagger$ is convex. ∎

**Definition 4.0.9.** [2] *Let $\mathscr{H}$ be a Hilbert space, $\mathfrak{B}(\mathscr{H})$ the set of all bounded operators on $\mathscr{H}$. Considering that $\mathfrak{B}(\mathscr{H})$ is imbued with the usual addition, scalar multiplication, product and adjoint operations, then the following families of seminorms generate the topological bases that define five vector space topologies on $\mathfrak{B}(\mathscr{H})$:*

---

[2] This exposition of the five families of seminorms that are used to define the five usual operator topologies is heavily inspired by the excellent exposition in [26].

(i) $\varrho_{\text{op}}(A) \equiv \|A\|_{\text{op}} \equiv \|A\| := \sup_{x} \{\|Ax\|_{\mathcal{H}} \,|\, x \in \mathcal{H},\, \|x\|_{\mathcal{H}} \leq 1\}$    *(operator or uniform topology).*

(ii) $\varrho_x^s(A) := \|Ax\|_{\mathcal{H}},\, x \in \mathcal{H}$    *(strong operator topology).*

(iii) $\varrho_{x,y}^w(A) := |\langle x, Ay \rangle|,\, x,y \in \mathcal{H}$    *(weak operator topology).*

(iv) $\varrho_{(x_n)_{n \in \mathbb{N}}}^{\sigma\text{-s}}(A) := \left(\sum_{n=1}^{\infty} \|Ax_n\|^2\right)^{\frac{1}{2}},\, (x_n)_{n \in \mathbb{N}} \in \ell_2(\mathcal{H})$    *($\sigma$-strong operator topology).*

(v) $\varrho_{(x_n,y_n)_{n \in \mathbb{N}}}^{\sigma\text{-w}}(A) := \left(\sum_{n=1}^{\infty} |\langle x_n, Ay_n \rangle|\right)^{\frac{1}{2}},\, \{x_n, y_n\}_{n \in \mathbb{N}} \subseteq \ell_2(\mathcal{H})$    *($\sigma$-weak operator topology).*

*We shall use the acronyms* WOT *for weak operator topology and* SOT *for strong operator topology, in the literature one might find the following notations for taking the closure with respect to each of the previous topologies $\overline{(\,\cdot\,)}^{\|\cdot\|}$, $\overline{(\,\cdot\,)}^{\text{WOT}}$, $\overline{(\,\cdot\,)}^{\text{SOT}}$, ...; we instead opt for the following notation $\overline{(\,\cdot\,)}^{\text{op}}, \overline{(\,\cdot\,)}^{w}, \overline{(\,\cdot\,)}^{s}$, ...*

We can now define the concrete version of a von Neumann algebra:

**Definition 4.0.10.** *A concrete von Neumann algebra is a weak-operator closed $*$-invariant subspace of $\mathfrak{B}(\mathcal{H})$, where $\mathcal{H}$ is a Hilbert space.*

We can then enunciate the fundamental theorem of von Neumann algebras, which is so classic we shall not give a proof of this fact, just point the reader to the proofs in the following books: [13, 64, 8].

**Theorem 4.0.11** (Bicommutant)**.** *Let $\mathfrak{M}$ be a unital $*$-invariant sub-algebra of $\mathfrak{B}(\mathcal{H})$, then:*

$$\mathfrak{M}'' = \overline{\mathfrak{M}}^{w} = \overline{\mathfrak{M}}^{s} = \overline{\mathfrak{M}}^{\sigma\text{-w}} = \overline{\mathfrak{M}}^{\sigma\text{-s}}.$$

**Definition 4.0.12.** *Let $\mathcal{V}$ be a normed vector space with norm $\|\cdot\|_{\mathcal{V}}$, then we define the open ball of radius $r \in \mathbb{R}_+$ centered at $c \in \mathcal{V}$ by $\mathcal{B}_r^c(\mathcal{V}) := \{v \in \mathcal{V} \,|\, \|v - c\| < r\}$. In the case of a ball centered at $0 \in \mathcal{V}$ we denote $\mathcal{B}_r^0(\mathcal{V}) \equiv \mathcal{B}_r(\mathcal{V})$.*

**Lemma 4.0.13.** *considering a net of operators $\{T_\alpha\}_\alpha \subseteq \mathfrak{B}(\mathcal{H})$ and a operator $T \in \mathfrak{B}(\mathcal{H})$, then if the sequence $\|T_\alpha - T\| \to 0$ this implies $\forall x \in \mathcal{H}, \|T_\alpha x - Tx\|_{\mathcal{H}} \to 0$ which consequentially implies $\forall x, y \in \mathcal{H}, |\langle x, (T_\alpha - T)y \rangle| \to 0$.*

*These assertions are equivalent to saying that the uniform topology WOT $\subseteq$ SOT $\subseteq$ uniform topology.*

*Proof:* This follows trivially from the inequalities, where we consider $x, y \in \mathcal{H}$ such that $\|x\|_{\mathcal{H}} \leq 1$, $\|y\|_{\mathcal{H}} \leq 1$, but that are otherwise arbitrary:

$$\|A\| = \sup_u \{\|Au\|_{\mathcal{H}} \,|\, u \in \mathcal{H},\, \|u\|_{\mathcal{H}} \leq 1\} \geq \|Ax\|_{\mathcal{H}} \geq$$

$$\geq \|y\|_{\mathcal{H}} \|Ax\|_{\mathcal{H}} \geq |\langle y, Ax \rangle|$$

where the inequality in the last line is Cauchy–Schwarz. ∎

**Lemma 4.0.14.** *The maps $* : \mathfrak{B}(\mathcal{H}) \to \mathfrak{B}(\mathcal{H})$ and $\mathfrak{Re} : \mathfrak{B}(\mathcal{H}) \to \mathfrak{B}(\mathcal{H})$ are WOT-continuous.*

*Proof:* Suppose that $\{T_i\}_{i \in I} \subset \mathfrak{B}(\mathscr{H})$ is a net of operators WOT-converging to $T \in \mathfrak{B}(\mathscr{H})$, so that $\forall u, v \in \mathscr{H}, \langle u, T_i v \rangle \to \langle u, T v \rangle$ then considering the adjoint of the operator and taking the complex conjugation of the reflected inner product we get $\langle v, T_i^* u \rangle \to \langle v, T^* u \rangle$ and since this in valid for any $u, v \in \mathscr{H}$ this implies $T_i^* \xrightarrow{w} T^*$. From this it immediately follows that

$$\mathfrak{Re}(T_i) = \frac{T_i + T_i^*}{2} \xrightarrow{w} \frac{T + T^*}{2} = \mathfrak{Re}(T)$$

Since in the WOT we can use that the inner product is sesquilinear and the absolute value obeys the triangle inequality, therefore linear combinations of weakly converging operators also converge weakly, which makes the WOT a linear topology. ∎

**Corollary 4.0.15.** *If $K \subset X$ is convex with $X$ locally convex, then it is weakly closed if and only if is it strongly closed.*

*Proof:* Since by the last Lemma 4.0.13 WOT $\subseteq$ SOT, that is if $O \in$ WOT $\Rightarrow O \in$ SOT, this implies $C \notin$ SOT $\Rightarrow C \notin$ WOT; then $\overline{K}^s \subseteq \overline{K}^w$. It remains to be proven that for a convex $K$ that $\overline{K}^w \subseteq \overline{K}^s$. For that we can prove that $\overline{K}^w \setminus \overline{K}^s = \varnothing$.

Let $K = \overline{K}^s$ be a strongly closed set, $K \subseteq \overline{K}^w$ as is always true. If $x \in X \setminus K$ then by the geometric form of the Hahn-Banach Theorem B.0.24, $K$ that is convex and strongly closed and $\{x\}$ that is convex and strongly compact are such that the can be separated by a continuous linear functional $\mathfrak{Re}(f)$ and a number $\alpha \in \mathbb{R}$ such that:

$$\forall y \in K, \quad \mathfrak{Re}(f)(y) \leq \alpha \leq \mathfrak{Re}(f)(x)$$

Thus the set $\mathfrak{Re}(f)^{-1}((\alpha, +\infty))$, which is weakly open since the continuity of $f$ composed with $\mathfrak{Re}$ which is WOT-continuous by Lemma 4.0.14, therefore $\mathfrak{Re}(f)$ is WOT-continuous, that makes $\mathfrak{Re}(f)^{-1}((\alpha, +\infty))$ be a weakly open set that contains $x$ and does not intersect $K$, by definition the closure of a set is the set of its adherent points, that is, those points for which every open neighbourhood intersect the set whose closure is being taken, then $x$ is not in the weak closure of $K$. Since $x$ was a arbitrary point outside $K$ then the weak closure cannot add any new points to $K$ and therefore $\overline{K}^w \setminus \overline{K}^s = \varnothing \implies \overline{K}^w = \overline{K}^s$. ∎

**Theorem 4.0.16** (The Gel'fand homomorphism in $C^*$-algebras)**.** *Being $\mathscr{A}$ a $C^*$-algebra with unit, and letting $A \in \mathscr{A}_\dagger$ and $\Phi_A : C(\sigma(A)) \to \mathscr{A}$ defined as the extension by the Bounded Linear Transformation theorem B.0.15 of the map $\phi_A \equiv \phi : P(\sigma(A)) \to \mathscr{A}$ from the set $P(\sigma(A))$ of the complex polynomials defined in the spectrum of A, given by $\phi_A(p) = p(A)$ for $p \in P(\sigma(A))$ and by Theorem B.0.44 we have that $\|\phi_A(p)\| = \|p\|_\infty$. This is so because, by the Stone-Weierstrass theorem B.0.22, the set $P(\sigma(A))$ that trivially separates points by the degree $1$ monomial, is composed only of self-adjoint elements since $A \in \mathscr{A}_\dagger$, and by theorem B.0.41 the spectrum $\sigma(A)$ is compact, implies that $P(\sigma(A))$ is dense in $C(\sigma(A)) \subset C(\mathbb{R})$, real by theorem B.0.48, the Bounded Linear Transformation theorem also implies that $\|\Phi_A(f)\| = \|f\|_\infty$ for all $f \in C(\sigma(A))$. For this map $\Phi_A$ the following assertions are valid:*

*1) - The map $\Phi_A$ is a algebraic $*$-homomorphism:*

$$\Phi_A(\alpha f + \beta g) = \alpha \Phi_A(f) + \beta \Phi_A(g), \quad \Phi_A(fg) = \Phi_A(f)\Phi_A(g), \quad \Phi_A(f)^* = \Phi_A(\overline{f}), \quad \Phi_A(1) = \mathbb{1};$$
(4.1)

*for all $f, g \in C(\sigma(A))$ and $\forall \alpha, \beta \in \mathbb{C}$. Since $fg = gf$ it follows from the second equation that $\Phi_A(f)\Phi_A(g) = \Phi_A(g)\Phi_A(f), \forall f, g \in C(\sigma(A))$.*

2) - *If $F \geq 0$ then $\sigma(\Phi_A(f)) \subseteq [0, +\infty)$, for $\mathscr{A} = \mathfrak{B}(\mathscr{H})$, we have that in the operator ordering $\Phi_A(f) \geq 0$.*

3) - *If $(f_n)_{n \in \mathbb{N}} \subset C(\sigma(A))$ converges in the norm $\|\cdot\|_\infty$ to a function $f \in C(\sigma(A))$ then $\Phi_A(f_n)$ converges to $\Phi_A(f)$ in the norm of $\mathscr{A}$. Reciprocally, if $(\Phi_A(f_n))_{n \in \mathbb{N}}$ converges in the norm of $\mathscr{A}$, then there exists $f \in C(\sigma(A))$ such that $\lim_{n \to \infty} \Phi_A(f_n) = \Phi_A(f)$. This tells us that the set $\{\Phi_A(g) \mid g \in C(\sigma(A))\}$ is closed in the norm of $\mathscr{A}$. This fact in conjunction with assertion 1 of this list makes $\{\Phi_A(g) \mid g \in C(\sigma(A))\}$ a abelian sub-$C^*$-algebra of $\mathscr{A}$.*

4) - *$\sigma(\Phi_A(f)) = \{f(\lambda) \mid \lambda \in \sigma(A)\} =: f(\sigma(A)), \forall f \in C(\sigma(A))$. Considering $\mathscr{A} = \mathfrak{B}(\mathscr{H})$, for some Hilbert space $\mathscr{H}$, then if $\psi \in \mathscr{H}$ is a eigenvector of $A \in \mathfrak{B}(\mathscr{H})_\dagger$ with eigenvalue $\lambda_0$, then $\Phi_A(f)\psi = f(\lambda_0)\psi, \forall f \in C(\sigma(A))$.*

*The $*$-homomorphism $\Phi_A : C(\sigma(A)) \to \mathscr{A}$ is denominated as Gel'fand homomorphism. When the $C^*$-algebra is explicitly given by $\mathfrak{B}(\mathscr{H})$, the bounded operators in a Hilbert space $\mathscr{H}$, the Gel'fand homomorphism is denoted by the following abuse of notation $\Phi_A(f) \equiv f(A), \forall f \in C(\sigma(A))$ and is called the continuous functional calculus.*

*Proof:* Lets prove those assertions item by item:

1) - The map $\Phi_A : C(\sigma(A)) \to \mathscr{A}$ is bounded, since, as we mention in the statement of the theorem, $\sigma(A)$ is compact which makes the functions $f \in C(\sigma(A))$ bounded, and by the equality $\|\Phi_A(f)\| = \|f\|_\infty$ the boundedness of $\Phi_A$ is verified. The $*$-homomorphism properties in (4.1) are trivially observed in $P(\sigma(A))$ and are extended by continuity to $C(\sigma(A))$.

2) - If $f \geq 0$ then we can construct $g \in C(\sigma(A))$ such that $f = g^2$, therefore by the $*$-homomorphism properties 4.1 proved in the last item, $\Phi_A(f) = \Phi_A(g^2) = \Phi_A(g)^2$, since $g$ is a real and continuous function then by the third equality in (4.1) $\Phi_A(g)^2$ is self-adjoint and by theorem B.0.42 since $\Phi_A(g)^2$ is a square, and therefore a polynomial in $\mathscr{A}$, we are then mapping the real spectrum of $\Phi_A(g)$, again by theorem B.0.48, to its square which makes $\sigma(\Phi_A(f)) = \sigma(\Phi_A(g)^2) \subseteq [0, +\infty)$. For $\mathscr{A} = \mathfrak{B}(\mathscr{H})$ we make the same argument and arrive at $\Phi_A(g)^2 = \Phi_A(g)\Phi_A(g) = \Phi_A(g)^*\Phi_A(g)$ which is a positive operator.

3) - Since $\|\Phi_A(f) - \Phi_A(f_n)\| = \|\Phi_A(f - f_n)\| = \|f - f_n\|_\infty$ then if $\|f - f_n\|_\infty \to 0$ follows that $\|\Phi_A(f) - \Phi_A(f_n)\| \to 0$. Reciprocally, if $\Phi_A(f_n)$ converges in the norm of $\mathscr{A}$, then it is a Cauchy sequence in $\mathscr{A}$, that means that for every $\varepsilon > 0$ there is a $N \in \mathbb{N}$ such that $\|\Phi(f_n) - \Phi_A(f_m)\| < \varepsilon, \forall m, n \geq N$, but since $\|\Phi_A(f_n) - \Phi_A(f_m)\| = \|f_n - f_m\|_\infty$ and therefore the sequence $(f_n)_{n \in \mathbb{N}}$ is Cauchy in $C(\sigma(A))$ with the norm $\|\|_\infty$. As $C(\sigma(A))$ is complete with relation to $\|\cdot\|_\infty$, there exists $f \in C(\sigma(A))$ such that $f_n \xrightarrow{n \to \infty} f$ and therefore, using again the equality of the norms, $\lim_{n \to \infty} \Phi_A(f_n) = \Phi_A(f)$.

4) - If $\lambda$ isn't in the image of $\sigma(A)$ by $f$ then $r(x) := \frac{1}{f(x) - \lambda}$ is continuous and therefore $\Phi_A(r)$ is well defined and it holds by the second equality in (4.1) that $\Phi_A(r)\Phi_A(f - \lambda) = \Phi_A(f - \lambda)\Phi_A(r) = \mathbb{1}$ proving that $\Phi_A(f) - \lambda\mathbb{1}$ is invertible, and therefore $\lambda \in \rho(\Phi_A(f))$, the resolvent set of $\Phi_A(f)$. This shows that the complementary set of the image of $\sigma(A)$ by $f$, $\mathbb{C} \setminus \{f(\xi) \,|\, \xi \in \sigma(A)\}$ is a subset of $\rho(\Phi_A(f))$, this means in other words $\sigma(\Phi_A(f)) \subseteq \{f(\xi) \,|\, \xi \in \sigma(A)\}$. To prove the reverse inclusion we choose $\mu \in \{f(\lambda) \,|\, \lambda \in \sigma(A)\}$, this means that $\mu = f(\lambda_0)$ for some $\lambda_0 \in \sigma(A)$, if we suppose that $\mu \in \rho(\Phi_A(f))$, therefore $F := \Phi_A(f) - f(\lambda_0)\mathbb{1}$ is invertible, considering $p \in P(\sigma(A))$ a polynomial that approximates $f$, guaranteed by the Weierstrass approximation theorem, then $\|f-p\|_\infty \leq \varepsilon$, for a small $\varepsilon > 0$. Defining then $P := \Phi_A(p) - p(\lambda_0)\mathbb{1}$, we have that $F - P = \Phi_A(f-p) - (f(\lambda_0) - p(\lambda_0))\mathbb{1}$ and so:

$$\|F-P\| \leq \|\Phi_A(f-p)\| + |f(\lambda_0) - p(\lambda_0)|\|\mathbb{1}\| = \|f-p\|_\infty + |f(\lambda_0) - p(\lambda_0)| \leq 2\|f-p\|_\infty < 2\varepsilon.$$

Now, since by corollary B.0.29, choosing $\varepsilon$ small enough such that $\|F - P\| < \|F^{-1}\|^{-1}$, then $P$ will be invertible in $\mathscr{A}$, which implies $p(\lambda_0) \notin \sigma(\Phi(p)) = \sigma(p(A))$ when $\lambda_0 \in \sigma(A)$, but this contradicts theorem B.0.42, therefore our supposition was wrong and $\mu \notin \rho(\Phi(f)) \Rightarrow \mu \in \sigma(\Phi(f))$, which proves that $\{f(\lambda) \,|\, \lambda \in \sigma(A)\}$ establishing the equality between the two sets. For $\mathscr{A} = \mathfrak{B}(\mathscr{H})$, and $\psi \in \mathscr{H}$ being a eigenvector of $A \in \mathfrak{B}_\dagger$ such that $A\psi = \lambda_0\psi$, then to prove that $\Phi_A(f)\psi = f(\lambda_0)\psi$, $\forall f \in C(\sigma(A))$ we consider that for any polynomial $p$ we have $\Phi_A(p)\psi = p(A)\psi = p(\lambda_0)\psi$, so that by considering a sequence of polynomials that converge to $f$ in the norm $\|\cdot\|_\infty$, then by continuity the extension $\Phi_A$ is such that $\Phi_A(f)\psi = f(\lambda_0)\psi$.

∎

We can then get a version of this result for von Neumann algebras instead of general $C^*$-algebras, where we can extend the domain of functions from continuous to Borel functions such proof is made in the Appendix B in theorem B.0.56, denoting such extension by $\hat{\Phi}_A$ for each self-adjoint $A \in \mathscr{A}$.

The Riesz-Markov representation theorem B.0.52 for complex measures allows us to speak about spectral measures $\mu_{\psi,A} : \mathscr{B}(\sigma(A)) \to \mathbb{R}$ for each $\psi \in \mathscr{H}$ and $A \in \mathscr{A}_\dagger$, such that $\forall f \in C(\sigma(A))$, $\langle \psi, f(A)\psi \rangle = \int_{\sigma(A)} f d\mu_{\psi,A}$, as is discussed in Appendix B, equation (B.16).

Considering then the spectral projectors given by $P_{(\cdot)} := \hat{\Phi}(\chi_{(\cdot)}) : \mathscr{B}(\sigma(A)) \to \mathfrak{B}(\mathscr{H})$, whose properties are better commented in Appendix B, theorem B.0.57. We can then define, analogously to the definition of integration with respect to a measure in definition C.0.24, integration with respect to a (orthogonal) projection-valued measure, or PVM in short, in this specific scenario of spectral measures.

Being $s = \sum_{n=1}^{N} \alpha_n \chi_{E_n}$ a arbitrary simple function with a finite number $N \in \mathbb{N}$ of terms, $\alpha_i \in \mathbb{C}$, and for a finite collection of measurable sets $\{E_n\}_{n=1,\ldots,N} \subset \mathscr{B}(\sigma(A))$, then defining the integral of this simple function to be the operator:

$$\int_{\sigma(A)} s(x) dP(x) := \sum_{n=1}^{N} \alpha_n P_{E_n}$$

Then it is obvious this map is linear on simple functions and it can be easily seen that the resulting operator is bounded, since for $\psi \in \mathscr{H}$:

$$\left\| \int_{\sigma(A)} s(x) dP(x) \, \psi \right\|_\mathscr{H}^2 = \left\langle \sum_{i=1}^{N} \alpha_i P_{E_i} \psi, \sum_{j=1}^{N} \alpha_j P_{E_j} \psi \right\rangle = \sum_{i,j=1}^{N} \left\langle \alpha_i \psi, \alpha_j P_{E_i}^* P_{E_j} \psi \right\rangle,$$

We can consider, without loss of generality, that the $\{E_n\}_{n=1,...,N}$ are disjoint, for if they aren't we can consider the collection formed by all the $E_n$'s subtracted from all the possible intersections between them and attaching to the collection these intersections themselves subtracted with all other intersections, and so on until we get to the intersection of all sets, since the initial collection is finite this new collection will also be, and the coefficients of the simple function in this new representation by disjoint sets will be equal to the old coefficients for the original sets subtracted the intersections and for each intersection set it will be the sum off all coefficients associated with each set of the intersection before the subtraction with the other intersections. Then for these disjoint sets $P_{E_i}P_{E_j} = P_{E_i \cap E_j} = \delta_{ij}P_{E_i}$, then continuing from the last equation line:

$$\sum_{i=1}^{N} |\alpha_i|^2 \langle \psi, P_{E_i}\psi\rangle \leq \left(\sup_{i\in\{1,...,N\}} |\alpha_i|\right)^2 \sum_{i=1}^{N} \langle \psi, P_{E_i}\psi\rangle. \tag{4.2}$$

By additivity and monotonicity:

$$\sum_{i=1}^{N} \langle \psi, P_{E_i}\psi\rangle \leq \left\langle \psi, P_{\bigcup_{i=1,...,N} E_i}\psi \right\rangle \leq \langle \psi, P_{\sigma(A)}\psi\rangle = \langle \psi, \psi\rangle = \|\psi\|^2.$$

Substituting this back to (4.2) and considering that $\|s\|_\infty = \sup_{i=1,...,N} |\alpha_i|$ for this disjoint set representation, we get:

$$\left\|\int_{\sigma(A)} s(x)dP(x)\ \psi\right\|_{\mathcal{H}}^2 \leq \|\psi\|^2 \|s\|_\infty^2.$$

Taking the square root on both sides and the least upper bound over unit vectors $\psi \in \mathcal{H}$: $\|\int_{\sigma(A)} s(x)dP(x)\| \leq \|s\|_\infty$. Hence integration with respect to a PVM is a Bounded linear transformation, and by the BLT theorem B.0.15 there exists a unique linear extension of this integration to the complete set of bounded Borel measurable functions:

$$\int_{\sigma(A)} f(x)dP(x), \qquad \forall f \in B^{\mathscr{B}}(\sigma(A)).$$

Having established this we can prove the Spectral decomposition theorem for self-adjoint operators in $\mathfrak{B}(\mathcal{H})$:

**Theorem 4.0.17** (Spectral decomposition of self-adjoint operators). *Let $\mathcal{H}$ be a Hilbert space and $A \in \mathfrak{B}(\mathcal{H})$ a self-adjoint operator, then:*

(i) *There exists a **unique** bounded PVM, $P^{(A)} : \mathscr{B}(\sigma(A)) \to \mathfrak{B}(\mathcal{H})$, given by theorem B.0.57 with $\mathscr{B}(\sigma(A)) \ni B \mapsto P_B := \hat{\Phi}_A(\chi_B) \equiv \chi\chi_B(A)$, such that:*

$$A = \int_{\sigma(A)} \lambda dP^{(A)}(\lambda).$$

*In a unique way we also have:*

$$g(A) = \int_{\sigma(A)} g(\lambda)dP^{(A)}(\lambda), \qquad \forall g \in B^{\mathscr{B}}(\sigma(A)).$$

(ii) *If $f \in B^{\mathscr{B}}(\sigma(A))$, the operator $\int_{\sigma(A)} f(\lambda)dP^{(A)}(\lambda)$ commutes with every operator in $\mathfrak{B}(\mathcal{H})$ that commutes with A.*

*Proof:* For a arbitrary simple function $s = \sum_{n=1}^{N} \alpha_n \chi_{E_n}$ where $\{E_n\}_{n=1,\dots,N} \subset \mathscr{B}(\sigma(A))$ is a collection of disjoint sets, we have by theorem B.0.57 item *(iii)* that for every $\psi \in \mathscr{H}$:

$$\left\langle \psi, \left(\int_{\sigma(A)} s(x)dP(x)\right)\psi \right\rangle = \left\langle \psi, \sum_{n=1}^{N}\alpha_n P_{E_n}\psi\right\rangle = \sum_{n=1}^{N}\alpha_n\langle \psi, P_{E_n}\psi\rangle = \sum_{n=1}^{N}\alpha_n \mu_{\psi,A}(E_n) =$$

$$= \sum_{n=1}^{N}\alpha_n \int_{\sigma(A)} \chi_{E_n} d\mu_{\psi,A} = \int_{\sigma(A)} \sum_{n=1}^{N}\alpha_n \chi_{E_n} d\mu_{\psi,A} = \int_{\sigma(A)} s\, d\mu_{\psi,A} = \langle \psi, s(A)\psi\rangle$$

Hence, by the continuity of integration with respect to a PVM, guaranteed by the boundedness brought by the BLT theorem B.0.15 and the boundedness-continuous correspondence for Linear operators acting between normed vector spaces in proposition B.0.6, we have that:

$$g(A) = \int_{\sigma(A)} g(\lambda) dP(\lambda), \qquad \forall g \in B^{\mathscr{B}}(\sigma(A)),$$

and therefore:

$$\mathbb{1}_{\mathbb{R}}(A) = A = \int_{\sigma(A)} \mathbb{1}_{\mathbb{R}}(\lambda) dP(\lambda) = \int_{\sigma(A)} \lambda\, dP(\lambda).$$

The uniqueness can be ascertained by considering that if $A = \int_{\sigma(A)} \lambda dP(\lambda) = \int_{\sigma(A)} \lambda dE(\lambda)$, then using linearity and the homomorphism property of the functional calculus, we get that $p(A) = \int_{\sigma(A)} p(\lambda) dP(\lambda) = \int_{\sigma(A)} p(\lambda) dE(\lambda)$ and therefore, $\forall \psi \in \mathscr{H}$:

$$\left\langle \psi, \left(\int_{\sigma(A)} p(\lambda) dP(\lambda)\right)\psi\right\rangle = \left\langle \psi, \left(\int_{\sigma(A)} p(\lambda) dE(\lambda)\right)\psi\right\rangle$$

and therefore $\int_{\sigma(A)} p(\lambda) d\mu_{\psi,A} = \int_{\sigma(A)} p(\lambda) d\nu_{\psi,\mathfrak{Q}}$ for the measure $\nu_{\psi,\mathfrak{Q}}(B) := \langle \psi, E_B \psi\rangle$, $\forall B \in \mathscr{B}(K)$ were $K \subseteq \mathbb{R}$ is a compact set and $\mathfrak{Q} : \mathscr{B}(K) \to \mathfrak{B}(\mathscr{H})$ is the PVM associated with the projectors $E$. By the Stone-Weierstrass theorem B.0.22 we conclude that $\int_{\sigma(A)} f(\lambda) d\mu_{\psi,A} = \int_{\sigma(A)} f(\lambda) d\nu_{\psi,\mathfrak{Q}}$ for every continuous function $f \in C(\sigma(A))$. Using Lusin's theorem B.0.54 and corollary B.0.55 we get that $\int_{\sigma(A)} g(\lambda) d\mu_{\psi,A} = \int_{\sigma(A)} g(\lambda) d\nu_{\psi,\mathfrak{Q}}$ for every limited Borel function $g \in B^{\mathscr{B}}(\sigma(A))$, in particular, for every Borel set $B \subset \sigma(A)$ we have $\int_{\sigma(A)} \chi_B(\lambda) d\mu_{\psi,A} = \int_{\sigma(A)} \chi_B(\lambda) d\nu_{\psi,\mathfrak{Q}} \implies \mu_{\psi,A}(B) = \nu_{\psi,\mathfrak{Q}}(B)$, by theorem B.0.57 item *(iii)* then $\langle \psi, P_B \psi\rangle = \langle \psi, E_B \psi\rangle$, which by the polarization Identity, prop. B.0.7, implies that $P_B = E_B$. As the choice of $B$ was arbitrary, this means that the PVM $\mathfrak{P}$ of the spectral projectors coincides with $\mathfrak{Q}$, when $A = \int_{\sigma(A)} \lambda dP(\lambda) = \int_{\sigma(A)} \lambda dE(\lambda)$.

To prove item *(ii)*, we note that if $F \in \mathfrak{B}$ is a arbitrary operator that commutes with $A$, then assuming by induction that $F$ commutes with $A^n$ this implies that $FA^{n+1} = FAA^n = AFA^n = AA^nF = A^{n+1}F$, hence $F$ commutes with arbitrary powers of $A$ and therefore, by linearity, to every polynomial $p(A)$, by the continuity of the map that gives integration with respect to a PVM and the Stone-Weierstrass theorem B.0.22, which tells us we can approximate any continuous function with polynomials, we get that for $h \in C(\sigma(A))$, $Fh(A) = h(A)F$. Considering then that by the Riesz-Markov theorem B.0.52, there exists a unique measure $\upsilon_{y,x;A}$ such that $\langle y, h(A)x\rangle = \int_{\sigma(A)} h\, d\upsilon_{y,x;A}$ for every $h \in C(\sigma(A))$, then calculating:

$$\int_{\sigma(A)} h\, d\upsilon_{y,Fx;A} = \langle y, h(A)Fx\rangle = \langle y, Fh(A)x\rangle = \langle F^*y, h(A)x\rangle = \int_{\sigma(A)} h\, d\upsilon_{F^*y,x;A}$$

but since the measure $v_{y,Fx;A}$ is unique, then $v_{F^*y,x;A} = v_{y,Fx,A}$, hence:

$$\langle y, \hat{\Phi}_A(f)Fx \rangle = \int_{\sigma(A)} f dv_{y,Fx;A} = \int_{\sigma(A)} f dv_{F^*y,x;A} = \langle F^*y, \hat{\Phi}_A(f)x \rangle = \langle y, F\hat{\Phi}_A(f)x \rangle,$$

for any $x, y \in \mathscr{H}$, $f \in B^{\mathscr{B}}(\sigma(A))$. As the vectors $x, y$ are arbitrary, $f(A)F \equiv \hat{\Phi}_A(f)F = F\hat{\Phi}_A(f) \equiv Ff(A)$ if $f \in B^{\mathscr{B}}(\sigma(A))$. Ultimately this means that, since $F$ was a arbitrary operator that commutes with $A$, then the operator $f(A) = \int_{\sigma(A)} g(\lambda) dP^{(A)}(\lambda)$ commutes with every operator in $\mathfrak{B}(\mathscr{H})$ that commutes with $A$. ∎

The following fundamental theorems of $C^*$-algebras will be needed in our exposition of all matters involving AQFT, the first theorem is available in every $C^*$-algebra book and even in some AQFT books, so we will omit its proof, since it becomes rather long, technical and involved, I particularly appreciate the proof in [8] (in portuguese) since it elegantly makes references to elementary notions directly instead of compressing them all in a convoluted, overly technical terminology, the proof in [77] is also really good, there being also a generalization of the usual GNS representation theorem for unital $*$-algebras instead of the usual $C^*$-algebras, this has some interest in the following discussion since the Fewster-Verch *framework* is given also in terms of general $*$-algebras. The, for the second theorem we do give a proof, but this proof has to rely on some propositions of the Appendix B that make references to definitions that will only be made in chapter 7, so we consider that to be more like a sketch of the proof.

**Theorem 4.0.18** (Gel'fand-Naĭmark-Segal (GNS representation)). *Let $\mathscr{A}$ be a $C^*$-algebra, and $\omega : \mathscr{A} \to \mathbb{C}$ a positive linear functional that is normalized by $\|\omega\|_{\mathscr{A}'} = 1$ (as we will see lather these are called the states of the $C^*$-algebra). Therefore, there exists a Hilbert space $\mathscr{H}_\omega$ and a representation $\pi_\omega$ of the $C^*$-algebra $\mathscr{A}$ by bounded operators acting on the Hilbert space $\mathscr{H}_\omega$, such that $\pi(a^*) = \pi(a)^*, \forall a \in \mathscr{A}$.*

*If, beyond that, $\mathscr{A}$ is a unital $C^*$-algebra then there exists a vector $\Omega \in \mathscr{H}_\omega$ such that $\omega(a) = \langle \Omega, \pi_\omega(a)\Omega \rangle_{\mathscr{H}_\omega}$, and the set $\{\pi_\omega(a)\Omega \in \mathscr{H}_\omega \,|\, a \in \mathscr{A}\}$ is dense in $\mathscr{H}_\omega$, this density property of a vector in a given representation is the definition of what is called a **cyclic vector** for the representtion $\pi_\omega$.*

**Theorem 4.0.19** (Noncommutative Gel'fand-Naĭmark). *Any $C^*$-algebra $\mathscr{A}$ possesses a faithful representation $\pi : \mathscr{A} \to \mathfrak{B}(\mathscr{H})$ on some Hilbert space $\mathscr{H}$. Thus, $\mathscr{A}$ is isomorphic to a sub-$C^*$-algebra of $\mathfrak{B}(\mathscr{H})$.*

*Proof:* Let $a \in \mathscr{A}$, with $a \neq 0$. Then as $a^*a$ is self-adjoint, it is in particular normal, hence by theorem B.0.50, we find a state $\varphi : \mathscr{A} \to \mathbb{C}$ with $\varphi(a^*a) = \|a^*a\| = \|a\| \neq 0$. By the GNS construction, theorem 4.0.18, we obtain a Hilbert space $\mathscr{H}_\varphi$ and a representation $\pi_\varphi : \mathscr{A} \to \mathfrak{B}(\mathscr{H})$ with a cyclic vector $\Omega_\varphi \in \mathscr{H}_\varphi$ such that:

$$\|\pi_\varphi(a)\Omega_\varphi\|^2 = \langle \Omega_\varphi, \pi_\varphi(a)\Omega_\varphi \rangle = \varphi(a^*a) \neq 0.$$

Hence, $\pi_\varphi(a) \neq 0$.

We then put $\mathscr{H} := \bigoplus_{\varphi \in \mathfrak{S}(\mathscr{A})} \mathscr{H}_\varphi$ and $\pi := \bigoplus_{\varphi \in \mathfrak{S}(\mathscr{A})} \pi_\varphi$, where $\mathfrak{S}(\mathscr{A})$ is the set of all states in $\mathscr{A}$. By the previous consideration, we have $\pi(a) \neq 0, \forall a \in \mathscr{A}$ with $a \neq 0$, i.e., $\pi$ is faithful. By proposition B.0.51, $\mathscr{A}$ is then isomorphic to $\pi(\mathscr{A}) \subseteq \mathfrak{B}(\mathscr{H})$. ∎

## 4.1 Haag-Kastler Axioms

1) **Existence of Algebras:** for each QFT we use a given symbol like for example $\mathscr{A}$ as a label for that theory defined in some collection of globally hyperbolic spacetimes.

For each spacetime $\mathbb{M}$ in this collection there will be a unital $*$-algebra $\mathscr{A}(\mathbb{M})$.

There will also be a collection of sub-$*$-algebras $\mathscr{A}(\mathbb{M}; N)$ labeled by causally convex open subsets $N$ of $\mathbb{M}$, all of which contain the unit of $\mathscr{A}(\mathbb{M})$, so the $\mathscr{A}(\mathbb{M}; N)$ collectively yields $\mathscr{A}(\mathbb{M})$.

2) **Isotony:** If $N_1 \subset N_2$ then $\mathscr{A}(\mathbb{M}; N_1) \subset \mathscr{A}(\mathbb{M}; N_2)$.

Before going on with the axioms, we recall briefly the definition of causal convexity.[3] Every spacetime we consider is a Lorentzian spacetime, that is, a smooth, Hausdorff, paracompact, manifold $\mathbb{M}$ with at most finitely many connected components, equipped with a smooth Lorentzian metric $g$, and a choice of time-orientation, which allows us to classify every nonzero causal vectors as pointing to the future or to the past.

If $x \in \mathbb{M}$, $J^+(x)$ is the *causal future* of $x$, which is the set of points reached starting from $x$ by smooth causal curves directed to the future, equivalently $J^-(x)$ is the *causal past* of $x$, where instead of the future, the smooth causal curves directed to the past. For a subset $K \subset \mathbb{M}$ we identify $J^{\pm}(K) = \bigcup_{x \in K} J^{\pm}(x)$ and also $J(K) = J^+(K) \cup J^-(K)$. The **causal hull**, or sometimes called causal envelope, of $K$ is $J^+(K) \cap J^-(K)$; that is, the set of all points that lie on causal curves with both endpoints in $S$, and a subset is **causally convex** if it is its own **causal hull**, that is if it contains every causal curve that begins and ends in itself.

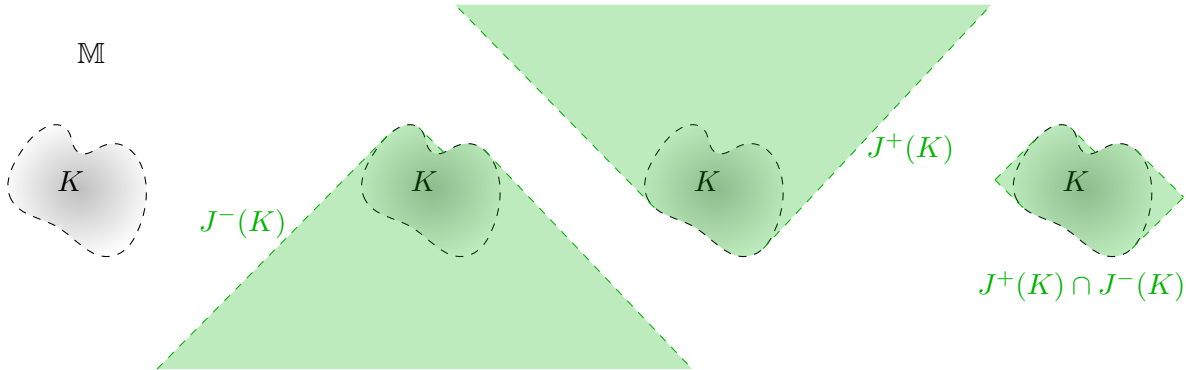

Figure 4 — Representations in spacetime diagrams of a subset $K \subseteq \mathbb{M}$, its causal past $J^-(K)$, causal future $J^+(K)$ and causal hull $J^+(K) \cap J^-(K)$. In these spacetime diagrams the arrow of time is oriented pointing up the page and light rays are inclined at 45° with relation to the vertical axis.

A Cauchy surface is a set intersected exactly once by every inextendible smooth timelike curve. The causal complement of a set $K$ is defined as $K^\perp = \mathbb{M} \setminus J(K)$, and sets $K$ and $S$ are causally disjoint if $K \subseteq S^\perp$ or equivalently $S \subseteq K^\perp$, *i.e.*, if there is no causal curve joining $K$ and $S$. In a globally hyperbolic spacetime, the causal future and past of an open set are open, while those of a compact set are closed; accordingly, if $K$ is compact then $K^\perp$ is open and $K^{\perp\perp}$ is closed, not necessarily compact, and contains $K$.

---

[3] The reader that is either unfamiliar, or forgetful, of these notions can find a more detailed description on classic textbooks such as [99], or for a slightly more compact discussion, but more tuned for the upcoming *framework* proposed by Fewster and Verch in AQFT in curved space times, the articles [40, 39].

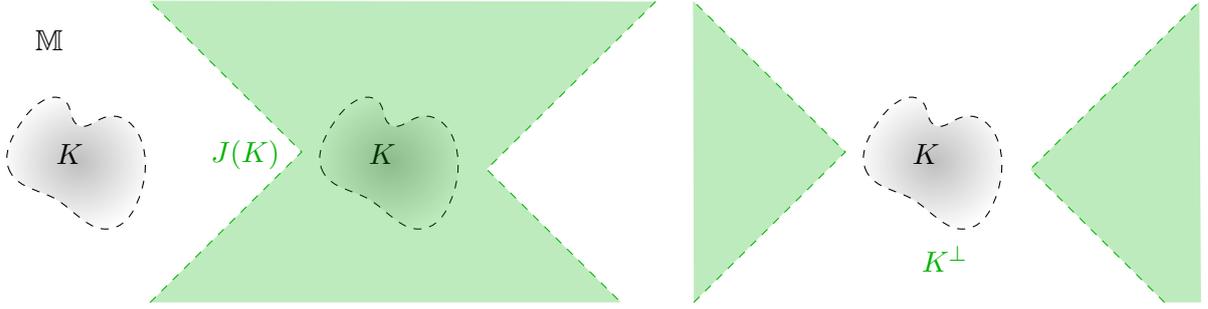

Figure 5 — Representations in spacetime diagrams of a subset $K \subseteq \mathbb{M}$, $J(K)$ and the causal complement $K^\perp = \mathbb{M} \setminus J(K)$.

The *future Cauchy development* $D^+(K)$ of a set $K \subseteq \mathbb{M}$ is the set of points $p$ so that every past-inextendible piecewise smooth causal curve through $p$ meets $K$, analogously $D^-(K)$ is the *past Cauchy development* of $K$ and contains the points though which every future-inextendible piecewise smooth causal curve meets $K$; also $D(K) = D^+(K) \cup D^-(K)$.

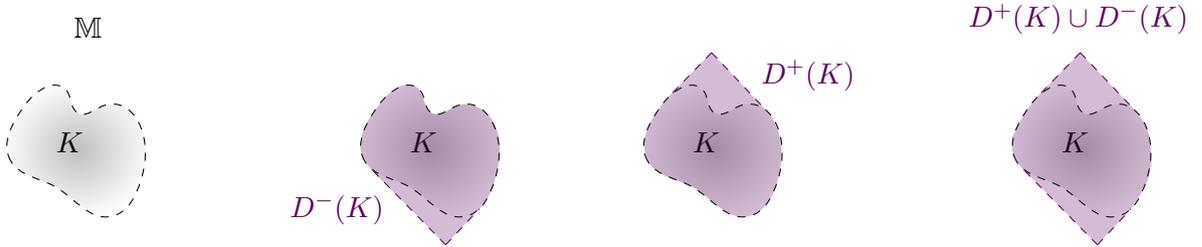

Figure 6 — Representations in spacetime diagrams of a subset $K \subseteq \mathbb{M}$, its past Cauchy development $D^-(K)$, future Cauchy development $D^+(K)$ and $D(K) = D^+(K) \cup D^-(K)$.

3) **Compatibility:** If $D$ is a causally convex open subset of the spacetime $\mathbb{M}$ on which $\mathscr{A}$ is defined, then $\mathscr{A}$ is also defined on $\mathbb{D}$, the globally hyperbolic spacetime composed of $D$ with the metric and temporal orientation inherited from $\mathbb{M}$, and there exists an injective $*$-unital homomorphism $\alpha_{\mathbb{M};\mathbb{D}} : \mathscr{A}(\mathbb{M}) \to \mathscr{A}(\mathbb{D})$ whose image is $\mathscr{A}(\mathbb{M}; D)$ and for which the composition rule applies:

$$\alpha_{\mathbb{M}_1;\mathbb{M}_2} \circ \alpha_{\mathbb{M}_2;\mathbb{M}_3} = \alpha_{\mathbb{M}_1;\mathbb{M}_3} \,, \tag{4.3}$$

for $\mathbb{M}_3 \subset \mathbb{M}_2 \subset \mathbb{M}_1$.

4) **Time slicing property or Existence of dynamics:**[4] If $N$ contains a Cauchy surface of $\mathbb{M}$ then $\mathscr{A}(\mathbb{M}, N) = \mathscr{A}(\mathbb{M})$.

5) **Einstein causality:** If $N_1$ and $N_2$ are causally disjoint regions then:

$$[A_1, A_2] = 0 \quad , \forall \ A_1 \in \mathscr{A}(\mathbb{M}, N_1), A_2 \in \mathscr{A}(\mathbb{M}, N_2).$$

6) **Haag property:** Let $K$ be a compact subset of $\mathbb{M}$, further suppose that a given element $A \in \mathscr{A}(\mathbb{M})$ commutes with every element of $\mathscr{A}(\mathbb{M}; N)$, $\forall\ N \in K^\perp$, where $K^\perp$ is the causal complement of $K$ given by $K^\perp = \mathbb{M} \setminus (J^+(K) \cup J^-(K))$. Then $A \in \mathscr{A}(\mathbb{M}; L)$ whenever $L \subset \mathbb{M}$ is a *connected* open causally convex subset containing $K$.[5]

---

[4] Axioms 4) and 6) are not usually considered to be part of the original Haag-Kastler axioms, although they are adjacent to them, and always included in AQFT's, so for simplicity of exposition we will abuse the nomenclature and include them as part of the Haag-Kastler axioms.

## 4.2 Observables and States

The usual physical interpretation in AQFT is that the self-adjoint elements of $\mathscr{A}(\mathbb{M})$ are the local observables of the theory $\mathscr{A}$, among these those belonging to $\mathscr{A}(\mathbb{M}; N)$ are the observables that can subsequently be located in $N$.

In retrospect, usual quantum mechanics makes the association between observables and spectral measurements also called projector-valued measurements (PVM's), however several considerations arising from the problem of quantum measurements (also familiar to Quantum Information) suggest that it is more beneficial in descriptions general flexibility of this association between observables and PVM's for an association between observables and measures valued with positive operators (POVM's) [21].

Taking this into account, we should then think of $\mathscr{A}(\mathbb{M}; N)$ as including all POVM valuations localizable in $N$. It then becomes instructive to differentiate sharp measurements associated with PVM's from more general unsharp measurements associated with POVM's.

**Definition 4.2.1** (States). *In AQFT the states of a theory in $\mathbb{M}$ are linear functionals $\omega$ of $\mathscr{A}(\mathbb{M})$ on $\mathbb{C}$ which are positive, $\omega(A^*A) \geq 0$, $\forall A \in \mathscr{A}(\mathbb{M})$ and normalized $\|\omega\|_{\mathscr{A}(\mathbb{M})'} = 1$. We denote the set of states on $\mathscr{A}(\mathbb{M})$ by $\mathfrak{S}(\mathscr{A}(\mathbb{M}))$.*

$\omega(A)$ is then the expected value of $A$ in $\omega$.

**Lemma 4.2.2** (Cauchy-Schwarz inequality for states). *Let $\mathscr{A}$ be a $C^*$-algebra and $\omega$ a state. Then for all $A, B \in \mathscr{A}$:*

$$\omega(A^*B) = \overline{\omega(B^*A)},$$
$$|\omega(A^*B)|^2 \leq \omega(A^*A)\omega(B^*B).$$

*Proof:* From the positivity of $\omega$, we get that for all $\lambda \in \mathbb{C}$:

$$\omega\left((A+\lambda B)^*(A+\lambda B)\right) = \omega(A^*A) + \bar{\lambda}\omega(B^*A) + \lambda(A^*B) + |\lambda|^2\omega(B^*B) \geq 0$$

If we take, in particular, $\lambda = 1$, then as the result of the above sum must be a positive real number, this forces the imaginary parts of $\omega(A^*B)$ and $\omega(B^*A)$ to cancel, hence $\mathfrak{Im}(\omega(A^*B)) = -\mathfrak{Im}(\omega(B^*A))$. Making also the particular choice $\lambda = i$, by the same reasoning as above the new imaginary parts, that are the old real parts multiplied by $-i$ and $i$, must cancel, hence $\mathfrak{Re}(\omega(A^*B)) = \mathfrak{Re}(\omega(B^*A))$, therefore $\omega(A^*B) = \overline{\omega(B^*A)}$ and the inequality becomes:

$$\omega(A^*A) + \overline{\lambda\omega(A^*B)} + \lambda(A^*B) + |\lambda|^2\omega(B^*B) \geq 0.$$

---

[5] This is a weakened form of the usual AQFT axiom that receives the name of *Haag duality*, and that is usually defined for von Neumann algebras instead of general unital $*$-algebras, in which case it is expressed as $\mathfrak{A}(\mathbb{M}, N)' = \mathfrak{A}(\mathbb{M}, N^\perp)$ for QFT's given by collections of von Neumann algebras $\mathfrak{A}$. The usual Haag duality axiom can be seen as a completeness requirement, since Einstein causality only requires that $\mathfrak{A}(\mathbb{M}, N^\perp) \subseteq \mathfrak{A}(\mathbb{M}, N)'$, were equality not to hold this would mean that there would be a observable that was not localizable within any of the causally convex regions determined by $N$, as being outside the influence region of $N$, but that still commutes with every element of the local algebra $\mathfrak{A}(\mathbb{M}, N)$.

If $\omega(B^*B) = 0$ then $|\omega(A^*B)|^2 \leq 0 \Rightarrow \omega(A^*B) = 0$ and the inequality is trivial. If $\omega(B^*B) \neq 0$, we take $\lambda = -\frac{\overline{\omega(A^*B)}}{\omega(B^*B)}$, and hence:

$$\omega(A^*A) - \frac{|\omega(A^*B)|^2}{\omega(B^*B)} \geq 0,$$

$$|\omega(A^*B)|^2 \leq \omega(A^*A)\omega(B^*B).$$

∎

**Proposition 4.2.3.** *Let $\omega$ be a linear functional over a $C^*$-algebra $\mathscr{A}$. The following conditions are equivalent:*

(i) *$\omega$ is positive;*

(ii) *$\omega$ is continuous and $\|\omega\| = \omega(\mathbb{1})$.*

*Proof:* We begin by proving that $(i) \Rightarrow (ii)$. We note that the set $\{\omega(A) \in \mathbb{R} \,|\, A \in \mathscr{A},\, A \geq 0, \|A\| = 1\}$ is bounded by some constant $M > 0$. In fact, if this is not true, there exists a sequence of positive elements $A_n$ with $\|A_n\| = 1$, such that $\omega(A_n) > n2^n$, this can be obtained by considering a divergent sequence $\{\omega(\widetilde{A}_i)\}_{i \in \mathbb{N}}$ and choosing a subsequence of this sequence, where each element of the subsequence is such that their value is greater than their sequence index times two to the power of the index. Nevertheless, because of the norm limitation of the $\{A_n\}_{n \in \mathbb{N}}$, the sequence of elements $B_k = \sum_{n=1}^{k} 2^{-n} A_n$ are norm convergent to some positive element $B$ and it follows from positivity that:

$$\sum_{n=1}^{k} < \omega(B_k) \leq \omega(B), \qquad \forall\, k \in \mathbb{N}.$$

But this is a contradiction, since $B$ is a positive element of $\mathscr{A}$ and hence $\omega$ takes it to a limited real number. Hence, we can define:

$$M = \sup\{\omega(A) \in \mathbb{R} \,|\, A \in \mathscr{A},\, A \geq 0, \|A\| = 1\} < +\infty,$$

but it follows easily from the polarization identity (B.2), that for each $A \in \mathscr{A}$ with $\|A\| = 1$, there exists $B_n \in \mathscr{A}_+$, $n = 0, 1, \ldots, 3$; with $\|B_n\| \leq 1$ and:

$$A = \sum_{n=0}^{3} i^n B_n.$$

Thus $|\omega(A)| \leq 4M < +\infty$, which means that $\omega$ is bounded and hence continuous, by proposition B.0.6.

We then note that $\omega(\mathbb{1}) \leq \|\omega\|\|\mathbb{1}\| = \|\omega\|$, to prove the reverse, we first note that: If $a \in \mathscr{A}$ is self-adjoint then $\omega(a) \in \mathbb{R}$ and $|\omega(a)| \leq \|a\|\omega(\mathbb{1})$, this can be easily verified since $\|a\|\mathbb{1} \geq a \geq -\|a\|\mathbb{1}$, since both $\|a\|\mathbb{1} - a$ and $a + \|a\|\mathbb{1}$ are self-adjoint and by the continuous functional calculus theorem 4.0.16, we can use the square root function to get the self adjoint elements $(\|a\|\mathbb{1} - a)^{\frac{1}{2}}$ and $(a + \|a\|\mathbb{1})^{\frac{1}{2}}$, hence $\|a\|\mathbb{1} - a = \left((\|a\|\mathbb{1} - a)^{\frac{1}{2}}\right)^* (\|a\|\mathbb{1} - a)^{\frac{1}{2}}$ and $a + \|a\|\mathbb{1} = \left((a + \|a\|\mathbb{1})^{\frac{1}{2}}\right)^* (a + \|a\|\mathbb{1})^{\frac{1}{2}}$, where it is used both the fact that these new elements are self-adjoint and the homomorphism property of the continuous functional calculus applied to the square root function, therefore both $\|a\|\mathbb{1} - a$ and $a + \|a\|\mathbb{1}$ are positive and $\|a\|\mathbb{1} \geq a \geq -\|a\|\mathbb{1}$.

We can then find, by the linearity and positivity of $\omega$, that $\|a\|\omega(\mathbb{1}) \geq \omega(a) \geq -\|a\|\omega(\mathbb{1})$. Going back to the general case, for a arbitrary $c \in \mathscr{A}$:

$$|\omega(c)|^2 = |\omega(\mathbb{1}^*c)|^2 \leq \omega(\mathbb{1})\omega(c^*c) \leq \omega(\mathbb{1})\|c^*c\|\omega(\mathbb{1}) \leq \omega(\mathbb{1})^2\|c\|^2,$$

where we used the fact that $\omega(c^*c) \leq \|c^*c\|\,\omega(\mathbb{1})$ in the second inequality, from this we arrive at $\|\omega\| \leq \omega(\mathbb{1})$. Therefore $\|\omega\| = \omega(\mathbb{1})$.

To prove the converse we can assume without loss of generality that $\|\omega\| = 1$, since we can just consider the new linear functional $\omega_n = \frac{1}{\omega(\mathbb{1})}\omega$, hence from $(ii)$, $\omega(\mathbb{1}) = 1$. We take $a \in \mathscr{A}$ to be positive, hence it is self-adjoint, and again by the continuous functional calculus theorem 4.0.16, we can write $a = (a)^{\frac{1}{2}}(a)^{\frac{1}{2}}$, where $b = (a)^{\frac{1}{2}}$ is also self-adjoint, by the $C^*$-algebra version of the spectral mapping theorem B.0.44, $\sigma(a) = \sigma(b^2) = \sigma(b)^2 \subseteq [0, \|b\|^2] = [0, \|a\|]$, where theorem B.0.48 was used together with the $B^*$-property of the norm. By the same argument $\sigma\left(\mathbb{1} - \frac{a}{\|a\|}\right) \subseteq [0,1]$, as $\mathbb{1} - \frac{a}{\|a\|}$ is clearly self-adjoint, and is the image of $a$ through the polynomial $p(x) = 1 - \frac{x}{\|a\|}$ and $\left\|\mathbb{1} - \frac{a}{\|a\|}\right\| = 1$.

Since $\left\|\mathbb{1} - \frac{a}{\|a\|}\right\| \leq 1$ then $\omega\left(\mathbb{1} - \frac{a}{\|a\|}\right) \leq \|\omega\|\left\|\mathbb{1} - \frac{a}{\|a\|}\right\| \leq 1$, and therefore we can conclude that $\omega(a) \geq 0, \forall\, a \in \mathscr{A}_+$. ∎

**Definition 4.2.4** (Vector state). *In any given nondegenerate representation $(\mathscr{H}, \pi)$ of a $C^*$-algebra $\mathscr{A}$, the functional given by:*

$$\omega_\psi(A) = \langle \psi,\, \pi(A)\psi \rangle_\mathscr{H}, \qquad \forall\, A \in \mathscr{A};$$

*is a state for $\|\psi\|_\mathscr{H} = 1$, since it is linear in $\mathscr{A}$ as it is the composition of the linear $\pi$ with the linear entry of the sesquilinear product $\langle\,\cdot\,,\,\cdot\,\rangle_\mathscr{H}$, it is also positive, because:*

$$\omega_\psi(A^*A) = \langle \psi,\, \pi(A^*A)\psi \rangle_\mathscr{H} = \langle \pi(A)\psi,\, \pi(A)\psi \rangle_\mathscr{H} = \|\pi(A)\psi\|_\mathscr{H}^2 \geq 0.$$

*The functional is normalized since in a unital $C^*$-algebra $\|\omega_\psi\|_{\mathscr{A}(\mathbb{M})'} = \omega_\psi(\mathbb{1}) = \langle \psi, \pi(\mathbb{1})\psi \rangle_\mathscr{H} = \langle \psi, \psi \rangle = \|\psi\|_\mathscr{H}^2 = 1$. For non-unital $C^*$-algebras we can consider the approximants of identity, that always exist in $C^*$-algebras, to get that same equation in the limit of the approximants, for more information on this we recommend [13].*

**Lemma 4.2.5.** *Let $\mathscr{A} \subseteq \mathfrak{B}(\mathscr{H})$ be a $C^*$-algebra. Then the set of $2\times 2$ matrix of bounded operators $M_2(\mathscr{A})$ acting on $\mathscr{H} \oplus \mathscr{H} \equiv \mathscr{H}^{\oplus 2}$ is a $C^*$-algebra, being $T \equiv (T_{ij})_{i,j \in \{1,2\}}$ and for all $i,j \in \{1,2\}$:*

$$\|T_{ij}\|_{\mathrm{op}(\mathfrak{B}(\mathscr{H}))} \leq \|T\|_{\mathrm{op}(\mathfrak{B}(\mathscr{H}^{\oplus 2}))}.$$

*Proof:* That $M_2(\mathscr{A})$ is a $C^*$-algebra is clear from the structure of matrix algebra, the involution $* : M_2(\mathscr{A}) \to M_2(\mathscr{A})$ operation is given by the usual transpose of the hermitian conjugate of the component operators, and the norm in this space is given by the usual operator norm $\|T\|_{\mathrm{op}(\mathfrak{B}(\mathscr{H}^{\oplus 2}))} := \sup_h \left\{\|Th\|_{\mathscr{H}^{\oplus 2}} \,\big|\, h \in \mathscr{H}^{\oplus 2},\, \|h\|_{\mathscr{H}^{\oplus 2}} \leq 1\right\}$, which then fulfills all $C^*$-algebra axioms.

The inequality follows from the ensuing quick calculations, for any $T \in M_2(\mathscr{A})$ and any vector $\begin{pmatrix} h_1 \\ h_2 \end{pmatrix} \in \mathscr{H}^{\oplus 2}$,

$$\left\| T \begin{pmatrix} h_1 \\ h_2 \end{pmatrix} \right\|_{\mathscr{H}^{\oplus 2}}^2 = \left( \sqrt{\|T_{11}h_1 + T_{12}h_2\|_{\mathscr{H}}^2 + \|T_{21}h_1 + T_{22}h_2\|_{\mathscr{H}}^2} \right)^2$$

$$= \|T_{11}h_1 + T_{12}h_2\|_{\mathscr{H}}^2 + \|T_{21}h_1 + T_{22}h_2\|_{\mathscr{H}}^2$$

Putting $h_1 = 0$, then clearly $\left\| T \begin{pmatrix} 0 \\ h_2 \end{pmatrix} \right\|_{\mathscr{H}^{\oplus 2}}^2 \geq \|T_{12}h_2\|_{\mathscr{H}}^2$ and $\left\| T \begin{pmatrix} 0 \\ h_2 \end{pmatrix} \right\|_{\mathscr{H}^{\oplus 2}}^2 \geq \|T_{22}h_2\|_{\mathscr{H}}^2$, analogously if we put $h_2 = 0$ then $\left\| T \begin{pmatrix} h_1 \\ 0 \end{pmatrix} \right\|_{\mathscr{H}^{\oplus 2}}^2 \geq \|T_{11}h_1\|_{\mathscr{H}}^2$ and $\left\| T \begin{pmatrix} h_1 \\ 0 \end{pmatrix} \right\|_{\mathscr{H}^{\oplus 2}}^2 \geq \|T_{21}h_1\|_{\mathscr{H}}^2$, completing our proof by considering that $\sup_{\begin{pmatrix} h_1 \\ h_2 \end{pmatrix}} \left\| T \begin{pmatrix} h_1 \\ h_2 \end{pmatrix} \right\|_{\mathscr{H}^{\oplus 2}}^2 \geq \left\| T \begin{pmatrix} 0 \\ h_2 \end{pmatrix} \right\|_{\mathscr{H}^{\oplus 2}}^2$ and analogously for $\begin{pmatrix} h_1 \\ 0 \end{pmatrix}$, so by considering the supremum on both sides and joining all these results we get that $\forall i,j \in \{1,2\}\ \|T_{ij}\|_{\text{op}(\mathfrak{B}(\mathscr{H}))} \leq \|T\|_{\text{op}(\mathfrak{B}(\mathscr{H}^{\oplus 2}))}$. ∎

**Lemma 4.2.6.** *Let $\mathscr{A} \subseteq \mathfrak{B}(\mathscr{H})$ be a sub-$C^*$-algebra and $\mathscr{B} := \overline{\mathscr{A}}^s$. Then we have that $M_2(\mathscr{B}) = \overline{M_2(\mathscr{A})}^s$.*

*Proof:* Being $b \equiv (b_{ij})_{i,j \in \{1,2\}} \in M_2(\mathscr{B})$, then for each $i, j \in \{1, 2\}$, there is a net $\{a_{ij}^\kappa\}_\kappa \subseteq \mathscr{A}$ which SOT-converges to $b_{ij}$. For any $\begin{pmatrix} h_1 \\ h_2 \end{pmatrix} \in \mathscr{H}^{\oplus 2}$, we have:

$$\left\| (a_{ij}^\kappa - b_{ij})_{i,j} \begin{pmatrix} h_1 \\ h_2 \end{pmatrix} \right\|_{\mathscr{H}^{\oplus 2}}^2 = \|(a_{11}^\kappa - b_{11}^\kappa)h_1 + (a_{12}^\kappa - b_{12}^\kappa)h_2\|_{\mathscr{H}}^2 + \|(a_{21}^\kappa - b_{21}^\kappa)h_1 + (a_{22}^\kappa - b_{22}^\kappa)h_2\|_{\mathscr{H}}^2$$

$$\leq (\|(a_{11}^\kappa - b_{11}^\kappa)h_1\|_{\mathscr{H}} + \|(a_{12}^\kappa - b_{12}^\kappa)h_2\|_{\mathscr{H}})^2 + (\|(a_{21}^\kappa - b_{21}^\kappa)h_1\|_{\mathscr{H}} + \|(a_{22}^\kappa - b_{22}^\kappa)h_2\|_{\mathscr{H}})^2 \to 0$$

Thus $a_{ij}^\kappa \xrightarrow[\kappa \to \infty]{s} b_{ij}$. ∎

**Definition 4.2.7.** *Considering $X$ a topological space, we define:*

- $C(X)$ *the set of all continuous functions from $X$ to $\mathbb{C}$.*

- $C_b(X)$ *the set of all bounded continuous functions from $X$ to $\mathbb{C}$.*

- $C_c(X) := \{f \in C(\Omega) \mid \text{supp}(f) \text{ is compact}\}$, *this is the set of continuous functions with compact support[6] from $X$ to $\mathbb{C}$.*

- *If $X = \Omega$ is a locally Hausdorff space, $C_0(\Omega) := \left\{ f \in C(\Omega) \,\middle|\, \overline{\{x \in \Omega \mid |f(x)| \geq \varepsilon\}} \text{ is compact } \forall \varepsilon \in \mathbb{R}_+ \right\}$ is the set of continuous functions from $\Omega$ to $\mathbb{C}$ that **vanish at infinity**.*

**Lemma 4.2.8.** *If $f : \mathbb{R} \to \mathbb{C}$ is a bounded continuous function, then $f$ is **strongly continuous**, that is, for every Hilbert space $\mathscr{H}$ and every net $\{T_\alpha\}_\alpha \subseteq \mathfrak{B}(\mathscr{H})$ of self-adjoint operators converging strongly to a self-adjoint operator $T \in \mathfrak{B}(\mathscr{H})$, we have that $f(T_\alpha) \xrightarrow{s} f(T)$.*

*Proof:* Letting $S = \{f : \mathbb{R} \to \mathbb{C} \mid f \text{ is strongly continuous}\}$, $S$ is trivially a complex vector space, additionally, if $f, g \in S$ and $f$ is bounded, then $fg \in S$, since supposing $\{T_\alpha\}_\alpha \subseteq \mathfrak{B}(\mathscr{H})$ a net of self-adjoint operators converging strongly to the self-adjoint operator $T \in \mathfrak{B}(\mathcal{H})$, and letting $h \in \mathcal{H}$ then:

$$\|f(T_\alpha)g(T_\alpha)h - f(T)g(T)h\| \leq \|f(T_\alpha)g(T_\alpha)h - f(T_\alpha)g(T)h\| + \|f(T_\alpha)g(T)h - f(T)g(T)h\|$$
$$\leq \|f\|_\infty \|g(T_\alpha)h - g(T)h\| + \|f(T_\alpha)g(T)h - f(T)g(T)h\| \xrightarrow{s} 0$$

so that $(fg)(T_\alpha) \xrightarrow{s} (fg)(T)$.

---

[6] Remembering that $\text{supp}(f) := \overline{\{x \in X \mid f(x) \neq 0\}}$, where the closure is taken with relation to the topology of $X$.

Setting now $S_0 := C_0(\mathbb{R}) \cap S$, we shall then prove that $S_0$ is a closed self-adjoint sub-$C^*$-algebra of $C_0(\mathbb{R})$ which separates the points of $\mathbb{R}$, and that trivially vanishes nowhere, since we can always consider a test function centered at any given point $x$, so that $f(x) \neq 0$. Then by the Stone-Weierstrass Theorem B.0.22 we have that $C_0(\mathbb{R}) = S_0 \subseteq S$.

To prove that $S_0$ is closed, first we note that $S_0$ is trivially linear and closed under multiplication by the above discussion, now considering a net $\{g_\alpha\}_\alpha$ in $S_0$ such that for $f \in C_0(\mathbb{R})$, $\|g_\alpha - f\|_\infty \to 0$, considering as before that $T_i, T$ are self-adjoint in $\mathfrak{B}(\mathcal{H})$ with $T_i \xrightarrow{s} T$. Let $\varepsilon > 0$, there is $N$ such that for all $\alpha \geq N$, $\|g_\alpha - f\|_\infty \leq \varepsilon$. Fixing an arbitrary $h \in \mathcal{H}$, then there is a set $I$ such that for all $i \geq I$, $\|g_N(T_i)h - g_N(T)h\| < \varepsilon$. Then considering $i \geq I$ and $\alpha \geq N$:

$$\|f(T_i)h - f(T)h\| \leq \|f(T_i)h - g_\alpha(T_i)h\| + \|g_\alpha(T_i)h - g_\alpha(T)h\| + \|g_\alpha(T)h - f(T)h\| < (2\|h\| + 1)\varepsilon.$$

So $f(T_i) \xrightarrow{s} f(T)$, and therefore $f \in S_0$.

To prove that $S_0$ is self-adjoint, we shall prove that if $f \in S_0$ then $\overline{f} \in S_0$. Considering now:

$$\left\|\overline{f}(T_i)h - \overline{f}(T)h\right\|^2 = \left\|\overline{f}(T_i)h\right\|^2 + \left\|\overline{f}(T)h\right\|^2 - \langle\overline{f}(T_i)h, \overline{f}(T)h\rangle - \langle\overline{f}(T)h, \overline{f}(T_i)h\rangle$$
$$= \|f(T_i)h\|^2 - \|f(T)h\|^2 + \langle\overline{f}(T)f(T)h, h\rangle + \langle h, \overline{f}(T)f(T)h\rangle - \langle h, f(T_i)\overline{f}(T)h\rangle - \langle f(T_i)\overline{f}(T)h, h\rangle$$
$$= \|f(T_i)h\|^2 - \|f(T)h\|^2 + \langle f(T)\overline{f}(T)h, h\rangle + \langle h, f(T)\overline{f}(T)h\rangle - \langle h, f(T_i)\overline{f}(T)h\rangle - \langle f(T_i)\overline{f}(T)h, h\rangle$$
$$= \|f(T_i)h\|^2 - \|f(T)h\|^2 + \langle h, (f(T) + f(T_i))\overline{f}(T)h\rangle + \langle (f(T_i) - f(T))\overline{f}(T)h, h\rangle \to 0$$

Hence $\overline{f}(T_i) \xrightarrow{s} \overline{f}(T) \implies \overline{f} \in S_0$.

To prove that $S_0$ separates the points of $\mathbb{R}$, we define:

$$f(x) = \frac{1}{1+x^2} \quad \text{and} \quad g(x) = \frac{x}{1+x^2}$$

Certainly, $f, g \in C_0(\mathbb{R})$ and $\|f\|_\infty, \|g\|_\infty \leq 1$. It is not hard to see that the collection $\{f, g\}$ separates the points of $\mathbb{R}$, because $f$ is a even function monotonically decreasing in $\mathbb{R}_+$ so $f$ can differentiate any two different points, both either in $[0, \infty)$ or both in $(-\infty, 0]$; and $g$ is a odd function so it can differentiate two points such that one of them is in $[0, \infty)$ and the other is in $(-\infty, 0]$, this exhausts all possibilities of different points, so it remains to show that $f, g \in S_0$. Considering then:

$$g(T_i) - g(T) = (1 + T_i^2)^{-1} T_i - T(1 + T^2)^{-1}$$
$$= (1 + T_i^2)^{-1} \left(T_i(1 + T^2) - (1 + T_i^2)T\right)(1 + T^2)^{-1}$$
$$= (1 + T_i^2)^{-1} \left(T_i - T + T_i(T - T_i)T\right)(1 + T^2)^{-1}$$
$$= f(T_i)(T_i - T)f(T) + g(T_i)(T - T_i)g(T)$$

As we can see, both $f$ and $g$ are contractions, this implies by definition that in the operator norm, $\|f(T_i)\| \leq 1$, $\|g(T_i)\| \leq 1$, and therefore

$$\|g(T_i) - g(T)\| \leq \|f(T_i)\|\|(T_i - T)f(T)h\| + \|g(T_i)\|\|(T - T_i)g(T)h\|$$
$$\leq \|(T_i - T)f(T)h\| + \|(T - T_i)g(T)h\| \to 0$$

Therefore, $g(T_i) \xrightarrow{s} g(T)$, so $g \in S_0$. Since $g$ is bounded, and the inclusion map $S_0 \ni x \mapsto x \in S$ is trivially strongly continuous, we have, by the beginning discussion, that $gx \in S$. Whence $f + gx = 1 \implies f = 1 - gx \in S$, concluding that $f \in S_0$.

So $C_0(\mathbb{R}) = S_0 \subseteq S$ by the Stone-Weierstrass Theorem B.0.22, but more than that, we can now prove that $C_b(\mathbb{R}) \subseteq S$. Let $f, g \in S_0$ be the same functions from the last paragraph, and being $h \in C_b(\mathbb{R})$, as $h$ is bounded $hf, hg \in C_0(\mathbb{R})$, then by what was proved before $hf, hg \in S_0$, again considering the inclusion $S_0 \ni x \mapsto x \in S$, $hgx \in S$, and as before $f + gx = 1$, then $h = h(f + gx) = hf + hgx \in S$, as desired. ∎

The following theorem is a application of spectral theory, and as a standard result of functional analysis, we shall not give a proof of this theorem, since it would be too involved, but only point to the wonderful exposition of a proof in the reference [77].

**Theorem 4.2.9** (Stone's Theorem). *Let $\mathcal{H}$ be a separable Hilbert space, if $A: \mathcal{H} \supset D(A) \to \mathcal{H}$ is a, possibly unbounded, self-adjoint operator on $D(A)$, a dense subspace of $\mathcal{H}$, then defining:*

$$U(t) = e^{itA} \quad , t \in \mathbb{R}$$

*by the functional calculus, then $U(t)$ is bounded on $D(A)$ and extends by continuity to a unitary operator on $\mathcal{H}$, that has the following properties:*

(i) $U(t + s) = U(t)U(s) \quad , \forall s, t \in \mathbb{R}$,

(ii) $U$ *is strongly continuous*, $\forall t_0 \in \mathbb{R}$, $s\text{-}\lim_{t \to t_0} U(t) = U(t_0)$,

(iii) $D(A) = \{\psi \in \mathcal{H} \mid \exists \lim_{t \to 0} t^{-1}(U(t)\psi - \psi)\}$ *and the map $t \mapsto U(t)\psi$ is a solution to what can be interpreted as a normalized Schrödinger equation:*

$$\frac{d}{dt}U(t)\psi := \lim_{s \to t} \frac{U(s)\psi - U(t)\psi}{s - t} = iA(U(t)\psi) \quad , \forall \psi \in D(A) \text{ and } t \in \mathbb{R},$$

*which can be rewritten as:*
$$A\psi = \frac{1}{i} \lim_{\tau \to 0} \frac{U(\tau)\psi - \psi}{\tau}.$$

*Conversely, given a strongly continuous one-parameter unitary group $U: \mathbb{R} \to \mathfrak{B}(\mathcal{H})$, that is $U(t)$ is a unitary for every $t \in \mathbb{R}$ and has properties (i) and (ii), then defining:*

$$D(A) := \{\psi \in \mathcal{H} \mid \exists \lim_{t \to 0} t^{-1}(U(t)\psi - \psi)\}$$

$$A\psi := \frac{1}{i} \lim_{\tau \to 0} \frac{U(\tau)\psi - \psi}{\tau} \quad , \psi \in D(A).$$

*Then $D(A)$ is dense in $\mathcal{H}$, $A$ is a self-adjoint operator in $D(A)$, and we have $U(t) = e^{itA}$, $\forall t \in \mathbb{R}$.*

**Theorem 4.2.10** (Kaplansky Density Theorem). *Let $\mathscr{H}$ be a Hilbert space, $\mathscr{A}$ a sub-$C^*$-algebra of $\mathfrak{B}(\mathscr{H})$, and $\mathscr{B} := \overline{\mathscr{A}}^s$ the SOT-closure of $\mathscr{A}$ in $\mathfrak{B}(\mathscr{H})$. Then:*

- *$\mathscr{A}_\dagger$ is SOT-dense in $\mathscr{B}_\dagger$.*

- *The closed unit ball of $\mathscr{A}_\dagger$ is SOT-dense in the closed unit ball of $\mathscr{B}_\dagger$.*

- *The closed unit ball of $\mathscr{A}_+$ is SOT-dense in the closed unit ball of $\mathscr{B}_+$.*

- *The closed unit ball of $\mathscr{A}$ is SOT-dense in the closed unit ball of $\mathscr{B}$.*

- *If $\mathbb{1}_{\mathscr{H}} \in \mathscr{A}$ then the unitaries of $\mathscr{A}$ are strongly dense in the unitaries of $\mathscr{B}$.*

*Note what this theorem guarantees: given an operator $b$ in the SOT-closure of $\mathscr{A}$, we can find a net of operators $\{a_\kappa\}_\kappa \subseteq \mathscr{A}$ such that $a_\kappa \xrightarrow{s} b$ and $\|a_\kappa\| \leq \|b\|$, $\forall \kappa$. Moreover, if $b$ is self-adjoint (or positive, or unitary), we can take all the $a_\kappa$ to be self-adjoint (or positive, or unitary, respectively).*

*Proof:*

- Proving first that $\mathscr{A}_\dagger$ is $SOT$-dense in $\mathscr{B}_\dagger$: Let $b \in \mathscr{B}_\dagger$, then there is a net $\{a_\kappa\}_\kappa \subset \mathscr{A}$ such that $a_\kappa \xrightarrow{s} b$, then by Lemma 4.0.13 $a_\kappa \xrightarrow{w} b$ and it follows easily from Lemma 4.0.14 that $\mathfrak{Re}(a_\kappa) \xrightarrow{w} \mathfrak{Re}(b) = b$. Thus, since by Lemma 4.0.8 $\mathscr{A}_\dagger$ is a convex space, then
$$b \in \overline{\mathscr{A}_\dagger}^w = \overline{\mathscr{A}_\dagger}^s$$

- Proving that the closed unit ball of $\mathscr{A}$ is $SOT$-dense in the closed unit ball of $\mathscr{B}$: Letting $b \in \mathcal{B}_1(\mathscr{B}_\dagger)$, by the first item there is a net $\{a_\kappa\}_\kappa \subset \mathscr{A}_\dagger$ such that $a_\kappa \xrightarrow{s} b$. Considering the function $f : \mathbb{R} \to \mathbb{C}$ defined by $f(t) := \min\{\max\{-1, t\}, 1\}$ that has the following graph:

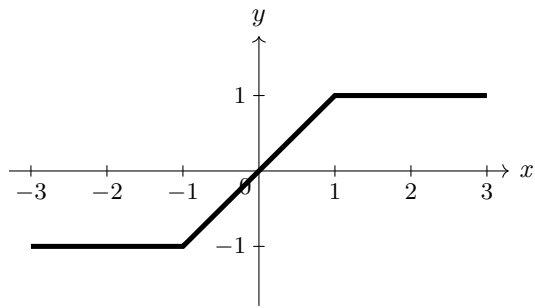

Figure 7 — $f(t) = \min\{\max\{-1, t\}, 1\}$

Then $f$ is continuous and bounded, so by Lemma 4.2.8, $f$ is strongly continuous. Thus, $f(a_\kappa) \xrightarrow{s} f(b)$. However, $b$ is self-adjoint and since it is inside the unit ball $\|b\| \leq 1$, so by Proposition B.0.48 we have $\sigma(b) \subseteq [-1, 1]$, so $f\Big|_{\sigma(b)}(t) = t$ and by the functional calculus $f(b) = b$. Moreover since the function $f$ is such that $\overline{f} = f$ and $\|f\|_\infty \leq 1$, these imply $f(a_\kappa)^* = f(a_\kappa)$ and $\|f(a_\kappa)\| \leq 1$ for all $\kappa$, that implies $f(a_\kappa) \in \mathcal{B}_1(\mathscr{A}_\dagger)$ and $f(a_\kappa) \xrightarrow{s} f(b) = b$.

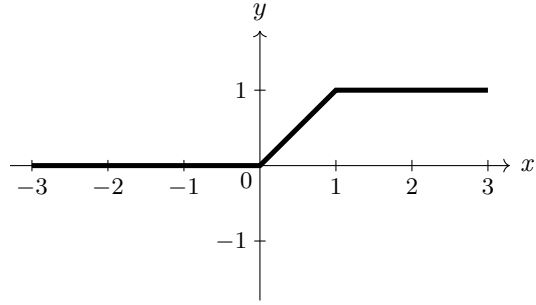

Figure 8 — $g(t) = \min\{\max\{0,t\},1\}$

- Proving that the closed unit ball of $\mathscr{A}_+$ is $SOT$-dense in the closed unit ball of $\mathscr{B}_+$: Considering that $b \in \mathcal{B}_1(\mathscr{B}_+)$, by the second item, there is a net $\{a_\kappa\}_\kappa \subset \mathcal{B}_1(\mathscr{A}_\dagger)$ such that $a_\kappa \xrightarrow{s} b$. Defining the function $f : \mathbb{R} \to \mathbb{C}$ by $g(t) := \min\{\max\{0,t\},1\}$ that has the following graph:

  Then again since $f$ is continuous and bounded by Lemma 4.2.8, $f$ is strongly continuous. Thus, $f(a_\kappa) \xrightarrow{s} f(b)$. Since $b$ is positive and is in the unit ball $\sigma(b) \subseteq [0,1]$, so $f(b) = b$. Moreover $\overline{f} = f$, $\|f\|_\infty \leq 1$ and $f \geq 0$, so $f(a_\kappa)$ is self-adjoint, of norm at most 1, and is positive $\forall \kappa$. Hence $f(a_\kappa) \in \mathcal{B}_1(\mathscr{A}_+)$ for all $\kappa$ and $f(a_\kappa) \xrightarrow{s} f(b) = b$.

- Proving that the closed unit ball of $\mathscr{A}$ is $SOT$-dense in the closed unit ball of $\mathscr{B}$: Let $b \in \mathcal{B}_1(\mathscr{B})$, then constructing:

$$\tilde{b} := \begin{pmatrix} 0 & b \\ b^* & 0 \end{pmatrix} \in M_2(\mathscr{B})$$

which is by construction self-adjoint and we have that, for $\mathbf{f} = \begin{pmatrix} f_1 \\ f_2 \end{pmatrix} \in \mathscr{H} \oplus \mathscr{H} \equiv \mathscr{H}^{\oplus 2}$:

$$\|\tilde{b}\| = \sup_{(\|\mathbf{f}\|_{\mathscr{H}^{\oplus 2}}) \leq 1} \left\| \begin{pmatrix} 0 & b \\ b^* & 0 \end{pmatrix} \begin{pmatrix} f_1 \\ f_2 \end{pmatrix} \right\|_{\mathscr{H}^{\oplus 2}} = \sup_{(\|\mathbf{f}\|_{\mathscr{H}^{\oplus 2}}) \leq 1} (\|bf_2\|_{\mathscr{H}}^2 + \|b^* f_1\|_{\mathscr{H}}^2)^{\frac{1}{2}} \leq$$

$$\leq \sup_{(\|\mathbf{f}\|_{\mathscr{H}^{\oplus 2}}) \leq 1} (\|f_2\|_{\mathscr{H}}^2 + \|f_1\|_{\mathscr{H}}^2)^{\frac{1}{2}} = 1$$

Then $\tilde{b} \in \mathcal{B}_1(M_2(\mathscr{B})_\dagger)$. Since by Lemma 4.2.6 we have:

$$M_2(\mathscr{B}) = \overline{M_2(\mathscr{A})}^s$$

it follows by the second item that there is a net $\{\tilde{a}^\kappa\}_\kappa$ in the unit ball $\mathcal{B}_1(M_2(\mathscr{A})_\dagger)$ which SOT-converges to $\tilde{b} \in \mathcal{B}_1(M_2(\mathscr{B})_\dagger)$. The following inequality follows from considering only one component from the Hilbert sum, for a arbitrary $h \in \mathscr{H}$:

$$\|\tilde{a}_{12}^\kappa h - bh\|_{\mathscr{H}}^2 \leq \left\| (\tilde{a}^\kappa - \tilde{b}) \begin{pmatrix} 0 \\ h \end{pmatrix} \right\|_{\mathscr{H}^{\oplus 2}}^2 \to 0$$

so $\tilde{a}_{12}^\kappa \xrightarrow[\kappa \to \infty]{s} b$. Because of Lemma 4.2.5 $\|\tilde{a}_{12}^\kappa\| \leq \|\tilde{a}^\kappa\| \leq 1$, so $\tilde{a}_{12}^\kappa \in \mathcal{B}_1(\mathscr{A}), \forall \kappa$.

- Proving that if $\mathbb{1}_{\mathcal{H}} \in \mathcal{A}$ then the unitaries of $\mathcal{A}$ are strongly dense in the unitaries of $\mathcal{B}$: Considering a unitary $u \in \mathcal{B}$, follows from Stone's theorem 4.2.9 that there is a self-adjoint operator $b \in \mathcal{B}$, such that $u = e^{ib}$, by the first item, there is a net $\{a_\kappa\}_\kappa$ of self-adjoint operators in $\mathcal{A}$ such that $a_\kappa \xrightarrow{s} b$, and since by lemma 4.2.8 the map $t \mapsto e^{it}$ is strongly continuous, then $e^{ia_\kappa} \xrightarrow{s} e^{ib} = u$, where, as we have seen in theorem 4.2.9, $e^{ia_\kappa}$ is unitary for every $\kappa$, since each $a_\kappa$ is self-adjoint.

∎

**Proposition 4.2.11.** *If $K$ is a $*$-invariant subset of $\mathfrak{B}(\mathcal{H})$, i.e. if $A \in K$ then $A^* \in K$, then $K'$ is a von Neumann algebra.*

*Proof:* If $T \in K'$ and $B \in K$, which implies $B^* \in K$, then $TB^* = B^*T$, by taking the adjoint on both sides $BT^* = T^*B$, as $T^*$ commutes with any element of $K$ then $T^* \in K'$, this proves that $K'$ is also a $*$-invariant set, since $\mathbb{1}_{\mathcal{H}}$ commutes with all elements of $\mathfrak{B}(\mathcal{H})$, then it also commutes with all elements of $K$, hence $\mathbb{1}_{\mathcal{H}} \in K'$. More than that, $K'$ is a unital $*$-algebra, since for any $T_1, T_2 \in K'$, $\lambda \in \mathbb{C}$ and $B \in K$, $T_1B = BT_1 \Rightarrow \lambda T_1 B = B\lambda T_1$, beyound that since $T_2B = BT_2$ then $B(T_1 + T_2) = (T_1 + T_2)B$ and $(T_1T_2)B = T_1BT_2 = B(T_1T_2)$ follow, this proves $K'$ is closed under sum, scalar multiplication and multiplication so it's a unital $*$-algebra.

Finally, we assume that $T_\alpha$ is in a net in $K'$ that converges weakly to some $T \in \mathfrak{B}(\mathcal{H})$. Then for all $u, v \in \mathcal{H}$ and $B \in K$, $\langle u, (T_\alpha B - BT_\alpha)v \rangle_{\mathcal{H}} = 0$. Now, $\langle u, T_\alpha Bv \rangle_{\mathcal{H}} \xrightarrow{\alpha} \langle u, TBv \rangle_{\mathcal{H}}$, because $T_\alpha$ converges weakly to $T$, similarly $\langle u, BT_\alpha v \rangle_{\mathcal{H}} = \langle B^*u, T_\alpha v \rangle_{\mathcal{H}} \xrightarrow{\alpha} \langle B^*u, Tv \rangle_{\mathcal{H}} = \langle u, BT_\alpha v \rangle_{\mathcal{H}}$, it then follows that $\langle u, (TB - BT)v \rangle_{\mathcal{H}} = 0$, since this is so for all $u, v \in \mathcal{H}$, $TB = BT, \forall B \in K$; then this proves that $T \in K'$, making $K'$ a weak-operator closed $*$-invariant subspace of $\mathfrak{B}(\mathcal{H})$, hence a von Neumann algebra. ∎

The following Theorem has been originally proven by J. Glimm and R. Kadison in [49], and has useful consequences for characterizing unitarily inequivalent pure states.

**Theorem 4.2.12.** *If $\mathcal{A}$ is a $C^*$-algebra acting irreducibly upon the Hilbert space $\mathcal{H}$ and $\{x_i\}_i, \{y_i\}_i$ are two finite sets of vectors with $i = 1, ..., n$, the first set consisting of linearly independent vectors, then there exists an operator $A$ in $\mathcal{A}$ such that $Ax_i = y_i$. In particular, A acts algebraically irreducibly upon $\mathcal{H}$. If $Bx_i = y_i$ for some self-adjoint operator B, then a self-adjoint operator A in $\mathcal{A}$ may be chosen such that $Ax_i = y_i$.*

*Proof:* Choosing an orthonormal basis for the space generated by $\{x_i\}_i$, there will be a set of $n$ vectors such that the class of operators mapping the basis onto the original set of linearly independent vectors, coincides with the class that maps the $x$'s onto the $y$'s. It suffices, therefore, to deal with the case where $\{x_i\}$ is an orthonormal set.

Noting that if $z_1, ..., z_n$ are vectors such that $\|z_i\| \leq r$, there is an operator $B$ of norm not exceeding $(2n)^{\frac{1}{2}}r$ (and self-adjoint, in case $z_i = Tx_i$, for some self-adjoint operator $T$) such that $Bx_i = z_i$. In fact, let $x_1, ..., x_n, x_{n+1}, ..., x_m$ be an extension of the set $x_1, ..., x_n$ to an orthonormal basis of the space $\mathcal{V}$ generated by $x_1, ..., x_n, z_1, ..., z_n$.

With $T$ and $S$ linear transformations on $\mathcal{V}$, the mapping $(T, S) \mapsto \text{tr}(S^*T)$ defines an inner product and hence a norm: $\|T\|_{\mathcal{V}} = (\text{tr}(T^*T))^{\frac{1}{2}}$, of $T$ is $\left(\sum_{i,j}^m |\alpha_{ij}|^2\right)^{\frac{1}{2}}$, where the $\alpha_{ij}$ are

the matrix coordinates of $T$ relative to any orthonormal basis of $\mathcal{V}$. The matrix whose first $n$ columns (and conjugate first $n$ rows, in the self-adjoint case) are $z_1, \ldots, z_n$; or, rather, their co-ordinate representation, relative to $x_1, \ldots, x_m$, and zeros elsewhere, represents an operator $B$ such that $Bx_i = z_i$, and $\|B\| \leq (2n)^{\frac{1}{2}}$. It can be proven that $\|B\| \leq \|B\|_{\mathcal{V'}}$, where, as always $\|\cdot\|$ denotes to usual operator norm, the proof of this fact is done by considering the $B^*$ condition, $\|A\|^2 = \|A^*A\| = \||A|^2\| = \||A|\|^2 = \left( \sup_{\substack{\varphi \in \mathcal{H} \\ \|\varphi\|_{\mathcal{H}} = 1}} |\langle \varphi, |A| \varphi \rangle| \right)^2 \leq \operatorname{tr}(|A|)^2$, hence $\|A^*A\| \leq \operatorname{tr}(|A^*A|) = \operatorname{tr}(A^*A) = \|A\|_{\mathcal{V'}}^2$. Defining $B$ to be 0 on the orthogonal complement of $\mathcal{V}$, we have an operator with the desired properties.

We proceed now to the construction of an operator $A$ in $\mathscr{A}$ such that $Ax_i = y_i$. Choose $B_0$ such that $B_0 x_i = y_i$, and $A_0$ in $\mathscr{A}$ such that $\|A_0 x_i - B_0 x_i\| = \|A_0 x_i - y_i\| < \left(2(2n)^{\frac{1}{2}}\right)^{-1}$. (Recall that since $\mathcal{H}$ is irreducible, its strong closure is the set of all bounded operators on $\mathcal{H}$). Next choose $B_1$, such that $B_1 x_i = y_i - A_0 x_i$, with $\|B_1\| < \frac{1}{2}$. Kaplansky's theorem 4.2.10 guarantees the existence of an operator $A_1$ in $\mathscr{A}$ with $\|A_1\| < \frac{1}{2}$. such that $\|A_1 x_i - B_1 x_i\| < \left(4(2n)^{\frac{1}{2}}\right)^{-1}$.

Suppose now that $B_k$ has been constructed with $\|B_k\| < \frac{1}{2^k}$, $B_k x_i = y_i - A_0 x_i - A_1 x_i - \ldots - A_{k-1} x_i$. Choosing $A_k \in \mathscr{A}$ with $\|A_k\| \leq \frac{1}{2^k}$ such that $\|A_k x_i - B_k x_i\| < \frac{1}{2^{k+1}(2n)^{\frac{1}{2}}}$, and $B_{k+1}$ with $\|B_{k+1}\| \leq \frac{1}{2^{k+1}}$, with $B_{k+1} x_i = y_i - A_0 x_i - A_1 x_i - \ldots - A_k x_i$. Note that if $Tx_i = y_i$, for some self-adjoint operator $T$, the results of the third paragraph of this proof allows us to choose self-adjoint $B_k$'s and hence, by Kaplansky's theorem 4.2.10, self-adjoint $A_k$'s. The sum $\sum_{k=0}^{\infty} A_k$ converges in norm to an operator $A$ in $\mathscr{A}$, and:

$$y_i - Ax_i = y_i - \sum_{k=0}^{\infty} A_k x_i = \lim_{k \to \infty} (y_i - A_0 x_i - \ldots - A_k x_i) = \lim_{k \to \infty} B_{k+1} x_i = 0.$$

∎

**Theorem 4.2.13.** *If $\mathscr{A}$ is a $C^*$-algebra, whose representation $\pi$ acts irreducibly on $\mathcal{H}$ and $V$ is a unitary operator on $\mathcal{H}$ such that $Vx_k = y_k$, for vectors $\{x_k\}_k$ and $\{y_k\}_k$ of $\mathcal{H}$ with $k = 1, \ldots, n$; then there is a unitary operator $U \in \mathscr{A}$ such that $Ux_k = y_k$ and $\sigma(U) \neq \mathbb{S}(\mathbb{C}) := \{\lambda \in \mathbb{C} \mid |\lambda| = 1\}$.*

*Proof:* We can pass to a orthonormal basis for the finite-dimensional space generated by $\{x_1, \ldots, x_n\}$ since if $x_k$ is linearly dependent to the rest of the $\{x_j\}_j$ so will be $y_k$ by $V$ to the rest of $\{y_j\}_j$, so we can, without loss of generality, assume that $\{x_k\}_k$ is a orthonormal set of vectors, and so is $\{y_k\}_k$. Considering a unitary extension of $V$ acting on $\{x_1, y_1, \ldots, x_n, y_n\}$, which by the same argument as above can be assumed to be a orthonormal set of vectors, by $\tilde{V} x_k = V x_k = y_k$ and $\tilde{V} y_k = V^* y_k = x_k$, then considering a diagonalizing basis of $\{x_1, y_1, \ldots, x_n, y_n\}$ for the unitary operator $\tilde{V}$, we see that it suffices to consider $\tilde{V} z_k = \beta_k z_k$, with $|\beta_k| = 1$ and $z_k$, in this diagonalizing basis. Choosing real $\alpha_j$ in the half open interval $(-\pi, \pi]$ such that $e^{i\alpha_j} = \beta_j$ and letting $A$ be a self-adjoint operator in $\mathscr{A}$ such that $Az_j = \alpha z_j$, such a operator exists because of theorem 4.2.12.

Defining $g : \mathbb{R} \to \mathbb{R}$ as:

$$g(x) := \begin{cases} \min(\{\alpha_j\}_j) & \text{, for } x < \min(\{\alpha_j\}_j) \\ x & \text{, for } \min(\{\alpha_j\}_j) \leq x \leq \max(\{\alpha_j\}_j) \\ \max(\{\alpha_j\}_j) & \text{, for } x > \max(\{\alpha_j\}_j) \end{cases}$$

since $\overline{g} = g$, $\|g\|_\infty \leq \max(\{\alpha_j\}_j)$ then as we already argued in the Kaplansky density theorem 4.2.10, then $g(A)$ is a self-adjoint operator in $\mathscr{A}$ with $\sigma(g(A)) \subseteq [\min(\{\alpha_j\}_j), \max(\{\alpha_j\}_j)]$, and by construction $g(A)z_j = \alpha_j z_j$. It follows that $U := e^{ig(A)}$ is a unitary operator in $\mathscr{A}$, with $\sigma(U) \neq \mathbb{S}(\mathbb{C})$ and $Uz_j = \beta_j z_j$. ∎

**Lemma 4.2.14.** *If $\{\pi_\alpha\}_\alpha$ are $*$-representations of a self-adjoint operator algebra $\mathscr{A}$, then $\{\pi_\alpha\}_\alpha$ consists of mutually disjoint representations if and only if $\overline{\pi(\mathscr{A})}^s = \bigoplus_\alpha \overline{\pi_\alpha(\mathscr{A})}^s$, where $\pi = \bigoplus_\alpha \pi_\alpha$.*

*Proof:* We suppose that $\pi_\alpha(\mathscr{A})$ acts on a Hilbert space $\mathscr{H}_\alpha$, that $\mathscr{H} = \bigoplus_\alpha \mathscr{H}_\alpha$ and that $P_\alpha$ is the orthogonal projector of $\mathscr{H}$ onto $\mathscr{H}_\alpha$.

If $\overline{\pi(\mathscr{A})}^s = \bigoplus_\alpha \overline{\pi_\alpha(\mathscr{A})}^s$, and $U$ is a partial isometry, *i.e.* a operator such that $UU^*$ and $U^*U$ are projectors, $U$ maps a initial subspace $I$ into a final subspace $\mathcal{F}$ and $\ker(U) = I^\perp$, with $U^*U = P_I$ and $UU^* = P_\mathcal{F}$, $\|U\psi\|_\mathscr{H} = \|\psi\|_\mathscr{H}$ for $\psi \in I \subset \mathscr{H}$; of a subspace $E_\alpha(\mathscr{H}_\alpha)$ into another subspace $E_\beta(\mathscr{H}_\beta)$, with $\alpha \neq \beta$ and such that $E_\alpha$ and $E_\beta$ are projectors into exactly those subspaces that make $U$ intertwine $\pi_\alpha(\mathscr{A})$ with $\pi_\beta(\mathscr{A})$, so that $U\pi_\alpha(A)U^* = \pi_\beta(A)E_\beta$, for each $A \in \mathscr{A}$, as a consequence $U$ commutes with $\pi(\mathscr{A})$. In fact, $U\pi(A) = U\pi_\alpha(A) = \pi_\beta(A)U = \pi(A)U$. Thus $U$ commutes with $\bigoplus_\alpha \pi_\alpha(\mathscr{A})$, and, in particular with each $P_\alpha$. But then $UP_\alpha = U = P_\alpha U = 0$; so that $0 = E_\alpha = E_\beta$ and $\{\pi_\alpha\}_\alpha$ consists of mutually disjoint representations.

Now we assume instead that $\{\pi_\alpha\}_\alpha$ are mutually disjoint representations, and $V$ is a partial isometry in the commutant of $\pi$, with initial space given by the projector $E\alpha$ within $P_\alpha$ and final space $E_\beta$ within $P_\beta$; then $VE_\alpha\pi(A)E_\alpha V^* = VV^*V\pi(A)V^*VV^* = E_\beta\pi(A)VV^*E_\beta = E_\beta\pi_\beta(A)E_\beta$ for each $A \in \mathscr{A}$. That would mean, for non-zero subspaces $E_\alpha$, $E_\beta$; unitarily equivalent subrepresentations, thus, by disjointedness, $E_\alpha = E_\beta = 0$. It follows that the central carrier of $P_\alpha$, given by $C(P_\alpha) = \bigwedge\{Q \in \mathcal{Z}(\pi(\mathscr{A})') \,|\, F \geq P_\alpha\}$, where $\mathcal{Z}(\pi(\mathscr{A})') = \pi(\mathscr{A})' \cap \pi(\mathscr{A})''$ is the center of $\pi(\mathscr{A})'$ and the operation $\wedge$ is such that for any two projectors $G$ and $H$, $G \wedge H$ is the projector onto the subspace $\text{Ran}(G) \cap \text{Ran}(H)$, as seen previously; resuming, the central carrier of $P_\alpha$ needs to be orthogonal to the central carrier of $P_\beta$, for $\alpha \neq \beta$, otherwise we could take $E_\alpha$ to be $C(P_\alpha) \wedge C(P_\beta) \neq 0$ which would violate disjointedness.

Since the central carrier of $P_\alpha$ is orthogonal to the central carrier of $P_\beta$, $\alpha \neq \beta$, implies in particular that it is orthogonal to $P_\beta$. Since $\sum_\alpha P_\alpha = \mathbb{1}_\mathscr{H}$, and the central carrier of $P_\alpha$ projects into a subspace that contains $\text{Ran}(P_\alpha)$, then $P_\alpha$ must be its own central carrier, as it cannot include any of the rages of the other $P_\beta$'s ($\beta \neq \alpha$) that together complete the space $\mathscr{H}$. In particular, $P_\alpha$ lies in the center of the commutant, $\mathcal{Z}(\pi(\mathscr{A})')$, by considering that $\mathcal{Z}(\pi(\mathscr{A})') = \pi(\mathscr{A})' \cap \pi(\mathscr{A})''$, then $P_\alpha \in \pi(\mathscr{A})''$ and by the bicommutant theorem 4.0.11, $P_\alpha \in \overline{\pi(\mathscr{A})}^s$. It is immediate from this that $\overline{\pi(\mathscr{A})}^s = \bigoplus_\alpha \overline{\pi_\alpha(\mathscr{A})}^s$. ∎

**Corollary 4.2.15.** *If $\{\pi_\alpha\}_\alpha$ is a family of irreducible $*$-representations of a self-adjoint operator algebra $\mathscr{A}$, no two of which are unitarily equivalent, then $\overline{\pi(\mathscr{A})}^s = \bigoplus_\alpha \mathfrak{B}(\mathscr{H}_\alpha)$ where $\pi = \bigoplus_\alpha \pi_\alpha$ and $\mathfrak{B}(\mathscr{H}_\alpha)$ is the algebra of all bounded operators on the representation space of $\pi_\alpha$.*

*Proof:* If we consider $\pi_\alpha$ to be an irreducible representation, then for a projector $P$ that projects into the subspace $\pi_\alpha(\mathscr{A})'\mathscr{H}$, then $P\mathscr{H}$ is a space invariant under $\pi(\mathscr{A})$, as every $\psi \in P\mathscr{H}$ is such that $\psi = F\phi$, for some $F \in \pi(\mathscr{A})'$ and $\phi \in \mathscr{H}$, then $\pi(A)\psi = \pi(A)F\phi = F\pi(A)\phi \in P\mathscr{H}$. Since $\pi$ is irreducible $P\mathscr{H}$ must be either $\mathbf{0}$ or $\mathscr{H}$, which in turn implies that $P = 0$ or $P = \mathbb{1}$. Hence, the only projectors in $\pi(\mathscr{A})'$ are 0 and $\mathbb{1}$, and since $\pi(\mathscr{A})'$ is a von Neumann algebra by proposition 4.2.11, and as such it is generated by the double commutant of its projector lattice by theorem 6.1.8, hence $\pi(\mathscr{A})' = \mathbb{C}\mathbb{1}$.

As projectors in the commutant of $\pi((A))$ correspond to invariant subspaces, since by proposition A.1.4 any projection into an invariant subspace $P$ commutes with $\pi(A), \forall A \in \mathscr{A}$, then $P \in \pi((A))'$, these invariant subspaces give rise to subrepresentations as by definition A.1.3, therefore any two such irreducible representations are either unitarily equivalent or disjoint, since the only projectors in $\pi(\mathscr{A})' = \mathbb{C}\mathbb{1}$ are $\mathbf{0}$ and $\mathbb{1}$. This result combined with lemma 4.2.14 gives immediately a proof of the corollary. ∎

**Lemma 4.2.16.** *If $\pi$ is a $*$-representation of the $C^*$-algebra $\mathscr{A}$ and $U$ is a unitary operator in $\pi(\mathscr{A})$ with $\sigma(U) \neq \mathbb{S}(\mathbb{C})$, there is a unitary element $U_0$ in $\mathscr{A}$ such that $\pi(U_0) = U$.*

*Proof:* Considering the function $\arg : \mathbb{C} \to (-\pi, \pi)$ that gives the angle of complex numbers in the Gauss-Argand plane representation of complex numbers (the polar representation), such that for a $z \in \mathbb{C}$, $|z| = 1$, $z = e^{i\arg(z)}$, then letting $f$ be a continuous extension of arg to $\mathbb{S}(\mathbb{C})$, then for a unitary operator $U \in \pi(\mathscr{A})$, $U = e^{if(U)}$ by the continuous functional calculus for normal operators (ref. [77]), and since $f$ has image only in the real numbers, then $f(U)$ is self-adjoint (since $\widehat{\Phi}_U(f)^* = \widehat{\Phi}_U(\overline{f})$).

Let $A$ be a self-adjoint element in $\mathscr{A}$ such that $\pi(A) = f(U)$ (Recall that $\pi(\mathscr{A})$ is a $C^*$-algebra, so that $f(U)$ lies in $\pi(\mathscr{A})$ by the continuous functional calculus theorem 4.0.16. If it is $\pi(B) = f(U)$, then $A$ may be chosen as $\frac{1}{2}(B^* + B))$. If we take $U_0 := e^{iA}$, then $\pi(U_0) = e^{if(U)} = U$, by the uniform continuity of $\pi$, proposition A.1.12. ∎

**Corollary 4.2.17.** *If $\{\pi_\alpha\}$ is a family of unitarily inequivalent irreducible $*$-representations of the $C^*$-algebra $\mathscr{A}$ on Hilbert spaces $\{\mathscr{H}_\alpha\}_\alpha$, $\pi$ is the direct sum of $\{\pi_\alpha\}_\alpha$, $\mathscr{H}$ the direct sum of $\{\mathscr{H}_\alpha\}_\alpha$, $\{x_1, ..., x_n\}$ and $\{y_1, ..., y_n\}$ are two finite sets of vectors with $\{x_1, ..., x_n\}$ linearly independent and each $X_j$ and corresponding $y_j$ in some $\mathscr{H}_\alpha$; then there is an $A$ in $\mathscr{A}$ such that $\pi(A)X_j = y_j$. If $Bx_j = y_j$ for some self-adjoint or unitary operator $B$ on $\mathscr{H}$ then $A$ may be chosen self-adjoint or unitary, respectively.*

*Proof:* The argument of theorem 4.2.12 applies directly to the first assertion once we note that the general constructions and norm estimates of that theorem can be performed on each $\mathscr{H}_\alpha$, since each $x_j, y_j$ lie in some $\mathscr{H}_\alpha$ and the strong approximations are valid by virtue of corollary 4.2.15. With $B$ self-adjoint, each $P_\alpha B P_\alpha$ is trivially self-adjoint and $P_\alpha B P_\alpha x_j = y_j$ (for $x_j, y_j \in \mathscr{H}_\alpha$), where $P_\alpha$ is the projection of $\mathscr{H}$ onto $\mathscr{H}_\alpha$, so that the argument of 4.2.12, in the self-adjoint case, applies to give a self-adjoint operator $\pi(A)$ such that $\pi(A)x_j = y_j, j = 1, ..., n$. Of course, $A$ may be chosen self-adjoint in this case. If $B$ is unitary it can be replaced by one which maps each $\mathscr{H}_\alpha$ onto itself and acts in the same way on $\{x_j, ..., x_n\}$ (extend the mappings of $x_j$ onto $y_j$ on each $\mathscr{H}_\alpha$).

Having the self-adjoint result, in this case, the argument of theorem 4.2.13 now applies to give a unitary operator $\pi(U)$ such that $\pi(U)x_j = y_j, j = 1, ..., n$ and $\sigma(\pi(U)) \neq \mathbb{S}(\mathbb{C})$. From lemma 4.2.16, $U$ may be chosen as a unitary operator in $\mathscr{A}$. ∎

**Lemma 4.2.18.** *If $\varrho, \varsigma$ are two pure states of a $C^*$-algebra $\mathscr{A}$, such that $\varrho$ and $\varsigma$ do not have associated unitarily equivalent representations of $\mathscr{A}$, then $\|\varrho - \varsigma\| = 2$.*

*Proof:* Considering that $\pi_\varrho$ and $\pi_\varsigma$ are unitarily inequivalent and $\pi = \pi_\varrho \oplus \pi_\varsigma$, their direct sum, represents $\mathscr{A}$ on the direct sum $\mathscr{H} = \mathscr{H}_\varrho \oplus \mathscr{H}_\varsigma$, then there are unit vectors $x$ and $y$ respectively in $\mathscr{H}_\varrho$ and $\mathscr{H}_\varsigma$, such that $\omega_x \circ \pi = \langle x, \pi(\cdot)x \rangle = \varrho$ and $\omega_y \circ \pi = \langle y, \pi(\cdot)y \rangle = \varsigma$ (i.e. $x$ and $y$ are respectively the cyclic vectors of the GNS representations for the states $\varrho$ and $\varsigma$). According to corollary 4.2.17, we can find $U$ in $\mathscr{A}$ such that $\pi(U)x = x$ and $\pi(U)y = -y$. Then $|(\varrho - \varsigma)(U)| = |\langle x, \pi(U)x \rangle - \langle y, \pi(U)y \rangle| = |\langle x, x \rangle + \langle y, y \rangle| = 2$, so that $\|\varrho - \varsigma\| = 2$. ∎

**Proposition 4.2.19.** *Let $\pi : \mathscr{A} \to \mathfrak{B}(\mathscr{H})$ be a cyclic representation of $\mathscr{H}$. If $\psi \in \mathscr{H}$ is a cyclic unit vector for $\pi$, then, for each $a \in \mathscr{A}$*

$$\omega(a) = \langle \psi, \pi(a)\psi \rangle \tag{4.4}$$

*is a state on $\mathscr{A}$, whose GNS representation $\pi_\omega$ is unitarily equivalent to $\pi$.*

*Proof:* Define $u : \mathscr{H}_\omega \to \mathscr{H}$ first on $\pi_\omega(\mathscr{A})\varphi_\omega$ (which is a dense subspace of $\mathscr{H}_\omega$) by

$$u\pi_\omega(a)\varphi_\omega := \pi(a)\psi \qquad , a \in \mathscr{A}$$

Using then the GNS representation and (4.4), we obtain:

$$\|\pi_\omega(a)\varphi_\omega\|^2 = \omega(a^*a) = \langle \psi, \pi(a^*a)\psi \rangle = \|\pi(a)\psi\|^2$$

This shows that $u$ is well defined as well as isometric, so that it extends to $\mathscr{H}_\omega$ by continuity. Its image is then the closure of $\pi(\mathscr{A})\psi$, which is $\mathscr{H}$, since $\psi$ is cyclic by assumption. Thus $u$ is surjective and hence unitary. Finally, we compute, for $a, b \in \mathscr{A}$

$$u\pi_\omega(a)\pi_\omega(b)\varphi_\omega = \pi(a)\pi(b)\psi = \pi(a)u\pi_\omega(b)\varphi_\omega$$

so that $u\pi_\omega(a) = \pi(a)u$ on the dense space $\pi_\omega(\mathscr{A})\varphi_\omega$, and hence everywhere. ∎

## 4.3 Measurement Scheme

For the measurement scheme we consider the following notation for the system being measured, and the probe that measures:

- System: $S$.
- Probe: $P$.

We must also consider the types of system and Probe Interactions, in the Fewster-Verch measurement scheme (sometimes referred to as the FV *framework*) measurements are required to take place in a compact region of spacetime, this restriction is both a simplifying condition and a expected requirement for exactness in the measurements, as will be commented in the conclusion with the results of chapter 7.

Then measurements of observables in the probe are representative of measurements of observables induced in the system, these two assertions collectively imply that the system and probe interactions occur only in a compact region of spacetime.

Of particular interest in a measurement scheme are the following parts of the measurement process.

- Probe.

- Interaction dynamics.

- Induced observables in the system.

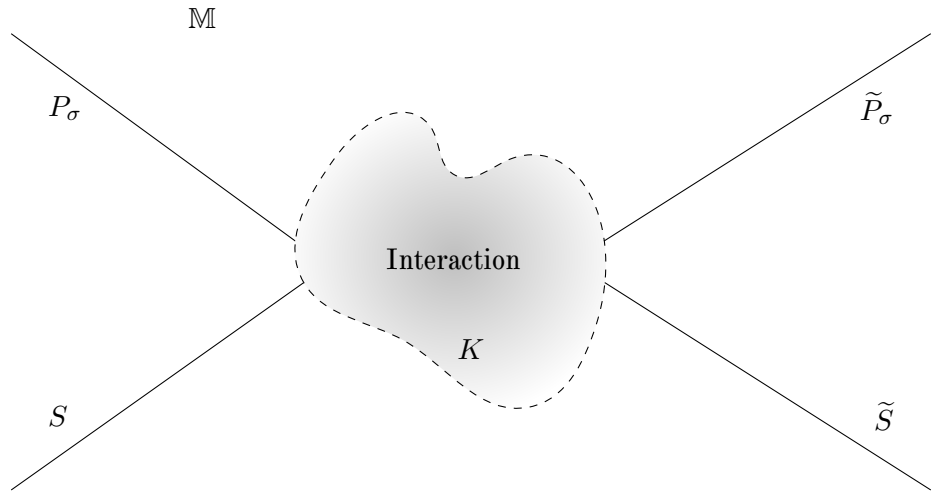

Figure 9 — Depiction of a measurement realized on a compact interaction region $K$ as a scattering of a probe, prepared in a state $\sigma$, $P_\sigma$ and the quantum system to be measured $S$. After the measurement interaction they become $\widetilde{P}_\sigma$ and $\widetilde{S}$.

We should expect that long before the interaction and long after the interaction, the fields of the system and the probe will be independent of each other, so observables associated with one will have to switch with observables of the other.

A general framework would be useful in which relations of local observables were made explicit and which already had results that could be used in the analysis of the behavior of local measurement schemes.

Considering two QFT's $\mathscr{A}$ representing the system and $\mathscr{B}$ the probe, being $\alpha_{\mathbb{M};D}$ and $\beta_{\mathbb{M};D}$ the respective $*$-homomorphisms acting as inclusion maps.

The combination of these theories in a non-interacting way can then be performed by $\mathscr{U} = \mathscr{A} \otimes \mathscr{B}$, which every time space $\mathbb{M}$ receives the algebra $\mathscr{U}(\mathbb{M}) = \mathscr{A}(\mathbb{M}) \otimes \mathscr{B}(\mathbb{M})$ and the local subalgebras $\mathscr{U}(\mathbb{M};D) = \mathscr{A}(\mathbb{M};D) \otimes \mathscr{B}(\mathbb{M};D)$ with inclusion maps given by $\alpha_{\mathbb{M};D} \otimes \beta_{\mathbb{M};D}$.

A measurement experiment will then be described by another theory $\mathscr{C}$ in which system and probe are in interaction, this theory in turn with inclusion maps $\gamma_{\mathbb{M};D}$.

We assume that the interaction occurs only in a compact spacetime region $K$. Which means that $\mathscr{C}$ must reduce to $\mathscr{A} \otimes \mathscr{B}$ outside the causal envelope $J^+(K) \cap J^-(K)$. That is, for every causally convex open set $\mathbb{L} \subseteq \mathbb{M} \setminus (J^+(K) \cap J^-(K))$ there is an isomorphism:

$$\chi_\mathbb{L} : \mathscr{A}(\mathbb{L}) \otimes \mathscr{B}(\mathbb{L}) \to \mathscr{C}(\mathbb{L}),$$

which has a compatibility with the local structures of both theories through the following relationship:

If $\mathbb{O} \subset \mathbb{L}$ and both open sets are causally convex in $\mathbb{M} \setminus (J^+(K) \cap J^-(K))$ then the following diagram commutes:

$$\begin{array}{ccc} \mathscr{A}(\mathbb{O}) \otimes \mathscr{B}(\mathbb{O}) & \xrightarrow{\alpha_{\mathbb{L};\mathbb{O}} \otimes \beta_{\mathbb{L};\mathbb{O}}} & \mathscr{A}(\mathbb{L}) \otimes \mathscr{B}(\mathbb{L}) \\ \chi_\mathbb{O} \downarrow & & \downarrow \chi_\mathbb{L} \\ \mathscr{C}(\mathbb{O}) & \xrightarrow{\gamma_{\mathbb{L};\mathbb{O}}} & \mathscr{C}(\mathbb{L}) \end{array}$$

this means:

$$\chi_\mathbb{L} \circ \alpha_{\mathbb{L};\mathbb{O}} \otimes \beta_{\mathbb{L};\mathbb{O}} = \gamma_{\mathbb{L};\mathbb{O}} \circ \chi_\mathbb{O} \tag{4.5}$$

Compact sets $K$ of interaction regions naturally define input $(-)$ and output $(+)$ regions: $\mathbb{M}^\pm = \mathbb{M} \setminus J^\mp(K)$ which are causally convex open regions that together cover the exterior of $J^+(K) \cap J^-(K)$. As you can see these regions are defined covariantly once the interaction region is specified.

It is expected from intuition that the regions $\mathbb{M}^\pm$ contain Cauchy surfaces of $\mathbb{M}$, and this is in fact true with a proof given in lemma A.4 of [39], therefore by the Haag-Kastler axioms $\alpha_{\mathbb{M};\mathbb{M}^\pm}$, $\beta_{\mathbb{M};\mathbb{M}^\pm}$, $\gamma_{\mathbb{M};\mathbb{M}^\pm}$ and $\chi_{\mathbb{M}^\pm}$ are isomorphisms, just like the compositions:

$$\kappa_{\mathbb{M};\mathbb{M}^\pm} := \gamma_{\mathbb{M};\mathbb{M}^\pm} \circ \chi_{\mathbb{M}^\pm} : \mathscr{A}(\mathbb{M}^\pm) \otimes \mathscr{B}(\mathbb{M}^\pm) \to \mathscr{C}(\mathbb{M}) \tag{4.6}$$

Using these maps we can define delayed $(-)$ and advanced $(+)$ response maps, $\tau^\pm : \mathscr{A}(\mathbb{M}) \otimes \mathscr{B}(\mathbb{M}) \to \mathcal{C}(\mathbb{M})$:

$$\tau^\pm \equiv \tau_{\mathbb{M};\mathbb{M}^\pm} := \kappa_{\mathbb{M};\mathbb{M}^\pm} \circ \left(\alpha_{\mathbb{M};\mathbb{M}^\pm} \otimes \beta_{\mathbb{M};\mathbb{M}^\pm}\right)^{-1}$$

which identifies the uncoupled system with the coupled system at earlier $(-)$ or later $(+)$ times.

And then combining these maps we can form the scattering morphism $\Theta : \mathscr{A}(\mathbb{M}) \otimes \mathscr{B}(\mathbb{M}) \to \mathscr{A}(\mathbb{M}) \otimes \mathscr{B}(\mathbb{M})$:

$$\Theta := \left(\tau_{\mathbb{M};\mathbb{M}^-}\right)^{-1} \circ \tau_{\mathbb{M};\mathbb{M}^+},$$

which is an automorphism of $\mathscr{A}(\mathbb{M}) \otimes \mathscr{B}(\mathbb{M})$ and note that $\Theta$ maps elements of the algebra of later $(+)$ times into elements of previous $(-)$ times.

Fewster and Verch go on to prove the following regularity properties of the scattering morphism in an appendix:

**Proposition 4.3.1.** *The following assertions are true about the scattering morphism $\Theta$:*

*(a) Let $\hat{K}$ be a compact set containing the matching region $K$ and $\hat{\Theta}$ the scattering morphism obtained by replacing $K$ by $\hat{K}$ in the definition of $\Theta$. So $\hat{\Theta} = \Theta$.*

*(b) Let $\mathbb{L}$ be a causally convex open subset of $K^\perp = \mathbb{M}^+ \cap \mathbb{M}^-$, then $\Theta$ acts trivially on $\mathscr{A}(\mathbb{M};\mathbb{L}) \otimes \mathscr{B}(\mathbb{M};\mathbb{L})$. (i.e. $\Theta \circ (\alpha_{\mathbb{M};\mathbb{L}} \otimes \beta_{\mathbb{M};\mathbb{L}}) = \alpha_{\mathbb{M};\mathbb{L}} \otimes \beta_{\mathbb{M};\mathbb{L}}$).*

*(c) Suppose that $\mathbb{L}^+$ (respectively $\mathbb{L}^-$) is a causally convex open subset of $\mathbb{M}^+$ (resp. $\mathbb{M}^-$) and that $\mathbb{L}^+ \subset D(\mathbb{L}^-)$ (resp. $L^- \subset D(L^+)$), where $D(\mathbb{L}^-)$ is the Cauchy development of $\mathbb{L}^-$. Then $\Theta\left(\mathscr{A}(\mathbb{M};\mathbb{L}^+) \otimes \mathscr{B}(\mathbb{M};\mathbb{L}^+)\right) \subset \mathscr{A}(\mathbb{M};\mathbb{L}^-) \otimes \mathscr{B}(\mathbb{M};\mathbb{L}^-)$.*

*Proof:* In the proofs we shall use the following short hand notation for the morphisms $\alpha^\pm \equiv \alpha_{\mathbb{M};\mathbb{M}^\pm}$, $\beta^\pm \equiv \beta_{\mathbb{M};\mathbb{M}^\pm}, \ldots$; $\kappa^\pm \equiv \kappa_{\mathbb{M};\mathbb{M}^\pm}$ and $\chi^\pm \equiv \chi_{\mathbb{M}^\pm}$, and $\widehat{\alpha}^\pm \equiv \alpha_{\widehat{\mathbb{M}};\widehat{\mathbb{M}}^\pm}$, $\widehat{\beta}^\pm \equiv \beta_{\widehat{\mathbb{M}};\widehat{\mathbb{M}}^\pm}, \ldots$; $\widehat{\kappa}^\pm \equiv \kappa_{\widehat{\mathbb{M}};\widehat{\mathbb{M}}^\pm}$ and $\widehat{\chi}^\pm \equiv \chi_{\widehat{\mathbb{M}}^\pm}$

(a) Note that $\widehat{\mathbb{M}}^\pm = \mathbb{M} \setminus J^\mp(\widehat{K})$ are subsets of $\mathbb{M}^\pm$, giving $(\alpha^\pm)^{-1} \circ \widehat{\alpha}^\pm = \alpha_{\mathbb{M}^\pm;\widehat{\mathbb{M}}^\pm}$ by the rule for consecutive inclusions (4.3) and the time-slice property (4). We may then calculate:

$$\kappa^\pm \circ (\alpha^\pm \otimes \beta^\pm)^{-1} \circ (\widehat{\alpha}^\pm \otimes \widehat{\beta}^\pm) \stackrel{(4.6)}{=} \gamma^\pm \circ \chi^\pm \circ (\alpha_{\mathbb{M}^\pm;\widehat{\mathbb{M}}^\pm} \otimes \beta_{\mathbb{M}^\pm;\widehat{\mathbb{M}}^\pm}) \stackrel{(4.5)}{=} \gamma^\pm \circ \gamma_{\mathbb{M}^\pm;\widehat{\mathbb{M}}^\pm} \circ \widehat{\chi}^\pm = \widehat{\gamma}^\pm \circ \widehat{\chi}^\pm \stackrel{(4.6)}{=} \widehat{\kappa}^\pm.$$

were the relation $(\nu^\pm)^{-1} \circ \widehat{\nu}^\pm = \nu_{\mathbb{M}^\pm;\widehat{\mathbb{M}}^\pm}$ was used in the first and third equalities with the morphisms $\alpha \otimes \beta$ and $\gamma$, in this last case in the form $\widehat{\gamma}^\pm = \gamma^\pm \circ \gamma_{\mathbb{M}^\pm;\widehat{\mathbb{M}}^\pm}$.

It follows that $\tau^\pm = \widehat{\tau}^\pm$ and hence $\widehat{\Theta} = \Theta$.

(b) First suppose that $\mathbb{L} \subset \mathbb{M}^\pm$ is an open causally convex set. Then:

$$\kappa^\pm \circ (\alpha_{\mathbb{M}^\pm;\mathbb{L}^\pm} \otimes \beta_{\mathbb{M}^\pm;\mathbb{L}^\pm}) = \gamma^\pm \circ \chi_{\mathbb{M}^\pm} \circ (\alpha_{\mathbb{M}^\pm;\mathbb{L}} \otimes \beta_{\mathbb{M}^\pm;\mathbb{L}}) = \gamma^\pm \circ \chi_{\mathbb{M}^\pm;\mathbb{L}} \circ \chi_\mathbb{L} = \chi_{\mathbb{M};\mathbb{L}} \circ \chi_\mathbb{L}.$$

where it was again used (4.6), (4.5) and the property $(\nu^\pm)^{-1} \circ \widehat{\nu}^\pm = \nu_{\mathbb{M}^\pm;\widehat{\mathbb{M}}^\pm}$. Then the definition of $\tau^\pm$ gives:

$$\tau^\pm \circ (\alpha_{\mathbb{M};\mathbb{L}} \otimes \beta_{\mathbb{M};\mathbb{L}}) = \kappa^\pm \circ (\alpha^\pm \circ \beta^\pm)^{-1} \circ (\alpha_{\mathbb{M},\mathbb{L}} \otimes \beta_{\mathbb{M},\mathbb{L}}) = \kappa^\pm \circ (\alpha_{\mathbb{M}^\pm;\mathbb{L}^\pm} \otimes \beta_{\mathbb{M}^\pm;\mathbb{L}^\pm}) = \gamma_{\mathbb{M};\mathbb{L}} \circ \chi_\mathbb{L}. \quad (4.7)$$

Since in this item $\mathbb{L} \subset \mathbb{M}^+ \cap \mathbb{M}^-$, so we have:

$$\tau^+ \circ (\alpha_{\mathbb{M};\mathbb{L}} \otimes \beta_{\mathbb{M};\mathbb{L}}) = \tau^- \circ (\alpha_{\mathbb{M};\mathbb{L}} \otimes \beta_{\mathbb{M};\mathbb{L}}),$$

and hence

$$\Theta \circ (\alpha_{\mathbb{M};\mathbb{L}} \otimes \beta_{\mathbb{M};\mathbb{L}}) = (\alpha_{\mathbb{M};\mathbb{L}} \otimes \beta_{\mathbb{M};\mathbb{L}}),$$

*i.e.*, $\Theta$ acts trivially on $\mathcal{A}(\mathbb{M}; L) \otimes \mathcal{B}(\mathbb{M}; L)$.

(c) Finally we apply (4.7) to $L^+ \subset \mathbb{M}^+$ and $L^- \subset \mathbb{M}^-$ giving:

$$\tau^+ \circ (\alpha_{\mathbb{M};\mathbb{L}^+} \otimes \beta_{\mathbb{M};\mathbb{L}^+}) = \gamma_{\mathbb{M};\mathbb{L}^+} \circ \chi_{\mathbb{L}^+}, \qquad \tau^- \circ (\alpha_{\mathbb{M};\mathbb{L}^-} \otimes \beta_{\mathbb{M};\mathbb{L}^-}) = \gamma_{\mathbb{M};\mathbb{L}^-} \circ \chi_{\mathbb{L}^-},$$

the first of which asserts that $\tau^+ \circ (\alpha_{\mathbb{M};\mathbb{L}^+} \otimes \beta_{\mathbb{M};\mathbb{L}^+})$ factors through $\gamma_{\mathbb{M};\mathbb{L}^+}$, while the second implies that $(\tau^-)^{-1} \circ \gamma_{\mathbb{M},\mathbb{L}^-}$ factors through $\alpha_{\mathbb{M},\mathbb{L}^-} \otimes \beta_{\mathbb{M},\mathbb{L}^-}$. As $\gamma_{\mathbb{M};\mathbb{L}^+}$ factors via $\gamma_{\mathbb{M};\mathbb{L}^-}$ due to the assumption $\mathbb{L}^+ \subset D(\mathbb{L}^-)$ and the timeslice property (4), the two observations combine to show that $\Theta \circ (\alpha_{\mathbb{M};\mathbb{L}^+} \otimes \beta_{\mathbb{M};\mathbb{L}^+})$ factors through $\alpha_{\mathbb{M};\mathbb{L}^-} \otimes \beta_{\mathbb{M};\mathbb{L}^-}$, as required.

∎

Isomorphisms allow mapping coupled states to uncoupled states. So we can consider a state $\varpi$ of $\mathscr{C}(\mathbb{M})$ as a decoupled state in earlier times if $(\kappa_{\mathbb{M};\mathbb{M}^-})^*\varpi$ is a product state of $\mathcal{A}(\mathbb{M}^-) \otimes \mathcal{B}(\mathbb{M}^-)$, or equivalently if $(\tau^-)^*\varpi$ is a product state of $\mathcal{A}(\mathbb{M}) \otimes \mathcal{B}(\mathbb{M})$.[7]

Reciprocally, probe observables at later times are precisely those that can be expressed as $\kappa_{\mathbb{M},\mathbb{M}^+}(\mathbb{1} \otimes B)$ for $B \in \mathcal{B}(\mathbb{M}^+)$ or equivalently $\tau^+(\mathbb{1} \otimes B)$ to $B \in \mathcal{B}(\mathbb{M})$. Analogously we can identify states that are uncorrelated at late times and probe observables that are measured at early times.

---

[7] Remembering that the pullback is defined by $(\xi^*\varpi)(X) = \varpi(\xi(X))$

With this in mind, we can then conclude that if we prepare the probe in a state $\sigma$ of $\mathscr{B}(\mathbb{M})$ and consider a system in a state $\omega$ of $\mathscr{A}(\mathbb{M})$ then the posterior coupled state will be given by $\omega_\sigma := \left((\tau^-)^{-1}\right)^* (\omega \otimes \sigma)$ of $\mathscr{C}(\mathbb{M})$. While assuming $B \in \mathscr{B}(\mathbb{M})$ an observable of the probe at later times, which in turn can be identified with the observable $\widetilde{B} := \tau^+ (\mathbb{1} \otimes B) \in \mathscr{C}(\mathbb{M})$. This allows us to make the following definition:

Then we can calculate the following expected value:

$$\omega_\sigma(\widetilde{B}) = \left((\tau^-)^{-1}\right)^* (\omega \otimes \sigma) \left(\tau^+ (\mathbb{1} \otimes B)\right) = (\omega \otimes \sigma) \left((\tau^-)^{-1} \circ \tau^+ (\mathbb{1} \otimes B)\right) =$$

$$= (\omega \otimes \sigma) (\Theta (\mathbb{1} \otimes B))$$

That is:

$$\omega_\sigma(\widetilde{B}) = (\omega \otimes \sigma) (\Theta (\mathbb{1} \otimes B)). \tag{4.8}$$

We can now ask if there is any correspondence between probe observables and system observables, so that measurements on the probe can be interpreted as measurements on the system conditioned by the state of probe preparation $\sigma$. That is, for each probe observable $B \in \mathscr{B}(\mathbb{M})$ we want to find an observable of the system $A \in \mathscr{A}(\mathbb{M})$ dependent on $B$ and $\sigma$, such that for all $\omega$ states of $\mathscr{A}(\mathbb{M})$:

$$\omega_\sigma(\widetilde{B}) = \omega(A).$$

This can be accomplished by the map $\eta_\sigma : \mathscr{A}(\mathbb{M}) \otimes \mathscr{B}(\mathbb{M}) \to \mathscr{A}(\mathbb{M})$ that linearly extends the relation $\eta_\sigma(A \otimes B) = \sigma(B)A$ (this can be strengthened to a continuously linear map, if needed). Considering then $C = \sum_{i=1}^n A_i \otimes B_i \in \mathscr{A}(\mathbb{M}) \otimes \mathscr{B}(\mathbb{M})$, we get that, for a $G \in \mathscr{A}(\mathbb{M})$, $G\eta_\sigma(C) = \sum_{i=1}^n \sigma(B_i) G A_i = \eta_\sigma((G \otimes \mathbb{1}) \sum_{i=1}^n A_i \otimes B_i)$ and $\eta_\sigma(C)G = \sum_{i=1}^n \sigma(B_i) A_i G = \eta_\sigma(\sum_{i=1}^n (A_i \otimes B_i)(G \otimes \mathbb{1}))$ and therefore $\eta_\sigma$ obeys the following relations, for $C \in \mathscr{A}(\mathbb{M}) \otimes \mathscr{B}(\mathbb{M})$ and $A \in \mathscr{A}(\mathbb{M})$:

$$A\eta_\sigma(C) = \eta_\sigma((A \otimes \mathbb{1})C), \qquad \eta_\sigma(C)A = \eta_\sigma(C(A \otimes \mathbb{1})).$$

The map $\eta_\sigma$ is also completely positive, definition A.0.13, in fact, let us consider a element $C \in M_n(\mathbb{C}) \otimes (\mathscr{A}(\mathbb{M}) \otimes \mathscr{B}(\mathbb{M}))$ given as a finite sum:

$$C = \sum_{r=1}^N M_r \otimes (A_r \otimes B_r),$$

for some finite $N$, hence to show complete positivity of $\eta_\sigma$ we must show that $(\mathbb{1}_{M_n(\mathbb{C})} \otimes \eta_\sigma)(C^*C)$ is positive in $M_n(\mathbb{C}) \otimes \mathscr{A}(\mathbb{M})$ for any such $C$ and arbitrary $n \in \mathbb{N}$. Computing:

$$(\mathbb{1}_{M_n(\mathbb{C})} \otimes \eta_\sigma)(C^*C) = \sum_{\substack{r=1 \\ s=1}}^N (\mathbb{1}_{M_n(\mathbb{C})} \otimes \eta_\sigma)(M_r^* M_s \otimes A_r^* A_s \otimes B_r^* B_s) = \sum_{\substack{r=1 \\ s=1}}^N \sigma(B_r^* B_s) M_r^* M_s \otimes A_r^* A_s.$$

As $\sigma$ is a state, $X_{rs} \equiv \sigma(B_r^* B_s)$ are the components of a positive matrix $X \in M_N(\mathbb{C})$, this can be proven by considering an arbitrary vector $y \equiv (y_r)_{r=1}^N \in \mathbb{C}^N$, then $\langle y, Xy \rangle =$ $\sum_{\substack{r=1\\s=1}}^{N} \overline{x_r}\sigma(B_r^* B_s)x_s = \sigma\left(\sum_{\substack{r=1\\s=1}}^{N} \overline{x_r} B_r^* x_s B_s\right)$, setting $B = \sum_{r=1}^{N} x_r B_r$, then $\sigma\left(\sum_{\substack{r=1\\s=1}}^{N} \overline{x_r} B_r^* x_s B_s\right) = \sigma(B^* B) \geq 0$, hence $\langle y, Xy \rangle \geq 0, \ \forall\, y \in \mathbb{C}^N$.

These components can be decomposed as:

$$\sigma(B_r^* B_s) = \sum_{i=1}^{N} \overline{\nu_r^{(i)}} \nu_s^{(i)},$$

for suitable mutually orthogonal vectors $\{\nu^{(i)}\}_{i=1}^{N}$, where $\nu_k^{(i)}$ stands for the $k$th component of $\nu^{(i)}$. Whereupon we find:

$$(\mathbb{1}_{M_n(\mathbb{C})} \otimes \eta_\sigma)(C^* C) = \sum_{i=1}^{N} W_i^* W_i, \qquad \text{where } W_i = \sum_{s=1}^{N} \nu_s^{(i)} M_s \otimes A_s. \qquad (4.9)$$

This proves that $\eta_\sigma$ is completely positive. Which, then allows us to then define:

**Definition 4.3.2** (Induced system observables). *A map $\varepsilon_\sigma : \mathscr{B}(\mathbb{M}) \to \mathscr{A}(\mathbb{M})$ is called a induced system observable when it is given by:*

$$\varepsilon_\sigma(B) := ((\eta_\sigma \circ \Theta)(\mathbb{1} \otimes B)) \qquad , B \in \mathscr{B}(\mathbb{M}).$$

*With this induced system observable we can then calculate the expected value of $\varepsilon_\sigma(B)$:*

$$\omega(\varepsilon_\sigma(B)) = \omega((\eta_\sigma \circ \Theta)(\mathbb{1} \otimes B)) = (\omega \otimes \sigma)(\Theta(\mathbb{1} \otimes B)) \stackrel{(4.8)}{=} \omega_\sigma(\widetilde{B})$$

*where we used the definition of $\eta_\sigma$ in the second equality, where the second entries of $\Theta(\mathbb{1} \otimes B)$ are sent to be evaluated by $\sigma$, leaving the first entries of $\Theta(\mathbb{1} \otimes B)$ to be evaluated by the linear functional $\omega$. That is, from the previous discussion we see that the theory corresponding to the probe $\mathscr{B}$, the one corresponding to the coupling $\mathcal{C}$, the identification maps $\chi$ and the state of preparation of the probe $\sigma$ constitute a **measurement scheme** of the observable induced in the system $\varepsilon_\sigma(B)$. (Note that this is a abuse of nomenclature, as it was never required for $B$ to be self-adjoint).*

**Theorem 4.3.3.** *For each probe preparation state $\sigma$, $A = \varepsilon_\sigma(B)$ is the only solution to $\omega_\sigma(\widetilde{B}) = \omega(A)$ if $\mathscr{A}(\mathbb{M})$ is separated by their states. Also, in general, $\varepsilon_\sigma$ is a completely positive linear map and has the properties:*

$$\varepsilon_\sigma(\mathbb{1}) = \mathbb{1}, \quad \varepsilon_\sigma(B^*) = \varepsilon_\sigma(B)^*, \quad \varepsilon_\sigma(B)^* \varepsilon_\sigma(B) \leq \varepsilon_\sigma(B^* B) \qquad (4.10)$$

*For fixed $B$, the map $\sigma \mapsto \varepsilon_\sigma(B)$ is weak-$*$continuous, i.e. continuous in the weak-$*$ topology, definition B.0.16.*

*Proof:* By the foregoing discussion we established that indeed $\omega(\varepsilon_\sigma(B)) = \omega_\sigma(\widetilde{B})$, since it is an hypothesis that $\mathscr{A}(\mathbb{M})$ is separated by its states, this means that if $A, B \in \mathscr{A}(\mathbb{M})$, $A \neq B$, there is at least one state $\varpi$ of $\mathscr{A}(\mathbb{M})$, such that $\varpi(A) \neq \varpi(B)$; then if there was any other $G : \mathscr{B}(\mathbb{M}) \to \mathscr{A}(\mathbb{M})$, $G \neq \varepsilon_\sigma$, for which $\omega_\sigma(\widetilde{B}) = \omega(G(B))$, but since $G \neq \varepsilon_\sigma$, there is at least one $B_0 \in \mathscr{B}(\mathbb{M})$ for which $G(B_0) \neq \varepsilon_\sigma(B_0)$, let $\omega_0$ be the separating state for these elements of $\mathscr{A}(\mathbb{M})$, but then $\varpi_\sigma(B_0) = \varpi(G(B_0)) \neq \varpi(\varepsilon_\sigma(B_0)) = \varpi_\sigma(B_0)$, absurd! therefore $\varepsilon_\sigma$ is unique.

In particular, we note that the vector states in a common dense domain within a faithful Hilbert space representation provide a separating set of states, since by the faithfulness of the representation $\pi(C) = \mathbf{0} \Leftrightarrow C = 0$, and since these vector states are associated with a dense subset of the representation Hilbert space, this means that for a $\pi(C) \neq \mathbf{0}$ there is at least one vector state $\omega_\psi$ for which $\omega_\psi(\pi(C)) \neq 0$, consider $\omega_\psi$ to be such a vector state for $\pi(A - B)$, $A \neq B$, $A, B \in \mathscr{A}(\mathbb{M})$, then $\omega_\psi \circ \pi(A) = \omega_\psi \circ \pi(B) \Leftrightarrow \omega_\psi \circ \pi(A - B) = 0$, but since $\pi$ is faithful and $A - B \neq 0 \Rightarrow \pi(A - B) \neq \mathbf{0}$ and as $\omega_\psi$ was chosen as to be different from zero for $\pi(A - B)$, this would give a absurd, hence $\omega_\psi \circ \pi(A) \neq \omega_\psi \circ \pi(B)$ and therefore $\omega_\psi \circ \pi$ separates the elements $A$ and $B$. For $C^*$-algebras, a faithful representation can always be found by the noncommutative Gel'fand-Naĭmark theorem 4.0.19 and hence $\mathscr{A}(\mathbb{M})$ is separated by the vector states in that representation, provided that $\mathscr{A}(\mathbb{M})$ is a $C^*$-algebra.

We then compute:
$$\varepsilon_\sigma(\mathbb{1}) = \eta_\sigma(\Theta(\mathbb{1} \otimes \mathbb{1})) \stackrel{4.3.1(\mathbf{b})}{=} \eta_\sigma(\mathbb{1} \otimes \mathbb{1}) = \mathbb{1},$$

this proves $\varepsilon_\sigma(\mathbb{1}) = \mathbb{1}$ and we also note that:
$$\eta_\sigma((A \otimes B)^*) = \eta_\sigma(A^* \otimes B^*) = \sigma(B^*)A^* = (\sigma(B)A)^* = (\eta_\sigma(A \otimes B))^*,$$

since $\Theta$ is a $*$-homomorphism, as it is defined by compositions of $*$-homomorphisms, then $\varepsilon(B^*) = (\eta_\sigma \circ \Theta)(\mathbb{1} \otimes B^*) = (\eta_\sigma \circ \Theta)((\mathbb{1} \otimes B)^*) = \eta_\sigma((\Theta(\mathbb{1} \otimes B))^*) = ((\eta_\sigma \circ \Theta)(\mathbb{1} \otimes B))^* = \varepsilon(B)^*$.

To prove that $\varepsilon_\sigma$ is completely positive, we use the previously proven fact that $\eta_\sigma$ is completely positive, then for any $C = \sum_{r=1}^{N} M_r \otimes B_r \in M_N(\mathbb{C}) \otimes \mathscr{B}$,
$$(\mathbb{1}_{M_N(\mathbb{C})} \otimes \varepsilon_\sigma)(C^*C) = (\mathbb{1}_{M_N(\mathbb{C})} \otimes \eta_\sigma)(D^*D) \stackrel{(4.9)}{\geq} 0,$$

where $D = \sum_{r=1}^{N} M_r \otimes \Theta(\mathbb{1} \otimes B_r)$.

That $\sigma \mapsto \varepsilon_\sigma(B)$ is weak-$*$ continuous for a fixed $B \in \mathscr{B}(\mathbb{M})$ follows from the definition of $\eta_\sigma$, since close functionals to $\sigma$ are mapped closed to the same element as $\varepsilon_\sigma(B)$ ($B$ is fixed), in fact it is just the same element scaled by the difference of the evaluation of $\sigma$ and the close functional considered.

Finally to prove $\varepsilon_\sigma(B)^* \varepsilon_\sigma(B) \leq \varepsilon_\sigma(B^*B)$, we let $B \in \mathscr{B}(\mathbb{M})$ and $C := \Theta(\mathbb{1} \otimes B) \in \mathscr{A}(\mathbb{M}) \otimes \mathscr{B}(\mathbb{M})$, which, therefore, can be decomposed as:
$$C = \sum_{r=1}^{K} A_r \otimes B_r,$$

for some $K \in \mathbb{N}$.

Calculating the difference we get:
$$\varepsilon_\sigma(B^*B) - \varepsilon_\sigma(B)^* \varepsilon_\sigma(B) = \sum_{\substack{r=1\\s=1}}^{K} \left[\sigma(B_r^*B_s) - \overline{\sigma(B_r)}\sigma(B_s)\right] A_r^* A_s,$$

and the factors in the square brackets determine a positive matrix $T$, by $T_{rs} = \left(\sigma(B_r^*B_s) - \overline{\sigma(B_r)}\sigma(B_s)\right)$, since for any $x \equiv (x_r)_{r=1}^{K}$:
$$\langle x, Tx \rangle = \sum_{\substack{r=1\\s=1}}^{K} \overline{x_r}[\sigma(B_r^*B_s) - \overline{\sigma(B_r)}\sigma(B_s)]x_s = \sum_{\substack{r=1\\s=1}}^{K} \sigma(\overline{x_r}B_r^* x_s B_s) - \overline{\sigma(x_r B_r)}\sigma(x_s B_s) =$$

$$= \sigma\left(\sum_{\substack{r=1\\s=1}}^{K} \overline{x_r} B_r^* x_s B_s\right) - \overline{\sigma\left(\sum_{r=1}^{K} x_r B_r\right)} \sigma\left(\sum_{r=1}^{K} x_r B_r\right),$$

writing $B = \sum_{r=1}^{K} x_r B_r$ we get:

$$\sigma\left(\sum_{\substack{r=1\\s=1}}^{K} \overline{x_r} B_r^* x_s B_s\right) - \overline{\sigma\left(\sum_{r=1}^{K} x_r B_r\right)} \sigma\left(\sum_{r=1}^{K} x_r B_r\right) = \sigma(B^* B) - \overline{\sigma(B)}\sigma(B) =$$
$$= \sigma(B^* B) - |\sigma(B)|^2,$$

as $B = \mathbb{1}_{\mathscr{B}(\mathbb{M})} B$ due to the Cauchy–Schwarz inequality and the fact that $\sigma$ is a state, we have:

$$|\sigma(B)|^2 \leq \sigma(B^* B)\sigma(\mathbb{1}_{\mathscr{B}(\mathbb{M})}^* \mathbb{1}_{\mathscr{B}(\mathbb{M})}) = \sigma(B^* B).$$

Thus $\sigma(B^* B) - |\sigma(B)|^2 \geq 0 \Rightarrow \langle x, Tx \rangle \geq 0$. Therefore we can use the same type of decomposition as we used when proving the complete positivity of $\eta_\sigma$, namely, since $\sigma(B_r^* B_s) - \overline{\sigma(B_r)}\sigma(B_s)$ are the components of a positive matrix, then:

$$\sigma(B_r^* B_s) - \overline{\sigma(B_r)}\sigma(B_s) = \sum_{i=1}^{K} \overline{w_r^{(i)}} w_s^{(i)},$$

for suitable mutually orthogonal vectors $\{w^{(i)}\}_{i=1}^{K}$, whereupon we find:

$$\varepsilon_\sigma(B^* B) - \varepsilon_\sigma(B)^* \varepsilon_\sigma(B) = \sum_{i=1}^{K} Y_i^* Y_i, \qquad \text{where } Y_i = \sum_{s=1}^{K} w_s^{(i)} A_s.$$

Therefore $\varepsilon_\sigma(B^* B) - \varepsilon_\sigma(B)^* \varepsilon_\sigma(B)$ is a positive element and $\varepsilon_\sigma(B^* B) \geq \varepsilon_\sigma(B)^* \varepsilon_\sigma(B)$. ∎

Using that $\tau^+$ is a homomorphism, then:

$$\widetilde{B}^2 = \tau^+(\mathbb{1} \otimes B)\tau^+(\mathbb{1} \otimes B) = \tau^+((\mathbb{1} \otimes B)^2) = \tau^+(\mathbb{1} \otimes B^2) = \widetilde{B^2}, \tag{4.11}$$

and we also verify that:

$$\operatorname{Var}\left(\widetilde{B}; \omega_\sigma\right) = \omega_\sigma\left(\widetilde{B}^2\right) - \omega_\sigma\left(\widetilde{B}\right)^2 \stackrel{(4.11)}{=} \omega_\sigma\left(\widetilde{B^2}\right) - \omega_\sigma\left(\widetilde{B}\right)^2 =$$
$$= \omega\left(\varepsilon_\sigma\left(B^2\right)\right) - \omega\left(\varepsilon_\sigma\left(B\right)\right)^2 \stackrel{(4.10)}{\geq} \omega\left(A^2\right) - \omega\left(A\right)^2 = \operatorname{Var}(A; \omega)$$

Therefore, measurements in practice are less accurate than a hypothetical direct measurement of induced observables in the state of the system (where we implicitly assume that the outcomes are distributed according to a spectral measure for a representation of $\widetilde{B}$, so that the standard formula for the quantum mechanical variance applies).

**Theorem 4.3.4.** *For each observable of probe $B \in \mathscr{B}(\mathbb{M})$ the induced observable in the system $\varepsilon_\sigma(B)$ can be located at any causally convex connected open set containing the matching region $K$. If $B$ can be located in $K^\perp$ then $\varepsilon_\sigma(B) = \sigma(B)\mathbb{1}$.*

*Proof:* Suppose that $L$ is a (possibly disconnected) open causally convex subset of $\mathbb{M}$ contained in $K^\perp$, so in particular $L \subset \mathbb{M}^+ \cap \mathbb{M}^-$. We have already shown that $\Theta$ acts trivially on $\mathscr{A}(\mathbb{M}; L) \otimes \mathscr{B}(\mathbb{M}; L)$, and now use this fact:

Suppose that $A \in \mathscr{A}(\mathbb{M}; L)$ is a system observable localised in $L$ and let $B \in \mathscr{B}(\mathbb{M})$ be any probe observable. Then $(A \otimes \mathbb{1})$ is invariant over $\Theta$, and we may compute:

$$[\varepsilon_\sigma(B), A] = [\eta_\sigma(\Theta(\mathbb{1} \otimes B)), A] = \eta_\sigma([\Theta(\mathbb{1} \otimes B), A \otimes \mathbb{1}]) =$$

$$= \eta_\sigma(\Theta([\mathbb{1} \otimes B, A \otimes \mathbb{1}])) = 0.$$

This shows that the induced observable $\varepsilon_\sigma(B)$ commutes with all system observables localised in the causal complement of $K$. Therefore, all field observables induced by probe observables are localisable in any connected open causally convex set containing the coupling region $K$.

Whereas, if we let $B \in \mathscr{B}(\mathbb{M}; L)$ be a probe observable localisable in $L$. Then $\Theta$ leaves $\mathbb{1} \otimes B$ invariant, and hence:

$$\varepsilon_\sigma(B) = \eta_\sigma(\Theta(\mathbb{1} \otimes B)) = \eta_\sigma(\mathbb{1} \otimes B) = \sigma(B)\mathbb{1}.$$

This shows that the system observable induced by a probe observable belonging to the causal complement of the coupling region is a fixed multiple of the identity (determined by the probe preparation state $\sigma$ and $B$) and provides no information about the field. ∎

We then consider maps $E : \mathcal{X} \to \mathscr{A}$, where $\mathcal{X}$ is a given $\sigma$-algebra with a total space $\Omega_\mathcal{X}$, so that $E$ has the properties of a measure, *i.e.* is at least finitely additive for the purely $*$-algebraic setting and $E(\varnothing) = \mathbf{0}$, also has the particular imposition $E(\Omega_\mathcal{X}) = \mathbb{1}_\mathscr{A}$, and takes values in the so called *effects of* $\mathscr{A}$, that is, the elements $A \in \mathscr{A}$ such that $A$ and $\mathbb{1} - A_\mathscr{A}$ are both positive.

We then interpret $\mathcal{X}$ as the set of potential measurement results and for each $X \in \mathcal{X}$, $\omega(E(X))$ is the probability that a value contained in $X$ is measured in state $\omega$.

**Definition 4.3.5** (Effect valued measure (EVM)). *Let $(\Omega_\mathcal{X}, \mathcal{X})$ be a measurable space, and $\mathscr{A}$ a $*$-algebra, consider a map $E : \mathcal{X} \to \mathscr{A}$, such that for any finite collection $\{X_k\}_{k=1}^n \subset \mathcal{X}$ of pairwise disjoint sets:*

$$E(\Omega_\mathcal{X}) = \mathbb{1}_\mathscr{A}, \quad E(\varnothing) = \mathbf{0}, \quad E\left(\bigcup_{k=1}^n X_k\right) = \sum_{k=1}^n E(X_k) \quad \text{and} \quad \{E(X), \mathbb{1}_\mathscr{A} - E(X)\} \subset \mathscr{A}_+, \, \forall X \in \mathcal{X}.$$

*Any $E$ map of this type will be called a **effects valued measure** (EVM) (a kind of POVM).*

It is a simple consequence of the definitions that a EVM $E$ taking values in the effects of the probe induces a corresponding EVM $\mathcal{X} \ni X \mapsto \varepsilon_\sigma(E(X)) \in \mathscr{A}(\mathbb{M})$ taking values in the effects of the system. Due to 4.10, we have that $\varepsilon_\sigma(E(X))^2 \leq \varepsilon_\sigma(E(X)^2)$, meaning that even if the EVM $E$ happens to be sharp, $E(X)^2 = E(X)$ (*i.e.* $E$ is a PVM), the induced EVM $\varepsilon_\sigma \circ E(X)$ will generally not be sharp with $\varepsilon_\sigma(E(X))^2 \leq \varepsilon_\sigma(E(X))$. This shows how the incorporation of the probe, with its own fluctuations, increase variance in measurement outcomes, even though the induced system observable is typically less sharp than the probe observable, but (as seen in the previous variance estimate) sharper than the actual measurement made on the coupled system.

Having introduced the notion of a effect valued measure, we naturally consider the case of a joint EVM, since, even though induced observables can be localised in a suitable open, connected and causally convex neighbourhood of the causal hull $J^+(K) \cap J^-(K)$ of the coupling region $K$, as we have seen; this would seem to suggest that two experimenters in causally disjoint spacetime regions would be able to measure incompatible system observables, as differently from the induced probe observables that they are measuring there is no reason to suppose that the system observables associated with each of these induced probe observables have also causally disjoint localisation, or are otherwise compatible. To elucidate this situation the concept of a *joint* EVM is introduced.

**Definition 4.3.6** (Joint EVM). *Two EVM's, $E_i : \mathcal{X}_i \to \mathscr{A}$, $i = 1, 2$; are said to have a joint EVM if they are the marginals of a EVM $E : \mathcal{X}_1 \otimes \mathcal{X}_2 \to \mathscr{A}$, that is, if $E_1(X_1) = E(X_1 \times \Omega_{\mathcal{X}_2})$ and $E_2(X_2) = E(\Omega_{\mathcal{X}_1} \times X_2)$, $\forall X_1 \in \mathcal{X}_1, X_2 \in \mathcal{X}_2$.*

Therefore, if $E$ is a a joint EVM valued in the effects of the probe, then $\varepsilon_\sigma \circ E$ is a joint EVM for the system EVM's $\varepsilon_\sigma \circ E_i$, $i \in I$. As $\varepsilon_\sigma$ is not in general a homomorphism, the $E_i$ being commuting EVM's does not imply that the induced system EVM's $\varepsilon_\sigma \circ E_i$ will commute. The same point about the sharpness of $E_i$ not implying sharpness of $\varepsilon_\sigma \circ E_i$, made three paragraphs ago, also applies here. Therefore, the information the experimenters can obtain concerning incompatible system observables is limited to what can be provided by an unsharp joint EVM.

Being $A$ and $B$ effects of the system and the probe respectively, and considering a joint measurement of both at later times, we then have:

$$\text{Prob}_\sigma\left(A\&B;\omega\right) = \omega(\eta_\sigma \circ \Theta\left(A \otimes B\right))$$

And a conditional probability:

$$\text{Prob}_\sigma\left(A|B;\omega\right) = \frac{\text{Prob}_\sigma\left(A\&B;\omega\right)}{\text{Prob}_\sigma\left(B;\omega\right)}$$

Defining then a map $\mathfrak{I}_\sigma(B) : \mathscr{A}\left(\mathbb{M}\right)^*_+ \to \mathscr{A}\left(\mathbb{M}\right)^*_+$ corresponding to the effect $B$ and probe preparation state $\sigma$ by:

$$\forall A \in \mathscr{A}\left(\mathbb{M}\right), \quad \left(\mathfrak{I}_\sigma(B)\left(\omega\right)\right)(A) := (\omega \otimes \sigma)\left(\Theta\left(A \otimes B\right)\right) = \left(\Theta^*\left(\omega \otimes \sigma\right)\right)(A \otimes B)$$

So we have that:
$$\text{Prob}_\sigma\left(A|B;\omega\right) = \frac{\left(\mathfrak{I}_\sigma(B)\left(\omega\right)\right)(A)}{\left(\mathfrak{I}_\sigma(B)\left(\omega\right)\right)(\mathbb{1})}$$

So we can interpret $\mathfrak{I}_\sigma(B)\left(\omega\right)$ as being a non-normalized update of the state of the system conditioned on the observation of the effect $B$ on the probe. Then the normalized state:

$$\omega` = \frac{\mathfrak{I}_\sigma(B)\left(\omega\right)}{\left(\mathfrak{I}_\sigma(B)\left(\omega\right)\right)(\mathbb{1})} = \frac{\mathfrak{I}_\sigma(B)\left(\omega\right)}{\omega(\varepsilon_\sigma(B))},$$

is the post-selected state of the system after a selective probe measurement. To check that $\omega`$ is indeed positive, recall that the effect $B$ is positive and therefore takes the form $B = \sum_{i=1}^K C_i^* C_i$ for some finite set of elements $\{C_i\}_{i=1}^K \subset \mathscr{B}(\mathbb{M})$, then:

$$\left(\mathfrak{I}_\sigma(B)\left(\omega\right)\right)(A^*A) = \sum_{i=1}^K \left(\Theta^*(\omega \otimes \sigma)\right)(A^*A \otimes C_i^*C_i) = \sum_{i=1}^K \left(\Theta^*(\omega \otimes \sigma)\right)((A \otimes C_i)^*(A \otimes C_i)) \geq 0$$

and it follows that $\omega`(A^*A) \geq 0$ for all $A \in \mathscr{A}(\mathbb{M})$. It is also evident that $\omega`$ is normalized as $\|\omega`\|_{\mathscr{A}(\mathbb{M})^*} = \omega`(\mathbb{1})$.

We then synthesize these conclusions in the definition:

**Definition 4.3.7** (Pre-instrument). *The map $\mathfrak{I}_\sigma : \mathscr{B}(\mathbb{M}) \times \mathscr{A}(\mathbb{M})^*_+ \to \mathscr{A}(\mathbb{M})^*_+$ given by:*

$$(\mathfrak{I}_\sigma(B)(\omega))(A) := (\omega \otimes \sigma)(\Theta(A \otimes B)) = (\Theta^*(\omega \otimes \sigma))(A \otimes B), \qquad (4.12)$$

$\forall B \in \mathscr{B}(\mathbb{M}), \omega \in \mathscr{A}(\mathbb{M})^*_+$ *and* $A \in \mathscr{A}(\mathbb{M})$ *is called a **pre-instrument** and is the unnormalized updated system state conditioned on the probe effect $B$ being observed in a system on a state $\omega$, for a probe prepared in a state $\sigma$.*

An *instrument* will be the composition of a pre-instrument with an EVM, *i.e.* a measure $\mathcal{X} \ni X \mapsto \mathfrak{I}_\sigma(E(X)) \in \mathscr{A}(\mathbb{M})^{**}_+$ on the $\sigma$-algebra of measurement outcomes valued in positive maps on the state space, this is the original notion of a quantum instrument, as introduced by Davies and Lewis in [27].

A **non-selective probe measurement** will then be the case where there is no conditional on the measurement result. That is, considering an EVM $E : \mathcal{X} \to \mathscr{B}(\mathbb{M})$, then a non-selective probe measurement will correspond to the pre-instrument $\mathfrak{I}_\sigma(E(\Omega_\mathcal{X})) = \mathfrak{I}_\sigma(\mathbb{1})$.

A justification for this definition stems from the case where $\Omega_\mathcal{X}$ is finite and using the additivity property of instruments:

$$\sum_{a \in \Omega_\mathcal{X}} \mathfrak{I}_\sigma(E(\{a\})) = \mathfrak{I}_\sigma(E(\Omega_\mathcal{X}))$$

So the updated state resulting from a non-selective measure:

$$\omega'_{ns} = \mathfrak{I}_\sigma(\mathbb{1})(\omega)$$

Or more explicitly:

$$\omega'_{ns}(A) = (\mathfrak{I}_\sigma(\mathbb{1})(\omega))(A) = (\Theta^*(\omega \otimes \sigma))(A \otimes \mathbb{1}) \qquad , \forall A \in \mathscr{A}(\mathbb{M})$$

And we check that if $A$ is located in $K^\perp$ then $\Theta(A \otimes \mathbb{1}) = (A \otimes \mathbb{1})$ and therefore $\omega'_{ns}(A) = \omega(A)$.

**Theorem 4.3.8.** *Consider a measurement of a probe effect $B$ in which the effect is observed. For each $A \in \mathscr{A}(\mathbb{M}; K^\perp)$, the expectation value of $A$ is unchanged in the post-selected state:*

$$\omega'(A) = \frac{\omega(A\varepsilon_\sigma(B))}{\varepsilon_\sigma(B)} = \omega(A),$$

*if and only if $A$ is uncorrelated with $\varepsilon_\sigma(B)$ in the original system state $\omega$.*

*Proof:* Since $A$ is localisable in $K^\perp$, then $\Theta(A \otimes \mathbb{1}) = A \otimes \mathbb{1}$, and so:

$$A\varepsilon_\sigma(B) = A\eta_\sigma(\Theta(\mathbb{1} \otimes B)) = \eta_\sigma((A \otimes \mathbb{1})\Theta(\mathbb{1} \otimes \mathbb{1})) = \eta_\sigma(\Theta(A \otimes \mathbb{1})\Theta(\mathbb{1} \otimes \mathbb{1})) = \eta_\sigma(\Theta(A \otimes B)),$$

by the definition of pre-instrument (4.12):

$$\mathfrak{I}_\sigma(B)(\omega)(A) = \eta_\sigma(\Theta(A \otimes B)) = A\varepsilon_\sigma(B),$$

then the normalized post-selected state, conditioned on the effect being observed, is:

$$\omega'(A) = \frac{\mathfrak{I}_\sigma(B)(\omega)(A)}{\mathfrak{I}_\sigma(B)(\omega)(\mathbb{1})} = \frac{A\varepsilon_\sigma(B)}{\varepsilon_\sigma(B)}.$$

Evidently $\omega'(A) = \omega(A)$ if and only if $\omega(A\varepsilon_\sigma(B)) = \omega(A)\omega(\varepsilon_\sigma(B))$. ∎

**Theorem 4.3.9.** *If two measurements of the field system are made in compact interaction regions $K_1$ and $K_2$, with $K_2 \cap J^-(K_1) = \varnothing$, for two probe systems $\mathscr{B}_i(\mathbb{M})$ and coupled systems $\mathscr{C}_i(\mathbb{M})$, each with its own scattering morphism $\Theta_i$ on $\mathscr{A}(\mathbb{M}) \otimes \mathscr{B}_i(\mathbb{M})$, $i = 1, 2$. On the three-fold tensor product $\mathscr{B}_1(\mathbb{M}) \otimes \mathscr{A}(\mathbb{M}) \otimes \mathscr{B}_2(\mathbb{M})$, we define $\widehat{\Theta}_1 := \overset{\leftrightarrow}{\Theta}_1 \otimes \mathbb{1}_{\mathscr{B}_2}$, where $\overset{\leftrightarrow}{\Theta}_1(x,y) := \Theta_1(y,x)$, $\forall\, x \in \mathscr{B}_1(\mathbb{M})$, $y \in \mathscr{A}(\mathbb{M})$ and $\widehat{\Theta}_2 := \mathbb{1}_{\mathscr{B}_1} \otimes \Theta_2$.*

*The two probes may be considered as a single probe with algebra $\mathscr{B}_1(\mathbb{M}) \otimes \mathscr{B}_2(\mathbb{M})$, a coupling region $K_1 \cup K_2$ and a combined scattering morphism $\widehat{\Theta}$ on $\mathscr{B}_1(\mathbb{M}) \otimes \mathscr{A}(\mathbb{M}) \otimes \mathscr{B}_2(\mathbb{M})$, such that they obey a natural causal factorisation formula:*

$$\widehat{\Theta} = \widehat{\Theta}_1 \circ \widehat{\Theta}_2. \tag{4.13}$$

*For all probe preparations $\sigma_i$ of $\mathscr{B}_i(\mathbb{M})$ and all probe observables $B_i \in \mathscr{B}_i(\mathbb{M})$, the following identity for the pre-instruments holds:*

$$\mathfrak{I}_{\sigma_2}(B_2) \circ \mathfrak{I}_{\sigma_1}(B_1) = \mathfrak{I}_{\sigma_1 \otimes \sigma_2}(B_1 \otimes B_2). \tag{4.14}$$

*If, in fact, K1 and K2 are causally disjoint, we have:*

$$\mathfrak{I}_{\sigma_2}(B_2) \circ \mathfrak{I}_{\sigma_1}(B_1) = \mathfrak{I}_{\sigma_1 \otimes \sigma_2}(B_1 \otimes B_2) = \mathfrak{I}_{\sigma_1}(B_1) \circ \mathfrak{I}_{\sigma_2}(B_2).$$

*Proof:* The composition of the pre-instruments is computed as follows. For any system state $\omega$ and system observable $A$, we have:

$$\mathfrak{I}_{\sigma_2}(B_2)(\mathfrak{I}_{\sigma_1}(B_1)(\omega))(A) = (\mathfrak{I}_{\sigma_1}(B_1)(\omega) \otimes \sigma_2)(\Theta_2(A \otimes B_2)) =$$
$$\overset{\text{def. } \eta_{\sigma_2}}{=} \mathfrak{I}_{\sigma_1}(B_1)(\omega)(\eta_{\sigma_2}(\Theta_2(A \otimes B_2))) =$$
$$= (\Theta_1^*(\omega \otimes \sigma_1))(\eta_{\sigma_2}(\Theta_2(A \otimes B_2)) \otimes B_1) =$$
$$= \left( \overset{\leftrightarrow}{\Theta}_1{}^*(\sigma_1 \otimes \omega) \otimes \sigma_2 \right) (B_1 \otimes (\Theta_2(A \otimes B_2))) =$$
$$= (\sigma_1 \otimes \omega \otimes \sigma_2) \left( \left( \overset{\leftrightarrow}{\Theta}_1 \otimes \mathbb{1}_{\mathscr{B}_2} \right) (B_1 \otimes \Theta_2(A \otimes B_2)) \right) =$$
$$= (\sigma_1 \otimes \omega \otimes \sigma_2) \left( \left( \widehat{\Theta}_1 \circ \widehat{\Theta}_2 \right) (B_1 \otimes A \otimes B_2) \right) =$$
$$= (\sigma_1 \otimes \omega \otimes \sigma_2)(\widehat{\Theta}(B_1 \otimes A \otimes B_2)) = \tag{4.15}$$
$$= \mathfrak{I}_{\sigma_1 \otimes \sigma_2}(B_1 \otimes B_2)(\omega)(A).$$

This last expression can be better justified formally, by changing from the algebra $\mathscr{B}_1(\mathbb{M}) \otimes \mathscr{A}(\mathbb{M}) \otimes \mathscr{B}_2(\mathbb{M})$ to the algebra $\mathscr{A}(\mathbb{M}) \otimes \mathscr{B}_1(\mathbb{M}) \otimes \mathscr{B}_2(\mathbb{M})$, if we consider the morphism given by $\breve{\Theta}(x,y,z) := \widehat{\Theta}(y,x,z)$, then (4.15) becomes:

$$(\omega \otimes \sigma_1 \otimes \sigma_2)(\breve{\Theta}(A \otimes B_1 \otimes B_2)),$$

and hence we get more clearly, at least formally, that $(\omega \otimes (\sigma_1 \otimes \sigma_2))(\breve{\Theta}(A \otimes (B_1 \otimes B_2))) = \mathfrak{I}_{\sigma_1 \otimes \sigma_2}(B_1 \otimes B_2)(\omega)(A)$, by simple application of the definition of $\mathfrak{I}$. This proves the first statement; the second is an immediate consequence of the possibility of factoring $\breve{\Theta}$ either as we just did or in the inverse order, since both are possible as $K_1$ and $K_2$ are causally disjoint. ∎

**Corollary 4.3.10.** *Consider two probes as described above, with $K_2 \cap J^-(K_1) = \varnothing$, effects $B_i \in \mathscr{B}_i(\mathbb{M})$ and probe preparation states $\sigma_i$, $i = 1, 2$. Suppose $B_1$ has nonzero probability of being observed in system state $\omega$, and that $B_2$ has nonzero probability of being observed in system state $\omega_1^\backprime$, the post-selected system state conditioned on $B_1$ being observed in state $\omega$. Then the post-selected state $\omega_{12}^\backprime\backprime$ conditioned on $B_2$ being observed in state $\omega_1^\backprime$ coincides with the post-selected state $\omega_{12}^\backprime$ conditioned on $B_1 \otimes B_2$ being observed in state $\omega$.*

*Proof:* We may compute:
$$\omega_1^\backprime = \frac{\mathfrak{I}_{\sigma_1}(B_1)(\omega)}{\mathfrak{I}_{\sigma_1}(B_1)(\omega)(\mathbb{1}_{\mathscr{A}})},$$

and:
$$\omega_{12}^{\backprime\backprime} = \frac{\mathfrak{I}_{\sigma_1}(B_2)(\omega_1^\backprime)}{\mathfrak{I}_{\sigma_1}(B_2)(\omega_1^\backprime)(\mathbb{1}_{\mathscr{A}})}$$

conditioned on both effects being observed, post-selecting on the $B_1$ measurement at the intermediate step. The normalisation factors applied to $\omega_1^\backprime$ cancel in the formula for $\omega_{12}^{\backprime\backprime}$ and so we also have:

$$\omega_{12}^{\backprime\backprime} = \frac{\mathfrak{I}_{\sigma_1}(B_2)(\mathfrak{I}_{\sigma_1}(B_1)(\omega))}{\mathfrak{I}_{\sigma_1}(B_2)(\mathfrak{I}_{\sigma_1}(B_1)(\omega))(\mathbb{1}_{\mathscr{A}})} = \frac{\mathfrak{I}_{\sigma_1 \otimes \sigma_2}(B_1 \otimes B_2)(\omega)}{\mathfrak{I}_{\sigma_1 \otimes B_2}(B_1 \otimes B_2)(\omega)(\mathbb{1}_{\mathscr{A}})} = \omega_{12}^\backprime,$$

where the denominators are equal by setting $A = \mathbb{1}_{\mathscr{A}}$ in 4.14. ∎

Having these definitions, and theorems, we can then enunciate and comment on the result obtained by Fewster et al. with regard to impossible measurements, this result is a rigorous generalization of previous results obtained by Sorkin in 1993 and later synthesized in the so-called Sorkin protocol [91], this generalization was obtained using the *framework* of covariant schemes of local measurements proposed by Fewster and Verch and exposed so far.

Considering then that three observers $A$, $B$ and $C$ respectively perform operations on individually connected regions $O_1$, $O_2$ and $O_3$. We then assume that:

$$O_2 \cap J^-(O_1) = \varnothing \ ;$$

$$O_3 \cap J^-(O_2) = \varnothing \ ;$$

$O_3$ has a spacelike separation of $O_1$ ;

$O_3$ has a compact closure $\overline{O}_3$.

Let $\mathcal{S}$ be the theory that describes the system, $\mathcal{P}_1$ and $\mathcal{P}_2$ two probe theories, one for $A$ and one for $B$, with compact matching regions $K_1$ and $K_2$ contained respectively in $O_1$ and $O_2$.

Denoting the corresponding input and output regions by $\mathbb{M}_1^{\mp}$ and $\mathbb{M}_2^{\mp}$, the initial states of the probes by $\sigma_1$ and $\sigma_2$ and the associated scatter maps by $\Theta_i : \mathcal{S} \otimes \mathcal{P}_i \to \mathcal{S} \otimes \mathcal{P}_i$ to $i \in \{1, 2\}$.

So $\hat{\Theta}_1 := \Theta_1 \otimes_3 \mathbb{1}$ and $\hat{\Theta}_2 := \Theta_2 \otimes_2 \mathbb{1}$ where the subscript of the tensor product represents the position where the second factor will be inserted.

Let $C$ be a system observable localizable in $O_3$.

Due to the assumptions about the regions $O_1$, $O_2$ and $O_3$ we have that these regions admit a causal ordering in which the region $O_1$ precedes the region $O_2$ which in turn precedes the region $O_3$.

From the previous discussions we see that if $A$ and $B$ perform a measurement each in their respective regions, then the expected value of $C$'s observable will be given by:

$$\omega_{AB}(C) = (\omega \otimes \sigma_1 \otimes \sigma_2)\left(\left(\hat{\Theta}_1 \circ \hat{\Theta}_2\right)(C \otimes \mathbb{1} \otimes \mathbb{1})\right)$$

On the other hand, if $A$ does not perform it's experiment, we will have:

$$\omega_B(C) = (\omega \otimes \sigma_2)\left(\Theta_2(C \otimes \mathbb{1})\right)$$

Considering the situation described above with the observations that $K_2 \cap J^-(K_1) = \varnothing$; $O_3 \cap J^-(K_2)$; $\overline{O}_3$ has a space-like separation of $K_1$. Then:

$$\forall C \in \mathcal{S}(O_3), \quad \left(\hat{\Theta}_1 \circ \hat{\Theta}_2\right)(C \otimes \mathbb{1} \otimes \mathbb{1}) = \hat{\Theta}_2(C \otimes \mathbb{1} \otimes \mathbb{1})$$

Which immediately implies:

$$\omega_{AB}(C) = (\omega \otimes \sigma_1 \otimes \sigma_2)\left(\left(\hat{\Theta}_1 \circ \hat{\Theta}_2\right)(C \otimes \mathbb{1} \otimes \mathbb{1})\right) =$$

$$= (\omega \otimes \sigma_1 \otimes \sigma_2)\left(\hat{\Theta}_2(C \otimes \mathbb{1} \otimes \mathbb{1})\right) = (\omega \otimes \sigma_2)\left(\Theta_2(C \otimes \mathbb{1})\right) = \omega_B(C)$$

The result of $C$'s measurement is independent of whether or not $A$ performed any experiments, i.e., there is no superluminal transmission of signals.

Continuing with the most recent publications, we see that Fewster and Verch's framework [41], yielded its own version of Stinespring's Dilation Theorem:

**Theorem 4.3.11** (Stinespring dilation)**.** *Let $A$ be a $C^*$-unital algebra, $\mathscr{H}$ a Hilbert space and $\mathfrak{B}(\mathscr{H})$ the bounded operators on $\mathscr{H}$. For every completely positive map*

$$\Phi : A \to \mathfrak{B}(\mathscr{H})$$

*There is a Hilbert space $\mathscr{G}$ and a $*$-unital homomorphism $\pi : A \to \mathfrak{B}(\mathscr{G})$, such that:*

$$\Phi(a) = V^* \pi(a) V \ ,$$

*Where $V : \mathscr{H} \to \mathscr{G}$ is a bounded operator. We have to:*

$$\|\Phi(\mathbb{1})\| = \|V\|^2$$

We know that as a consequence of this theorem we have that "for each observable of the system we have a measurement scheme" [21], and as it appears, this result can be generalized to Fewster and Verch's framework for measurement schemes in quantum field theories, using the formalism of AQFT [36], which gives a great *a posteriori* justification for this *framework* in particular.

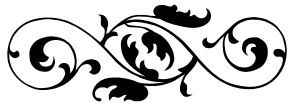

# CHAPTER 5

# Decoherence Model

*Klaus Hepp's [56] paper on a model of a quantum measurement apparatus that reproduces the wavefunction collapse effect through an infinite chain of spins is directly related to the so-called "measurement problem", which many physicists disagree if it can really be called a problem, since the theory does not present discrepant predictions with the experiments, either way this is the name present in the literature [50, 101]. The understanding of the article involves heavy use of the theory of representations of algebras $C^*$, in which good references are Dixmier's book [30] and Mackey's preceding article [75].*

Within the AQFT formalism we assume that the set of observables of the system generates a $C^*$-algebra $\mathscr{A}$ with unit. The set $\mathfrak{S}(\mathscr{A})$ of all positive linear functionals $\omega$ on $\mathscr{A}$ with $\omega(\mathbb{1}) = 1$ contains all states of the system. Also, in this chapter we implicitly consider that we have that every $\omega \in \mathfrak{S}(\mathscr{A})$ gives rise through the GNS construction, theorem 4.0.18, to a $*$-representation $\pi_\omega$ of $\mathscr{A}$ in a Hilbert space $\mathscr{H}_\omega$ with a cyclic vector $\varphi_\omega$, we consider this GNS triple $(\mathscr{H}_\omega, \pi_\omega, \varphi_\omega)$ associated with $(\mathscr{A}, \omega)$ as given, and may evoke them at any time without making reference to their origins.

The main result in this chapter is the analysis of a specific measurement model, the so called "Coleman-Hepp" model, with the theoretical resources that are developed throughout the chapter. The intuition for considering that model in particular is that when a apparatus measures a quantum system, this measurement establishes a correlation between the quantum system and the measurement apparatus, for quantum systems the principal way in which they can become correlated is through entanglement, so by considering a extensive apparatus each of the interactions of the components of the apparatus with the quantum system being measured increases the entanglement between these two systems, in this way increasing their correlation, until these correlations coalesce into a agreement of physical values.

In this way, this coherence between a observable value in the measured quantum system and their respective representation in the apparatus allows for a description of measurement that doesn't involve any type of "global collapse of the wave functions of probability amplitudes" (as this isn't well defined in relativistic theories either way), but a cumulative transition from independent, *a priori* undefined states to a classically probabilistic ensemble of, effectively, totally correlated states, this is the essence of decoherence.

In the particular case of the Coleman-Hepp model, the quantum system being measured is a $\frac{1}{2}$ spin particle, such as a electron, that moves in a one dimensional motion interacting locally with a infinite array of $\frac{1}{2}$ spins, where each local interaction between the electron and those $\frac{1}{2}$ spin sites in the apparatus array reinforces the total correlation between those systems, that develop a certain equivalence with regards to measurement values, in this specific case, of spins.

We may then establish a operational equivalence with respect to measurements in the sense proposed by Jauch [62] and exposed in Chapter 3.

In view of this equivalence, it could happen that for vector states, with a given representation

$\pi$, $\omega_1$, $\omega_2 \in \mathfrak{S}(\mathscr{A})$, can be in a incoherent mixture given by $\mathfrak{S}(\mathscr{A}) \ni \omega = \lambda_1 \omega_1 + \lambda_2 \omega_2$, with both $\lambda_1, \lambda_2 > 0$ and $\lambda_1 + \lambda_2 = 1$; since they are vector states they can be written as $\omega_i = \omega_{\Psi_i} \circ \pi$, $i = 1, 2$ by the vectors $\Psi_1, \Psi_2 \in \mathscr{H}_\pi$, and then one may consider the vector given by:

$$\Psi = \frac{\Psi_1 + \lambda \Psi_2}{\|\Psi_1 + \lambda \Psi_2\|} \, , \, \forall \lambda \in \mathbb{C}, \, \Psi_1 + \lambda \Psi_2 \neq 0.$$

And then the pure state $\omega_\Psi$ associated with the vector $\Psi$ on $\mathscr{A}$ could possibly always be such that:

$$\omega_\Psi(A) = \frac{1}{1 + |\lambda|^2} \omega_1(A) + \frac{|\lambda|^2}{1 + |\lambda|^2} \omega_2(A), \, \forall A \in \mathscr{A}.$$

This will only be true if there is no possible overlapping expectancy of values for any element $A \in \mathscr{A}$ (observable in a algebra of local observables), so as to not double count the contributions from each state. They have to be **disjoint**, or in other words:

**Definition 5.0.1** (Disjoint states). *Two states $\omega_1, \omega_2 \in \mathfrak{S}(\mathscr{A})$ are said to be* disjoint *if **no** subrepresentation of $\pi_{\omega_1}$ is unitarily equivalent to any subrepresentation of $\pi_{\omega_2}$, in symbols:*

$$\nexists U \text{ unitary operator such that } \quad U \pi_{\omega_1}\bigg|_{\mathscr{S}_{\omega_1}}(A) = \pi_{\omega_2}\bigg|_{\mathscr{S}_{\omega_2}}(A) U \, , \, \forall A \in \mathscr{A} \text{ and}$$

*for any fixed $\mathscr{S}_{\omega_1}, \mathscr{S}_{\omega_2}$ that are respectively vector subspaces of $\mathscr{H}_{\omega_1}$ and $\mathscr{H}_{\omega_2}$*

*That is if for their GNS representations there exists no nonzero partial isometry, which is the definition of disjointness for representations in general. If $\omega_1, \omega_2$ are not disjoint, they are called relatively coherent.*

**Definition 5.0.2.** *Two states $\omega_1, \omega_2 \in \mathfrak{S}(\mathscr{A})$ are said to be **quasi-equivalent** if every subrepresentation of $\pi_1$ has a subrepresentation that is unitarily equivalent to some subrepresentation of $\pi_2$, and vice versa. That is equivalent to say that the non-zero subrepresentations of $\pi_1$ are not disjoint from $\pi_2$, and the non-zero subrepresentations of $\pi_2$ are not disjoint from $\pi_1$.*

**Lemma 5.0.3.** *$\omega_1, \omega_2 \in \mathfrak{S}(\mathscr{A})$ are disjoint, if and only if for every representation $\pi$ of $\mathscr{A}$ with $\omega_i = \omega_{\Psi_i} \circ \pi$ for some $\Psi_i$, $i = 1, 2$; one has:*

$$\langle \Psi_1, \pi(A) \Psi_2 \rangle = 0 \, , \, \forall A \in \mathscr{A} \tag{5.1}$$

*Proof:* Considering $\pi_{\omega_i}$ as the subrepresentations of $\pi$ and $E_i \in \pi(\mathscr{A})'$(the algebraic dual) are the orthogonal projectors on $\pi(\mathscr{A})\Psi_i$ with central supports $F_i \in \mathcal{Z}(\pi(\mathscr{A})')$ that verify $F_i E_i = E_i$ and since the $\pi_{\omega_i}$ are disjoint there can be no unitary transformation (isometric isomorphism) that maps $\pi_{\omega_1}(\mathscr{A})$ to $\pi_{\omega_2}(\mathscr{A})$ therefore, since both are orthogonal projections and come from the center of $\pi(\mathscr{A})'$ then $F_1 F_2 = F_2 F_1 = (F_1 F_2)^*$ would be such a unitary transformation for a non null image of $F_1 F_2$, since for this subspace $F_1 F_2 = \mathbb{1}$, therefore we have that $F_1 F_2 = 0$ and so:

$$\langle \Psi_1, \pi(A) \Psi_2 \rangle = \langle \Psi_1, F_2 \pi(A) \Psi_2 \rangle = \langle \Psi_1, \pi(A) F_2 \Psi_2 \rangle = \langle \pi(A)^* \Psi_1, F_2 \Psi_2 \rangle = \langle \pi(A^*) \Psi_1, F_2 \Psi_2 \rangle =$$

$$= \langle E_1 \pi(A^*) \Psi_1, F_2 \Psi_2 \rangle = \langle \pi(A^*) \Psi_1, E_1 F_2 \Psi_2 \rangle = \langle \pi(A^*) \Psi_1, E_1 F_1 F_2 \Psi_2 \rangle = 0 \, , \, \forall A \in \mathscr{A}$$

The converse follows by the contrapositive, that is if $\omega_1, \omega_2$ are not disjoint then $\langle \Psi_1, \pi(A) \Psi_2 \rangle \neq$ $\neq 0$ for some $A \in \mathscr{A}$ (which is logically the same as the converse $C \to D \Leftrightarrow \neg D \to \neg C$), but as we have shown in the other case this implies $F_1 F_2 = \mathbb{1}$ for some subspace $\implies \pi(\mathscr{A})F_1\Psi_1 \cap \pi(\mathscr{A})F_2\Psi_2 \neq \varnothing$ and therefore:

$$\exists\, \pi(A), \pi(B) \in \pi(\mathscr{A}),\, 0 \neq \langle \pi(A)F_1\Psi_1, \pi(B)F_2\Psi_2\rangle = \langle \Psi_1, F_1\pi(A)^*\pi(B)F_2\Psi_2\rangle = \langle \Psi_1, F_1F_2\pi(A^*)\pi(B)\Psi_2\rangle$$

Since $F_1F_2\pi(A^*)\pi(B)\Psi_2 \neq 0$ then this vector is invariant by $F_1F_2$ and is in the associated subspace $\pi(\mathscr{A})F_1\Psi_1 \cap \pi(\mathscr{A})F_2\Psi_2 \implies F_1F_2\pi(A^*)\pi(B)\Psi_2 = F_1\pi(A^*B)\Psi_2 \in \pi(\mathscr{A})F_2\Psi_2$ therefore $\exists C \in \mathscr{A}$ such that $\pi(C)\Psi_2 = F_1\pi(A^*B)\Psi_2$ and for this $C$ we have that:

$$\langle \Psi_1, \pi(C)\Psi_2\rangle \neq 0.$$

∎

From this Lemma we have the following corollary:

**Corollary 5.0.4.** *If $\omega_1, \omega_2 \in \mathfrak{S}(\mathscr{A})$ are disjoint, and $\alpha \in \mathrm{Aut}(\mathscr{A})$, then $\omega_1 \circ \alpha$ and $\omega_2 \circ \alpha$ are disjoint.*

It is easy to see that, since an automorphism is a bijective map that preserves the algebraic structure, then $\forall A \in \mathscr{A}$, $\omega_i(\alpha(A)) = \omega_i(\widetilde{A})$ for some $\widetilde{A} \in \mathscr{A}$, follows that $\langle \Psi_1, \pi(\alpha(A))\Psi_2\rangle = \langle \Psi_1, \pi(\widetilde{A})\Psi_2\rangle = 0$ by Lemma 5.0.3, since it is so for any element of the algebra and therefore $\omega_1 \circ \alpha$ and $\omega_2 \circ \alpha$ are disjoint. The notion of disjointedness is important for models of measurement processes as it gives a notion of discernability, that is, the different pointer positions of an apparatus must be disjoint states for a big natural algebra of observables most of the times as is the case in § 3.2 of this dissertation.

## 5.1 Weakly converging sequence of states

Since we have from Corollary 5.0.4 that disjointedness is invariant by automorphisms, that means that coherence cannot be destroyed by an automorphic time evolution during the measurement process. Nevertheless we may still have sequences of relatively coherent states which converge weakly in $\mathfrak{S}(\mathscr{A})$ towards disjoint states. Intuitively that means that the cross-therms were would lie the interference must all converge to zero.

**Lemma 5.1.1.** *Considering sequences $\omega_{i,n} \xrightarrow{w} \omega_i$, for $i = 1, 2$; with $\omega_1, \omega_2$ disjoint. Let $\pi_n$ be representations of $\mathscr{A}$ and $\Psi_{i,n} \in \mathscr{H}_{\pi_n}$ with $\omega_{i,n} = \omega_{\Psi_{i,n}} \circ \pi_n$, $i = 1, 2$. Then, $\forall A \in \mathscr{A}$:*

$$\lim_{n \to \infty} \langle \Psi_{1,n}, \pi_n(A)\Psi_{2,n}\rangle = 0 \tag{5.2}$$

*Proof:* If $\omega_2(A^*A) = 0$, then by Cauchy-Schwarz:

$$|\langle \Psi_{1,n}, \pi_n(A)\Psi_{2,n}\rangle|^2 \leq \langle \Psi_{1,n}, \Psi_{1,n}\rangle\langle \pi_n(A)\Psi_{2,n}, \pi_n(A)\Psi_{2,n}\rangle = 1\langle \Psi_{2,n}, \pi_n(A^*A)\Psi_{2,n}\rangle = \omega_{2,n}(A^*A)$$

$$\implies |\langle \Psi_{1,n}, \pi_n(A)\Psi_{2,n}\rangle|^2 \leq \omega_{2,n}(A^*A) \xrightarrow{w} 0 \implies \lim_{n \to \infty}\langle \Psi_{1,n}, \pi_n(A)\Psi_{2,n}\rangle = 0$$

Otherwise $\omega_{2,n}(A^*A) \neq 0$ for any sufficiently large $n$ and this sequence does not converge weakly to zero. That being so, we may then define:[1]

$$\omega_{2,n}^A(\,\cdot\,) = \frac{\omega_{2,n}(A^*(\,\cdot\,)A)}{\omega_{2,n}(A^*A)}$$

$\omega_{2,n}^A$ trivially belongs to $\mathfrak{S}(\mathscr{A})$. So $\omega_{2,n}^A \xrightarrow{w} \omega_2^A := \frac{\omega_2(A^*(\,\cdot\,)A)}{\omega_2(A^*A)}$ that is well defined.

Since $\omega_1$ and $\omega_2$ are disjoint, then there exists $\Psi_1, \Psi_2$ such that:

$$\forall A \in \mathscr{A},\, \langle \Psi_1, \pi(A)\Psi_2 \rangle = 0 \text{ yet for } B \in \mathscr{A},\, \omega_2^A(B) = \frac{\omega_2(A^*BA)}{\omega_2(A^*A)} = \frac{\langle \Psi_2, \pi(A^*BA)\Psi_2 \rangle}{\langle \Psi_2, \pi(A^*A)\Psi_2 \rangle} =$$

$$= \left\langle \pi(A)\Psi_2,\, \pi(B)\frac{\pi(A)}{\omega_2(A^*A)}\Psi_2 \right\rangle = \left\langle \frac{\pi(A)}{(\omega_2(A^*A))^{\frac{1}{2}}}\Psi_2,\, \pi(B)\frac{\pi(A)}{(\omega_2(A^*A))^{\frac{1}{2}}}\Psi_2 \right\rangle$$

Consequently $\omega_2^A = \omega_{\frac{\pi(A)\Psi_2}{\sqrt{\omega_2(A^*A)}}}$ and then if we calculate the inner product:

$$\text{for } C \in \mathscr{A},\, \left\langle \Psi_1,\, \pi(C)\frac{\pi(A)}{(\omega_2(A^*A))^{\frac{1}{2}}}\Psi_2 \right\rangle = \frac{1}{(\omega_2(A^*A))^{\frac{1}{2}}}\langle \Psi_1, \pi(C)\pi(A)\Psi_2 \rangle =$$

$$= \frac{1}{(\omega_2(A^*A))^{\frac{1}{2}}}\langle \Psi_1, \pi(CA)\Psi_2 \rangle = 0\,,\text{ by hypothesis and lemma 5.0.3}$$

Therefore $\omega_1$ and $\omega_2^A$ are disjoint for all $A \in \mathscr{A}$, i.e., there are no subrepresentation $\pi_{\omega_1}$ that is unitarily equivalent to any other subrepresentation $\pi_{\omega_2^A}$. To continue we will need the lemma 4.2.18 by which since both $\omega_1$ and $\omega_2^A$ fulfill the hypothesis, we have $\|\omega_1 - \omega_2^A\| = 2$.

Considering that $\omega_{2,n}^A$ is a vector state with respect to the vector $\frac{\pi_n(A)\Psi_{2,n}}{(\omega_{2,n}(A^*A))^{\frac{1}{2}}} = \frac{\pi_n(A)\Psi_{2,n}}{\|\pi_n(A)\Psi_{2,n}\|}$

Proving the following proposition we arrive one step closer to proving what we wish:

**Proposition 5.1.2.** *Let $\alpha, \beta$ be vector states with vectors $\psi$ and $\phi$ respectively, that is $\alpha(Q) = \langle \psi, Q\psi \rangle$ and $\beta(Q) = \langle \phi, Q\phi \rangle$ with $\|\psi\| = \|\phi\| = 1$. Then the following inequality is true:*

$$\|\alpha - \beta\| \leq 2\sqrt{1 - |\langle \psi, \phi \rangle|}$$

*Proof:* Considering that $P_\psi, P_\phi \in \mathfrak{B}(\mathscr{A})$ are orthogonal projectors in the subspaces generated by the vectors $\psi, \phi \in \mathscr{H}$ respectively on a Hilbert space $\mathscr{H}$. Calculating:

$$|\alpha(A) - \beta(A)| = |\operatorname{tr}(AP_\psi) - \operatorname{tr}(AP_\phi)| = |\operatorname{tr}(A(P_\psi - P_\phi))| \leq \|A\|\|P_\psi - P_\phi\|_1 \quad (5.3)$$

Where the norm $\|\cdot\|_1$ shows up because $P_\psi, P_\phi$ are rank 1 projectors, the trace norm $\|\cdot\|_1$ given by $\|A\|_1 = \operatorname{tr}\left(\sqrt{A^*A}\right)$ is unitarily invariant, so considering:

---

[1] This definition of $\omega_{2,n}^A$ is a instance of, what was called in the literature [71], the "reduced" state of the **folium** determined by $\omega_{2,n}$ by the operation $\omega \mapsto \frac{\omega(A^*(\,\cdot\,)A)}{\omega(A^*A)}$. The algebras $\mathscr{A}(\mathbb{M}; N)$ were not initially considered to be observables, originally the observables were thought to arise only as limits of local algebra elements, with the elements of $\mathscr{A}(\mathbb{M}; N)$ seen to be the result of operations performed within $N$ on the states of $\mathscr{A}(\mathbb{M})$ [41, 53]. The name "folium" of $\omega$, for the set of states reachable by operations in a given state $\omega$ is actually poorly chosen, as it does not match the mathematical definition of a *foliation*, since $\mathfrak{S}(\mathscr{A}(\mathbb{M}; N))$ is by no means foliated by its "folia", for any such "folium" may contain sub-"folia" [71].

$$(P_\psi - P_\phi)^2 = P_\psi^2 + P_\phi^2 - P_\psi P_\phi - P_\phi P_\psi$$

Since $P_\phi$ is a rank 1 orthogonal projector then $P_\psi = \langle \psi, (\cdot)\rangle \psi$, not only that but $(P_\psi - P_\phi)$ has a maximum rank of 2 with 2 associated eigenvalues $k_1$ and $k_2$, yet:

$$\text{tr}(P_\psi - P_\phi) = \text{tr}(P_\psi) - \text{tr}(P_\phi) = 1 - 1 = 0 \implies k_2 = -k_1$$

Since $P_\psi - P_\phi$ is still a projector and is self-adjoint:

$$\|P_\psi - P_\phi\|_1 = \lambda_1 + \lambda_2 = ((\lambda_1 + \lambda_2)^2)^{\frac{1}{2}} = (\lambda_1^2 + \lambda_2^2 + 2\lambda_1\lambda_2)^{\frac{1}{2}} = (2\lambda_1^2 + 2\lambda_2^2)^{\frac{1}{2}} = \sqrt{2}(\lambda_1^2 + \lambda_2^2)^{\frac{1}{2}} =$$

$$\sqrt{2}\left(\text{tr}((P_\psi - P_\phi)^2)^{\frac{1}{2}}\right) = \sqrt{2}\left(\text{tr}(P_\psi^2 + P_\phi^2 - P_\psi P_\phi - P_\phi P_\psi)^{\frac{1}{2}}\right) = \sqrt{2}(1 + 1 - 2|\langle\psi,\phi\rangle|^2)^{\frac{1}{2}} =$$

$$= 2\sqrt{1 - |\langle\psi,\phi\rangle|^2}.$$

Where it was used that $\text{tr}(P_\psi P_\phi) = \sum_{i\in I}\langle e_i, \langle \psi, \phi\rangle \psi \langle \phi, e_i\rangle\rangle = \sum_{i\in I}\langle \psi, \phi\rangle\langle \phi, e_i\rangle\langle e_i, \psi\rangle = |\langle\psi,\phi\rangle|^2$ for any basis $\{e_i\}_{i\in I}$ of $\mathscr{H}$.

And then substituting this in 5.3:

$$|\alpha(A) - \beta(A)| \leq \|A\|\|P_\psi - P_\phi\|_1 = \|A\|(2\sqrt{1 - |\langle\psi,\phi\rangle|^2})$$

$$\implies \sup_{A\in\mathscr{A}, \|A\|\neq 0} \frac{|\alpha(A) - \beta(A)|}{\|A\|} \leq 2\sqrt{1 - |\langle\psi,\phi\rangle|^2}$$

$$\therefore \|\alpha - \beta\| \leq 2\sqrt{1 - |\langle\psi,\phi\rangle|^2}$$

∎

Having proved that we then have that:

$$\|\omega_{1,n} - \omega_{2,n}^2\|^2 \leq 4 - 4\left|\left\langle \Psi_{1,n}, \frac{\pi_n(A)\Psi_{2,n}}{\|\pi_n(A)\Psi_{2,n}\|^2}\right\rangle\right|^2 \leq 4$$

By the weak continuity of the norm we have that:

$$4 \leftarrow \|\omega_{1,n} - \omega_{2,n}^2\|^2 \leq 4 - 4\frac{|\langle\Psi_{1,n}, \pi_n(A)\Psi_{2,n}\rangle|^2}{\|\pi_n(A)\Psi_{2,n}\|^2} \leq 4$$

$$\downarrow$$

$$0 \leftarrow \|\omega_{1,n} - \omega_{2,n}^2\|^2 - 4 \leq -4\frac{|\langle\Psi_{1,n}, \pi_n(A)\Psi_{2,n}\rangle|^2}{\|\pi_n(A)\Psi_{2,n}\|^2} \leq 0$$

And by the theorem:

$$|\langle\Psi_{1,n}, \pi_n(A)\Psi_{2,n}\rangle|^2 \xrightarrow{n\to\infty} 0 \implies \lim_{n\to\infty}\langle\Psi_{1,n}, \pi_n(A)\Psi_{2,n}\rangle = 0, \forall A \in \mathscr{A}$$

∎

We also have the following result about mixtures that have each of its composing states disjoint with all the composing states of another mixture:

**Proposition 5.1.3.** *Let $\omega_i = \sum_{n=1}^{\infty} \lambda_{i,n} \omega_{i,n}$, $\lambda_{i,n} \leq 0$, $\sum_{n=1}^{\infty} \lambda_{i,n} = 1$, for $i = 1, 2$; and let $\omega_{1,n}, \omega_{2,n} \in \mathfrak{S}(\mathscr{A})$ be disjoint $\forall m, n$. Then $\omega_1$ and $\omega_2$ are disjoint.*

*Proof:* We can speak of the representations $\pi_{\omega_{i,n}}$ acting on $\mathscr{H}_{i,n}$ each associated with one of the $\omega_{i,n}$, whence the representation $\bigoplus_{n=1}^{\infty} \pi_{\omega_{i,n}}$ can be constructed by Hilbert sums and we must have that the representation $\pi_{\omega_i}$ is unitarily equivalent to some subrepresentation of $\bigoplus_{n=1}^{\infty} \pi_{\omega_{i,n}}$, additionally any of such $\bigoplus_{k \in K \subseteq \mathbb{N}} \pi_{\omega_{1,n}}$ is disjoint to any of $\bigoplus_{k \in K \subseteq \mathbb{N}} \pi_{\omega_{2,n}}$, since any unitary operator acts individually at each $\pi_{\omega_{1,n}}$ and since each of these cannot be transformed to any of the $\pi_{\omega_{2,m}}$, for any subspace of the domains of the $\pi_{\omega_{1,n}}$, and by any unitary operator and this implies that there is no unitary transformation that acts on any $\pi_{\omega_1}$ and transforms it in any $\pi_{\omega_2}$, for any subspace, and are hence disjoint. ∎

## 5.2 Limits on quasilocal algebras

Considering now a *quasilocal algebra of observables* $\mathscr{A}$, if we think about the nature of classical observables, the classical observables of the system do not necessarily belong to $\mathscr{A}$, for they are supposed to to correspond to operations which can be made outside of any bounded set. Considering then any given bounded region $\Lambda$ from the set of bounded regions $\mathscr{D}$, we let $\widetilde{\mathscr{A}}(\Lambda)$ be the $C^*$-algebra of $\mathscr{A}$ generated by $\{\mathscr{A}(M) \in \mathscr{A} \mid M \in \mathscr{D}, M \cap \Lambda = \varnothing\}$, that is, $\widetilde{\mathscr{A}}(\Lambda)$ is to be interpreted as the algebra of observables measurable outside $\Lambda$. By the Haag-Kastler axioms then $\widetilde{\mathscr{A}}(\Lambda)$ and $\mathscr{A}(\Lambda)$ commute. Considering $\pi$ a $*-$representation of $\mathscr{A}$. Then:

$$\mathscr{L}_\pi = \bigcap_{\Lambda \in \mathscr{D}} \pi\left(\widetilde{\mathscr{A}}(\Lambda)\right)'' \tag{5.4}$$

where $''$ denotes the bicommutant, $\mathscr{L}_\pi$ is called the **algebra of observables at infinity** in the representation $\pi$, that is, it is the algebra of observables measurable outside any given bounded region [73].

**Proposition 5.2.1.** *For any $*-$representation $\pi$ of $\mathscr{A}$, the algebra $\mathscr{L}_\pi$ is contained in the center of $\pi(\mathscr{A})'$.*

*Proof:* $\mathscr{L}_\pi$ is trivially contained in $\pi(\mathscr{A})''$, and it suffices by the fact that $\bigcup_{\Lambda \in \mathscr{D}} \mathscr{A}(\Lambda)$ is dense with relation to the norm of the $C^*$ algebra $\mathscr{A}$ by the first of the Haag-Kastler axioms, to show that given any $B \in \mathscr{L}_\pi$ and any $A \in \mathscr{A}(\Lambda)$ for some $\Lambda \in \mathscr{D}$, that $[B, A] = 0$. Since $B \in \mathscr{L}_\pi \Rightarrow B \in \pi(\widetilde{\mathscr{A}}(\Lambda))$ in particular, and since $\widetilde{\mathscr{A}}(\Lambda)$ commutes with $\mathscr{A}(\Lambda)$, and the $*-$representation preserves the algebra, then $[B, A] = 0 \Rightarrow \mathscr{L}_\pi \subseteq \mathcal{Z}(\pi(\mathscr{A})'') = \mathcal{Z}(\pi(\mathscr{A})')$ and the proposition is proved. ∎

It is a consequence of the fact that $\mathscr{L}_\pi \in \mathcal{Z}(\pi(\mathscr{A})')$ that $\mathscr{L}_\pi$ is abelian, which is a necessary prerequisite for a set of classical observables.

Considering states that could be distinguished only if they differ "far away" in space, that is, if they differ macroscopically and not just by local fluctuations, D. Ruelle and O. E. Lanford introduced in [73] the concept of having only *short-range correlations* for representations and states:

**Definition 5.2.2.** *We say that a representation $\pi$ has **short-range correlations** if the corresponding algebra $\mathscr{L}_\pi$ contains only the scalars $\mathbb{C}\mathbb{1}$, and that a state $\varrho$ on $\mathscr{A}$ has **short-range correlations** if the corresponding GNS cyclic representation does.*

This makes sense, since $\mathbb{C}\mathbb{1}$ always commutes with any set of representation elements, so every $\pi\left(\widetilde{\mathscr{A}}(\Lambda)\right)''$ contains $\mathbb{C}\mathbb{1}$ and therefore $\mathscr{L}_\pi$ also always does. If $\mathscr{L}_\pi$ contains any other element $g$, that means $g \in \pi\left(\widetilde{\mathscr{A}}(\Lambda)\right)''$ for every $\Lambda \in \mathscr{D}$, that means that there is a observable at infinity that is present in the closure of every $\widetilde{\mathscr{A}}(\Lambda)$ and therefore this would be a kind of *global correlation* or *long-range correlation*.

**Proposition 5.2.3.** *Consider that the state $\omega$ has short-range correlations then this is equivalent to the assertion that for every $A \in \mathscr{A}$ and $\varepsilon > 0$ there exists a region $\Lambda \in \mathscr{D}$ such that:*

$$|\omega(AB) - \omega(A)\omega(B)| \leq \|B\|\varepsilon \quad, \forall B \in \widetilde{\mathscr{A}}(\Lambda)$$

*Proof:* The direct implication can be proven by a proof by contradiction, consider that $\omega$ has short-range correlations but that the inequality is not valid, then there exists $A \in \mathscr{A}$, an increasing net $\{M_\alpha\}_\alpha \subseteq \mathscr{D}$ of bounded open sets whose union is the whole space, eventually avoiding any given region of $\mathscr{D}$, and operators $B_\alpha \in \widetilde{\mathscr{A}}(M_\alpha)$, $\|B_\alpha\| \leq 1$, such that:

$$\lim_\alpha \left(\omega(AB_\alpha) - \omega(A)\omega(B_\alpha)\right) \neq 0$$

Since it is a increasing net, we may pass to a subnet were it can be assumed that $\pi_\omega(B_\alpha)$ converges in the weak operator topology, since by the Banach—Alaoglu theorem B.0.18 the set is compact and nets on a compact set always have a convergent subnet. This limit will have to be in $\bigcap_\alpha \pi(\widetilde{\mathscr{A}}(M_\alpha))''$ and by the hypothesis of $\omega$ having short-range correlations then it must be of the form $b\mathbb{1}$, $b \in \mathbb{C}$. Then:

$$\lim_\alpha \left(\omega(AB_\alpha) - \omega(A)\omega(B_\alpha)\right) = \lim_\alpha \left(\langle \varphi_\omega, \pi_\omega(A)\pi_\omega(B_\alpha)\varphi_\omega \rangle - \langle \varphi_\omega, \pi_\omega(A)\varphi_\omega \rangle \langle \varphi_\omega, \pi_\omega(B_\alpha)\varphi_\omega \rangle\right)$$
$$= b\langle \varphi_\omega, \pi_\omega(A)\varphi_\omega \rangle - b\langle \varphi_\omega, \pi_\omega(A)\varphi_\omega \rangle = 0$$

which is a contradiction and thus proves that short-range correlations implies the inequality. Considering now the converse, the inequality is valid and considering $C \in \mathscr{L}_{\pi_\omega}$, and $\varepsilon > 0$ we have:

$$|\langle \varphi_\omega, \pi_\omega(A)C\varphi_\omega \rangle - \langle \varphi_\omega, \pi_\omega(A)\varphi_\omega \rangle \langle \varphi_\omega, C\varphi_\omega \rangle| \leq \|B\|\varepsilon \quad, \forall A \in \mathscr{A}$$

Replacing $A$ by $\lambda A$ we get that the LHS gets multiplied by $|\lambda|$ and leaves the RHS unchanged, henceforth the LHS must be zero. Letting $b = \langle \varphi_\omega, C\varphi_\omega \rangle$, and using Proposition 5.2.1 we get:

$$\langle \varphi_\omega, C\pi_\omega(A_1 A_2)\varphi_\omega \rangle - \langle \varphi_\omega, \pi_\omega(A_1 A_2)\varphi_\omega \rangle \langle \varphi_\omega, C\varphi_\omega \rangle = 0$$

$C \in \mathcal{Z}(\pi(\mathscr{A})'') \implies \langle \varphi_\omega, \pi_\omega(A_1)C\pi_\omega(A_2)\varphi_\omega \rangle = b\langle \varphi_\omega, \pi_\omega(A_1)\pi_\omega(A_2)\varphi_\omega \rangle$
$$\langle \pi_\omega(A_1)\varphi_\omega, C\pi_\omega(A_2)\varphi_\omega \rangle = b\langle \pi_\omega(A_1)\varphi_\omega, \pi_\omega(A_2)\varphi_\omega \rangle \quad, \forall A_1, A_2 \in \mathscr{A}.$$

And hence:
$$C = b\mathbb{1}$$

∎

A class of states of interest is the class of primary states.

**Definition 5.2.4.** *A **primary state** $\omega$ is such that $\pi_\omega(\mathscr{A})'' \cap \pi_\omega(\mathscr{A})' = \mathbb{C}\mathbb{1}$.*

Therefore by Proposition 5.2.1 and Definition 5.2.2 every *primary state* trivially has short-range correlations.

As it was probably perceived by the reader by now, a subrepresentation $\pi\big|_\mathscr{G}$ in a subspace $\mathscr{G} \subseteq \mathscr{H}$ of the Hilbert space where the bounded operators mapped by $\pi$ act correspond to having projectors $P_\mathscr{G}$ such that when writing $\pi = \pi\big|_\mathscr{G} + \pi\big|_{\mathscr{G}^\complement}$ then $P_\mathscr{G}\pi = \pi\big|_\mathscr{G} = \pi P_\mathscr{G}$, so that $P_\mathscr{G} \in \pi(\mathscr{A})'$

Which leads us to the following proposition:

**Proposition 5.2.5.** *For a state $\omega$, if $\omega = t\omega_1 + (1-t)\omega_2$ for some $t \in (0,1)$, then $\omega_1$ and $\omega_2$ are disjoint if, and only if there is a projection $K \in \mathcal{Z}(\pi_\omega(\mathscr{A})) = \pi_\omega(\mathscr{A})' \cap \pi_\omega(\mathscr{A})''$ such that:*

$$\pi_\omega\bigg|_{K\mathscr{H}_\omega}(\mathscr{A}) \simeq \pi_{\omega_1}(\mathscr{A}), \tag{5.5}$$

$$\pi_\omega\bigg|_{K^\perp\mathscr{H}_\omega}(\mathscr{A}) \simeq \pi_{\omega_1}(\mathscr{A}). \tag{5.6}$$

*Since, as we just established, $K \in \pi(\mathscr{A})'$ the key assumption being made is that $K \in \pi_\omega(\mathscr{A})''$.*

*Proof:* Consider the reverse order statement as true, that is consider that the equations (5.5) and (5.6) hold, we then argue by contradiction that equivalent subrepresentations $\pi_1(\mathscr{A}) \subseteq \pi_{\omega_1}(\mathscr{A})$ and $\pi_2(\mathscr{A}) \subseteq \pi_{\omega_2}(\mathscr{A})$ are given by projections $K_1\mathscr{H}_\omega \subseteq K\mathscr{H}_\omega (K_1 \leq K)$ and $K_2\mathscr{H}_\omega \subseteq K^\perp\mathscr{H}_\omega = (\mathbb{1} - K)\mathscr{H}_\omega(K_2 \leq K^\perp)$, respectively, so that:

$$\pi_i(A) = \pi_\omega\bigg|_{K_i\mathscr{H}_\omega}(A) \quad , i = 1, 2\,;\, \forall A \in \mathscr{A}.$$

and the partial isometry $\mathbf{w}\colon K\mathscr{H}_\omega \to K^\perp\mathscr{H}_\omega$ on $\mathscr{H}_\omega$ that implements the unitary equivalence between the subrepresentations, whose restriction to $K_1\mathscr{H}_\omega$ implements a unitary equivalence between $\pi_1(\mathscr{A})$ and $\pi_2(\mathscr{A})$ by definition satisfies $\mathbf{w}^*\mathbf{w} = K_1$ and $\mathbf{w}\mathbf{w}^* = K_2$, moreover $K_1 \leq K$ implies $\mathbf{w}K = \mathbf{w}$ and analogously $K_2 \leq K^\perp$ implies $K^\perp\mathbf{w} = \mathbf{w}$, combining both we get $K^\perp\mathbf{w}K = \mathbf{w}$. If then $K \in \pi_\omega(\mathscr{A})''$, then $\mathbf{w}K = K\mathbf{w}$ this in $K^\perp\mathbf{w} = \mathbf{w}$ implies

$$\mathbf{w} = K^\perp K\mathbf{w} = 0$$

which gives a contradiction since it implies that there are no equivalent subrepresentations $\pi_1$ and $\pi_2$ and therefore $\omega_1$ and $\omega_2$ are disjoint.

To prove the direct assertion, first we will need the following lemma:

**Lemma 5.2.6.** *For any functional $\hat{\omega} \in \mathscr{A}^*$ such that $0 \leq \hat{\omega} \leq \omega$, where $\omega \in \mathfrak{S}(\mathscr{A})$, there is an operator $c \in \pi_\omega(\mathscr{A})'$ on $\mathscr{H}_\omega$ such that $0 \leq c \leq \mathbb{1}_\mathscr{H}$ and*

$$\hat{\omega}(A) = \langle \varphi_\omega, c\pi_\omega(A)\varphi_\omega \rangle \quad , A \in \mathscr{A}$$

*In particular there is a vector $\xi \in \mathscr{H}_\omega$ such that:*

$$\hat{\omega}(A) = \langle \xi, \pi_\omega(A)\xi \rangle \tag{5.7}$$

*Proof:* Using the Cauchy–Schwarz for the positive semidefinite form $\langle a, b \rangle_\wedge = \hat{\omega}(a^*b)$ gives:

$$|\hat{\omega}(a^*b)|^2 \leq \hat{\omega}(a^*a)\hat{\omega}(b^*b) \leq \omega(a^*a)\omega(b^*b) = \|\pi_\omega(a)\varphi_\omega\|^2 \|\pi_\omega(b)\varphi_\omega\|^2 \tag{5.8}$$

Hence we obtain a well-defined positive bilinear form $B$ on $\mathscr{H}_\omega$, initially defined in the dense domain $\pi_\omega(\mathscr{A})\varphi_\omega \times \pi_\omega(\mathscr{A})\varphi_\omega$ by the formula:

$$B\left(\pi_\omega(a)\varphi_\omega,\, \pi_\omega(b)\varphi_\omega\right) = \hat{\omega}(a^*b)$$

and extended to $\mathscr{H}_\omega \times \mathscr{H}_\omega$ by continuity, the inequality (5.8) immediately gives $|B(\psi, \phi)| \leq \|\psi\|\|\phi\|$, and hence by proposition B.0.14 we have an operator $0 \leq c \leq \mathbb{1}_\mathscr{H}$, as $\|c\| = \|B\| \leq \|\omega\| = 1$ such that $B(\psi, \phi) = \langle \psi, c\phi \rangle$, that implies $B(\pi_\omega(\mathbb{1})\varphi_\omega, \pi_\omega(b)\varphi_\omega) = \hat{\omega}(\mathbb{1}b) = \hat{\omega}(b) = \langle \varphi_\omega, c\,\pi_\omega(b)\varphi_\omega \rangle$, $\forall b \in \mathscr{A}$. We now compute:

$$\hat{\omega}(a^*b^*d) = B(\pi_\omega(ba)\varphi_\omega,\, \pi_\omega(d)\varphi_\omega) = \langle \pi_\omega(a)\varphi_\omega,\, \pi_\omega(b^*)\,c\,\pi_\omega(d)\varphi_\omega \rangle$$
$$B(\pi_\omega(a)\varphi_\omega,\, \pi_\omega(b^*d)\varphi_\omega) = \langle \pi_\omega(a)\varphi_\omega,\, c\,\pi_\omega(b^*)\pi_\omega(d)\varphi_\omega \rangle,$$

so that $[c, \pi_\omega(b^*)] = 0$ for each $b \in \mathscr{A}$, i.e., $c \in \pi_\omega(\mathscr{A})'$. Using B.0.27 and writing $c = c_1^2$ with $c_1^* = c_1$, and then $\xi = c_1\varphi_\omega$, completes the proof. ∎

Continuing the proof of the direct assertion of Proposition 5.2.5. Assume:

$$\omega = t\omega_1 + (1-t)\omega_2 = \tilde{\omega}_1 + \tilde{\omega}_2 \qquad, t \in (0,1)$$

with $\tilde{\omega}_1 = t\omega_1$ and $\tilde{\omega}_2 = (1-t)\omega_2$, so that $0 \leq \tilde{\omega}_1 \leq \omega$ and $0 \leq \tilde{\omega}_2 \leq \omega$. It follows from the first claim in Lemma 5.2.6 that there is $c \in \mathfrak{B}(\mathscr{H}_\omega)$ such that $\forall a \in \mathscr{A}$:

$$\tilde{\omega}_1(a) = \langle \varphi_\omega,\, c\,\pi_\omega(a)\varphi_\omega \rangle \tag{5.9}$$

$$\tilde{\omega}_2(a) = \langle \varphi_\omega,\, (\mathbb{1}_{\mathscr{H}_\omega} - c)\,\pi_\omega(a)\varphi_\omega \rangle \tag{5.10}$$

where the equation (5.10) comes from considering that by the GNS construction $\langle \varphi_\omega, \pi_\omega(a)\varphi_\omega \rangle = \omega(a) = \tilde{\omega}_1(a) + \tilde{\omega}_2(a) \implies \tilde{\omega}_2(a) = \omega(a) - \tilde{\omega}_1(a) = \langle \varphi_\omega, (\mathbb{1}_{\mathscr{H}_\omega} - c)\pi_\omega(a)\varphi_\omega \rangle$. Define $\tilde{\omega} \in \mathscr{A}^*$ by

$$\tilde{\omega}(a) := \langle \varphi_\omega,\, c\,(\mathbb{1}_{\mathscr{H}_\omega} - c)\,\pi_\omega(a)\varphi_\omega \rangle$$

We have $0 \leq \tilde{\omega} \leq \tilde{\omega}_1$, since $0 \leq c \leq \mathbb{1}_{\mathscr{H}_\omega} \implies c(\mathbb{1}_{\mathscr{H}_\omega} - c) \leq c$, as well as $0 \leq \tilde{\omega} \leq \tilde{\omega}_2$, where also $c(\mathbb{1}_{\mathscr{H}_\omega} - c) \leq \mathbb{1}_{\mathscr{H}_\omega} - c$. Assuming now that $\omega_1$ and $\omega_2$ are disjoint. Applying (5.7) with the substitution in the lemma of $\omega \rightsquigarrow \tilde{\omega}_i$ which implies that there is unitarily equivalent subrepresentations of $\pi_{\omega_1}$ and $\pi_{\omega_2}$ because of the existence of $\tilde{\omega}$, but since $\omega_1$ and $\omega_2$ are assumed to be disjoint therefore $\tilde{\omega} = 0$. Since $\varphi_\omega$ is cyclic for $\pi_\omega(\mathscr{A})$ by the GNS construction, this implies $c(\mathbb{1}_{\mathscr{H}_\omega} - c) = 0 \implies c^2 = c$. Since $c \geq 0$, which implies $c^* = c$, from these characteristics follows that $c$ is a projection, henceforth called $K$. Therefore,

$$\omega_1(a) = \frac{\langle \varphi_\omega,\, K\pi_\omega(a)\varphi_\omega \rangle}{\|K\varphi_\omega\|^2} \tag{5.11}$$

$$\omega_2(a) = \frac{\langle \varphi_\omega,\, K^\perp \pi_\omega(a)\varphi_\omega \rangle}{\|K^\perp \varphi_\omega\|^2} \tag{5.12}$$

where $t = \|K\varphi_\omega\|^2$, and $1 - t = \|K^\perp \varphi_\omega\|^2$. We see from these formulae and Proposition 4.2.19 that $\pi_{\omega_1}$ and $\pi_{\omega_2}$ are equivalent to the restrictions of $\pi_\omega$ to $K\mathcal{H}_\omega$ and $K^\perp \mathcal{H}_\omega$ respectively; under this equivalence, the cyclic vectors $\varphi_{\omega_1}$ and $\varphi_{\omega_2}$ correspond with $\frac{K\varphi_\omega}{\|K\varphi_\omega\|}$ and $\frac{K^\perp \varphi_\omega}{\|K^\perp \varphi_\omega\|}$, respectively. Since $K \in \pi_\omega(\mathscr{A})'$ by Lemma 5.2.6, it only remains to be shown that $K \in \pi_\omega(\mathscr{A})''$. To this effect, to any $b \in \pi_\omega(\mathscr{A})'$, $a \in \mathscr{A}$ and $\psi \in \mathcal{H}_\omega$, define $\hat\omega \in \mathscr{A}^*$ such that:

$$\hat\omega(a) := \langle K^\perp b K \psi, \pi_\omega(a) K^\perp b K \psi \rangle$$

Then $\hat\omega$ is positive, as well as quasiequivalent to $\omega_2$ since the resulting vector $bK\psi$ is projected by $K^\perp$ and in equation (5.12) $K^\perp \varphi_\omega$ maps to this same subspace. Yet, for $a \in \mathscr{A}^+$ we have the inequalities:

$$0 \leq \hat\omega(a) \leq \|K^\perp b\|^2 \langle K\psi, \pi_\omega(a) K\psi \rangle,$$

So that assuming that $K\psi$ is a unit vector, $0 \leq \hat\omega \leq \hat\omega_1$ for the state:

$$\hat\omega_1(a) = \langle \psi, K\pi_\omega(a) K\psi \rangle.$$

Since $K\psi \in K\mathcal{H}_\omega$ implies that $\hat\omega_1$ is quasiequivalent to $\omega_1$, so that $\hat\omega$ itself is quasiequivalent to $\omega_1$ by Lemma 5.2.6. Again invoking disjointness of $\omega_1$ and $\omega_2$, it follows that $\hat\omega = 0$, which since, at least the direction, of $\psi$ was arbitrary, in turn yields $K^\perp bK = 0$ for any $b \in \pi_\omega(\mathscr{A})'$. This forces $K \in \pi_\omega(\mathscr{A})''$. ∎

**Remark:** This Proposition can also prove Proposition 5.0.3 for $\omega = \frac{1}{2}(\omega_1 + \omega_2)$.

**Corollary 5.2.7.** *If $\omega_1, \omega_2 \in \mathfrak{S}(\mathscr{A})$ are primary, then either $\omega_1$ and $\omega_2$ are disjoint or quasiequivalent.*

*Proof:* The $\omega_1$ and $\omega_2$ could be disjoint, in which case Proposition 5.2.5 would be valid, if they were not disjoint then because the centers of $\pi_{\omega_1}(\mathscr{A})$ and $\pi_{\omega_2}(\mathscr{A})$ are $\mathbb{C}\mathbb{1}$ then this implies that one can always find to every subrepresentation of $\pi_1$ a subrepresentation only of the center in such a way that this is unitarily equivalent to a subrepresentation of the center of $\pi_2$. ∎

**Corollary 5.2.8.** *A state is primary if, and only if it has no convex decomposition into disjoint states.*

*Proof:* If a primary state can be decomposed into disjoint states, by Proposition 5.2.5 $K \in \mathcal{Z}(\pi_\omega(\mathscr{A})) = \pi_\omega(\mathscr{A})' \cap \pi_\omega(\mathscr{A})'' = \mathbb{C}\mathbb{1} \implies K = \mathbb{1}$ therefore:

$$\pi_\omega\bigg|_{\mathbb{1}\mathcal{H}_\omega}(\mathscr{A}) = \pi_\omega(\mathscr{A}) \simeq \pi_{\omega_1}(\mathscr{A}),$$

$$\pi_\omega\bigg|_{\mathbb{1}^\perp \mathcal{H}_\omega}(\mathscr{A}) = \pi_\omega\bigg|_{\mathbf{0}}(\mathscr{A}) = \mathbf{0} \simeq \pi_{\omega_2}(\mathscr{A})$$

Therefore $\omega_2$ would not be a state, absurd. ∎

Let $\Psi_1, \Psi_2$ be orthogonal vector states in $\mathcal{H}_\omega$ with projectors $E_1, E_2$ and with $\omega$ primary. Then $\omega_{\Psi_1}$ and $\omega_{\Psi_2}$ are coherent, and the measurement problem for $A = a_1 E_1 + a_2 E_2$ is well-posed and non-trivial.

Another special class of observables at infinity are the macroscopic observables. Considering successive approximations of the region at infinity with bounded regions, that is a sequence $\{\Lambda_n\}_{n\in\mathbb{N}} \subseteq \mathscr{D}$ such that for any given bounded region $K \in \mathscr{D}$, $\Lambda_n \cap K = \varnothing$ for almost all $n \in \mathbb{N}$. Let $A_n \in \mathscr{A}(\Lambda_n)$ with $\|A_n\| \leq b$, uniformly in $n$ for some $b \in \mathbb{R}$, and letting $\pi$ be a representation of $\mathscr{A}$. If

$$w\text{-}\lim_{N\to\infty} \frac{1}{N} \sum_{n=1}^{N} \pi(A_n) = A \tag{5.13}$$

exists, then $A \in \mathscr{L}_\pi$.

**Lemma 5.2.9.** *Consider that $\omega$ has short range correlations and let $\{A_n \in \mathscr{A}(\Lambda_n)\}$ be such that $w\text{-}\lim_{N\to\infty} \frac{1}{N} \sum_{n=1}^{N} \pi(A_n) = A \in \mathscr{L}_\pi$. If*

$$\lim_{N\to\infty} \frac{1}{N} \sum_{n=1}^{N} \omega(A_n) = a \qquad , a \in \mathbb{C}$$

*exists, then in $\mathscr{H}_\omega$ one has:*

$$w\text{-}\lim_{N\to\infty} \frac{1}{N} \sum_{n=1}^{N} \pi_\omega(A_n) = a \tag{5.14}$$

*Proof:* Since $\frac{1}{N} \sum_{n=1}^{N} \omega(A_n)$ is uniformly limited, by the Bounded Linear Transformation Theorem B.0.15 it suffices to prove (5.14) for a dense set of states. If $w\text{-}\lim_{N\to\infty} \frac{1}{N} \sum_{n=1}^{N} \pi_\omega(A_n) = a$ is true, then that means:

$$\left\langle \pi_\omega(A)\Psi_\omega, w\text{-}\lim_{N\to\infty} \frac{1}{N} \sum_{n=1}^{N} \pi_\omega(A_n) \pi_\omega(A)\Psi_\omega \right\rangle = \langle \pi_\omega(A)\Psi_\omega, a\pi_\omega(A)\Psi_\omega \rangle$$

$$\lim_{N\to\infty} \frac{1}{N} \sum_{n=1}^{N} \langle \Psi_\omega, \pi_\omega(A^*A_n A)\Psi_\omega \rangle = \lim_{N\to\infty} \frac{1}{N} \sum_{n=1}^{N} \omega(A^*A_n A) = a\omega(A^*A)$$

And hence we can limit ourselves to prove that for any $\Lambda_1 \in \mathscr{D}$ and $\forall A \in \mathscr{A}(\Lambda_1)$ we have that $\lim_{N\to\infty} \frac{1}{N} \sum_{n=1}^{N} \omega(A^*A_n A) = a\omega(A^*A)$.

Given $A \in \mathscr{A}(\Lambda_1)$ and $\varepsilon > 0$, we choose $\Lambda_2 \in \mathscr{D}$ in such a way that by the providence that $\omega$ has short-range correlations then by the validity of Proposition 5.2.3 we have:

$$|\omega(AB) - \omega(A)\omega(B)| \leq \|B\|\varepsilon \qquad , \forall B \in \widetilde{\mathscr{A}}(\Lambda_2).$$

Since $A_n$ belongs to the algebra of a region converging to infinity, then for a $M \in \mathbb{N}$ sufficiently big, since $A_n \in \mathscr{A}(\Lambda_n)$, in such a way that by this convergence for $n \geq M$ then $\Lambda_1 \cap \Lambda_n = \varnothing$ and therefore the $A_n$ commute with the elements of $\mathscr{A}(\Lambda_1)$, analogously $M$ can be made big enough such that $\Lambda_n$ also doesn't intersect $\Lambda_2$ in such a way that $A_n \in \widetilde{\mathscr{A}}(\Lambda_2) \cap \mathscr{A}(\Lambda_1)'$ for $n \leq M$.

Therefore by locality, for $N \leq M$:

$$\left| \frac{1}{N} \sum_{n=1}^{N} \omega(A^*A_n A) - a\omega(A^*A) \right| = \left| \frac{1}{N} \sum_{n=1}^{M} (\omega(A^*A_n A) - a\omega(A^*A)) + \frac{1}{N} \sum_{n=1+M}^{N} (\omega(A^*A_n A) - a\omega(A^*A)) \right| \leq$$

$$\leq \left| \frac{1}{N} \sum_{n=1}^{M} \omega(A^*A_nA) - \frac{M}{N} a\omega(A^*A) \right| + \left| \frac{1}{N} \sum_{n=1+M}^{N} (\omega(A^*A_nA) - a\omega(A^*A)) \right| \leq$$

$$\leq \frac{1}{N} \left| \sum_{n=1}^{M} \omega(A^*A_nA) \right| + \frac{M}{N}|a|\omega(A^*A) + \left| \frac{1}{N} \sum_{n=1+M}^{N} \omega(A^*A_nA) - \frac{(N-M)}{N} a\omega(A^*A) \right| \leq$$

$$\leq \frac{1}{N} \sum_{n=1}^{M} |\omega(A^*A_nA)| + \frac{M}{N}|a|\omega(A^*A) + \left| \frac{1}{N} \sum_{n=1+M}^{N} \omega(A^*A_nA) - a\omega(A^*A) + \frac{(M)}{N} a\omega(A^*A) \right| \leq$$

$$\leq \frac{1}{N} \sum_{n=1}^{M} |\omega(A^*A_nA)| + 2\frac{M}{N}|a|\omega(A^*A) + \left| \frac{1}{N} \sum_{n=1+M}^{N} \omega(A^*A_nA) - a\omega(A^*A) \right| \quad (5.15)$$

Considering that $\omega$ has short-range correlations by hypothesis and the choice of $M$:

$$\omega(A^*A_nA) = \omega(A^*AA_n)$$

And

$$|\omega(A^*AA_n) - \omega(A^*A)\omega(A_n) \leq \|A_n\|\varepsilon = b\varepsilon$$
$$\implies -b\varepsilon \leq \omega(A^*AA_n) - \omega(A^*A)\omega(A_n) \leq b\varepsilon$$
$$\implies \omega(A^*AA_n) \leq \omega(A^*A)\omega(A_n) + b\varepsilon$$

Therefore continuing from the inequality at (5.15):

$$\leq \frac{1}{N} \sum_{n=1}^{M} |\omega(A^*A_nA)| + \frac{M}{N}|a|\omega(A^*A) + \left| \frac{1}{N} \sum_{n=1+M}^{N} \omega(A_n) - a \right| |\omega(A^*A)| + b\varepsilon$$

Taking a limit of this inequality, we have that the first two terms, the first being a finite sum, go to zero with $\frac{1}{N}$ resulting in:

$$\lim_{N\to\infty} \left| \frac{1}{N} \sum_{n=1}^{N} \omega(A^*A_nA) - a\omega(A^*A) \right| \leq \lim_{N\to\infty} \left( \left| \frac{1}{N} \sum_{n=1+M}^{N} \omega(A_n) - a \right| |\omega(A^*A)| + b\varepsilon \right) =$$

$$\leq \lim_{N\to\infty} \left| \frac{1}{N} \sum_{n=1}^{N} \omega(A_n) - a \right| |\omega(A^*A)| + \lim_{N\to\infty} \left| \frac{1}{N} \sum_{n=1}^{M+1} \omega(A_n) \right| |\omega(A^*A)| + b\varepsilon$$

The first term is equal to zero because of the hypothesis of the convergence of the weighted sum to $a$, the second term is a finite sum that again goes to zero as $\frac{1}{N}$, which leaves:

$$\lim_{N\to\infty} \left| \frac{1}{N} \sum_{n=1}^{N} \omega(A^*A_nA) - a\omega(A^*A) \right| \leq b\varepsilon$$

Since $\varepsilon > 0$ can be taken arbitrarily close to zero this implies:

$$\therefore \lim_{N\to\infty} \frac{1}{N} \sum_{n=1}^{N} \omega(A^*A_nA) = a\omega(A^*A)$$

Which as we had show in the beginning is equivalent to:

$$w\text{-}\lim_{N\to\infty} \frac{1}{N} \sum_{n=1}^{N} \pi_\omega(A_n) = a$$

∎

As consequence of Lemma 5.2.9 we have that for primary states $\omega_1$, $\omega_2$ the existence of a macroscopic observable with different expectation values for $\omega_1$, $\omega_2$ implies the disjointness of $\omega_1$, $\omega_2$:

**Lemma 5.2.10.** *Let $\omega_1, \omega_2 \in \mathfrak{S}(\mathscr{A})$ be primary states, and $\{A_n \in \mathscr{A}(\Lambda_n)\}_n$ a sequence of elements from algebras contained in a sequence of bounded regions that are converging to infinity with* $w\text{-}\lim_{N \to \infty} \frac{1}{N} \sum_{n=1}^{N} \pi_{\omega_i}(A_n) = A$. *If:*

$$\lim_{N \to \infty} \frac{1}{N} \sum_{n=1}^{N} \omega_i(A_n) = a_i \quad , \quad i = 1, 2;$$

*with $a_1 \neq a_2$, then $\omega_1$ and $\omega_2$ are disjoint.*

*Proof:* By Lemma 5.2.9, $\frac{1}{N} \sum_{n=1}^{N} \pi_{\omega_i}(A_n) \xrightarrow[N \to \infty]{w} a_i$, if $\omega_1$ and $\omega_2$ are not disjoint then by Proposition 5.2.7 they are quasiequivalent and therefore or $\pi_{\omega_1} \leq \pi_{\omega_2}$ or $\pi_{\omega_2} \leq \pi_{\omega_1}$. Choosing, without loss of generality, the second case, then there exists a projector $E \in \pi_{\omega_1}(\mathscr{A})$ such that $\pi_{\omega_2}(A) = \pi_{\omega_1}(A)E$, $\forall A \in \mathscr{A}$. Hence $a_2 E = a_1$ and since those are constants that do not depend on subspaces then $a_2 = a_1$, completing the proof by contrapositive. ∎

Thus we see how differences in the expectation values of macroscopic pointers lead to disjointness. Only with a measurement apparatus with infinitely many degrees of freedom can one have a non-trivial quasilocal algebra.

## 5.3 The quasilocal algebra $\mathscr{A}_{\text{spin}}$

We can then consider applying these propositions to the specific case of quasilocal algebras $\mathscr{A}_{\text{spin}}$ of infinitely many spin $^1/_2$ systems on $\mathbb{C}^2$ at lattice sites $n = 1, 2, 3, \dots$; a pure state $|e\rangle \in \mathbb{C}^2$ is characterized by a unit vector $e \in \mathbb{R}^3$ with $\vec{\sigma} \cdot e |e\rangle = |e\rangle$. That is, considering these eigenvectors for $|e\rangle = \binom{a}{b}$:

$$\vec{\sigma} \cdot e |e\rangle = \begin{pmatrix} e_z & e_x - ie_y \\ e_x + ie_y & -e_z \end{pmatrix} \begin{pmatrix} a \\ b \end{pmatrix} = \begin{pmatrix} a \\ b \end{pmatrix} \iff (e_z - 1)a + (e_x - ie_y)b = 0$$

Therefore a $\lambda = +1$ eigenstate of $\vec{\sigma} \cdot e$ is given by:

$$|e\rangle_\theta = \frac{1}{\sqrt{2(1 - e_z)}} \begin{pmatrix} e_x - ie_y \\ 1 - e_z \end{pmatrix} e^{i\theta} \quad , \quad \theta \in \mathbb{R} \text{ a phase.}$$

therefore simplifying

$$|_\vartheta\langle e^1 | e^2 \rangle_\theta|^2 = \frac{2(e_z^1 - 1)(e_z^2 - 1)(1 + e^1 \cdot e^2)}{4(1 - e_z^1)(1 - e_z^2)} \left| e^{i(\theta + \vartheta)} \right|^2 = \frac{1 + e_1 \cdot e_2}{2}$$

Two product states in $\bigotimes_{n=1}^{N} \mathbb{C}_n^2$ given by $|e^i\rangle_N = \bigotimes_{n=1}^{N} |e_n^i\rangle$ have a scalar product such that:

$$|_N\langle e^1 | e^2 \rangle_N|^2 = \prod_{n=1}^{N} \frac{(1 + e_n^1 \cdot e_n^2)}{2}$$

This can be seen as a random distribution of $1+e_n^1 \cdot e_n^2$ values that gives rise to a scalar product that looks like a unnormalized Dirichlet distribution, but with the caveat that $\sum_{n=1}^{N} \frac{(1+e_n^1 \cdot e_n^2)}{2} \neq 1$ even though $\frac{(1+e_n^1 \cdot e_n^2)}{2} \in [0,1] \, \forall n < N$ for which we can consider a normalization coefficient for the sum that can be taken out of the finite product as part of the (missing) normalization.

We may then, very roughly, majorize the Dirichlet distribution by a binomial distribution considering that whenever $e_n^1 \cdot e_n^2 \neq 0$ we set it equal to 1 as to have a upper bound binomial distribution for the renormalized values $\frac{1/2}{3/2} = \frac{1}{3}$ and $\frac{1}{3/2} = \frac{2}{3}$. Since for the binomial distribution we have the *de Moivre—Laplace theorem*, that is a special case of the central limit theorem for binomial distributions, which implies $\prod_{n=1}^{N} \frac{(1+e_n^1 \cdot e_n^2)}{2} \approx e^{-\frac{(k-\frac{2}{3}N)^2}{2N\frac{2}{9}}}$ where k is the number of $e_n^1$ equal to $e_n^2$, the difference $(k - \frac{2}{3}N)$ between the number of times the components are different and the expected number of times this would happen considering the probability 2/3 in $N$ terms can be approximated by the norm distance between the original vectors $\sqrt{\sum_{n=1}^{N} \|e_n^1 - e_n^2\|^2}$ so that:

$$\prod_{n=1}^{N} \frac{(1+e_n^1 \cdot e_n^2)}{2} \approx e^{-\frac{\sum_{n=1}^{N} \|e_n^1-e_n^2\|^2}{\frac{4}{9}N}} \lesssim e^{-\frac{\sum_{n=1}^{N} \|e_n^1-e_n^2\|^2}{4N}} \leq e^{-\frac{\left\|\sum_{n=1}^{N}(e_n^1-e_n^2)\right\|^2}{4N}}$$

And since:

$$e^{-\frac{\left\|\sum_{n=1}^{N}(e_n^1-e_n^2)\right\|^2}{4N}} = \exp\left(-\frac{1}{4N} \left\|\left(\sum_{n=1}^{N} \frac{e_n^1}{N^{\frac{1}{2}+\varepsilon}} - \sum_{n=1}^{N} \frac{e_n^2}{N^{\frac{1}{2}+\varepsilon}}\right) N^{\frac{1}{2}+\varepsilon}\right\|^2\right) = \exp\left(-N^{2\varepsilon} \frac{\|f_{N\varepsilon}^1 - f_{N\varepsilon}^2\|^2}{4}\right)$$
(5.16)

Where

$$f_{N\varepsilon}^i = \sum_{n=1}^{N} \frac{e_n^i}{N^{\frac{1}{2}+\varepsilon}} \quad , i=1,\,2\,;$$

This upper bound leads us directly, in conjunction with the lemmas and propositions of the last section, to the following lemma:

**Lemma 5.3.1.** *Let* $|e^i\rangle = \otimes_{n=1}^{\infty} |e_n^i\rangle$, $i=1,\,2$; *product states in* $\mathscr{A}_{spin}$, *such that for some* $\varepsilon > 0$:

$$\lim_{N \to \infty} \sum_{n=1}^{N} \left(\frac{e_n^1 - e_n^2}{N^{\frac{1}{2}+\varepsilon}}\right) = d \neq 0$$

*Then* $|e^1\rangle$ *and* $|e^2\rangle$ *are weakly inequivalent and therefore disjoint.*

*Proof:* The $|e^i\rangle_N$ are tending to $|e^i\rangle$ as $N \to \infty$ and $0 \leq |_N\langle e^1 | e^2 \rangle_N|^2 \leq e^{-N^{2\varepsilon} \frac{\|f_{N\varepsilon}^1 - f_{N\varepsilon}^2\|^2}{4}}$ taking the limit were we have $0 \leq |\langle e^1 | e^2 \rangle|^2 \leq e^{-\lim_{N \to \infty} N^{2\varepsilon} \frac{\|d\|^2}{4}} = 0 \implies |\langle e^1 | e^2 \rangle|^2 = 0$, and the same is valid for $|_N\langle e^1 | \pi(A) e^2 \rangle_N| = |_N\langle e^1 | \widetilde{e}^2 \rangle_N|$ their inner product going weakly to zero means that they are weakly inequivalent and by Lemma 5.0.3 disjoint. ∎

As we can see from 5.16 it is possible to arrive at a result of the correlation on finite approximations of the infinite system:

$$|_N\langle e^1|A|e^2\rangle_N|^2 = \mathcal{O}\left(e^{-\frac{N^{2\varepsilon}||d||^2}{4}}\right) \tag{5.17}$$

For every extension of a operator of $\mathscr{B}\left(\otimes_{n=1}^{M}\mathbb{C}_n^2\right)$ in $\mathscr{B}\left(\otimes_{n=1}^{N}\mathbb{C}_n^2\right) \ni A$, where $M < N$, and $A$ acts on $\otimes_{n=M}^{N}\mathbb{C}_n^2$ as a identity operator, $||A|| \leq 1$ and fixed $M$, this goes uniformly as $N \to \infty$. Hence macroscopic differences between product states imply a rapidly decreasing overlap for all finite approximations.

## 5.4 Coleman-Hepp Model

The so-called Coleman Model (or as Bell [9] called it: the Coleman-Hepp model) is thus composed of the operators of $L^2(\mathbb{R}) \otimes \left(\otimes_{n=1}^{\infty}\mathbb{C}_n^2\right)$:

$$H_0 = p, \quad H = H_0 + \widehat{V}, \quad \widehat{V} = \sum_{n=1}^{\infty} V(x-n)\sigma_n^1$$

Where the function $V$ is real, continuous and of compact support $\text{supp}(V) = [0,r]$, $r \in \mathbb{R}$ with $\int_{\mathbb{R}} V(x)dx = \frac{\pi}{2}$

Considering then the Dyson iterated equation for the time evolution operator $U(t) = e^{iH_0\frac{t}{\hbar}}e^{-iH\frac{t}{\hbar}}$ is:

$$U(t) = \mathbb{1} - i\int_0^t V(s)U(s)ds \tag{5.18}$$

with

$$V(s) = e^{iH_0\frac{s}{\hbar}}\widehat{V}e^{-iH_0\frac{s}{\hbar}} = \sum_{n=1}^{\infty} V(x+s-n)\sigma_n^1$$

The solution to (5.18) is then given by:

$$U(t) = \mathcal{T}\text{-}\left\{e^{-i\int_0^t \sum_{n=1}^{\infty} V(x+s-n)\sigma_n^1 ds}\right\} \tag{5.19}$$

That is, when considering the time ordering operation $\mathcal{T}$-:

$$U(t) = \sum_{v=0}^{\infty}(-i)^v \int_0^t \int_0^{t_1} \ldots \int_0^{t_{v-1}} \prod_{k=1}^{v}\sum_{n=1}^{\infty} V(x+s_k-n)\sigma_n^1 dk$$

Where the integral $\int_0^{t_{-1}} \prod_{k=1}^{0} \ldots dk$ is understood to be the identity $\mathbb{1}$ and $t_0 \equiv t$ for simplicity of expression.

As one can see equation (5.19) leaves $L^2(I) \otimes \left(\bigotimes_{n=1}^{\infty}\mathbb{C}_n^2\right)$ invariant if $I$ is any bounded open interval in $\mathbb{R}$. Considering that the Hilbert space of the system being measured is given by $\mathscr{H}_S = L^2(I) \otimes \mathbb{C}^2$ with projectors $P^{\pm}$ associated with the eigenvectors $\Psi_{\pm} \in \mathbb{C}^2$, $\sigma^3 \Psi_{\pm} = \pm\Psi_{\pm}$, and considering the algebra of observables of the apparatus that measures the system be $\mathscr{A}_A = \mathscr{A}_{\text{spin}}$, and were the algebra of the combined system plus apparatus is $\mathscr{A}$.

Since the eigenvalues of $\sigma^3$ are $\lambda_1 = 1$, $\lambda_2 = -1$ with eigenvectors given by $\Psi_+ = \binom{1}{0}e^{i\theta}$ and $\Psi_- = \binom{0}{1}e^{i\varrho}$

Ignoring the phases and considering that $\sigma^1$ acts on the eigenvectors $\Psi_+$, $\Psi_-$:

$$\sigma^1 \Psi_+ = \begin{pmatrix} 0 & 1 \\ 1 & 0 \end{pmatrix} \begin{pmatrix} 1 \\ 0 \end{pmatrix} = \begin{pmatrix} 0 \\ 1 \end{pmatrix} = \Psi_-$$

$$\sigma^1 \Psi_- = \begin{pmatrix} 0 & 1 \\ 1 & 0 \end{pmatrix} \begin{pmatrix} 0 \\ 1 \end{pmatrix} = \begin{pmatrix} 1 \\ 0 \end{pmatrix} = \Psi_+$$

That is, the actions of the $\sigma_n^1$ flip the eigenvectors of $\sigma^3$ from those of eigenvalue $+1$ to $-1$ and vice-versa.

Altering the potential of the interaction so as to make the potential affect only the positive eigenvalue eigenvectors of $\sigma^3$ (that we will call spin up, and analogously negative eigenvalue eigenvectors of spin down) of the subsystems of the apparatus:

$$\widetilde{V} = \sum_{n=1}^{\infty} V(x-n)\sigma_n^1 \left( \frac{\mathbb{1}}{2} - \frac{\sigma_n^3}{2} \right)$$

So that the interaction picture evolution $\alpha_t \circ \alpha_{-t} = W(t)(\,\cdot\,)W(t)^*$, where it can be easily verified that:

$$W(t) = P^+ + P^- U(t) \in \mathscr{A}$$

Which is automorphic for all $t$. Then choosing the initial states $\chi \otimes \Psi_\pm \otimes \varphi_+$, where $\chi \in L^2(I)$, for a bounded open interval $I \subset \mathbb{R}$ is the wave function of the measured system with spin given by $\psi_\pm$ and the apparatus state $\varphi_\pm$ has all spins up or all spins down, where in particular we prepare the apparatus as having only spins up. Since $I$ and $\mathrm{supp}(V)$ are bounded the interaction between the system $L^2(I) \otimes \mathbb{C}^2$ and apparatus $\bigotimes_{n=2}^{\infty} \mathbb{C}_n^2$ occurs at most for a finite number $N$ of sites at each time, so that for a given unitary $U \in \mathscr{A}$:

$$U(\infty) = U \prod_{n=1}^{\infty} \sigma_n^1$$

$$U = \left( \prod_{n=1}^{N} \sigma_n^1 \right) \exp\left( -i \int_0^\infty \sum_{n=1}^{N} V(x+s-n) \right) \sigma_n^1$$

Hence as states on $\mathscr{A}$ we get that:

$$W(t)\, \chi \otimes \Psi_+ \otimes \varphi_+ = \chi \otimes \Psi_+ \otimes \varphi_+$$

$$W(t)\, \chi \otimes \Psi_- \otimes \varphi_+ \xrightarrow{w} U\left( \chi \otimes \Psi_- \otimes \varphi_- \right)$$

As we know by Lemma 5.1.3 it was not actually necessary to begin with the pointer state of the apparatus in a pure state, that is, instead of starting from $\omega_{\varphi_+}$ we could consider states that differ from $\varphi_+$ at finitely many $n$ positions given by $\varphi_n = K_n \varphi_+$ for local unitary operators $K_n \in \mathscr{A}_{\mathrm{spin}}$, then taking $\omega = \sum_{m=1}^{M} p_m \omega_{\varphi_m}$ for a finite $M$ as the initial prepared pointer state, then, for example, the evolution of $\omega_{\chi \otimes \psi_-} \otimes \omega$ would converge weakly to $\omega_{\chi \otimes \psi_-} \otimes \sum_{m=1}^{M} p_m \omega_{V_m \varphi_-}$, with local unitary $V_m \in \mathscr{A}$. By Lemma 5.1.3, this pointer position would be disjoint from $\omega_{\chi \otimes \psi_-} \otimes \omega$, as long as the number of sites that were not originally aligned in the apparatus is finite by Lemma 5.3.1.

In this model we see that we adopt $p$ as being the free Hamiltonian, this is made as a simplifying choice for the model, that although not physical can be substituted by the realistic free Hamiltonian $H_0 = \frac{p^2}{2m}$ without many differences in the argument beyond the diffusion of the wave packet $\chi$ of the measured system. As we can see in the limit $N \to \infty$, $t \to \infty$, and exactly only in this limit, that we get a evolution from a superposition of a spin up state and a spin down state $\Psi = c_+ \Psi_+ + c_- \Psi_-$ so that the vector $\phi = \chi \otimes \Psi \otimes \varphi_+$ whose evolution by $W(t)$ leads to $W(t)\phi = \Phi = c_+ \chi \otimes \Psi_+ \otimes \varphi_+ + \tilde{c}_- \chi \otimes \Psi_- \otimes \varphi_-$ where $|\tilde{c}_-| = c_-$, seizing the opportunity to introduce the denotations $\phi_\pm \equiv \chi \otimes \Psi_\pm \otimes \varphi_\pm$.

Considering the finite approximation $\Phi_n = c_+ \chi \otimes \Psi_+ \otimes \left( \bigotimes_{k=1}^{n} \binom{1}{0} \right) + \tilde{c}_- \chi \otimes \Psi_- \otimes \left( \bigotimes_{k=1}^{n} \binom{0}{1} \right)$ the corresponding sequence of states $\omega_{\Phi_n}$ that tends weakly as $n \to \infty$ (which also implies $t \to \infty$, since the amount of time also increases to flip all the spins in $\bigotimes_{k=1}^{n} \binom{1}{0}$) by Lemma 5.1.1 to the mixed state $\omega = |c_+|^2 \omega_{\phi+} + |c_-|^2 \omega_{\phi-}$ composed of two macroscopically distinct states, these being disjoint. Therefore we get a evolution from a pure state to a mixture that can be understood to have a classical probability of being the actual evolution from the system by measurement.

We reserve this space to note that several approximations were made in the process to obtain the solution of the evolution operator for the Colemann-Hepp model, not least of which was the necessity of using the interaction picture, which is known in the AQFT literature to not exist in general, since differently from the usual QM case, in which the Stone-von Neumann guarantees unitary equivalence of representations with relation to the CCR, in QFT we have a infinite number of degrees of freedom, for which not only are unitarily inequivalent representations possible; in fact there exist uncountably many unitarily inequivalent representations of the equal-time canonical commutation relations [47], Haag's theorem then establishes that interaction representations, with some expected requirements, are always unitarily inequivalent in the CCR with the free field representation [93, 33]. This has posed somewhat of a problem for the description of interacting theories within AQFT.

Although, very recently, Buchholz and Fredenhagen in [19] have proposed a way to try to circumvent this difficulty, by using the so-called dynamical $C^*$-algebras that are much inspired in scattering theory with the inclusion of insights gained in the study of perturbative algebraic quantum field theory (pAQFT), in doing so this approach substitutes the problem of inequivalent CCR representations to a problem about the admittance in these $C^*$-algebras of physically necessary states such as a stable ground state. This gives hope that eventually it will be possible to adapt the Coleman-Hepp model for these dynamical $C^*$-algebras, and, in doing so, possibly obtain a result that may also be valid for AQFT in curved spacetimes, as this approach can also be made locally covariant by requiring the generally covariant locality principle [17].

About this result from Hepp, John Bell made the criticism in [9] that "the reduction of the wave packet" would only occur after an infinite time and that the observable:

$$Z_{m=\lfloor t-r-w \rfloor} = \sigma_0^1 \prod_{n=1}^{\lfloor t-r-w \rfloor} \sigma_n^2$$

where $t$ is the time, $r$ is the interaction radius and $w$ is the support length of the considered wave packet, he affirms that this would not obey the result of Lemma 5.3.1. Bell says this because the calculation of the matrix elements:

$$\langle \psi_{+,t},\, Z_m \psi_{-,t}\rangle = \left\langle W(t)\left(\chi \otimes \Psi_+ \otimes \varphi_+\right),\, \sigma_0^1 \prod_{n=1}^{\lfloor t-r-w \rfloor} \sigma_n^2\, W(t)\left(\chi \otimes \Psi_- \otimes \varphi_+\right)\right\rangle$$

Where the $\sigma_n^2$ would unflip the spins that spins that were flipped by the Pauli matrix $\sigma_n^1$ in $W(t) = P^+ + P^- \mathcal{T}\left\{e^{-i\int_0^t \sum_{n=1}^{\infty} V(x+s-n)\sigma_n^1 ds}\right\}$ and so these matrix elements would not tend to 0 in the limit, since the operations that would lead to that were being undone and for every increasing time the inner product increases. The problem with that is, using a argument present in [78], although $Z_m = \pi_m(A_m)$ for some $A_m \in \mathscr{A}$, in fact $A_m \in \mathscr{A}(\Lambda_m)$ in the quasilocal algebra for some compact region $\Lambda_m \in \mathscr{D}$, for each finite $M$, there is no element $A \in \mathscr{A}$ such that the infinite sequence $\{\lambda_m\}_m$ of inner products satisfies:

$$\lim_{m\to\infty} \langle \psi_{+,m+r+w},\, \pi_m(A)\psi_{-,m+r+w}\rangle = \lim_{m\to\infty} \lambda_m = c \neq 0 \tag{5.20}$$

This assertion can be shown in an elementary way as follows, since $\mathscr{A}$ is a quasilocal algebra, then for every $A \in \mathscr{A} = \overline{\bigcup_{\Lambda_\alpha \in \mathscr{D}} \mathscr{A}(\Lambda_\alpha)}^w$, given any $\varepsilon > 0$, there exists $A_L \in \bigcup_{\Lambda_\alpha \in \mathscr{D}} \mathscr{A}(\Lambda_\alpha)$ such that $\|A - A_L\| \leq \varepsilon$, then:

$$\langle \psi_{+,m+r+w},\, \pi_m(A)\psi_{-,m+r+w}\rangle = \langle \psi_{+,m+r+w},\, \pi_m(A-A_L)\psi_{-,m+r+w}\rangle + \langle \psi_{+,m+r+w},\, \pi_m(A_L)\psi_{-,m+r+w}\rangle$$

Considering that $\langle \psi_{+,m+r+w},\, \pi_m(A_L)\psi_{-,m+r+w}\rangle = 0$ for $m$ sufficiently big as to make $\pi_m(A_L)$ alter only the limited amount of sites contained in its localization region, leaving the rest unaltered and therefore rendering the inner product equal to zero because of the remaining orthogonalities. Using also that $\|\pi_m(A - A_L)\| \leq \|A - A_L\|$ then:

$$\langle \psi_{+,m+r+w},\, \pi_m(A)\psi_{-,m+r+w}\rangle \leq \|A - A_L\| = \varepsilon$$

Which is in contradiction with equation (5.20). So that such a element $A = \lim_{m\to\infty} Z_m$ is not a part of the algebra of observables $\mathscr{A}$. The other criticism that *only when* the apparatus is infinite that it is possible to actually say that "the reduction of the wave packet" occurred, not when speaking about any finite approximation, but as we had seen with the WAY theorem, there exists scenarios for which no exact measurement is possible within a finite measuring apparatus and where the full disjointness of the pointer states only exists in the limit $N \to \infty$.

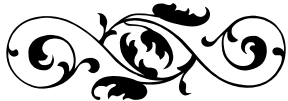

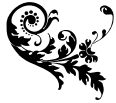 **CHAPTER 6** 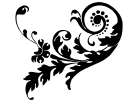

# Contextuality

*In this chapter we will explore the notion of contextuality which has its origin with the Kochen-Specker Theorem proven originally [67] in 1967. Beginning the chapter by giving two distinct and not completely logically equivalent proofs of the Kochen-Specker Theorem, one more technically involved which exposes the connections of this theorem with results from operator theory and one more geometrically involved which manifests the logical aspects of contextuality. Then we consider two [55, 45] proposed physical situations in which the notion of contextuality allows one to resolve possible ambiguities that the measurement setups are arranged to create.*

## 6.1 Kochen-Specker Theorem

The Kochen-Specker theorem is the first result that brings contextuality as a physically relevant concept. This now sometimes called Kochen-Specker contextuality, usually so addressed to differentiate from other, more recent, conceptions of contextuality such as Spekken's contextuality [92], at a first moment one can think that contextuality only makes sense when considering classical *hidden variables* that must display contextuality so as to make the description with these hidden variables match experiment results, but more than this contextuality can be faced as a high level delimitation between quantum aspects and classical aspects, at least to what pertains to information, this will be more precisely put in chapter 7 in which the formalism of category theory will be used to arrive at a abstract formulation of contextuality that is applicable in a variety of physical descriptions.

We beguin with a first proof of the Kochen-Specker theorem that relies on Gleason's theorem and hence depends more explicitly on results from operator theory and in some sense measure theory when considering frame functions that can be thought as a inherent and intrinsic characteristic of any one Hilbert space, in the same way that curvature is a intrinsic characteristic of any space-time.

**Theorem 6.1.1** (Gleason). *Let $\mathcal{H}$ be a Hilbert space of finite dimension $\neq 2$, or infinite dimensional and separable. For any map $\mu$ from the lattice of projectors of the Hilbert space to non-negative values, $\mu : \mathcal{P}(\mathcal{H}) \to [0, +\infty]$, such that $\mu(\mathbb{1}) < +\infty$ and that, like the (ii) axiom of quantum states, $\mu$ has the property of $\sigma$-additivity, there exists a positive trace-class operator $T \in \mathcal{J}_1(\mathcal{H})$ such that:*

$$\mu(P) = \mathrm{tr}(TP) \ , \textit{ for any } P \in \mathcal{P}(\mathcal{H}).$$

*Proof:* As per the hypothesis we consider separable infinite dimensional, or finite dimensional Hilbert spaces, for $\dim(\mathcal{H}) = 1$, $\mathcal{H}$ is isomorphic to $\mathbb{C}$, since the convex set of positive trace-class operators in this case is composed of multiplicative operators $S_m : \mathbb{C} \ni z \mapsto mz \in \mathbb{C}$, where $\mathrm{tr}(S_m) = m \geq 0$. Clearly $\mathcal{P}(\mathcal{H}) = \{0, 1\}$, these seen as multiplicative operators; maps $\mu : \mathcal{P}(\mathbb{C}) \to [0, +\infty]$ that obey $\sigma$-additivity are such that $\mu(0 + 1) = \mu(0) + \mu(1) = \mu(1) \Rightarrow \mu(0) = 0$ and considering that $\mu(1) < +\infty$, so that $\mu(1)$ can be any arbitrary positive finite value, we see that $T := S_{\mu(1)}$ satisfies uniquely Gleason's theorem for each given $\mu$.

Let us now consider the case $\dim(\mathscr{H}) \geq 2$, we begin by defining a non-negative *frame function* from $\mathscr{H}$ to be a mapping from the sphere on $\mathscr{H}$ denoted by $\mathbb{S}(\mathscr{H}) \equiv \{x \in \mathscr{H} \mid \|x\| = 1\}$:

**Definition 6.1.2** (Frame function). *A **frame function** is a map $f : \mathbb{S}(\mathscr{H}) \to [0, +\infty)$ such that for every Hilbert basis (orthonormal basis) $\{x_i\}_{i \in K}$ of $\mathscr{H}$ there exists a weight $W \in [0, +\infty)$ such that:*

$$\sum_{i \in K} f(x_i) = W.$$

We then use the following lemma, which we will prove in appendix C, lemma C.0.31:

**Lemma 6.1.3.** *On any Hilbert space $\mathscr{H}$ that is either separable or of finite dimension $> 2$, for any non-negative frame function $f$ there exists a bounded, self-adjoint operator $T \in \mathfrak{B}(\mathscr{H})$ such that $f(x) = \langle x, Tx \rangle$, for every $x \in \mathbb{S}(\mathscr{H})$.*

Considering then the projectors $P_x(\,\cdot\,) := \langle x, \cdot \rangle x$, $x \in \mathbb{S}(\mathscr{H})$. Considering the hypothesis made about $\mu$ in the theorem, it is straightforward to see that $f(x) := \mu(P_x)$, $x \in \mathbb{S}(\mathscr{H})$ is a non-negative frame function, since $\mu \geq 0$ and for any Hilbert basis $\{x_i\}_{i \in K}$:

$$\sum_{i \in K} f(x_i) = \sum_{i \in K} \mu(P_{x_i}) = \mu\left(\text{s-}\sum_{i \in K} P_{x_i}\right) = \mu(\mathbb{1}) \leq +\infty.$$

By lemma 6.1.3 there is a self-adjoint operator $T$ such that $\mu(P_x) = \langle x, Tx \rangle$, $\forall x \in \mathbb{S}(\mathscr{H})$. Since $\langle x, Tx \rangle = \mu(P_x) \geq 0$, $\forall x \in \mathbb{S}(\mathscr{H})$ then for a arbitrary vector $\xi \in \mathscr{H}$, as we have that $\xi = \|\xi\| \frac{\xi}{\|\xi\|}$ then $\langle \xi, T\xi \rangle = \|\xi\|^2 \left\langle \frac{\xi}{\|\xi\|}, T \frac{\xi}{\|\xi\|} \right\rangle \geq 0$, then $T$ is positive, *i.e.* $T = |T|$, and therefore for a Hilbert basis $\{x_i\}_{i \in K}$:

$$+\infty > \mu(\mathbb{1}) = \sum_{i \in K} f(x_i) = \sum_{i \in K} \langle x_i, Tx_i \rangle = \sum_{i \in K} \langle x_i, |T|x_i \rangle.$$

By definition B.0.12 then $T = |T|$ is of trace class. Considering now $P \in \mathcal{P}(\mathscr{H})$ and picking a Hilbert basis $\{y_i\}_{i \in J}$ of the subspace $P(\mathscr{H})$ and completing it with a Hilbert basis $\{y_i\}_{i \in J'}$ of $P(\mathscr{H})^\perp$. Then by proposition B.0.5 $J$ is countable (enumerable or finite), whence:

$$P = \text{s-}\sum_{i \in J} P_{y_i},$$

by proposition B.0.9. By the orthogonality of the basis:

$$P_{y_i} P_{y_j} = \delta_{ij} P_{y_i}.$$

Since we have that $Py_i = y_i$ if $i \in J$, and $Py_i = 0$ if $i \in J'$, this allows us to write:

$$\mu(P) = \sum_{i \in J} \mu(P_{y_i}) = \sum_{i \in J} \langle y_i, Ty_i \rangle = \sum_{i \in J \cup J'} \langle y_i, TPy_i \rangle = \text{tr}(TP).$$

∎

**Theorem 6.1.4** (Kochen-Specker, measure theory version). *If $\mathscr{H}$ is a Hilbert space, separable or of finite dimension $n > 2$, there is no function $\mu : \mathcal{P}(\mathscr{H}) \to [0,1]$ that is a $\sigma$-additive probability measure on $\mathcal{P}(\mathscr{H})$ taking values only on $\{0,1\}$.*

*Proof:* Considering a unit vector $x \in \mathbb{S}(\mathscr{H})$ and the associated orthogonal projector given by $P_x(\,\cdot\,) := \langle x, \cdot \rangle x$, any function $\mu : \mathcal{P}(\mathscr{H}) \to [0,1]$ that fulfills the properties given in the statement of the theorem gives, by Gleason's theorem 6.1.1, a map $\mathbb{S}(\mathscr{H}) \ni x \mapsto \mu(P_x) = \text{tr}(T P_x) = \langle x, Tx \rangle$, where each $\mu$ determines a unique $T \in \mathcal{J}_1(\mathscr{H})$ with $T \geq 0$ and with $\mu(\mathbb{1}) = \text{tr}(T) = 1$ since $\mu$ is a probability measure. This map is continuous in $\mathbb{S}(\mathscr{H})$, for $T$ is bounded and hence continuous in $\mathscr{H}$ and the inner product is automatically bicontinuous in the topology of $\mathscr{H}$, therefore both and their composition is continuous in $\mathbb{S}(\mathscr{H})$ for the $\mathscr{H}$ induced topology.

$\mathbb{S}(\mathscr{H})$ is path-connected, *i.e.* for any two elements $\phi, \psi \in \mathbb{S}(\mathscr{H})$ there is a continuous path $\gamma : [0,1] \to \mathbb{S}(\mathscr{H})$ starting at $\gamma(0) = \phi$ and ending at $\gamma(1) = \psi$; which can be seen by the fact that, if $t\psi + (1-t)\phi \neq 0, \forall t \in [0,1]$ the path given by:

$$\gamma_1(t) = \frac{t\psi + (1-t)\phi}{\|t\psi + (1-t)\phi\|} = \frac{t\psi + (1-t)\phi}{\sqrt{\langle t\psi + (1-t)\phi, t\psi + (1-t)\phi \rangle}}, \qquad t \in [0,1];$$

is a continuous path since it is formed by the composition of continuous functions, and if $t\psi + (1-t)\phi = \mathbf{0}$ for some $t_0 \in [0,1]$ then $t_0 \psi = (t_0 - 1)\phi$, as $\psi \neq \mathbf{0} \neq \phi$ since they are in $\mathbb{S}(\mathscr{H})$, hence $t_0 \in (0,1)$ and $\psi = \frac{(t_0-1)}{t_0}\phi$ which implies further that $\left|\frac{(t_0-1)}{t_0}\right| = 1$ and therefore $t_0 = \frac{1}{2}$ and $\psi = -\phi$ in which case we can use the path given by $\gamma_2(t) = e^{i\pi t}\psi$. Then, since $\mathbb{S}(\mathscr{H})$ is path-connected and the map $\mathbb{S}(\mathscr{H}) \ni x \mapsto \mu(P_x) = \langle x, Tx \rangle$ is continuous then the map image must also be path-connected, as it can be seen to be constituted of the composition of paths in $\mathbb{S}(\mathscr{H})$ with the measure $\mu$.

Since the image of $\mu$ must be contained in $\{0,1\}$, the possibilities of the image are either $\{0,1\}$ or $\{0\}$ or $\{1\}$; but there is no continuous path for $\{0,1\}$ in the standard topology of $\mathbb{R} \cap [0,1]$, which must be the one considered since $\mu$ is a probability measure $\mu : \mathcal{P}(\mathscr{H}) \to [0,1]$, which eliminates this possibility. If the image is $\{0\}$ then $\mu(P_x) = \langle x, Tx \rangle = 0$ for any $x \in \mathbb{S}(\mathscr{H})$, hence $\text{tr}(T) = 0$, violating the consequence of the probability measure hypothesis $\text{tr}(T) = 1$; whereas if the image is $\{1\}$, then $\mu(P_x) = \langle x, Tx \rangle = 1$ for any $x \in \mathbb{S}(\mathscr{H})$, which would imply that, for any Hilbert basis $\{e_k\}_{0 < k \leq \dim(\mathscr{H})}$:

$$\text{tr}(T) = \sum_{k=1}^{\dim(\mathscr{H})} \langle x_k, T x_k \rangle = \dim(\mathscr{H}),$$

which is again different from 1 for any Hilbert space with dimension $> 2$.

For the case with dimension $n = 2$ Gleason's theorem 6.1.1 is not valid, hence the previous argument doesn't work. Indeed since:

$$P_{-x}(\,\cdot\,) = \langle -x, \cdot \rangle(-x) = \langle x, \cdot \rangle x = P_x(\,\cdot\,),$$

then $\mu(P_{-x}) = \mu(P_x)$ and for every orthonormal basis of a Hilbert space of dimension 2, $\{e_1, e_2\}$, we have:

$$\mu(P_{e_1}) + \mu(P_{e_2}) = 1,$$

but we can re-characterize these vectors by using their angles to a third unit vector $x \in \mathscr{H}$, then if $\theta = \arccos |\langle x, e_1 \rangle|$ and the angle between $x$ and $e_2$ is $\phi$, by the norm of $x$ we conclude:

$$\cos^2 \theta + \cos^2 \phi = 1 \Rightarrow \cos \phi = \sqrt{1 - \cos^2 \theta} = \sin \theta.$$

By the positivity of $\cos \theta = |\langle x, e_1 \rangle|$, then:

$$\sin(\theta) = \sin(\arccos |\langle x, e_1 \rangle|) = \cos\left(\frac{\pi}{2} - \arccos |\langle x, e_1 \rangle|\right)$$

and

$$\phi = \arccos\left(\cos\left(\frac{\pi}{2} - \arccos |\langle x, e_1 \rangle|\right)\right),$$

then either $\phi = \frac{\pi}{2} - \theta$ or $\phi = \theta - \frac{\pi}{2}$, hence if $P_{v(\theta)}$ is a projector associated with a vector $v(\theta)$ that makes an angle $\theta$ with the reference unit vector $x$, we get that:

$$\mu(P_{v(\theta)}) + \mu\left(P_{v(\pm(\frac{\pi}{2}-\theta))}\right) = 1$$

If we then consider an completely arbitrary function $g : \left[0, \frac{\pi}{2}\right) \to \{0, 1\}$, and define the measure $\mu : \mathfrak{B}(\mathscr{H}) \to \{0, 1\}$ to be:

$$\mu(P_{v(\theta)}) := \begin{cases} g(\theta), & \text{for } \theta \in \left[0, \frac{\pi}{2}\right); \\ 1 - g\left(\theta - \frac{\pi}{2}\right), & \text{for } \theta \in \left[\frac{\pi}{2}, \pi\right); \\ g(\theta - \pi), & \text{for } \theta \in \left[\pi, \frac{3\pi}{2}\right); \\ 1 - g\left(\theta - \frac{3\pi}{2}\right), & \text{for } \theta \in \left[\frac{3\pi}{2}, 2\pi\right). \end{cases}$$

Then the existence of a measure $\mu$ that is a $\sigma$-additive probability measure on $\mathcal{P}(\mathscr{H})$ taking values only on $\{0, 1\}$ is verified for $\dim(\mathscr{H}) = 2$. For $\dim(\mathscr{H}) = 1$ the probability measure also exists and is trivially equal to 1 for any projector.

∎

To clarify the logical consequences of the Kochen-Specker theorem, we continue to give a second proof of the theorem, that requires a different set of hypothesis, arguably a weaker set of hypothesis, that also clarify the logical considerations and consequences of the Kochen-Specker theorem.

A vizualization resource which we will make use in proving the Kochen-Specker theorem for the second time, with a proof that is very close to the original, will be the use of graphs to represent sets of vectors, in those, some relations between the vectors are represented graphically in the drawing, the specific rules for drawing the graphs which we will be using will result in our graphs being Greechie diagrams:

**Definition 6.1.5.** *(Greechie diagrams) A graph in which vectors are represented by nodes and mutually orthogonal vectors lie on the same line, that is usually a straight line, but can more generally be a smooth curve, in which case we generally refer to those as generalized Greechie diagrams.*

We then enunciate and prove the usual Kochen-Specker theorem:

**Theorem 6.1.6** (Kochen-Specker). *Let $\mathfrak{B}(\mathscr{H})$ be the algebra of bounded linear operators on a Hilbert space $\mathscr{H}$, let $\mathfrak{B}(\mathscr{H})_\dagger$ be the real vector space of bounded and self-adjoint operators in $\mathscr{H}$ and considering **valuation functions** $v : \mathfrak{B}(\mathscr{H})_\dagger \to \mathbb{R}$ such that $\forall\, a \in \mathfrak{B}(\mathscr{H})_\dagger : v(a) \in \sigma(a)$ and that $v(f(a)) := f(v(a))$ is well defined for all Borel measurable functions $f : \mathbb{R} \to \mathbb{R}$ where $f(a)$ is the operator given by the Borel functional calculus from theorem B.0.56. So, for $\dim(\mathscr{H}) \geq 3$, there are no valuation functions $v : \mathfrak{B}(\mathscr{H})_\dagger \to \mathbb{R}$.*

*Proof:* We begin by considering that if the valuation function is such that $v(a) \in \sigma(a)$ then for projectors such as $P$ either $v(P) = 0$ or $v(P) = 1$, and $v(\mathbb{1}) = 1$ whereas $v(\mathbf{0}) = 0$.

Obviously two projectors that are mutually orthogonal commute. By the von Neumann-Varadarajan theorem A.0.11 we have that for any collection $\{B_i\}_{i \in I}$ of two by two commuting self-adjoint operators, there is a self-adjoint operator $B$ and Borel functions $f_{B_i} : \mathbb{R} \to \mathbb{R}$ such that $f_{B_i}(B) = B_i$, $\forall i \in I$, then for any set of mutually orthogonal projectors $\{P_i\}_{i \in J}$, there will exist a self-adjoint operator $P$ and Borel functions $f_{P_i} : \mathbb{R} \to \mathbb{R}$ given by the previously commented theorem and we will have then that:

$$v\left(\sum_{i \in J} P_i\right) = v\left(\sum_{i \in J} f_{P_i}(P)\right) = v\left(\left(\sum_{i \in J} f_{P_i}\right)(P)\right) = \left(\sum_{i \in J} f_{P_i}\right)(v(P)) =$$
$$= \sum_{i \in J} f_{P_i}(v(P)) = \sum_{i \in J} v(f_{P_i}(P)) = \sum_{i \in J} v(P_i).$$

Considering two projectors $P_1$ and $P_2$ that are orthogonal to each other, such that $v(P_1) = 1$, then considering any completion of $\{P_1, P_2\}$ with other projectors into a partition of unity, let's say $\sum_{i \in K} P_i = \mathbb{1}$, then:

$$1 = v(\mathbb{1}) = v\left(\sum_{i \in K} P_i\right) = \sum_{i \in K} v(P_i),$$

and as $v(P_1) = 1$ and $0 \leq v(P) \leq 1$, for any projector $P$, implies that

$$v(P_i) = 0, \;\; \forall\, i \neq 1,$$

in particular $v(P_2) = 0$. That is, if $P \perp Q$ and $v(P) = 1$ this implies $v(Q) = 0$.

Therefore for any set of orthogonal projectors that is a partition of unity we have that necessarily the valuation of one of these projectors is 1 and the rest has valuation 0.

This proof which follows the original[1] also uses a set of 117 vectors of $\mathbb{R}^3$ and their respective projectors in a closed loop that leads to a projector being forced to receive two distinct valuations.

---

[1] The author knows that there are other equivalent proofs that use less directions than the original 117, such as Peres' 33 directions proof [84] and the 31 directions of the Kochen and Conway proof [85][2], that is, as of the time of the defense of this dissertation, the proof with the smallest number of directions within 3 dimensions (Cabello, Estebaranz and García-Alcaine have a weaker proof using 18 directions [22], but this proof is only valid for $\dim(\mathscr{H}) \geq 4$), nevertheless these proofs are usually made in a more combinatorial way, although they do have cube representations that could make the proofs more geometrical, whereas the original proof gives more intuition on how this double valuation happens using simple cycles and epicycles.

[2] Kochen and Conway have never published their proof with 31 vectors but they have communicated their proof to Peres that has put some information of the proof in his book on Quantum Theory [85], more specifically in page 114 a cube with a pattern of dots informs the 31 directions by joining the dots in the surface of the cube passing through the center of the cube and connecting to other dots on the other side, in page 197 there is a brief commentary on the nature of these directions and on page 211 there is an exercise to prove that those directions indeed form 71 orthogonal pairs, which belong to 17 orthogonal triads.

We begin by considering the following piece of the construction with a particular easy realization in therms of projections within the directions of the shown vectors in the following Greechie diagram:

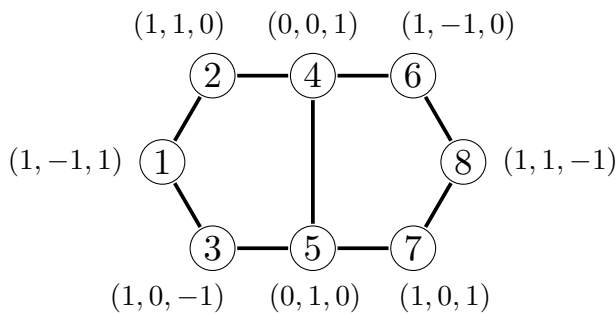

Figure 10 — Greechie diagram for a base piece of the structure that will be used to prove the Kochen-Specker theorem, the triples represent the coordinates of the vectors in $\mathbb{R}^3$, this diagram in particular has received the nickname of "bug".

As we can see form this Greechie diagram, if we assign the value 1 to the valuation of the node ① of the graph, which we represent with the color green in the following image, and consider the color red for the nodes that receive the value 0, then considering the previous rules about the valuations of projectors, the following Greechie diagrams represent the only possible configurations where node ① has value 1:

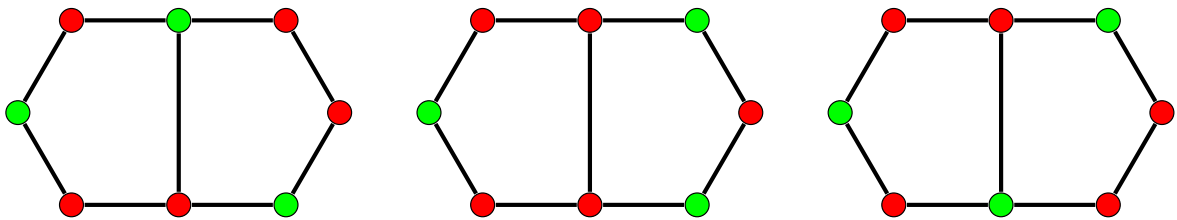

Figure 11 — All possible colorings of the "bug" diagram in which node ① has received the value 1, represented by the color green, and where the value 0 is represented with the color red.

As we can see in all possibilities, the imposition of value 1 to node ①, implies in the value 0 to node ⑧, analogously if we were to impose the value 1 to node ⑧ then node ① would receive 0, as can be seen by a simple rotation of the previous figure.

To prove kochen-Specken we then assemble a new graph using this old one as a building block, in such a way that we begin by considering a basis and for each element of this basis we create one of such blocks in such a way as to make the oposite point of the block orthogonal to the previous vector of the basis in a cyclic fashion. We then complete each of these pairs with another vector forming a basis, from this last vector we construct another blocks such that the opposite point of the block is also orthogonal to the same previous vector in the original basis. We repeat this process in such away as to use 5 of those blocks for each vector of the original basis in such a way that the last opposite point of the last block is the initial vector of the original basis from which the first block of that sequence originated.

We arrive at the following Greechie diagram to represent this collection of vectors, were the generalized rule of continuous curves in Greechie diagrams is used for the original basis:

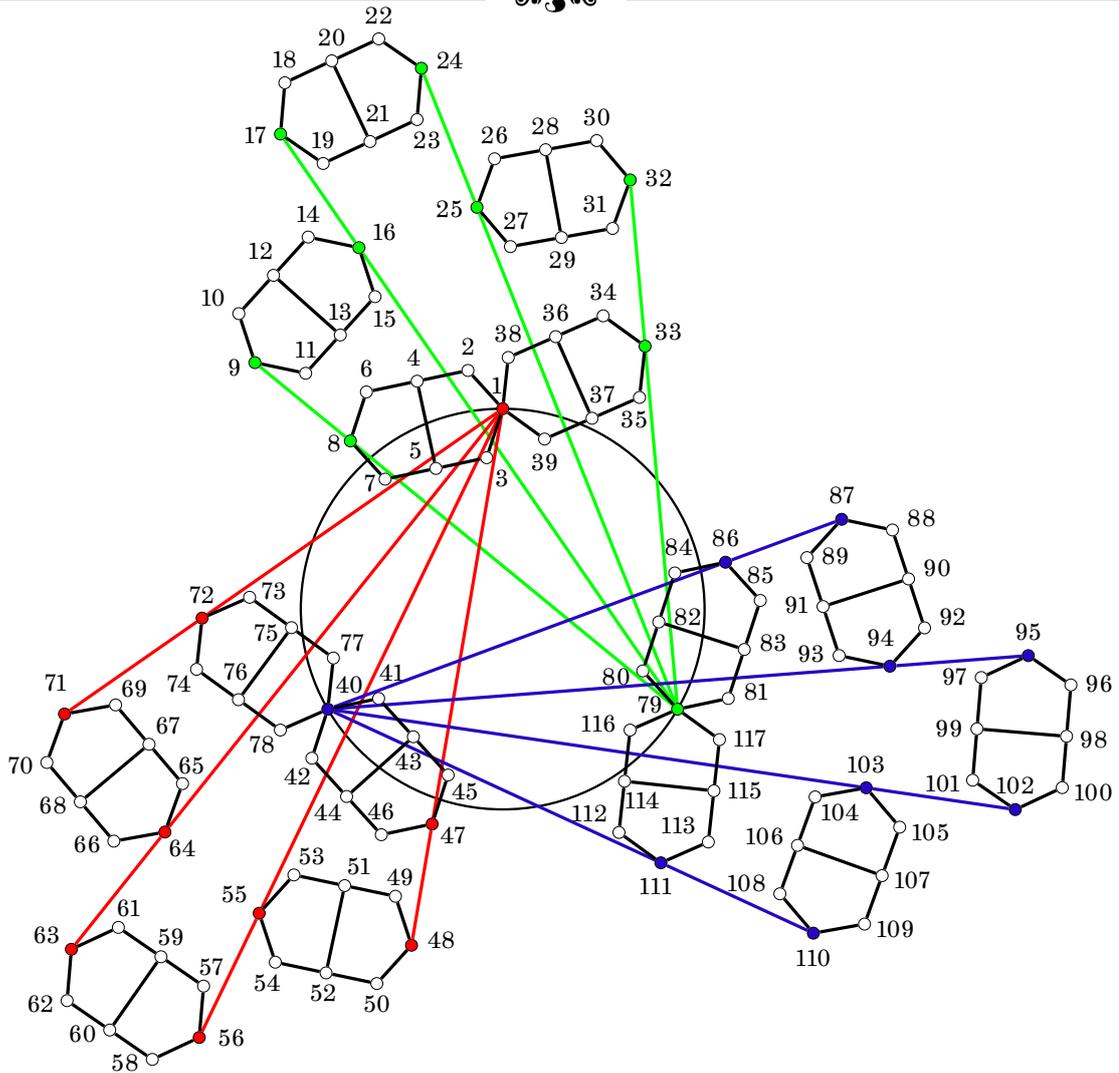

Figure 12 — Complete (generalised) Greechie diagram of the 117 vectors of the original Kochen-Specker proof.

Where the vectors are given by the following table:

Table 1 — Listing of vector coordinates for vectors 1-9 of Figure 12.

| | |
|---|---|
| 1 | $(1, 0, 0)$ |
| 2 | $\left(0, \sqrt{\frac{25-8\sqrt{5}-\sqrt{80\sqrt{5}-175}}{10}}, \sqrt{\frac{8\sqrt{5}-15+\sqrt{80\sqrt{5}-175}}{10}}\right)$ |
| 3 | $\left(0, \frac{\sqrt{75+5\sqrt{80\sqrt{5}-175}}}{10}, -\sqrt{\frac{25-5\sqrt{80\sqrt{5}-175}}{10}}\right)$ |
| 4 | $\left(\frac{1-\sqrt{5}+\sqrt{14\sqrt{5}-10}}{8}, -\frac{\sqrt{-47+\frac{113}{\sqrt{5}}+\sqrt{26\sqrt{5}-50}}}{4}, \frac{2}{\sqrt{5+2\sqrt{5}}}\right)$ |
| 5 | $\left(-\frac{\sqrt{34-6\sqrt{5}+2\sqrt{26\sqrt{5}-50}}}{8}, \sqrt{\frac{5-\sqrt{80\sqrt{5}-175}}{30+2\sqrt{80\sqrt{5}-175}}}, \frac{1}{\sqrt{2}}\right)$ |
| 6 | $\left(\frac{\sqrt{17-3\sqrt{5}+\sqrt{26\sqrt{5}-50}}}{4}, \frac{7\sqrt{5}-15+\sqrt{14\sqrt{5}-10}}{8\sqrt{5}}, \frac{\sqrt{80\sqrt{5}-175}-5}{10}\right)$ |
| 7 | $\left(\frac{1}{\sqrt{\frac{3}{2}+\sqrt{\frac{4}{\sqrt{5}}-\frac{7}{4}}}}, \sqrt{\frac{2}{15+6\sqrt{5}+\sqrt{5+4\sqrt{5}}}}, \frac{\sqrt{25+5\sqrt{80\sqrt{5}-175}}}{10}\right)$ |
| 8 | $\left(\frac{\sqrt{5}-1}{4}, -\frac{\sqrt{2(5+\sqrt{5})}}{4}, 0\right)$ |
| 9 | $\left(\frac{\sqrt{2(5+\sqrt{5})}}{4}, \frac{\sqrt{5}-1}{4}, 0\right)$ |

Table 2 — Listing of vector coordinates for vectors 10-32 of Figure 12.

| | |
|---|---|
| 10 | $\left(\dfrac{7-3\sqrt{5}-2\sqrt{\frac{7\sqrt{5}-5}{10}}}{8},\ \dfrac{\sqrt{17-3\sqrt{5}-\sqrt{26\sqrt{5}-50}}}{4},\ \dfrac{5+\sqrt{80\sqrt{5}-175}}{10}\right)$ |
| 11 | $\left(-\dfrac{\sqrt{5}-1}{8}\sqrt{3+\sqrt{\frac{16}{\sqrt{5}}-7}},\ \dfrac{1}{\sqrt{\frac{3}{2}-\sqrt{\frac{4}{\sqrt{5}}-\frac{7}{4}}}},\ -\dfrac{\sqrt{25-5\sqrt{80\sqrt{5}-175}}}{10}\right)$ |
| 12 | $\left(\sqrt{-\dfrac{7}{8}+\dfrac{3}{\sqrt{5}}-\dfrac{\sqrt{\sqrt{5}-\frac{35}{16}}}{2}},\ \dfrac{-1-\sqrt{1+\frac{4}{-9-4\sqrt{5}}}}{4},\ \dfrac{2}{\sqrt{5+2\sqrt{5}}}\right)$ |
| 13 | $\left(-\dfrac{\sqrt{7+\sqrt{16\sqrt{5}-35}}}{4},\ \dfrac{\sqrt{1-\sqrt{1+\frac{4}{-9-4\sqrt{5}}}}}{4},\ \dfrac{1}{\sqrt{2}}\right)$ |
| 14 | $\left(\dfrac{3-\sqrt{5}+2\sqrt{\frac{17\sqrt{5}+35}{10}}}{8},\ \dfrac{\sqrt{33-13\sqrt{5}+\sqrt{166\sqrt{5}-370}}}{4},\ \dfrac{\sqrt{80\sqrt{5}-175}-5}{10}\right)$ |
| 15 | $\left(\dfrac{\sqrt{45+15\sqrt{5}-\sqrt{70\sqrt{5}-50}}}{4\sqrt{10}},\ \dfrac{\sqrt{15-3\sqrt{5}-\sqrt{166\sqrt{5}-370}}}{4\sqrt{2}},\ \dfrac{\sqrt{25+5\sqrt{80\sqrt{5}-175}}}{10}\right)$ |
| 16 | $\left(\dfrac{\sqrt{2(5-\sqrt{5})}}{4},\ \dfrac{-1-\sqrt{5}}{4},\ 0\right)$ |
| 17 | $\left(\dfrac{1+\sqrt{5}}{4},\ \dfrac{\sqrt{2(5-\sqrt{5})}}{4},\ 0\right)$ |
| 18 | $\left(-\dfrac{\sqrt{33-13\sqrt{5}-\sqrt{166\sqrt{5}-370}}}{4},\ \dfrac{-3+\sqrt{5}+2\sqrt{\frac{35+17\sqrt{5}}{10}}}{8},\ \dfrac{5+\sqrt{80\sqrt{5}-175}}{10}\right)$ |
| 19 | $\left(\dfrac{1-\sqrt{5}}{\sqrt{6-2\sqrt{\frac{16}{\sqrt{5}}-7}}},\ \dfrac{1+\sqrt{5}}{8}\sqrt{3+\sqrt{\frac{16}{\sqrt{5}}-7}},\ -\dfrac{\sqrt{25-5\sqrt{80\sqrt{5}-175}}}{10}\right)$ |
| 20 | $\left(\sqrt{\dfrac{13\sqrt{5}-25}{8}},\ -\sqrt{\dfrac{5-\sqrt{5}}{10}}\big/2,\ \dfrac{2}{\sqrt{5+2\sqrt{5}}}\right)$ |
| 21 | $\left(-\dfrac{\sqrt{1+3\sqrt{5}}}{4},\ -\dfrac{\sqrt{7-3\sqrt{5}}}{4},\ \dfrac{1}{\sqrt{2}}\right)$ |
| 22 | $\left(\dfrac{\sqrt{33-13\sqrt{5}+\sqrt{166\sqrt{5}-370}}}{4},\ \dfrac{3-\sqrt{5}+2\sqrt{\frac{17\sqrt{5}+35}{10}}}{8},\ \dfrac{\sqrt{80\sqrt{5}-175}-5}{10}\right)$ |
| 23 | $\left(\dfrac{15-3\sqrt{5}-\sqrt{166\sqrt{5}-370}}{4\sqrt{2}},\ \dfrac{\sqrt{45+15\sqrt{5}-\sqrt{70\sqrt{5}-50}}}{4\sqrt{10}},\ \dfrac{\sqrt{25+5\sqrt{80\sqrt{5}-175}}}{10}\right)$ |
| 24 | $\left(\dfrac{1+\sqrt{5}}{4},\ -\dfrac{\sqrt{2(5-\sqrt{5})}}{4},\ 0\right)$ |
| 25 | $\left(\dfrac{\sqrt{2(5-\sqrt{5})}}{4},\ \dfrac{1+\sqrt{5}}{4},\ 0\right)$ |
| 26 | $\left(\dfrac{3-\sqrt{5}-2\sqrt{\frac{35+17\sqrt{5}}{10}}}{8},\ \dfrac{\sqrt{33-13\sqrt{5}-\sqrt{166\sqrt{5}-370}}}{4},\ \dfrac{5+\sqrt{80\sqrt{5}-175}}{10}\right)$ |
| 27 | $\left(-\dfrac{\sqrt{\frac{9+3\sqrt{5}+\sqrt{\frac{14}{\sqrt{5}}-2}}{2}}}{4},\ \dfrac{\sqrt{5}-1}{\sqrt{6-2\sqrt{\frac{16}{\sqrt{5}}-7}}},\ -\dfrac{\sqrt{25-5\sqrt{80\sqrt{5}-175}}}{10}\right)$ |
| 28 | $\left(\sqrt{-\dfrac{7}{8}+\dfrac{3}{\sqrt{5}}+\dfrac{\sqrt{\sqrt{5}-\frac{35}{16}}}{2}},\ \dfrac{\sqrt{1+\frac{4}{-9-4\sqrt{5}}}-1}{4},\ \dfrac{2}{\sqrt{5+2\sqrt{5}}}\right)$ |
| 29 | $\left(-\dfrac{\sqrt{7-\sqrt{16\sqrt{5}-35}}}{4},\ -\dfrac{\sqrt{1+\sqrt{1+\frac{4}{-9-4\sqrt{5}}}}}{4},\ \dfrac{1}{\sqrt{2}}\right)$ |
| 30 | $\left(\dfrac{7-3\sqrt{5}+2\sqrt{\frac{7\sqrt{5}-5}{10}}}{8},\ \dfrac{\sqrt{17-3\sqrt{5}+\sqrt{26\sqrt{5}-50}}}{4},\ \dfrac{\sqrt{80\sqrt{5}-175}-5}{10}\right)$ |
| 31 | $\left(\dfrac{\sqrt{45-15\sqrt{5}-\sqrt{2170\sqrt{5}-4850}}}{4\sqrt{10}},\ \dfrac{15+3\sqrt{5}-\sqrt{26\sqrt{5}-50}}{4\sqrt{2}},\ \dfrac{\sqrt{25+5\sqrt{80\sqrt{5}-175}}}{10}\right)$ |
| 32 | $\left(\dfrac{\sqrt{2(5+\sqrt{5})}}{4},\ \dfrac{1-\sqrt{5}}{4},\ 0\right)$ |

Table 3 — Listing of vector coordinates for vectors 33-55 of Figure 12.

| | |
|---|---|
| 33 | $\left(\frac{\sqrt{5}-1}{4}, \frac{\sqrt{2(5+\sqrt{5})}}{4}, 0\right)$ |
| 34 | $\left(-\frac{\sqrt{17-3\sqrt{5}-\sqrt{26\sqrt{5}-50}}}{4}, \frac{-7+3\sqrt{5}+2\sqrt{\frac{7\sqrt{5}-5}{10}}}{8}, \frac{5+\sqrt{80\sqrt{5}-175}}{10}\right)$ |
| 35 | $\left(-\frac{\sqrt{\frac{15+3\sqrt{5}+\sqrt{26\sqrt{5}-50}}{2}}}{4}, \frac{\sqrt{9-3\sqrt{5}+\sqrt{\frac{434}{\sqrt{5}}-194}}}{4\sqrt{2}}, -\frac{\sqrt{25-5\sqrt{80\sqrt{5}-175}}}{10}\right)$ |
| 36 | $\left(\frac{-1+\sqrt{5}+2\sqrt{\frac{7\sqrt{5}-5}{2}}}{8}, \frac{\sqrt{-47+\frac{113}{\sqrt{5}}-\sqrt{26\sqrt{5}-50}}}{4}, \frac{2}{\sqrt{5+2\sqrt{5}}}\right)$ |
| 37 | $\left(-\frac{\sqrt{17-3\sqrt{5}-\sqrt{26\sqrt{5}-50}}}{4\sqrt{2}}, -\frac{\sqrt{\frac{-1+3\sqrt{5}+\sqrt{26\sqrt{5}-50}}{2}}}{4}, \frac{1}{\sqrt{2}}\right)$ |
| 38 | $\left(0, \frac{-2+\sqrt{5}+\sqrt{1+\frac{4}{\sqrt{5}}}}{2}, \frac{\sqrt{80\sqrt{5}-175}-5}{10}\right)$ |
| 39 | $\left(0, \frac{\sqrt{3-\sqrt{\frac{16\sqrt{5}-35}{5}}}}{2}, \frac{\sqrt{25+5\sqrt{80\sqrt{5}-175}}}{10}\right)$ |
| 40 | $(0, 1, 0)$ |
| 41 | $\left(\sqrt{\frac{8\sqrt{5}-15+\sqrt{80\sqrt{5}-175}}{10}}, 0, \sqrt{\frac{25-8\sqrt{5}-\sqrt{80\sqrt{5}-175}}{10}}\right)$ |
| 42 | $\left(-\frac{\sqrt{25-5\sqrt{80\sqrt{5}-175}}}{10}, 0, \frac{\sqrt{75+5\sqrt{80\sqrt{5}-175}}}{10}\right)$ |
| 43 | $\left(\frac{2}{\sqrt{5+2\sqrt{5}}}, \frac{1-\sqrt{5}+\sqrt{14\sqrt{5}-10}}{8}, -\frac{\sqrt{-47+\frac{113}{\sqrt{5}}+\sqrt{26\sqrt{5}-50}}}{4}\right)$ |
| 44 | $\left(\frac{1}{\sqrt{2}}, -\frac{\sqrt{34-6\sqrt{5}+2\sqrt{26\sqrt{5}-50}}}{8}, \sqrt{\frac{5-\sqrt{80\sqrt{5}-175}}{30+2\sqrt{80\sqrt{5}-175}}}\right)$ |
| 45 | $\left(\frac{\sqrt{80\sqrt{5}-175}-5}{10}, \frac{\sqrt{17-3\sqrt{5}+\sqrt{26\sqrt{5}-50}}}{4}, \frac{7\sqrt{5}-15+\sqrt{14\sqrt{5}-10}}{8\sqrt{5}}\right)$ |
| 46 | $\left(\frac{\sqrt{25+5\sqrt{80\sqrt{5}-175}}}{10}, \frac{1}{\sqrt{\frac{3}{2}+\sqrt{\frac{4}{\sqrt{5}}-\frac{7}{4}}}}, \sqrt{\frac{2}{15+6\sqrt{5}+\sqrt{5+4\sqrt{5}}}}\right)$ |
| 47 | $\left(0, \frac{\sqrt{5}-1}{4}, -\frac{\sqrt{2(5+\sqrt{5})}}{4}\right)$ |
| 48 | $\left(0, \frac{\sqrt{2(5+\sqrt{5})}}{4}, \frac{\sqrt{5}-1}{4}\right)$ |
| 49 | $\left(\frac{5+\sqrt{80\sqrt{5}-175}}{10}, \frac{7-3\sqrt{5}-2\sqrt{\frac{7\sqrt{5}-5}{10}}}{8}, \frac{\sqrt{17-3\sqrt{5}-\sqrt{26\sqrt{5}-50}}}{4}\right)$ |
| 50 | $\left(-\frac{\sqrt{25-5\sqrt{80\sqrt{5}-175}}}{10}, -\frac{\sqrt{5}-1}{8}\sqrt{3+\sqrt{\frac{16}{\sqrt{5}}-7}}, \frac{1}{\sqrt{\frac{3}{2}-\sqrt{\frac{4}{\sqrt{5}}-\frac{7}{4}}}}\right)$ |
| 51 | $\left(\frac{2}{\sqrt{5+2\sqrt{5}}}, \sqrt{-\frac{7}{8}+\frac{3}{\sqrt{5}}-\frac{\sqrt{\sqrt{5}-\frac{35}{16}}}{2}}, \frac{-1-\sqrt{1+\frac{4}{-9-4\sqrt{5}}}}{4}\right)$ |
| 52 | $\left(\frac{1}{\sqrt{2}}, -\frac{\sqrt{7+\sqrt{16\sqrt{5}-35}}}{4}, \frac{\sqrt{1-\sqrt{1+\frac{4}{-9-4\sqrt{5}}}}}{4}\right)$ |
| 53 | $\left(\frac{\sqrt{80\sqrt{5}-175}-5}{10}, \frac{3-\sqrt{5}+2\sqrt{\frac{17\sqrt{5}+35}{10}}}{8}, \frac{\sqrt{33-13\sqrt{5}+\sqrt{166\sqrt{5}-370}}}{4}\right)$ |
| 54 | $\left(\frac{\sqrt{25+5\sqrt{80\sqrt{5}-175}}}{10}, \frac{\sqrt{45+15\sqrt{5}-\sqrt{70\sqrt{5}-50}}}{4\sqrt{10}}, \frac{\sqrt{15-3\sqrt{5}-\sqrt{166\sqrt{5}-370}}}{4\sqrt{2}}\right)$ |
| 55 | $\left(0, \frac{\sqrt{2(5-\sqrt{5})}}{4}, \frac{-1-\sqrt{5}}{4}\right)$ |

Table 4 — Listing of vector coordinates for vectors 56-78 of Figure 12.

| | |
|---|---|
| 56 | $\left(0, \frac{1+\sqrt{5}}{4}, \frac{\sqrt{2(5-\sqrt{5})}}{4}\right)$ |
| 57 | $\left(\frac{5+\sqrt{80\sqrt{5}-175}}{10}, -\frac{\sqrt{33-13\sqrt{5}-\sqrt{166\sqrt{5}-370}}}{4}, \frac{-3+\sqrt{5}+2\sqrt{\frac{35+17\sqrt{5}}{10}}}{8}\right)$ |
| 58 | $\left(-\frac{\sqrt{25-5\sqrt{80\sqrt{5}-175}}}{10}, \frac{1-\sqrt{5}}{\sqrt{6-2\sqrt{\frac{16}{\sqrt{5}}-7}}}, \frac{1+\sqrt{5}}{8}\sqrt{3+\sqrt{\frac{16}{\sqrt{5}}-7}}\right)$ |
| 59 | $\left(\frac{2}{\sqrt{5+2\sqrt{5}}}, \sqrt{\frac{13\sqrt{5}-25}{8}}, -\frac{\sqrt{\frac{5-\sqrt{5}}{10}}}{2}\right)$ |
| 60 | $\left(\frac{1}{\sqrt{2}}, -\frac{\sqrt{1+3\sqrt{5}}}{4}, -\frac{\sqrt{7-3\sqrt{5}}}{4}\right)$ |
| 61 | $\left(\frac{\sqrt{80\sqrt{5}-175}-5}{10}, \frac{\sqrt{33-13\sqrt{5}+\sqrt{166\sqrt{5}-370}}}{4}, \frac{3-\sqrt{5}+2\sqrt{\frac{17\sqrt{5}+35}{10}}}{8}\right)$ |
| 62 | $\left(\frac{\sqrt{25+5\sqrt{80\sqrt{5}-175}}}{10}, \frac{15-3\sqrt{5}-\sqrt{166\sqrt{5}-370}}{4\sqrt{2}}, \frac{\sqrt{45+15\sqrt{5}-\sqrt{70\sqrt{5}-50}}}{4\sqrt{10}}\right)$ |
| 63 | $\left(0, \frac{1+\sqrt{5}}{4}, -\frac{\sqrt{2(5-\sqrt{5})}}{4}\right)$ |
| 64 | $\left(0, \frac{\sqrt{10-2\sqrt{5}}}{4}, \frac{1+\sqrt{5}}{4}\right)$ |
| 65 | $\left(\frac{5+\sqrt{80\sqrt{5}-175}}{10}, \frac{3-\sqrt{5}-2\sqrt{\frac{35+17\sqrt{5}}{10}}}{8}, \frac{\sqrt{33-13\sqrt{5}-\sqrt{166\sqrt{5}-370}}}{4}\right)$ |
| 66 | $\left(-\frac{\sqrt{25-5\sqrt{80\sqrt{5}-175}}}{10}, -\frac{\sqrt{\frac{9+3\sqrt{5}+\sqrt{\frac{14}{\sqrt{5}}-2}}{2}}}{4}, \frac{\sqrt{5}-1}{\sqrt{6-2\sqrt{\frac{16}{\sqrt{5}}-7}}}\right)$ |
| 67 | $\left(\frac{2}{\sqrt{5+2\sqrt{5}}}, \sqrt{-\frac{7}{8}+\frac{3}{\sqrt{5}}+\frac{\sqrt{\sqrt{5}-\frac{35}{16}}}{2}}, \frac{\sqrt{1+\frac{4}{-9-4\sqrt{5}}}-1}{4}\right)$ |
| 68 | $\left(\frac{1}{\sqrt{2}}, -\frac{\sqrt{7-\sqrt{16\sqrt{5}-35}}}{4}, -\frac{\sqrt{1+\sqrt{1+\frac{4}{-9-4\sqrt{5}}}}}{4}\right)$ |
| 69 | $\left(\frac{\sqrt{80\sqrt{5}-175}-5}{10}, \frac{7-3\sqrt{5}+2\sqrt{\frac{7\sqrt{5}-5}{10}}}{8}, \frac{\sqrt{17-3\sqrt{5}+\sqrt{26\sqrt{5}-50}}}{4}\right)$ |
| 70 | $\left(\frac{\sqrt{25+5\sqrt{80\sqrt{5}-175}}}{10}, \frac{\sqrt{45-15\sqrt{5}-\sqrt{2170\sqrt{5}-4850}}}{4\sqrt{10}}, \frac{15+3\sqrt{5}-\sqrt{26\sqrt{5}-50}}{4\sqrt{2}}\right)$ |
| 71 | $\left(0, \frac{\sqrt{2(5+\sqrt{5})}}{4}, \frac{1-\sqrt{5}}{4}\right)$ |
| 72 | $\left(0, \frac{\sqrt{5}-1}{4}, \frac{\sqrt{2(5+\sqrt{5})}}{4}\right)$ |
| 73 | $\left(\frac{5+\sqrt{80\sqrt{5}-175}}{10}, -\frac{\sqrt{17-3\sqrt{5}-\sqrt{26\sqrt{5}-50}}}{4}, \frac{-7+3\sqrt{5}+2\sqrt{\frac{7\sqrt{5}-5}{10}}}{8}\right)$ |
| 74 | $\left(-\frac{\sqrt{25-5\sqrt{80\sqrt{5}-175}}}{10}, -\frac{\sqrt{\frac{15+3\sqrt{5}+\sqrt{26\sqrt{5}-50}}{2}}}{4}, \frac{\sqrt{9-3\sqrt{5}+\sqrt{\frac{434}{\sqrt{5}}-194}}}{4\sqrt{2}}\right)$ |
| 75 | $\left(\frac{2}{\sqrt{5+2\sqrt{5}}}, \frac{-1+\sqrt{5}+2\sqrt{\frac{7\sqrt{5}-5}{2}}}{8}, \frac{\sqrt{-47+\frac{113}{\sqrt{5}}-\sqrt{26\sqrt{5}-50}}}{4}\right)$ |
| 76 | $\left(\frac{1}{\sqrt{2}}, -\frac{\sqrt{17-3\sqrt{5}-\sqrt{26\sqrt{5}-50}}}{4\sqrt{2}}, -\frac{\sqrt{\frac{-1+3\sqrt{5}+\sqrt{26\sqrt{5}-50}}{2}}}{4}\right)$ |
| 77 | $\left(\frac{\sqrt{80\sqrt{5}-175}-5}{10}, 0, \frac{-2+\sqrt{5}+\sqrt{1+\frac{4}{\sqrt{5}}}}{2}\right)$ |
| 78 | $\left(\frac{\sqrt{25+5\sqrt{80\sqrt{5}-175}}}{10}, 0, \frac{\sqrt{3-\sqrt{\frac{16\sqrt{5}-35}{5}}}}{2}\right)$ |

Table 5 — Listing of vector coordinates for vectors 79-102 of Figure 12.

| | |
|---|---|
| 79 | $(0,0,1)$ |
| 80 | $\left(\sqrt{\frac{25-8\sqrt{5}-\sqrt{80\sqrt{5}-175}}{10}}, \sqrt{\frac{8\sqrt{5}-15+\sqrt{80\sqrt{5}-175}}{10}}, 0\right)$ |
| 81 | $\left(\frac{\sqrt{75+5\sqrt{80\sqrt{5}-175}}}{10}, -\frac{\sqrt{25-5\sqrt{80\sqrt{5}-175}}}{10}, 0\right)$ |
| 82 | $\left(-\frac{\sqrt{-47+\frac{113}{\sqrt{5}}+\sqrt{26\sqrt{5}-50}}}{4}, \frac{2}{\sqrt{5+2\sqrt{5}}}, \frac{1-\sqrt{5}+\sqrt{14\sqrt{5}-10}}{8}\right)$ |
| 83 | $\left(\sqrt{\frac{5-\sqrt{80\sqrt{5}-175}}{30+2\sqrt{80\sqrt{5}-175}}}, \frac{1}{\sqrt{2}}, -\frac{\sqrt{34-6\sqrt{5}+2\sqrt{26\sqrt{5}-50}}}{8}\right)$ |
| 84 | $\left(\frac{7\sqrt{5}-15+\sqrt{14\sqrt{5}-10}}{8\sqrt{5}}, \frac{\sqrt{80\sqrt{5}-175}-5}{10}, \frac{\sqrt{17-3\sqrt{5}+\sqrt{26\sqrt{5}-50}}}{4}\right)$ |
| 85 | $\left(\sqrt{\frac{2}{15+6\sqrt{5}+\sqrt{5+4\sqrt{5}}}}, \frac{\sqrt{25+5\sqrt{80\sqrt{5}-175}}}{10}, \frac{1}{\sqrt{\frac{3}{2}+\sqrt{\frac{4}{\sqrt{5}}-\frac{7}{4}}}}\right)$ |
| 86 | $\left(-\frac{\sqrt{2(5+\sqrt{5})}}{4}, 0, \frac{\sqrt{5}-1}{4}\right)$ |
| 87 | $\left(\frac{\sqrt{5}-1}{4}, 0, \frac{\sqrt{2(5+\sqrt{5})}}{4}\right)$ |
| 88 | $\left(\frac{\sqrt{17-3\sqrt{5}-\sqrt{26\sqrt{5}-50}}}{4}, \frac{5+\sqrt{80\sqrt{5}-175}}{10}, \frac{7-3\sqrt{5}-2\sqrt{\frac{7\sqrt{5}-5}{10}}}{8}\right)$ |
| 89 | $\left(\frac{1}{\sqrt{\frac{3}{2}-\sqrt{\frac{4}{\sqrt{5}}-\frac{7}{4}}}}, -\frac{\sqrt{25-5\sqrt{80\sqrt{5}-175}}}{10}, -\frac{\sqrt{5}-1}{8}\sqrt{3+\sqrt{\frac{16}{\sqrt{5}}-7}}\right)$ |
| 90 | $\left(\frac{-1-\sqrt{1+\frac{4}{-9-4\sqrt{5}}}}{4}, \frac{2}{\sqrt{5+2\sqrt{5}}}, \sqrt{-\frac{7}{8}+\frac{3}{\sqrt{5}}-\frac{\sqrt{\sqrt{5}-\frac{35}{16}}}{2}}\right)$ |
| 91 | $\left(\frac{\sqrt{1-\sqrt{1+\frac{4}{-9-4\sqrt{5}}}}}{4}, \frac{1}{\sqrt{2}}, -\frac{\sqrt{7+\sqrt{16\sqrt{5}-35}}}{4}\right)$ |
| 92 | $\left(\frac{\sqrt{33-13\sqrt{5}+\sqrt{166\sqrt{5}-370}}}{4}, \frac{\sqrt{80\sqrt{5}-175}-5}{10}, \frac{3-\sqrt{5}+2\sqrt{\frac{17\sqrt{5}+35}{10}}}{8}\right)$ |
| 93 | $\left(\frac{\sqrt{15-3\sqrt{5}-\sqrt{166\sqrt{5}-370}}}{4\sqrt{2}}, \frac{\sqrt{25+5\sqrt{80\sqrt{5}-175}}}{10}, \frac{\sqrt{45+15\sqrt{5}-\sqrt{70\sqrt{5}-50}}}{4\sqrt{10}}\right)$ |
| 94 | $\left(\frac{-1-\sqrt{5}}{4}, 0, \frac{\sqrt{2(5-\sqrt{5})}}{4}\right)$ |
| 95 | $\left(\frac{\sqrt{2(5-\sqrt{5})}}{4}, 0, \frac{1+\sqrt{5}}{4}\right)$ |
| 96 | $\left(\frac{-3+\sqrt{5}+2\sqrt{\frac{35+17\sqrt{5}}{10}}}{8}, \frac{5+\sqrt{80\sqrt{5}-175}}{10}, -\frac{\sqrt{33-13\sqrt{5}-\sqrt{166\sqrt{5}-370}}}{4}\right)$ |
| 97 | $\left(\frac{1+\sqrt{5}}{8}\sqrt{3+\sqrt{\frac{16}{\sqrt{5}}-7}}, -\frac{\sqrt{25-5\sqrt{80\sqrt{5}-175}}}{10}, \frac{1-\sqrt{5}}{\sqrt{6-2\sqrt{\frac{16}{\sqrt{5}}-7}}}\right)$ |
| 98 | $\left(-\frac{\sqrt{\frac{5-\sqrt{5}}{10}}}{2}, \frac{2}{\sqrt{5+2\sqrt{5}}}, \sqrt{\frac{13\sqrt{5}-25}{8}}\right)$ |
| 99 | $\left(-\frac{\sqrt{7-3\sqrt{5}}}{4}, \frac{1}{\sqrt{2}}, -\frac{\sqrt{1+3\sqrt{5}}}{4}\right)$ |
| 100 | $\left(\frac{3-\sqrt{5}+2\sqrt{\frac{17\sqrt{5}+35}{10}}}{8}, \frac{\sqrt{80\sqrt{5}-175}-5}{10}, \frac{\sqrt{33-13\sqrt{5}+\sqrt{166\sqrt{5}-370}}}{4}\right)$ |
| 101 | $\left(\frac{\sqrt{45+15\sqrt{5}-\sqrt{70\sqrt{5}-50}}}{4\sqrt{10}}, \frac{\sqrt{25+5\sqrt{80\sqrt{5}-175}}}{10}, \frac{15-3\sqrt{5}-\sqrt{166\sqrt{5}-370}}{4\sqrt{2}}\right)$ |
| 102 | $\left(-\frac{\sqrt{2(5-\sqrt{5})}}{4}, 0, \frac{1+\sqrt{5}}{4}\right)$ |

Table 6 — Listing of vector coordinates for vectors 103-117 of Figure 12.

| 103 | $\left(\frac{1+\sqrt{5}}{4}, 0, \frac{\sqrt{2(5-\sqrt{5})}}{4}\right)$ |
|---|---|
| 104 | $\left(\frac{\sqrt{33-13\sqrt{5}-\sqrt{166\sqrt{5}-370}}}{4}, \frac{5+\sqrt{80\sqrt{5}-175}}{10}, \frac{3-\sqrt{5}-2\sqrt{\frac{35+17\sqrt{5}}{10}}}{8}\right)$ |
| 105 | $\left(\frac{\sqrt{5}-1}{\sqrt{6-2\sqrt{\frac{16}{\sqrt{5}}-7}}}, -\frac{\sqrt{25-5\sqrt{80\sqrt{5}-175}}}{10}, -\frac{\sqrt{\frac{9+3\sqrt{5}+\sqrt{\frac{14}{\sqrt{5}}-2}}{2}}}{4}\right)$ |
| 106 | $\left(\frac{\sqrt{1+\frac{4}{-9-4\sqrt{5}}}-1}{4}, \frac{2}{\sqrt{5+2\sqrt{5}}}, \sqrt{-\frac{7}{8}+\frac{3}{\sqrt{5}}+\frac{\sqrt{\sqrt{5}-\frac{35}{16}}}{2}}\right)$ |
| 107 | $\left(-\frac{\sqrt{1+\sqrt{1+\frac{4}{-9-4\sqrt{5}}}}}{4}, \frac{1}{\sqrt{2}}, -\frac{\sqrt{7-\sqrt{16\sqrt{5}-35}}}{4}\right)$ |
| 108 | $\left(\frac{\sqrt{17-3\sqrt{5}+\sqrt{26\sqrt{5}-50}}}{4}, \frac{\sqrt{80\sqrt{5}-175}-5}{10}, \frac{7-3\sqrt{5}+2\sqrt{\frac{7\sqrt{5}-5}{10}}}{8}\right)$ |
| 109 | $\left(\frac{15+3\sqrt{5}-\sqrt{26\sqrt{5}-50}}{4\sqrt{2}}, \frac{\sqrt{25+5\sqrt{80\sqrt{5}-175}}}{10}, \frac{\sqrt{45-15\sqrt{5}-\sqrt{2170\sqrt{5}-4850}}}{4\sqrt{10}}\right)$ |
| 110 | $\left(\frac{1-\sqrt{5}}{4}, 0, \frac{\sqrt{2(5+\sqrt{5})}}{4}\right)$ |
| 111 | $\left(\frac{\sqrt{2(5+\sqrt{5})}}{4}, 0, \frac{\sqrt{5}-1}{4}\right)$ |
| 112 | $\left(\frac{-7+3\sqrt{5}+2\sqrt{\frac{7\sqrt{5}-5}{10}}}{8}, \frac{5+\sqrt{80\sqrt{5}-175}}{10}, -\frac{\sqrt{17-3\sqrt{5}-\sqrt{26\sqrt{5}-50}}}{4}\right)$ |
| 113 | $\left(\frac{\sqrt{9-3\sqrt{5}+\sqrt{\frac{434}{\sqrt{5}}-194}}}{4\sqrt{2}}, -\frac{\sqrt{25-5\sqrt{80\sqrt{5}-175}}}{10}, -\frac{\sqrt{\frac{15+3\sqrt{5}+\sqrt{26\sqrt{5}-50}}{2}}}{4}\right)$ |
| 114 | $\left(\frac{\sqrt{-47+\frac{113}{\sqrt{5}}-\sqrt{26\sqrt{5}-50}}}{4}, \frac{2}{\sqrt{5+2\sqrt{5}}}, \frac{-1+\sqrt{5}+2\sqrt{\frac{7\sqrt{5}-5}{2}}}{8}\right)$ |
| 115 | $\left(-\frac{\sqrt{\frac{-1+3\sqrt{5}+\sqrt{26\sqrt{5}-50}}{2}}}{4}, \frac{1}{\sqrt{2}}, -\frac{\sqrt{17-3\sqrt{5}-\sqrt{26\sqrt{5}-50}}}{4\sqrt{2}}\right)$ |
| 116 | $\left(\frac{-2+\sqrt{5}+\sqrt{1+\frac{4}{\sqrt{5}}}}{2}, \frac{\sqrt{80\sqrt{5}-175}-5}{10}, 0\right)$ |
| 117 | $\left(\frac{\sqrt{3-\sqrt{\frac{16\sqrt{5}-35}{5}}}}{2}, \frac{\sqrt{25+5\sqrt{80\sqrt{5}-175}}}{10}, 0\right)$ |

These vectors can be obtained by rotating the first block in the construction in the following directions in sequence, where $0 \leq n \leq 4$:

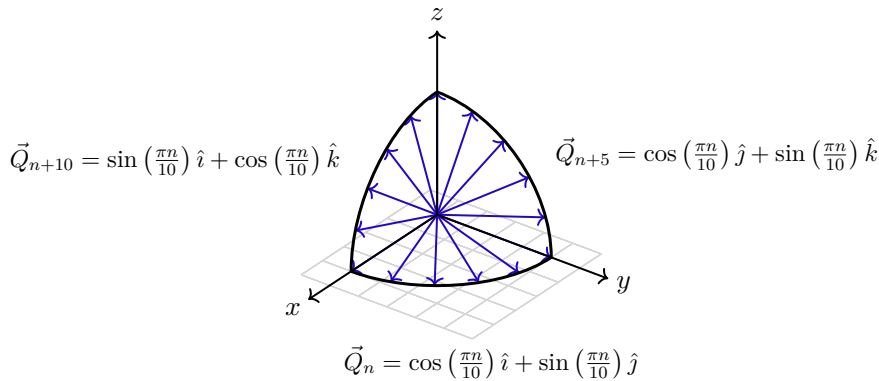

Figure 13 — Rotated vectors associated with each first node of the inner "bug" diagrams.  ∎

We note that in the Kochen-Specker proof, sets of mutually orthogonal projectors that form a basis were the main concept utilized, since mutually orthogonal projectors automatically commute, we can consider the generalization of this property specifically from projectors to elements of a von Neumann algebra that can be broadly used to describe quantum systems, into what are then called contexts:

**Definition 6.1.7** (Context). *Let $\mathscr{H}$ be a Hilbert space of a certain quantum system, and let $\mathcal{N} \subseteq \mathfrak{B}(\mathscr{H})$ denote the von Neumann algebra of observables from the system, then a context is an abelian von Neumann sub-algebra of $\mathcal{N}$.*

This generalization can be more easily understood by considering the theorem 6.1.8 that states that the set of orthogonal projectors of a von Neumann algebra do in fact generate the full von Neumann algebra. Better yet, by von Neumann's bicommutant theorem 4.0.11 the process of taking a bicommutant is equivalent to taking either the weak or strong operator closures of that set, hence the von Neumann algebra generated can be more clearly understood to be the completion of the set of projectors:

**Theorem 6.1.8.** *Let $\mathfrak{M}$ be a (concrete) von Neumann algebra on the complex Hilbert space $\mathscr{H} \neq \{\mathbf{0}\}$ and $\mathcal{P}_{\mathfrak{M}}(\mathscr{H}) := \mathcal{P}(\mathscr{H}) \cap \mathfrak{M}$ be the set of orthogonal projectors $P \in \mathfrak{M}$. Then The von Neumann algebra generated by $\mathcal{P}_{\mathfrak{M}}(\mathscr{H})$ is $\mathfrak{M}$ itself:*

$$\mathcal{P}_{\mathfrak{M}}(\mathscr{H})'' = \mathfrak{M}.$$

*Proof:* As $\mathcal{P}_{\mathfrak{M}}(\mathscr{H}) \subseteq \mathfrak{M}$, then we have $\mathfrak{M}' \subseteq \mathcal{P}_{\mathfrak{M}}(\mathscr{H})'$ and finally $\mathcal{P}_{\mathfrak{M}}(\mathscr{H})'' \subseteq \mathfrak{M}'' = \mathfrak{M}$. We go on to prove $\mathfrak{M} \subseteq \mathcal{P}_{\mathfrak{M}}(\mathscr{H})''$, since the von Neumann algebra given by $\mathcal{P}_{\mathfrak{M}}(\mathscr{H})''$ is closed in the strong operator topology by von Neumann's bicommutant theorem 4.0.11 , it is sufficient to prove that if $A \in \mathfrak{M}$, there exists a sequence of elements $\{A_n\}_{n\in\mathbb{N}} \subset \mathcal{P}_{\mathfrak{M}}(\mathscr{H})''$ such that $A_n \to A$ strongly. As $A = \frac{1}{2}(A + A^*) + i\frac{(A-A^*)}{2i}$, where both terms are self-adjoint elements of $\mathfrak{M}$, we may prove the convergence individually for each self-adjoint part and hence limit ourselves to prove our claim for self-adjoint elements $A \in \mathfrak{M}$.

Since we are talking about self-adjoint operators, we can apply the spectral theorem for bounded self-adjoint operators 4.0.17, and write $A = \int_K x dP^{(A)}(x)$ where $\{P^{(A)}(E)\}_{E \in \mathscr{B}(\mathbb{R})}$ is the PVM associated to $A$, and $K \subset \mathbb{R}$ is a sufficiently large compact set containing the spectrum of $A$. By item *(ii)* of theorem 4.0.17 we have $P^{(A)}(E) = \int_K \chi_E(x) dP^{(A)}(x)$ commutes with every bounded operator commuting with $A$ (and $A^* = A$) where $\chi_E$ is the characteristic function of the Borel set $E \subset \mathbb{R}$.

As $A \in \mathfrak{M} = (\mathfrak{M}')'$, $P^{(A)}(E)$ commutes with every element of $\mathfrak{M}'$ and thus $P^{(A)}(E) \in \mathfrak{M}'' = \mathfrak{M}$. In particular, $P^{(A)}(E) \in \mathcal{P}_{\mathfrak{M}}(\mathscr{H})$ because it is an orthogonal projector. Finally, consider a bounded sequence of simple functions $\{s_n\}_{n\in\mathbb{N}}$ tending pointwise to $\mathbb{1}_{\mathbb{R}} : K \ni x \mapsto x \in \mathbb{R}$ (such a sequence exists as a consequence of proposition C.0.23). Then theorem B.0.58 yields $\int_K s_n(x) dP^{(A)}(x) \to A$ strongly as $n \to \infty$.

On the other hand

$$A_n = \int_K s_n(x) dP^{(A)}(x) = \sum_{k=1}^{N_n} s_{kn} P^{(A)}(E_{kn})$$

commutes with every operator which commutes with the elements of $\mathcal{P}_\mathfrak{M}(\mathscr{H})$ because we saw $P^{(A)}(E_{kn}) \in \mathcal{P}_\mathfrak{M}(\mathscr{H})$. We conclude that $A_n \in \mathcal{P}_\mathfrak{M}(\mathscr{H})''$. Summarising, we have found that $\mathcal{P}_\mathfrak{M}(\mathscr{H})'' \ni A_n \to A$ strongly, as we wanted. ∎

We can also generalize the notion of context to a abstract context:

**Definition 6.1.9.** *Let $\mathscr{A}$ be a $*$-algebra, a abelian unital $*$-sub-algebra of $\mathscr{A}$ is said to be a abstract context of $\mathscr{A}$.*

Having said all that, we can also consider contexts generated by a self-adjoint set of commuting elements:

**Definition 6.1.10** (Generated context)**.** *Let $\mathscr{A}$ be a unital $*$-algebra, and $\{A_n\}_{n\in I}$ a collection of self-adjoint commuting elements, then there is at least one context containing $\{A_n\}_{n\in I}$, namely the space of all polynomials of $\{A_n\}_{n\in I}$, then we can consider the smallest of all such contexts $\bigcap_C \{C \subseteq \mathscr{A} \mid C \text{ is a context containing } \{A_n\}_{n\in I}\}$, and this context is said to be the* context generated *by $\{A_n\}_{n\in I}$, denoted by $\mathcal{C}(\{A_n\}_{n\in I})$.*

In particular, all resolutions of the identity generate a context. If $\{\mathbf{e}_i\}_{i=1}^n \subseteq \mathbb{C}^n$ is an orthogonal basis, then $\mathcal{C}(\{P_{\mathbf{e}_i}\}_{i=1}^n)$ are the diagonal matrices in this basis.

As we repeatedly deal with compatible sets of observables both in the case of the Kochen-Specker theorem and also in the Frauchinger-Renner Gedankenexperiment, it is to be expected that the notion of context naturally enters both situations [32]. We then follow on to prove a related result that will motivate a step in the Frauchinger-Renner argument in their Gedankenexperiment.

## 6.2 Hardy's Paradox

Lucien Hardy in his paper [55] considers the situation in which a couple of Mach-Zehnder interferometers are arranged in such a way as to have a possible annihilation between one of the beams of each interferometer. This Gedankenexperiment is originally used as a argument against the existence of classical hidden variables that can be used to explain quantum phenomena while maintaining the classical characteristics of *realism* and *locality*.

This example is important because, beyond the fact that it serves to show that contextuality can have physically feasible consequences, but also because the type of argument used that is dependent on one given possible configuration of the experiment will be the basis on which the following Frauchinger-Renner Gedankenexperiment will be formulated.

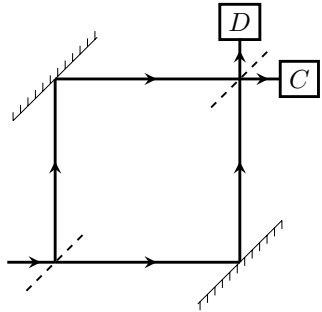

Figure 14 — Mach-Zehnder interferometer.

Considering two Mach-Zehnder interferometers one for a beam of electrons $\mathbf{e}^-$ and one for a beam of positrons $\mathbf{e}^+$ in such a way that they are juxtaposed as to have a point $P$ being a intersection of a electron beam and a positron beam, in this point electrons and positrons annihilate themselves.

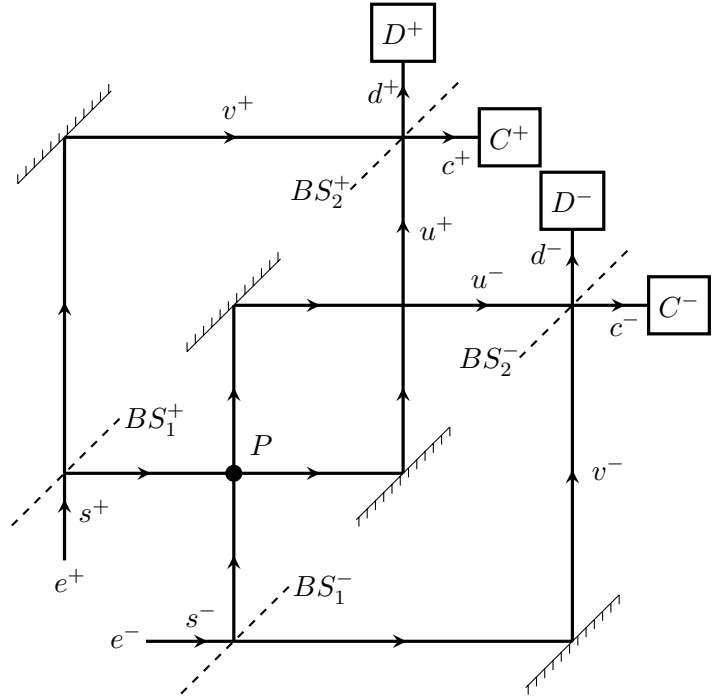

Figure 15 — Experimental setup for Hardy's Gedankenexperiment.

The action of the first beam splitters of each Mach-Zehnder interferometers are such that:

$$BS_1^\pm s^\pm = \frac{1}{\sqrt{2}}(v^\pm + iu^\pm),$$

and of the second beam splitters of each interferometers are:

$$BS_2^\pm u^\pm = \frac{1}{\sqrt{2}}(c^\pm + id^\pm), \qquad BS_2^\pm v^\pm = \frac{1}{\sqrt{2}}(d^\pm + ic^\pm).$$

The point $P$ is the locus of any possible annihilation of beams, which we will represent by the acting of $A_P : \mathscr{H}^+ \otimes \mathscr{H}^- \to \mathscr{H}^+ \otimes \mathscr{H}^-$, where:

$$A_p(u^+ \otimes u^-) = \gamma \oplus \gamma \in \mathscr{H}^+ \oplus \mathscr{H}^-.$$

In the situation were a annihilation occurs, if the beam splitters $BS_2^\pm$ were beforehand removed we would have that:

$$c^\pm = u^\pm \quad \text{and} \quad d^\pm = v^\pm.$$

Considering then that the initial state of the system is:

$$s^+ \otimes s^- \in \mathscr{H}^+ \otimes \mathscr{H}^-.$$

After each Beam passes respectively through $BS_1^+$ and $BS_1^-$:

$$BS_1^+ \otimes BS_1^- (s^+ \otimes s^-) = \frac{1}{2}(-u^+ \otimes u^- + iu^+ \otimes v^- + iv^+ \otimes u^- + v^+ \otimes v^-).$$

After the beams cross in point $P$:

$$A_P\left(\frac{1}{2}(-u^+ \otimes u^- + iu^+ \otimes v^- + iv^+ \otimes u^- + v^+ \otimes v^-)\right) = \frac{1}{2}(-\gamma \oplus \gamma + iu^+ \otimes v^- + iv^+ \otimes u^- + v^+ \otimes v^-).$$

Considering now the following different configurations:

(a) In case the beam splitters $BS_2^\pm$ had been removed, the resulting state shall be:

$$\implies \frac{1}{2}(-\gamma \oplus \gamma + ic^+ \otimes d^- + id^+ \otimes c^- + d^+ \otimes d^-), \tag{6.1}$$

(b) If only $BS_2^+$ had been placed, the state would be:

$$\implies \frac{1}{2\sqrt{2}}(-\sqrt{2}\gamma \oplus \gamma + 2ic^+ \otimes d^- + id^+ \otimes c^- - c^+ \otimes c^-), \tag{6.2}$$

(c) If only $BS_2^-$ had been placed, the state would be:

$$\implies \frac{1}{2\sqrt{2}}(-\sqrt{2}\gamma \oplus \gamma + 2id^+ \otimes c^- + ic^+ \otimes d^- - c^+ \otimes c^-), \tag{6.3}$$

(d) If both beam splitters had been placed, we would have:

$$\implies \frac{1}{4}(-2\gamma \oplus \gamma + id^+ \otimes c^- + ic^+ \otimes d^- - 3c^+ \otimes c^- - d^+ \otimes d^-). \tag{6.4}$$

In his considerations, Hardy then imposes a *realism condition* in which classical hidden variables, the complete dta of which shall be here collectively denoted by $\lambda$, whose value would allow to determine with absolute certainty which results the measurements in the detectors would be.

Each measurement can be realized in two different situations, as we have seen above, the measurement of each interferomenter can be realized with or without the respective beam splitters $BS_2^\pm$, the configuration with the beam splitter will be denoted by the value 0, and the configuration without the beam splitter will be denoted by $\infty$.

Then Hardy imposes a *locality condition* in the form of the consideration that the result of the measurement in one interferometer does not depend on the configuration of the measurement apparatus of the other interferometer, namely the measurement on the + indexed interferometer is independent on the presence or absence of the $BS_2^-$ beam splitter, equally the − interferometer is independent of $BS_2^+$.

We shall then write $D^+(0, \lambda) = 1$ if a positron is detected at $D^+$, where the zero in the first entry represents the presence of the beam splitter $BS_2^+$, and $\lambda$ represents the dependency on classical hidden variables, and write $D^+(0, \lambda) = 0$ if one isn't detected; completely analogously $D^-(0, \lambda) = 1$ means that a electron is detected at $D^-$ in the presence of $BS_2^-$ with hidden variables $\lambda$, and $D^-(0, \lambda) = 0$ if not. If the electron or positron are respectively detected at $C^\pm$ with $BS_2^\pm$ removed then $C^\pm(\infty, \lambda) = 1$ otherwise $C^\pm(\infty, \lambda) = 0$, where the $\infty$ on the first entry represents the absence of $BS_2^\pm$.

From scenario (6.1) we conclude, as there is no $c^+ \otimes c^-$ term, that:

$$C^+(\infty, \lambda)C^-(\infty, \lambda) = 0, \tag{6.5}$$

and from scenario (6.2) we deduce that:

$$\text{if } D^+(0,\lambda) = 1 \Rightarrow C^-(\infty,\lambda) = 1, \tag{6.6}$$

as the only term with $d^+$ has $c^-$. Similarly from (6.3) we get:

$$\text{if } D^-(0,\lambda) = 1 \implies C^+(\infty,\lambda) = 1, \tag{6.7}$$

and from (6.4) one has:

$$D^+(0,\lambda)D^-(0,\lambda) = 1 \text{ in } \frac{1}{16} \text{ of the experiments.} \tag{6.8}$$

Considering one of these experiments in which $D^+(0,\lambda)D^-(0,\lambda) = 1$ (we stress this dependency on a specific configuration) by (6.7) and (6.6) we conclude that $C^+(\infty,\lambda)C^-(\infty,\lambda) = 1$, but (6.5) says that $C^+(\infty,\lambda)C^+(\infty,\lambda) = 0$, hence we have a contradiction between the system and the assumptions of *locality* and *classical realism* that were supposed upon the system. This means that classical hidden variables are incompatible with sensible conditions of locality and realism on these hidden variables.

This result receives the name of Hardy's paradox since isolated each interferometer works in such a way that the state $BS_1^{\pm}s^{\pm} = \frac{1}{\sqrt{2}}(v^{\pm} + iu^{\pm})$ would be exactly the one to arrive respectively at $BS_2^{\pm}$ and hence:

$$BS_2^{\pm}\left(\frac{1}{\sqrt{2}}(v^{\pm} + iu^{\pm})\right) = \frac{1}{\sqrt{2}}\left(\frac{1}{\sqrt{2}}(d^{\pm} + ic^{\pm}) + i\frac{1}{\sqrt{2}}(c^{\pm} + id^{\pm})\right) = ic^{\pm}$$

Therefore each interferometer is arranged in such a way that the beams interfere in a completely destructive way in the direction of detectors $D^{\pm}$ and also interfere constructively in the direction of detectors $C^{\pm}$. The paradox lies in the situation in which a single electron and a single positron are sent into each respective interferometer, in the configuration we originally considered in which a annihilation can occur at point $P$ and with the beam splitters $BS_2^{\pm}$ placed, basically situation (6.4). If it is measured either particle in at least one of the detectors $D^{\pm}$, which is possible by the previous calculations, then as the only influence placed on either interferometer is the possible annihilation at point $P$, this means that a annihilation is influencing the measurement outcomes, but in a situation in which only one electron and positron are present in each interferometer beam, a annihilation means that no detection should occur in either detector since the particles had become gamma rays, this establishes a paradox of sorts when trying to force classical notions on this quantum system.

The actual detection of a particle in either $D^{\pm}$ would then indicate that a annihilation interaction was being perceived by the system, but without any such annihilation actually taking place, this type of contemplation upon this Gedankenexperiment has lead some to consider that the annihilation itself can be detected in this system without any such annihilation interaction happening, making it a **interaction-free measurement**.

Hardy's proof relies on a specific set of vector states, each of which are related by being the results of different apparatus configurations on the experiment, and the ability to assign classical data to each such vector state, in such a way that this classical data determines globally all the results of measurements of the system. In this sense Hardy's proof can be seen as a state-dependent version of the Kochen-Specker theorem [23].

## 6.3 Application: Frauchinger-Renner Gedankenexperiment

The Gedankenexperiment proposed by Daniela Frauchinger and Renato Renner [45] is a generalization of the gedankenexperiment proposed by Wigner [102] and now known as the paradox of Wigner's friend, that was discussed in the introduction in chapter 1. In a very similar way to the way the Kochen-Specker theorem 6.1.6 was originally demonstrated, the proof of the Impossibility Theorem involved in the Frauchinger-Renner gedankenexperiment scenario also involves a double valuation, in this case, a state that has measurements conditioned to other measurements in a chained manner in order to allow two ways of assigning the valuation of the measurement of this state by two different paths and which leads to the Impossibility Theorem:

**Theorem 6.3.1** (Frauchinger-Renner impossibility). *Any theory that satisfies the assumptions given by **Born's rule (Q)**, the ability to **generalize an assertion (C)** about the information available to an agent, with the possibility of simplifying information without any obstructions related to the receptacle of this information; and the **univaluation (S)** of measurement results of any one system, these assumptions lead to a contradiction in the scenario described by the Gedankenexperiment.*

To clarify what the conditions **(Q), (C)** and **(S)** mean we will elaborate on them in what follows:

— Condition **(Q)** is plain old Born's rule, which for all intents and purposes here just means that for a normalized vector $\psi \in \mathcal{H}$, if $P_\xi \psi = \psi$ then $\xi = \psi$ and therefore $\langle \psi, P_\xi \psi \rangle = 1$.

— To what pertains condition **(C)**, the ability to generalize an assertion about the available information we must consider a type of metalogic of the system to describe logical relations between the experimenter agents themselves. To make the exposition clear we shall follow [80] and adapt the Kripke semantics structure exposed in chapter 2 to substitute the 5th axiom, know as the Knowledge axiom in definition 2.3.6, to the following Trust axiom:

**Definition 6.3.2** (Trust axiom[3]). *We say that an agent $i$ "trusts"(or more precisely is amenable to the information of) $j$, and denote it by $i \leftharpoonup j$, if and only if:*

$$(\mathcal{M}, s) \models \mathcal{K}_i \mathcal{K}_j \phi \implies \mathcal{K}_i \phi, \quad \forall s \in \Sigma, \phi \in \Phi$$

This Trust axiom, although usually employed in situations in which a rational agent tries to filter information from all available sources, only relying upon what each agent considers to be a reliable source. This has the effect that two different agents may hold conflicting pieces of information as long as neither one considers the other to be a trustful source of information, and that no two chains of rational agents that trust the following agent in the chain can be formed such that these chains link the same rational agent to both agents that hold mutually incompatible information. In this way one can realize a sort of paraconsistent logic.

---

[3] The author has chosen to employ this formalism of Kripke semantics because it allows for a more simple and clear exposition of the Frauchinger-Renner Gedankenexperiment, in which the **semantic contents** *are contained in the **logical form*** of the expressions, whence no analytic—synthetic distinction is present. If the reader is skeptical, unconvinced or otherwise uncomfortable with this type of extremely formal exposition, a more casual exposition that tries to argue directly from a number of cumbersome physically motivated calculations (not presenting the explicit logical structure of the inferences made along them) is available in the article [18]. Nevertheless, this does not mean that the exposition made in this dissertation is any less physically motivated, just, in some sense, more organized by taking advantage of the expressive potential of Kripke semantics for explicitly showing in this situation the, otherwise implicit, logical structure of the argument.

That is to say, if rational agent $A$ doesn't trust rational agent $B$, and vice-versa, also if $A$ knows $q$ and $B$ knows $\neg q$, then the only conclusions any one of them can directly make about the other is that the other "thinks" he knows the opposite (perhaps a better describing word would be "believe" he knows the opposite, but it is better not to mix notions so freely, since for anyone of these agents it is valid that what they know, they know with 100% of certainty, it is not a likeliness evaluation in which there is some room for doubt, which could be the case for a belief, and it is also implied that this knowledge is grounded in some evidences, whereupon a belief might be so in spite of evidence). In symbols:

$$A \not\leadsto B \wedge B \not\leadsto A, \quad \text{then if } (\mathscr{K}_A q) \wedge (\mathscr{K}_B \neg q)$$

this makes what each of they can directly conclude about the other to be:

$$\mathscr{K}_A[\mathscr{K}_B \neg q] \quad \text{and} \quad \mathscr{K}_B[\mathscr{K}_A q]$$

and no more, meaning that one cannot reduce these formulae as $\mathscr{K}_A[\mathscr{K}_B \neg q] \Rightarrow \mathscr{K}_A \neg q$ or $\mathscr{K}_B[\mathscr{K}_A q] \Rightarrow \mathscr{K}_B q$, meaning that they are kept as $\mathscr{K}_A[\mathscr{K}_B \neg q]$ and $\mathscr{K}_B[\mathscr{K}_A q]$, where there is not necessarily any logical contradiction in this Kripke semantics. To be able to keep this information system consistent we must also ask the second condition about there being no agent that links two chains of trust leading to agents that hold mutually contradictory information, since this would lead to a contradiction by the application of the trust axiom, in symbols:

$$(A_1 \leadsto ... \leadsto A_k \leftsquigarrow ... \leftsquigarrow A_n) \wedge \mathscr{K}_{A_1} q \wedge \mathscr{K}_{A_n} \neg q \implies \mathscr{K}_{A_k}[q] \wedge \mathscr{K}_{A_k}[\neg q].$$

This is enough of a pure logic digression, what this brings to physics, more specifically to quantum mechanics is the possibility of making an adaptation of this formalism of trust relations to represent the information about the outcomes of measurements realized by agents. Since in quantum mechanics measurements almost always alter the state of the system being measured it would not make sense to consider that the original fifth knowledge axiom would make any sense, *e.g.* if for a given electron the assertion: $q \equiv$ "the projection on the $z$ direction of the spin of the electron is in the positive $z$ direction" is to be considered in a situation in which an experimenter $A$ has realized the measurement, if we resignify the meaning of $\mathscr{K}_A q$ to have the semantic meaning of "$A$ has realized a measurement obtaining as result, or otherwise acquired the knowledge that $q$", then while it would make sense to write $\mathscr{K}_A q$, axiom 5 would make this imply that assertion $q$ by itself would be true independently of any particular measurement experiment, but this makes no sense since if another experimenter $B$ decided to measure the spin projection in the $x$ direction then this would perturb the electron and make uncertain any subsequent measurement in the $z$ direction, making it possible to not be anymore in the positive $z$ direction as $q$ would imply.

The previous discussion motivates us to also resignify the trust relation $A \leftsquigarrow B$ to mean a trust that if $A$ were to repeat the same measurement as $B$ did, then $A$ would always get the same result as $B$, in other words $A$ trusts that $B$ has not made a destructive measurement, or failed to keep the system isolated, or immediately after taking that measurement has made, or allowed another person to make, a incompatible measurement that would perturb the system in such a way as to make his measurement not repeatable. From this considerations one can clearly see that the Trust axiom $A \leftsquigarrow B \iff \mathscr{K}_A \mathscr{K}_B q \Rightarrow \mathscr{K}_A q$ makes sense quantum mechanically, since as $A$ could in principle repeat the same measurement as $B$ and get the same result, then it makes as much sense for $A$ to know $q$ as does for $B$.

— Lastly, condition **(S)** means that a agent cannot measure two different results for the same value at the same time, that is, if $P_\xi \psi = \psi$ at a time $t$, then $(\mathbb{1} - P_\xi)\psi = \mathbf{0}$ also at time $t$. Equivalently in the notation of Kripke semantics, if $\mathscr{K}_A\big[(P_\xi \psi \neq 0) \wedge (P_\xi \psi = 0)\big]$ then $\mathscr{K}_A(P_\xi \psi \neq 0) \wedge (\neg \mathscr{K}_A(P_\xi \psi \neq 0))$, which is a usual, global, contradiction $A \wedge \neg A$, whereas the $\mathscr{K}_B[A \wedge \neg A]$ could be seen as a local, *i.e.* to agent $B$, contradiction.

Having made all of these important semantic considerations, we can go on to prove the Frauchinger-Renner no-go result, where the formal language of Kripke semantics with the substitution of axiom 5 for the Trust axiom, which we shall from now own number as $5'$, will allow us to simplify the exposition of the argument, providing also a formulaic backbone that makes explicit and more mechanically verifiable the logical reasoning behind the argument itself.

*Proof:* The Frauchinger-Renner Gedankenexperiment is formulated considering a generalization of Wigner's friend Gedankenexperiment, this time we shall have two laboratories $L$ and $\widetilde{L}$, each with its own two-level quantum system $d$ and respectively $\widetilde{d}$, each of which can be expressed in a orthonormal basis $\{a, b\}$ of a Hilbert space $\mathscr{H}_S$, experimenters $F$ and $\widetilde{F}$, and respective outside observers $W$ and $\widetilde{W}$.

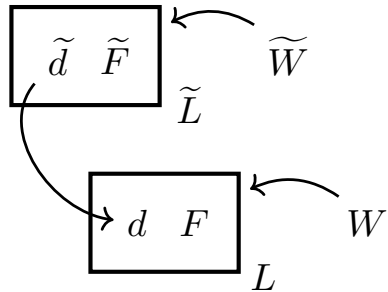

Figure 16 — Representation of the Frauchinger-Renner Gedankenexperiment as a variation in the Wigner's friend Gedankenexperiment.

It is considered, just as in the Wigner's friend Gedankenexperiment, that each experimenter $\widetilde{F}$ and $F$ can have their states measured to determine the measured outcome of the two level system that they perceived, as a simplifying assumption we shall consider this to be representable with the orthogonal basis $\{u, w\} \in \mathscr{H}_P$, where we have chosen other letters to better distinguish the states of the system and the states of the experimenters, of course one can criticize this simplification as non-representative of the complexity that is a rational agent, in which case the same experiment can be carried out mechanically with a second two-level quantum system being used as a replacement for the state of the experimenter memory where the measurement result is held.

As occurs in QM , each experimenter has to consider the evolution of all other quantum systems, that he is not measuring, as unitary (all constituent parts of the experimental setup are fundamentally quantum systems and even though each of the experimenters know that the other experimenters are going to perform measurements, he can only add the effects of these measurements is with correlations on the states implemented by isometries, since the state of that experimenter himself isn't getting entangled with the sistem being measured by another agent.)

The Gedankenexperiment is arranged in steps following this procedure table:

Table 7 — Frauchinger-Renner Gedankenexperiment experimental protocol.

| Step $^{(n)}t=0$ | Agent $\widetilde{F}$ prepares a quantum system $\widetilde{d}$ to be in a state $\sqrt{\frac{1}{3}}a + \sqrt{\frac{2}{3}}b$, subsequently $\widetilde{F}$ measures the system $\widetilde{d}$ in the orthogonal basis $\{a,b\} \subset \mathscr{H}_S$ and if the result of this measurement is $a$ then $F$ sends a quantum system $d$ prepared in state $a \in \mathscr{H}_S$ to lab $L$, otherwise if the result of the measurement is $b$, then $F$ sends a quantum system $d$ prepared in state $\sqrt{\frac{1}{2}}(a+b)$ to lab $L$. |
|---|---|
| Step $^{(n)}t=1$ | Agent $F$ measures $d$ with respect to the basis $\{a,b\}$ and preserves the resulting state as a recording. |
| Step $^{(n)}t=2$ | Agent $\widetilde{W}$ measures the entirety of the lab $\widetilde{L}$ in a orthogonal basis given by $\left\{\sqrt{\frac{1}{2}}(a\otimes u + b\otimes w), \sqrt{\frac{1}{2}}(a\otimes u - b\otimes w), \sqrt{\frac{1}{2}}(b\otimes u + a\otimes w), \sqrt{\frac{1}{2}}(b\otimes u - a\otimes w)\right\}$, more specifically with the projector $P_{\sqrt{\frac{1}{2}}(a\otimes u - b\otimes w)} \equiv P_{\tilde{c}}$ and if it obtains a non-null outcome he then announces the information "$P_{\tilde{c}}\widetilde{L} \neq 0$" publicly, otherwise he announces "$P_{\tilde{c}}\widetilde{L} = 0$". |
| Step $^{(n)}t=3$ | Agent $W$ measures the entirety of the lab $L$ in a orthogonal basis given by $\left\{\sqrt{\frac{1}{2}}(a\otimes u + b\otimes w), \sqrt{\frac{1}{2}}(a\otimes u - b\otimes w), \sqrt{\frac{1}{2}}(b\otimes u + a\otimes w), \sqrt{\frac{1}{2}}(b\otimes u - a\otimes w)\right\}$, more specifically with the projector $P_{\sqrt{\frac{1}{2}}(a\otimes u - b\otimes w)} \equiv P_c$ and if it obtains a non-null outcome he then announces the information "$P_c L \neq 0$" publicly, otherwise he announces "$P_c L = 0$". |
| Step $^{(n)}t=4$ | If "$P_c L \neq 0$" and "$P_{\tilde{c}}\widetilde{L} \neq 0$" then the experiment is halted, otherwise the index of $^{(n)}t$ is increased to $^{(n+1)}t$ and the procedure is repeated beginning from step 0, this is repeated until the experiment halts. |

We can then consider which agents receive relevant information from which other agents, or in other words what are the trust relations between the measuring agents. Clearly since $\widetilde{W}$ announces his sole measurement result for all the other agents at time $t=2$, then we must have that all other agents trust $\widetilde{W}$ at $t=2$, in particular $W_{t=3} \curvearrowleft \widetilde{W}_{t\geq 2}$. Analogously, agent $W$ announces his sole measurement result for all the other agents at time $t=3$, then we must have that all other agents trust $W$ at $t=3$, in particular $\widetilde{F}_4 \curvearrowleft W_{t=3}$.

If $\widetilde{W}$ can indirectly find out the information of the measurement of $F$ at time $t=1$, then $\widetilde{W}$ can trust this information since the quantum system $d$ that $F$ is measuring is and remains after its measurement isolated form $\widetilde{W}$, that is $\widetilde{W}_{t\geq 2} \curvearrowleft F_{t=2}$.

And finally since the measurements that $F$ and $\widetilde{F}$ make are compatible then any information one gets about the other they can trust as long as those systems have not been disturbed at those times by someone else, in particular $F_t = 2 \curvearrowleft \widetilde{F}_{t=0,1,2}$.

Synthesizing these particular relations into a trust hierarchy one gets:

$$\widetilde{F}_{t=4} \curvearrowleft W_{t=3} \curvearrowleft \widetilde{W}_{t\geq 2} \curvearrowleft F_{t=2} \curvearrowleft \widetilde{F}_{t=0,1,2}.$$

From the outset the following assertions are true because of the way the Gedankenexperiment was arranged and because every agent knows the arrangement of the experiment:

If $\widetilde{F}_{t=1}$ measures $b$ from system $\widetilde{d}$, then $\widetilde{F}$ knows that after $F$ measures $d = \sqrt{\frac{1}{2}}(a+b)$ in the basis $\{a,b\}$ it has as much chance of measuring $a$ as it has for measuring $b$ therefore $\widetilde{F}$ arrives at the conclusion that $W$ will be measuring the lab $L$ in the state $\sqrt{\frac{1}{2}}(a \otimes u + b \otimes w)$, which is orthogonal to $c = \sqrt{\frac{1}{2}}(a \otimes u - b \otimes w)$, hence $P_c L = 0$. In symbols:

$$\models \mathscr{K}_{\widetilde{F}_{t<3}}\big[\mathscr{K}_{\widetilde{F}_{t=1}}(\widetilde{d}=b) \Rightarrow \mathscr{K}_{W_{t=4}}(P_c L = 0)\big] \tag{6.9}$$

Since $F$ knows from the outset of the experiment that $\widetilde{F}$ will send him either a $d = a$ state if $\widetilde{F}$ measures $a$, or a $d = \sqrt{\frac{1}{2}}(a+b)$ if $\widetilde{F}$ measures $b$, then if $F$ measures $d = b$ he will know that $\widetilde{F}$ had measured $\widetilde{d} = b$. Therefore with the formula:

$$\models \mathscr{K}_{F_{t<3}}\big[\mathscr{K}_{F_{t=1}}(d=b) \Rightarrow \mathscr{K}_{\widetilde{F}_{t=0,1,2}}(\widetilde{d}=b)\big] \tag{6.10}$$

Since $\widetilde{W}$ knows that when he is going to make a measurement of the lab $\widetilde{L}$, he considers that the global state is at that time equal to $\frac{1}{\sqrt{3}}(a_{\widetilde{d}} \otimes u_{\widetilde{W}} \otimes a_d \otimes u_W + b_{\widetilde{d}} \otimes w_{\widetilde{W}} \otimes a_d \otimes u_W + b_{\widetilde{d}} \otimes w_{\widetilde{W}} \otimes b_d \otimes w_W)$ hence he can factor this state in the basis that he is measuring with to get $\sqrt{\frac{2}{3}}\frac{1}{\sqrt{2}}(a_{\widetilde{d}} \otimes u_{\widetilde{W}} + b_{\widetilde{d}} \otimes w_{\widetilde{W}}) \otimes a_d \otimes u_W + \frac{1}{\sqrt{3}} b_{\widetilde{d}} \otimes w_{\widetilde{W}} \otimes b_d \otimes w_W$, since the first part of the global state is perpendicular to the vector $\widetilde{c} \equiv a_{\widetilde{d}} \otimes u_{\widetilde{W}} - b_{\widetilde{d}} \otimes w_{\widetilde{W}}$ associated with the projector $P_{\widetilde{c}}$ that $\widetilde{W}$ uses to make his measurement, then if he gets $P_{\widetilde{c}}\widetilde{L} \neq 0$ that means that the only part of the global state that overlaps with this observation is the part $\frac{1}{\sqrt{3}} b_{\widetilde{d}} \otimes w_{\widetilde{W}} \otimes b_d \otimes w_W$, which makes $\widetilde{W}$ certain that $F$ measured $b$ at time $t = 2$, that is:

$$\models \mathscr{K}_{\widetilde{W}_{t<4}}\big[\mathscr{K}_{\widetilde{W}_{t=3}}(P_{\widetilde{c}}\widetilde{L} \neq 0) \Rightarrow \mathscr{K}_{F_{t=1,2}}(d=b)\big] \tag{6.11}$$

Since before the beginning of the experiment the participants must communicate amongst themselves to set up the experiment in the first place, then they can also inform each other of the implications that each deduced based on a possible measurement outcome, in particular the previous tree assertions can be made know to $F$, $\widetilde{W}$ and $W$ respectively:

$$\models \mathscr{K}_{F_{t=2}}\mathscr{K}_{\widetilde{F}_{t<3}}\big[\mathscr{K}_{\widetilde{F}_{t=1}}(\widetilde{d}=b) \Rightarrow \mathscr{K}_{W_{t=4}}(P_c L = 0)\big] \tag{6.12}$$

$$\models \mathscr{K}_{\widetilde{W}_{t=3}}\mathscr{K}_{F_{t<3}}\big[\mathscr{K}_{F_{t=1}}(d=b) \Rightarrow \mathscr{K}_{\widetilde{F}_{t=0,1,2}}(\widetilde{d}=b)\big] \tag{6.13}$$

$$\models \mathscr{K}_{W_{t=4}}\mathscr{K}_{\widetilde{W}_{t<4}}\big[\mathscr{K}_{\widetilde{W}_{t=3}}(P_{\widetilde{c}}\widetilde{L} \neq 0) \Rightarrow \mathscr{K}_{F_{t=1,2}}(d=b)\big] \tag{6.14}$$

These in part can be communicated again to other agents and so on, up to third- and fourth-order statements like:

$$\models \mathscr{K}_{\widetilde{W}_{t=3}}\mathscr{K}_{F_{t=2}}\mathscr{K}_{\widetilde{F}_{t<3}}\big[\mathscr{K}_{\widetilde{F}_{t=1}}(\widetilde{d}=b) \Rightarrow \mathscr{K}_{W_{t=4}}(P_c L = 0)\big] \tag{6.15}$$

$$\models \mathscr{K}_{W_{t=4}}\mathscr{K}_{\widetilde{W}_{t=3}}\mathscr{K}_{F_{t=2}}\mathscr{K}_{\widetilde{F}_{t<3}}\big[\mathscr{K}_{\widetilde{F}_{t=1}}(\widetilde{d}=b) \Rightarrow \mathscr{K}_{W_{t=4}}(P_c L = 0)\big] \tag{6.16}$$

$$\models \mathscr{K}_{W_{t=4}}\mathscr{K}_{\widetilde{W}_{t=3}}\mathscr{K}_{F_{t<3}}\big[\mathscr{K}_{F_{t=1}}(d=b) \Rightarrow \mathscr{K}_{\widetilde{F}_{t=0,1,2}}(\widetilde{d}=b)\big] \tag{6.17}$$

As we can only reduce the statements made about trusted agents, we shall begin by considering all of the previous statements that are within $W$'s knowledge,

$$\models \mathscr{K}_{W_{t=4}}\mathscr{K}_{\widetilde{W}_{t<4}}\big[\mathscr{K}_{\widetilde{W}_{t=3}}(P_{\widetilde{c}}\widetilde{L} \neq 0) \Rightarrow \mathscr{K}_{F_{t=1,2}}(d=b)\big] \tag{6.18}$$

$$\models \mathscr{K}_{W_{t=4}}\mathscr{K}_{\widetilde{W}_{t=3}}\mathscr{K}_{F_{t<3}}\big[\mathscr{K}_{F_{t=1}}(d=b) \Rightarrow \mathscr{K}_{\widetilde{F}_{t=0,1,2}}(\widetilde{d}=b)\big] \tag{6.19}$$

$$\models \mathcal{K}_{W_{t=4}}\mathcal{K}_{\widetilde{W}_{t=3}}\mathcal{K}_{F_{t=2}}\mathcal{K}_{\widetilde{F}_{t<3}}\big[\mathcal{K}_{\widetilde{F}_{t=1}}(\widetilde{d}=b) \Rightarrow \mathcal{K}_{W_{t=4}}(P_cL=0)\big] \quad (6.20)$$

First we combine the bottom two statements:

$$\models \mathcal{K}_{W_{t=4}}\mathcal{K}_{\widetilde{W}_{t=3}}\mathcal{K}_{F_{t=2}}\bigg[\big[\mathcal{K}_{F_{t=1}}(d=b) \Rightarrow \mathcal{K}_{\widetilde{F}_{t=0,1,2}}(\widetilde{d}=b)\big] \wedge$$
$$\wedge \big[\mathcal{K}_{\widetilde{F}_{t<3}}\big[\mathcal{K}_{\widetilde{F}_{t=1}}(\widetilde{d}=b) \Rightarrow \mathcal{K}_{W_{t=4}}(P_cL=0)\big]\big]\bigg], \quad (6.21)$$

and now we apply the trust condition $F_{t=2} \curvearrowleft \widetilde{F}_{t=0,1,2}$:

$$\models \mathcal{K}_{W_{t=4}}\mathcal{K}_{\widetilde{W}_{t=3}}\mathcal{K}_{F_{t=2}}\bigg[\big[\mathcal{K}_{F_{t=1}}(d=b) \Rightarrow \mathcal{K}_{\widetilde{F}_{t=0,1,2}}(\widetilde{d}=b)\big] \wedge$$
$$\wedge \big[\mathcal{K}_{\widetilde{F}_{t=1}}(\widetilde{d}=b) \Rightarrow \mathcal{K}_{W_{t=4}}(P_cL=0)\big]\bigg].$$

Using now the distribution axiom 2.3.2 for $F$:

$$\models \mathcal{K}_{W_{t=4}}\mathcal{K}_{\widetilde{W}_{t=3}}\mathcal{K}_{F_{t=2}}\big[\mathcal{K}_{F_{t=1}}(d=b) \Rightarrow \mathcal{K}_{W_{t=4}}(P_cL=0)\big]. \quad (6.22)$$

Now we combine this with the remaining statement about $W$'s knowledge:

$$\models \mathcal{K}_{W_{t=4}}\mathcal{K}_{\widetilde{W}_{t=3}}\bigg[\big[\mathcal{K}_{\widetilde{W}_{t=3}}(P_{\widetilde{c}}\widetilde{L} \neq 0) \Rightarrow \mathcal{K}_{F_{t=1,2}}(d=b)\big] \wedge$$
$$\wedge \big[\mathcal{K}_{F_{t=2}}\big[\mathcal{K}_{F_{t=1}}(d=b) \Rightarrow \mathcal{K}_{W_{t=4}}(P_cL=0)\big]\big]\bigg]. \quad (6.23)$$

Again, we apply the trust condition $\widetilde{W}_{t\geq2} \curvearrowleft F_{t=2}$, followed by the distribution axiom for $\widetilde{W}$, obtaining:

$$\models \mathcal{K}_{W_{t=4}}\mathcal{K}_{\widetilde{W}_{t=3}}\big[\mathcal{K}_{\widetilde{W}_{t=3}}(P_{\widetilde{c}}\widetilde{L} \neq 0) \Rightarrow \mathcal{K}_{W_{t=4}}(P_cL=0)\big]. \quad (6.24)$$

Note that the above holds for all possible outcome scenarios, and that $\widetilde{W}$ and $W$ reach the conclusion $\mathcal{K}_{\widetilde{W}_{t=3}}(P_{\widetilde{c}}\widetilde{L} \neq 0) \Rightarrow \mathcal{K}_{W_{t=4}}(P_cL=0)$ before the experiment even starts. Additionally, at the end of the experiment, after time $t=3$, $\widetilde{W}$ and $W$ learn of each other's outcomes, in particular:

$$\forall s \in \Sigma: \quad \big[(\mathcal{M},s) \models \mathcal{K}_{\widetilde{W}_{t=2}}(P_{\widetilde{c}}\widetilde{L} \neq 0)\big] \implies \big[(\mathcal{M},s) \models \mathcal{K}_{W_{t=3}}\mathcal{K}_{\widetilde{W}_{t=2,3}}(P_{\widetilde{c}}\widetilde{L} \neq 0)\big]. \quad (6.25)$$

Now we run the experiment. As in the Hardy's paradox Gedankenexperiment, we analyze a specific scenario within all possible configurations of outcomes that exist in the experiment $\Sigma$, namely, we follow the scenario $s$ in which both $\widetilde{W}$ reports "$P_{\widetilde{c}}\widetilde{L} \neq 0$" and $W$ reports "$P_cL \neq 0$", by considering from the very beginning the global state of the experiment:

$$\frac{1}{\sqrt{3}}(a_{\widetilde{d}} \otimes u_{\widetilde{W}} \otimes a_d \otimes u_W + b_{\widetilde{d}} \otimes w_{\widetilde{W}} \otimes a_d \otimes u_W + b_{\widetilde{d}} \otimes w_{\widetilde{W}} \otimes b_d \otimes w_W) =$$

$$= \frac{1}{\sqrt{3}}\bigg[(a_{\widetilde{d}} \otimes u_{\widetilde{W}} + b_{\widetilde{d}} \otimes w_{\widetilde{W}})\bigg(a_d \otimes u_W + \frac{1}{2}b_d \otimes w_W\bigg)\bigg] -$$
$$- \frac{1}{4\sqrt{3}}(a_{\widetilde{d}} \otimes u_{\widetilde{W}} - b_{\widetilde{d}} \otimes w_{\widetilde{W}})(a_d \otimes u_W + b_d \otimes w_W) +$$
$$+ \frac{1}{2\sqrt{3}}\bigg[\bigg(\sqrt{\frac{1}{2}}(a_{\widetilde{d}} \otimes u_{\widetilde{W}} - b_{\widetilde{d}} \otimes w_{\widetilde{W}})\bigg)\bigg(\sqrt{\frac{1}{2}}(a_d \otimes u_W - b_d \otimes w_W)\bigg)\bigg]$$

therefore the chance of this happening is $\left|\frac{1}{2\sqrt{3}}\right|^2 = \frac{1}{12}$, hence:

$$\exists s \in \Sigma: \quad (\mathcal{M}, s) \models \mathcal{K}_{\widetilde{W}_{t=2,3}}(P_{\tilde{c}}\widetilde{L} \neq 0) \wedge \mathcal{K}_{W_{t=3}}(P_c L \neq 0) \tag{6.26}$$

Given that $\widetilde{W}$ communicates his result to all other agents in the experiment, we must also have that $W$ knows the result of $\widetilde{W}$ at time $t = 3$:

$$\exists s \in \Sigma: \quad (\mathcal{M}, s) \models \mathcal{K}_{W_{t=3}}\left[\mathcal{K}_{\widetilde{W}_{t=2,3}}(P_{\tilde{c}}\widetilde{L} \neq 0) \wedge \mathcal{K}_{W_{t=3}}(P_c L \neq 0)\right]. \tag{6.27}$$

Now we can combine this with the assertion (6.24) to get:

$$\exists s \in \Sigma: \quad (\mathcal{M}, s) \models \mathcal{K}_{W_{t=3}}\Big[\left[\mathcal{K}_{\widetilde{W}_{t=2}}(P_{\tilde{c}}\widetilde{L} \neq 0) \wedge \mathcal{K}_{W_{t=3}}(P_c L \neq 0)\right] \wedge$$
$$\wedge \mathcal{K}_{\widetilde{W}_{t=3}}\left[\mathcal{K}_{\widetilde{W}_{t=2}}(P_{\tilde{c}}\widetilde{L} \neq 0) \Rightarrow \mathcal{K}_{W_{t=4}}(P_c L = 0)\right]\Big] \tag{6.28}$$

applying the trust relation $W_{t=3} \curvearrowleft \widetilde{W}_{t=2}$:

$$\exists s \in \Sigma: \quad (\mathcal{M}, s) \models \mathcal{K}_{W_{t=3}}\Big[\left[\mathcal{K}_{W_{t=3}}(P_c L \neq 0) \wedge \mathcal{K}_{\widetilde{W}_{t=2}}(P_{\tilde{c}}\widetilde{L} \neq 0)\right] \wedge$$
$$\wedge \left[\mathcal{K}_{\widetilde{W}_{t=2}}(P_{\tilde{c}}\widetilde{L} \neq 0) \Rightarrow \mathcal{K}_{W_{t=4}}(P_c L = 0)\right]\Big]$$

using now the distribution axiom for $\widetilde{W}$:

$$\exists s \in \Sigma: \quad (\mathcal{M}, s) \models \mathcal{K}_{W_{t=3}}\left[\mathcal{K}_{W_{t=3}}(P_c L \neq 0) \wedge \mathcal{K}_{W_{t=4}}(P_c L = 0)\right]. \tag{6.29}$$

Finally, we can apply the trivial trust relation $W_{t=3} \curvearrowleft W_{t=3}$ to obtain:

$$\exists s \in \Sigma: \quad (\mathcal{M}, s) \models \mathcal{K}_{W_{t=3}}\left[(P_c L \neq 0) \wedge (P_c L = 0)\right],$$

we now use condition **(S)** to get a global contradiction from a local one:

$$\mathcal{K}_{W_{t=3}} \wedge \neg \mathcal{K}_{W_{t=3}}$$

As a matter of fact this contradiction can be achieved by all agents at he end of the experiment by considering the emulation of the thought process of $W$ and his final announcement at $t = 3$. ∎

The author, having read the original paper [45] of D. Frauchiger and R. Renner couldn't avoid to notice a striking similarity between this process of contradictory two-valuation of an assertion in this Gedankenexperiment and the also contradictory two-valuation of the Kochen-Specker theorem 6.1.6 (or 6.1.4), even though this was not present in any way in the original paper.

Immersed in the thought that this could be a, small but nevertheless novel, contribution to the problem, the author got on to write and try to more naturally relate these two results. It was only then that he decided to search if there were any papers available that sought to expose this relationship, and after some search, through a lot of papers that were even just trying to understand the argument in the first place (it is after all a somewhat polemic paper with a extraordinary claim serving as the title, and the original exposition of the argument was so convoluted that it was difficult for anyone to call that "extraordinary evidence"), the author came across [80] that used Kripke semantics to simplify and organize the original argument and that also considered contextuality as a solutution to the problem.

The notion of contextuality that is used in [80], called "two-dimensional semantics", differs from our definition using abelian von Neuman sub-algebras, it instead uses the entire list of observers that have made measurements and provided information so that a conclusion could be reached, and at each measurement a new observer was appended to the list, in this sense being a context of under which measurements were those assertions derived.

The usual notion of contextuality simplifies this description by only looking at the orthogonal basis under which any of the measurements were performed, hence we can change the trust axiom to reflect this constraint:

**Definition 6.3.3** (Contextual trust axiom). *We say that an agent $i$ "trusts"(or more precisely is amenable to the information of) $j$, and denote it by $i \curvearrowleft j$, if and only if the context $\mathscr{C}$ of $i$ is the same as the context of $j$ and:*

$$(\mathcal{M}, s) \models \mathscr{K}_i \Big|_{\mathscr{C}} \mathscr{K}_j \Big|_{\mathscr{C}} \phi \implies \mathscr{K}_i \Big|_{\mathscr{C}} \phi, \quad \forall\, s \in \Sigma, \phi \in \Phi$$

Using this new axiom it becomes clear that $\widetilde{W}_{t \geq 2} \not\curvearrowleft F_{t=2}$ and $\widetilde{F}_{t=4} \not\curvearrowleft W_{t=3}$, since the contexts in which each of those makes their measurements is different, hence one cannot arrive at the contradiction since the assertion (6.23) now given by $\models \mathscr{K}_{W_{t=4}}\Big|_{\mathscr{C}_2} \mathscr{K}_{\widetilde{W}_{t=3}}\Big|_{\mathscr{C}_2} \Big[ \mathscr{K}_{\widetilde{W}_{t=3}}\Big|_{\mathscr{C}_2} (P_{\tilde{c}} \widetilde{L} \neq 0) \Rightarrow \mathscr{K}_{F_{t=1,2}}\Big|_{\mathscr{C}_1} (d = b) \Big] \wedge \Big[ \mathscr{K}_{F_{t=2}}\Big|_{\mathscr{C}_1} \Big[ \mathscr{K}_{F_{t=1}}\Big|_{\mathscr{C}_1} (d = b) \Rightarrow \mathscr{K}_{W_{t=4}}\Big|_{\mathscr{C}_2} (P_c L = 0) \Big] \Big]$ cannot be reduced anymore to an equivalent of (6.24).

The impossibility of such reductions is implicitly assumed by [80] to mean that there is a hard limitation on the reasoning of the agents that is imposed on them without their noticing which makes them perceive the information received by measurements in other contexts as independent from their own. A much simpler interpretation is that the agents *do, **in fact**, know* quantum mechanics, hence they must also know the Kochen-Specker theorem 6.1.6 (or 6.1.4), hence they know they cannot use the measurement results from another context in their own without excluding the possibility of arriving at contradictory results.

A more standard way to rephrase that is to say: For the agents to actually be considered to be using QM in their deductions, they have to carefully consider the compatibility, or lack thereof, of the measurement results when those are being logically combined.

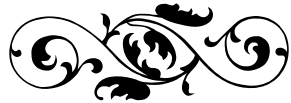

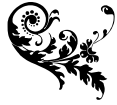 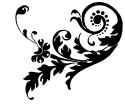

# CHAPTER 7

# AQFT and Contextuality in Sheaf Theory

Another interesting connection that emerged was between contextuality and Sheaf Theory [3], which in addition to unifying all expressions of contextuality [67, 55] and being a way of representing Kripke's semantics, such as As it happens, the Algebraic Quantum Field Theory itself can be seen as a specific construction of Sheaf Theory, where the [14] bundles arise naturally from the condition of locality of the observables in TQCA. So the natural question that arises is whether it is natural then to translate all these contextuality issues naturally into AQFT.

Sheaf Theory seems to be the natural language in which recent postulations of quantum contextuality and the theorems that involve it have been reformulated [32, 58, 59, 61, 60, 3], in addition, it could also led to a deepening of the connections between Algebraic Quantum Field Theory [4, 51, 37, 46, 16] and Quantum Measurement Theory [20, 21, 71], since they all could be formulated in this same language of *sewing* local data (Algebras, Contexts, Instruments) in a global structure.

**Definition 7.0.1** (Presheaves)**.** *Let $(X, \tau)$ be a topological space. A presheaf of sets F on X consists of the following structure:*

- *For each open set $V \in \tau$, a set $F(V)$ whose elements are called the* sections *of F over V, the sections of F over X are called* ***global sections***.

- *For each inclusion of open sets $V \subseteq W$, a function $res_{V,W} : F(W) \to F(V)$ which is called a restriction morphism. If $s \in F(W)$, then its restriction $res_{V,W}(s)$ is denoted $s\big|_V$. These restriction morphisms are required to satisfy two additional functorial properties:*

    - *$\forall\, V \in \tau$, the restriction morphism $res_{V,V} : F(V) \to F(V)$ is the identity morphism $\mathrm{id}_{F(V)} : F(V) \to F(V)$ on $F(V)$.*

    - *The composition of compatible restrictions is a restriction, that is, for $W \subseteq V \subseteq U$ then $res_{W,V} \circ res_{V,U} = res_{W,U}$.*

**Definition 7.0.2** (Sheaves)**.** *A* sheaf *is a presheaf that satisfies both of the following two additional axioms:*

- *(Locality) Suppose $U \in \tau$, and $\{V_i\}_{i \in I}$ is a open cover of U, and $s, t \in F(U)$ are sections. If $s\big|_{V_i} = t\big|_{V_i},\ \forall i \in I$, then $s = t$.*

- *(Gluing)* Suppose $U \in \tau$, $\{V_i\}_{i \in I}$ is a open cover of $U$, and $\{s_i \in F(V_i)\}_{i \in I}$ is a family of sections. If all pairs of sections agree on the overlap of their domains, that is, if $s_i\big|_{V_i \cap V_j} = s_j\big|_{V_i \cap V_j}$, $\forall i, j \in I$, then there exists a section $s \in F(U)$ such that $s\big|_{V_i} = s_i$, $\forall i \in I$.

Since a von Neumann algebra has a topology induced by its various equivalent norms, we can consider the presheaves of a von Neumann algebra, we will call these **observable presheaves**.

## 7.1   Sheaf Theory within category theory

We can actually generalize all these notions using category theory, which we will do in the following. First we define a category:

**Definition 7.1.1** (Category). *A category C is composed of **objects**, which we will say that are of type[1] $\mathbf{Obj}_C$, and **morphisms**, which are of type $\mathbf{Mor}_C$.*

*Every morphism $f$ is related to two (not necessarily distinct) objects, a domain object $\mathrm{Dom}(f)$ and a codomain object $\mathrm{Cod}(f)$ and we write $\mathrm{Dom}(f) \xrightarrow{f} \mathrm{Cod}(f)$. If $A \xrightarrow{f} B$ we say that $f$ is of type $\mathbf{Obj}_C(A, B)$.*

*For every pair of morphisms $f, g$ from a category C, if $\mathrm{Cod}(f) = \mathrm{Dom}(g)$ then there exists a morphism $g \circ f$ which is named their composite, with $\mathrm{Dom}(g \circ f) = \mathrm{Dom}(f)$ and $\mathrm{Cod}(g \circ f) = \mathrm{Cod}(g)$. Composition is associative, that is, the morphism $f \circ (g \circ h)$ is exactly the same as $(f \circ g) \circ h$, for all morphisms $f, g, h$ in the category C such that $\mathrm{Cod}(h) = \mathrm{Dom}(g)$ and $\mathrm{Cod}(g) = \mathrm{Dom}(f)$.*

*For every obeject $L$ in a category, there exists a morphism $L \xrightarrow{\mathrm{id}_L} L$ such that, whenever $\mathrm{Dom}(f) = L$ and $\mathrm{Cod}(f) = M$ for a morphism $f$, then $f \circ \mathrm{id}_L = f = \mathrm{id}_M \circ f$.*

*For more complicated relations between morphisms, such as $A \xrightarrow{f} B \xrightarrow{g} D$ and $A \xrightarrow{h} D$ we can write them all in the form of a diagram:*

$$\begin{array}{ccc} A & \xrightarrow{f} & B \\ & \searrow h & \downarrow g \\ & & D \end{array}$$

*and we say that the **diagram commutes** whenever all composition paths that go to the same object are equal, in this case if $h = g \circ f$. In fact it is common practice in Category Theory to use, for the most part, diagrams instead of equations.*

There is also a assisting definition that will be useful for writing the next assertions with shorthand.

**Definition 7.1.2** (Opposite category). *If C is a category, its opposite category $C^{\mathrm{ops}}$ is given by:*

$$\mathbf{Obj}_{C^{\mathrm{ops}}} := \mathbf{Obj}_C \qquad \text{and} \qquad \mathbf{Mor}_{C^{\mathrm{ops}}}(A, B) := \mathbf{Mor}_C(B, A)$$

---

[1] we use the word **type** here because this must **not** be seen as a set, in category theory we don't presuppose that the ZFC axioms are valid for the constituents of a category, this allows for more liberty about the things we can consider, such as the category of all sets **Set**, the category of all groups **Grp**, the category of all vector spaces **Vect**, the category of all algebras **Alg** amongst other such collections that do not form sets since they would be "too big"(this is a technical term); this liberty comes at the expense of not being able to manipulate these types as we can manipulate sets, *i.e.* by considering unions, intersections, set differences, cartesian products, considerations about subsets, etc. For more information about category theory we reccoment the book [72].

*such that, if* $\mathrm{Cod}_{C^{\mathrm{ops}}}(f) = \mathrm{Dom}_{C^{\mathrm{ops}}}(g)$ *then:*

$$g \circ_{C^{\mathrm{ops}}} f := f \circ_C g.$$

Also of particular importance are the initial objects and terminal objects in a category:

**Definition 7.1.3.** *We say that a object $I$ in the category $C$ is a initial object if for all objects $A$ in this category, there is a **unique** morphism $I \xrightarrow{\iota_A} A$. Whereas we say that a object $T$ in the category $C$ is terminal if for all objects $A$ in this category, there is a **unique** morphism $A \xrightarrow{\tau_A} T$.*

We then introduce the category of **Set**, which is the type of all sets from set theory.

**Definition 7.1.4** (**Set** Category)**.** **Set** *is the category whose objects are sets and the morphisms, such as those of type $\mathbf{Mor}_{\mathbf{Set}}(A, B)$, are the functions between sets, in the particular case the functions $f : A \to B$. The composition of morphisms is the function composition.*

*The empty set $\varnothing$ is the initial object of this category, and every set that contains only one element are the terminal objects of the category.*

**Definition 7.1.5** (The $*$-**UAlg** and $*$-**UCAlg** Categories)**.** *The category of unital $*$-algebras is denoted by $*$-**UAlg**. Its objects are the unital $*$-algebras, its morphisms are the unital $*$-homomorphisms, and its composition rule is the standard function composition. The category of commutative unital $*$-algebras is denoted by $*$-**UCAlg**, which is defined analogously to the $*$-**UAlg** category by changing "unital $*$-algebras" to "commutative unital $*$-algebras" everywhere.*

Contexts also form categories:

**Definition 7.1.6** (Categories of contexts)**.** *Consider the family of contexts of a unital $*$-algebra $\mathscr{A}$, for contexts $\mathcal{C}, \mathcal{D}$ the relation of $\mathcal{D}$ being a context of $\mathcal{C}$ ($\mathcal{D} \preceq \mathcal{C}$), i.e. $\mathcal{D}$ being a abelian $*$-sub-algebra of $\mathcal{C}$, is a pre-order, that is, it is a binary relation that is reflexive ($\forall$ contexts $\mathcal{D}$ of $\mathscr{A}$, $\mathcal{D} \preceq \mathcal{D}$) and transitive (if $\mathcal{E} \preceq \mathcal{D}$ and $\mathcal{D} \preceq \mathcal{C}$ then $\mathcal{E} \preceq \mathcal{C}$).*

*We can then consider the category $\mathfrak{C}(\mathscr{A})$ that has as objects the contexts of a unital $*$-algebra $\mathscr{A}$, and that has as morphisms the inclusion maps ($\iota : \mathcal{D} \hookrightarrow \mathcal{C}$, $x \mapsto x$) between contexts $(\mathcal{C}, \mathcal{D})$ that are pre-ordered ($\mathcal{D} \preceq \mathcal{C}$) by one being a context of the other.*

**Proposition 7.1.7.** *The initial object of $\mathfrak{C}(\mathscr{A})$ is the object context $\{\alpha \mathbb{1} \,|\, \alpha \in \mathbb{C}\}$, and with the exception of $\mathfrak{C}(\mathscr{A})$ coming from a already abelian algebra $\mathscr{A}$, there is no terminal object on $\mathfrak{C}(\mathscr{A})$.*

*Proof:* Since the objects of $\mathfrak{C}(\mathscr{A})$ are unital $*$-sub-algebras of $\mathscr{A}$, then they all contain $\mathbb{1}$ the identity of $\mathscr{A}$, hence the linearity property of the algebra implies they all contain a copy of $\{\alpha \mathbb{1} \,|\, \alpha \in \mathbb{C}\}$, which implies that there is always the inclusion morphism $\{\alpha \mathbb{1}\} \xrightarrow{\iota_{\alpha \mathbb{1}}^{(\cdot)}} (\,\cdot\,)$ between the set $\{\alpha \mathbb{1} \,|\, \alpha \in \mathbb{C}\}$, seen as a $*$-sub-algebra of $\mathscr{A}$, and every other $*$-sub-algebra of $\mathscr{A}$ that has to also contain $\{\alpha \mathbb{1} \,|\, \alpha \in \mathbb{C}\}$, this morphism is unique because it has to be a unital $*$-homomorphism and the domain is the multiples of the identity, hence for any other $*$-homomorphism $\{\alpha \mathbb{1}\} \xrightarrow{\varphi} (\,\cdot\,)$, $\varphi(\alpha \mathbb{1}) = \alpha \varphi(\mathbb{1}) = \alpha \mathbb{1}$ which is exactly the inclusion morphism.

Considering now that there were to exist a terminal object on $\mathfrak{C}(\mathscr{A})$, then all contexts of $\mathscr{A}$ would be contexts of this terminal object, which we shall call the terminal context. Since all elements of $\mathscr{A}$ generate contexts, this implies that the totality of $\mathscr{A}$ must be contained in this terminal context, which would in turn imply that $\mathscr{A}$ is abelian, since by hypothesis it isn't then this creates a contradiction, hence there is no such terminal object. ∎

We must also consider how to transform a category into another category, we do this initially by functors:

**Definition 7.1.8** (Functor). *For categories $\mathcal{C}, \mathcal{D}$; a **covariant functor** $F : \mathcal{C} \to \mathcal{D}$ is a transformation for which:*

(i) *For all objects $A$ in $\mathbf{Obj}_\mathcal{C}$, the functor assigns a object $F(A)$ in $\mathbf{Obj}_\mathcal{D}$.*

(ii) *For any two objects $A, B$ in $\mathbf{Obj}_\mathcal{C}$ and morphism $A \xrightarrow{f} B$ in $\mathbf{Mor}_\mathcal{C}(A, B)$, the functor assigns a morphism $F(A) \xrightarrow{F(f)} F(B)$ in $\mathbf{Mor}_\mathcal{D}(F(A), F(B))$ such that:*

$$F(\mathrm{id}_A) = \mathrm{id}_{F(A)} \qquad \text{and} \qquad F(g \circ f) = F(g) \circ F(f).$$

*Similarly, a **contravariant functor** $G : \mathcal{C} \to \mathcal{D}$ is a transformation in which:*

(i) *For all objects $A$ in $\mathbf{Obj}_\mathcal{C}$, the functor assigns a object $G(A)$ in $\mathbf{Obj}_\mathcal{D}$.*

(ii) *For any two objects $A, B$ in $\mathbf{Obj}_\mathcal{C}$ and morphism $A \xrightarrow{f} B$ in $\mathbf{Mor}_\mathcal{C}(A, B)$, the functor assigns a morphism $G(B) \xrightarrow{G(f)} G(A)$ in $\mathbf{Mor}_\mathcal{D}(G(B), G(A))$ such that:*

$$G(\mathrm{id}_A) = \mathrm{id}_{G(A)} \qquad \text{and} \qquad G(g \circ f) = G(f) \circ G(g).$$

We can understand contravariant functors $G : \mathcal{C} \to \mathcal{D}$ as as covariant functors $G : \mathcal{C}^{\mathrm{ops}} \to \mathcal{D}$ or $G : \mathcal{C} \to \mathcal{D}^{\mathrm{ops}}$.

We can then finally define the categorical notion of a presheaf:

**Definition 7.1.9** (Categorical presheaf). *Let $\mathcal{C}$ be a category. A contravariant functor $F : \mathcal{C} \to \mathbf{Set}$ is said to be a presheaf on $\mathcal{C}$.*

We see that this definition clearly includes the topological presheaves originally defined, since the $F(A)$ are sets for $A$ a object in $\mathcal{C}$, and if there is a morphism $A \xrightarrow{\iota_{A,B}} B$ between any two objects $A$ and $B$ there will be a function $F(f) : F(B) \to F(A)$ (a morphism in the category of **Set**), such that if:

$$K \xrightarrow{s} L$$
$$q \searrow \quad \downarrow r$$
$$M$$

this diagram commutes, that is, if $q = r \circ s$ then $F(q) = F(r \circ s) = F(s) \circ F(r)$. Therefore the morphisms that are images under the functor $F$ obey the same relations as the restriction morphisms of the topological presheaves.

**Definition 7.1.10** (Hom functor). *Let $\mathcal{C}$ be a locally small[2] category. For each $A$ in $\mathbf{Obj}_\mathcal{C}$ the contravariant functor $\mathrm{Hom}_\mathcal{C}(\,\cdot\,, A) : \mathcal{C}^{\mathrm{ops}} \to \mathbf{Set}$ is the presheaf over $\mathcal{C}$ that transforms $B$ in $\mathbf{Obj}_\mathcal{C}$ to the set:*

$$\mathrm{Hom}_\mathcal{C}(B, A) \equiv \mathbf{Mor}_\mathcal{C}(B, A)$$

*and transforms the morphism $f$ of $\mathbf{Mor}_\mathcal{C}(B, C)$ in the morphism:*

$$\mathbf{Mor}_\mathcal{C}(C, A) \xrightarrow{\mathrm{Hom}_\mathcal{C}(f, A)} \mathbf{Mor}_\mathcal{C}(B, A) \qquad \text{given by} \qquad \mathrm{Hom}_\mathcal{C}(f, A)(g) = g \circ f.$$

We can also define transformations between functors themselves:

**Definition 7.1.11** (Natural transformations). *Let $\mathcal{C}$ and $\mathcal{D}$ be any two categories and let $F : \mathcal{C} \to \mathcal{D}$ and $G : \mathcal{C} \to \mathcal{D}$ be covariant functors. A covariant **natural transformation** $N : F \to G$ is a collection of morphisms $(N_A)_{A \text{ in } \mathbf{Obj}_\mathcal{C}}$ such that $N_A$ is of type $\mathbf{Mor}_\mathcal{D}(F(A), G(A))$, with:*

$$\begin{array}{ccc} F(A) & \xrightarrow{F(f)} & F(B) \\ N_A \downarrow & & \downarrow N_B \\ G(A) & \xrightarrow{G(f)} & G(B) \end{array}$$

*commuting for all objects $A, B$ of type $\mathbf{Obj}_\mathcal{C}$ and all morphisms $f$ of respective types $\mathbf{Mor}_\mathcal{C}(A, B)$, in other words, if $N_B \circ F(f) = G(f) \circ N_A$.*

*For contravariant functors $F : \mathcal{C} \to \mathcal{D}$ and $G : \mathcal{C} \to \mathcal{D}$, we can define a contravariant **natural transformation** $M : F \to G$ which is also a collection of morphisms $(M_A)_{A \text{ in } \mathbf{Obj}_\mathcal{C}}$ such that $M_A$ is of type $\mathbf{Mor}_\mathcal{D}(F(A), G(A))$, such that:*

$$\begin{array}{ccc} F(A) & \xleftarrow{F(f)} & F(B) \\ M_A \downarrow & & \downarrow M_B \\ G(A) & \xleftarrow{G(f)} & G(B) \end{array}$$

*commutes for all objects $A, B$ of type $\mathbf{Obj}_\mathcal{C}$ and all morphisms $f$ of respective types $\mathbf{Mor}_\mathcal{C}(A, B)$, in equations, $M_A \circ F(f) = G(f) \circ M_B$.*

*We usually denote natural transformations, like $N \equiv (N_A)_{A \text{ in } \mathbf{Obj}_\mathcal{C}}$, by the following type of diagram:*

$$\mathcal{C} \underset{G}{\overset{F}{\rightrightarrows}} \Downarrow N \;\; \mathcal{D}$$

The notion of elements of a object actually exist in Category theory, to arrive at such a definition we must first define certain types of morphisms, introduce the notion of subobjects, and by taking inspiration from what happens in the **Set** category, define elements of a category.

**Definition 7.1.12** (Types of morphisms). *Let $\mathcal{C}$ be a category and $A \xrightarrow{f} B$ be a morphism in $\mathcal{C}$. We say $f$ is:*

(i) *a monomorphism if, and only if, for any $g, h$ of type $\mathbf{Mor}_\mathcal{C}(C, A)$, $f \circ g = f \circ h$ implies $g = h$. In this case we write $A \xhookrightarrow{f} B$;*

---
[2] This means that for any $A, B$ in $\mathbf{Obj}_\mathcal{C}$, $\mathbf{Mor}_\mathcal{C}(A, B)$ is in **Set**.

(ii) a **epimorphism** *if, and only if, for any $g, h$ of type $\mathbf{Mor}_\mathcal{C}(B,C)$, $g \circ f = h \circ f$ implies $g = h$. In this case we write $A \xrightarrow{f} \!\!\!\!\to B$;*

(iii) a **bimorphism** *if, and only if, $f$ is both a monomorphism and a epimorphism;*

(iv) a **isomorphism** *if, and only if, there is a morphism $B \xrightarrow{f^{-1}} A$ such that $f^{-1} \circ f = \mathrm{id}_A$ and $f \circ f^{-1} = \mathrm{id}_B$. We then say that $A$ and $B$ are isomorphic, and since this isomorphism relation is trivially an equivalence relation, we can consider the equivalence class $[A]_{\mathrm{iso}}$ of objects isomorphic to $A$;*

(v) a **endomorphism** *if, and only if, $A = B$;*

(vi) a **automorphism** *if, and only if, $f$ is both an isomorphism and an endomorphism.*

We note in a proposition that a bimorphism need not be a isomorphism, even though an isomorphism always is a bimorphism.

**Proposition 7.1.13.** *An isomorphism is always a bimorphism, but a bimorphism isn't necessarily a isomorphism.*

*Proof:* Let $\mathcal{C}$ be a category and $A \xrightarrow{f} B$ be an isomorphism. Then if $g, h$ are two morphisms of type $\mathbf{Mor}_\mathcal{C}(C, A)$ and if $f \circ g = f \circ h$ then we can compose this in both sides with $f^{-1}$ on the left and get $f^{-1} \circ f \circ g = g = f^{-1} \circ f \circ h = h$, hence $f$ is a epimorphism, equally if $r, s$ are two morphisms of type $\mathbf{Mor}_\mathcal{C}(B, C)$ and if $r \circ f = s \circ f$ then we can compose this in both sides with $f^{-1}$ on the right and get $r \circ f \circ f^{-1} = r = s \circ f \circ f^{-1} = s$ and therefore $f$ is also a epimorphism, making it a bimorphism.

Now we consider a special case, a category formed by two objects $A$ and $B$, each with their respective identity morphisms $\mathrm{id}_A$, $\mathrm{id}_B$ and a single other morphism from $A \xrightarrow{f} B$, but no morphism on the other direction. Then $f$ is a bimorphism since $f \circ \mathrm{id}_A = f = \mathrm{id}_B \circ f$, but $f$ is not a isomorphism since there is no $B \xrightarrow{f^{-1}} A$ morphism. ∎

**Definition 7.1.14.** *Let $\mathcal{C}$ be a category, $A, B, C$ objects of type $\mathbf{Obj}_\mathcal{C}$, and let $f$ in $\mathbf{Mor}_\mathcal{C}(A, C)$ and $g$ in $\mathbf{Mor}_\mathcal{C}(B, C)$ be monomorphisms. If there is $h$ in $\mathbf{Mor}_\mathcal{C}(A, B)$ such that $f = g \circ h$, or in other words, if:*

$$\begin{array}{ccc} A & \xrightarrow{h} & B \\ & \searrow_{f} & \downarrow_{g} \\ & & C \end{array}$$

*commutes, then we say that $f$ **factors through** $g$.*

This leads to the quite natural notion of equivalence of monomorphisms:

**Definition 7.1.15** (Equivalence of monomorphisms)**.** *Let $\mathcal{C}$ be a category and $B$ a object of $\mathbf{Obj}_\mathcal{C}$. Let $f$ and $g$ be monomorphisms on $\mathcal{C}$ with codomain $B$. We'll say $f \sim_{\mathrm{mono}} g$ if, and only if, $f$ factors through $g$ and $g$ factors through $f$.*

**Proposition 7.1.16.** *Let $\mathcal{C}$ be a category and $A, B, C$ in $\mathbf{Obj}_\mathcal{C}$. Let $f$ in $\mathbf{Mor}_\mathcal{C}(A, B)$ and $g$ in $\mathbf{Mor}_\mathcal{C}(C, B)$. Then $f \sim_{\mathrm{mono}} g$ if, and only if, $A \sim_{\mathrm{iso}} C$.*

*Proof:* Suppose that $f \sim_{\text{mono}} g$, then there are morphisms $h$ and $k$ such that $f = g \circ h$ and $g = f \circ k$, using these expressions $f \circ \text{id}_A = f = g \circ h = (f \circ k) \circ h = f \circ (k \circ h)$, but since f is a monomorphism it is left-canceling and we conclude $k \circ h = \text{id}_A$. Analogously, since $g$ is also a monomorphism, we get $h \circ k = \text{id}_C$. Hence, $h$ is a isomorphism with $k = h^{-1}$, proving $A \sim_{\text{iso}} C$.

Now supposing $A \sim_{\text{iso}} C$, then we can use the isomorphism to factor $f$ through $g$ and vice-versa. ∎

**Proposition 7.1.17.** *Subsets on set theory are equivalence classes of monomorphisms.*

*Proof:* For the **Set** category, for $B$ in $\mathbf{Obj_{Set}}$, and each $C \subseteq B$, let $\iota_C$ in $\mathbf{Mor_{Set}}(C, B)$ be the inclusion map $C \ni x \mapsto x \in B$. $\iota_C$ is trivially a monomorphism for every subset $C$. Furthermore, $C \mapsto [\iota_C]_{\text{mono}}$ defines a bijection between subsets of B and equivalence classes of monomorphisms with codomain B. It is straightforward to see that $C \mapsto [\iota_C]_{\text{mono}}$ is surjective, by taking the image of any representative $\iota_C$ as the input set for the map. To see that it is also injective, notice that if $D \subsetneq C$, then $\iota_D$ and $\iota_C$ have different ranges, so they can't be equivalent. Hence Subsets are isomorphic to equivalence classes of monomorphisms. ∎

Due to this example, we define subobjects in a category in the following manner:

**Definition 7.1.18** (Subobjects). *Let $\mathcal{C}$ be a category. If $B$ in $\mathbf{Obj}_\mathcal{C}$, then the **subobjects** of $B$ are the equivalence classes of monomorphisms in $\mathcal{C}$ with codomain $B$. We shall denote the collection of subobjects of $B$ as $\mathbf{Sub}(B)$.*

We also note that some categories admit that we identify subobjects with objects in the category (as it happens with **Set**, for example), but this doesn't always hold.

As it happens, points in set theory are exactly those subobjects with a terminal object as the domain:

**Proposition 7.1.19.** *Points in a set are subobjects with Terminal Domains.*

*Proof:* We first note that if $T$ is a terminal object in a category $\mathcal{C}$, and $A, B$ are objects of type $\mathbf{Obj}_\mathcal{C}$, if $T \xrightarrow{f} B$ is a morphism, as are $g, h$ in $\mathbf{Mor}_\mathcal{C}(A, T)$, then we can clearly see that $f$ is a monomorphism, for which it is required that $f \circ g = f \circ h$ implies $g = h$, but $\mathbf{Mor}_\mathcal{C}(A, T)$ hasa unique morphism by the terminal object property of $T$, hence it is impossible for $g \neq h$, which makes all morphisms $f$ with $\text{Dom}(f) = T$ be monomorphisms.

Furthermore, given a $f$ of type $\mathbf{Mor}_\mathcal{C}(T, B)$, which isa monomorphism, since $T$ is terminal, $\mathbf{Mor}_\mathcal{C}(T, T) \equiv \mathbf{End}_\mathcal{C}(T)$ has a single element, so the only morphism of type $\mathbf{Mor}_\mathcal{C}(T, B)$ that $f$ can factor through is itself, *i.e.* the unique morphism in $\mathbf{End}_\mathcal{C}(T)$ must be the identity morphism $\tau_T \equiv \text{id}_T$, hence for any morphism $T \xrightarrow{g} B$, which is also a monomorphism, for $f$ to factor through $g$ we must have $f = g \circ \text{id}_T = g$. In other words, the the only element of type $\mathbf{Mor}_\mathcal{C}(T, B)$ in the equivalence class $[f]_{\text{mono}}$ of the subobjects $\mathbf{Sub}(B)$ is $f$ itself.

More than that, the domain of all other morphisms in $[f]_{\text{mono}}$ must also be terminal, to understand why we consider another monomorphism $A \xrightarrow{h} B$, from another object $A$ of $\mathcal{C}$ such that $f \sim_{\text{mono}} g$, and therefore there is another morphism $T \xrightarrow{s} A$ that factors $f$ through $h$; together with another terminal object $D$, which is isomorphic to $T$ since if $\tau_D$ in $\mathbf{Mor}_\mathcal{C}(D, T)$ and $\widetilde{\tau}_T$ in $\mathbf{Mor}_\mathcal{C}(T, D)$ are the respective unique terminal morphisms, then $\text{id}_T = \widetilde{\tau}_T \circ \tau_D$ and $\text{id}_D = \tau_D \circ \widetilde{\tau}_T$ which makes $\widetilde{\tau}_T = \tau_D^{-1}$, in a diagram:

$$B \xleftarrow{f} T \underset{\tau_D}{\overset{\tau_D^{-1}}{\rightleftarrows}} D$$
$$\searrow_h \quad \tau_A \Big\Updownarrow s$$
$$A$$

Then any arbitrary object $R$ has a morphism to $T$ given by $\tau_R$ and hence a morphism to $A$ by $s \circ \tau_R$, to prove that this morphism is unique we consider that $f \circ \tau_A$ is a monomorphism, since $f \circ \tau_A = h$ and $h$ is a monomorphism, this also means that $\tau_A$ is a monomorphism for if $\tau_A \circ w = \tau_A \circ v$ then we can compose this with $f$ on the left to get $w = v$, making $\tau_A$ a monomorphism, hence if there are two morphisms $a, b : E \to A$, we must have $\tau_A \circ a = \tau_A \circ b = \tau_E$ by the terminal property of $T$, then by the monomorphism property of $\tau_A$, $a = b$. Also, since $h = f \circ \tau_A = (h \circ s) \circ \tau_A$, then by the monomorphism property of $h$ we have that $s \circ \tau_A = \mathrm{id}_A$, from the terminal property of $T$ we also get $\tau_A \circ s = \mathrm{T}$ from which we see that $s = \tau_A^{-1}$, therefore $A \sim_{\mathrm{iso}} T$ just as $D \sim_{\mathrm{iso}} T$.

In **Set** the terminal objects are the singleton sets (sets with a single element), since any function from a singleton can only be the constant function equal to the image of that one element, then, for functions $f : \{x\} \to \{y_f\} \in B$ we can identify the equivalence classes $[f]_{\mathrm{mono}} \in \mathbf{Sub}(B)$ as being the points, or elements, of $B$. ∎

As the points, or in other words the elements, of a set in **Set** are in a bijection with those subobjects of this set with terminal domain, then this motivates the abstraction of the definition of *elements of a object* in a more general category.

**Definition 7.1.20** (Elements of a object). *Let $\mathcal{C}$ be a category and $B$ in $\mathbf{Obj}_\mathcal{C}$. The elements of $B$ are the subobjects of $B$ corresponding to morphisms with terminal domain*

Having gone through all these considerations, we can observe that presheaves themselves form a category:

**Definition 7.1.21** (Category of presheaves). *For a small[3] category $\mathcal{C}$, the category of presheaves on $\mathcal{C}$, denoted by $\mathbf{Set}^{\mathcal{C}^{\mathrm{ops}}}$ is defined by:*

- ***Objects** given by functors of type $\mathcal{C}^{\mathrm{ops}} \to \mathbf{Set}$, i.e. presheaves.*

- ***Morphisms** given by natural transformations between presheaves.*

**Proposition 7.1.22.** *Let $F$ and $G$ be presheaves over some category $\mathcal{C}$, a natural transformation $\phi : F \to G$ is a monomorphism if, and only if, all of the component functions $(\phi_A)_{A \text{ in } \mathbf{Obj}_\mathcal{C}}$ are monomorphisms (in particular, they are injective functions).*

*Proof:* Let us first prove that if $F \xrightarrow{\phi} G$ is a monomorphism in the category of presheaves, then, for each $U$ in $\mathcal{C}$, $\phi(U) : F(U) \to G(U)$ is a injective function, that means that if $\phi(U)(x) = \phi(U)(y)$, for $x, y \in F(U)$ then $x = y$. We define a presheaf $H$ by:

$$H(V) := \mathrm{Hom}_\mathcal{C}(V, U)$$

For each $V$ in $\mathcal{C}$, independently of they being objects or morphisms. Now let $H \xRightarrow{\alpha} F$ be defined by $\alpha_V(f) = F(f)(x)$, for each $f$ in $\mathbf{Mor}_\mathcal{C}(V, U)$ which is clearly a contravariant natural transformation, since by the following diagram:

---

[3] This means that $\mathbf{Obj}_\mathcal{C}$ and $\mathbf{Mor}_\mathcal{C}$ are of type **Set**.

$$\begin{array}{ccc} H(V) & \xleftarrow{H(h)} & H(W) \\ \alpha_V \downarrow & & \downarrow \alpha_W \\ F(V) & \xleftarrow{F(h)} & F(W) \end{array}$$

Then calculating $\alpha_V \circ H(h)$ in a arbitrary $W \xrightarrow{g} U$ morphism, we get $\alpha_V \circ H(h)(g) = \alpha_V \circ \text{Hom}_{\mathcal{C}}(h,U)(g) = \alpha_V(g \circ h) = F(g \circ h)(x) = (F(h) \circ F(g))(x) = F(h) \circ (\alpha_W(g)) = F(h) \circ \alpha_W(g)$, which makes $H \xRightarrow{\alpha} F$ a contravariant natural transformation. Analogously $H \xRightarrow{\beta} F$ defined by $\beta_V(f) = F(f)(y)$, for each $f$ in $\textbf{Mor}_{\mathcal{C}}(V,U)$, is also a contravariant natural transformation. We now show that $\phi \circ \alpha = \phi \circ \beta$, since:

$$(\phi \circ \alpha)(V)(f) = \phi(V)(F(f)(x)) = G(f)(\phi(f)(x)) = G(f)(\phi(f)(y)) = \phi(V)(F(f)(y)) = (\phi \circ \beta)(V)(f)$$

For any $f$ in $\textbf{Mor}_{\mathcal{C}}(V,U)$, where we used in the second equal sign that $\phi$ is a natural transformation, and in the third equal sign we used the hypothesis $\phi(U)(x) = \phi(U)(y)$. Since $\phi$ is also a monomorphism by hypothesis then we can cancel it on the left to get $\alpha = \beta$, and so:

$$x = F(\text{id}_U)(x) = \alpha_U(\text{id}_U) = \beta_U(\text{id}_U) = F(\text{id}_U)(y) = y,$$

as required.

Now we prove the converse, by taking as hypothesis that each of the component morphisms $\phi_V = \phi(V) : F(U) \to G(U)$, for each $V$ in $\mathcal{C}$, which makes each $\phi_V$ a injective function, since it is a monomorphism between sets. Letting $\alpha, \beta : H \to F$ be morphisms in the category of sheaves, such that $\phi \circ \alpha = \phi \circ \beta$, then for $\phi$ to be a monomorphism, we have to prove that $\alpha = \beta$. This equality is equivalent to the equality of the functions $\alpha(U)$ and $\beta(U)$ for each $U$ in $\mathcal{C}$. Hence:

$$(\phi \circ \alpha)(U)(x) = (\phi \circ \beta)(U)(x) \implies \phi_U(\alpha(U)(x)) = \phi_U(\beta(U)(x))$$

and then the injectivity of $\phi_U$ implies that $\alpha(U)(x) = \beta(U)(x)$, as needed. ∎

This has a important corollary in the form of:

**Corollary 7.1.23.** *Let $F$ and $G$ be presheaves over some category $\mathcal{C}$, two monomorphisms in the category of sheaves $f, g : F \to G$ are equivalent if, and only if $\text{Ran} f_A = \text{Ran} g_A \subseteq G(A)$, for each component of $f$ and $g$ associated with a object $A$ from $\textbf{Obj}_{\mathcal{C}}$.*

*Proof:* First supposing that $f \sim_{\text{mono}} g$, then $f = g \circ h$ and $g = f \circ s$ for morphisms $h, s$ in the category of sheaves, this means that, for each component morphism (in these cases functions) associated with a object $A$ of $\mathcal{C}$, $f_A = (g \circ h)_A$ and $g_A = (f \circ s)_A$, using that these are functions from the set $F(A)$ to $G(A)$, then they can only be equal as functions if every image of $f_A$ is reached by $(g \circ h)_A = g_A \circ h_A$ and hence is reached by $g_A$, equally every image of $g_A$ needs to be reached by $f_A \circ s_A$ and hence by $f_A$, this implies $\text{Ran} f_A = \text{Ran} g_A$ both as subsets of $G(A)$.

Considering now the converse, $\text{Ran} f_A = \text{Ran} g_A \subseteq G(A)$ for each object $A$ of the category $\mathcal{C}$, and by proposition 7.1.22 $f_A$ and $g_A$ are monomorphisms in the category $\textbf{Set}$, since the natural transformations $f$ and $g$ are taken to be monomorphisms in the category of sheaves. As $\text{Ran} f_A = \text{Ran} g_A$ we can take in particular the automorphism $\alpha_A$ of this range given by $\alpha_A(f_A(x)) := g_A(x)$ with a inverse $\alpha_A^{-1}(g_A(x)) = f_A(x)$, then $\alpha_A$ factors $g_A$ through $f_A$ and $\alpha_A^{-1}$ factors $f_A$ through $g_A$. Promoting this automorphism of sets $\alpha_A$ to a automorphism of sheaves $\alpha$ taken to be the collection $(\alpha_A)_{A \text{ in } \textbf{Obj}_{\mathcal{C}}}$, then $\alpha$ factors $g$ through $f$ and $\alpha^{-1}$ factors $f$ through $g$, making $f \sim_{\text{mono}} g$. ∎

Which, in part, has the following corollary:

**Corollary 7.1.24** (Subobjects of presheaves). *Let $G$ be a presheaf over some category $\mathcal{C}$. The subobjects of $G$ can be identified with the presheaves $F$ in $\mathbf{Obj}_{\mathbf{Set}^{\mathcal{C}^{\text{ops}}}}$ such that, for all $A$ in $\mathbf{Obj}_{\mathcal{C}}$, $F(A) \subseteq G(A)$ and such that there is a natural transformation $\iota = (\iota_A)_A$ from $F$ to $G$ such that $\iota_A : F(A) \to G(A)$ is the inclusion function $x \overset{\iota_A}{\hookrightarrow} x$ for each $A$ in $\mathbf{Obj}_{\mathcal{C}}$.*

*Proof:* The subobjects of a presheaf $G$ are exactly the equivalence classes of monomorphisms of sheaves in $\mathbf{Sub}(G)$, but by corollary 7.1.23 if two monomorphisms of sheaves are equivalent if and only if the ranges of the components associated with each $A$ in $\mathbf{Obj}_{\mathcal{C}}$ are equal and they are subsets of $G(A)$. Therefore for each subobject $[f_A]_{\text{mono}}$ of a presheaf $G$, we can associate another presheaf $F$ in $\mathbf{Obj}_{\mathbf{Set}^{\mathcal{C}^{\text{ops}}}}$, with $F(A) = \text{Ran } f_A \subseteq G(A)$ and for morphisms $g$ in $\mathbf{Mor}_{\mathcal{C}}(B, A)$ to be so that $F(g) \subseteq G(g)$, where the inclusion is taken in the sense of the graphs of the functions $F(g)$ and $G(g)$ (i.e. $\{(x, F(g)(x)) \mid x \in F(A)\} \subseteq \{(x, G(g)(x)) \mid x \in G(A)\}$), or in other words $G(g)\Big|_{F(A)} = F(g)$.

Since both $F(A)$ and $G(A)$ are sets, there is trivially a inclusion function $\iota_A : F(A) \hookrightarrow G(A)$ with $F(A) \ni x \overset{\iota_A}{\hookrightarrow} x \in G(A)$, this function is a natural transformation since, considering objects $A$ and $B$ in $\mathbf{Obj}_{\mathcal{C}}$ and any morphism $g$ in $\mathbf{Mor}_{\mathcal{C}}(B, A)$, with sheaves $F, G$ with $F(D) \subseteq G(D)$ for every object $D$ in $\mathbf{Obj}_{\mathcal{C}}$, then $G(g) \circ \iota_A = G(g)\Big|_{F(A)} = F(g) = \iota_B \circ F(g)$, that is, the following diagram commutes:

$$\begin{array}{ccc} F(A) & \xrightarrow{F(g)} & F(B) \\ \iota_A \downarrow & & \downarrow \iota_B \\ G(A) & \xrightarrow{G(g)} & G(B) \end{array}$$

∎

So subobjects of presheaves are basically other presheaves, this already shows a great regularity and a good behavior when we are treating presheaves in particular instead of general categories, in which a subobject of a object need not itself be a object. To be able to speak about elements we use the fact that presheaves transport a category to sets, then we can use our results for elements of sets being the equivalence classes of monomorphisms composed of functions parting from a singleton and also having to arrive at the element of a specific singleton, a point of the subobject (also subset in this case), hence the equivalence between points in sets and these equivalence classes of monomorphisms parting from the terminal elements of the **Set** category, *i.e.* the singletons.

Hence:

**Lemma 7.1.25.** *Let $\mathcal{C}$ be a arbitrary small category, the initial object of the presheaf category $\mathbf{Set}^{\mathcal{C}^{\text{ops}}}$ is the contravariant functor $\mathbf{0}_{\mathcal{C}} : \mathcal{C} \to \mathbf{Set}$ that maps all objects of $\mathcal{C}$ to the empty set $\varnothing$ and all morphisms of $\mathcal{C}$ to the empty function $\varnothing \mapsto \varnothing$ (also seen as the identity morphism of $\varnothing$).*

*The terminal object of the presheaf category $\mathbf{Set}^{\mathcal{C}^{\text{ops}}}$ is the contravariant functor $\mathbf{1}_{\mathcal{C}} : \mathcal{C} \to \mathbf{Set}$ that maps all objects of $\mathcal{C}$ to a singleton set, and all morphisms of $\mathcal{C}$ to the identity function on this singleton.*

*Proof:* We first note that the only initial object of **Set** is the empty set $\varnothing$, this can be easily seen by the fact that any function from the empty set to another set $X$, like $l : \varnothing \to X$, would need to have its graph be a subset of $\varnothing \times X \equiv \varnothing$, hence it would be an empty function, unique for each set $X$ in the codomain.

For a natural transformation between presheaves, seen as a morphism, to be unique would imply all of its component functions to be unique between any two sets, in particular, a initial morphism would imply the existence of a collection of functions parting from a single set for each object, of which, the totality of this collection, would arrive at every set, in particular to the empty set, which would make the domain of such a function need to be empty for it to be a function, hence the only possible initial object in the category of presheaves is a contravariant functor that maps all objects of $\mathcal{C}$ to the empty set, and, for reasons of compatibility, would also need to map all morphisms of $\mathcal{C}$ to the empty function $\varnothing \mapsto \varnothing$, *i.e.* it would have to be the $\mathbf{0}_\mathcal{C}$ functor.

On the other hand, for a presheaf to be terminal any other presheaf would need to arrive in it by natural transformations, in particular a presheaf that maps all objects to a singleton set would need to have a component function of the natural transformation mapping this singleton set to the set mapped by the terminal presheaf for each object in $\mathcal{C}$, hence these sets need to have the same cardinality, *i.e.* one element, for those component functions to be functions, therefore the terminal object of the presheaf category needs to be a contravariant functor that maps all objects of $\mathcal{C}$ to a singleton set, and all morphisms, for compatibility's sake, to the identity function on this singleton, making it the $\mathbf{1}_\mathcal{C}$ functor. ∎

We are now able to conjoin the association of subobjects of presheaves to other presheaves given by the corollary 7.1.24 with the observation that the terminal objects of the category of presheaves is exactly the functor that maps objects to singleton sets, from lemma 7.1.25, which allows us to talk about the elements of presheaves, concordant with definition 7.1.20, indirectly through the following definition:

**Definition 7.1.26** (Global sections of presheaves)**.** *Let $\mathcal{C}$ be a small category and let $G$ be a presheaf over $\mathcal{C}$. We say that a presheaf $F$ over $\mathcal{C}$ is a **global section** of $G$ if, and only if, for each $A$ in* $\mathbf{Obj}_\mathcal{C}$, $F(A) = \{p_A\}$, *for some point $p_A \in G(A)$, and $\iota : F \Longrightarrow G$ with $\iota_A(p_A) = p_A$ (the inclusion function) is a natural transformation.*

Notice that a global section $F$ must transform any morphism $A \xrightarrow{g} B$ to the unique map $\{p_B\} \xmapsto{F(g)} \{p_A\}$, hence, a global section is completely determined by the collection $(p_A)_{A \text{ in } \mathbf{Obj}_\mathcal{C}}$. The naturality condition for $\iota$, $\iota_A \circ F(f) = G(f) \circ \iota_B$, with the fact that $F(B) = \{p_B\}$, then implies:

$$(\iota_A \circ F(f))(p_B) = (G(f) \circ \iota_B)(p_B),$$

$$\iota_A(F(f)(p_B)) = G(f)(\iota_B(p_B)),$$

$$\iota_A(p_A) = G(f)(p_B),$$

$$p_A = G(f)(p_B).$$

This then implies we can identify global sections on a presheaf with the presheaf's elements, since a global section is a subobject by corollary 7.1.24 and each of these global sections map objects to a single point, which by lemma 7.1.25 makes them a terminal object, hence the global section functor applied to any morphism of $\mathcal{C}$ gives us a function that has only one point as image, this spells out the presheaf's element.

We see that this definition abstracts the initial definition of global sections, by considering that the global section $F$ of any subobject $K$ of a presheaf $G$, is defined in such a way that we can always map elements associated with the global section back to the presheaf $G$, by the way of the composition of the natural transformations, whose components are the inclusion functions, of $F$ to $K$ with $K$ to $G$, in the same way as the restriction morphisms $res$ were to be composed.

We can now enunciate a definition of the **observable presheaf** within category theory, from which we will associate the global sections of this presheaf with the valuation functions present in Kochen-Specker.

**Definition 7.1.27** (Category of quantum observables). *Considering a von Neumann algebra $\mathfrak{R}$ of self-adjoint elements, if $A, B \in \mathfrak{R}$ are such that there is some Borel measurable function $f : \sigma(B) \to \mathbb{R}$ such that, using the Borel functional calculus from theorem B.0.56, $A = f(B)$, then we can say that such a relation establishes a pre-ordering $A \preceq B$, since $A = \mathbb{1}_{\mathbb{R}}(A)$ hence $A \preceq A$ and if $A \preceq B \preceq C$ this means that there are Borel measurable functions $f$ and $g$ such that $A = f(B)$ and $B = g(C)$, hence $A = f \circ g(C)$ and $A \preceq C$.*

*Then a category $\mathfrak{Obs}$ of quantum observables can be formulated such that $\mathbf{Obj}_{\mathfrak{Obs}} = \mathfrak{R}$ and if $A \preceq B$, i.e. $A = f(B)$ for objects $A$ and $B$ and a Borel measurable function $f$, then there is a unique morphism in $\mathbf{Mor}_{\mathfrak{Obs}}(A, B)$, namely $f$, where uniqueness is guaranteed by theorem B.0.53 used in the Borel functional calculus theorem B.0.56.*

*Since this category was defined over a pre-order $\preceq$, it is categorized as a pre-order category.*

**Definition 7.1.28** (Gel'fand spectra and Gel'fand transform). *Let $\mathfrak{V} \in \mathfrak{C}(\mathfrak{M})$ be an abelian sub-von Neumann algebra of the von Neumann algebra $\mathfrak{M}$. We denote by $\Xi_{\mathfrak{V}}$ the so called **Gel'fand spectrum** of $\mathfrak{V}$, the set composed of positive linear functionals, e.g. $\lambda : \mathfrak{V} \to \mathbb{C}$, these functionals are called **characters** and are of norm 1, i.e. $\|\lambda\|_{\mathfrak{V}'} = 1$, that are also $*$-algebra homomorphisms, that is:*

$$\forall A, B \in \mathfrak{V}, \quad \lambda(AB^*) = \lambda(A)\overline{\lambda(B)}, \quad \forall \lambda \in \Xi_{\mathfrak{V}}.$$

*For each element $A \in \mathfrak{V}$ we then define its **Gel'fand transform** as the function $\hat{\tilde{A}} : \Xi_{\mathfrak{V}} \to \mathbb{R}$ defined by $\hat{\tilde{A}}(\lambda) := \lambda(A)$.*

**Proposition 7.1.29.** *Every character $\lambda \in \Xi_{\mathfrak{V}}$, $\lambda$ is continuous.*

*Proof:* Since $\|\lambda\|_{\mathfrak{V}'} = 1$ for every character $\lambda \in \Xi_{\mathfrak{V}}$, this means that $|\lambda(A)| \leq \|A\|$, $\forall A \in \mathfrak{V}$, then by proposition B.0.6, $\lambda$ is continuous. ∎

**Proposition 7.1.30.** *The set $\Xi_{\mathfrak{V}}$ is non empty and the map $\lambda \mapsto \ker(\lambda)$ defines a bijection from $\Xi_{\mathfrak{V}}$ onto the set of all maximal ideals of $\mathfrak{V}$.*

*Proof:* We begin by noting that $\lambda(\mathbb{1}A) = \lambda(A\mathbb{1}) = \lambda(\mathbb{1})\lambda(A) = \lambda(A)\lambda(\mathbb{1}) = \lambda(A)$ implies that $\lambda(\mathbb{1}) = 1$, since $\lambda$ cannot be the trivial zero homomorphism as it doesn't have a norm of 1.

Let $I$ denote the closed ideal $\ker(\lambda)$, for $\lambda \in \Xi_{\mathfrak{V}}$, which is an ideal since if $A \in \ker(\lambda)$, then $\lambda(A) = 0$ and by the homomorphism property of $\lambda$, the product of $A$ with any other element of $\mathfrak{V}$ will also be zero; and $\ker(\lambda)$ is trivially closed, since if $A \in \overline{\ker(\lambda)}^{\text{op}}$, but $A \notin \ker(\lambda)$, then there would be a sequence $\{A_n\}_{n\in\mathbb{N}} \subset \ker(\lambda)$ such that, for every $\delta > 0$, we could find $N(\delta) \in \mathbb{N}$ with $\|A_n - A\| < \delta$ whenever $n > N(\delta)$. Considering the continuity of $\lambda$, established on the previous proposition, then for every $\varepsilon > 0$ there exists $\delta(\varepsilon) > 0$, with $|\lambda(B_n) - \lambda(B)| < \varepsilon$ whenever $\|B_n - B\| < \delta$, using the previous sequence sequence $\{A_n\}_{n\in\mathbb{N}} \subset \ker(\lambda)$ and $\delta$ adjusted for the $\varepsilon$ inequality, we get:

$$|\lambda(A_n) - \lambda(A)| = |\lambda(A)| < \varepsilon$$

for every $\varepsilon > 0$, which makes $\lambda(A) = 0$, which is an absurd for all $A \in \overline{\ker(\lambda)}^{\mathrm{op}}$ and $A \notin \ker(\lambda)$, hence this set is empty and $\overline{\ker(\lambda)}^{\mathrm{op}} = \ker(\lambda)$.

The ideal $I = \ker(\lambda)$ is proper since $\lambda \neq 0$ as $\|\lambda\|_{\mathfrak{V}'} = 1$, and since $A - \lambda(A)\mathbb{1} \in \ker(\lambda) = I$, $\forall A \in \mathfrak{V}$, then $\mathfrak{V} = I + \mathbb{C}\mathbb{1}$, where the sum is a sum of subspaces, from this it clearly follows that $I$ is a maximal ideal of $\mathfrak{V}$.

If $\lambda_1, \lambda_2 \in \Xi_{\mathfrak{V}}$ and $\ker(\lambda_1) = \ker(\lambda_2)$, then as $A - \lambda_2(A)\mathbb{1} \in \ker(\lambda_2)$, then $\lambda_1(A - \lambda_2(A)\mathbb{1}) = 0$ which implies $\lambda_1(A) = \lambda_2(A)$, thus $\lambda_1 = \lambda_2$ which proves injectivity of the map $\lambda \mapsto \ker(\lambda)$.

Now we consider the elementary fact that the closure of a normed algebra ideal is an ideal, that is, for an arbitrary proper ideal $I$, if $\kappa \in \overline{I}^{\mathrm{op}}$ then there is a sequence $\{\kappa_n\}_{n\in\mathbb{N}}$ such that $\forall \varepsilon > 0$ there is an $N \in \mathbb{N}$, where for $n > N$:

$$\|\kappa_n - \kappa\| < \varepsilon,$$

this closure is also closed under linear combinations of sequences, since:

$$\|(\alpha\kappa_n + \beta\nu_n) - (\alpha\kappa + \beta\nu)\| \leq |\alpha|\|\kappa_n - \kappa\| + |\beta|\|\nu_n - \nu\| < (|\alpha| + |\beta|)\varepsilon.$$

If we now take $\kappa \in \overline{I}^{\mathrm{op}}$, a sequence tending to $\kappa$ such as $\{\kappa_n\}_{n\in\mathbb{N}} \subset I$, an arbitrary $x \in \mathfrak{V}$ and take $N \in \mathbb{N}$ for which $\|\kappa_n - \kappa\| < \frac{\varepsilon}{\|x\|}$, whenever $n > N$, then for $n > N$:

$$\|x\kappa_n - x\kappa\| \leq \|x\|\|\kappa_n - \kappa\| \leq \varepsilon,$$

then $x\kappa \in \overline{I}^{\mathrm{op}}$, and therefore $\overline{I}^{\mathrm{op}}$ is an ideal. More than that, it is an proper ideal, since if we consider a element $u \in \mathfrak{V}$, such that $A - Au = A - uA \in I$, for all $A \in \mathfrak{V}$, for $B \in I$ with $\|u - B\| < 1$, then the element $v = \mathbb{1} - u + B$ is invertible in $\mathfrak{V}$ by proposition B.0.28, then for $A \in \mathfrak{V}$, $Av = A - Au + AB \in I$, as $v$ is invertible we can take $Av^{-1}v = A \in I$, making $A = Av \subseteq I$, but this contradicts the assumption that $I$ is proper, as a consequence we must have that $\|u - B\| \geq 1$ for all $B \in I$, hence $u \notin \overline{I}^{\mathrm{op}}$, making it a proper subset of $\mathfrak{V}$, and therefore a proper ideal.

If we further consider $I$ to be a maximal ideal, then $\overline{I}^{\mathrm{op}} \supseteq I$ and $\overline{I}^{\mathrm{op}}$ is an ideal, makes the fact that $I$ is a maximal ideal imply $I = \overline{I}^{\mathrm{op}}$, as there can be no other proper ideal containing $I$, this is what it pertains to be maximal.

The quotient $\mathfrak{V}/I$ is composed of the equivalence classes $[A] = A + I := \{A + v \mid v \in I\}$ (where the equivalence relation was $A \sim B \Leftrightarrow A - B \in I$), which is also a algebra by its own right when we define sums as $(A+I)+(B+I) = (A+B)+I$ and products as $(A+I)(B+I) = (AB)+I$, where multiplication by scalars would also be such that $\alpha(A + I) = \alpha A + I$. If $J$ is an ideal in $\mathfrak{V}/I$ and $\pi(\,\cdot\,) = [\,\cdot\,]$ be seen as the quotient map from $\mathfrak{V}$ to $\mathfrak{V}/I$, then considering that if $v \in J$, $av \in J$ for all $a \in \mathfrak{V}/I$ in particular $(a - a)v \in J \Rightarrow \mathbf{0}_{\mathfrak{V}/I} \in J$ and $\mathbf{0}_{\mathfrak{V}/I} = [0] = [u], \forall u \in I$, obviously, hence $\pi^{-1}(J)$ is an ideal of $\mathfrak{V}$ containing $I$, this makes $\pi^{-1}(J)$ equal to either $\mathfrak{V}$ or $I$ by the maximality of $I$. Therefore, either $J = \mathfrak{V}/I$ or $J = \{\mathbf{0}_{\mathfrak{V}/I}\}$, which make them the only two ideals of $\mathfrak{V}/I$.

Now suppose $\pi(a) \equiv [a]$ is a non-zero element of $\mathfrak{V}/I$. Then as $J = \pi(a)\mathfrak{V}/I$ must be a non-zero ideal of $\mathfrak{V}/I$, and therefore $J = \mathfrak{V}/I$. In other words this means that there is an element $b$ of $\mathfrak{V}$ such that $(a+I)(b+I) = \mathbb{1} + I$, where $\mathbb{1} + I$ is the identity of $\mathfrak{V}/I$, so $a + I$ is invertible. This makes $\mathfrak{V}/I$ a unital Banach algebra in which every non-zero element is invertible.

Using theorem B.0.38 we get that $\mathfrak{V}/I = \mathbb{C}(\mathbb{1} + I)$, from this it follows trivially that $\mathfrak{V} = I + \mathbb{C}\mathbb{1}$. If we define then the linear functional $\hat{\lambda} : \mathfrak{V} \to \mathbb{C}$ such that $\hat{\lambda}(a + \alpha\mathbb{1}) = \alpha$, whenever $a \in I$, $\alpha \in \mathbb{C}$. Then $\hat{\lambda}$ is trivially a character and $\ker(\hat{\lambda}) = I$. Therefore for each maximal ideal $I$ of $\mathfrak{V}$ we can find a character $\hat{\lambda}$, such that $I = \ker(\hat{\lambda})$, making the map $\lambda \mapsto \ker(\lambda)$ is then surjective onto the set of all maximal ideals of $\mathfrak{V}$.

Thus, we have shown that the map $\lambda \mapsto \ker(\lambda)$ is a bijection from the characters onto the maximal ideals of $\mathfrak{V}$.

Since $\mathfrak{V}$ is unital, then if $(\mathbb{1} - u) \in I$, a ideal of $\mathfrak{V}$, then for all $A \in \mathfrak{V}$, $A(\mathbb{1}-u) = A - Au \in I$ and $(\mathbb{1} - u)A = A - uA \in I$, for a $u \in \mathfrak{V}$, a property of an ideal that is called modular, hence all ideals of a unital algebra are modular. Therefore proper modular ideals, or in this case just proper ideals, form a chain under inclusion, since for all of them $u$ is such that $(\mathbb{1} - u)$ is an element of each ideal on the chain, hence in this totally ordered chain, for each finite sub-chain we always have a upper bound in the form of the last set of the sub-chain, we then have by Zorn's lemma a maximal element in this chain which is a maximal proper ideal of $\mathfrak{V}$. Hence there is at least one maximal proper ideal in $\mathfrak{V}$, since $\{0_{\mathfrak{V}}\}$ is a proper ideal of $\mathfrak{V}$ and we can use the previous argument to construct a maximal proper ideal, using then the construction of the $\hat{\lambda} : \mathfrak{V} \to \mathbb{C}$ character, used for proving surjectivity of the map $\lambda \mapsto \ker(\lambda)$, applied to that one maximal proper ideal, we obtain that $\Xi_{\mathfrak{V}} \neq \varnothing$. ∎

In the next important theorem we follow, somewhat loosely, [28]:

**Theorem 7.1.31.** *Let $\mathfrak{A}$ be an abelian von Neumann algebra. Then the Gel'fand spectrum $\Xi_{\mathfrak{A}}$ is homeomorphic to the Stone space $S(\mathcal{P}(\mathfrak{A}))$ of the lattice $\mathcal{P}(\mathfrak{A})$ (def. C.0.14).*

*Proof:* Let $\lambda \in \Xi_{\mathfrak{A}}$, then:
$$\beta_\lambda := \{P \in \mathcal{P}(\mathfrak{A}) \setminus \{\mathbf{0}\} \,|\, \lambda(P) = 1\}$$

is such that $\mathbf{0} \notin \beta_\lambda$, obviously, and if $P_1, P_2 \in \beta_\lambda$, clearly there exists a $P \in \beta_\lambda$ such that $P \leq P_1$ and $P \leq P_2$, since $\lambda(P_1 P_2) = \lambda(P_1)\lambda(P_2) = 1$, hence $P_1 P_2 \in \beta_\lambda$, as they commute, and we can take $P = P_1 P_2$, since $P_1 P_2 \leq P_1$ and $P_1 P_2 \leq P_2$. Moreover $\lambda(\mathbb{1} - P) = 1 - \lambda(P)$, so that $\forall P \in \mathcal{P}(\mathfrak{A})$, either $P \in \beta_\lambda$ or $\mathbb{1} - P \in \beta_\lambda$. Let $P \in \beta_\lambda$ and $K \in \mathcal{P}(\mathfrak{A})$, such that $P \leq K$, then as $PK = P = KP$ we can calculate $\lambda(KP) = \lambda(K)\lambda(P) = \lambda(P) = 1$ making $\lambda(K) = 1$ and hence $K \in \beta_\lambda$, so $\beta_\lambda$ is upwardly closed, and since the meets operation in the lattice of orthogonal projectors is given by the product of projectors, as we have already seen, $\beta_\lambda$ is closed under finite products of projectors, hence $\beta_\lambda$ is a filter.

More than that, since $\mathbf{0} \notin \beta_\lambda$ and if $P \in \beta_\lambda$ then $\mathbb{1} - P \notin \beta_\lambda$, this makes $\beta_\lambda$ a proper filter, even more, if we suppose there is another proper filter $\mathcal{R}$ of $\mathcal{P}(\mathfrak{A})$ such that $\beta_\lambda \subsetneq \mathcal{R} \subsetneq \mathcal{P}(\mathfrak{A})$, then there would be a projector $R \in \mathcal{R}$ and $R \notin \beta_\lambda$, but then $(\mathbb{1} - R) \in \beta_\lambda \subsetneq \mathcal{R}$, but for $\mathcal{R}$ to be a filter it must be closed under finite meets, or in this case products, hence $R(\mathbb{1} - R) = R - R = \mathbf{0} \in \mathcal{R}$, and beyond that $\mathcal{R}$ must be upwardly closed, but since $\mathbf{0} \leq P$, $\forall P \in \mathcal{P}(\mathfrak{A})$, then $\mathcal{R} = \mathcal{P}(\mathfrak{A})$, so that $\mathcal{R}$ is not a proper filter, making $\beta_\lambda$ necessarily a ultrafilter, def. C.0.13, and so a element of the Stone space $S(\mathcal{P}(\mathfrak{A}))$.

Let $\lambda, \kappa \in \Xi_\mathfrak{A}$ such that $\beta_\lambda = \beta_\kappa$, then for all $P \in \mathcal{P}(\mathfrak{A}), \lambda(P) = 1 \Leftrightarrow \kappa(P) = 1$. Hence $\lambda$ and $\kappa$ agree on $\mathcal{P}(\mathfrak{A})$ and therefore also on the finite linear combinations of elements of $\mathcal{P}(\mathfrak{A})$. By the continuity of $\lambda$ and $\kappa$, proven in proposition 7.1.29, we conclude from the spectral decomposition theorem 4.0.17, that $\lambda = \kappa$. The mapping:

$$\begin{array}{rcl} \Xi_\mathfrak{A} & \to & S(\mathcal{P}(\mathfrak{A})) \\ \lambda & \mapsto & \beta_\lambda \end{array}$$

is therefore injective.

Conversely, let $\beta \in S(\mathcal{P}(\mathfrak{A}))$ be given. We define a mapping:

$$\lambda_\beta : \mathcal{P}(\mathfrak{A}) \to \{0, 1\}$$

by

$$\lambda_\beta(P) = \chi_\beta(P) := \begin{cases} 1, & \text{if } P \in \beta; \\ 0, & \text{if } P \notin \beta. \end{cases}$$

Since $\beta$ is an ultrafilter $P, Q \in \beta$ if and only if $P \wedge Q = PQ \in \beta$, as they commute in this abelian algebra, this makes:

$$\forall P, Q \in \mathcal{P}(\mathfrak{A}), \quad \lambda_\beta(PQ) = \lambda_\beta(P)\lambda_\beta(Q).$$

In the next step we extend $\lambda_\beta$ to a linear functional on the finite linear combinations of elements of $\mathcal{P}(\mathfrak{A})$ (which we will also denote by $\lambda_\beta$). Let $A = \sum_{j=1}^{m} b_j P_j$ be an orthogonal representation of an element of the finite linear combinations of elements of $\mathcal{P}(\mathfrak{A})$. If $P_1 + ... + P_m = \mathbb{1}$ and $b_1, ..., b_m$ are all different from zero, then we call such a representation a complete orthogonal representation. For the basis elements of the topology of $S(\mathcal{P}(\mathfrak{A}))$, we write $S_P(\mathcal{P}(\mathfrak{A})) := \{\mathcal{F} \in S(\mathcal{P}(\mathfrak{A})) \mid P \in \mathcal{F}\}$. Then clearly:

$$S(\mathcal{P}(\mathfrak{A})) = \bigcup_{j \leq m} S_{P_j}(\mathcal{P}(\mathfrak{A})),$$

we have that $\beta \in S_{P_j}(\mathcal{P}(\mathfrak{A}))$ for exactly one $j \leq m$, since the $\{P_n\}_{n=1,...,m}$ are mutually orthogonal, then if any ultrafilter were to contain any two of these projectors, say $P_n, P_k, n \neq k$, then it would need to contain their meet also, which is $P_n P_k = \mathbf{0}_\mathfrak{A}$, but since any filter is upwardly closed this filter would contain every element of $\mathcal{P}(\mathfrak{A})$, since $\mathbf{0}_\mathfrak{A}$ is under every projector, but then this filter wouldn't be proper and hence wouldn't be a ultrafilter. Therefore each ultrafilter in $S(\mathcal{P}(\mathfrak{A}))$ must contain at most one of the mutually orthogonal projectors $\{P_n\}_{n=1,...,m}$, and it must contain at least one because the $\{P_n\}_{n=1,...,m}$ partition $\mathfrak{A}$ and hence every projector is in at least one of these subspaces.

We denote by $j_\beta$ the unique index $j$ for which $\beta \in S_{P_j}(\mathcal{P}(\mathfrak{A}))$, then we define:

$$\lambda_\beta \left( \sum_{j=1}^{m} b_j P_j \right) := b_{j_\beta}.$$

We must show that the functional $\lambda_\beta$ from the finite linear combinations of $\mathcal{P}(\mathfrak{A})$ to $\mathbb{C}$ is well defined. Let $\sum_{j=1}^{m} b_j P_j$ and $\sum_{k=1}^{n} c_k Q_k$ be two complete orthogonal representations of $A$, a element of the finite linear combinations of $\mathcal{P}(\mathfrak{A})$. If we let $V_j := \operatorname{Ran}(P_j)$ and $W_k := \operatorname{Ran}(Q_k)$, $j \leq m, k \leq n$ and let $v_{j_0} \in V_{j_0}$, then:

$$b_{j_0} v_{j_0} = \left(\sum_{j=1}^{m} b_j P_j\right) v_{j_0} = \left(\sum_{k=1}^{n} c_k Q_k\right) v_{j_0} \in W_1 + ... + W_n.$$

Hence $v_{j_0} \in W_1 + ... + W_n$, this shows $V_1 + ... + V_m \subseteq W_1 + ... + W_n$. By the symmetry of the argument we get $W_1 + ... + W_n = V_1 + ... + V_m$ and therefore $P_1 + ... + P_m = Q_1 + ... + Q_n$.

If for a ultrafilter $\alpha \in S(\mathcal{P}(\mathfrak{A}))$, $P_1 + ... + P_m \notin \alpha$, and consequently $Q_1 + ... + Q_n \notin \alpha$, then both $b_{j_\alpha}$ and $c_{k_\alpha}$ would need to be equal to zero. If $P_1 + ... + P_m \in \alpha$ then, due to orthogonality, there are uniquely determined indices $j_0, k_0$, such that $P_{j_0}, Q_{k_0} \in \alpha$, but then $P_{j_0} Q_{k_0} \in \alpha$ and in particular $P_{j_0} Q_{k_0} \neq \mathbf{0}_{\mathfrak{A}}$, then $Q_{k_0} \sum_{j=1}^{m} b_j P_j = Q_{k_0} \sum_{k=1}^{n} c_k Q_k = Q_{k_0} c_{k_0}$ and multiplying this by $P_{j_0}$ on both sides we get $P_{j_0} Q_{k_0} \sum_{j=1}^{m} b_j P_j = P_{j_0} Q_{k_0} b_{j_0} = P_{j_0} Q_{k_0} c_{k_0}$ this implies $b_{j_0} = c_{k_0}$.

By the previous argument, as $b_{j_\beta}, c_{k_\beta}$ imply that $\beta \in S_{P_{j_\beta}}(\mathcal{P}(\mathfrak{A}))$ and $\beta \in S_{Q_{k_\beta}}(\mathcal{P}(\mathfrak{A}))$, we arrive at $\lambda_\beta \left(\sum_{j=1}^{m} b_j P_j\right) = b_{j_\beta} = c_{k_\beta} = \lambda_\beta \left(\sum_{k=1}^{n} c_k Q_k\right)$, this proves that $\lambda_\beta$ is well defined on the finite linear combinations of $\mathcal{P}(\mathfrak{A})$.

Now let $\sum_{k=1}^{n} a_k E_k$ be an arbitrary element of the finite linear combinations of $\mathcal{P}(\mathfrak{A})$, we see that:

$$\lambda_\beta \left(\sum_{k=1}^{n} a_k E_k\right) = a_{j_1} + ... + a_{j_s},$$

where $j_1, ..., j_s$ are precisely those indices for which $E_{j_1}, ..., E_{j_s}$ are elements of $\beta$. This shows that $\lambda_\beta$ is linear. Multiplicativity follows from linearity and the fact that $\lambda_\beta$ is multiplicative on projections. $\lambda_\beta$ is continuous in norm because for orthogonal representations we have that:

$$\left\|\sum_{k=1}^{n} a_k P_k\right\| = \max_{k \leq n} |a_k|.$$

The BLT theorem B.0.15 then assures us that $\lambda_\beta$ has a unique extension to a multiplicative linear functional on $\mathcal{P}(\mathfrak{A})$, this extension can be further extended to $\mathfrak{A}$, by noting that since $\mathfrak{A}$ is abelian, then $AA^* = A^*A, \forall A \in \mathfrak{A}$ and then every element of $\mathfrak{A}$ is normal, using now the generalization of the spectral decomposition theorem for normal elements proven in chapter 8 section 8.3.3, theorem 8.54(d) of [77], we can extend $\lambda_\beta$ to the entirety of $\mathfrak{A}$, this extension we shall also denote by $\lambda_\beta$, it is clear that this last one is a character. By construction we have:

$$\beta_{\lambda_\beta} = \beta,$$

This establishes the surjectivity of the map $\beta_{(\cdot)} : \Xi_{\mathfrak{A}} \to S(\mathcal{P}(\mathfrak{A}))$. And therefore the mapping $\lambda \mapsto \beta_\lambda$ is a bijection from $\Xi_{\mathfrak{A}}$ to $S(\mathcal{P}(\mathfrak{A}))$.

Considering that from the get go $S(\mathcal{P}(\mathfrak{A}))$ is a Hausdorff compact space, and $\Xi_{\mathfrak{A}}$ is also compact, since any character lies within $\mathcal{B}_1(\mathfrak{A}')$ then by the Banach-Alaoglu theorem B.0.18, $\Xi_{\mathfrak{A}}$ is compact in the weak-$^*$ topology which is also Hausdorff by proposition B.0.17; then by proposition B.0.26 we only need to show that the mapping $\lambda \mapsto \beta_\lambda$ is continuous for it to be a homeomorphism.

Let $\lambda_0 \in \Xi_\mathfrak{A}$, $0 < \varepsilon < 1$ and $P \in \mathcal{P}(\mathfrak{A})$ be such that $\lambda_0(P) = 1$. Then:

$$N_\varepsilon^P(\lambda_0) := \{\lambda \in \Xi_\mathfrak{A} \,|\, |\lambda(P) - \lambda_0(P)| < \varepsilon\}$$

is an open neighborhood of $\lambda_0$ and from $\varepsilon < 1$, we conclude:

$$\lambda \in N_\varepsilon^P(\lambda_0) \xLeftrightarrow{\varepsilon<1,\ \mathrm{Ran}(\lambda)=\{0,1\}} \lambda(P) = \lambda_0(P)$$

$$\Leftrightarrow \lambda(P) = 1 \iff P \in \beta_\lambda$$

$$\Leftrightarrow \beta_\lambda \in S_P(\mathcal{P}(\mathfrak{A})).$$

This means that $N_\varepsilon^P(\lambda_0)$ is mapped bijectively onto the open neighborhood $S_P(\mathcal{P}(\mathfrak{A}))$ that also contains $\beta_{\lambda_0}$. As we have already alluded to, the $S_P(\mathcal{P}(\mathfrak{A}))$ with $P \in \beta_{\lambda_0}$ form a neighborhood base of $\beta_{\lambda_0}$ in the topology of the Stone space $S(\mathcal{P}(\mathfrak{A}))$. Hence $\lambda \mapsto \beta_\lambda$ is continuous and is therefore a homeomorphism. ∎

**Proposition 7.1.32.** *For each $A \in \mathfrak{V}$ and $\lambda \in \Xi_\mathfrak{V}$, it holds that:*

$$\lambda(A) \in \sigma(A).$$

*More than that, it is true that:*

$$\sigma(A) = \{\lambda(A) \,|\, \lambda \in \Xi_\mathfrak{V}\}.$$

*Proof:* If we calculate:
$$\lambda(A - \lambda(A)\mathbb{1}) = \lambda(A) - \lambda(A) = 0,$$

and so $A - \lambda(A)\mathbb{1} \in \ker(\lambda)$. For $A - \lambda(A)\mathbb{1}$ to be invertible there would need to exist a element $B$ such that $(A - \lambda(A)\mathbb{1})B = B(A - \lambda(A)\mathbb{1}) = \mathbb{1}$, but then $\lambda(A - \lambda(A)\mathbb{1})\lambda(B) = \lambda(B)\lambda(A - \lambda(A)\mathbb{1}) = \lambda(\mathbb{1}) = 1$, as $\lambda(A - \lambda(A)\mathbb{1}) = 0$, there can be no $B$ such that $\lambda(A - \lambda(A)\mathbb{1})\lambda(B) = \lambda(B)\lambda(A - \lambda(A)\mathbb{1}) \neq 0$. Therefore $A - \lambda(A)\mathbb{1}$ is not invertible and hence $\lambda(A) \in \sigma(A)$.

On the other hand, as we have already shown that $\{\lambda(A) \,|\, \lambda \in \Xi_\mathfrak{V}\} \subseteq \sigma(A)$, we must show that $\sigma(A) \subseteq \{\lambda(A) \,|\, \lambda \in \Xi_\mathfrak{V}\}$, if $\nu \in \sigma(A)$, then $A - \nu\mathbb{1}$ isn't invertible and $I = (A - \nu\mathbb{1})\mathfrak{V}$ is an ideal, since for every $(A - \nu\mathbb{1})V \in I$, where $V \in \mathfrak{V}$, $(A - \nu\mathbb{1})V$ isn't invertible, we can see this by contradiction, if there was $W \in \mathfrak{V}$ such that $(A - \nu\mathbb{1})VW = W(A - \nu\mathbb{1})V = VW(A - \nu\mathbb{1}) = \mathbb{1}$, then $(A - \nu\mathbb{1})$ would be invertible with inverse equal to $VW$. So, since the ideal $I$ is clearly a proper ideal, by proposition 7.1.30 $I$ is contained in a maximal ideal $\ker(\lambda)$, where $\lambda \in \Xi_\mathfrak{V}$, then in particular $0 = \lambda((A - \nu\mathbb{1})\mathbb{1}) = \lambda(A) - \nu\lambda(\mathbb{1}) = \lambda(A) - \nu$, hence $\lambda(A) = \nu$, proving that for every $\nu \in \sigma(A)$, we can find $\lambda \in \Xi_\mathfrak{V}$ such that $\lambda(A) = \nu$, which proves the converse containment $\sigma(A) \subseteq \{\lambda(A) \,|\, \lambda \in \Xi_\mathfrak{V}\}$. ∎

**Proposition 7.1.33.** *Given a self-adjoint $A \in \mathfrak{V}_\dagger$ and any Borel measurable function $f \in B^\mathscr{B}(\sigma(A))$ and $\lambda \in \Xi_{\mathcal{C}(\{A\})}$, for the context $\mathcal{C}(\{A\})$ generated by $A$, it holds that $f(A) \in \mathcal{C}(\{A\})$ and that $\lambda(f(A)) = f(\lambda(A))$. Furthermore:*

$$\mathcal{C}(\{A\}) = \{f(A) \,|\, f \in B^\mathscr{B}(\sigma(A))\}.$$

*Also, every context* $\mathfrak{W} \in \mathfrak{C}(\mathfrak{M})$ *is such that:*

$$\mathfrak{W} = \mathcal{C}(\{A\}),$$

*for some* $A \in \mathfrak{M}$.

*Proof:* We begin by noting that by the spectral decomposition theorem 4.0.17, more specifically item *(ii)*, for every Borel measurable function $f \in B^{\mathscr{B}}(\sigma(A))$, $f(A)$ commutes with every element that commutes with $A$, and since these elements commute, $Bf(A) = f(A)B$

We begin by proving that $f(A) \in \mathcal{C}(\{A\})$, as $\mathcal{C}(\{A\})$ is a von Neumann algebra $\mathcal{C}(\{A\})'' = \mathcal{C}(\{A\})$, by the Borel functional calculus theorem B.0.56, since each $f(A)$ commutes with every element that commutes with $A$, then $f(A) \in \mathcal{C}(\{A\})'' = \mathcal{C}(\{A\})$.

To prove then that $\mathcal{C}(\{A\}) = \{f(A) \mid f \in B^{\mathscr{B}}(\sigma(A))\}$, we notice that the previous argument $\{f(A) \mid f \in B^{\mathscr{B}}(\sigma(A))\} \subseteq \mathcal{C}(\{A\})$, to prove the inverse containment we notice that by item $(c)$ of theorem B.0.56, $\{f(A) \mid f \in B^{\mathscr{B}}(\sigma(A))\}$ is closed in the strong operator topology, but since $\{f(A) \mid f \in B^{\mathscr{B}}(\sigma(A))\}$ is a unital $*$-invariant sub-algebra of $\mathfrak{B}(\mathscr{H})$, we get by von Neumann's bicommutant theorem that $\overline{\{f(A) \mid f \in B^{\mathscr{B}}(\sigma(A))\}}^w = \overline{\{f(A) \mid f \in B^{\mathscr{B}}(\sigma(A))\}}^s = \{f(A) \mid f \in B^{\mathscr{B}}(\sigma(A))\}$, hence $\{f(A) \mid f \in B^{\mathscr{B}}(\sigma(A))\}$ is a context, but since $\mathcal{C}(\{A\})$ must be the smallest context containing $A$, $\mathcal{C}(\{A\}) = \{f(A) \mid f \in B^{\mathscr{B}}(\sigma(A))\}$.

For us to prove that $\forall \mathfrak{W} \in \mathfrak{C}(\mathfrak{M})$, $\mathfrak{W} = \mathcal{C}(\{A\})$, for some $A \in \mathfrak{M}$, we consider that every element of $\mathfrak{W}$ commutes with every other element of $\mathfrak{W}$, then by the von Neumann-Varadarajan theorem A.0.11 there exists a self-adjoint element $A \in \mathfrak{M}$, such that for every operator $R \in \mathfrak{W}$, $\exists f$ a bounded Borel measurable function such that $R = f(A)$, hence $\mathfrak{W} \subseteq \{f(A) \mid f \in B^{\mathscr{B}}(\sigma(A))\}$, also $A \in \mathfrak{W}$ since $\mathfrak{W}$ is a von Neumann algebra and as a consequence of the Borel functional calculus B.0.56, $A \in \mathfrak{W}'' = \mathfrak{W}$, but as we saw in the last paragraph the set $\{f(A) \mid f \in B^{\mathscr{B}}(\sigma(A))\}$ is the smallest context that contains $A$, hence $\{f(A) \mid f \in B^{\mathscr{B}}(\sigma(A))\} \subseteq \mathfrak{W}$, and the equality $\mathfrak{W} = \{f(A) \mid f \in B^{\mathscr{B}}(\sigma(A))\}$ is proved.

Finally we prove that $\lambda(f(A)) = f(\lambda(A))$, for every character $\lambda \in \Xi_{\mathcal{C}(\{A\})}$ and bounded Borel measurable function $f$. By the spectral decomposition theorem 4.0.17 we know that we can decompose the operator $f(A) = \int_{\sigma(A)} f(\lambda) dP^{(A)}(\lambda)$, for any $f \in B^{\mathscr{B}}(\sigma(A))$ and self-adjoint $A$. The main idea now is to redo the steps used to get the spectral decomposition theorem, that is, firstly considering continuous functions, then using Lusin's theorem B.0.54 and its corollary B.0.55 to arrive at the case for a Borel measurable function.

Therefore, by corollary B.0.55 we know that any Borel measurable function has a integral that can be approximated through the integral of a continuous function with compact support, we consider that such a function can be in part approximated by a sequence simple functions, by proposition C.0.23, lets say that $\{s_n\}_{n \in \mathbb{N}}$ is such a sequence of simple functions, where $s_n(x) = \sum_{k=1}^{n} f(c_{n,k}) \chi_{E_{n,k}}(x)$ where we assume, without loss of generality, that the sets $\{E_{n,k}\}_{k=1}^{n}$ are disjoint, we can then use this approximation to calculate:

$$\lambda(f(A)) = \lambda\left(\lim_{n \to \infty} s_n(A)\right) = \lambda\left(\lim_{n \to \infty} \int_{\sigma(A)} s_n(\lambda) dP^{(A)}(\lambda)\right) = \lambda\left(\lim_{n \to \infty} \sum_{k=1}^{n} f(c_{n,k}) P_{E_{n,k}}\right),$$

and by the continuity of the characters $\lambda$, proved in proposition 7.1.29, we get:

$$\lim_{n \to \infty} \lambda\left(\sum_{k=1}^{n} f(c_{n,k}) P_{E_{n,k}}\right) = \lim_{n \to \infty} \sum_{k=1}^{n} f(c_{n,k}) \lambda(P_{E_{n,k}}) = \lim_{n \to \infty} f\left(\sum_{k=1}^{n} c_{n,k} \lambda(P_{E_{n,k}})\right) = f(\lambda(A)),$$

where we used that since $\lambda(P_{E_{n,k}}) \in \sigma(P_{E_{n,k}}) = \{0,1\}$, and since the $\{E_{n,k}\}_{k=1}^{n}$ are disjoint, $P_{E_{n,i}} P_{E_{n,j}} = 0$ for $i \neq j$, so at most one of the $\lambda(P_{E_{n,k}})$ in the sum are going to be equal to 1, which allows us to take the function $f$ to outside the sum, then, as $P_{E_{n,k}}$ projects into consecutively smaller subspaces, and since $s_n \xrightarrow{n\to\infty} f$ this makes $s_n(x) = \sum_{k=1}^{n} f(c_{n,k})\chi_{E_{n,k}}(x) \xrightarrow{n\to\infty} f(x)$ as $E_{n,k} \xrightarrow{n\to\infty} \{x\}$ then $c_{n,k} \xrightarrow{n\to\infty} x$ for some sequence of $k \leq n$, in particular there will be a subsequence of $c_{n,k} \to \lambda(A) \in \sigma(A)$ for such a subsequence $\lim_{n\to\infty} \lambda(\sum_{k=1}^{n} c_{n,k} P_{E_{n,k}}) = \lambda(A)$, by the approximation property of $\{s_n\}_{n\in\mathbb{N}}$ that has to be at least pointwise, considering this if it were to return a different value there would be a mismatch between values and hence $f\left(\sum_{k=1}^{n} c_{n,k} P_{E_{n,k}}\right) = \sum_{k=1}^{n} f(c_{n,k}) P_{E_{n,k}}$ wouldn't approximate $f(A)$. ∎

**Definition 7.1.34** (Gel'fand presheaf). *For a von Neumann algebra $\mathfrak{M}$, we define the **Gel'fand presheaf** $\underline{\Xi}$ as the contravariant functor $\underline{\Xi} : \mathfrak{C}(\mathfrak{M}) \to \mathbf{Set}$, that maps each context $V$ (object) in $\mathfrak{C}(\mathfrak{M})$ to $\underline{\Xi}(V) \equiv \underline{\Xi}_V := \Xi_V$, seen as a set, and for every inclusion $\iota_{W,V} : V \hookrightarrow W$ (morphism) where $V, W$ in $\mathfrak{C}(\mathfrak{M})$ are such that $V \subseteq W$, we assign $\underline{\Xi}(\iota_{V,W}) := res_{V,W}$ where $res_{V,W} : \Xi_W \to \Xi_V$ is the restriction in the Gel'fand spectrum:*

$$res_{V,W} : \begin{array}{ccc} \Xi_W & \to & \Xi_V \\ \lambda & \mapsto & \lambda\big|_V \end{array}$$

**Definition 7.1.35** (Presheaves of observables). *Considering the category of presheaves over the quantum observables, $\mathbf{Set}^{\mathfrak{Obs}^{\mathrm{ops}}}$, we define a **spectral presheaf** $\widehat{S} : \mathfrak{C}(\mathfrak{Obs}) \to \mathbf{CH}$, where $\mathfrak{C}(\mathfrak{Obs})$ stands for the category of the contexts of the quantum observables, whose objects are the contexts of the original von Neumann algebra of self-adjoint elements, and whose morphisms are lifted from functional dependencies to set inclusions (as it is a consequence of the spectral decomposition theorem 4.0.17 that if $A = f(B)$ then $A$ commutes with $B$ and on the other hand if $A$ and $B$ commute, we have by the von Neumann-Varadajan theorem A.0.11 that there exists a $f$ for which $A = f(B)$); and $\mathbf{CH}$ is the category of compact Hausdorff spaces.*

*The spectral presheaf is then defined by:*

- *The transformation of elements $\widehat{S}(\mathcal{A}) := S(\mathcal{P}(\mathcal{A}))$, for all objects $\mathcal{A}$ in $\mathfrak{C}(\mathfrak{Obs})$ that are abelian von Neumann sub-algebras of the original total von Neumann algebra in $\mathfrak{Obs}$.*

- *The transformation of morphisms, in this case inclusion morphisms $\iota_{\mathcal{A},\mathcal{B}} : \mathcal{A} \hookrightarrow \mathcal{B}$, $x \mapsto x$ for $\mathcal{A} \subseteq \mathcal{B}$, then $\widehat{S}(\iota_{\mathcal{A},\mathcal{B}}) := \eta_{\mathcal{A}}^{\mathcal{B}}$, where the mapping $\eta_{\mathcal{A}}^{\mathcal{B}} : S(\mathcal{P}(\mathcal{B})) \to S(\mathcal{P}(\mathcal{A}))$ is defined by $\beta \xmapsto{\eta_{\mathcal{A}}^{\mathcal{B}}} \beta \cap \mathcal{A}$.*

*We know from theorem 7.1.31 that there is a canonical homeomorphism $\pi_{\mathcal{A}} : S(\mathcal{P}(\mathcal{A})) \to \Xi_{\mathcal{A}}$. This homeomorphism can be composed with $\widehat{S}$, to get a new contravariant functor $\mathscr{O}(\,\cdot\,) \equiv \pi_{(\,\cdot\,)} \circ \widehat{S}(\,\cdot\,) : \mathfrak{C}(\mathfrak{Obs}) \to \underline{\Xi}$, where $\pi_{\mathcal{A}} \circ \widehat{S}(\mathcal{A}) = \pi_{\mathcal{A}}(S(\mathcal{P}(\mathcal{A}))) = \Xi_{\mathcal{A}}$, and that takes the inclusion morphisms $\iota_{\mathcal{A},\mathcal{B}}$ of $\mathfrak{C}(\mathfrak{Obs})$ to $\pi_{\mathcal{A}} \circ \eta_{\mathcal{A}}^{\mathcal{B}}$, this new presheaf is called the **observable presheaf**.*

**Proposition 7.1.36.** *The homeomorphism $\pi_{\mathcal{A}} : S(\mathcal{P}(\mathcal{A})) \to \Xi_{\mathcal{A}}$ intertwines the ordinary restriction $res_{\mathcal{B},\mathcal{A}} : \Xi_{\mathcal{A}} \to \Xi_{\mathcal{B}}$ on $\underline{\Xi}$:*

$$res_{\mathcal{A},\mathcal{B}}: \Xi_{\mathcal{B}} \to \Xi_{\mathcal{A}}$$
$$\lambda \mapsto \lambda\big|_{\mathcal{A}}$$

with $\eta^{\mathcal{B}}_{\mathcal{A}}$:

$$res_{\mathcal{A},\mathcal{B}} \circ \pi_{\mathcal{B}} = \pi_{\mathcal{A}} \circ \eta^{\mathcal{B}}_{\mathcal{A}}.$$

*Proof:* By considering the full diagram:

$$\begin{array}{ccc} S(\mathcal{P}(\mathcal{A})) & \xleftarrow{\eta^{\mathcal{B}}_{\mathcal{A}}} & S(\mathcal{P}(\mathcal{B})) \\ \pi_{\mathcal{A}} \downarrow & & \downarrow \pi_{\mathcal{B}} \\ \Xi_{\mathcal{A}} & \xleftarrow[res_{\mathcal{A},\mathcal{B}}]{} & \Xi_{\mathcal{B}} \end{array}$$

Then since $\widehat{\mathcal{S}}$ and $\underline{\Xi}$ are presheaves in the category of presheaves, the morphisms between them are necessarily natural transformations, as is the case with $\pi_{(\cdot)}$ since it is a canonical homeomorphism, hence it must obey the naturality condition for contravariant natural transformations, *i.e.* $\pi_{\mathcal{A}} \circ \eta^{\mathcal{B}}_{\mathcal{A}} = res_{\mathcal{A},\mathcal{B}} \circ \pi_{\mathcal{B}}$. ∎

If there was a a global section $s$ of $\mathcal{O}$, the following diagram would commute:

$$\begin{array}{ccc} \mathcal{M} & \xrightarrow{\iota_{\mathcal{M},\mathcal{N}}} & \mathcal{N} \\ s \downarrow & & \downarrow s \\ \Xi_{\mathcal{M}} & \xleftarrow[\mathcal{O}(\iota_{\mathcal{M},\mathcal{N}})]{} & \Xi_{\mathcal{N}} \end{array}$$

For $\mathcal{M}, \mathcal{N} \in \mathfrak{C}(\mathfrak{Obf})$, with $\mathcal{M} \subseteq \mathcal{N}$, $s(\mathcal{M}) \in \Xi_{\mathcal{M}}$ and $s(\mathcal{M}) = \mathcal{O}(\iota_{\mathcal{M},\mathcal{N}})(s(\iota_{\mathcal{M},\mathcal{N}}(\mathcal{M})))$.

Such a choice of one element $s(\mathcal{M})$ of the Gel'fand spectrum $\Xi_{\mathcal{M}}$ per abelian subalgebra $\mathcal{M}$ of $\mathfrak{Obf}$, compatible with the spectral presheaf mappings, *i.e.* with restrictions $\Xi_{\mathcal{N}} \to \Xi_{\mathcal{M}}$, would give a valuation function when restricted to the self-adjoint elements: for all $A \in \mathcal{M}_{\dagger}, s(\mathcal{M})(A) \in \sigma(A)$ and $s(\mathcal{M})(f(A)) = f(s(\mathcal{M})(A))$, for every Borel measurable function $f: \sigma(A) \to \mathbb{R}$. Notwithstanding, the Kochen-Specker theorem 6.1.6 (or 6.1.4) gives us clear limitations for the existence of such valuations when the subjacent Hilbert space of these von Neumann algebras have dimension greater than 2, hence that also means the inexistence of global sections for the respective observable presheaves, or in other words the inexistence of points in those presheaves.

We can make a further refinement to this argument by considering that:

**Lemma 7.1.37.** *Let $v: \mathfrak{B}(\mathcal{H})_{\dagger} \to \mathbb{R}$ be a valuation function in $\mathfrak{B}(\mathcal{H})_{\dagger}$. For every context $\mathbf{C}$ in $\mathfrak{C}(\mathfrak{B}(\mathcal{H}))$, there is a unique character $\varphi^v_{\mathbf{C}} \in \Xi_{\mathbf{C}}$ that coincides with $v$ on the self-adjoint elements of $\mathbf{C}$.*

*Proof:* Let $\mathbf{C}$ be a context in $\mathfrak{C}(\mathfrak{B}(\mathcal{H}))$, and suppose there exists at least one such character $\varphi^v_{\mathbf{C}}$. Then, due to linearity, we have that for all $A \in \mathbf{C}$:

$$\varphi^v_{\mathbf{C}}(A) = \varphi^v_{\mathbf{C}}\left(\frac{1}{2}(A + A^*)\right) + i\varphi^v_{\mathbf{C}}\left(\frac{i}{2}(A^* - A)\right) = \tag{7.1}$$

$$= v\left(\frac{1}{2}(A + A^*)\right) + iv\left(\frac{i}{2}(A^* - A)\right) \tag{7.2}$$

as $\frac{1}{2}(A+A^*)$ and $\frac{i}{2}(A^*-A)$ are self-adjoint. Hence, if $\varphi_{\mathbf{C}}^v$ exists, it is forced by $v$ to be unique, since any other character would have to give the same values as $\varphi_{\mathbf{C}}^v$. To prove existence we use equation (7.2) as a definition of $\varphi_{\mathbf{C}}^v$, only remaining to show that $\varphi_{\mathbf{C}}^v$ is indeed a character.

The definition of valuation function implies that $v(\mathbf{0}) = 0$, by picking $f(x) = 0$ and using $v(f(A)) = f(v(A))$. Similarly, by picking $g(x) = 1$, $v(\mathbb{1}) = 1$. Hence, given $A, B \in \mathbf{C} \cap \mathfrak{B}(\mathscr{H})_\dagger$, it follows directly from equation (7.2) that:

$$\varphi_{\mathbf{C}}^v(A + iB) = \varphi_{\mathbf{C}}^v(A) + i\varphi_{\mathbf{C}}^v(B).$$

Using again the definition of valuation for the function $h(x) = ax$, $\forall\, a \in \mathbb{R}$, $v(aA) = v(h(A)) = h(v(A)) = av(A)$, hence for $A \in \mathbf{C} \cap \mathfrak{B}(\mathscr{H})_\dagger$, $\varphi_{\mathbf{C}}^v(aA) = v(aA) = av(A) = a\varphi_{\mathbf{C}}^v(A)$. Using these last two results, we find that for $\alpha \in \mathbb{C}$, we have:

$$\varphi_{\mathbf{C}}^v(\alpha A) = \varphi_{\mathbf{C}}^v(\mathfrak{Re}(\alpha)A + i\mathfrak{Im}(\alpha)A) =$$
$$= \mathfrak{Re}(\alpha)\varphi_{\mathbf{C}}^v(A) + i\mathfrak{Im}(\alpha)\varphi_{\mathbf{C}}^v(A) =$$
$$= \alpha\varphi_{\mathbf{C}}^v(A).$$

Using the same argument as in equation (7.1), we get that for $A \in \mathbf{C} \cap \mathfrak{B}(\mathscr{H})_\dagger$, $A = \frac{1}{2}(A+A^*) + i\left(\frac{i}{2}(A^*-A)\right)$, then $A^* = \frac{1}{2}(A+A^*) - i\left(\frac{i}{2}(A^*-A)\right)$, hence by equation (7.2) which we are using as the definition of $\varphi_{\mathbf{C}}^v$, we get:

$$\varphi_{\mathbf{C}}^v(A^*) = v\left(\frac{1}{2}(A+A^*)\right) + iv\left(-\frac{i}{2}(A^*-A)\right) =$$
$$= v\left(\frac{1}{2}(A+A^*)\right) - iv\left(\frac{i}{2}(A^*-A)\right),$$

but $v\left(\frac{1}{2}(A+A^*)\right)$ and $v\left(\frac{i}{2}(A^*-A)\right)$ are real values, hence $\overline{v\left(\frac{1}{2}(A+A^*)\right) + iv\left(\frac{i}{2}(A^*-A)\right)} = v\left(\frac{1}{2}(A+A^*)\right) - iv\left(\frac{i}{2}(A^*-A)\right)$, which makes $\varphi_{\mathbf{C}}^v(A^*) = \overline{\varphi_{\mathbf{C}}^v(A)}$.

By the von Neumann-Varadajan theorem A.0.11, we know that for $A, B \in \mathbf{C} \cap \mathfrak{B}(\mathscr{H})_\dagger$, there is a $C \in \mathbf{C}$ for which $A = f(C)$ and $B = g(C)$, for real functions $f : \sigma(C) \to \sigma(A), g : \sigma(C) \to \sigma(B)$, using this we then get:

$$\varphi_{\mathbf{C}}^v(A+B) = v(A+B) = v(f(C)+g(C)) = v((f+g)(C)) = (f+g)v(C) = f(v(C)) + g(v(C)) =$$
$$= v(f(C)) + v(g(C)) = v(A) + v(B) = \varphi_{\mathbf{C}}^v(A) + \varphi_{\mathbf{C}}^v(B).$$

Using then the decomposition of any $K \in \mathbf{C}$ into a sum of self-adjoint elements by equation (7.1), and using the above equalities to recombine these self-adjoint components, and recombining them at the end again with equation (7.1), we get:

$$\varphi_{\mathbf{C}}^v(A+B) = \varphi_{\mathbf{C}}^v(A) + \varphi_{\mathbf{C}}^v(B) \quad \text{and} \quad \varphi_{\mathbf{C}}^v(\alpha A) = \alpha\varphi_{\mathbf{C}}^v(A),$$

for any $A, B \in \mathbf{C}$ and $\alpha \in \mathbb{C}$. Therefore we have proven that $\varphi_{\mathbf{C}}^v : \mathbf{C} \to \mathbb{C}$ is a linear functional.

Considering again the von Neumann-Varadajan theorem A.0.11, for $A, B \in \mathbf{C} \cap \mathfrak{B}(\mathscr{H})_\dagger$, there is a $C \in \mathbf{C}$ for which $A = f(C)$ and $B = g(C)$, for real functions $f : \sigma(C) \to \sigma(A), g : \sigma(C) \to \sigma(B)$, hence:

$$\varphi_{\mathbf{C}}^v(AB) = v(f(C)g(C)) = v((fg)(C)) = (fg)(v(C)) = f(v(C))g(v(C)) =$$
$$= v(f(C))v(g(C)) = v(A)v(B) = \varphi_{\mathbf{C}}^v(A)\varphi_{\mathbf{C}}^v(B).$$

Hence using again the (7.1) decomposition:

$$\varphi_{\mathbf{C}}^v(MN) = \varphi_{\mathbf{C}}^v(M)\varphi_{\mathbf{C}}^v(N),$$

for any $M, N \in \mathbf{C}$, making $\varphi_{\mathbf{C}}^v$ a homomorphism, which together with its compatibility with involution already proven, makes it a $*$-homomorphism, and therefore a character. ∎

A character (multiplicative linear functional) of an abelian $C^*$-algebra $\mathscr{A}$ is a pure state of $\mathscr{A}$. So the above lemma shows that a valuation function induces a pure state on every abelian subalgebra $\mathscr{A} \subseteq \mathfrak{M}$, of a von Neumann algebra $\mathfrak{M}$.

From which we get the following theorem:

**Theorem 7.1.38.** *Let $v$ be a valuation on $\mathfrak{Obs}$ and $\varphi_{\mathbf{C}}^v$ be the unique character on the context $\mathbf{C}$ in $\mathfrak{C}(\mathfrak{Obs})$, induced by means of lemma 7.1.37. Then defining $X_v(\mathbf{C}) := \varphi_{\mathbf{C}}^v$ for each $\mathbf{C}$ in $\mathfrak{C}(\mathfrak{Obs})$ defines a global section of $\underline{\Xi}$. Furthermore, this assignment is bijective.*

*Proof:* For a valuation $v$ on $\mathfrak{Obs}$, we first show that $X_v : \mathfrak{C}(\mathfrak{Obs}) \to \underline{\Xi}(\mathfrak{C}(\mathfrak{Obs}))$ is a presheaf, firstly it does send contexts to characters, whose collection form a set for each valuation, and it will send inclusion morphisms through the contravariant Gel'fand presheaf $\underline{\Xi}$, which makes $X_v$ also contravariant and hence a presheaf.

We shall now prove that $X_v(\mathbf{C})$ is indeed a global section, to do so we consider that for each morphism $\mathbf{C} \xrightarrow{f} \mathbf{D}$ the expression:

$$\varphi_{\mathbf{C}}^v = \underline{\Xi}(f)(\varphi_{\mathbf{D}}^v) = \varphi_{\mathbf{D}}^v\Big|_{\mathbf{C}}$$

holds. It must hold since both $\varphi_{\mathbf{C}}^v$ and $\varphi_{\mathbf{D}}^v\Big|_{\mathbf{C}}$ are characters on $\mathbf{C}$ that extend $v$ to $\mathbf{C}$, and by lemma 7.1.37 such extensions are unique.

We will now prove that this assignment is a bijection. Beginning by proving injectivity, we suppose $v \neq w$ are valuations. Since they are not equal, there is at least one element $A$ in $\mathbf{Obj}_{\mathfrak{Obs}}$ such that $v(A) \neq w(A)$, hence $X_v(\mathcal{C}(A)) \neq X_w(\mathcal{C}(A))$, where $\mathcal{C}(A)$ is the context generated by $A$.

For proving surjectivity we assume that $X$ is a a global section of $\underline{\Xi}$. For each object $A$ in $\mathfrak{Obs}$ and considering characters $\varphi_{\mathbf{C}}$ obtained by $X(\mathbf{C}) = \{\varphi_{\mathbf{C}}\}$, we define $v_X(A) := \varphi_{\mathcal{C}(A)}(A)$.

To show that $v_X$ is indeed a a valuation, we use proposition 7.1.33, that tells us that $\mathcal{C}(\{A\}) = \{f(A) \mid f : \sigma(A) \to \mathbb{R}\}$, as a consequence $\mathcal{C}(\{f(A)\}) \subseteq \mathcal{C}(\{A\})$, meaning there is a unique morphism $f$ in $\mathbf{Mor}_{\mathfrak{C}(\mathfrak{Obs})}(\mathcal{C}(f(A)), \mathcal{C}(A))$ (an inclusion morphism). As by hypothesis $X$ is a global section on $\underline{\Xi}$, that means:

$$\varphi_{\mathcal{C}(\{f(A)\})} = \underline{\Xi}(f)(\varphi_{\mathcal{C}(\{A\})}) = \varphi_{\mathcal{C}(\{A\})}\Big|_{\mathcal{C}(\{f(A)\})}$$

With these remarks in mind, notice that given an object $A$ in $\mathfrak{Obs}$ and a Borel measurable function $f : \mathbb{R} \to \mathbb{R}$,

$$v_X(f(A)) = \varphi_{\mathcal{C}(\{f(A)\})}(f(A)) = \varphi_{\mathcal{C}(\{A\})}\Big|_{\mathcal{C}(\{f(A)\})}(f(A)) = \varphi_{\mathcal{C}(\{A\})}(f(A)),$$

again by proposition 7.1.33, using that $\varphi(f(A)) = f(\varphi(A))$ for a Borel measurable function $f : \mathbb{R} \to \mathbb{R}$ and a character $\varphi \in \Xi_{\mathcal{C}(\{A\})}$, $A \in \mathfrak{V}_\dagger$, we get the following equation continuing from the last:

$$\varphi_{\mathcal{C}(\{A\})}(f(A)) = f(\varphi_{\mathcal{C}(\{A\})}(A)) = f(v_X(A)),$$

Hence $v_X(f(A)) = v_X(f(A))$, making $v_X$ a valuation. If we now use $v_X$ to induce a global section, we'll get back to $X$ due to the fact that lemma 7.1.37 ensures uniqueness of characters extending valuations. Indeed, notice that given an object $A$ in $\mathfrak{Obs}$, we have:

$$\varphi^{v_X}_{\mathcal{C}(\{A\})}(A) = v_X(A) = \varphi_{\mathcal{C}(\{A\})}(A),$$

which ensures $\varphi^{v_X}_{\mathcal{C}(\{A\})}$ and $\varphi_{\mathcal{C}(\{A\})}$ agree on $\mathcal{C}(\{A\})$, since characters commute with the functional calculus and $\mathcal{C}(\{A\}) = \{f(A) \mid f \in B^{\mathscr{B}}(\sigma(A))\}$ as given by proposition 7.1.33. Nonetheless, proposition 7.1.33 also ensures that each context in $\mathfrak{C}(\mathfrak{Obs})$ is generated as $\mathcal{C}(\{B\})$ for some $B \in \mathfrak{Obs}$, and therefore $\varphi^{v_X}_{\mathbf{C}} = \varphi_{\mathbf{C}}$ holds for all contexts $\mathbf{C}$ in $\mathfrak{C}(\mathfrak{Obs})$. ∎

We conclude this chapter by noting that the Kochen-Specker theorem 6.1.6 therefore implies by this last theorem that for von Neumann algebras $\mathfrak{M}$ associated with (by being a closed $*$-invariant subspace of the bounded operators of) a Hilbert space $\mathscr{H}$ with dimension greater than 2, then the Gel'fand presheaf of this von Neumann algebra of observables $\underline{\Xi} : \mathfrak{C}(\mathfrak{M}) \to \mathbf{Set}$ admits no global sections, or, in other words, has no points.

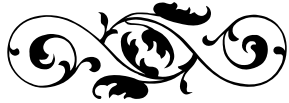

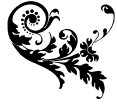 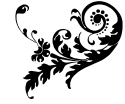

# CHAPTER 8

# Conclusion

Throughout this dissertation we explored several aspects of QMT, departing from aspects that originate from standard non-relativistic QM, such as the interpretative difficulties of the process of measurement, as were originally raised by Wigner in his Gedankenexperiment, for example. This justified the study of the logic of assertions present in QM, which can be obtained by the standard semantic approach of the Tarski-Lindenbaum algebra of the propositional system determined by the elementary assertions present in quantum mechanics. This gives as a result a kind of logic that has several unfamiliar properties from those that would be expected from logic systems used for reasoning, this fact has profound implications when considering quantum systems that contain agents capable of reasoning, as we have seen in the Frauchinger-Renner Gedankenexperiment.

When naïvely incorporating reasoning based on classical logic into quantum systems, we usually obtain contradictions, as we know from the history of quantum mechanics, but to use the constructed quantum logic into the description of Gedankenexperiments would pose several difficulties, the first of which is that the Birkhoff-von Neumann quantum logic can represent only the reasoning of yes-no questions, that is the logic of *probability one* quantum propositions, whereas, in classical logic we have seen that the generalization to assertions about the probability of any event was quite natural, in quantum logic there is no currently known propositional system to deal with assertions that incorporate probabilities within the assertions themselves and how new such assertions could be composed or in other way combined [90].

The study of the contextuality of quantum systems then presents a way to deal with this difficulty with quantum logic, namely that within any context we can reason about the quantum system with classical logic, this is one of the conclusions we can take from the results exposed in chapter 7, where the presheaf structure allows one to think about quantum systems as nontrivial sewings of classical systems, where the non-triviality comes from the fact that this combination, in general, isn't globally a classical system, a consequence of the Kochen-Specker theorem. From this insight we conclude that the supposed inconsistency brought about by the Frauchinger-Renner Gedankenexperiment is actually obstructed by the contextual structure present in the logic that must be used by the agents present in the Gedankenexperiment.

This rephrasing of the compatibility relations of observables into the concept of quantum contexts, ties in with the measurement problem in quantum theories, be it they QM or QFT, as the concept of a "wave function collapse", or reduction of wave packet (whichever way one may prefer this to be referenced), can be seen as an imposition of realism into quantum systems, hence the limitations imposed by the contextual theorems of Kochen-Specker and Kochen-Conway, unexplored in this dissertation, seem to provide strong evidence for concluding that measurement processes really must be achieved by decoherence processes such as those studied in chapter 5, whose main chosen example was in the Coleman-Hepp model for $\frac{1}{2}$-spin measurements.

Of great importance in the analysis of the subject of this dissertation, was the mathematically rigorous framework of AQFT, which allowed several of these problems to be postulated into a, mostly, self-consistent form, and whose relative ease, when compared with other approaches, to be adapted into more general scenarios, such as QFT's in curved spacetimes, likely justify,*a posteriori* the choice of the author in adopting this formalism for the majority of the sections in this dissertation. A point can also be made about the requirement in the FV framework, where the measurement is supposed to occur by a interaction on a compact region, as by the consequence of Haag's theorem, mentioned when solving the Coleman-Hepp model, the descriptions after a measurement is performed cannot be obtained as an isomorphism from the interaction region, only from a equivalent interaction free region before the measurement process actually begins, hence the necessity of the interaction region be compact. We also note that those measurements can only be substantiated when the considered observable and induced probe observable lie in causally disjoint regions, therefore they commute by Einstein causality, and can therefore also be thought to be present within the same context.

We also note in passing that, although more popular when it comes to the description of measurements in QFT's, the Unruh—DeWitt detector models that are usually considered, present several problems to what relates to their well definedness, with examples where the only fields that are non-singular solutions for the equations of motion for the combined system being those in which there is no interaction between the measured system and probe, we refer to [42] for a more precise stipulation of this fact.

It is in the best hopes of the author that this dissertation may have contributed in being a accessible presentation of relatively old results and newer approaches to the theory of quantum measurement, in the AQFT framework, establishing connections between they, and possibly serving to bring closure for the polemic Frauchinger-Renner no-go result, that can be seen as adjunct to the whole project of getting a completely well posed description os measurements in quantum theories.

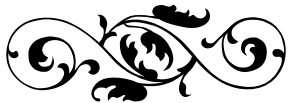

# Appendix A

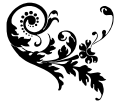 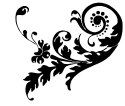

# Algebra and Representation Theory

We shall here prove the equivalence of the $C^*$ condition $\|AA^*\| = \|A\|\|A^*\|$ and the $B^*$ condition $\|AA^*\| = \|A\|^2$. We follow the exposition of [31] of these results proven originally by Araki and Elliott in [5].

**Proposition A.0.1.** *The involution that obeys the $C^*$ condition $\|AA^*\| = \|A\|\|A^*\|$ in a $C^*$-algebra is continuous.*

*Proof:* First we show that the set $\mathscr{A}_\dagger = A \in \mathscr{A} \mid A = A^*$ of hermitian elements in a $C^*$-algebra $\mathscr{A}$ is closed. Let $\{h_n\}_{n \in \mathbb{N}}$ be a convergent sequence in $\mathscr{A}_\dagger$ whose limit is $h + ik$, with $h, k \in \mathscr{A}_\dagger$. For $\widetilde{h}_n = h_n - h$ we have that $\widetilde{h}_n$ converges to $ik$, and also we may assume that $\|\widetilde{h}_n\| \leq 1$. The spectral mapping theorem B.0.42 for polynomials gives:

$$\sigma_{\mathscr{A}}(\widetilde{h}_n^2 - \widetilde{h}_n^4) = \{\lambda^2 - \lambda^4 \mid \lambda \in \sigma_{\mathscr{A}}(\widetilde{h}_n)\},$$

for $x \in \mathscr{A}_\dagger$ we have by theorem B.0.41 that the spectral radius $r(x) = \|x\|$ and by theorem B.0.48 $\sigma_{\mathscr{A}}(x)$ is contained in $\mathbb{R}$. Considering that :

$$\|\widetilde{h}_n^2 - \widetilde{h}_n^4\| = \sup_\lambda\{\lambda^2 - \lambda^4 \mid \lambda \in \sigma_{\mathscr{A}}(\widetilde{h}_n)\} \leq$$

$$\leq \sup_\lambda\{\lambda^2 \mid \lambda \in \sigma_{\mathscr{A}}(\widetilde{h}_n)\} = \|\widetilde{h}_n^2\|$$

Letting $n \to \infty$, we obtain $\| - k^2 - k^4\| \leq \|k^2\|$, and hence:

$$\sup_\lambda\{\lambda^2 + \lambda^4 \mid \lambda \in \sigma_{\mathscr{A}}(k)\} \leq \sup_\lambda\{\lambda^2 \mid \lambda \in \sigma_{\mathscr{A}}(k)\}.$$

Choose $\mu \in \sigma_{\mathscr{A}}(k)$ such that $\mu^2 = \sup_\lambda\{\lambda^2 \mid \lambda \in \sigma_{\mathscr{A}}(\widetilde{h}_n)\}$, then $\mu^2 + \mu^4 \leq \mu^2 \Rightarrow \mu = 0$. It follows that $\|k\| = r(k) = 0$ and so $k = 0$, this shows that $\mathscr{A}_\dagger$ is closed.

Now we show that the graph of the map $x \mapsto x^*$ of $\mathscr{A}$ onto $\mathscr{A}$ is closed. Consider the following argument: Suppose $x_n \to x$ and $x_n^* \to y$, then $x_n + x_n^* \to x + y$ and $\frac{1}{i}(x_n - x_n^*) \to \frac{1}{i}(x - y)$. Since $\mathscr{A}_\dagger$ is closed, $x + y$ and $\frac{1}{i}(x - y)$ have to be hermitian and so $x + y = x^* + y^*$ and $x - y = y^* - x^*$, whence $y = x^*$ and by the closed graph theorem B.0.25, for real Banach spaces, the involution in $\mathscr{A}$ is continuous. ∎

The following proposition gives a motivation for why these two conditions must be equivalent, by showing that they agree on a equivalent norm to that which is immanent to the $C^*$ algebra,the full on equivalence of the conditions for the original norms involves a lengthy and intricate proof, which for brevity's sake we will omit in the assertion immediately the next, only giving the statement of the proposition, a bibliographic reference to the proof, and then utilizing it to establish the equality between those conditions.

**Proposition A.0.2.** *Let $\mathscr{A}$ be a $C^*$-algebra. Then*
$$\|x\|_o = \|x^*x\|^{\frac{1}{2}}$$
*is a equivalent $C^*$-norm on $\mathscr{A}$ such that $\|x^*\|_o = \|x\|_o$ for all $x \in \mathscr{A}$, and $\|h\|_o = \|h\|$ for all self-adjoint $h \in \mathscr{A}$.*

*Proof:* By proposition A.0.1 there exists a constant $M > 1$ such that $\|x^*\| \leq M\|x\|$ for all $x \in \mathscr{A}$, in particular it is true for $x^* \in \mathscr{A}$, that is $\|(x^*)^*\| = \|x\| \leq M\|x^*\|$. Then $M^{-\frac{1}{2}}\|x\|^{\frac{1}{2}} \leq \|x^*\|$, and we have that:
$$M^{-\frac{1}{2}}\|x\| \leq \|x^*\|^{\frac{1}{2}}\|x\|^{\frac{1}{2}} = \|x\|_o \leq M^{\frac{1}{2}}\|x\|$$
so that $\|\cdot\|_o$ and $\|\cdot\|$ are equivalent. Clearly $\|\cdot\|_o$ is homogeneous since $\|\alpha x\|_o = \|\overline{\alpha}x^*\alpha x\|^{\frac{1}{2}} = |\alpha|\|x^*x\|^{\frac{1}{2}}$ and is submultiplicative since $\|xy\|_o = \|y^*x^*xy\|^{\frac{1}{2}} \leq \|y^*\|^{\frac{1}{2}}\|x^*x\|^{\frac{1}{2}}\|y\|^{\frac{1}{2}} = \|x\|_o(\|y^*\|\|y\|)^{\frac{1}{2}} = \|x\|_o\|y^*y\|^{\frac{1}{2}} = \|x\|_o\|y\|_o$. To prove the triangle inequality, let $x, y \in \mathscr{A}$, then:
$$\|x+y\|_o^2 = \|(x+y)^*(x+y)\| \leq \|x^*x\| + \|y^*y\| + \|x^*y + y^*x\|,$$
so it is enough to prove that $\|x^*y + y^*x\| \leq 2\|x\|_o\|y\|_o$. For any positive integer $n$:
$$\left\|(x^*y)^{2^{(n-1)}} + (y^*x)^{2^{(n-1)}}\right\|^2 =$$
$$= \left\|(x^*y)^{2^n} + (y^*x)^{2^n} + (x^*y)^{2^{(n-1)}}(y^*x)^{2^{(n-1)}} + (y^*x)^{2^{(n-1)}}(x^*y)^{2^{(n-1)}}\right\|,$$
and since
$$\left\|(x^*y)^{2^{(n-1)}}(y^*x)^{2^{(n-1)}} + (y^*x)^{2^{(n-1)}}(x^*y)^{2^{(n-1)}}\right\| \leq \left\|(x^*y)^{2^{(n-1)}}(y^*x)^{2^{(n-1)}}\right\| + \left\|(y^*x)^{2^{(n-1)}}(x^*y)^{2^{(n-1)}}\right\| \leq$$
$$\leq 2(\|x^*\|\|x\|\|y^*\|\|y\|)^{2^{n-1}} = 2(\|x^*x\|\|y^*y\|)^{2^{n-1}}$$
then:
$$\left\|(x^*y)^{2^{(n-1)}} + (y^*x)^{2^{(n-1)}}\right\|^2 \leq \left\|(x^*y)^{2^n} + (x^*y)^{2^n}\right\| + 2(\|x^*x\|\|y^*y\|)^{2^{n-1}} \tag{A.1}$$

By the spectral radius theorem B.0.41 $\lim_{n \to \infty} \|u^n\|^{\frac{1}{n}} = r(u)$, that is, for any $\varepsilon > 0$ there is a $n_0(\varepsilon) \in \mathbb{N}$ such that $\|u^{n_0}\|^{\frac{1}{n_0}} \leq r(u) + \varepsilon$, using the same logic there is a big enough $n_0 \in \mathbb{N}$ such that $\left\|(u^2)^{2^{n-1}}\right\|^{\frac{1}{2^{n-1}}} \leq r(a)^2 + \varepsilon$ applying this to $\left\|(x^*y)^{2^n}\right\|$ and $\left\|(y^*x)^{2^n}\right\|$ we get:
$$\left\|(x^*y)^{2^n}\right\| \leq (r(x^*y)^2 + \varepsilon)^{2^{n-1}} \quad \text{and} \quad \left\|(y^*x)^{2^n}\right\| \leq (r(y^*x)^2 + \varepsilon)^{2^{n-1}}.$$

Then by theorem B.0.31 the spectral radii $r(kq) = r(qk)$ and remembering that $(u^*)^{-1} = (u^{-1})^*$, which makes $\sigma(u^*) = \{\overline{\lambda} \in \mathbb{C} \mid \lambda \in \sigma(u)\}$, and the definition B.0.39 of spectral radius implies $r(u^*) = r(u)$, then all of this together gives $r(x^*y) = r(yx^*) = r(xy^*) = r(y^*x)$ and substituting this into the inequality:
$$\left\|(x^*y)^{2^n}\right\| \leq (r(x^*y)r(y^*x) + \varepsilon)^{2^{n-1}} \leq (\|x^*y\|\|y^*x\| + \varepsilon)^{2^{n-1}} \leq$$

$$\leq (\|x^*x\|\|y^*y\| + \varepsilon)^{2^{n-1}}$$

and analogously:

$$\|(y^*x)^{2^n}\| \leq (\|x^*x\|\|y^*y\| + \varepsilon)^{2^{n-1}},$$

so that:

$$\|(x^*y)^{2^n} + (y^*x)^{2^n}\| \leq 2(\|x^*x\|\|y^*y\| + \varepsilon)^{2^{n-1}} \tag{A.2}$$

Beginning with (A.2) and applying (A.1) recursively we obtain:

$$\left\|(x^*y)^{2^{k-1}} + (y^*x)^{2^{k-1}}\right\|^2 \leq 4(\|x^*x\|\|y^*y\| + \varepsilon)^{2^{k-1}}$$

for any $k$, $1 \leq k \leq n+1$. Thus in particular for an arbitrary $\varepsilon > 0$:

$$\|x^*y + y^*x\|^2 \leq 4\left(\|x^*x\|\|y^*y\| + \varepsilon\right)$$

Hence $\|x*y + y*x\|^2 \leq 2\|x\|_o\|y\|_o$. So we have seen that $\|\cdot\|_o$ is an equivalent algebra norm on $\mathscr{A}$. Furthermore, $\|h\|_o = \|h^*h\|^{\frac{1}{2}} = \|h\|$ for all hermitian $h \in \mathscr{A}$ and so $\|x\|_o^2 = \|x*x\| = \|x*x\|_o$; i.e., $\|\cdot\|_o$ is a $C^*$-norm on $\mathscr{A}$ with $\|x^*\|_o = \|x\|_o$ for all $x \in \mathscr{A}$. ∎

The following proof is rather extensive, including a version of the Russo-Dye theorem and therefore we shall only enunciate the statement of the proposition and refer the reader to [31] proposition (15.1) pages 42-45 for a proof.

**Proposition A.0.3.** *Let $\mathscr{A}$ be a $C^*$-algebra with identity $\mathbb{1}$. Considering $U(\mathscr{A})$ denote the group of unitary elements in $\mathscr{A}$. Then every element $x$ in $\mathscr{A}$ is a linear combination of unitary elements and $\|x\| = \|x\|_u$, where*

$$\|x\|_u = \inf\left\{\sum_{i=1}^N |\lambda_i| \,\bigg|\, \sum_{i=1}^N \lambda_i u_i,\, \lambda_i \in \mathbb{C},\, u_i \in U(\mathscr{A}),\, \forall i = 1, \ldots, N\right\}.$$

**Theorem A.0.4.** *The involution in a $C^*$-algebra is isometric. This makes $\|x^*\| = \|x\|$ and therefore $\|x^*\|\|x\| = \|x\|^2$ which makes the $C^*$ condition equivalent to the $B^*$ condition.*

*Proof:* This follows directly form proposition A.0.3 that establishes $\|x\| = \|x\|_u, \forall x \in \mathscr{A}$ and the trivial observation that $\|x^*\|_u = \|x\|_u$, i.e. the involution in a $C^*$-algebra is isometric. Therefore $\|xx^*\| = \|x\|\|x^*\| = \|x\|^2$. ∎

In the following we shall establish the required results to prove the von Neumann-Varadarajan theorem A.0.11, the generally know theorem ws proved by von Neumann in [95] in german, in the year 1931 for operators in a Hilbert space, the result following from a application of the spectral theorem, although after that it seems to have become somewhat of a folk theorem not being proved in many references but only being alluded to, in the case of [67] it cites a difficult to find paper of Naĭmark [79] from 1954 also in german of which the desired result is theorem number 6, apparently; Mackey in [76] just cites this result as von Neumann's theorem, and by all means it should be called that, but as since V. S. Varadarajan published a generalization of this theorem in 1962 on the paper [97], this allows we to increase specificity of the theorem being used by calling it the von Neumann-Varadarajan theorem, as did Moretti in [77].

The approach that we used to prove the von Neumann-Varadarajan theorem follows the simplified version and exposition in [63].

**Definition A.0.5** ($\sigma$-homomorphism). *A $\sigma$-homomorphism is a mapping from $\mathscr{B}(\mathbb{R})$, the Borel sets of the real line, into a bounded orthocomplemented lattice $\mathcal{L}$, being $t : \mathscr{B}(\mathbb{R}) \to \mathcal{L}$ any such homomorphism, then:*

(i) $t(\varnothing) = \mathbf{0}, \quad t(\mathbb{R}) = \mathbf{1}$;

(ii) $\Delta_1 \cap \Delta_2 = \varnothing$ *implies* $t(\Delta_1) \perp t(\Delta_2)$;

(iii) *for any collection $\{\Delta_i\}_{i \in I}$ such that $\Delta_i \cap \Delta_j = \varnothing$, $\forall i \neq j$, then:*

$$t\left(\bigcup_{i \in I} \Delta_i\right) = \bigvee_{i \in I} t(\Delta_i).$$

*The range of a $\sigma$-homomorphism $t : \mathscr{B}(\mathbb{R}) \to \mathcal{L}$ is given by $\mathrm{Ran}(t) = \{b \in \mathcal{L} \,|\, b = t(\Delta), \Delta \in \mathscr{B}(\mathbb{R})\}$.*

**Proposition A.0.6.** *The range of a $\sigma$-homomorphism is a Boolean sub-lattice of its codomain lattice $\mathcal{L}$.*

*Proof:* By construction the range of any such $\sigma$-homomorphism $t$ is bounded since $t(\varnothing) = \mathbf{0} \in \mathrm{Ran}(t)$ and $t(\mathbb{R}) = \mathbf{1} \in \mathrm{Ran}(t)$. It is orthocomplemented since it is bounded and we have a orthogonal complement map $\neg : \mathcal{L} \to \mathcal{L}$ given by $\neg t(\Delta) := t(\mathbb{R} \setminus \Delta), \forall \Delta \in \mathscr{B}(\mathbb{R})$, then $t(\Delta) \vee \neg t(\Delta) = t(\Delta) \vee t(\mathbb{R} \setminus \Delta) \stackrel{(iii)}{=} t(\Delta \cup \mathbb{R} \setminus \Delta) = t(\mathbb{R}) = \mathbf{1}$, and $t(\Delta) \wedge \neg t(\Delta) = \neg(\neg t(\Delta) \vee t(\Delta)) = \neg(t(\mathbb{R} \setminus \Delta) \vee t(\Delta)) \stackrel{(iii)}{=} \neg t(\mathbb{R}) = t(\varnothing) = \mathbf{0}$.

Clearly $\neg(\neg t(\Delta)) = \neg t(\mathbb{R} \setminus \Delta) = t(\mathbb{R} \setminus (\mathbb{R} \setminus \Delta)) = t(\Delta)$.

Considering then that if $t(\Delta_1) \preccurlyeq t(\Delta_2) \Rightarrow t(\Delta_1) \vee t(\Delta_2) = t(\Delta_2)$, then by the absorption property $t(\Delta_1) = t(\Delta_1) \wedge (t(\Delta_1) \vee t(\Delta_2)) = t(\Delta_1) \wedge t(\Delta_2)$, therefore we have that $\neg t(\Delta_1) \vee \neg t(\Delta_2) = \neg(t(\Delta_1) \wedge t(\Delta_2)) = \neg t(\Delta_1)$, which implies that $\neg t(\Delta_2) \preccurlyeq \neg t(\Delta_1)$.

$\mathrm{Ran}(t)$ is distributive since we have that $t(A) \vee t(B) = t((A \setminus B) \cup (A \cap B)) \vee t((B \setminus A) \cup (A \cap B)) \stackrel{(iii)}{=} (t(A \setminus B) \vee t(A \cap B)) \vee (t(B \setminus A) \vee t(A \cap B)) = t(A \setminus B) \vee (t(A \cap B) \vee t(A \cap B)) \vee t(B \setminus A) = t(A \setminus B) \vee t(A \cap B) \vee t(B \setminus A) = t((A \setminus B) \cup (B \setminus A)) \vee t(A \cap B) = t(A \cup B), \forall A, B \in \mathscr{B}(\mathbb{R})$ and since by De Morgan's Law $t(A) \wedge t(B) = \neg(\neg t(A) \vee \neg t(B)) = \neg(t(\mathbb{R} \setminus A) \vee t(\mathbb{R} \setminus B)) = \neg(t((\mathbb{R} \setminus A) \cup (\mathbb{R} \setminus B))) = \neg(t(\mathbb{R} \setminus (A \cap B))) = t(A \cap B)$ then $\forall \Delta_1, \Delta_2, \Delta_3 \in \mathscr{B}$ :

$$t(\Delta_1) \wedge (t(\Delta_2) \vee t(\Delta_3)) = t(\Delta_1) \wedge t(\Delta_2 \cup \Delta_3) = t(\Delta_1 \cap (\Delta_2 \cup \Delta_3)) =$$

$$= t((\Delta_1 \cap \Delta_2) \cup (\Delta_1 \cap \Delta_3)) = t(\Delta_1 \cap \Delta_2) \vee t(\Delta_1 \cap \Delta_3) = (t(\Delta_1) \wedge t(\Delta_2)) \vee (t(\Delta_1) \wedge t(\Delta_3)).$$

∎

**Definition A.0.7** (Separable $\sigma$-algebra). *A $\sigma$-algebra $\Sigma$ is said to be **separable** if there exists a countable set $\mathcal{D} \subset \Sigma$ such that the smallest sub $\sigma$-algebra of $\Sigma$ containing $\mathcal{D}$ is $\Sigma$ itself. $\mathcal{D}$ is said to generate $\Sigma$.*

*This makes $\Sigma$ be **separable** in the topological sense if for any finite measure $\mu$ of $(X, \Sigma)$ we consider the metric topology generated by the metric $d(A, B) = \mu((A \setminus B) \cup (B \setminus A)) \,\forall A, B \in \Sigma$.*

**Proposition A.0.8.** *Let $U$ be a space and $\mathcal{U}$ be a separable $\sigma$-algebra of subsets of $U$. Then there exists a real valued $\mathcal{U}$-measurable function $f : (U, \mathcal{U}) \to \mathbb{R}$ on $U$ such that:*

$$\mathcal{U} = \{f^{-1}(E) \,|\, E \in \mathscr{B}(\mathbb{R})\}$$

*Proof:* Let $\{D_n\}_{n\in\mathbb{N}} \subset \wp(\mathcal{U})$ generate $\mathcal{U}$ and let $\chi_{D_n}: U \to \{0,1\}$ be the characteristic functions associated with each $D_n$. Then the map $\gamma: U \to \bigtimes_{n\in\mathbb{N}}[0,1]$ given by $U \ni u \xmapsto{\gamma} (\chi_{D_1}(u), \chi_{D_2}(u), ...) \in \bigtimes_{n\in\mathbb{N}}\{0,1\}$ is a $\mathcal{U}$-measurable map of $U$. If $\mathscr{B}\left(\bigtimes_{n\in\mathbb{N}}[0,1]\right)$ is the class of Borel subsets of $\bigtimes_{n\in\mathbb{N}}[0,1]$, it is obvious that $\mathcal{U} = \left\{\gamma^{-1}(E) \,|\, E \in \mathscr{B}\left(\bigtimes_{n\in\mathbb{N}}[0,1]\right)\right\}$. Since by the pointed out implication of theorem [69] there is a isomorphism between $\bigtimes_{n\in\mathbb{N}}[0,1]$ and $\mathbb{R}$, say $\beta: \bigtimes_{n\in\mathbb{N}}[0,1] \to \mathbb{R}$ which preserves the Borel structures on the two spaces, it follows that $f \equiv \beta \circ \gamma$ is a real valued $\mathcal{U}$-measurable function on $U$ such that $\mathcal{U} = \{f^{-1}(E) \,|\, E \in \mathscr{B}(\mathbb{R})\}$. ∎

**Theorem A.0.9.** *A Boolean sub-lattice $\mathcal{R}$ of a lattice $\mathcal{L}$ is separable if and only if there exists an $\sigma$-homomorphism $t$ such that $\mathcal{R} = \mathrm{Ran}(t)$.*

*Proof:* Let $t: \mathscr{B}(\mathbb{R}) \to \mathcal{R}$ be a $\sigma$-homomorphism of $\mathscr{B}(\mathbb{R})$ onto $\mathcal{R}$ and consider the set $\mathcal{D} = \{t((-\infty,q)) \,|\, q \in \mathbb{Q}\}$. Suppose $\mathcal{R}_0$ is the sub $\sigma$-algebra of $\mathcal{R}$ generated by $\mathcal{D}$, since the generating set $\mathcal{D}$ is composed of elements that can be indexed by rational numbers, which are countable, then these are dense in $\mathcal{R}_0$ which is therefore separable. Moreover since $\{E \in \mathscr{B}(\mathbb{R}) \,|\, t(E) \in \mathcal{R}_0\}$ is a $\sigma$-algebra of Borel sets of $\mathbb{R}$ including all intervals $(-\infty,q)$, for $q \in \mathbb{Q}$, it then includes all Borel sets and hence $\mathcal{R}_0$ is the range of $t$. This proves that $\mathcal{R} = \mathcal{R}_0 = \mathrm{Ran}(t)$.

Conversely, let $\mathcal{R}$ be separable and $\mathcal{D}_0 \subset \mathcal{R}$ any countable collection that generates $\mathcal{R}$. By the Loomis–Sikorski representation theorem C.0.17, there is a measurable space $(X, \Sigma)$ and a $\sigma$-ideal $\mathcal{N}$ such that there is a $\sigma$-algebra isomorphism $u: \Sigma/_{\sim_\mathcal{N}} \to \mathcal{R}$. Let $C_1, C_2, ...$ be subsets of $\Sigma/_{\sim_\mathcal{N}}$ such that $\mathcal{D}_0 = \{u(C_j)\}_{j=1,2,...}$ and let $\Sigma_0$ be the $\sigma$-algebra of subsets of $X$ generated by the $\{C_j\}_{j\in\mathbb{N}}$. $\Sigma_0$ is separable and hence by proposition A.0.8 there exists a real valued $\Sigma_0$-measurable function $f$ such that $\Sigma_0 = \{f^{-1}(E) \,|\, e \in \mathscr{B}(\mathbb{R})\}$.

If now $h(E) := u(f^{-1}(E))$ for any $E \in \mathscr{B}(\mathbb{R})$, then $h$ is obviously a $\sigma$-homomorphism of $\mathscr{B}(\mathbb{R})$ into $\mathcal{R}$. Since the range of $h: \mathscr{B}(\mathbb{R}) \to \mathcal{R}$ includes all the $u(C_j)$, $h$ is onto $\mathcal{R}$. ∎

**Theorem A.0.10.** *If there are two $\sigma$-homomorphisms $x$ and $y$ such that $\mathrm{Ran}(y) \subseteq \mathrm{Ran}(x)$, then there exists a Borel function $u: \mathbb{R} \to \mathbb{R}$ such that $y(\Delta) = x(u^{-1}(\Delta))$ for all $\Delta \in \mathscr{B}(\mathbb{R})$.*

*Proof:* Let $(r_n)_{n\in\mathbb{N}}$ be an enumeration of the rational numbers. Set

$$b_n = y((-\infty, r_n)).$$

We shall show below that it is possible to construct Borel sets $\Delta_n$ with the properties:

(i) $b_n = x(\Delta_n)$;

(ii) $\Delta_i \subseteq \Delta_j$ whenever $i < j$;

(iii) $x\left(\bigcup_{n\in\mathbb{N}} \Delta_n\right) = \mathbf{1}$.

These sets will be instrumental in proving the relation $y(\Delta) = x(u^{-1}(\Delta))$, as will be seen later.

We shall construct these sets by induction. The base case for induction is trivial by the hypothesis, and therefore $\exists \Delta_1 \in \mathscr{B}(\mathbb{R})$. Let $\Delta_1, \Delta_2, ..., \Delta_n$ be so constructed that they satisfy conditions *(i)* and *(ii)*, and let $\Pi(1), \Pi(2), \ldots, \Pi(n)$ be a permutation of $1, 2, \ldots, n$ such that $r_{\Pi(1)} < r_{\Pi(2)} < \ldots < r_{\Pi(n)}$. We now use the hypothesis of the theorem. Since $b_{n+1} \in \text{Ran}(y)$ it follows from the hypothesis $\text{Ran}(y) \subseteq \text{Ran}(x)$ that $b_{n+1} \in \text{Ran}(x)$ also. Thus there exists a $\Delta \in \mathscr{B}(\mathbb{R})$ such that $b_{n+1} = x(\Delta)$. We now distinguish three cases:

1) $r_{n+1} < r_{\Pi(n)}$;

2) there exists a $k \in \mathbb{N}$ such that $1 \leq k \leq n$ and $r_{\Pi(k)} < r_{n+1} < r_{\Pi(k+1)}$;

3) $r_{\Pi(n)} < r_{n+1}$

We define,

- In case 1), $\Delta_{n+1} := \Delta \cap \Delta_{\Pi(n)}$,

- in case 2), $\Delta_{n+1} := \Delta_{\Pi(k)} \cup (\Delta \cap \Delta_{\Pi(k+1)})$,

- in case 3), $\Delta_{n+1} := \Delta_{\Pi(n)} \cup \Delta$,

so that properties $(i)$ and $(ii)$ are also satisfied for the sets $\Delta_1, \Delta_2, \ldots, \Delta_{n+1}$ We find

$$x\left(\bigcup_{n \in \mathbb{N}} \Delta_n\right) = \bigvee_{n \in \mathbb{N}} x(\Delta_n) = \bigvee_{n \in \mathbb{N}} b_n = \bigvee_{n \in \mathbb{N}} y((-\infty, r_n)) = \mathbf{1}$$

so that property $(iii)$ is also verified. This finishes the construction of the sets of interest.

Now, to prove the theorem:

Denote by $X_0 \equiv \bigcup_{n \in \mathbb{N}} \Delta_n$ the union of all these sets $\Delta_n$. For any $t \in X_0$ define

$$\tilde{u}(t) = \inf \{r_n \mid t \in \Delta_n\}$$

We have then $\{t \mid \tilde{u}(t) < s\} = \bigcup_{\substack{n \in \mathbb{N} \\ r_n < s}} \Delta_n$, so that $\tilde{u}(t)$ is a Borel function. If we define:

$$u(t) = \begin{cases} \tilde{u}(t) & \text{, for } t \in X_0, \\ 0 & \text{, for } t \in X_0^{\complement}. \end{cases}$$

then

$$\{t \mid u(t) < s\} = \begin{cases} \{t \mid \tilde{u}(t) < s\} & \text{, for } s \leq 0, \\ \{t \mid \tilde{u}(t) < s\} \cup X_0^{\complement} & \text{, for } s > 0. \end{cases}$$

so that $u(t)$ is also a Borel function. Furthermore

$$x(\{t \mid u(t) < s\}) = \bigvee_{\substack{n \in \mathbb{N} \\ r_n < s}} x(\Delta_n) = \bigvee_{\substack{n \in \mathbb{N} \\ r_n < s}} b_n = \bigvee_{\substack{n \in \mathbb{N} \\ r_n < s}} y((-\infty, r_n)) =$$

$$= y((-\infty, s)) \text{, for any real } s.$$

Since the sets $(-\infty, s)$ generate all the Borel sets, we conclude

$$x(u^{-1}(\Delta))) = y(\Delta), \text{ for all } \Delta \in \mathscr{B}(\mathbb{R}).$$

Thus the function $u(t)$ does everything that is required of it, and thereby completes the proof of the theorem. ∎

**Theorem A.0.11** (von Neumann-Varadarajan). *Let $\{x_i\}_{i \in I}$ be a family of compatible observables (i.e. $\sigma$-homomorphisms) in a separable lattice $\mathcal{L}$ so that $x_i$ is compatible (i.e. commutes) with $x_j$ for all $i, j \in I$. Then there exist an observable $x$ and Borel functions $u_i : \mathbb{R} \to \mathbb{R}$ such that $x_i = u_i(x)$ functionally.*

*Proof:* Let $\mathfrak{B}(\operatorname{Ran}(x_i)) := \bigcap \{R \in \wp(\mathcal{L}) \,|\, \operatorname{Ran}(x_i) \subseteq R, R \text{ is a Boolean subalgebra of } \mathcal{L}\}$ be the smallest Boolean sub-lattice generated by the ranges $\operatorname{Ran}(x_i)$ of the observables $x_i$. Since $\mathcal{L}$ is separable, $\mathfrak{B}(\operatorname{Ran}(x_i))$ is separable too. By the theorem A.0.9 there exists an observable $x$ such that $\mathfrak{B}(\operatorname{Ran}(x_i)) = \operatorname{Ran}(x)$.

Clearly $\operatorname{Ran}(x_i) \subseteq \operatorname{Ran}(x)$ and the theorem A.0.10 then tells us that there exist Borel functions $\tilde{u}_i : \mathbb{R} \to \mathbb{R}$ such that $x_i = x \circ \tilde{u}_i^{-1}$. This proves everything we needed since by defining the expression $u_i(x)$ in the statement of the theorem to be defined as $u_i(x) := x \circ \tilde{u}_i^{-1}$, since if an observable $x$ yields the value $\xi$, then it's composition with a Borel function $u(x)$ would need to yield $u(\xi)$ and therefore the value $u(\xi)$ can only be a element of a Borel set $\Delta$ if $\xi \in u^{-1}(\Delta)$ which makes the natural definition of the observable $u(x)$ to be the homomorphism which sends the set $\Delta \in \mathscr{B}(\mathbb{R})$ into $x(u^{-1}(\Delta))$. ∎

**Definition A.0.12** (Positive map). *Let $\mathcal{A}$ and $\mathcal{B}$ be $C^*$-algebras and $\phi : \mathcal{A} \to \mathcal{B}$ a linear map. $\phi$ is said to be a **positive map** if it maps positive elements $a_+ \in \mathcal{A}$ into positive elements $b_+ \in \mathcal{B}$, where positive elements of a $C^*$-algebra are defined as those elements $c_+$ for which there exists a finite collection $\{c_k\}_{k=1}^n \subsetneq \mathscr{C}$ for which $c_+ = \sum_{k=1}^n c_k^* c_k$.*

**Definition A.0.13** (Complete positive map). *Let $\mathcal{A}$ and $\mathcal{B}$ be $C^*$-algebras and $\phi : \mathcal{A} \to \mathcal{B}$ a linear map. Then the induced map $\mathbb{1}_{M_n(\mathbb{C})} \otimes \phi : M_n(\mathbb{C}) \otimes \mathcal{A} \to M_n(\mathbb{C}) \otimes \mathcal{B}$ defined by:*

$$M_n(\mathbb{C}) \otimes \mathcal{A} \ni \begin{pmatrix} a_{11} & \ldots & a_{1n} \\ \vdots & \ddots & \vdots \\ a_{n1} & \ldots & a_{nn} \end{pmatrix} \xmapsto{\mathbb{1}_{M_n(\mathbb{C})} \otimes \phi} \begin{pmatrix} \phi(a_{11}) & \ldots & \phi(a_{1n}) \\ \vdots & \ddots & \vdots \\ \phi(a_{n1}) & \ldots & \phi(a_{nn}) \end{pmatrix} \in M_n(\mathbb{C}) \otimes \mathcal{B}.$$

*$\phi$ is then called n-positive if $\mathbb{1}_{M_n(\mathbb{C})} \otimes \phi$ is a positive map, and **completely positive** if $\phi$ is n-positive for all $n \in \mathbb{N}$.*

## A.1　Representation Theory

To count as a repository of definitions for the uninitiated, in here we define the basic elements of representation theory:

**Definition A.1.1** (Magma). *A magma $(\mathbf{M}, \cdot)$ is a a set $\mathbf{M}$ matched with an binary operation $\cdot : \mathbf{M} \times \mathbf{M} \to \mathbf{M}$, such that $\forall\, a, b \in \mathbf{M}$, $a \cdot b \in \mathbf{M}$.*

**Definition A.1.2** (Representation). *A representation $(\mathcal{V}, \pi)$, is composed of a vector space $\mathcal{V}$ and a map $\pi : \mathbf{M} \to \operatorname{End}_{\mathbb{E}}(\mathcal{V})$ that sends elements $m$ from a magma $(\mathbf{M}, \cdot)$ to a linear transformation $\pi(m) : \mathcal{V} \to \mathcal{V}$, i.e. an endomorphism of the vector space $\mathcal{V}$ over some algebra subfield $\mathbb{E}$ of the algebra field over which $\mathcal{V}$ is defined, such that:*

$$\pi(m \cdot n) = \pi(m) \circ \pi(n).$$

*Virtually, any useful representation lies in at least one of the following:*

- *A **group representation** $(\mathcal{V}, \pi)$, is such that $\pi: G \to \mathrm{GL}(\mathcal{V}, \mathbb{F})$ is a representation from a group $G$ to the General linear group of the vector space $\mathcal{V}$ over the algebra field $\mathbb{F}$, $\mathrm{GL}(\mathcal{V}, \mathbb{F})$, and is formed by the set of all bijective linear transformations over the algebra subfield $\mathbb{F}$ of $\mathcal{V}$ to itself, together with functional composition as the group operation, making the group representation an group homomorphism.*

- *A **associative algebra representation** $(\mathcal{V}, \pi)$, is such that $\pi: \mathcal{A} \to \mathrm{End}_\mathbb{F}(\mathcal{V})$ is a representation from a associative algebra $\mathcal{A}$ to the endomorphism of $\mathcal{V}$ over $\mathbb{F}$, a algebra subfield of the algebra field over which $\mathcal{A}$ is an algebra, making $\mathrm{End}_\mathbb{F}(\mathcal{V})$ itself a associative algebra under linear functional composition seen as a bilinear product, making the associative algebra representation an algebra homomorphism. A recurring subcase of associative algebra representations in this dissertation is the following:*

    - *A **Banach $*$-algebra representation** $(\mathscr{H}, \pi)$ (also called a $*$-representation for short), is such that $\pi: \mathscr{A} \to \mathfrak{B}(\mathscr{H})$ is a $*$-morphism[1] between some Banach $*$-algebra $\mathscr{A}$ and $\mathfrak{B}(\mathscr{H})$ such that, $\forall a, b \in \mathscr{A}$: $\pi(a^*) = \pi(a)^*$, $\pi(ab) = \pi(a)\pi(b)$ and $\pi(\alpha a + \beta b) = \alpha \pi(a) + \beta \pi(b)$.*

- *A **Lie algebra representation** $(\mathcal{V}, \pi)$, is such that $\pi: \mathfrak{a} \to \mathfrak{gl}(\mathcal{V}, \mathbb{F})$ is a representation from a Lie algebra $\mathfrak{a}$ to the General linear Lie algebra of a vector space $\mathcal{V}$ over $\mathbb{F}$, a algebra subfield of the algebra field over which $\mathcal{A}$ is an algebra, $\mathfrak{gl}(\mathcal{V}, \mathbb{F})$, and is formed by the set of all bijective linear transformations over the algebra subfield $\mathbb{F}$ of $\mathcal{V}$ to itself, such that, $\forall a, b \in \mathfrak{a}$:*

$$\pi([a,b]) = [\pi(a), \pi(b)] = \pi(a)\pi(b) - \pi(b)\pi(a), \quad \text{and} \quad \pi(\alpha a + \beta b) = \alpha \pi(a) + \beta \pi(b),$$

*making the Lie algebra representation an Lie algebra homomorphism.*

*At times the map $\pi$ is by itself called a representation leaving the vector space $\mathcal{V}$ implicitly understood, this can be somewhat confusing when treating subrepresentations, but since it is common practice we also adopt it in the body of this dissertation, for clarity in this appendix we opt to be explicit about the vector spaces.*

**Definition A.1.3** (Subrepresentation). *If $(\mathcal{V}, \pi)$ is a representation of a magma $\mathbf{M}$, and $\mathcal{W}$ is a linear subspace of $\mathcal{V}$ **stable under** $\mathbf{M}$, i.e. $\forall m \in \mathbf{M}, \forall w \in \mathcal{W}, \pi(m)w \in \mathcal{W}$; then $(\mathcal{W}, \pi\!\upharpoonright_\mathcal{W})$ is a **subrepresentation** of $(\mathcal{V}, \pi)$, where the map $\pi\!\upharpoonright_\mathcal{W}: \mathbf{M} \to \mathrm{End}_\mathbb{E}(\mathcal{W})$ is such that for any $m \in \mathbf{M}$, $\pi\!\upharpoonright_\mathcal{W}(m) := \pi(m)\Big|_\mathcal{W}$.*

**Proposition A.1.4.** *A subspace $\mathcal{W} \subset \mathcal{V}$ is invariant under the map $\pi: \mathbf{M} \to \mathrm{End}_\mathbb{F}(\mathcal{V})$, if and only if, $\forall A \in \mathbf{M}$, and for a projector $P_\mathcal{W}: \mathcal{V} \to \mathcal{W}$ onto $\mathcal{W}$:*

$$\pi(A) P_\mathcal{W} = P_\mathcal{W} \pi(A).$$

*Proof:* Clearly if $\mathcal{W}$ is invariant under $\pi$:

$$P_\mathcal{W} \pi(A) P_\mathcal{W} = \pi(A) P_\mathcal{W},$$

---

[1] We call this only a $*$-morphism instead of a full $*$-homomorphism, even though it is a algebra homomorphism, because it doesn't include any information about the norm in $\mathscr{A}$ being equivalent to the operator norm in $\mathfrak{B}(\mathscr{H})$.

for all $A \in \mathbf{M}$, since the $\pi(A)$'s cannot map vectors outside $\mathcal{W}$ by hypothesis, hence $\forall A \in \mathbf{M}$:

$$\pi(A)P_{\mathcal{W}} = (P_{\mathcal{W}}\pi(A)^*P_{\mathcal{W}})^* = (\pi(A^*)P_{\mathcal{W}})^* = P_{\mathcal{W}}\pi(A).$$

∎

**Definition A.1.5** (Irreducible and reducible representations). *If the representation $(\mathcal{V}, \pi)$ has exactly two subrepresentations, namely $(\{\mathbf{0}\}, \pi\!\restriction_{\{\mathbf{0}\}})$, where $\{\mathbf{0}\}$ is the trivial null subspace for which $\pi\!\restriction_{\{\mathbf{0}\}}$ is a null map, and $(\mathcal{V}, \pi)$ itself, then the representation is said to be irreducible; if $(\mathcal{V}, \pi)$ has a proper nontrivial subrepresentation, the representation is said to be reducible.*

**Definition A.1.6** (Faithful representation). *A faithful representation is a representation $(\mathcal{V}, \pi)$, for which the map $\pi$ is injective, for algebras, this is equivalent to $\ker \pi = \{\mathbf{0}\}$, since the map $\pi$ is then taken to be linear on the algebra.*

**Definition A.1.7** (Direct sum representation). *If $\{(\mathcal{V}_\alpha, \pi_\alpha)\}_{\alpha \in I}$ is a family of representations of the same magma $(\mathbf{M}, \cdot)$, then there is a representation $\left(\bigoplus_\alpha \mathcal{V}_{\alpha \in I}, \bigoplus_{\alpha \in I} \pi_\alpha\right)$ acting on the direct sum $\bigoplus_{\alpha \in I} \mathcal{V}_\alpha$, whose map is denoted by $\pi \equiv \bigoplus_{\alpha \in I} \pi_\alpha$ and defined by:*

$$\pi(m)\left(\underset{\alpha \in I}{\times} v_\alpha\right) := \underset{\alpha \in I}{\times} \pi_\alpha(m)v_\alpha, \qquad \forall m \in \mathbf{M},$$

*where $\underset{\alpha \in I}{\times}$ is a notation to represent each of the components of elements of the direct sum $\bigoplus_\alpha \mathcal{V}_\alpha$.*

**Definition A.1.8** (Cyclic representation of a $C^*$-algebra). *A cyclic representation of a $C^*$-algebra $\mathscr{A}$ is defined to be a triple $(\mathscr{H}, \pi, \Omega)$, where $(\mathscr{H}, \pi)$ is a $*$-representation of $\mathscr{A}$ and $\Omega$ is a vector in $\mathscr{H}$ which is cyclic for $\pi$, in $\mathscr{H}$, i.e. $\mathrm{span}\{\pi(A)\Omega \,|\, A \in \mathscr{A}\}$ is norm dense in $\mathscr{H}$.*

**Definition A.1.9** (Non-degenerate representation). *A representation $(\mathcal{V}, \pi)$ of a magma $(\mathbf{M}, \cdot)$ is said to be nondegenerate if:*

$$\{v \in \mathcal{V} \,|\, \pi(A)v = 0, \ \forall A \in \mathbf{M}\} = \{\mathbf{0}\}.$$

**Definition A.1.10** (Unitarily equivalent $C^*$-algebra representations). *For a $C^*$-algebra $\mathscr{A}$, and two representations $(\mathscr{H}_1, \pi_1)$, $(\mathscr{H}, \pi_2)$, as $\pi_{(\_)} : \mathscr{A} \to \mathfrak{B}(\mathscr{H}_{(\_)})$, if there exists a unitary operator $U : \mathscr{H}_1 \to \mathscr{H}_2$ that intertwines $\pi_1$ and $\pi_2$:*

$$U\pi_1(a) = \pi_2(a)U, \qquad \forall a \in \mathscr{A};$$

*then the representations $(\mathscr{H}_1, \pi_1)$, $(\mathscr{H}, \pi_2)$ are said to be **unitarily equivalent**.*

**Proposition A.1.11.** *Let $(\mathscr{H}, \pi)$ be a nondegenerate representation of the $C^*$-algebra $\mathscr{A}$. It follows that $\pi$ is the direct sum of a family of cyclic subrepresentations.*

*Proof:* Let $\{\Omega_\alpha\}_{\alpha \in I}$ denote a maximal family of nonzero vectors in $\mathscr{H}$ such that:

$$\langle \pi(A)\Omega_\alpha, \pi(B)\Omega_\beta \rangle_\mathscr{H} = 0,$$

for all $A, B \in \mathscr{A}$, whenever $\alpha \neq \beta$. The existence of such a family can be deduced with the aid of Zorn's lemma. Next define $\mathscr{H}_\alpha$ as the Hilbert subspace formed by closing the linear subspace $\{\pi(A)\Omega_\alpha \,|\, A \in \mathscr{A}\}$. This is an invariant subspace so we can introduce $\pi_\alpha$ by $\pi_\alpha(A) := P_{\mathscr{H}_\alpha}\pi(A)P_{\mathscr{H}_\alpha}$ and it follows that each $(\mathscr{H}_\alpha, \pi_\alpha, \Omega_\alpha)$ is a cyclic representation of $\mathscr{A}$. But the maximality of the $\{\Omega_\alpha\}_{\alpha \in I}$ and the nondegeneracy of $\pi$ imply that there is no nonzero $\Omega$ which is orthogonal to each subspace $\mathscr{H}_\alpha$, and hence:

$$\mathscr{H} = \bigoplus_{\alpha \in I} \mathscr{H}_\alpha, \qquad \pi = \bigoplus_{\alpha \in I} \pi_\alpha.$$

∎

**Proposition A.1.12.** *Let $\mathscr{A}$ be a Banach $*$-algebra with identity, and $\pi$ a $*$-morphism of $\mathscr{A}$ into $\mathfrak{B}(\mathscr{H})$. Then $\pi$ is continuous and*

$$\|\pi(A)\| \leq \|A\|, \tag{A.3}$$

*for all $A \in \mathscr{A}$.*

*Also, if $\mathscr{A}$ is a unital $C^*$-algebra, then $\pi$ is injective if, and only if $\pi$ is isometric,* i.e. $\|\pi(a)\| = \|a\|$, *for any $a \in \mathscr{A}$.*

*Proof:* First assume $A = A^*$. Then since $\mathfrak{B}(\mathscr{H})$ is a $C^*$-algebra and $\pi(A) \in \mathfrak{B}(\mathscr{H})$, one has:

$$\|\pi(A)\| = \sup\{|\lambda| \in \mathbb{R} \,|\, \lambda \in \sigma(\pi(A))\},$$

by theorem B.0.41. Next define $P = \pi(\mathbb{1}_{\mathscr{A}})$, it follows from the definition of $\pi$ that $P$ is a projection in $\mathfrak{B}(\mathscr{H})$. Hence replacing $\mathfrak{B}(\mathscr{H})$ by the $C^*$-algebra $\mathscr{G} := P\mathfrak{B}(\mathscr{H})P$ the projection $P$ becomes the identity $\mathbb{1}_{\mathscr{G}}$, of the new algebra $\mathscr{G}$.

Moreover, $\pi(\mathscr{A}) \subset \mathscr{G}$. Now it follows from the definitions of a $*$-morphism and of the spectrum that $\sigma_{\mathscr{G}}(\pi(A)) \subset \sigma_{\mathscr{A}}(A)$. Therefore:

$$\|\pi(A)\| \leq \sup\{|\lambda| \in \mathbb{R} \,|\, \lambda \in \sigma_{\mathscr{A}}(A)\} \leq \|A\|$$

by definition B.0.39. Finally, if $A$ is not self-adjoint one can combine this inequality with the $C^*$-norm property and the product inequality to deduce that

$$\|\pi(A)\|^2 = \|\pi(A)^*\pi(A)\| = \|\pi(A^*A)\| \leq \|A^*A\| \leq \|A\|^2.$$

Thus $\|\pi(A)\| \leq \|A\|$ for all $A \in \mathscr{A}$ and $\pi$ is continuous.

For the particular case of a unital $C^*$-algebra $\mathscr{A}$, if $\pi$ is isometric then if $0 = \|\pi(a) - \pi(b)\| = \|\pi(a - b)\| = \|a - b\|$, but since $\|\cdot\|$ is a norm, then $\|a - b\| = 0 \Leftrightarrow a - b = 0 \Rightarrow a = b$, therefore $\pi$ is injective. To prove that injectiveness implies that $\pi$ is isometric, we use the previous general result for unital Banach $*$-algebras $\|\pi(a)\| \leq \|a\|$, $\forall A \in \mathscr{A}$, therefore we must only prove that $\|\pi(a)\| \geq \|a\|$. We observe that, for $C^*$-algebras, if $\|\pi(a)\| \leq \|a\|$ is true for any self-adjoint element $a$, then this inequality is true for **any** element $b$ of $\mathscr{A}$, since:

$$\|\pi(b)\|^2 = \|\pi(b)^*\pi(b)\| = \|\pi(b^*b)\| \stackrel{\text{hyp.}}{\geq} \|b^*b\| = \|b\|^2, \tag{A.4}$$

so that $\|\pi(b)\| \geq \|b\|$, $\forall b \in \mathscr{A}$.

By contradiction, assume that there is a self-adjoint element $a \in \mathscr{A}$, such that $\|\pi(a)\| < \|a\|$, by theorem B.0.48 $\sigma(a) \subseteq [-\|a\|, \|a\|]$ and $r(a) = \|a\|$ theorem B.0.41, hence, by the definition of the spectral radius, either $\|a\| \in \sigma(a)$ or $-\|a\| \in \sigma(a)$. Similarly $\sigma(\pi(a)) \subseteq [-\|\pi(a)\|, \|\pi(a)\|] \stackrel{(A.3)}{\leq}$ $[-\|a\|, \|a\|]$. Choose a continuous map $f : [-\|a\|, \|a\|] \to \mathbb{R}$ that vanishes on $[-\|\pi(a)\|, \|\pi(a)\|]$ and such that, in the borders, $f(-\|a\|) = f(\|a\|) = 1$. We have that by linearity and multiplicity, $p(\pi(a)) = \pi(p(a))$ for any polynomial $p$, but by the just proved continuity of $\pi$, we can exchange the limits when considering Stone-Weierstrass approximations, theorem B.0.22, to get that $g(\pi(a)) = \pi(g(a))$ for continuous functions $g$, hence for our function $f$, $\pi(f(a)) = f(\pi(a)) = 0$, for $f\big|_{\sigma(\pi(a))} = 0$ and by the continuous functional calculus, theorem 4.0.16, $\|f(a)\| = \|f\|_\infty \geq 1$, this makes $f(a) \neq 0$, hence, by the last equation, contradicting the hypothesis of injectivity of $\pi$, which is absurd, therefore there is no self-adjoint element $a$ in $\mathscr{A}$ for which $\|\pi(a)\| < \|a\|$, hence $\|\pi(a)\| \geq \|a\|$ and by (A.4) $\|\pi(b)\| \geq \|b\|$, $\forall b \in \mathscr{A}$, making $\|\pi(b)\| = \|b\|$, $\forall b \in \mathscr{A}$. ∎

**Lemma A.1.13.** *Let $\mathscr{A}$ and $\mathscr{B}$ be $C^*$-algebras and let $\pi : \mathscr{A} \to \mathscr{B}$ be an isometric $*$-homomorphism. Then $\pi(\mathscr{A}) \subseteq \mathscr{B}$ is a closed $*$-sub-algebra; in particular $\pi(\mathscr{A})$ is a $C^*$-algebra.*

*Proof:* Since $\pi$ is a $*$-homomorphism, it is clear that $\pi(\mathscr{A}) \subseteq \mathscr{B}$ is a $*$-sub-algebra. Moreover, $\pi(\mathscr{A})$ is closed by a general fact for isometric maps: Let $\{\pi(x_n)\}_{n\in\mathbb{N}}$ be a Cauchy sequence in $\pi(\mathscr{A})$. Then $\|x_n - x_m\| = \|\pi(x_n) - \pi(x_m)\|$, since $\pi$ is isometric. Hence, $\{x_n\}_{n\in\mathbb{N}}$ is a Cauchy sequence in $\mathscr{A}$ and we find a limit $x_n \to x$ since $\mathscr{A}$ is complete. Then $\pi(x_n) \to \pi(x) \in \pi(\mathscr{A})$, i.e. $\pi(\mathscr{A})$ is complete and hence closed. ∎

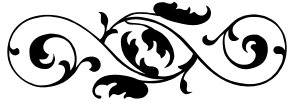

# Appendix B

# Functional Analysis

**Definition B.0.1.** *A Hilbert space $\mathscr{H}$ is a complex vector space that has a sesquilinear inner product $\langle \cdot, \cdot \rangle_{\mathscr{H}} : \mathscr{H} \times \mathscr{H} \to \mathbb{C}$ that is also a complete metric space with respect to the metric $d : \mathscr{H} \times \mathscr{H} \to \mathbb{R}$ induced by this inner product:*

$$d(x, y) = \|x - y\| = \sqrt{\langle x - y, x - y \rangle_{\mathscr{H}}}$$

**Theorem B.0.2** (Bessel's inequality)**.** *For any orthonormal system $N \subset X$ in an inner product space $(X, \langle \cdot, \cdot \rangle)$, i.e. not necessarily complete with respect to the norm induced by the inner product, we have that:*

$$\sum_{z \in N} |\langle z, x \rangle|^2 \leq \|x\|^2 \ , \ \forall\, x \in X. \tag{B.1}$$

*In particular, only a countable number of the products $\langle z, x \rangle$ are non-zero, at most.*

*Proof:* By the definition of the sum of an indexed set of positive numbers $\{\alpha_i\}_{i \in I} \subset \mathbb{R}_+$ with an index set $I$ with arbitrary cardinality we have that:

$$\sum_{i \in I} \alpha_i := \sup_{J} \left\{ \sum_{j \in J} \alpha_j \,\middle|\, J \subseteq I,\ J \text{ is finite} \right\}$$

If this sum is finite the subset of $I$ for which $\alpha_i \neq 0$ is finite or countable. By restricting to this subset, this sum equals either the sum of finitely many terms or the sum of the series $\sum_{n=0}^{+\infty} \alpha_{i_n}$ for a enumerable subsequence of indexes $\{i_n\}_{n \in \mathbb{N}_0} \subseteq I$. this can be seen by supposing that $0 \leq \sum_{i \in I} \alpha_i < +\infty$, the case $\sum_{i \in I} \alpha_i = 0$ implies $\alpha_i = 0$, $\forall\, i \in I$ since the $\alpha_i$ are non-negative. Assuming then that $\sum_{i \in I} \alpha_i > 0$, then any $\alpha_i$ must be contained in the interval $\left[0, \sum_{i \in I} \alpha_i\right]$, for otherwise the sum would be larger than the sum itself. In particular if $\alpha_i \neq 0$ then $\alpha_i \in \left(0, \sum_{i \in I} \alpha_i\right]$, defining then $S_n := \frac{1}{n} \sum_{i \in I} \alpha_i$ for $n \in \mathbb{N}$, letting $N_k$ be the number of indexes $i \in I$ such that $\alpha_i \in (S_{k+1}, S_k]$ this makes the estimate $S_{k+1} N_k \leq \sum_{i \in I} \alpha_i < +\infty$ be valid, hence $N_k$ is a finite set for any $k$. But $\bigcup_{k=1}^{+\infty} (S_{k+1}, S_k] = \left(0, \sum_{i \in I} \alpha_i\right]$, so all $\alpha_i \neq 0$ are accounted for, therefore there can be at most enumerably many of these values, since there are enumerably many intervals $(S_{k+1}, S_k]$ and each interval contains a finite number of $\alpha_i = 0$.

From these consideration we conclude that for inequality (B.1) to hold, it must hold for all finite subsets $F \subset N$. So, for $F = \{z_1, \ldots, z_n\}$, $x \in X$ and $\alpha - 1, \ldots, \alpha_n \in \mathbb{C}$. Expanding then $\left\|x - \sum_{k=1}^{n} \alpha_k z_k\right\|^2$ in terms of the inner product of $X$, taking into consideration its linearity and the orthogonality between the $z_k$'s, we have:

$$\left\| x - \sum_{k=1}^{n} \alpha_k z_k \right\|^2 = \|x\|^2 + \sum_{k=1}^{n} |\alpha_k|^2 - \sum_{k=1}^{n} \alpha_k \langle x, z_k \rangle - \sum_{k=1}^{n} \overline{\alpha_k \langle x, z_k \rangle}.$$

By summing and subtracting $\sum_{k=1}^{n} |\langle x, z_k \rangle|^2$:

$$\|x\|^2 - \sum_{k=1}^{n} |\langle x, z_k \rangle|^2 + \sum_{k=1}^{n} \left( |\langle x, z_k \rangle|^2 - \alpha_k \langle x, z_k \rangle - \overline{\alpha_k \langle x, z_k \rangle} + |\alpha_k|^2 \right),$$

so

$$\left\| x - \sum_{k=1}^{n} \alpha_k z_k \right\|^2 = \|x\|^2 - \sum_{k=1}^{n} |\langle x, z_k \rangle|^2 + \sum_{k=1}^{n} |\langle z_k, x \rangle - \alpha_k|^2.$$

Taking the minimum point $\alpha_k = \langle z_k, x \rangle$ for $k = 1, \ldots, n$, then:

$$0 \leq \left\| x - \sum_{k=1}^{n} \alpha_k z_k \right\|^2 = \|x\|^2 - \sum_{k=1}^{n} |\langle x, z_k \rangle|^2,$$

and finally:

$$\sum_{k=1}^{n} |\langle x, z_k \rangle|^2 \leq \|x\|^2.$$

■

**Definition B.0.3.** *A Hilbert space $\mathscr{H}$ is **separable** if it is separable as a topological space with relation to the metric topology induced by the inner product $\langle \cdot, \cdot \rangle$, that is if it admits a **countable dense subset**.*

**Lemma B.0.4.** *If $\mathscr{H}$ is a Hilbert space imbued with the inner product $\langle \cdot, \cdot \rangle : \mathscr{H} \times \mathscr{H} \to \mathbb{C}$, and $N \subset \mathscr{H}$ is an Hilbert space orthonormal basis, then if $x \in \mathscr{H}$, we have that at most countably many products $\langle z, x \rangle$, where $z \in N$, are non-zero; and:*

$$x = \sum_{z \in N} \langle x, z \rangle \, z,$$

*where the series converges in the sense that partial sums converge in the inner product topology.*

*Proof:* By theorem B.0.2 only countably many coefficients $\langle z, x \rangle$ are non-null, at most, indicate these numbers by $\langle z_n, x \rangle$, $n \in \mathbb{N}$ and fix partial sums $S_N := \sum_{n=1}^{N} \langle z_n, x \rangle z_n$. The system $\{\langle z_n, x \rangle z_n\}_{n \in \mathbb{N}}$ is by construction orthogonal, and because $\|z_n\| = 1, \forall n \in \mathbb{N}$, by Bessel's inequality (B.1) implies $\sum_{n=1}^{+\infty} \|\langle z_n, x \rangle z_n\|^2 < +\infty$. Since $\mathscr{H}$ is complete, then the expression $\left\| \sum_{k=1}^{n} x_k - \sum_{k'=1}^{m} x_{k'} \right\|^2 = \sum_{k=m+1}^{n} \|x_k\|^2 < \sum_{k=m+1}^{+\infty} \|x_k\|^2 < +\infty$ for $m < n$, which is valid for orthogonal $x_k$'s with convergence of the sum of the square of their norms, this establishes a convergence between the partial sums when $m \to +\infty$, therefore these partial sums form a Cauchy sequence and hence converge to a unique element $x' \in \mathscr{H}$, specifying this result to the system $\{\langle z_n, x \rangle z_n\}_{n \in \mathbb{N}}$ we get $x' = \sum_{n=1}^{+\infty} \langle z_n, x \rangle z_n$. Moreover it is an elementary result from mathematical analysis that rearranging a Cauchy sequence leads to the same limit (Proved for example in [77], Lemma 3.25(b).), so we can rearrange that previous sequence and get the same limit.

We claim that $x' = x = \sum_{z \in N} \langle x, z \rangle z$, to prove this we consider that the linearity and continuity of the inner product and a $z' \in N$, will force the inner product of the difference between $x$ and $x'$ together with the $z'$ to equal:

$$\langle x - x', z \rangle = \langle x, z' \rangle - \sum_{z \in N} \langle x, z \rangle \langle z, z' \rangle = \langle x, z' \rangle - \langle x, z' \rangle = 0$$

where we have used the fact that the set $N$ is a orthonormal basis of $\mathscr{H}$. Since $z' \in N$ is arbitrary, $x - x' \in N^\perp$ and so $x - x' = \mathbf{0}$, as $N^\perp = \{\mathbf{0}\}$ by assumption. ∎

**Proposition B.0.5.** *Let $\mathscr{H} \neq \{\mathbf{0}\}$ be a Hilbert space.*

(a) *$\mathscr{H}$ is separable if and only if either $\dim(\mathscr{H}) < \infty$ or it has a enumerable orthonormal basis.*

(b) *if $\mathscr{H}$ is separable then every basis is either finite, with cardinality equal to $\dim(\mathscr{H})$, or enumerable.*

(c) *if $\mathscr{H}$ is separable then it is isomorphic either to $\ell_2(\mathbb{C})$, or to $\mathbb{C}^n$ with the standard topology, where $n = \dim(\mathscr{H})$.*

*Proof:* (a) If the Hilbert space has a finite or enumerable orthonormal basis $N$, then by lemma B.0.4 and any $x \in \mathscr{H}$, we have $x = \sum_{\xi \in N} \langle \xi, x \rangle \xi$, were this sum converges converges in the sense that the partial sums converge in the inner product topology; the possibility of writing any element of the Hilbert space in this way ensures that a countable dense set exists, because rational numbers are dense in the reals. This set consists clearly of finite linear combinations of basis elements with complex coefficients having rational real and rational imaginary parts, since this is clear and easy to prove we shall omit this part of the proof.

Conversely, suppose a Hilbert space is separable. Since every Hilbert space different from $\{\mathbf{0}\}$ admits a complete orthonormal basis, a fact that comes from considering the collection of orthonormal systems $\mathcal{V}$ in $\mathscr{H}$ and the partial order given by the set inclusion relation, obviously any subset $E \subset \mathcal{V}$ is bounded above by the set formed by the union of all its constituent systems $\bigcup E \in \mathcal{V}$ this clears the hypothesis of Zorn's lemma that every chain $E$ of the partial ordered set $\mathcal{V}$ has a upper bound in $\mathcal{V}$, follows as implication of the lemma that there is a maximal element $M \in \mathcal{V}$, since it is maximal there is no other non-zero vector that is orthogonal to every element in $M$ but is not in $M$ itself, this makes $M$ a complete orthonormal system, *i.e.* a orthonormal basis, because of this we know bases exist, and we want to show that any basis must be enumerable at most.

Suppose, by contradiction, that $N$ is an uncountable orthonormal basis for the separable Hilbert space $\mathscr{H}$. For any chosen $z, z' \in N$, $z \neq z'$, any point $x \in \mathscr{H}$ satisfies $\|z-z'\| \leq \|x-z'\| + \|z-x\|$, by the triangle inequality induced by the inner product. At the same time $\{z, z'\}$ is an orthonormal set of vectors, so $\|z-z'\|^2 = \langle z-z', z-z' \rangle = \|z\|^2 + \|z'\|^2 = 1+1 = 2$. Hence $\|x-z\| + \|x-z'\| \geq \sqrt{2}$. This implies that two open balls of radius $\varepsilon < \frac{\sqrt{2}}{2}$ centered at $z$ and $z'$ are disjoint, irrespective of how we pick $z, z' \in N$ with $z \neq z'$. Call $\{\mathcal{B}_\varepsilon^z(\mathscr{H})\}_{z \in N}$ a family of such balls parametrised by their centres $z \in N$. If $D \subset \mathcal{H}$ is a countable dense set (the space is separable), then for any $z \in N$ there exists $x \in D$ with $x \in \mathcal{B}_\varepsilon^z(\mathcal{H})$. The balls are pairwise disjoint, so there will be one $x$ for each ball, all different from one another. But the cardinality of $\{\mathcal{B}_\varepsilon^z(\mathcal{H})\}_{z \in N}$ is not countable, hence neither $D$ can be countable, a contradiction.

(b) From the basic theory, if a (orthonormal or algebraic) basis is finite, the cardinality of any other basis equals the dimension of the space. Moreover, any linearly independent set (viewed as a basis) cannot contain a number of vectors exceeding the dimension. From this, if a Hilbert space is separable and one of its bases is finite, then all bases are finite and have cardinality $\dim(\mathcal{H})$. Under the same hypotheses, if a basis is enumerable then any other is enumerable by item (a).

(c) Fix a orthonormal basis $N$. Using again lemma B.0.4 that for any vector $x \in \mathcal{H}$ it is possible to write $x = \sum_{u \in N} \langle u, x \rangle u$; one verifies quickly that the map sending $\mathcal{H} \ni x = \sum_{u \in N} \alpha_u u$ to the (finite or infinite) family $\{\alpha_u\}_{u \in N}$, i.e. $\alpha_u = \langle u, x \rangle$, is an isometric isomorphism sending $\mathcal{H}$ to $\mathbb{C}^n$ (if $\dim(\mathcal{H})$ is finite) or to $\ell_2(\mathbb{C})$ (if $\dim(\mathcal{H})$ is infinite).

∎

**Proposition B.0.6.** *Let $(\mathcal{V}, \|\cdot\|_\mathcal{V})$ and $(\mathcal{W}, \|\cdot\|_\mathcal{W})$ be normed vector spaces. A linear operator $T: \mathcal{V} \to \mathcal{W}$ is bounded if and only if it is continuous.*

*Proof:* If $T$ is bounded then $\exists M > 0$ such that $\|Tu\|_\mathcal{W} \leq M\|u\|_\mathcal{V}$, $\forall u \in \mathcal{V}$. Let $\varepsilon > 0$, and $u, v \in \mathcal{V}$ be such that $\|u - v\|_\mathcal{V} \leq \frac{\varepsilon}{M}$. Then:

$$\|Tu - Tv\|_\mathcal{W} = \|T(u - v)\|_\mathcal{W} \leq M\|u - v\|_\mathcal{V} \leq M\frac{\varepsilon}{M} = \varepsilon.$$

Therefore considering $\delta = \frac{\varepsilon}{M}$, we get that $T$ fulfills the definition of a continuous linear transformation.

If $T$ is continuous, then for every $\varepsilon > 0$ and $u \in \mathcal{V}$, $\exists \delta > 0$ such that $\|Tu - Tv\|_\mathcal{W} \leq \varepsilon$ for every $v \in \mathcal{V}$ with $\|u - v\|_\mathcal{V} \leq \delta$. Taking $u = \mathbf{0}$ and fixating $\varepsilon$, we have then that $\|Tv\|_\mathcal{W} \leq \varepsilon$, whenever $\|v\|_\mathcal{V} \leq \delta$. Considering then an arbitrary non-null vector $u \in \mathcal{V}$, and choosing $v$ such that:

$$v = \frac{\delta}{\|u\|_\mathcal{V}} u,$$

we see that:

$$\|v\|_\mathcal{V} = \left\|\frac{\delta}{\|u\|_\mathcal{V}} u\right\| = \delta.$$

Therefore, for such $v$, it is true that $\|Tv\|_\mathcal{W} \leq \varepsilon$, consequently:

$$\frac{\delta}{\|u\|_\mathcal{V}} \|Tu\|_\mathcal{W} = \left\|T\left(\frac{\delta}{\|u\|_\mathcal{V}} u\right)\right\| = \|Tv\|_\mathcal{W} \leq \varepsilon,$$

which implies:

$$\|Tu\|_\mathcal{W} \leq \frac{\varepsilon}{\delta} \|u\|_\mathcal{V}.$$

Defining $M := \frac{\varepsilon}{\delta}$, we have shown that $\|Tu\|_\mathcal{W} \leq M\|u\|_\mathcal{V}$ for every $u \neq \mathbf{0}$, but for $u = \mathbf{0}$ this relation is trivially satisfied, and hence the relation is valid for all $u \in \mathcal{V}$, showing that $T$ is bounded. ∎

**Proposition B.0.7** (Sesquilinear form polarization identity). *Let $\mathcal{V}$ be a complex vector space and $A : \mathcal{V} \to \mathcal{V}$ be a linear operator in $\mathcal{V}$, then $\forall\, u, v \in \mathcal{V}$:*

$$\langle u,\, Av \rangle = \frac{1}{4} \sum_{n=0}^{3} i^{-n} \left\langle (u + i^n v),\, A(u + i^n v) \right\rangle$$

$$\langle u,\, Av \rangle = \frac{1}{4} \sum_{n=0}^{3} i^{n} \left\langle (u + i^{-n} v),\, A(u + i^{-n} v) \right\rangle$$

*Proof:* Trivial expansion of the expression and canceling of terms with opposite signs. ∎

**Proposition B.0.8** ($*$-algebra element polarization identity). *Let $\mathscr{A}$ be a $*$-algebra and $A, B \in \mathscr{A}$, then:*

$$AB^* = \frac{1}{4} \sum_{n=0}^{3} i^n (A + i^n B)(A + i^n B)^* \tag{B.2}$$

*Proof:* Trivial expansion of the expression and canceling of terms with opposite signs. ∎

**Proposition B.0.9.** *Let $\mathscr{H}$ be a Hilbert space and $P \in \mathfrak{B}(\mathscr{H})$ an orthogonal projector onto a subspace $\mathscr{M} \subset \mathscr{H}$, then if $N$ is a orthonormal basis on $\mathscr{M}$ we have that $\forall\, x \in \mathscr{H}$:*

$$Px = \sum_{u \in N} \langle u,\, x \rangle u$$

*Proof:* We may extend $N$ to a orthonormal basis of $\mathscr{H}$ by adding a orthonormal basis $N'$ of $M^\perp$. We can immediately verify that, varying $x \in \mathscr{H}$:

$$Rx = \sum_{u \in N} \langle u,\, x \rangle u$$

and

$$R'x = \sum_{v \in N'} \langle v,\, x \rangle v$$

define bounded operators (at most countably many products $\langle u,\, x \rangle$ are non-zero for every fixed $x \in \mathscr{H}$, by lemma B.0.4), and they satisfy $RR = R$, $R(\mathscr{H}) = M$, $R'R' = R'$, $R'(\mathscr{H}) = M^\perp$ and also $R'R = RR' = 0$ and $R + R' = \mathbb{1}$. This implies that $R$ and $R'$ are projectors associated to $M \oplus M^\perp$. By uniqueness of the decomposition of any vector we must have $R = P$ (and $R' = \mathbb{1} - P$). ∎

**Proposition B.0.10** (Hölder's inequality). *Let $(X, \Sigma, \mu)$ be a measure space, then for $p, q \in \mathbb{R}$ such that $p > 1, q > 1$ and $\frac{1}{p} + \frac{1}{q} = 1$, then for any measurable functions $f, g : X \to \mathbb{C}$ such that $\int_X |f(x)|^p d\mu(x) < +\infty$ and $\int_X |g(x)|^q d\mu(x) < +\infty$ then:*

$$\int_X |f(x)||g(x)| d\mu(x) \leq \left[ \int_X |f(x)|^p d\mu(x) \right]^{\frac{1}{p}} \left[ \int_X |g(x)|^q d\mu(x) \right]^{\frac{1}{q}}. \tag{B.3}$$

*Proof:* The logarithm function is concave in the interval $(0, +\infty)$, therefore for any $a, b \in (0, +\infty)$, we have for the convex combination $\ln\left(\frac{1}{p}a + \frac{1}{q}b\right) \geq \frac{1}{p}\ln(a) + \frac{1}{q}\ln(b) = \ln\left(a^{\frac{1}{p}} b^{\frac{1}{q}}\right)$, the first combination is convex since by the hypothesis of the proposition $\frac{1}{p}, \frac{1}{q} \in (0,1)$ and $\frac{1}{p}+\frac{1}{q}=1$; taking the exponential of both sides of the logarithm inequality we get the so called Young's inequality:

$$a^{\frac{1}{p}}b^{\frac{1}{q}} \leq \frac{a}{p} + \frac{b}{q}. \tag{B.4}$$

We then note that Hölder's inequality is trivial for either $\int_X |f(x)|^p d\mu(x) = 0$ or $\int_X |g(x)|^q d\mu(x)$, since either $|f|=0$ or $|g|=0$ $\mu$-almost everywhere. Therefore for $\int_X |f(x)|^p d\mu(x) \neq 0 \neq \int_X |g(x)|^q d\mu(x)$ and $x \in X$ we take:

$$a = \frac{|f(x)|^p}{\int_X |f(x)|^p d\mu(x)} \quad b = \frac{|g(x)|^q}{\int_X |g(x)|^q d\mu(x)}$$

Substituting $a$ and $b$ in Young's inequality (B.4):

$$\frac{|f(x)|}{\left[\int_X |f(x)|^p d\mu(x)\right]^{\frac{1}{p}}} \frac{|g(x)|}{\left[\int_X |g(x)|^q d\mu(x)\right]^{\frac{1}{q}}} \leq \frac{1}{p} \frac{|f(x)|^p}{\int_X |f(x)|^p d\mu(x)} + \frac{1}{q} \frac{|g(x)|^q}{\int_X |g(x)|^q d\mu(x)}$$

Taking the integral in $X$ with respect to $\mu$ of both sides of this expression gives Hölder's inequality:

$$\frac{\int_X |f(x)||g(x)| d\mu(x)}{\left[\int_X |f(x)|^p d\mu(x)\right]^{\frac{1}{p}} \left[\int_X |g(x)|^q d\mu(x)\right]^{\frac{1}{q}}} \leq \frac{1}{p} \frac{\int_X |f(x)|^p d\mu(x)}{\int_X |f(x)|^p d\mu(x)} + \frac{1}{q} \frac{\int_X |g(x)|^q d\mu(x)}{\int_X |g(x)|^q d\mu(x)} = \frac{1}{p} + \frac{1}{q} = 1$$

$$\implies \int_X |f(x)||g(x)| d\mu(x) \leq \left[\int_X |f(x)|^p d\mu(x)\right]^{\frac{1}{p}} \left[\int_X |g(x)|^q d\mu(x)\right]^{\frac{1}{q}}.$$

∎

**Corollary B.0.11** (Riesz-Fischer)**.** *Let $\{\phi_i\}_{i \in I} \subseteq L^2(X, \mu)$ be a enumerable orthonormal basis, then there is an isometric isomorphism between $L^2(X, d\mu)$ and $\ell_2(\mathbb{C})$ given by:*

$$\Phi: L^2(X, d\mu) \to \ell_2(\mathbb{C}), \ L^2(X, d\mu) \ni \psi \xmapsto{\Phi} \left(\int_X \psi(x)\overline{\phi_i(x)} d\mu(x)\right)_{i \in I} \in \ell_2(\mathbb{C}).$$

*Proof:* First we prove that $L^2(X, \mu)$ is a Hilbert space with the inner product defined by $\langle f, g \rangle := \int_X f(x)\overline{g(x)} d\mu(x)$. Since for any $f, g \in L^2(X, \mu)$ we have $\int_X |f(x)|^2 d\mu < +\infty$ and $\int_X |g(x)|^2 d\mu < +\infty$, then by Holder's inequality, prop. B.0.10:

$$\int_X |f(x)||g(x)| d\mu \leq \left[\int_X |f(x)|^2 d\mu(x)\right]^{\frac{1}{2}} \left[\int_X |g(x)|^2 d\mu(x)\right]^{\frac{1}{2}} < +\infty$$

Also since $\left|\int_X f(x)\overline{g(x)} d\mu(x)\right| \leq \int_X |f(x)||g(x)| d\mu(x)$ it follows that:

$$\left|\int_X f(x)\overline{g(x)} d\mu(x)\right| \leq \left[\int_X |f(x)|^2 d\mu(x)\right]^{\frac{1}{2}} \left[\int_X |g(x)|^2 d\mu(x)\right]^{\frac{1}{2}} < +\infty$$

And hence $\langle f, g \rangle = \int_X f(x)\overline{g(x)} d\mu(x)$ is well defined for every $f, g \in L^2(X, \mu)$. Sesquilinearity follows from the linearity of the integral, the conjugate symmetry is self-evident from the definition. Positive definiteness holds by construction: If $\int_X |f(x)| d\mu(x) = 0$, then $|f|^2$ and hence $f$ is zero almost everywhere, thus the equivalence class of $f$, which is the actual element of $L^2(X, \mu)$, is the equivalence class of the zero function,the additive neutral element of the space. We shall prove the completeness for $L^p(X,\mu)$ since the proof for $L^2(X,\mu)$ is no less complicated, this also proves as bonus the second version of the Riesz-Fischer theorem in which "For $p \geq 1$ the spaces $L^p(X, d\mu)$ are complete metric spaces with relation to the distance $d_p(f,g) := \|f - g\|_p = \left[\int_X |f-g|^p d\mu\right]^{\frac{1}{p}}$".

Let $\{f_n\}_{n\in\mathbb{N}}$ be a Cauchy sequence in $L^p(X,\mu)$ for the norm $\|\cdot\|_p$, that is, for every $\varepsilon > 0$ there exists $N(\varepsilon) \in \mathbb{N}$ such that $\|f_n - f_m\|_p < \varepsilon$, $\forall n, m > N(\varepsilon)$. Obviously for any such sequence $\{f_n\}_{n\in\mathbb{N}}$ there is a subsequence $\{g_n\}_{n\in\mathbb{N}}$ such that:

$$\|g_{k+1} - g_k\|_p < \frac{1}{2^k}, \tag{B.5}$$

since we can take $g_k := f_{n_k}$ where $n_k > N\left(\frac{1}{2^{k+1}}\right)$, $k \in \mathbb{N}$. From Hölder's inequality we can prove Minkowski's inequality which shall be instrumental in what follows. To prove Minkowski's inequality we first note that the application $x \mapsto x^p$ is continuous and convex for $p \geq 1$ and $x > 0$, hence:

$$\left(\frac{(|f(x)| + |g(x)|)}{2}\right)^p \leq \frac{1}{2}(|f(x)|^p + |g(x)|^p)$$

Since $|f - g| \leq |f| + |g|$, whence:

$$\left(\frac{(|f - g|)}{2}\right)^p \leq \frac{1}{2}(|f|^p + |g|^p)$$

This makes us conclude that if $f, g \in L^p(X, \mu)$ then $f - g \in L^p(X, \mu)$. Then for $0 < \int_X |f-g|^p < +\infty$, we write:

$$|f - g|^p = |f - g||f - g|^{p-1} \leq (|f| + |g|)|f - g|^{p-1} = |f||f - g|^{p-1} + |g||f - g|^{p-1},$$

which makes:

$$\int_X |f - g|^p d\mu = \int_X |f||f - g|^{p-1} d\mu + \int_X |g||f - g|^{p-1} d\mu. \tag{B.6}$$

Considering then Holder's inequality (B.3) for $q = \frac{p}{p-1}$ applied to each of the elements of the sum in the RHS of the previous line:

$$\int_X |f||f - g|^{p-1} d\mu \leq \left[\int_X |f|^p d\mu\right]^{\frac{1}{p}} \left[\int_X |f - g|^{(p-1)q} d\mu\right]^{\frac{1}{q}} = \left[\int_X |f|^p d\mu\right]^{\frac{1}{p}} \left[\int_X |f - g|^p d\mu\right]^{\frac{1}{q}},$$

and analogously:

$$\int_X |g||f - g|^{p-1} d\mu \leq \left[\int_X |g|^p d\mu\right]^{\frac{1}{p}} \left[\int_X |f - g|^p d\mu\right]^{\frac{1}{q}}.$$

Substituting both these expressions back into (B.6):

$$\int_X |f - g|^p d\mu \leq \left(\left[\int_X |f|^p d\mu\right]^{\frac{1}{p}} + \left[\int_X |g|^p d\mu\right]^{\frac{1}{p}}\right) \left[\int_X |f - g|^p d\mu\right]^{\frac{1}{q}}$$

Since we supposed $0 < \int_X |f - g|^p d\mu < +\infty$ then $0 < \left[\int_X |f-g|^p d\mu\right]^{\frac{1}{q}} < +\infty$ and therefore we can divide both sides of the previous expression by $\left[\int_X |f-g|^p d\mu\right]^{\frac{1}{q}}$, remembering that $1 - \frac{1}{q} = \frac{p}{p} - \frac{p-1}{p} = \frac{1}{p}$ then we get Minkowski's inequality:

$$\frac{\int_X |f - g|^p d\mu}{\left[\int_X |f - g|^p d\mu\right]^{\frac{1}{q}}} \leq \left[\int_X |f|^p d\mu\right]^{\frac{1}{p}} + \left[\int_X |g|^p d\mu\right]^{\frac{1}{p}}$$

$$\implies \left[\int_X |f-g|^p d\mu\right]^{\frac{1}{p}} \leq \left[\int_X |f|^p d\mu\right]^{\frac{1}{p}} + \left[\int_X |g|^p d\mu\right]^{\frac{1}{p}}, \tag{B.7}$$

which is also clearly true for $\int_X |f-g|^p d\mu = 0$. From Minkowski's inequality and relation (B.5) we get:

$$\left\|\sum_{k=1}^n |g_{k+1} - g - k|\right\|_p \leq \sum_{k=1}^n \|g_{k+1} - g - k\|_p \leq \sum_{k=1}^n \frac{1}{2^k},$$

which makes:

$$\int_X \left(\sum_{k=1}^n |g_{k+1} - g - k|\right)^p d\mu \leq \left(\sum_{k=1}^n \frac{1}{2^k}\right)^p.$$

By Fatou's lemma C.0.25 we have that:

$$\int_X \liminf_{n\to\infty} \left(\sum_{k=1}^n |g_{k+1} - g - k|\right)^p d\mu \leq \liminf_{n\to\infty} \int_X \left(\sum_{k=1}^n |g_{k+1} - g - k|\right)^p d\mu \leq \liminf_{n\to\infty} \left(\sum_{k=1}^n \frac{1}{2^k}\right)^p = 1$$

Now since $\left\{\sum_{k=1}^n |g_{k+1} - g - k|\right\}_{n\in\mathbb{N}}$ is a non-decreasing sequence, so is $\left\{\left(\sum_{k=1}^n |g_{k+1} - g - k|\right)^p\right\}_{n\in\mathbb{N}}$ and this sequence converges to $\sum_{k=1}^\infty |g_{k+1} - g - k|$. Therefore $\liminf_{n\to\infty} \left(\sum_{k=1}^n |g_{k+1} - g - k|\right)^p = \left(\sum_{k=1}^\infty |g_{k+1} - g - k|\right)^p$ and we conclude that:

$$\int_X \left(\sum_{k=1}^\infty |g_{k+1} - g - k|\right)^p d\mu \leq 1,$$

which implies that $\left\|\sum_{k=1}^\infty |g_{k+1} - g - k|\right\|_p \leq 1$, whence $\sum_{k=1}^\infty |g_{k+1} - g - k| < +\infty$ $\mu$-almost everywhere, and so $g_1(x) + \sum_{k=1}^n (g_{k+1} - g - k)$ converges absolutely for its inputs with the exception of a set of null $\mu$-measure. In noting that:

$$g_1(x) + \sum_{k=1}^{n-1}(g_{k+1} - g - k) = g_n(x),$$

and so we conclude that $\lim_{n\to\infty} g_n(x)$ exists $\mu$-almost everywhere.

Denoting by $G$ the set in which this limit exists, we have $\mu(X \setminus G) = 0$, and defining $f : X \to \mathbb{C}$ to be:

$$f = \begin{cases} \lim_{n\to\infty} g_n(x), & \text{for } x \in G, \\ 0, & \text{for } x \in X \setminus G. \end{cases}$$

We want to prove that $f$ is the limit in $L^p(X, \mu)$ of the sequence $\{f_n\}_{n\in\mathbb{N}}$, i.e. $\|f - f_m\|_p \to 0$ when $m \to \infty$. Fixating $\varepsilon > 0$, we know from the hypothesis that for $n, m > N(\varepsilon)$, $\|f_n - f_m\|_p < \varepsilon$, then Fatou's lemma C.0.25 for $m > N(\varepsilon)$ will imply that:

$$\int_X |f - f_m|^p d\mu \leq \int_X \liminf_{n\to\infty} |g_n - f_m|^p d\mu \leq \liminf_{n\to\infty} \int_X |g_n - f_m|^p d\mu = \liminf_{n\to\infty} (\|g_n - f_m\|_p)^p \leq \varepsilon^p$$

This proves that $f - f_m \in L^p(X, \mu)$, and since $f = (f - f_m) + f_m$ this implies that $f \in L^p(X, \mu)$ since $L^p(X, \mu)$ is closed under addition. At the same time this last inequality implies that $\|f - f_m\| \to 0$ when $m \to \infty$. We have shown that the Cauchy sequence $\{f_n\}_{n\in\mathbb{N}} \subset L^p(X,\mu)$ has a limit $f \in L^p(X,\mu)$ in the norm $\|\cdot\|_p$, this makes $L^p(X,\mu)$ a complete metric space in this norm, and since $\sqrt{\langle f, f \rangle} = \left(\int_X f(x)\overline{f(x)}d\mu(x)\right)^{\frac{1}{2}} = \left(\int_X |f(x)|^2 d\mu(x)\right)^{\frac{1}{2}} = \|f\|_2, \forall f \in L^2(X,\mu)$, this completes the proof that $L^2(X,\mu)$ is a Hilbert space. Since $L^2(X,\mu)$ is separable and has a enumerable orthonormal basis, by hypothesis, then by proposition B.0.5 item (c) it is isometrically isomorphic $\ell_2(\mathbb{C})$. ∎

**Definition B.0.12** (Trace-class). *Let $\mathscr{H}$ be a Hilbert space and $A \in \mathfrak{B}(\mathscr{H})$, we say that $A$ is a **trace-class** operator, in symbols $A \in \mathcal{J}_1(\mathscr{H})$, if for a Hilbert basis $\{x_i\}_{i\in K}$ of $\mathscr{H}$ we have that $\sum_{i \in K} \langle x_i, |A|x_i \rangle < +\infty$.*

The following theorem is a fundamental theorem for functional analysis, we shall only enunciate the statement of the theorem, the proof is simple and can be found in any book of functional analysis, such as [68, 15, 105]

**Theorem B.0.13** (Riesz representation theorem). *Let $l \in \mathscr{H}'$ be a continuous linear functional in a Hilbert space $\mathscr{H}$, then there exists a unique $\phi \in \mathscr{H}$ such that:*

$$l(x) = \langle \phi, x \rangle \quad , \forall x \in \mathscr{H}, \text{ and } \|l\|_{\mathscr{H}'} = \|\phi\|_{\mathscr{H}}.$$

**Corollary B.0.14.** *Let $\mathcal{S} : \mathscr{H}_2 \times \mathscr{H}_1 \to \mathbb{C}$ be a bicontinuous sesquilinear form. Then, there exists a unique bounded operator $S : \mathscr{H}_2 \to \mathscr{H}_1$ such that:*

$$\mathcal{S}(u,v) = \langle Su, v \rangle_{\mathscr{H}_1} \quad , \forall\, u \in \mathscr{H}_2, v \in \mathscr{H}_1.$$

*Proof:* For each fixed $u \in \mathscr{H}_2$, the map $v \mapsto \mathcal{S}(u,v)$ is a continuous linear functional in $\mathscr{H}_1$. By the Riesz representation theorem B.0.13 for each $u \in \mathscr{H}_2$ there exists a vector $\eta_u \in \mathscr{H}_1$ such that $\mathcal{S}(u,v) = \langle \eta_u, v \rangle_{\mathscr{H}_1}$. Being $S : \mathscr{H}_2 \to \mathscr{H}_1$ a map that takes $u$ to $\eta_u$, $S(u) = \eta_u$. We write $\mathcal{S}(u,v) = \langle S(u), v \rangle_{\mathscr{H}_1}$ for every $u \in \mathscr{H}_2$ and $v \in \mathscr{H}_1$.

Since $\mathcal{S}$ is sesquilinear we have that:

$$\langle S(\alpha_1 u_1 + \alpha_2 u_2), v \rangle = \mathcal{S}(\alpha_1 u_1 + \alpha_2 u_2, v) = \overline{\alpha}_1 \mathcal{S}(u_1, v) + \overline{\alpha}_2 \mathcal{S}(u_2, v) =$$

$$= \langle \alpha_1 S(u_1), v \rangle + \langle \alpha_2 S(u_2), v \rangle = \langle \alpha_1 S(u_1) + \alpha_2 S(u_2), v \rangle,$$

for all $u_1, u_2 \in \mathscr{H}_1$, $v \in \mathscr{H}_2$ and $\alpha_1, \alpha_2 \in \mathbb{C}$ which implies that $S$ is linear $S(\alpha_1 u_1 + \alpha_2 u_2) = \alpha_1 S(u_1) + \alpha_2 S(u_2)$. By the hypothesis of bicontinuity of $\mathcal{S}$ we have that $\|Su\|^2_{\mathscr{H}_1} = |\langle Su, Su \rangle_{\mathscr{H}_1}| = |\mathcal{S}(u, Su)| \leq M\|u\|_{\mathscr{H}_2}\|Su\|_{\mathscr{H}_2}$. Since for $\|Su\|_{\mathscr{H}_1} \neq 0$, $\|Su\|_{\mathscr{H}_1} \leq M\|u\|_{\mathscr{H}_2}$, and for $\|Su\|_{\mathscr{H}_1} = 0$ this inequality is trivially true, then $S : \mathscr{H}_2 \to \mathscr{H}_1$ is a limited linear operator $S \in \mathfrak{B}(\mathscr{H}_2, \mathscr{H}_1)$. The unicity of $S$ is proven by considering that if there were another such linear map $\tilde{S} : \mathscr{H}_2 \to \mathscr{H}_1$ then since for any $u \in \mathscr{H}_2$ we have that $\langle Su, v \rangle - \langle \tilde{S}u, v \rangle = 0, \forall v \in \mathscr{H}_1$ this implies $0 = \langle (S - \tilde{S})u, (S - \tilde{S})u \rangle = \|(S - \tilde{S})u\|^2 \Rightarrow (S - \tilde{S})u = 0 \Rightarrow Su = \tilde{S}u$, but $u$ is arbitrary, then $S = \tilde{S}$, since they map the same inputs to the same outputs. ∎

**Theorem B.0.15** (Bounded Linear Transformation Theorem). *Considering a normed vector space $\mathcal{V}$ with norm $\|\cdot\|_{\mathcal{V}}$ and another normed vector space $\mathcal{W}$ with norm $\|\cdot\|_{\mathcal{W}}$, that is complete in this norm, i.e., $(\mathcal{W}, \|\cdot\|_{\mathcal{W}})$ is a Banach space, if $\mathcal{V}$ is itself a dense subspace of a complete normed vector space* $\tilde{\mathcal{V}}$, *with norm* $\|\cdot\|_{\tilde{\mathcal{V}}}$ *such that* $\|v\|_{\mathcal{V}} = \|v\|_{\tilde{\mathcal{V}}}$, $\forall v \in \mathcal{V}$. *Then for every bounded linear transformation* $T : \mathcal{V} \to \mathcal{W}$, $T \in \mathfrak{B}(\mathcal{V}, \mathcal{W})$, *there exists a unique extension* $\tilde{T} : \tilde{\mathcal{V}} \to \mathcal{W}$, $\tilde{T} \in \mathfrak{B}(\tilde{\mathcal{V}}, \mathcal{W})$ *such that* $\|\tilde{T}\|_{\mathfrak{B}(\tilde{\mathcal{V}},\mathcal{W})} := \sup_{\substack{\tilde{v} \in \tilde{\mathcal{V}} \\ \tilde{v} \neq 0}} \frac{\|\tilde{T}\tilde{v}\|_{\mathcal{W}}}{\|\tilde{v}\|_{\tilde{\mathcal{V}}}} = \sup_{\substack{v \in \mathcal{V} \\ v \neq 0}} \frac{\|Tv\|_{\mathcal{W}}}{\|v\|_{\mathcal{V}}} =: \|T\|_{\mathfrak{B}(\mathcal{V},\mathcal{W})}$. *This theorem is usually called informally the BLT theorem in the literature.*

**Definition B.0.16** (Weak-* topology). *For a topological vector space $X$ the weak-* topology is a topology on $X^{\#}$, the continuous dual space of $X$, whose topological basis is generated by the following family of seminorms:*

$$\varrho^{w\text{-}*}_{\{x_k\}_{k=1}^n}(A) := \sup_{j \in \{1,\ldots,n\}} \left\{ |Ax_j| \,\Big|\, x_j \in \{x_k\}_{k=1}^n \right\}.$$

**Proposition B.0.17.** *The weak-* topology is Hausdorff.*

*Proof:* This is proven by showing that the set of seminorms that generate the topological basis is separating, *i.e.* $\forall y \in Y$, $y \neq \mathbf{0}_Y$ implies that there is a seminorm $\varrho$, of the family that generates the weak-* topology, such that $\varrho(y) \neq 0$. The usual notion of separability comes by considering the difference of points, like $x - y$.

Let $f \in X^{\#}$ be such that $f \neq 0$, then $\exists x \in X$, such that $f(x) \neq 0$, then $\varrho^{w\text{-}*}_{\{x\}}(f) = |f(x)| \neq 0$, hence there is a norm on the family that generates the weak-* topology that is separating for each $f \in X^{\#}$, hence the weak-* topology is Hausdorff. ∎

We now also state the Banach-Alaoglu theorem, whose proof can be found in [86, 15, 89].

**Theorem B.0.18** (Banach-Alaoglu). *If $(\mathcal{V}, \|\cdot\|_{\mathcal{V}})$ is a normed vector space then the closed unit ball in the continuous dual space $\mathcal{V}^{\#}$, i.e. the set of continuous linear functionals over $\mathcal{V}$ endowed with its usual operator norm, is compact with respect to the weak-* topology.*

In the following we enunciate the renowned Liouville theorem in complex analysis, we give no proof of such theorem since it is widely know and can be found in standard texts such as [88] theorem 10.23 in page 212, [24] theorem 3.4 page 77, and a generalized version for functions valued at Banach spaces in [105] corollary 4 in page 129.

**Theorem B.0.19** (Liouville). *If $f : \mathbb{C} \to \mathbb{C}$ is an entire function that is also limited, that is, there exists $M \in \mathbb{R}_+$ such that $|f(z)| \leq M$, $\forall z \in \mathbb{C}$; then $f$ is constant.*

In the following we will enunciate the Stone-Weierstrass Theorem, but before that we will introduce the notions of a separating set of functions and of a set of functions that vanish nowhere.

**Definition B.0.20.** *Let $\mathscr{F}$ be a family of functions on a set $X$, then $\mathscr{F}$ is said to **separate points** on $X$ if to every pair $x_1, x_2 \in X$ of distinct points $x_1 \neq x_2$, there exists a function $f \in \mathscr{F}$ such that $f(x_1) \neq f(x_2)$.*

**Definition B.0.21.** *Let $\mathscr{F}$ be a family of functions from a set $X$ either to $\mathbb{R}$ or $\mathbb{C}$, then $\mathscr{F}$ is said to **vanish nowhere** on $X$ if for every $x \in X$, $\exists f \in \mathscr{F}$ such that $f(x) \neq 0$.*

The Stone-Weierstrass Theorem is a standard result of mathematical analysis and can be found in any number of books [87, 105, 34, 86], the importance of the Stone-Weierstrass theorem lies in the fact that it yields a method of uniform approximation of all continuous functions defined on a compact space $X$ by special classes of functions.

**Theorem B.0.22** (Stone-Weierstrass)**.** *Suppose $X$ is a locally compact Hausdorff space and $\mathscr{C}$ is a is a subalgebra of $C_0(X, \mathbb{C})$. Then $\mathscr{C}$ is dense in $C_0(X, \mathbb{C})$ (in the topology of the supremum norm) if and only if $\mathscr{C}$ is a self-adjoint subalgebra of $C(X, \mathbb{C})$, it separates points and vanishes nowhere.*

**Theorem B.0.23** (Hahn-Banach)**.** *Suppose $\varrho : X \to \mathbb{R}$ is a seminorm on the vector space $X$ over the field $\mathbb{K} = \mathbb{R}$ or $\mathbb{C}$, if on a vector subspace $M \subseteq X$ a linear functional $f : M \to \mathbb{K}$ is such that:*

$$|f(m)| \leq \varrho(m) \quad , \forall m \in M.$$

*Then there exists a linear functional $F : X \to \mathbb{K}$ that extends $f$ in such a way that:*

$$F(m) = f(m) \quad , \forall m \in M,$$

$$|F(x)| \leq \varrho(x) \quad , \forall x \in X.$$

**Theorem B.0.24** (Geometric Hahn-Banach)**.** *Let $A, B$ be be non-empty convex subsets of a locally convex topological vector space $X$. If the interior $\mathring{A} \neq \varnothing$ and $B \cap \mathring{A} = \varnothing$ then there exists a non-zero continuous linear functional $f$ on $X$ such that $\sup_{a \in A} \mathfrak{Re}(f)(A) \leq \inf_{b \in B} \mathfrak{Re}(f)(B)$ and also $\mathfrak{Re}(f)(a) < \inf_{b \in B} \mathfrak{Re}(f)(B)$ for all $a \in \mathring{A}$.*

We take advantage of having enunciated the previous fundamental theorems of functional analysis to enunciate another fundamental theorem, the Closed Graph theorem for Banach spaces, whose proof can be found in virtually any functional analysis book such as [68, 86, 105].

**Theorem B.0.25** (Closed Graph)**.** *If $T : X \to Y$ is an everywhere-defined linear operator between Banach spaces $X$ and $Y$, then the following are equivalent:*

*(a) $T$ is continuous;*

*(b) The graph of $T$ is closed in the product topology on $X \times Y$, i.e. $T$ is closed;*

*(c) If the sequence $\{x_n\}_{n \in \mathbb{N}}$ converges to $x \in X$, then the sequence $\{Tx_n\}_{n \in \mathbb{N}}$ converges to $Tx \in Y$.*

**Proposition B.0.26.** *If $X$ is a a compact space, $Y$ is a topological space whose topology is Hausdorff and $f : X \to Y$ is a continuous bijection, then $f$ is a homeomorphism.*

*Proof:* Since $f$ is a bijection, we can take $g = f^{-1}$, and to prove that $f$ is a homeomorphism we must show that $g : Y \to X$ is continuous. For any $V \subseteq X$, we have that $g^{-1}(V) = f(V)$. Then we need to show that if $V$ is closed in $X$, then $g^{-1}(V)$ is closed in $Y$.

We suppose $V$ is closed in $X$, and since $X$ is compact, $V$ is compact, and hence $f(V)$ is compact, as it is the image of a compact set by a continuous function.

For a Hausdorff topological space, compact sets are closed by the following argument: For a topological space $T$, whose topology is Hausdorff, we consider a compact subset $C \subset T$. Let $a \in T \setminus C$, for any point $x \in C$, the Hausdorff condition guarantees the existence of disjoint open sets $U_a^x$ that contains $a$ but not $x$, and $W_x^a$ that contains $x$ but doesn't contain $a$.

If $C$ is composed of a finite number of points $\{x_i\}_{i=1}^n$, then we can take $U_a = \bigcap_{i=1}^n U_a^{x_i}$, and get that $a \in U_a \subset T \setminus C$. If $C$ is not finite, then the set $\{W_x^a \subset T \,|\, x \in C\}$ is an open cover for $C$, since $C$ is compact it has a finite subcover, say $\{W^a_{x_j}\}_{j=1}^m$. Let $U_a = \bigcap_{j=1}^m U^{x_j}_a$, then $U_a$ is open because it is a finite intersection of open sets, also $a \in U_a$, as $a \in U^{x_j}_a$, $\forall j \in \{1, \ldots, m\}$.

Finally if $b \in U_a$, then $b \in U^{x_j}_a$, $\forall j \in \{1, \ldots, m\}$, and since $C \subseteq \bigcup_{j=1}^m W^a_{x_j}$, as $b \notin W^a_{x_j}$, $\forall j \in \{1, \ldots, m\}$, $b \notin C$, thus $U_a \subset T \setminus C$. Considering that $T \setminus C = \bigcup_{a \in T \setminus C} U_a$, this makes $T \setminus C$ be open, it is a arbitrary union of open sets, hence its complement $(T \setminus C)^\complement = C$ is closed.

Therefore as $f(V)$ is compact in a Hausdorff topology, $f(V)$ is closed, but $f(V) = g^{-1}(V)$, which implies that $g^{-1}(V)$ is closed. From this it follows that $g$ is continuous, and hence $f$ is a homeomorphism. ∎

The following proposition is a known proposition in Functional analysis and basically every book on the subject has a proof of it, because of the usual length of these proofs we will omit the proof, only enunciating the statement of the proposition and indicating the books [68, 86, 77] for proofs.

**Proposition B.0.27.** *Let $T : \mathscr{H} \to \mathscr{H}$ be a positive bounded self-adjoint linear operator on a complex Hilbert space $\mathscr{H}$, then there is a unique bounded self-adjoint linear operator $A$, for which:*

$$T = A^2.$$

*This operator $A$ commutes with every bounded linear operator that commutes with $T$.*

**Proposition B.0.28.** *Let $\mathscr{B}$ be a Banach algebra with unity, for every $w \in \mathscr{B}$ with $\|w\| < 1$ then there exists $(\mathbb{1} - w)^{-1} \in \mathscr{B}$ given by:*

$$(\mathbb{1} - w)^{-1} = \mathbb{1} + \sum_{k=1}^{\infty} w^k, \tag{B.8}$$

*where the RHS converges in the $\mathscr{B}$ norm, this series is called the Neumann series.*

*Proof:* Firstly we will prove that the Neumann series converges, if $s_n := \mathbb{1} + \sum_{k=1}^n w^k$, then for $m < n$ we have that $s_n - s_m = \sum_{k=m+1}^n w^k$ which implies:

$$\|s_n - s_m\| \leq \sum_{k=m+1}^n \|w^k\| \leq \sum_{k=m+1}^n \|w\|^k = \|w\|^{m+1} \sum_{k=0}^{n-m-1} \|w\|^k \leq \|w\|^{m+1} \sum_{k=0}^{\infty} \|w\|^k = \frac{\|w\|^{m+1}}{1 - \|w\|}.$$

where we used that $\sum_{k=0}^{\infty} \|w\|^k$ is the infinite sum of a geometric progression with $\|w\| < 1$ which converges to $\frac{1}{1-\|w\|}$, and for this same reason $\|w\|^{m+1}$ can be made as small as needed by choosing $m$ big enough, this proves that $(s_n)_{n \in \mathbb{N}}$ is a Cauchy sequence in the norm of $\mathscr{B}$, and therefore converges. Being $v \in \mathscr{B}$ its limit, we have:

$$wv = w + w\left(\lim_{n \to \infty} \sum_{k=1}^n w^k\right) = w + \lim_{n \to \infty} \sum_{k=1}^n w^{k+1} = w + \lim_{n \to \infty}\left[\left(\sum_{k=1}^n w^k\right) + w^{k+1} - w\right] =$$

$$= \lim_{n \to \infty} w^{k+1} + \lim_{n \to \infty} \sum_{k=1}^n w^k = v - \mathbb{1}.$$

where it was used the continuity of the product in $\mathcal{B}$ and the fact that $\lim_{n\to\infty} w^{n+1} = 0$, because $\|w^{n+1}\| \leq \|w\|^{n+1} \xrightarrow{n\to\infty} 0$, since $\|w\| < 1$. This makes $(\mathbb{1} - w)v = v - (v - \mathbb{1}) = \mathbb{1}$. Analogously:

$$vw = w + \left(\lim_{n\to\infty} \sum_{k=1}^{n} w^k\right) w = w + \lim_{n\to\infty} \sum_{k=1}^{n} w^{k+1} = w + \lim_{n\to\infty} \left[\left(\sum_{k=1}^{n} w^k\right) + w^{k+1} - w\right] =$$

$$= \lim_{n\to\infty} w^{k+1} + \lim_{n\to\infty} \sum_{k=1}^{n} w^k = v - \mathbb{1}.$$

and from this we conclude that $v(\mathbb{1} - w) = v - (v - \mathbb{1}) = \mathbb{1}$. ∎

**Corollary B.0.29.** *Considering $\mathcal{B}$ to be a Banach algebra with unity, and $w \in \mathcal{B}$ a invertible element, if $v \in \mathcal{B}$ is such that $\|\mathbb{1} - vw^{-1}\| < 1$, a case of particular interest of this inequality occurs when $\|w - v\| < \|w^{-1}\|^{-1}$. Then $v$ is invertible and can be written with the following series that converges in the norm of $\mathcal{B}$:*

$$v^{-1} = w^{-1}\left(\mathbb{1} + \sum_{k=1}^{\infty} (\mathbb{1} - vw^{-1})^k\right).$$

*Proof:* Since $v = v - w + w = (\mathbb{1} - (w-v)w^{-1})w$, then $\mathbb{1} - (w-v)w^{-1}$ will be invertible by proposition B.0.28 if $\|(w-v)w^{-1}\| < 1$, since by the properties of a Banach algebra $\|(w-v)w^{-1}\| \leq \|(w-v)\|\|w^{-1}\|$, therefore the hypothesis of proposition B.0.28 will be fulfilled if $\|(w-v)\| < \|w^{-1}\|^{-1}$, and then:

$$v^{-1} = w^{-1}\left(\mathbb{1} - (w-v)w^{-1}\right)^{-1} \stackrel{(B.8)}{=} w^{-1}\left(\mathbb{1} + \sum_{k=1}^{\infty} \left[(w-v)w^{-1}\right]^k\right) = w^{-1}\left(\mathbb{1} + \sum_{k=1}^{\infty} \left(\mathbb{1} - vw^{-1}\right)^k\right).$$

∎

**Definition B.0.30** (Spectrum). *Let $x$ be an element of an algebra $\mathscr{A}$. If $\mathscr{A}$ has a identity element $\mathbb{1}$, the spectrum of $x$ is defined to be:*

$$\sigma_{\mathscr{A}}(x) = \{\lambda \in \mathbb{C} \,|\, \nexists\, (x - \lambda\mathbb{1})^{-1} \in \mathscr{A}\}.$$

*If $\mathscr{A}$ has no identity element then the spectrum of $x$ is defined to be the set:*

$$\sigma_{\mathscr{A}}(x) = \left\{\lambda \in \mathbb{C} \,\bigg|\, \lambda \neq 0 \text{ and } \nexists\, z, w \in \mathscr{A},\, \frac{x}{\lambda} + z - \frac{xz}{\lambda} = w + \frac{x}{\lambda} - \frac{wx}{\lambda} = 0\right\} \cup \{0\}.$$

*Clearly if $(x - \lambda\mathbb{1})$ is invertible and $\lambda \neq 0$ then $\left(\mathbb{1} - \frac{x}{\lambda}\right)$ is inverible with inverse given by $\left(\mathbb{1} - \frac{x}{\lambda}\right)^{-1} = -\lambda(x - \lambda\mathbb{1})^{-1}$, as $-\lambda(x - \lambda\mathbb{1})^{-1}\left(\mathbb{1} - \frac{x}{\lambda}\right) = (x - \lambda\mathbb{1})^{-1}(x - \lambda\mathbb{1}) = \mathbb{1}$; which implies that $\frac{x}{\lambda} + z - \frac{xz}{\lambda} = \frac{x}{\lambda} + \left(\mathbb{1} - \frac{x}{\lambda}\right)z$ is zero for $z = -\left(\mathbb{1} - \frac{x}{\lambda}\right)^{-1}\frac{x}{\lambda}$ and $w + \frac{x}{\lambda} - \frac{wx}{\lambda} = \frac{x}{\lambda} + w\left(\mathbb{1} - \frac{x}{\lambda}\right)$ which is also zero for $w = -\frac{x}{\lambda}\left(\mathbb{1} - \frac{x}{\lambda}\right)^{-1}$ and hence all spectrum values that are on the first definition are on the second. If in converse there are no $z, w \in \mathscr{A}$ such that $\frac{x}{\lambda} + z - \frac{xz}{\lambda} = w + \frac{x}{\lambda} - \frac{wx}{\lambda} = 0$ then there can be no inverse of $\left(\mathbb{1} - \frac{x}{\lambda}\right)$ for taking $z = -\left(\mathbb{1} - \frac{x}{\lambda}\right)^{-1}\frac{x}{\lambda}$ and $w = -\frac{x}{\lambda}\left(\mathbb{1} - \frac{x}{\lambda}\right)^{-1}$ would make the previous statement false, therefore all the spectrum values of the second definition, with the possible exception of the value $\lambda = 0$, are in the first definition.*

*For each element $x \in \mathscr{A}$ if there exists $z \in \mathscr{A}$ such that the first condition on the second definition of spectrum $x + z - xz = 0$ is fulfilled (we can multiply the entire equation by $\lambda \neq 0$ without*

*loss of generality), then we call x **right quasi-regular**, if there exists $w \in \mathscr{A}$ that fulfills the second condition on the second definition of spectrum $w + x - wx = 0$ then we call x **left quasi-regular**. x is **quasi-regular** if it is both **right quasi-regular** and **left quasi-regular**.*

*We define the resolvent set as being $\rho_{\mathscr{A}}(x) = \mathbb{C} \setminus \sigma_{\mathscr{A}}(x)$. We shall write $\sigma(x)$ and $\rho(x)$ when the underlying algebra is implicitly understood.*

**Theorem B.0.31.** *If $x, y$ are elements of an algebra $\mathscr{A}$, then the sets $\sigma_{\mathscr{A}}(xy)$ and $\sigma_{\mathscr{A}}(yx)$ differ at most by the number 0; i.e. $\sigma_{\mathscr{A}}(xy) \cup \{0\} = \sigma_{\mathscr{A}}(yx) \cup \{0\}$.*

*Proof:* Let $\lambda$ be a nonzero complex number, if $\lambda \notin \sigma_{\mathscr{A}}(xy)$, it follows from definition B.0.30 $\frac{xy}{\lambda}$ is quasi-regular and hence $\exists z, w \in \mathscr{A}$ such that $\frac{xy}{\lambda} + z - \frac{xyz}{\lambda} = w + \frac{xy}{\lambda} - \frac{wxy}{\lambda} = 0$, letting $u = -\frac{yx}{\lambda} + \frac{yzx}{\lambda}$ and $v = -\frac{yx}{\lambda} + \frac{ywx}{\lambda}$ calculating:

$$yx + u - yxu = yx + (-yx + yzx) - yx(-yx + yzx) = y(z + xy - xyz)x = y(0)x = 0$$

and

$$yx + v - vyx = yx + (-yx + ywx) - (-yx + ywx)yx = y(w + xy - wxy)x = y(0)x = 0,$$

we see then that $\frac{yx}{\lambda}$ is quasi-regular and hence $\lambda \notin \sigma_{\mathscr{A}}(yx)$. ∎

**Definition B.0.32** (Resolvent operator). *Let $\mathscr{B}$ be a Banach algebra with unity, $u \in \mathscr{B}$, if $\mu \in \rho(u)$, we define the **resolvent operator** of $u$ calculated at $\mu$ by:*

$$R_\mu(u) := (\mu \mathbb{1} - u)^{-1}.$$

*since $\mu \in \rho(u)$, $\mu \mathbb{1} - u$ is invertible and therefore $R_\mu(u) \in \mathscr{B}$.*

**Lemma B.0.33.** *Let $\mathscr{B}$ be a Banach algebra with unity, if $u \in \mathscr{B}$, $\mu \in \rho(u)$ and $\lambda \in \mathbb{C}$ is such that $|\lambda - \mu| < \|R_\mu(u)\|^{-1}$, then it is also true that $\lambda \in \rho(u)$, as are the following equalities:*

$$R_\lambda(u) = R_\mu(u) \left[ \mathbb{1} + \sum_{n=1}^{\infty} (\mu - \lambda)^n (R_\mu(u))^n \right] = \left[ \mathbb{1} + \sum_{n=1}^{\infty} (\mu - \lambda)^n (R_\mu(u))^n \right] R_\mu(u). \quad \text{(B.9)}$$

*Proof:* Since $|\lambda - \mu| < \|R_\mu(u)\|^{-1}$ then $\|(\lambda - \mu)R_\mu(u)\| < 1$ so that by the same reasoning as in proposition B.0.28, both series converge to elements of $\mathscr{B}$, and then the fact that the series commutes with $R_\mu(u)$ is evident. To prove the first equality we then calculate:

$$(\lambda \mathbb{1} - u)R_\mu(u) = ((\lambda - \mu)\mathbb{1} + (\mu \mathbb{1} - u)) R_\mu(u) = -(\mu - \lambda)R_\mu(u) + \mathbb{1}.$$

This makes:

$$(\lambda \mathbb{1} - u)R_\mu(u) \left[ \mathbb{1} + \sum_{n=1}^{\infty} (\mu - \lambda)^n (R_\mu(u))^n \right] =$$

$$= -(\mu - \lambda)R_\mu(u) \left[ \mathbb{1} + \sum_{n=1}^{\infty} (\mu - \lambda)^n (R_\mu(u))^n \right] + \left[ \mathbb{1} + \sum_{n=1}^{\infty} (\mu - \lambda)^n (R_\mu(u))^n \right] =$$

$$= -\sum_{n=1}^{\infty} (\mu - \lambda)^n (R_\mu(u))^n + \mathbb{1} + \sum_{n=1}^{\infty} (\mu - \lambda)^n (R_\mu(u))^n = \mathbb{1}.$$

∎

**Proposition B.0.34.** *Let $\mathcal{B}$ be a Banach algebra with unity, if $u \in \mathcal{B}$, then $\rho(u)$ is a open subset of $\mathbb{C}$, which implies, by the complement, that $\sigma(u)$ is a closed subset of $\mathbb{C}$.*

*Proof:* The lemma B.0.33 says that for every $\mu \in \rho(u)$ we have that, for the following open neighbourhoods $\{\lambda \in \mathbb{C} \,|\, |\lambda - \mu| < \|R_\mu(u)\|^{-1}\} \subseteq \rho(u)$, so $\rho(u)$ is open, and by the complement $\sigma(u)$ is a closed subset of $\mathbb{C}$. ∎

The proof of the following propositions are needed to prove the Spectral mapping theorem, they are standard results proven originally by Gel'fand.

**Proposition B.0.35.** *Let $\mathcal{B}$ be a Banach algebra with unity, for $u \in \mathcal{B}$, and every $F \in \mathcal{B}'$, a continuous linear functional in $\mathcal{B}$, then the function $g_F : \rho(u) \to \mathbb{C}$ given by $g_F(\lambda) := F(R_\lambda(u))$ is holomorphic in each connected component of $\rho(u)$.*

*Proof:* Let $\mu \in \rho(u)$ and $\lambda \in \mathbb{C}$ such that $|\lambda - \mu| < \|R_\mu(u)\|^{-1}$, then by lemma B.0.33 $\lambda \in \rho(u)$ and:

$$g_F(\lambda) := F(R_\lambda(u)) \stackrel{(B.9)}{=} F\left(R_\mu(u) + \sum_{n=1}^{\infty} (\mu - \lambda)^n (R_\mu(u))^{n+1}\right) =$$

$$\stackrel{\text{Cont.}}{=} F(R_\mu(u)) + \sum_{n=1}^{\infty} (\mu - \lambda)^n F\left((R_\mu(u))^{n+1}\right). \quad (B.10)$$

Considering then the estimate:

$$\left|F\left((R_\mu(u))^{n+1}\right)\right| \leq \|F\|_{\mathcal{B}'} \left\|(R_\mu(u))^{n+1}\right\| \leq \|F\|_{\mathcal{B}'} \|R_\mu(u)\|^{n+1},$$

where $\|\cdot\|_{\mathcal{B}'}$ is the usual supremum norm in $\mathcal{B}'$ given by $\|F\|_{\mathcal{B}'} := \sup_{x \in \mathcal{B}} \frac{|F(x)|}{\|x\|}$, and from $|\lambda - \mu| < \|R_\mu(u)\|^{-1}$ we see that B.10 is absolutely convergent, analogously to B.9, and therefore this series defines a holomorphic function on the open ball of radius $\|R_\mu\|^{-1}$ centered at $\mu$, $\mathcal{B}^\mu_{\|R_\mu\|^{-1}}(\mathbb{C})$, that can be extended by the standard method of analytic continuation to the connected component of $\rho(u)$ that contains $\mu$. ∎

**Lemma B.0.36.** *If $\mathcal{V}$ is a normed complex vector space and $x \in \mathcal{V}$ is such that $\forall F \in \mathcal{V}'$, $F(x) = 0$, then $x = 0$.*

*Proof:* Lets first construct a functional $l_{x_0} : \mathcal{V} \to \mathbb{C}$ such that $\|l_{x_0}\|_{\mathcal{V}'} = 1$ and $l_{x_0}(x_0) = \|x_0\|$, if $x_0 = 0$ any functional with norm 1 of $\mathcal{V}'$ suffices, if $x_0 \neq 0$ then considering the linear subspace $V_1 := \{\alpha x_0 \,|\, \alpha \in \mathbb{C}\}$, we can define the functional $\breve{l}_{x_0} : V_1 \to \mathbb{C}$ by $\breve{l}_{x_0}(\alpha x_0) := \alpha \|x_0\|$. We can extend $\breve{l}_{x_0}$ by the Hahn-Banach theorem B.0.23 to a functional $l_{x_0} : \mathcal{V} \to \mathbb{C}$ that extends $\breve{l}_{x_0}$ with $\|l_{x_0}\|_{\mathcal{V}'} = \|\breve{l}_{x_0}\|_{\mathcal{V}'}$ and $l_{x_0}(x_0) = \breve{l}_{x_0}(x_0) = \|x_0\|$, but more than that:

$$\|\breve{l}_{x_0}\|_{\mathcal{V}'} = \sup_{\substack{y \in V_1 \\ y \neq 0}} \frac{|\breve{l}_{x_0}(y)|}{\|y\|} = \sup_{\substack{\alpha \in \mathbb{C} \\ \alpha \neq 0}} \frac{|\breve{l}_{x_0}(\alpha x_0)|}{\|\alpha x_0\|} = \sup_{\substack{\alpha \in \mathbb{C} \\ \alpha \neq 0}} \frac{|\alpha| \|x_0\| \,|}{\|\alpha x_0\|} = 1$$

and so $\|l_{x_0}\|_{\mathcal{V}'} = 1$. To prove the lemma we consider the hypothesis that $\forall F \in \mathcal{V}'$, $F(x) = 0$ then in particular $l_x(x) = 0$, but $l_x(x) = \|x\| \Rightarrow \|x\| = 0$ and therefore $x = 0$. ∎

**Proposition B.0.37.** *Let $\mathcal{B}$ be a Banach algebra with unity, if $u \in \mathcal{B}$, then $\sigma(u)$ is a non empty set and it is contained in the closed ball of radius $\|u\|$ centered at 0, $\overline{\mathcal{B}}_{\|u\|}(\mathbb{C}) = \{z \in \mathbb{C} \,|\, |z| \leq \|u\|\}$.*

*Proof:* Lets suppose that $\rho(u) = \mathbb{C}$, then by proposition B.0.35, $\forall F \in \mathcal{B}'$ the defined $g_F(\lambda) := F(R_\lambda(u))$ would be an entire function, that is, analytic everywhere in $\mathbb{C}$. Now, for $|\lambda| > \|u\| \Rightarrow \|\lambda^{-1} u\| < 1$, the proposition B.0.28 gives us:

$$R_\lambda(u) = (\mathbb{1}\lambda - u)^{-1} = \lambda^{-1}(\mathbb{1} - \lambda^{-1} u)^{-1} \stackrel{(B.8)}{=} \lambda^{-1} \left[\mathbb{1} + \sum_{n=1}^{\infty} \lambda^{-n} u^n \right] \tag{B.11}$$

So,

$$\|R_\lambda(u)\| \leq \frac{1}{|\lambda|}\left[1 + \sum_{n=1}^{\infty}\left(\frac{\|u\|}{|\lambda|}\right)^n\right] = \frac{1}{|\lambda| - \|u\|}.$$

where in the last equality the infinite sum of a decreasing geometric sum was used. The last expression tells us that $\lim_{|\lambda| \to \infty} \|R_\lambda(u)\| = 0$, therefore by $|g_F(\lambda)| = |F(R_\lambda)(u)| \leq \|F\|_{\mathcal{B}'}\|R_\lambda(u)\|$, from this follows that $\lim_{|\lambda| \to \infty} |g_F(\lambda)| = 0$, and then $g_F$ is a entire function that is limited and converges to zero as $\lambda$ goes to infinity. By the famous Liouville theorem B.0.19 of complex analysis, these facts imply that $g_F$ is identically zero for all $\lambda \in \mathbb{C}$, so that $F(R_\lambda(u)) = 0, \forall F \in \mathcal{B}'$, then by lemma B.0.36 $R_\lambda(u) = 0$, absurd, since $R_\lambda(u)$ is by definition the inverse of an operator. We conclude that $\rho(u) \neq \mathbb{C}$ from which follows $\sigma(u) \neq \varnothing$. Since equation (B.11) is valid for $|\lambda| > \|u\|$ then $\{z \in \mathbb{C} \,|\, |z| > \|u\|\} \subset \rho(u)$ which implies, by taking complements, $\sigma(u) \subset \{z \in \mathbb{C} \,|\, |z| \leq \|u\|\}$. ∎

**Theorem B.0.38** (Gel'fand-Mazur)**.** *If $\mathcal{A}$ is a unital Banach algebra in which every non-zero element is invertible, then $\mathcal{A} = \mathbb{C}\mathbb{1}$.*

*Proof:* This follows immediately from proposition B.0.37, by the following argument: Let $a \in \mathcal{A}$ be a non-zero element, then $\sigma(a)$ is non empty by proposition B.0.37, lets say $\mathbb{C} \ni \lambda \in \sigma(a)$, then $a - \lambda\mathbb{1}$ is not invertible, but $a - \lambda\mathbb{1} \in \mathcal{A}$ and every non-zero element of $\mathcal{A}$ is invertible, hence $a - \lambda\mathbb{1} = 0 \Rightarrow a = \lambda\mathbb{1}$. This has as consequence that $\mathcal{A} = \mathbb{C}\mathbb{1}$. ∎

In the following we define the spectral radius.

**Definition B.0.39.** *Let $\mathcal{B}$ be a Banach algebra with unity, if $u \in \mathcal{B}$, we define the **spectral radius** of $u$ by:*

$$r(u) := \sup_{\lambda \in \sigma(u)} |\lambda|.$$

*By Proposition B.0.37 it is clear that $r(u) \leq \|u\|$.*

Then we enunciate the following fundamental theorem, initially proven by Beurling, called the Spectral Radius Theorem although it can easily be found in standard functional analysis texts, such as the aforementioned [68, 105, 8].

**Theorem B.0.40** (Spectral Radius Theorem)**.** *Let $\mathcal{B}$ be a Banach algebra with unity, if $u \in \mathcal{B}$, then*

$$r(u) = \inf_{n \geq 1} \|u^n\|^{\frac{1}{n}} = \lim_{n \to \infty} \|u^n\|^{\frac{1}{n}}.$$

**Theorem B.0.41.** *If $\mathscr{A}$ is a $C^*$-algebra with a unit, considering $a \in \mathscr{A}$ that is either self-adjoint $a = a^*$ or normal $aa^* = a^*a$, then the spectral radius has the following equality:*

$$r(a) := \sup_{\lambda \in \sigma(a)} |\lambda| = \|a\|. \tag{B.12}$$

*More generally if $\mathscr{C} \subset \mathscr{A}$ is a unital $C^*$-subalgebra, with the same unit, of the unital $C^*$-algebra $\mathscr{A}$ then $\sigma_{\mathscr{A}}(x) = \sigma_{\mathscr{C}}(x), \forall x \in \mathscr{C}$.*

*Proof:* By the $C^*$ property $\forall b \in \mathscr{A}, \|b^*b\| = \|b\|^2$, therefore for a self-adjoint element $a$ of the algebra $\|a^2\| = \|a\|^2$, and applying iteratively this property to $a^{2^{n-1}}$ that is also self-adjoint since it is the product of $2^{n-1}$ self-adjoint elements:

$$\|a^{2^n}\| = \|a^{2^{n-1}}\|^2 = \|a^{2^{n-2}}\|^{2^2} = ... = \|a\|^{2^n}. \tag{B.13}$$

This implies by Lemma B.0.40 that:

$$r(a) = \lim_{k \to \infty} \|a^k\|^{\frac{1}{k}} = \lim_{n \to \infty} \|a^{2^n}\|^{\frac{1}{2^n}} \stackrel{(B.13)}{=} \lim_{n \to \infty} \|a\| = \|a\|. \tag{B.14}$$

Proceeding to the proof of the normal element case, invoking again the $C^*$ property, $\forall b \in \mathscr{A}, \|(b^{2^n})^* b^{2^n}\| = \|b^{2^n}\|^2$, and for a normal operator $a$ we have that $(a^{2^n})^* a^{2^n} = (a^*a)^{2^n}$, which together make $\|a^{2^n}\|^2 = \|(a^*a)^{2^n}\|$. Considering now that $a^*a$ is self-adjoint then by equation (B.13) we have $\|(a^*a)^{2^n}\| = \|a^*a\|^{2^n}$ so that by the $C^*$ property $\|a^*a\|^{2^n} = \|a\|^{2^{n+1}} \Rightarrow \|a^{2^n}\|^2 = (\|a\|^{2^n})^2 \Rightarrow \|a^{2^n}\| = \|a\|^{2^n}$ from where we can then follow the chain of equalities in (B.14) again, completing the proof of the first part.

More generally we get that if $\mathscr{C} \subset \mathscr{A}$ is a unital $C^*$-subalgebra, with the same unit of the unital $C^*$-algebra $\mathscr{A}$, then $\sigma_{\mathscr{A}}(x) = \sigma_{\mathscr{C}}(x), \forall x \in \mathscr{C}$, the assertion $\sigma_{\mathscr{A}}(x) \subseteq \sigma_{\mathscr{C}}(x)$ is clear since the inverible elements of $\mathscr{C}$, Inv($\mathscr{C}$), are contained in Inv($\mathscr{A}$), hence Inv($\mathscr{A}$)$^\complement$ $\subseteq$ Inv($\mathscr{C}$)$^\complement$.

Considering a $x \in \mathscr{C}$ that is invertible in $\mathscr{A}$, then $x^*x$ is a self-adjoint element of $\mathscr{C}$ that is also invertible in $\mathscr{A}$, by theorem B.0.48 $\sigma_{\mathscr{C}}(x^*x) \subseteq \mathbb{R}$, in particular every point of $\sigma_{\mathscr{C}}(x^*x)$ is a boundary point. The useful fact that $\partial \sigma_{\mathscr{C}}(y) \subseteq \sigma_{\mathscr{A}}(y), \forall y \in \mathscr{C}$ is apparent from considering the contradiction of assuming that if $0 \notin \sigma_{\mathscr{A}}(y)$ and $0 \in \partial \sigma_{\mathscr{C}}(y)$, then $y \in \mathscr{B}$ is invertible in $\mathscr{A}$ and there is a sequence of complex numbers $\{\lambda_n\}_{n \in \mathbb{N}}$ with $\lambda_n \to 0$ such that $\lambda_n \notin \sigma_{\mathscr{C}}(y)$, thus $(y - \lambda_n \mathbb{1})^{-1}$ is a sequence of elements of $\mathscr{C}$ with the property that, since inversion is continuous in Inv($\mathscr{A}$), the sequence converges to $y^{-1}$ as $n \to \infty$. It follows that $y^{-1} = \lim_{n \to \infty}(y - \lambda_n \mathbb{1})^{-1} \in \overline{\mathscr{C}} = \mathscr{C}$, contradicting the hypothesis that $0 \in \sigma_{\mathscr{C}}(y)$, hence if $0 \in \partial \sigma_{\mathscr{C}}(y)$ then $0 \in \sigma_{\mathscr{A}}(y)$, we have proven that for 0, but the same argument can be applied for any $\lambda \in \partial \sigma_{\mathscr{C}}(y)$ by exchanging $y$ for $y - \lambda_n \mathbb{1}$ (while keeping the $y$ in the expressions $\sigma_{\mathscr{C}}(y)$ and $\sigma_{\mathscr{A}}(x)$), which in turn implies $\partial \sigma_{\mathscr{C}}(y) \subseteq \sigma_{\mathscr{A}}(y)$.

Whence applying the useful fact that $\partial \sigma_{\mathscr{C}}(y) \subseteq \sigma_{\mathscr{A}}(y), \forall y \in \mathscr{C}$; on $\sigma_{\mathscr{C}}(x^*x) = \partial \sigma_{\mathscr{C}}(x^*x) \subseteq \sigma_{\mathscr{A}}(x^*x)$, as $x^*x \in$ Inv($\mathscr{A}$) then $0 \notin \sigma_{\mathscr{A}}(x^*x)$ which by the useful fact implies $0 \notin \sigma_{\mathscr{C}}(x^*x)$, therefore $(x^*x)^{-1} \in \mathscr{C}$. As $x, (x^*x)^{-1} \in \mathscr{B}$, then $(x^*x)^{-1}x^* \in \mathscr{C}$ and we note that $(x^*x)^{-1}x^*x = \mathbb{1}$, hence $(x^*x)^{-1}x^*$ is a left inverse of $x$, but since in $\mathscr{A}$ $x$ can only have one unique inverse this implies that it is the complete inverse $x^{-1} = (x^*x)^{-1}x^* \in \mathscr{C}$.

Then any element $x \in \mathscr{C}$ which is invertible in $\mathscr{A}$ one has $x^{-1} \in \mathscr{C}$, making $\sigma_{\mathscr{C}}(x) \subseteq \sigma_{\mathscr{A}}(x), \forall x \in \mathscr{C}$, from which, finally, $\sigma_{\mathscr{A}}(x) = \sigma_{\mathscr{C}}(x), \forall x \in \mathscr{C}$, which means that for $C^*$-algebras we can speak of spectra without referring to which $C^*$-subalgebra, all with the same unit, the spectrum is withdrawn. ∎

**Theorem B.0.42** (Spectral mapping theorem)**.** *Let $\mathcal{B}$ be a Banach algebra with unit, being $u \in \mathcal{B}$, then for any polynomial $p$:*

$$\sigma(p(u)) = p(\sigma(u)) := \{p(\lambda) \in \mathbb{C} \,|\, \lambda \in \sigma(u)\}.$$

*Proof:* Consider that $\mu \in \sigma(p(u))$, which is not empty by Proposition B.0.37, and let $\alpha_1, ..., \alpha_n$ be the $n$ roots of the polynomial $p(z) - \mu$ in $\mathbb{C}$, then $p(u) - \mu\mathbb{1} = a_n(u - \alpha_1\mathbb{1})...(u - \alpha_n\mathbb{1})$. If these roots are all such that $\alpha_i \notin \sigma(u)$, then by definition of the spectrum, each of the factors $(u - \alpha_j\mathbb{1})$ would be invertible as would be the product $a_n(u - \alpha_1\mathbb{1})...(u - \alpha_n\mathbb{1})$, which is in contradiction with the assumption that $\mu \in \sigma(p(u))$. Therefore some $\alpha_i$ makes $(u - \alpha_i\mathbb{1})$ non-invertible and then $\alpha_i \in \sigma(u)$, and since $\alpha_i$ is a root of $p(z) - \mu$ which implies $p(\alpha_i) = \mu$, but $\mu$ was arbitrarily chosen in the spectrum of $p(u)$ so that $\sigma(p(u)) \subseteq \{p(\lambda) \,|\, \lambda \in \sigma(u)\}$.

To prove the inverse inclusion, as we already said $\sigma(u)$ is non-empty, for $\lambda \in \sigma(u)$ the polynomial $p(z) - p(\lambda)$ has clearly $\lambda$ as a root. Therefore we can factor $p(z) - p(\lambda) = (z - \lambda)q(z)$, where $q$ is a polynomial of degree $n-1$, $p(u) - p(\lambda)\mathbb{1} = (u - \lambda\mathbb{1})q(u)$ and we remember that $(u - \lambda\mathbb{1})$ is not invertible. To continue we then prove the following lemma:

**Lemma B.0.43.** *Let $\mathcal{B}$ be a Banach algebra with unit, for $u \in \mathcal{B}$, $\lambda \in \sigma(u)$ and a arbitrary polynomial $q$ then $(u - \lambda\mathbb{1})q(u) \notin \text{Inv}(\mathcal{B})$.*

*Proof:* Defining $p(z) := (z - \lambda)q(z)$ then adapting this polynomial for $\mathcal{B}$ we have $p(u) = (u - \lambda\mathbb{1})q(u) \in \mathcal{B}$, since they are polynomials $uq(u) = q(u)u$ and $up(u) = p(u)u$ and therefore $q(u)$ and $p(u)$ commute with $u$. We shall prove that $p(u) \notin \text{Inv}(\mathcal{B})$ through a proof by contradiction, let us suppose that there exists a $w \in \mathcal{B}$ such that $wp(u) = p(u)w = \mathbb{1}$.

Considering $[w, u] = wu - uw$, left multiplying by $p(u)$ we get $p(u)[w, u] = u - p(u)uw = u - up(u)w = u - u = 0 \Rightarrow p(u)[w, u] = 0$ and again left multiplying this time by $w$ we get $[w, u] = 0$ so that $wu = uw$ and therefore $wq(u) = q(u)w$.

By hypothesis $p(u)w = wp(u) \Rightarrow (u - \lambda\mathbb{1})q(u)w = w(u - \lambda\mathbb{1})q(u) = \mathbb{1}$, using now that $q(u)$ commutes with $u$ and $w$ we may write $w(u - \lambda\mathbb{1})q(u) = q(u)w(u - \lambda\mathbb{1})$, so that $(u - \lambda\mathbb{1})(q(u)w) = (q(u)w)(u - \lambda\mathbb{1}) = \mathbb{1}$, which means that $(u - \lambda\mathbb{1})$ has a inverse given by $(u - \lambda\mathbb{1})^{-1} = q(u)w$, but this contradicts the hypothesis that $\lambda \in \sigma(u)$, therefore $p(u) = (u - \lambda\mathbb{1})q(u)$ does not have a inverse. ∎

Continuing from where we left, since $(u - \lambda\mathbb{1})$ is not invertible, then by the above Lemma B.0.43 we have that $(u - \lambda\mathbb{1})q(u) = p(u) - p(\lambda)\mathbb{1}$ is also not invertible, implies $p(\lambda) \in \sigma(p(u))$ and since we had chosen an arbitrary $\lambda \in \sigma(u)$ we get $\{p(\lambda) \,|\, \lambda \in \sigma(u)\} \subseteq \sigma(p(u))$, this completes the proof that $\sigma(p(u)) = \{p(\lambda) \,|\, \lambda \in \sigma(u)\}$. ∎

Specifically for $C^*$-algebras this theorem can be made more precise when considering self-adjoint elements of the $C^*$-algebra, with the additional information that then $\|p(a)\| = \left\|p\Big|_{\sigma(a)}\right\|_\infty := \sup_{\lambda \in \sigma(a)} |p(\lambda)|$.

**Theorem B.0.44.** *Let $\mathscr{A}$ be a $C^*$-algebra with unit, being $a \in \mathscr{A}_\dagger$, and for any polynomial $p(x) = \sum_{k=0}^{n} c_k x^k$, the spectrum of $p(a)$ is equal to the image by $p$ of the spectrum of $a$:*

$$\sigma(p(a)) = p(\sigma(a)) := \{p(\lambda) \in \mathbb{C} \,|\, \lambda \in \sigma(u)\}.$$

*Moreover* $\|p(a)\| = \left\|p\big|_{\sigma(a)}\right\|_\infty := \sup_{\lambda \in \sigma(a)} |p(\lambda)|$.

*Proof:* Since every $C^*$-algebra with unit is a Banach algebra with unit, then by the Spectral mapping theorem B.0.42 the first assertion is immediately true. The equality of the norm follows from considering the $C^*$ property $\|p(a)\|^2 = \|p(a)^*p(a)\|$:

$$p(a)^*p(a) = \left(\sum_{k=0}^n c_k a^k\right)^* \left(\sum_{j=0}^n c_j a^j\right) \stackrel{a^*=a}{=} \left(\sum_{k=0}^n \overline{c_k} a^k\right)\left(\sum_{j=0}^n c_j a^j\right) = \sum_{k,j=0}^n \overline{c_k} c_j a^{k+j}$$

Since $p(a)^*p(a) \in \mathscr{A}_\dagger$, applying Theorem B.0.41 we get:

$$\|p(a)\|^2 = \|p(a)^*p(a)\| \stackrel{(B.0.41)}{\equiv} r(p(a)^*p(a)) \equiv \sup_{\mu \in \sigma(p(a)^*p(a))} |\mu| \stackrel{(B.0.42)}{=} \sup_{\mu \in \{\overline{p(\lambda)}p(\lambda)\,|\,\lambda \in \sigma(a)\}} |\mu| =$$
$$= \sup_{\lambda \in \sigma(a)} |\overline{p(\lambda)}p(\lambda)| = \sup_{\lambda \in \sigma(a)} |p(\lambda)|^2 = \left(\sup_{\lambda \in \sigma(a)} |p(\lambda)|\right)^2$$

Establishing the last equality. ∎

**Proposition B.0.45.** *Let $\mathcal{B}$ be a Banach algebra with unit, and $u \in \mathrm{Inv}(\mathcal{B})$ a invertible element of $\mathcal{B}$. Then $\sigma(u^{-1}) = \{\lambda^{-1} \in \mathbb{C} \,|\, \lambda \in \sigma(u)\}$.*

*Proof:* If $u$ is invertible then so is $0\mathbb{1} - u$, and therefore $0 \in \rho(u) \implies 0 \notin \sigma(u)$, so $\lambda^{-1}$ is always well defined for $\lambda \in \sigma(u)$. Whereas, for $\lambda \neq 0$, $(\lambda\mathbb{1} - u) = -\lambda u(\lambda^{-1}\mathbb{1} - u^{-1})$, which shows that $\lambda \in \sigma(u)$, that is $\lambda\mathbb{1} - u$ is non-invertible, if and only if $-\lambda u(\lambda^{-1}\mathbb{1} - u^{-1})$ is non-invertible, but since $u$ is invertible, $\lambda^{-1}\mathbb{1} - u^{-1}$ must not be and therefore $\lambda^{-1} \in \sigma(u^{-1})$. ∎

**Proposition B.0.46.** *Let $\mathcal{B}$ be a Banach $*$–algebra with unit, and $u \in \mathrm{Inv}(\mathcal{B})$ a invertible element of $\mathcal{B}$. Then $\sigma(u^*) = \{\overline{\lambda} \in \mathbb{C} \,|\, \lambda \in \sigma(u)\}$.*

*Proof:* Considering that if $u$ is invertible then $uu^{-1} = u^{-1}u = \mathbb{1}$, so that by taking the involution of these equalities we get $(u^{-1})^*u^* = u^*(u^{-1})^* = \mathbb{1}$, and therefore $(u^*)^{-1} = (u^{-1})^*$, that is $u^*$ is invertible if and only if $u$ is invertible, so that calculating $(\lambda\mathbb{1} - u)^* = \overline{\lambda}\mathbb{1} - u^*$ and so $\overline{\lambda}\mathbb{1} - u^*$ is invertible if and only if $\lambda\mathbb{1} - u$ is invertible, so that by the contrapositive $\lambda \in \sigma(u)$ if and only if $\overline{(\lambda)} \in \sigma(u^*)$. ∎

We shall make a abuse of notation and denote the set of the inverted elements of the spectrum of $u$ by $\sigma(u)^{-1} \equiv \{\lambda^{-1} \in \mathbb{C} \,|\, \lambda \in \sigma(u)\}$ and the set of the complex conjugate elements of the spectrum of $u$ by $\sigma(u)^* \equiv \{\overline{\lambda} \in \mathbb{C} \,|\, \lambda \in \sigma(u)\}$, so that the conclusions of the propositions B.0.45 and B.0.46 can be synthesized as $\sigma(u^{-1}) = \sigma(u)^{-1}$ and $\sigma(u^*) = \sigma(u)^*$, for $u \in \mathrm{Inv}(\mathcal{B})$.

**Proposition B.0.47.** *Let $\mathscr{A}$ be a $C^*$-algebra with unit, if $u \in \mathscr{A}$ is unitary, then $\sigma(u) \subseteq \mathbb{S}(\mathbb{C}) :=$*
$= \{\lambda \in \mathbb{C} \,|\, |\lambda| = 1\}$.

*Proof:* Since $u$ is a unitary then by the $C^*$ property $\|u\|^2 = \|u^*u\| = \|\mathbb{1}\| = 1 \Rightarrow \|u\| = 1$. Beyond that $u$ is unitary and therefore normal ($uu^* = u^*u (= \mathbb{1})$), so by theorem B.0.41, $r(u) = \|u\| = 1$. This shows that $\sigma(u)$ is a closed subset of the unit disc centered at 0, $\overline{\mathcal{B}_1}(\mathbb{C}) := \overline{\mathcal{D}_1} = \{\lambda \in \mathbb{C} \,|\, |\lambda| \leq 1\}$. By propositions B.0.45 and B.0.46 $\sigma(u) = \sigma(u^*)^* = \sigma(u^{-1})^* = \left((\sigma(u))^{-1}\right)^*$, from this we see that if $\lambda \in \sigma(u)$ and $|\lambda| < 1$ it is a easy calculation to show that, for $\lambda = x + iy;\, x, y \in \mathbb{R}$, $|\overline{\lambda^{-1}}| = \frac{1}{\sqrt{x^2+y^2}} = |\lambda|^{-1} > 1$ so that $\overline{\lambda^{-1}} \notin \left((\sigma(u))^{-1}\right)^*$ and therefore $\lambda \notin \sigma(u)$, absurd. So this leaves us with $|\lambda| = 1$ which implies $\sigma(u) \subseteq \mathbb{S}(\mathbb{C})$. ∎

**Theorem B.0.48.** *Let $\mathscr{A}$ be a $C^*$-algebra with unit, for $A \in \mathscr{A}_\dagger$. Then $\sigma(A) \subseteq \mathbb{R}$, and more precisely $\sigma(A)$ is contained in a compact subset of $[-\|A\|, \|A\|]$.*

*Proof:* For $A = 0$ this result is trivial, for $A \neq 0$ and let $\mathbb{R} \ni c > 0$ and $\lambda \in \mathbb{C}$, such that the imaginary part of $\lambda$ is not zero. The case $|\lambda| > \|A\|$ implies by proposition B.0.37 that $\lambda \notin \sigma(A)$, which makes it sufficient to consider that $|\lambda| \leq \|A\|$. Choosing $c < \|a\|^{-1}$, the norm of the elements $\pm icA$ will be $c\|A\| < 1$ and by proposition B.0.28 the elements $\mathbb{1} \pm icA$ are invertible. Beyond that, the choices we made make $c < \|A\|^{-1} \leq |\lambda|^{-1}$, which makes $1 \pm ic\lambda \neq 0$. We have then:

$$\lambda\mathbb{1} - A = \left(\frac{2ic\lambda}{2ic}\right)\mathbb{1} - \left(\frac{2ic}{2ic}\right)A$$

$$= \left(\frac{(1+ic\lambda) - (1-ic\lambda)}{2ic}\right)\mathbb{1} - \left(\frac{ic[(1-ic\lambda)+(1+ic\lambda)]}{2ic}\right)A$$

$$= \left(\frac{1}{2ic}\right)\left[(1+ic\lambda)(\mathbb{1}-icA) - (1-ic\lambda)(\mathbb{1}+icA)\right]$$

$$= \left(\frac{1-ip\lambda}{2ip}\right)\left[\left(\frac{1+ip\lambda}{1-ip\lambda}\right)(\mathbb{1}-icA) - (\mathbb{1}+icA)\right]$$

$$= \left(\frac{1-ip\lambda}{2ip}\right)\left[\left(\frac{1+ip\lambda}{1-ip\lambda}\right)\mathbb{1} - (\mathbb{1}+icA)(\mathbb{1}-icA)^{-1}\right](\mathbb{1}-icA). \quad (B.15)$$

The invertibility of $(\mathbb{1} - icA)$ has already been guaranteed, so from the last equality we see that $\lambda\mathbb{1} - A$ will be invertible if $\left(\frac{1+ip\lambda}{1-ip\lambda}\right)\mathbb{1} - (\mathbb{1}+icA)(\mathbb{1}-icA)^{-1}$ is invertible. We note that since $(\mathbb{1}+icA)^{-1}(\mathbb{1}+icA) = \mathbb{1} = (\mathbb{1}+icA)(\mathbb{1}+icA)^{-1} \Rightarrow (\mathbb{1}+icA)^{-1}a = a(\mathbb{1}+icA)^{-1}$, so they commute, and therefore:

$$\left((\mathbb{1}+icA)(\mathbb{1}-icA)^{-1}\right)^{-1} = (\mathbb{1}-icA)(\mathbb{1}+icA)^{-1} \stackrel{\text{comm. } A}{=} (\mathbb{1}+icA)^{-1}(\mathbb{1}-icA) =$$

$$\stackrel{A^*=A}{=} \left((\mathbb{1}-icA)^*\right)^{-1}(\mathbb{1}+icA)^* = \left((\mathbb{1}+icA)(\mathbb{1}-icA)^{-1}\right)^*.$$

which proves that $(\mathbb{1}+icA)(\mathbb{1}-icA)^{-1}$ is unitary. Writing $\lambda = x + iy$ with $x, y \in \mathbb{R}$ we have:

$$\left|\frac{1+ic\lambda}{1-ic\lambda}\right|^2 = \frac{(1-cy)^2 + (cx)^2}{(1+cy)^2 + (cx)^2} \neq 1, \text{ for } y \neq 0.$$

Since $(\mathbb{1}+icA)(\mathbb{1}-icA)^{-1}$ is unitary, by proposition B.0.47 then its spectrum is formed of complex numbers with absolute value equal to 1 and since $\left|\frac{1+ic\lambda}{1-ic\lambda}\right| \neq 1$ then there is no element $T \in \mathscr{A}$ such that $\left(\left(\frac{1+ip\lambda}{1-ip\lambda}\right)\mathbb{1} - (\mathbb{1}+icA)(\mathbb{1}-icA)^{-1}\right)T = 0$ or $T\left(\left(\frac{1+ip\lambda}{1-ip\lambda}\right)\mathbb{1} - (\mathbb{1}+icA)(\mathbb{1}-icA)^{-1}\right) = 0$, and hence $\left(\frac{1+ip\lambda}{1-ip\lambda}\right)\mathbb{1} - (\mathbb{1}+icA)(\mathbb{1}-icA)^{-1}$ is invertible. Then by equation (B.15) so is $\lambda\mathbb{1} - A$ with:

$$(\lambda\mathbb{1} - A)^{-1} = \left(\frac{2ic}{1-ic\lambda}\right)(\mathbb{1}-icA)^{-1}\left(\left(\frac{1+ip\lambda}{1-ip\lambda}\right)\mathbb{1} - (\mathbb{1}+icA)(\mathbb{1}-icA)^{-1}\right)^{-1}.$$

Therefore $\lambda\mathbb{1} - A$ has a inverse for every $\lambda$ such that $\mathfrak{Im}(\lambda) \neq 0$, and so every non purely real number is in the resolvent set $\rho(A)$, in other words $\sigma(A) \subseteq \mathbb{R}$. Since by theorem B.0.41 $r(A) = \|A\|$, we conclude that $\sigma(A) \subseteq [-\|A\|, \|A\|]$. That $\sigma(u)$ is closed is implied by proposition B.0.34. ∎

Now we prove the commutative version of the Gel'fand-Naĭmark theorem, also called the *1st fundamental theorem of $C^*$-algebras*, this involves definitions that appear in chapter 7, such as the Gel'fand transform $\tilde{A}$ for each element $A$ of a abelian von Neumann algebra. This theorem is useful for proving the noncommuatative Gel'fand-Naĭmark theorem, also called the *2nd fundamental theorem of $C^*$-algebras*, theorem 4.0.19, in chapter 4.

**Theorem B.0.49** (Commutative Gel'fand-Naĭmark). *The Gel'fand transform, defined for $C^*$-algebras analogously as it was defined for von Neumann algebras in definition 7.1.28, is an isometric $*$-isomorphism for commutative, unital $C^*$-algebras.*

*Proof:* We begin by proving that the Gel'fand transform is a continuous, $*$-morphism with $\|\tilde{A}\|_\infty \leq \|A\|$. We begin by noticing that by proposition 7.1.29 every character $\lambda$ is continuous, in particular, if $\lambda_n \xrightarrow{n\to\infty} \lambda$ then $\lambda_n(x) \xrightarrow{n\to\infty} \lambda(x)$, for each $x$ in the relevant abelian $C^*$-algebra $\mathscr{C}$; then this implies that the Gel'fand transform $\tilde{x}$ is continuous with respect to the topology of pointwise convergence on $\Xi_\mathscr{C}$, as $\tilde{x}(\lambda_n) = \lambda_n(x)$ and $\tilde{x}(\lambda) = \lambda(x)$. We then check that the properties $\widetilde{\alpha x + \beta y} = \alpha \tilde{x} + \beta \tilde{y}$ and $\widetilde{xy} = \tilde{x}\,\tilde{y}$, are valid $\forall x, y \in \Xi_\mathscr{C}$ and $\alpha, \beta \in \mathbb{C}$; as for any character $\lambda \in \Xi_\mathscr{C}$, $\widetilde{\alpha x + \beta y}(\lambda) = \lambda(\alpha x + \beta y) = \alpha\lambda(x) + \beta\lambda(y) = \alpha\tilde{x} + \beta\tilde{y}$, and $\widetilde{xy}(\lambda) = \lambda(xy) = \lambda(x)\lambda(y) = \tilde{x}(\lambda)\tilde{y}(\lambda)$.

By the observation in the proof of proposition 7.1.29 we have that $|\tilde{x}(\lambda)| = |\lambda(x)| \leq \|x\|$, which implies $\|\tilde{x}\|_\infty \leq \|x\|$, also the Gel'fand transform is unital since $\tilde{\mathbb{1}}(\lambda) = \lambda(\mathbb{1}) = 1$, $\forall \lambda \in \Xi_\mathscr{C}$; by the $*$-morphism property of every character, which also makes:

$$\widetilde{x^*}(\lambda) = \lambda(x^*) = \lambda(x)^* = \bigl(\tilde{x}(\lambda)\bigr)^*, \qquad \forall \lambda \in \Xi_\mathscr{C}.$$

Hence $\widetilde{x^*} = (\tilde{x})^*$ and the Gel'fand transform is a unital $*$-homomorphism. Then using proposition 7.1.32 $\tilde{x}(\Xi_\mathscr{C}) = \sigma(x)$, thus $\|\tilde{x}\|_\infty = \sup_{\lambda \in \Xi_\mathscr{C}} |\tilde{x}(\lambda)| = \sup_{\lambda \in \sigma(x)} |\lambda| = r(x)$, and since every element of $\mathscr{C}$ is normal, thanks to the abelianess, then by theorem B.0.41 $r(x) = \|x\|$; in conclusion $\|\tilde{x}\|_\infty = \|x\|$, and the Gel'fand transform is isometric, since the Gel'fand transform is a unital $*$-homomorphism it can also be seen as a $*$-representation and by proposition A.1.12, the Gel'fand transform is injective.

To prove surjectivity, we first note that the Gel'fand spectrum $\Xi_\mathscr{C}$ is a compact Hausdorff space, since, by considering the pointwise convergence of any net $\lambda_{\alpha_\alpha}$ that converges to a functional $\lambda$, this is true if and only if $\lambda_n(x) \xrightarrow{n\to\infty} \lambda(x)$, for each $x \in \mathscr{C}$, it is easy, though extensive, to to check that if $\lambda_\alpha \in \Xi_\mathscr{C}$ then $\lambda \in \Xi_\mathscr{C}$, hence we shall omit this step. Therefore $\Xi_\mathscr{C}$ is a closed subset of the closed unit ball $\overline{\mathcal{B}_1(\mathscr{C}^\#)} = \{\lambda \in \mathscr{C}^\# \mid \|\lambda\|_{\mathscr{C}'} \leq 1\}$ of the dual space $\mathscr{C}^\#$. We know this set is compact in the weak-$*$ topology by the Banach-Alaoglu theorem B.0.18, and by proposition B.0.17 this ball is a Hausdorff space, which makes the closed subset $\Xi_\mathscr{C}$ a compact Hausdorff space.

By lemma A.1.13, $\tilde{\mathscr{C}} := \{\tilde{x} \in \Xi_\mathscr{C} \mid x \in \mathscr{C}\}$ is a closed $*$-sub-algebra of $C(\Xi_\mathscr{C})$. We shall now show that $\tilde{\mathscr{C}}$ separates points, let $\lambda, \xi \in \Xi_\mathscr{C}$ be two points in our compact space with $\lambda \neq \xi$ in particular $\exists x \in \mathscr{C}$ such that $\lambda(x) \neq \xi(x)$, hence $\tilde{x} \in \tilde{\mathscr{C}}$ separates the points $\lambda$ and $\xi$, since $\tilde{x}(\lambda) = \lambda(x) \neq \xi(x) = \tilde{x}(\xi)$. Hence, we verified all conditions of the Stone-Weierstrass theorem B.0.22, and we conclude $\tilde{\mathscr{C}} = C(\Xi_\mathscr{C})$, *i.e.* the Gel'fand transform is also surjective. ∎

**Theorem B.0.50.** *Let $\mathscr{A}$ be a unital $C^*$-algebra and let $x \in \mathscr{A}$ be a normal element. There is a state $\varphi : \mathscr{A} \to \mathbb{C}$ with $|\varphi(x)| = \|x\|$.*

*Proof:* We consider the smallest $C^*$-algebra that contains $\{\mathbb{1}_{\mathscr{A}}, x\}$, we denote this by $C^*(\{\mathbb{1}_{\mathscr{A}}, x\})$, this will obviously be a unital Banach algebra and we consider the Gel'fand transform $\widehat{(\cdot)} : C^*(\{\mathbb{1}_{\mathscr{A}}, x\}) \to C(\Xi_{C^*(\{\mathbb{1}_{\mathscr{A}}, x\})})$. Now, for $\hat{x}$ we find some character $\lambda_0 \in \Xi_{C^*(\{\mathbb{1}_{\mathscr{A}}, x\})}$ such that $|\hat{x}(\lambda_0)| = \|\hat{x}\|_\infty$, as $\lambda_0$ is a character $\lambda_0(\mathbb{1}_{\mathscr{A}}) = 1$ and $\|\lambda_0\|_{C^*(\{\mathbb{1}_{\mathscr{A}}, x\})'}$.

Since $x$ is normal, $C^*(\{\mathbb{1}_{\mathscr{A}}, x\})$ is in fact a commutative, unital $C^*$-algebra, as $C^*(\{\mathbb{1}_{\mathscr{A}}, x\})$ must be formed of all polynomials of $x$. Hence, by the commutative Gel'fand-Naĭmark theorem B.0.49, the Gel'fand transform is an isometric $*$-isomorphism, i.e. $\|\hat{x}\|_\infty = \|x\|$. Forgetting some of the structure of $\lambda_0$, we conclude that we found a linear and continuous map $\lambda_0 : C^*(\{\mathbb{1}_{\mathscr{A}}, x\}) \to \mathbb{C}$ with $\lambda_0(\mathbb{1}_{\mathscr{A}}) = 1$ and:
$$|\lambda_0(x)| = |\hat{x}(\lambda_0)| = \|\hat{x}\|_\infty = \|x\|.$$

We may hence apply the Hahn-Banach theorem B.0.23, and find a linear and continuous extension $\lambda : \mathscr{A} \to \mathbb{C}$ with $\|\lambda\| = \|\lambda_0\|$. This extension $\lambda$ is still positive, as it agrees with $\lambda_0$ on $C^*(\{\mathbb{1}_{\mathscr{A}}, x\})$, we have:
$$\|\lambda\| = \|\lambda_0\| = \lambda_0(\mathbb{1}_{\mathscr{A}}) = \lambda(\mathbb{1}_{\mathscr{A}}).$$

and by proposition 4.2.3, we obtain that $\lambda : \mathscr{A} \to \mathbb{C}$ is a positive linear functional. Thus, a state with:
$$|\lambda(x)| = |\lambda_0(x)| = \|x\|.$$
∎

**Lemma B.0.51.** *Let $\mathscr{A}, \mathscr{B}$ be $C^*$-algebras and let $\pi : \mathscr{A} \to \mathscr{B}$ be a $*$-homomorphism. Then, $\pi(\mathscr{A})$ is a $C^*$-algebra which is isomorphic to $\mathscr{A}/\ker(\pi)$.*

*Proof:* As we have already proved in proposition 7.1.30, $\ker(\pi)$ is an ideal for a $*$-homomorphism $\pi$, it is trivial to ascertain that $\mathscr{A}/\ker(\pi)$ is a $C^*$-algebra, we just have to verify that the set $\mathscr{A}/\ker(\pi)$ is closed under the sum, products, multiplications by scalars and involutions defined canonically for $\mathscr{A}/\ker(\pi)$ and that the equivalent quotient norm defined on $\mathscr{A}/\ker(\pi)$ obeys the $C^*$ property, in case of doubt this is proposition 1.8.2. in [30].

Let us denote the quotient map by $\widetilde{\pi} : \mathscr{A} \to \mathscr{A}/\ker(\pi)$, $x \mapsto x + \ker(\pi)$. We define a map $\varphi : \mathscr{A}/\ker(\pi) \to \mathscr{B}$ by $\varphi(x + \ker(\pi)) := \pi(x)$. This is an injective $*$-homomorphism with range $\pi(\mathscr{A})$. Essentially by proposition A.1.12 we have that $\varphi$ is even isometric, hence, any Cauchy sequence in $\pi(\mathscr{A})$ is then also a Cauchy sequence in $\mathscr{A}/\ker(\pi)$. This implies that $\pi(\mathscr{A})$ is complete.

Concluding, we find that $\pi(\mathscr{A}) \subseteq \mathscr{B}$ is a closed $*$-sub-algebra and hence it is a $C^*$-algebra. We conclude that $\varphi : \mathscr{A}/\ker(\pi) \to \pi(\mathscr{A})$ is a $*$-isomorphism. ∎

We shall then extend the continuous functional calculus constructed in 4.0.16 to a Borel functional calculus, where instead of continuous complex functions $f \in C(\sigma(A))$, we can consider the bounded Borel measurable complex functions $g \in B^{\mathscr{B}}(\sigma(A))$. To do this we first note that by the Riesz-Markov representation theorem, that we will enunciate but not prove since it is a well known theorem of functional analysis proven in [89], [88] and among others, we can write the expectation values of the functional calculus of operators as integrals with relation to the underlying function.

**Theorem B.0.52** (Riesz-Markov representation with complex measures). *Let $X$ be a locally compact and Hausdorff topological space and $l : C_c(X) \to \mathbb{C}$ a continuous linear functional. Then there exists a unique regular complex Borel measure, as defined in C.0.30, $\mu_l$ such that:*

$$\forall f \in C_c(X), \quad l(f) = \int_X f d\mu_l$$

*Moreover $\|l\|_{\mathscr{H}'} = \|\mu_l\|_\Sigma := |\mu_l|(X)$.*

Since $f \mapsto \langle \psi, f(A)\psi \rangle \in \mathbb{C}$ is a continuous linear functional in $X = \sigma(A)$ that is compact and therefore $C_c(\sigma(A)) = C(\sigma(A))$. Then by the Riesz-Markov representation theorem $\forall f \in C(\sigma(A))$:

$$\langle \psi, f(A)\psi \rangle = \int_{\sigma(A)} f d\mu_{\psi, A}, \tag{B.16}$$

Where the measure $\mu_{\psi, A}$ is called the spectral measure of the $A \in \mathfrak{B}(\mathscr{H})$ operator associated with the vector $\psi \in \mathscr{H}$. But as it turns out the RHS of B.16 is defined not only for continuous complex functions but for any bounded Borel measurable function, so what remains to be shown is that there exists a operator $g(A) \in \mathfrak{B}(\mathscr{H})$ for each $g \in B^{\mathscr{B}}(\sigma(A))$ such that $\langle \psi, g(A)\psi \rangle = \int_{\sigma(A)} g d\mu_{\psi, A}, \forall \psi \in \mathscr{H}$.

**Proposition B.0.53.** *For each $g \in B^{\mathscr{B}}(\sigma(A))$, the application $S_g : \mathscr{H} \times \mathscr{H} \to \mathbb{C}$ defined by:*

$$S_g(u, v) := \frac{1}{4} \sum_{n=0}^{3} i^{-n} \int_{\sigma(A)} g d\mu_{(u+i^n v), A},$$

*is a sesquilinear and bicontinuous map in $\mathscr{H}$, with $|S_g(u,v)| \leq \|g\|_\infty \|u\| \|v\|$, $\forall u, v \in \mathscr{H}$. Therefore by the corollary B.0.14 there is a unique limited linear operator which we will denote by $g(A)$ such that:*

$$S(u, v) = \langle u, g(A)v \rangle \quad , \forall u, v \in \mathscr{H}.$$

*It follows also that $\|g(A)\|_{\mathfrak{B}(\mathscr{H})} = \|g\|_\infty$.*

*Proof:* For each continuous function $f \in C(\sigma(A))$ we have that by the polarization identity, prop. B.0.7:

$$S_f(u, v) = \frac{1}{4} \sum_{n=0}^{3} i^{-3} \int_{\sigma(A)} f d\mu_{(u+i^n v), A} = \frac{1}{4} \sum_{n=0}^{3} i^{-3} \langle (u+i^n v), f(A)(u+i^n v) \rangle = \langle u, f(A)v \rangle$$

This proves that $S_f$ is sesquilinear and it is bicontinuous since, by Cauchy-Schwarz, $|\langle u, f(A)v \rangle| \leq \|f(A)\|_{\mathfrak{B}(\mathscr{H})} \|u\| \|v\|$. To show that these properties extend to the forms $S_g$ obtained by bounded Borel measurable functions $g \in B^{\mathscr{B}}(\sigma(A))$, we must use the fact that bounded Borel measurable functions can be approximated by continuous functions, this is obtained by Lusin's theorem, which we will enunciate but not prove, the proof can be found in [88] or [43].

**Theorem B.0.54** (Lusin). *Considering $X$ a locally compact and Hausdorff topological space and let $\mu : \mathcal{M} \to \mathbb{R}$ be a positive measure over the $\sigma$-algebra $\mathcal{M}$ of $X$ that contains the Borel $\sigma$-algebra of $X$, in such a way that:*

(a) $\mu(K) < \infty$ *for every compact $K \subset X$;*

(b) $\mu$ is regular.

(c) the measure space $(X, \mathcal{M}, \mu)$ is complete, that is, if $E \in \mathcal{M}$ and $\mu(E) = 0$ then $\forall F \subset E$, $F \in \mathcal{M}$.

*Suppose now that $g : (X, \mathcal{M}) \to \mathbb{C}$ is a measurable complex function in $X$ with the property that for every $B \subset X$ such that $\mu(B) < \infty$ then $g(x) = 0$ if $x \notin B$. Therefore for every $\varepsilon > 0$, there exists a function $f \in C_c(X)$ such that:*

$$\mu\left(\{x \in X \mid g(x) \neq f(x)\}\right) \leq \varepsilon.$$

*Beyound that $\varepsilon$ can be chosen in such a way that:*

$$\sup_{x \in X} |f(x)| \leq \sup_{x \in X} |g(x)|.$$

Form this theorem we trivially obtain the following corollary, that will be finally used to show that $S_g$ for $g \in B^{\mathscr{B}}(\sigma(A))$ is also a bicontinuous sesquilinear form:

**Corollary B.0.55.** *Considering $X$ a locally compact and Hausdorff topological space and let $\mu_j$, $j = 1, \ldots, n$; be a finite collection of measures satisfying the hypothesis for Lusin's theorem B.0.54. Let $g$ be a complex Borel measurable function in $X$ with the property that for every $B \subset X$ such that $\mu_j(B) < \infty$, $i = 1, \ldots, n$; then $g(x) = 0$ if $x \notin B$. Therefore, for every $\varepsilon > 0$ there exists $f \in C_c(X)$ such that:*

$$\mu_j\left(\{x \in X \mid g(x) \neq f(x)\}\right) \leq \varepsilon \quad , \text{ for every } j = 1, \ldots, n.$$

*Beyound that, $f$ can be chosen in such a way that:*

$$\sup_{x \in X} |f(x)| \leq \sup_{x \in X} |g(x)|.$$

*For such a $f$, the following estimate will be valid:*

$$\int_X |f - g| d\mu_j \leq 2\|g\|_\infty \varepsilon \quad , \text{ for every } j = 1, \ldots, n. \tag{B.17}$$

*Proof:* Letting $D := \{x \in X \mid g(x) \neq f(x)\}$. By hypothesis the measures $\{\mu_j\}_{j=1}^n$ have in common the Borel $\sigma$-algebra of $X$, where it can be defined the measure $\mu := \mu_1 + \ldots + \mu_n$ that also fulfills all conditions of Lusin's theorem B.0.54. Therefore there exists a function $f \in C_c(X)$ with $\mu(D) \leq \varepsilon$, that is, $\mu_1(D) + \ldots + \mu_n(D) \leq \varepsilon \implies \mu_j(D) \leq \varepsilon$, for all $j = 1, \ldots, n$; since the measures are positive.

Continuing we have that for every $j = 1, \ldots, n$:

$$\int_X |f - g| d\mu_j = \int_D |f - g| d\mu_j \leq \|f - g\|_\infty \mu_j(D) \leq 2\|g\|_\infty \varepsilon,$$

Where it was used that since $\|f\|_\infty \leq \|g\|_\infty$, which implies that $\|g - f\|_\infty \leq \|g\|_\infty + \|f\|_\infty \leq 2\|g\|_\infty$.

Conditions 1), 2) and 3) of Lusin's theorem B.0.54 are guaranteed by the Riesz-Markov representation theorem B.0.52 and therefore are valid for the spectral measures $\mu_{(u+i^n v), A}$ defined on $X = \sigma(A)$.

Using then corollary B.0.55, we can get the following estimate:

$$|S_g(u,v) - S_f(u,v)| = \left| \frac{1}{4} \sum_{n=0}^{3} i^{-n} \int_{\sigma(A)} g - f d\mu_{(u+i^n v), A} \right| \leq 8\|g\|_\infty \varepsilon \tag{B.18}$$

To prove that $S_g$ is bicontinuous then since by (B.18) the difference $|S_g - S_f|$ can be made arbitrarily small by choosing $f$ sufficiently close to $g$, then $|S_g(u,v) - S_f(u,v)| \leq \varepsilon$ for a arbitrarily small $\varepsilon(f)$. Then:

$$|S_g(u,v)| = |S_g(u,v) - S_f(u,v) + S_f(u,v)| \leq |S_g(u,v) - S_f(u,v)| + |S_f(u,v)| \leq \varepsilon + \|f(A)\|_{\mathfrak{B}(\mathcal{H})}\|u\|\|v\|,$$

by the continuous functional calculus $\|f(A)\|_{\mathfrak{B}(\mathcal{H})} = \|f\|_\infty$ and by Lusin's theorem B.0.54 $f$ can be chosen so as to $\|f\|_\infty \leq \|g\|_\infty$ and therefore, $\|S_g(u,v)\| \leq \varepsilon + \|g\|_\infty \|u\|\|v\|$, as $\varepsilon$ can be made arbitrarily small, $\|S_g(u,v)\| \leq \|g\|_\infty \|u\|\|v\|$ and therefore $S_g$ is bicontinuous.

To prove that $S_g$ is sesquilinear we consider that, as already proved, if $f$ is continuous then $S_f$ is sesquilinear which implies $S_f(u, \alpha_1 v_1 + \alpha_2 v_2) - \alpha_1 S_f(u, v_1) - \alpha_2 S_f(u, v_2) = 0$ and $S_f(\beta_1 u_1 + \beta_2 u_2, v) - \overline{\beta_1} S_f(u_1, v) - \overline{\beta_2} S_f(u_2, v) = 0$, we shall prove that $S_g$ is linear in the second argument, the proof that $S_g$ is anti-linear in the first argument is completely analogous.

$$|S_g(u, \alpha_1 v_1 + \alpha_2 v_2) - \alpha_1 S_g(u, v_1) - \alpha_2 S_g(u, v_2)| =$$
$$= |S_g(u, \alpha_1 v_1 + \alpha_2 v_2) - \alpha_1 S_g(u, v_1) - \alpha_2 S_g(u, v_2) - (S_f(u, \alpha_1 v_1 + \alpha_2 v_2) - \alpha_1 S_f(u, v_1) - \alpha_2 S_f(u, v_2))| \leq$$
$$\leq |S_g(u, \alpha_1 v_1 + \alpha_2 v_2) - S_f(u, \alpha_1 v_1 + \alpha_2 v_2)| + |\alpha_1||S_g(u, v_1) - S_f(u, v_1)| + |\alpha_2||S_g(u, v_2) - S_f(u, v_2)|$$

These last three terms, though (B.18), can be made as small as desired by choosing the adequate $f \in C(\sigma(A))$, this establishes the linearity $S_f(u, \alpha_1 v_1 + \alpha_2 v_2) = \alpha_1 S_f(u, v_1) + \alpha_2 S_f(u, v_2)$. ∎

Having then proved the existence of $g(A)$, when $g$ is a Borel measurable function, so we can obtain a Functional Calculus for Borel measurable functions, we shall only enunciate the theorem for such a construction, but the proof is very similar to the proof of the continuous functional calculus, a excellent reference for the proof is in chapter 40 of [8] (in portuguese), or for a text in English [77].

**Theorem B.0.56** (Borel functional calculus). *Let $\mathcal{H}$ be a Hilbert space and $A \in \mathfrak{B}(\mathcal{H})_\dagger$ be a self-djoint operator, to each Borel measurable function $g \in B^{\mathcal{B}}(\sigma(A))$ we can construct the operator $g(A)$ by proposition B.0.53, this assignment for each $A \in \mathfrak{B}(\mathcal{H})_\dagger$ will be denoted by $\hat{\Phi}_A : B^{\mathcal{B}}(\sigma(A)) \to \mathfrak{B}(\mathcal{H})$ and be such that $\hat{\Phi}_A(g) \equiv g(A)$, by proposition B.0.53, specifically the first part of the proof, it becomes clear that $\hat{\Phi}_A$ is the unique extension of $\Phi_A : C(\sigma(A)) \to \mathfrak{B}(\mathcal{H})$ from the Continuous functional calculus 4.0.16, that is, $\hat{\Phi}_A(f) = \Phi_A(f), \forall f \in C(\sigma(A))$. By B.0.53, $\|\hat{\Phi}_A(g)\|_{\mathfrak{B}(\mathcal{H})} \leq \|g\|_\infty, \forall g \in B^{\mathcal{B}}(\sigma(A))$ and beyound that the following assertions are true:*

*(a) - The map $\hat{\Phi}_A$ is a algebraic $*$-homomorphism:*

$$\hat{\Phi}_A(\alpha g + \beta h) = \alpha \hat{\Phi}_A(g) + \beta \hat{\Phi}_A(h), \quad \hat{\Phi}_A(gh) = \hat{\Phi}_A(g)\hat{\Phi}_A(h), \quad \hat{\Phi}_A(g)^* = \hat{\Phi}(\overline{g}), \quad \hat{\Phi}_A(1) = \mathbb{1};$$

for all $g, h \in B^{\mathscr{B}}(\sigma(A))$ and $\alpha, \beta \in \mathbb{C}$. From the second equation in the previous line and the fact that $gh = hg$, $\forall g, h \in B^{\mathscr{B}}(\sigma(A))$ we get that it is always true that $\hat{\Phi}_A(g)\hat{\Phi}_A(h) = \hat{\Phi}_A(h)\hat{\Phi}_A(g)$.

(b) - $g \geq 0 \implies \hat{\Phi}_A(g) \geq 0$ in the operator ordering.

(c) - Let $g, g_n \in B^{\mathscr{B}}(\sigma(A))$, $\forall n \in \mathbb{N}$, such that $\lim_{n \to \infty} g_n(x) = g(x)$, $\forall x \in \sigma(A)$, and such that $\exists M > 0$ with $\|g_n\|_\infty < M$, $\forall n \in \mathbb{N}$. Then $g_n(A)$ converges to $g(A)$ in SOT, i.e., $\lim_{n \to \infty} g_n(A)\psi = g(A)\psi$, $\forall \psi \in \mathscr{H}$.

(d) - If $\varphi \in \mathscr{H}$ is a eigenvector of $A$ with eigenvalue $\lambda_0$, then the spectral measure $\mu_{\varphi, A}$ is exactly the Dirac measure centered at $\lambda_0$, also $\hat{\Phi}_A(g)\varphi = g(\lambda_0)\varphi$ for all $g \in B^{\mathscr{B}}(\sigma(A))$. In general we have $\sigma(\hat{\Phi}_A(g)) \subset \overline{\{g(\lambda) \mid \lambda \in \sigma(A)\}}$.

We then get the following theorem on spectral projectors:

**Theorem B.0.57.** Let $A \in \mathfrak{B}(\mathscr{H})$ be a self-adjoint operator, the map $\mathfrak{P} : \mathscr{B}(\sigma(A)) \to \mathfrak{B}(\mathscr{H})$, which associates a limited operator to each Borel subset of $\sigma(A) \subset \mathbb{R}$, which is given by $\mathscr{B}(\sigma(A)) \ni B \overset{\mathfrak{P}}{\mapsto} P_B := \hat{\Phi}_A(\chi_B) \equiv \chi_B(A) \in \mathfrak{B}(\mathscr{H})$, i.e. $\mathfrak{P}(\,\cdot\,) \equiv \hat{\Phi}_A \circ \chi_{(\,\cdot\,)}$, is a **(orthogonal) projector-valued measure**, where the $P_B$ are called the **spectral projectors** of the operator $A$, which means:

(i) Each $P_B$ is a orthogonal projector, whence $P_B^2 = P_B^* = P_B$;

(ii) $P_\varnothing = 0$ and $P_{\sigma(A)} = \mathbb{1}$;

(iii) For all $B_1, B_2 \in \mathscr{B}(\sigma(A))$, $P_{B_1} P_{B_2} = P_{B_1 \cap B_2}$;

(iv) If $\{B_n\}_{n \in \mathbb{N}} \subset \mathscr{B}(\sigma(A))$ is a sequence such that $B_k \cap B_j = \varnothing$ for $k \neq j$, then:

$$P_{\bigcup_{n \in \mathbb{N}} B_n} = s\text{-}\lim_{N \to \infty} \sum_{n=1}^{N} P_{B_n};$$

(v) If $\psi \in \mathscr{H}$, then:

$$\mu_{\psi, A}(B) = \langle \psi, P_B \psi \rangle, \qquad \forall B \in \mathscr{B}(\sigma(A)).$$

*Proof:* (i) Since the characteristic functions are such that $\chi_B^2 = \chi_B$ and $\overline{\chi_B} = \chi_B$ then by theorem B.0.56 item (a), $P_B^2 = \chi_B(A)\chi_B(A) = \chi_B(A) = P_B$ and $P_B^* = \overline{\chi_B(A)} = \chi_B(A) = P_B$.

(ii) $\chi_\varnothing = 0$, from which we get, again, by theorem B.0.56 item (a) that $P_\varnothing = \hat{\Phi}_A(\chi_\varnothing) = \hat{\Phi}_A(0) = 0$. $\chi_{\sigma(A)}$ is equal to the constant function 1 in $\sigma(A)$, hence by theorem B.0.56 item (a) $P_{\sigma(A)} = \hat{\Phi}(\chi_{\sigma(A)}) = \hat{\Phi}(1) = \mathbb{1}$.

(iii) Trivially $\chi_{B_1}\chi_{B_2} = \chi_{B_1 \cap B_2}$, therefore by the homomorphism property of $\hat{\Phi}_A$ in theorem B.0.56 item (a), we have $P_{B_1} P_{B_2} = \hat{\Phi}_A(\chi_{B_1})\hat{\Phi}_A(\chi_{B_2}) = \hat{\Phi}_A(\chi_{B_1}\chi_{B_2}) = \hat{\Phi}_A(\chi_{B_1 \cap B_2}) = P_{B_1 \cap B_2}$.

(iv) The sequence of Borel functions $g_N = \sum_{n=1}^{N} \chi_{B_n}$, fulfills $\|g_N\|_\infty = 1$, $\forall N \in \mathbb{N}$ since the $B_n$ are disjoint, hence each point $x \in \sigma(A)$ can be in at most one of the $B_n$'s. As there is no overlap between each $\chi_{B_n}$ we can write:

$$\chi_{\bigcup_{n\in\mathbb{N}} B_n}(x) = \lim_{N\to\infty} \sum_{n=1}^{N} \chi_{B_n}(x) = \lim_{N\to\infty} g_N(x), \quad \forall x \in \sigma(A).$$

Using now theorem B.0.56 items (a) and (c):

$$\hat{\Phi}_A\left(\chi_{\bigcup_{n\in\mathbb{N}} B_n}\right) = \hat{\Phi}_A\left(\lim_{N\to\infty} g_N\right) \overset{(c)}{=} s\text{-}\lim_{N\to\infty} \hat{\Phi}_A\left(\sum_{n=1}^{N} \chi_{B_n}\right) \overset{(a)}{=} s\text{-}\lim_{N\to\infty} \sum_{n=1}^{N} \hat{\Phi}_A(\chi_{B_n}).$$

In other words:

$$P_{\bigcup_{n\in\mathbb{N}} B_n} = s\text{-}\lim_{N\to\infty} \sum_{n=1}^{N} P_B.$$

(v) Considering the remark on definition C.0.24, we have that $\mu_{\psi,A}(B) = \int_{\sigma(A)} \chi_B d\mu_{\psi,A} = \langle \psi, \chi(A)\psi \rangle \equiv \langle \psi, P_B \psi \rangle$.

∎

**Theorem B.0.58.** *If $\{f_n\}_{n\in\mathbb{N}} \subset B^{\mathscr{B}}(X)$ is a bounded sequence that converges to $f : X \to \mathbb{C}$ pointwise, then the integral of $f$ with respect to a spectral measure $P$ (as was defined in Chapter 4) exists and equals:*

$$\int_X f(x) dP(x) = s\text{-}\lim_{n\to\infty} \int_X f_n(x) dP(x).$$

*Proof:* Firstly we note that $f \in B^{\mathscr{B}}(X)$, because $f$ is measurable as a limit of measurable functions, and bounded by the constant that bounds the sequence $\{f_n\}_{n\in\mathbb{N}}$. Calculating:

$$\left\|\int_X f(x)dP(x)\psi\right\|^2 = \left\langle \int_X f(x)dP(x)\psi, \int_X f(x)dP(x)\psi \right\rangle = \left\langle \psi, \left(\int_X f(x)dP(x)\right)^* \int_X f(x)dP(x)\psi \right\rangle, \quad \text{(B.19)}$$

From here on we prove that $\left(\int_X f(x)dP(x)\right)^* = \int_X \overline{f(x)}dP(x)$. Since $f \in B^{\mathscr{B}}(X)$, then by proposition C.0.23 there is a sequence of simple functions $\{s_n\}_{n\in\mathbb{N}}$ converging to $f$ uniformly. Taking $\psi, \phi \in \mathscr{H}$, then:

$$\left\langle \int_X \overline{s_n(x)} dP(x)\psi, \phi \right\rangle = \left\langle \sum_{n=1}^{N} \overline{\alpha_n} P_{E_n} \psi, \phi \right\rangle = \sum_{n=1}^{N} \alpha_n \langle \psi, P_{E_n} \phi \rangle =$$

$$= \left\langle \psi, \sum_{n=1}^{N} \alpha_n P_{E_n} \phi \right\rangle = \left\langle \phi, \int_X s_n(x) dP(x) \phi \right\rangle.$$

By the continuity of complex conjugation we have that $\overline{s_n} \to \overline{f}$ uniformly, as $N \to \infty$. Hence by the uniform convergence of $s_n$, the continuity and linearity of the integral with respect to a PVM, plus the continuity of the inner product, when we take the limit as $n \to \infty$, the above equation gives:

$$\left\langle \int_X \overline{f(x)} dP(x)\psi, \phi \right\rangle = \left\langle \psi, \int_X f(x) dP(x) \phi \right\rangle,$$

So:

$$\left\langle \left[ \int_X \overline{f(x)} dP(x) - \left(\int_X f(x) dP(x)\right)^* \right] \psi, \phi \right\rangle = 0.$$

As $\psi, \phi \in \mathscr{H}$ are arbitrary, then $\int_X \overline{f(x)} dP(x) = \left(\int_X f(x) dP(x)\right)^*$. Taking advantage of this detour we can also prove that $\left(\int_X f(x) dP(x)\right)\left(\int_X g(x) dP(x)\right) = \int_X f(x)g(x) dP(x)$ for all $f, g \in B^{\mathscr{B}}(X)$. Again by proposition C.0.23 there are sequences of simple functions $\{s_n\}_{n \in \mathbb{N}}$ and $\{t_n\}_{n \in \mathbb{N}}$ that converge uniformly respectively to $f$ and $g$, then computing the following product:

$$\int_X s_n(x) dP(x) \int_X t_n(x) dP(x) = \left(\sum_{n=1}^N \alpha_n P_{E_n}\right)\left(\sum_{m=1}^M \beta_m P_{F_m}\right) =$$

$$= \sum_{n=1}^N \sum_{m=1}^M \alpha_n \beta_m P_{E_n \cap F_m} = \int_X s_n(x) t_n(x) dP(x)$$

With $m$ fixed, it is clear that $s_n t_m$ tends to $f t_m$ uniformly, as $n \to \infty$, because $t_m$ is bounded. By continuity and linearity of the integral, the limit as $n \to \infty$ gives:

$$\int_X f(x) dP(x) \int_X t_m(x) dP(x) = \int_X f(x) t_m(x) dP(x),$$

where we used the fact that the composite of bounded operators is continuous in its arguments. Similarly, since $f t_m$ tends to $fg$ uniformly as $m \to \infty$, and so does the integral with respect to a PVM of this expression, hence $\left(\int_X f(x) dP(x)\right)\left(\int_X g(x) dP(x)\right) = \int_X f(x)g(x) dP(x)$, $\forall f, g \in B^{\mathscr{B}}(X)$.

Continuing then from (B.19):

$$= \left\langle \psi, \left(\int_X \overline{f(x)} dP(x)\right) \int_X f(x) dP(x) \psi \right\rangle = \left\langle \psi, \int_X |f(x)|^2 dP(x) \psi \right\rangle = \int_X |f(x)|^2 d\mu_{\psi, \left(\int_X \lambda dP(\lambda)\right)}(x)$$

Since by item *(ii)* of theorem B.0.57 $P_X = \mathbb{1}$, then $\mu_{\psi, \left(\int_X \lambda dP(\lambda)\right)}(X) = \langle \psi, \int_X \chi_X(x) dP(x) \psi \rangle = \langle \psi, P_X \psi \rangle = \langle \psi, \psi \rangle = \|\psi\|^2 < +\infty$, $\forall \psi \in \mathscr{H}$, therefore $\mu_{\psi, \left(\int_X \lambda dP(\lambda)\right)}$ is a finite measure.

Considering now the bounded sequence $\{f_n\}_{n \in \mathbb{N}}$ that converges pointwise to $f$ given on the statement of the theorem, we have:

$$\left\| \left(\int_X f(x) dP(x) - \int_X f_n(x) dP(x)\right) \psi \right\|^2 = \int_X |f(x) - f_n(x)|^2 d\mu_{\psi, \left(\int_X \lambda dP(\lambda)\right)}(x).$$

The measure $\mu_{\psi, \left(\int_X \lambda dP(\lambda)\right)}$ is finite, so constant maps are integrable. By assumption $\|f_n\|_\infty < K < +\infty$ for any $n \in \mathbb{N}$, so $\|f\|_\infty \leq K$ and then $\|f_n - f\|_\infty^2 \leq (\|f_n\|_\infty + \|f\|_\infty)^2 < 4K^2$. Since $|f_n - f|^2 \to 0$ pointwise, we can invoke the dominated convergence theorem C.0.26 to obtain, as $n \to \infty$:

$$\left\| \left(\int_X f(x) dP(x) - \int_X f_n(x) dP(x)\right) \psi \right\| = \sqrt{\int_X |f(x) - f_n(x)|^2 d\mu_{\psi, \left(\int_X \lambda dP(\lambda)\right)}(x)} \to 0.$$

Given the freedom in choosing $\psi \in \mathscr{H}$, this proves that $\int_X f(x) dP(x) = \text{s-}\lim_{n \to \infty} \int_X f_n(x) dP(x)$. ∎

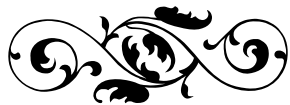

# Appendix C

# Measure Theory

We shall first prove the known *Measure Problem* of measure theory to differentiate this mathematical problem of the physical problem of measurement theory in **QM**.

Beginning by introducing basic measure theory notions, namely the *desiderata* for a volume measure $\mu$ in $\mathbb{R}^n$:

(a) **$\sigma$-additivity:** for every sequence $\{A_n\}_{n\in\mathbb{N}} \subset \mathbb{R}^n$ of two by two disjoint sets, i.e. $i,j \in \mathbb{N}, i \neq j, A_i \cap A_j = \varnothing$; then:
$$\mu\left(\bigcup_{n\in\mathbb{N}} A_n\right) = \sum_{n=1}^{\infty} A_n.$$

(b) **Translation invariance:** requires that the measure of a set $B \subset \mathbb{R}^n$ be equal to the measure of the translation of this set by any element $x \in \mathbb{R}^n$:
$$\mu(B+x) = \mu(B), \qquad B+x := \{b+x \in \mathbb{R}^n \,|\, b \in B\}.$$

(c) **Normalization:** $0 \leqslant \mu([0,1]^n) \leqslant +\infty$.

**Lemma C.0.1.** *For any set $X$ and collection $\mathcal{M} \subset \wp(X)$ that contains the empty set $\varnothing$, and a $\sigma$-additive function $\mu : \mathcal{M} \to [0,+\infty]$. If there is $A \in \mathcal{M}$ such that $\mu(A) \leqslant +\infty$ then $\mu(\varnothing) = 0$.*

*Proof:* Taking $A_1 = A$ and $A_k = \varnothing, \forall k > 1$, which makes the collection be constituted two by two disjoint sets. Since $A = A_1 \cup \left(\bigcup_{k>1} A_k\right)$, calculating the measure:
$$\mu(A) = \mu(A) + \sum_{k=2}^{\infty} \mu(\varnothing).$$

This implies $\mu(\varnothing) = 0$, since $\mu(A) < +\infty$ by hypothesis. ∎

**Corollary C.0.2.** *Let $A, B \in \mathcal{M}$, such that $A \subset B$ and $B \setminus A \in \mathcal{M}$. Given a $\sigma$-additive function $\mu : \mathcal{M} \to [0,+\infty]$ such that there exists a $C \in \mathcal{M}$ with $\mu(C) < +\infty$, then $\mu(A) \leqslant \mu(B)$.*

*Proof:* Note that $A$ and $B \setminus A$ are disjoint and that $B = A \cup (B \setminus A)$, for $A \subset B$; Then by the lemma we have that:
$$\mu(B) = \mu(A \cup (B \setminus A)) = \mu(A) + \mu(B \setminus A),$$
since $\mu(B \setminus A) \geqslant 0$, then $\mu(A) \leqslant \mu(B)$. ∎

**Theorem C.0.3** (The Measure Problem)**.** *There is no function $\mu : \mathcal{P}(X) \to [0,+\infty]$ that is concomitantly $\sigma$-additive, translation invariant and such that $\mu([0,1]) = 1$*

*Proof:* Let's start by constructing the **Vitali set** $\mathbb{V}$. Let $I = [0, 1]$; we can define in $I$ the relation

$$x, y \in I, \quad x \sim y \Leftrightarrow x - y \in \mathbb{Q},$$

which is clearly an equivalence relationship. Thus, as we know that $I$ is partitioned by these equivalence classes, applying the axiom of choice we can obtain a set formed by a single element from each equivalence class $[x]_\sim \in I/_\sim$, which we will call $\mathbb{V} \subset I$. Note that, given $r, s \in \mathbb{Q}$, if there exists $u \in (\mathbb{V} + r) \cap (\mathbb{V} + s)$ then, on the one hand, we have that $u = v + r$ for some $v \in \mathbb{V}$ and, on the other hand, that $u = v' + s$ for some $v' \in \mathbb{V}$. Now, this means that $v - v' = s - r \in \mathbb{Q}$, that is, that $v \sim v'$, which can only occur if $v = v'$ and, therefore, $r = s$, as we had constructed $\mathbb{V}$ so as to have exactly one element of each equivalence class in $I/_\sim$. In conclusion, we have $(\mathbb{V} + r) \cap (\mathbb{V} + s) = \emptyset$ if they are $r, s \in \mathbb{Q}$ with $r \neq s$.

Note also that, given $x \in I$ and the corresponding $y \in \mathbb{V}$ such that $y \in [x]_\sim$, we have $y + (x - y) \in \mathbb{V} + (x - y)$ (or, in other words, $x \in \mathbb{V} + (x - y)$), in addition to verifying that $x - y \in [-1, 1] \cap \mathbb{Q}$. Now, this implies the following containment order:

$$[0, 1] \subset \bigcup_{q \in \mathbb{Q} \cap [-1, 1]} (\mathbb{V} + q) \subset [-1, 2].$$

Thus, assuming the existence of a function $\mu$ according to the hypotheses of the statement, the Corollary C.0.2 holds and we obtain:

$$\mu([0, 1]) \leqslant \mu \left( \bigcup_{q \in \mathbb{Q} \cap [-1, 1]} (\mathbb{V} + q) \right) \leqslant \mu([-1, 2]).$$

Considering that $\mu$ is translation invariant and that $\mu([0, 1]) = 1$, we must have $\mu([-1, 2]) \leqslant 3$ and, finally, once that $\mathbb{Q}$ is enumerable and that, by the reasoning in the previous paragraph, the union above is disjoint, the hypothesis of $\sigma$-additivity of $\mu$ implies:

$$1 \leqslant \mu \left( \bigcup_{q \in \mathbb{Q} \cap [-1, 1]} (\mathbb{V} + q) \right) = \sum_{q \in \mathbb{Q} \cap [-1, 1]} \mu(\mathbb{V} + q) = \sum_{q \in \mathbb{Q} \cap [-1, 1]} \mu(\mathbb{V}) \leqslant 3.$$

One way or another, we get an absurd statement: either $\mu(\mathbb{V}) = 0$ and $1 \leqslant 0$, or $\mu(\mathbb{V}) > 0$ and $+\infty \leqslant 3$. Therefore, there cannot be any function $\mu$ with the characteristics specified in the statement, i.e. a finitely valued translationally invariant measure. ∎

**Definition C.0.4** ($\sigma$-algebra). *Let $X$ be a set and $\wp(X)$ be its power set. A subset $\Sigma \subseteq \wp(X)$ is called a $\sigma$-algebra if:*

(a) $X \in \Sigma$;

(b) $\Sigma$ is closed under complementation: $A \in \Sigma \Rightarrow (X \setminus A) \in \Sigma$;

(c) $\Sigma$ is closed under countable unions: $\{A_i\}_{i \in \mathbb{N}} \subset \Sigma \Rightarrow \bigcup_{i \in \mathbb{N}} A_i \in \Sigma$.

It follows that $\varnothing \in \Sigma$ since by (i) $X \in \Sigma$ and (ii) $X \setminus X = \varnothing \in \Sigma$, the $\sigma$-algebra is also closed under countable intersections by applying De Morgan's laws. Elements of the $\sigma$-algebra are called measurable sets, and $(X, \Sigma)$ is called a measurable space.

**Definition C.0.5** (Borel $\sigma$-algebra). *Given a topological space $(X, \tau)$, a **Borel set** $B \subset X$ is any set that can be formed by a given countable collection of open sets $\{V_i\}_{i \in I} \subset \tau$ through operations of countable union $\left(\bigcup_{\substack{j \in J \\ |J| < |\mathbb{N}|}}\right)$, countable intersection $\left(\bigcap_{\substack{j \in J \\ |J| < |\mathbb{N}|}}\right)$, and relative complement $(\backslash)$ of subcollections of those open sets together with the resulting sets of previous operations.*

*The **Borel $\sigma$-algebra** $\mathscr{B}_\tau(X)$ is the smallest $\sigma$-algebra containing all open sets, formed by all Borel sets of $(X, \tau)$, which can be generated considering the operations for a set $T$ given by: $T_\sigma$ which represents all countable unions of elements of $T$, $T_\delta$ that represents all countable intersections of elements of $T$ and $T_{\delta\sigma} = (T_\delta)_\sigma$. Then considering the sets $\tau$, $\tau_{\delta\sigma}$, $(\tau_{\delta\sigma})_{\delta\sigma}$, $((\tau_{\delta\sigma})_{\delta\sigma})_{\delta\sigma}$ and so forth iteratively, from the union of all those sets we get $\mathscr{B}_\tau(X)$.*

*We shall make the following abbreviation $\mathscr{B}(X)$ when the underlying topology is implicitly known.*

**Definition C.0.6** (Measurable functions and Borel measurable functions). *Considering two measurable spaces $(X, \Sigma)$ and $(Y, \Xi)$, a function $f : X \to Y$ is said to be **measurable** if:*

$$\forall E \in \Xi, \quad f^{-1}(E) \equiv \{x \in X \mid f(x) \in E\} \in \Sigma.$$

*and we represent it by $f : (X, \Sigma) \to (Y, \Xi)$. When $(Y, \Xi) = \mathbb{C}$ we say that the measurable functions are $\Sigma$-measurable.*

*If $(X, \tau)$ and $(Y, \kappa)$ are two topological spaces then a function $g : X \to Y$ is said to be **Borel measurable** if:*

$$\forall E \in \mathscr{B}_\kappa(Y), \quad g^{-1}(E) \equiv \{x \in X \mid f(x) \in E\} \in \mathscr{B}_\tau(X).$$

**Definition C.0.7** (Multiplicative class). *Considering the operations given by the countable unions $(\cdot)_\sigma$ and countable intersections $(\cdot)_\delta$ of elements of the underlying set to be operated upon, then if $\boldsymbol{G}$ is a family of open sets then considering the sequence: $\boldsymbol{G}_0 = \boldsymbol{G}, \boldsymbol{G}_1 = \boldsymbol{G}_\delta, \boldsymbol{G}_2 = \boldsymbol{G}_{\delta\sigma}, \boldsymbol{G}_3 = \boldsymbol{G}_{\delta\sigma\delta}, ...,$ we have that those $\boldsymbol{G}_n$'s where $n$ is a **odd** number are such that the intersection of any two elements of $\boldsymbol{G}_n$ is also in $\boldsymbol{G}_n$.*

*Similarly if $\boldsymbol{F}$ is a family of closed sets and considering the sequence $\boldsymbol{F}_0 = \boldsymbol{F}, \boldsymbol{F}_1 = \boldsymbol{F}_\sigma, \boldsymbol{F}_2 = \boldsymbol{F}_{\sigma\delta}, \boldsymbol{F}_3 = \boldsymbol{F}_{\sigma\delta\sigma}, ...,$ we have that those $\boldsymbol{F}_n$'s where $n$ is a **even** number are such that the intersection of any two elements of $\boldsymbol{F}_n$ is also in $\boldsymbol{F}_n$.*

*Then we say that a set $Y$ is of multiplicative class $\alpha \in \mathbb{N}_0$ if:*

- *in the case of $\alpha$ being a odd number $\alpha = 2n+1$, $n \in \mathbb{N}$, $Y \in \boldsymbol{G}_{2n+1}$, for a suitable collection of open sets $\boldsymbol{G}$, or,*

- *in the case of $\alpha$ being a even number $\alpha = 2n$, $n \in \mathbb{N}$, $Y \in \boldsymbol{F}_{2n}$, for a suitable collection of closed sets $\boldsymbol{F}$.*

**Definition C.0.8** (Class of Borel measurable functions). *A function $f : X \to Y$ is said to be a Borel measurable function of class $\alpha$, if for every closed Borel subset $F \in \mathscr{B}(Y)$, the set $f^{-1}(Y)$ is borelian of multiplicative class $\alpha$.*

**Definition C.0.9** (Borel Homeomorphism of class $(\alpha, \beta)$). *A function $f : X \to Y$ is a Borel homeomorphism of class $(\alpha, \beta)$ if $f$ is a bijection of $X$ and $Y$, and if $f$ is a Borel measurable function of class $\alpha$ and its inverse $f^{-1} : Y \to X$ is a Borel measurable function of class $\beta$.*

Given these definitions the following theorem was proved by Kuratowski in page 450 of [69]:

**Theorem C.0.10.** *For every pair of uncountable sets $A$ and $B$ of the multiplicative classes $(\alpha + l) > 2$ and $(\beta + l) > 2$, respectively, there exists a homeomorphism $f$ of $A$ onto $B$ of class $(\alpha, \beta)$.*

This theorem in particular implies that if $M_1$ and $M_2$ are complete separable metric spaces and if $E_1$ and $E_2$ are noncountable Borel subsets of $M_1$ and $M_2$ respectively. Then $E_1$ and $E_2$ are isomorphic as Borel spaces.

**Definition C.0.11** (Complex measure). *A complex measure on a measurable space $(X, \Sigma)$ is a map $\mu : \Sigma \to \mathbb{C}$ associating a complex number to every element in the $\sigma$-algebra $\Sigma$ on $X$ so that:*

*(a) $\mu(\varnothing) = 0$;*

*(b) For any collection $\{D_i\}_{i \in \mathbb{N}_0} \subset \Sigma$, such that $D_n \cap D_m = \varnothing$ if $n \neq m$, then:*

$$\mu\left(\bigcup_{i \in \mathbb{N}_0} D_n\right) = \sum_{i=0}^{+\infty} \mu(D_n)$$

*If $\mu(\Sigma) \subset \mathbb{R}$ then $\mu$ is called a signed measure or charge on $X$. The joining of a measurable space $(X, \Sigma)$ with a specific measure $\mu$ (that can be either a complex, positive or signed measure) is called a measure space $(X, \Sigma, \mu)$.*

We follow on, defining connected topological spaces, Stone spaces, and enunciating the Stone's representation theorem for Boolean algebras, the proof of which can be found in [54], the definition of a $\sigma$-ideal and the Loomis–Sikorski representation theorem.

**Definition C.0.12** (Connected topological space). *For a topological space $(X, \tau)$ to be connected we must have that whenever $U, V \in \tau$ are such that $U \cup V = X$ and $U \cap V = \varnothing$ then either $U = \varnothing$ or $V = \varnothing$.*

**Definition C.0.13** (Ultrafilter). *A ultrafilter $\mathfrak{F}$ is a proper filter (def. 2.2.9) of a lattice $\mathcal{L}$, such that there exists no proper filter $\mathfrak{G}$ of $\mathcal{L}$ with $\mathfrak{F} \subsetneq \mathfrak{G}$.*

**Definition C.0.14** (Stone space). *Considering $S$ to be a topological space, if $S$ is compact, Hausdorff, and such that all it's connected subspaces are singletons, which makes $S$ a totally disconnected space, then $S$ is a Stone space.*

*For a Boolean lattice $\mathcal{B}$, we define the Stone space of $\mathcal{B}$ to be the topological space $S(\mathcal{B}) := (U, \tau)$, where:*

*(a) $U$ is the set of ultrafilters in $\mathcal{B}$.*

*(b) $\tau$ is the topology generated by the basis consisting of all sets of the form $\{x \in U \mid b \in \mathcal{B} \text{ and } b \in x\}$.*

**Theorem C.0.15** (Stone representation). *If $\mathcal{B}$ is a Boolean algebra and $S(\mathcal{B})$ is its associated Stone space, then the set of concurrently open and closed sets (clopen sets) of $S(\mathcal{B})$ is a Boolean algebra under union, intersection, and complementation in $S(\mathcal{B})$, also this Boolean algebra of clopen sets is isomorphic to $\mathcal{B}$.*

We then follow this theorem with a comparable one for $\sigma$-algebras, one way to comprehend Stone's representation theorem is that one can always represent a abstract Boolean algebra defined on a lattice as a Boolean algebra of sets in which the meet, join and orthogonal complement are given respectively by the set operations of intersection, union and set complement on a collection of specific sets and that also has the minimum element **0** in the form of $\varnothing$ and its orthogonal complement $\neg \mathbf{0}$ as the set complement $\varnothing^\complement$.

Equally one can ask if it is possible to represent a abstract $\sigma$-algebra, understood to be a lattice closed under the usual operations of meet, join and orthogonal complement containing also a minimum element **0** and its complement that is the maximum element **1**, beyond that we also have closeness with respect to suprema of countable collections of elements of the lattice $\bigvee_{n=1}^{\infty}$ and infima $\bigwedge_{n=1}^{\infty}$; going to a usual concrete $\sigma$-algebra $\Sigma$ of a measurable space $(X, \Sigma)$, in which all the usual operations of countable intersection, countable union and set complement containing the empty set $\varnothing$ and the total set $X$. The answer to the possibility of this representation lies in the Loomis—Sikorski representation theorem, which we will enunciate and prove after defining a $\sigma$-ideal:

**Definition C.0.16** ($\sigma$-ideal). *Let $\mathcal{C}$ be a lattice closed under the operations of meet $\wedge$, join $\vee$, orthogonal complement $\neg$, also of suprema $\bigvee_{n=1}^{\infty}$ and infima $\bigwedge_{n=1}^{\infty}$ of countable collections of elements of the lattice, that is $\mathcal{C}$ is a abstract $\sigma$-algebra. A $\sigma$-ideal is a subset $\mathcal{N} \subset \mathcal{C}$ such that $\mathbf{0} \in \mathcal{N}$, $\mathcal{N}$ is closed under countable joins and $\mathcal{N}$ is downwardly closed, that is, if for a $N \in \mathcal{N}$ and a $A \in \mathcal{C}$ is such that $A \subset \mathcal{N}$, then $A \in \mathcal{N}$.*

*If $\mathcal{N}$ is a $\sigma$-ideal, then we say that two elements $A$ and $B$ of $\mathcal{C}$ are equivalent modulo $\mathcal{N}$ if their symmetric difference lies in $\mathcal{N}$, that is $A \sim_{\mathcal{N}} B \Leftrightarrow A \triangle B := (A \wedge (\neg B)) \vee (B \wedge (\neg A)) \in \mathcal{N}$.*

*The quotient of $\mathcal{C}$ by this equivalence relation is denoted by $\mathcal{C}/\sim_{\mathcal{N}}$.*

**Theorem C.0.17** (Loomis–Sikorski representation). *Let $\mathcal{C}$ be a lattice closed under the operations of meet $\wedge$, join $\vee$, orthogonal complement $\neg$, also of suprema $\left(\bigvee_{n=1}^{\infty}\right)$ and infima $\left(\bigwedge_{n=1}^{\infty}\right)$ of countable collections of elements of the lattice, that is $\mathcal{C}$ is a abstract $\sigma$-algebra, then there exists a concrete $\sigma$-algebra $\Sigma$ from a measurable space $(X, \Sigma)$ and a $\sigma$-ideal $\mathcal{N}$ of $\Sigma$ such that $\mathcal{C} \simeq \Sigma/\sim_{\mathcal{N}}$.*

*Proof:* This proof will follow the arguments originally presented by Loomis, applying Stone's representation theorem C.0.15, we can find a Stone space $X$ such that there is a Boolean algebra isomorphism $\varphi : \mathcal{C} \to \text{Clo}(X)$ from $\mathcal{C}$ seen as a Boolean algebra to the algebra of clopen sets of $X$, here denoted by $\text{Clo}(X)$. Letting $\Sigma$ be the Baire $\sigma$-algebra of $X$ generated by $\text{Clo}(X)$, that is the smallest $\sigma$-algebra such that all continuous real-valued functions are measurable.

The map $\varphi$ need not be a $\sigma$-algebra isomorphism, being merely a Boolean algebra isomorphism. In particular if $\{B_n\}_{n \in \mathbb{N}^0} \subset \mathcal{C}$ is such that $\bigwedge_{n=0}^{\infty} B_n = \mathbf{0}$ then $\bigcap_{n=0}^{\infty} \varphi(B_n) \in \Sigma$ is not necessarily the empty set, considering these sets $\bigcap_{n=0}^{\infty} \varphi(B_n)$ that are formed by $B_n$'s such that the countable intersection is empty, we shall call they the *basic null sets*. let $\mathcal{N}$ be the collection of sets in $\Sigma$ which can be covered by countably many basic null sets.

$\mathcal{N}$ is clearly a $\sigma$-ideal in $\Sigma$, since $\varnothing \in \mathcal{N}$ by vacuity in the forming intersection, it is clearly closed under countable unions since each of the elements in this union have each a countable covering by basic null sets and therefore their union will be a countable covering of the element formed by countable unions of elements from $\mathcal{N}$, it is also downwardly closed since if a $N \in \mathcal{N}$ and a $A \in \Sigma$ are such that $A \subset N$ then clearly if $\bigcup_{n=0}^{\infty} V_n$ is any countable cover of $N$ by the basic null sets $\{V_n\}_{n \in \mathbb{N}^0}$ then it is also a cover of $A$ by basic null sets, and therefore $\mathcal{N}$ is a $\sigma$-ideal in $\Sigma$.

Considering then the quotient map $[\,\cdot\,]_{\sim_\mathcal{N}} : \mathrm{Clo}(X) \to \Sigma/{\sim_\mathcal{N}}$, then the map $\varphi : \mathcal{C} \to \mathrm{Clo}(X)$ descends into the map $\widetilde{\varphi} : \mathcal{C} \to \Sigma/{\sim_\mathcal{N}}$, where $\widetilde{\varphi} = [\,\cdot\,]_{\sim_\mathcal{N}} \circ \varphi$:

$$\begin{array}{ccc} \mathcal{C} & \xrightarrow{\varphi} & \mathrm{Clo}(X) \\ & \searrow{\widetilde{\varphi}} & \downarrow{[\,\cdot\,]_{\sim_\mathcal{N}}} \\ & & \Sigma/{\sim_\mathcal{N}} \end{array}$$

Clearly $\widetilde{\varphi}$ is a Boolean Algebra homomorphism, since $\varphi$ was a Boolean Algebra isomorphism and $\widetilde{\varphi}$ still relates a given $A \in \mathcal{C}$ to a $\varphi(A) \in \mathrm{Clo}(X)$ even if this association is made by the equivalence class $[\varphi(A)]_{\sim_\mathcal{N}} \in \Sigma/{\mathcal{N}}$. More than that, if $\{B_n\}_{n \in \mathbb{N}^0} \subset \mathcal{C}$ is such that $\bigwedge_{n=0}^{\infty} B_n = \mathbf{0}$ then by construction $\bigcap_{n=0}^{\infty} \widetilde{\varphi}(B_n) = \varnothing$. From these two facts and De Morgan's laws it is clear that $\widetilde{\varphi}$ is in fact a $\sigma$-homomorphism.

Since by Stone's representation theorem C.0.15 $\varphi(\mathcal{C}) = \mathrm{Clo}(X)$ and this set generates $\Sigma$, then $\widetilde{\varphi}$ generates $\Sigma/{\mathcal{N}}$, and so $\widetilde{\varphi}$ is surjective.

The only remaining task is to show that $\widetilde{\varphi}$ is injective. As before, it suffices to show that $\widetilde{\varphi}(A) \neq \varnothing$ when $A \neq \mathbf{0}$. Suppose for contradiction that $A \neq \mathbf{0}$ and $\widetilde{\varphi}(A) = \varnothing$; then $\widetilde{\varphi}(A)$ can be covered by a countable family $\bigcap_{n=1}^{\infty} \widetilde{\varphi}(A_n^{(i)})$ of basic null sets, where $\bigwedge_{n=1}^{\infty} A_n^{(i)} = \mathbf{0}$ for each $i$. Since $A \neq \mathbf{0}$ and $\bigwedge_{n=1}^{\infty} A_n^{(1)} = \mathbf{0}$, we can find $n_1$ such that $A \wedge (\neg A_{n_1}^{(1)}) \neq \mathbf{0}$. Iterating this, we can find $n_2, n_3, n_4, \ldots$ such that $A \wedge \left(\neg(A_{n_1}^{(1)} \vee \ldots \vee A_{n_k}^{(k)})\right) \neq \mathbf{0}$ for all $k$. Since $\widetilde{\varphi}$ is a Boolean algebra isomorphism, we conclude that $\widetilde{\varphi}(A)$ is not covered by any finite subcollection of the $\widetilde{\varphi}(A_{n_1}^{(1)}), \widetilde{\varphi}(A_{n_2}^{(2)}), \ldots$ But all of these sets are clopen, so by compactness, $\widetilde{\varphi}(A)$ is not covered by the entire collection $\widetilde{\varphi}(A_{n_1}^{(1)}), \widetilde{\varphi}(A_{n_2}^{(2)}), \ldots$. But this contradicts the fact that $\widetilde{\varphi}(A)$ is covered by the $\bigcap_{n=1}^{\infty} \widetilde{\varphi}(A_n^{(i)})$.

∎

**Definition C.0.18** (Positive measure). *Let $X$ be a set and $\Sigma$ a $\sigma$-algebra over $X$. A set function $\mu : \Sigma \to \mathbb{R} \cup \{\infty\}$ is called a positive measure if:*

(i) *$\mu(E) \geq 0, \forall E \in \Sigma$;*

(ii) *$\mu(\varnothing) = 0$;*

(iii) *It is $\sigma$-additive, i.e. for $\{E_k\}_{k \in \mathbb{N}} \subset \Sigma$ a collection of pairwise disjoint sets:*

$$\mu\left(\bigcup_{k=1}^{\infty} E_k\right) = \sum_{k=1}^{\infty} \mu(E_k)$$

*If the first condition of positivity of $\mu$ is dropped and for $\mu$ to be well defined for every element of $\Sigma$, the regularity condition that $\mu$ takes on at most one of the values of $+\infty$ or $-\infty$, since this blocks situations in which the indeterminable form $\infty - \infty$ occurs, such as when we have two disjoint sets one being measured to $+\infty$ and the other to $-\infty$ and we consider the measure of the union of these sets, or when a set of the $\sigma$-algebra is measured to be one of the infinities and a subset of this set has a measure of the opposite sign infinity; then $\mu$ is called a **signed measure**.*

**Definition C.0.19** (Total Variation). *In a measure space $(X, \Sigma, \mu)$ the **total variation** of $\mu$, is by definition the $\sigma$-additive positive measure $|\mu|$ on $\Sigma$, such that for any $E \in \Sigma$, whose partitions are sets of the form $\{E_i\}_{i \in I}$ such that $\bigcup_{i \in I} E_i = E$ and $E_n \cap E_m = \varnothing$ if $n \neq m$, we then have:*

$$|\mu|(E) := \sup\left\{ \sum_{i \in I} |\mu(E_i)| \,\bigg|\, \{E_i\}_{i \in I} \text{ partition of } E \right\} \quad, \forall E \in \Sigma.$$

**Definition C.0.20** (Kinds of positive measures). *Measure spaces $(X, \Sigma, \mu)$ with positive measures $\mu$ are called respectively:*

(a) **Finite** *if $\mu(X) < +\infty$;*

(b) *A **probability space** and **probability measure**, if $\mu(X) = 1$;*

(c) *A **Borel space** and **Borel measure**, if $X$ is locally compact, Hausdorff and $\Sigma = \mathscr{B}(X)$, that is, $\Sigma$ is the Borel $\sigma$-algebra of $X$.*

*More generally if $X$ is locally compact, Hausdorff and $\Sigma \supset \mathscr{B}(X)$ then we may call the measure $\mu$:*

 a) **inner regular** *if:*

$$\mu(E) = \sup\{\mu(K) \,|\, K \subset E, K \text{ is compact }\} \quad, \forall E \in \Sigma;$$

 b) **outer regular** *if:*

$$\mu(E) = \inf\{\mu(V) \,|\, V \supset E, V \text{ is open }\} \quad, \forall E \in \Sigma;$$

 c) **regular** *when simultaneously inner and outer regular.*

**Proposition C.0.21.** *$|\mu|(X) < +\infty$, making $|\mu|$ a finite measure.*

*Proof:* Suppose first that some set $E \in \Sigma$ has $|\mu|(E) = +\infty$. Put $t = n(1 + |\mu(E)|)$. Since $|\mu|(E) > t$, as complex measures cannot diverge and then $|\mu(E)| < +\infty$, there is a partition $\{E_i\}_{i=1}^N$ of $E$ such that:

$$\sum_{i=1}^N |\mu(E_i)| > t$$

since the supremum over all partitions diverge. Considering the complex numbers $\{\mu(E_i)\}_{i=1}^N$, we write $\mu(E_k) = |\mu(E_k)|e^{i\alpha_k}$. Then for $-\pi \leq \theta \leq \pi$, we consider $V(\theta)$ to be the set of indexes such that $\cos(\alpha_k - \theta) > 0$, that is, $V(\theta) = \{k \in \mathbb{N} \,|\, \cos(\alpha_k - \theta) > 0\}$, then:

$$\left| \sum_{k \in V(\theta)} \mu(E_k) \right| = \left| \sum_{k \in V(\theta)} \mu(E_k) e^{-i\theta} \right| \geq \mathfrak{Re}\left[ \sum_{k \in V(\theta)} \mu(E_k) e^{-i\theta} \right].$$

But then, if we write the positive part of the cosine as $\cos^+(x) = \frac{\cos(x)+|\cos(x)|}{2}$, since the set $V(\theta)$ is made up of exactly those points in which this function differs from zero for $\alpha_k - \theta$, we can make the exchange:

$$\mathfrak{Re}\left[\sum_{k\in V(\theta)} \mu(E_k)e^{-i\theta}\right] = \sum_{k=1}^N |\mu(E_k)|\cos^+(\alpha_k - \theta),$$

choosing then the angle $\theta_0$ that maximizes the previous sum we get that this maximum must be at lest as large as the average of the sum over $[-\pi, \pi]$, this average is:

$$\frac{1}{2\pi}\int_{-\pi}^{\pi}\sum_{k=1}^N |\mu(E_k)|\cos^+(\alpha_k-\theta)d\theta = \frac{1}{2\pi}\sum_{k=1}^N |\mu(E_k)|\int_{-\pi}^{\pi}\frac{\cos(x)+|\cos(x)|}{2}d\theta = \frac{1}{\pi}\sum_{k=1}^N |\mu(E_k)|,$$

therefore $\sum_{k=1}^N |\mu(E_k)| \leq \pi\left|\sum_{k\in V(\theta_0)}\mu(E_k)\right|$, and hence:

$$\left|\sum_{k\in V(\theta_0)}\mu(E_k)\right| = \left|\mu\left(\bigcup_{k\in V(\theta_0)}E_k\right)\right| \geq \frac{1}{\pi}\sum_{i=1}^N |\mu(E_i)| > \frac{t}{\pi} > 1.$$

Setting $B = E \setminus \left(\bigcup_{k\in V(\theta_0)} E_k\right)$, we have that:

$$|\mu(B)| = \left|\mu(E) - \mu\left(\bigcup_{k\in V(\theta_0)}E_k\right)\right| \geq \left|\mu\left(\bigcup_{k\in V(\theta_0)}E_k\right)\right| - |\mu(E)| \geq \frac{t}{\pi} - |\mu(E)| = 1.$$

We have split the set $E$ into two disjoint sets, $B$ and $A = \bigcup_{k\in V(\theta_0)}E_k$, with $|\mu(B)| > 1$ and $|\mu(A)| > 1$. Evidently, at least one of $|\mu|(A)$ and $|\mu|(B)$ is $\infty$.

Now if $|\mu|(X) = \infty$, split $X$ into disjoint sets $A_1$ and $B_1$, following the construction above, with $|\mu(A_1)| > 1$ and $|\mu|(B_1) = \infty$. Split $B_1$ in the same way into $A_2$ and $B_2$ such that $|\mu(A_2)| > 1$ and $|\mu|(B_2) = \infty$, and continue with this process iteratively for every natural number, we then get a countably infinite disjoint collection $\{A_i\}_{i\in\mathbb{N}}$ such that $|\mu(A_i)| > 1$, $\forall i \in \mathbb{N}$; and the countable additivity of $\mu$ implies that

$$\mu\left(\bigcup_{i\in\mathbb{N}}A_i\right) = \sum_{i\in\mathbb{N}}\mu(A_i)$$

But this series cannot converge, since for all $i$ we have $|\mu(A_i)| > 1$ and therefore $\mu(A_i)$ does not tend to 0 as $i \to \infty$. Therefore the set $\bigcup_{i\in\mathbb{N}} A_i$ is not sent to a finite complex number by the measure $\mu$, this is a contradiction with the fact that $\mu$ is a complex measure, and therefore this shows that $|\mu|(X) < \infty$. ∎

**Definition C.0.22.** *A simple function $s : X \to \mathbb{C}$ is any function which can be written as $s = \sum_{n=1}^N \alpha_n \chi_{E_n}$, where $N$ is a finite natural number, $\alpha_n \in \mathbb{C}$ and $\chi_{E_n}$ are the characteristic functions for a sequence of disjoint measurable sets $\{E_n\}_{n=1,\ldots,N}$, i.e., for any $x \in X$, $\chi_{E_n}(x) = 1$ if $x \in E_n$ and $\chi_{E_n}(x) = 0$ if $x \notin E_n$. Hence a map is simple if its range is finite.*

**Proposition C.0.23.** *If $f : X \to [0, +\infty]$ is measurable in the measurable space $(X, \Sigma)$, there exists a sequence of simple functions $0 \leq s_1 \leq s_2 \leq \cdots \leq s_n \leq f$ with $s_n \to f$ pointwise. The convergence is uniform if $f$ is bounded. More generally, the space of complex valued simple functions with respect to $\Sigma$, denoted by $S(X)$, is dense in the space of $\mathbb{C}$-valued, $\Sigma$-measurable maps, with uniform convergence of these simple functions when the function to be approximated is in $B^{\mathscr{B}}(X)$.*

*Proof:* For given $0 < n \in \mathbb{N}$ and $1 \leq k \leq n2^n$ let us partition the real semi-axis $[0, +\infty)$ into Borel sets:
$$E_{n,k} := \left[\frac{k-1}{2^n}, \frac{k}{2^n}\right), \qquad E_n := [n, +\infty).$$

For each $n$ let:
$$\Upsilon^f_{n,k} := f^{-1}(E_{n,k}), \qquad \Upsilon^f_n := f^{-1}(E_n);$$

be subsets in $\Sigma$. Then define $s_0(x) := 0, \forall x \in X$, and:
$$s_n := \sum_{k=1}^{n2^n} \frac{k-1}{2^n} \chi_{\Upsilon^f_{n,k}} + n\chi_{\Upsilon^f_n}.$$

For any $n \in \mathbb{N}$. As the sets $\Upsilon^f_{n,k}$ and $\Upsilon^f_n$ represent a partition on the range of $f$, the functions $s_n$ are simple functions that assume the least value of the sets $E_{n,k}$ and $E_n$ for all inputs whose image in one of those respective sets, therefore, by construction, $0 \leq s_n \leq s_{n+1} \leq f$, $n \in \mathbb{N}$. Furthermore, for any given $x \in \bigcup_{k=1}^{n2^n} E_{n,k}$ we have $|f(x) - s_n(x)| \leq \frac{1}{2^n}$. Evidently $s_n \to f$ pointwise if $n \to \infty$, since any point $x$ is in a $E_{n,k}$ for a a $n$ big enough.

If $f$ is bounded, then we can take $n > \sup f$, then $|f(x) - s_n(x)| \leq \frac{1}{2^n}, \forall x \in X$ and hence $s_n \to f$ uniformly when $\sup f < n \to \infty$. For the more general case of complex valued functions, such as $g$, we can just separate the real $\mathfrak{Re}(g)$ and imaginary parts $\mathfrak{Im}(g)$, and hence get two real functions, which can in turn be divided in positive and negative parts by $(\cdot)^+ := \sup\{0, (\cdot)\}$ and $(\cdot)^- = -\inf\{0, (\cdot)\}$, in this way we get four non-negative real functions, each of which we can use the first part of the proof to prove this more general assertion. ∎

**Definition C.0.24** (Integration). *Let $f : (X, \Sigma) \to \mathbb{R}_+ \cup \{+\infty\}$ be a non-negative measurable function on a measure space $(X, \Sigma, \mu)$. We can then define:*
$$\int_E f d\mu = \sup_s \left\{ \sum_{i \in I} \alpha_i \mu(E_i) \in \mathbb{R}_+ \,\middle|\, s = \sum_{i \in I} \alpha_i \chi_{E_i} \text{ is a simple function}, 0 \leq s \leq f, \right.$$
$$\left. \{E_i\}_{i \in I} \text{ is a partition of } E \in \Sigma, I \text{ is a finite set} \right\}.$$

*If $f : (X, \Sigma) \to \overline{\mathbb{R}}$ isn't non-negative, but is still a measurable function, we can write $f = f^+ + f^-$, where $f^+(x) := \chi_{[0,+\infty]}(f(x))f(x)$ and $f^-(x) := -\chi_{[-\infty,0)}(f(x))f(x)$ for all $x \in X$, then if $\min\left(\int_E f^+ d\mu, \int_E f^- d\mu\right) < \infty$, we define:*
$$\int_E f d\mu = \int_E f^+ d\mu - \int_E f^- d\mu.$$

*We say that $f$ is Lebesgue integrable in this case if $\int_E |f| d\mu = \int_E f^+ + f^- d\mu \leq +\infty$.*

*For complex functions such as $g : (X, \Sigma) \to \mathbb{C}$, we can always write they like $g(z) = \mathfrak{Re}(g)(z) + i\mathfrak{Im}(g)(z)$ where the first term is a real function and the other term is a real function*

*multiplied by $i$. A sufficient and necessary condition for the integral of $g$ to be well defined is for $\int_E |g|d\mu$ to be Lebesgue integrable, then:*

$$\int_E g d\mu = \int_E \mathfrak{Re}(g) d\mu + i \int_E \mathfrak{Im}(g) d\mu.$$

Remark: *It is clear from these definitions that $\int_X \chi_E d\mu = \mu(E)$ for $E \in \Sigma$.*

We take a little diversion to enunciate Fatou's lemma which is used in the Riesz-Fischer theorem B.0.11 of Appendix B, a good proof of this lemma in portuguese can be found in Appendix G of chapter 31 of [8], or for a proof in english in [88].

**Lemma C.0.25** (Fatou)**.** *Let $(X, \Sigma, \mu)$ be a measure space with a positive measure $\mu$. Let $\{f_n\}_{n \in \mathbb{N}}$ be a sequence of measurable non-negative functions $f_n : X \to \mathbb{R}$. Then:*

$$\int \left( \liminf_{n \to \infty} f_n \right) d\mu \leq \liminf_{n \to \infty} \int_X f_n d\mu.$$

We then enunciate Lebesgue's dominated convergence theorem, whose proof can be found in any measure theory book that is worth its salt, in particular, the ones already cited [88, 43].

**Theorem C.0.26** (Lebesgue's dominated convergence)**.** *Suppose $\{f_n\}_{n \in \mathbb{N}}$ is a is a $L^1(\mathbb{C}, \mu)$ sequence of complex measurable functions on $X$ such that $\lim_{n \to \infty} f_n(x) = f(x)$ for almost every point $x \in X$. If there is a function $g \in L^1(\mathbb{R}, \mu)$ such that $|f_n(x)| \leq g(x)$ for almost every point $x \in X$ and $\forall\, n \in \mathbb{N}$. Then $f \in L^1(\mathbb{C}, \mu)$,*

$$\lim_{n \to \infty} \int_X |f - f_n| d\mu = 0,$$

*and:*

$$\lim_{n \to \infty} \int_X f_n d\mu = \int_X f d\mu.$$

**Proposition C.0.27.** *A $\sigma$-additive positive measure $\mu$ on a measurable space $(X, \Sigma)$, and hence also for $L^p(X, d\mu)$, **is separable** if the following conditions hold:*

*(i) $\mu$ is $\sigma$-finite (that is $\exists \{M_n\}_{n \in \mathbb{N}}$ such that $\forall n \in \mathbb{N}, \mu(M_n) < \infty$ and we have that $\bigcup_{n \in \mathbb{N}} M_n = X$).*

*(ii) the $\sigma$-algebra $\Sigma$ is generated by a countable collection of measurable sets at most.*

*Proof:* Let $\{E_i\}_{i \in I}$ be a countable sequence of sets in $\Sigma$, such that the $\sigma$-algebra $\Sigma = \sigma(\{E_i\}_{i \in I}) := \bigcap_F \{F \subseteq \wp(X) \,|\, \{E_i\}_{i \in I} \subseteq F, F \text{ is a } \sigma\text{-algebra}\}$, that is, the smallest $\sigma$-algebra that contains $\{E_i\}_{i \in I}$ is precisely $\Sigma$, *i.e.* $\{E_i\}_{i \in I}$ generates $\Sigma$. Since $\mu$ is $\sigma$-finite by hypothesis, being $\{M_n\}_{n \in \mathbb{N}}$ any covering of $X$ by sets of finite measure, then the sets $\{(E_i \cap M_n)\}_{i \in I, n \in \mathbb{N}}$ each have finite measure, are in a enumerable quantity since their cardinality is the product of the cardinalities of a countable set of indexes $I$ with the enumerable set $\mathbb{N}$ and is hence enumerable, and since the sets $\{E_i\}_{i \in I}$ already generated the $\sigma$-algebra $\Sigma$ then the refinement of these sets in the form of $\{(E_i \cap M_n)\}_{i \in I, n \in \mathbb{N}}$ must also generate the $\sigma$-algebra $\Sigma$.

　　The $\sigma$-algebra $\Sigma$ that is generated by $\{E_i\}_{i \in I}$ must also have a countable cardinality, since if we denote by $\breve{L}$ the set formed by all the countable intersections, countable unions and complements of the elements of $L$, then the hierarchy formed by $L_0 = \{E_i\}_{i \in I}$, $L_1 = \widetilde{\{E_i\}_{i \in I}} = \breve{L}_0$, $L_2 = \breve{L}_1, ...$ ; then:

$$\{E_i\}_{i \in I} \subseteq \bigcup_{n \in \mathbb{N}} \breve{L}_n \subseteq \sigma(\{E_i\}_{i \in I})$$

But since $\bigcup_{n \in \mathbb{N}} \breve{L}_n$ is clearly closed under countable intersections, countable unions and complements, then $\bigcup_{n \in \mathbb{N}} \breve{L}_n$ is a $\sigma$-algebra and hence $\sigma(\{E_i\}_{i \in I}) \subseteq \bigcup_{n \in \mathbb{N}} \breve{L}_n$ and therefore $\bigcup_{n \in \mathbb{N}} \breve{L}_n = \Sigma$, as each $L_n$ is formed by countable operations on countable elements then their cardinality is also countable and so is the cardinality of $\bigcup_{n \in \mathbb{N}} \breve{L}_n$, that is also $\Sigma$.

Having made all of these points explicit then we can consider without loss of generality that there are enumerable sequences such as $\{G_n\}_{n \in \mathbb{N}}$ with the properties that $\mu(G_n) < \infty, \forall n \in \mathbb{N}$, $\bigcup_{n \in \mathbb{N}} G_n = X$ and that $\{G_n\}_{n \in \mathbb{N}} = \bigcup_{n \in \mathbb{N}} \breve{L}_n$.

Considering then that for any given set $B \in \Sigma$ of finite measure:

$$\mu(B) = \inf_{\{B_i\}_{i \in \mathbb{N}}} \left\{ \sum_{i=1}^{\infty} \mu(B_i) \,\middle|\, B \subset \bigcup_{i \in \mathbb{N}} B_i,\ B_i \in \Sigma, \forall i \in \mathbb{N} \right\}$$

Consequently there exists a sequence $\{B_i\}_{i \in \mathbb{N}}$ of sets of $\Sigma$ such that:

$$B \subset \bigcup_{i \in \mathbb{N}} B_i \quad \text{and} \quad \mu\left(\bigcup_{i \in \mathbb{N}} B_i\right) \leq \mu(B) + \frac{\varepsilon}{2} \tag{C.1}$$

for a arbitrarily small value $\varepsilon > 0$. Since $\lim_{n \to \infty} \mu\left(\bigcup_{i=1}^{n} B_i\right) = \mu\left(\bigcup_{i=1}^{\infty} B_i\right)$ and this is a finite value, then we can evidently find a integer $n_0$ such that:

$$\mu\left(\bigcup_{i \in \mathbb{N}} B_i\right) \leq \mu\left(\bigcup_{i=1}^{n_0} B_i\right) + \frac{\varepsilon}{2}. \tag{C.2}$$

Clearly $\bigcup_{i=1}^{n_0} B_i =: B_0 \in \Sigma$ and we have that:

$$\mu(B \setminus B_0) \leq \mu\left(\bigcup_{i \in \mathbb{N}} B_i \setminus B_0\right) = \mu\left(\bigcup_{i \in \mathbb{N}} B_i\right) - \mu(B_0) \overset{(C.2)}{\leq} \frac{\varepsilon}{2}$$

and

$$\mu(B_0 \setminus B) \leq \mu\left(\bigcup_{i \in \mathbb{N}} B_i \setminus B\right) = \mu\left(\bigcup_{i \in \mathbb{N}} B_i\right) - \mu(B) \overset{(C.1)}{\leq} \frac{\varepsilon}{2}$$

And then $\mu(B \triangle B_0) \leq \varepsilon$. We conclude that for any set of finite measure $B \in \Sigma$ and any $\varepsilon > 0$ there is a set $B_0 \in \Sigma$ such that $\mu(B \triangle B_0) < \varepsilon$.

The operation $\mu(A \triangle B)$ has all the properties of a metric for sets with finite measure (the triangle inequality is a consequence of $(A \triangle B) \triangle (B \triangle C) = A \triangle C$, an identity easily proven using the associative property of $\triangle$, then $A \triangle C \subset (A \triangle B) \cup (B \triangle C)$ and hence $\mu(A \triangle C) \leq \mu(A \triangle B) + \mu(B \triangle C)$); which makes the subcollection of sets with a finite measure be a metric space with the metric $d(A, B) := \mu(A \triangle B)$, considering then any set with a finite measure $B \in \Sigma$ then, as we have shown, for any $\varepsilon > 0$ there is a set $B_0 \in \Sigma$ such that $d(B, B_0) = \mu(B \triangle B_0) < \varepsilon$, using now that the collection $\{G_n\}_{n \in \mathbb{N}} = \bigcup_{n \in \mathbb{N}} \breve{L}_n$ then $B_0 \in \{G_n\}_{n \in \mathbb{N}}$, lets say $B_0 = G_k$ for some $k$, then for a arbitrary $B \in \Sigma$ with a finite measure and for a $\varepsilon > 0$, there is always a $G_k \in \{G_n\}_{n \in \mathbb{N}}$ such that $d(B, G_k) = \mu(B \triangle G_k) < \varepsilon$, hence these sets $\{G_n\}_{n \in \mathbb{N}}$ are dense in $\Sigma$, this makes $\mu$ separable, since the sets with finite measure form a separable metric space. ∎

**Theorem C.0.28** (Liouville's volume). *For a $n$-parameter family of maps $F_t : \mathbb{R}^n \to \mathbb{R}^n$, given by:*

$$F_t(a_1, \ldots, a_n) = (x_1(a_1, \ldots, a_n, t), \ldots, x_n(a_1, \ldots, a_n, t)),$$

*and the $n$-parameter family of vector fields:*

$$\mathbf{X}(t)(a_1, \ldots, a_n) = \left(\frac{\partial x_1}{\partial t}(a_1, \ldots, a_n, t), \ldots, \frac{\partial x_n}{\partial t}(a_1, \ldots, a_n, t)\right) = (x'_1(a_1, \ldots, a_n, t), \ldots, x'_n(a_1, \ldots, a_n, t)).$$

*If $\nabla \cdot \mathbf{X}(t) = 0$ then $F_t$ is volume preserving.*

*Proof:* We shall first prove the following lemma:

**Lemma C.0.29.** $\frac{\partial}{\partial t}\left(\det \frac{\partial(x_1, \ldots, x_n)}{\partial(a_1, \ldots, a_n)}(t)\right) = \det\left(\frac{\partial(x_1, \ldots, x_n)}{\partial(a_1, \ldots, a_n)}(t)\right)(\nabla \cdot \mathbf{X}(t))$.

*Proof:* Consider $A$ to be a arbitrary $n \times n$ matrix, then its Schur decomposition allows one to write $A = UQU^{-1}$ where $U$ is a unitary matrix and $Q$ is a upper triangular matrix, writing $Q = D + N$ where $D$ is a diagonal matrix and $N$ is a strictly upper triangular and thus a nilpotent matrix, then $A = U(D + N)U^{-1}$.

Calculating $e^A := \sum_{k=0}^{\infty} \frac{A^k}{k!}$, and considering that since $A^k = (U(D+N)U^{-1})^k = U(D+//+N)^k U^{-1}$ and as $DN$ and $ND$ are also strictly upper triangular, then $A^k = U(D^k + N_k)U^{-1}$ where $N_k := DN_{K-1} + N_{K-1}D + N_{K-1}N$, $k = 2, 3, \ldots$ with $N_1 = N$ are each a strictly upper diagonal matrix; therefore $e^A = U\left(\sum_{k=0}^{\infty} \frac{D^k}{k!} + \sum_{k=0}^{\infty} \frac{N_k}{k!}\right)U^{-1} = U\left(e^D + \sum_{k=0}^{\infty} \frac{N_k}{k!}\right)U^{-1}$, $e^D$ is a diagonal matrix and the sum $\sum_{k=0}^{\infty} \frac{N_k}{k!}$ is strictly upper triangular, making $e^D + \sum_{k=0}^{\infty} \frac{N_k}{k!}$ be a upper triangular matrix.

Calculating $\det\left(e^A\right) = \det\left(U\left(e^D + \sum_{k=0}^{\infty} \frac{N_k}{k!}\right)U^{-1}\right) = \det\left(e^D + \sum_{k=0}^{\infty} \frac{N_k}{k!}\right)$, but the determinant of an upper triangular matrix is the product of the diagonal elements, letting $\lambda_1, \lambda_2, \ldots$ be such diagonal elements, then $\det\left(e^A\right) = \prod_{n \in \mathbb{N}} e^{\lambda_n} = e^{\left(\sum_{n \in \mathbb{N}} \lambda_n\right)} = e^{\operatorname{tr}(A)}$.

If the matrix $\frac{\partial(x_1, \ldots, x_n)}{\partial(a_1, \ldots, a_n)}$ is sufficiently close to the identity matrix, $\left\|\frac{\partial(x_1, \ldots, x_n)}{\partial(a_1, \ldots, a_n)} - \mathbb{1}\right\|_{H-S} < 1$ for the Hilbert–Schmidt norm $\|A\|_{H-S} := (\operatorname{tr}(A^*A))^{\frac{1}{2}}$, then we can define the natural logarithm of $\frac{\partial(x_1, \ldots, x_n)}{\partial(a_1, \ldots, a_n)}$ to be $\ln\left(\frac{\partial(x_1, \ldots, x_n)}{\partial(a_1, \ldots, a_n)}\right) := \sum_{k=1}^{\infty} \frac{(-1)^{k+1}}{k}\left(\frac{\partial(x_1, \ldots, x_n)}{\partial(a_1, \ldots, a_n)} - \mathbb{1}\right)^k$ and then:

$$e^{\operatorname{tr}\left(\ln\left(\frac{\partial(x_1, \ldots, x_n)}{\partial(a_1, \ldots, a_n)}(t)\right)\right)} = \det\left(\frac{\partial(x_1, \ldots, x_n)}{\partial(a_1, \ldots, a_n)}(t)\right). \tag{C.3}$$

If the matrix $\frac{\partial(x_1, \ldots, x_n)}{\partial(a_1, \ldots, a_n)}$ isn't sufficiently close to $\mathbb{1}$ but is diagonalizable then $\frac{\partial(x_1, \ldots, x_n)}{\partial(a_1, \ldots, a_n)} = VDV^{-1}$, where $D$ is a diagonal matrix $D = \operatorname{diag}(\lambda_1, \ldots, \lambda_n)$ then we can define the natural logarithm of $\frac{\partial(x_1, \ldots, x_n)}{\partial(a_1, \ldots, a_n)}$ to be $\ln\left(\frac{\partial(x_1, \ldots, x_n)}{\partial(a_1, \ldots, a_n)}\right) = V\operatorname{diag}(\ln(\lambda_1), \ldots, \ln(\lambda_n))V^{-1}$. If $\frac{\partial(x_1, \ldots, x_n)}{\partial(a_1, \ldots, a_n)}$ isn't diagonalizable then we can use the fact that the diagonalizable matrices are dense in $M_n(\mathbb{C})$ to justify that the formula $e^{\operatorname{tr}\left(\ln\left(\frac{\partial(x_1, \ldots, x_n)}{\partial(a_1, \ldots, a_n)}\right)\right)} = \det\left(\frac{\partial(x_1, \ldots, x_n)}{\partial(a_1, \ldots, a_n)}\right)$ is valid for all matrices.

That the diagonalizable matrices are dense in $M_n(\mathbb{C})$ is a consequence of considering again the Schur decomposition of a arbitrary matrix $A \in M_n(\mathbb{C})$, $A = U(\text{diag}(\lambda_1, \ldots, \lambda_n) + N)U^{-1}$ where again $N$ is a strictly upper triangular matrix, consider the sequence $\{\varepsilon_{ik}\}_{i \in \{1,\ldots,n\}, k \in \mathbb{N}}$ such that $\lambda_j + \varepsilon_{jk} \neq \lambda_l + \varepsilon_{lk}$ for $j \neq l$ and $\varepsilon_{ik} \xrightarrow{k \to \infty} 0$, $\forall i \in \{1, \ldots, n\}$. Then $U(\text{diag}(\lambda_1 + \varepsilon_{1k}, \ldots, \lambda_n + \varepsilon_{nk}) + N)U^{-1}$ is diagonalizable, since all their eigenvalues are distinct, and converges to $A$.

Hence from (C.3.3) we get:

$$\frac{\partial}{\partial t}\left(\det \frac{\partial(x_1, \ldots, x_n)}{\partial(a_1, \ldots, a_n)}(t)\right) = \frac{\partial}{\partial t} e^{\text{tr}\left(\ln\left(\frac{\partial(x_1, \ldots, x_n)}{\partial(a_1, \ldots, a_n)}(t)\right)\right)} =$$

$$= e^{\text{tr}\left(\ln\left(\frac{\partial(x_1, \ldots, x_n)}{\partial(a_1, \ldots, a_n)}\right)\right)} \text{tr}\left(\left(\frac{\partial(x_1, \ldots, x_n)}{\partial(a_1, \ldots, a_n)}\right)^{-1} \frac{\partial}{\partial t}\frac{\partial(x_1, \ldots, x_n)}{\partial(a_1, \ldots, a_n)}\right) =$$

$$= \det\left(\frac{\partial(x_1, \ldots, x_n)}{\partial(a_1, \ldots, a_n)}\right) \sum_{i,j=1}^{n} \left(\left(\frac{\partial(x_1, \ldots, x_n)}{\partial(a_1, \ldots, a_n)}\right)^{-1}_{ij} \frac{\partial}{\partial t}\left(\frac{\partial(x_1, \ldots, x_n)}{\partial(a-1, \ldots, a_n)}\right)_{ji}\right) =$$

$$= \det\left(\frac{\partial(x_1, \ldots, x_n)}{\partial(a_1, \ldots, a_n)}\right) \sum_{i,j=1}^{n} \frac{\partial a^i}{\partial x^j}\frac{\partial x'_j}{\partial a_i} = \det\left(\frac{\partial(x_1, \ldots, x_n)}{\partial(a_1, \ldots, a_n)}\right) \sum_{i,j=1}^{n}\sum_{k=1}^{n} \frac{\partial a^i}{\partial x^j}\frac{\partial x_k}{\partial a_i}\frac{\partial x'_j}{\partial x_k} =$$

$$= \det\left(\frac{\partial(x_1, \ldots, x_n)}{\partial(a_1, \ldots, a_n)}\right) \sum_{j,k=1}^{n}\sum_{k=1}^{n} \delta_{ik}\frac{\partial x'_j}{\partial x_k} = \det\left(\frac{\partial(x_1, \ldots, x_n)}{\partial(a_1, \ldots, a_n)}\right) \sum_{j=1}^{n} \frac{\partial x'_j}{\partial x_j} =$$

$$= \det\left(\frac{\partial(x_1, \ldots, x_n)}{\partial(a_1, \ldots, a_n)}(t)\right)(\nabla \cdot \mathbf{X}(t)).$$

∎

Now, for a bounded open set $R \subset \mathbb{R}^n$, let $R_t = F_t(R)$ then its volume is calculated as:

$$\text{Vol}(R_t) = \int_{R_t} dx_1 \ldots dx_n,$$

and by the change of variable formula for multiple integrals, we have:

$$\text{Vol}(R_t) = \int_{R_t} dx_1 \ldots dx_n = \int_R \det\left(\frac{\partial(x_1, \ldots, x_n)}{\partial(a_1, \ldots, a_n)}(t)\right) da_1 \ldots da_n,$$

so:

$$\text{Vol}'(R_t) = \int_R \frac{\partial}{\partial t} \det\left(\frac{\partial(x_1, \ldots, x_n)}{\partial(a_1, \ldots, a_n)}(t)\right) da_1 \ldots da_n \stackrel{C.0.29}{=} \int_R (\nabla \cdot \mathbf{X}(t)) \det\left(\frac{\partial(x_1, \ldots, x_n)}{\partial(a_1, \ldots, a_n)}(t)\right) da_1 \ldots da_n =$$

$$\text{Vol}'(R_t) = \int_{R_t} (\nabla \cdot \mathbf{X}(t)) \, dx_1 \ldots dx_n.$$

Therefore if $\nabla \cdot \mathbf{X}(t) = 0$ then the variation in volume by applying $F_t$ is zero. For the specific case of a Hamiltonian vector flow within generalized coordinates $(q^1, \ldots, q^n, p_1, \ldots, p_n)$ then $F_t(q^1, \ldots, q^n, p_1, \ldots, p_n) = \varphi_t(q^1, \ldots, q^n, p_1, \ldots, p_n)$, then:

$$\nabla \cdot \xi_H = \sum_{i=1}^{n} \frac{\partial \dot{q}^i}{\partial q^i} + \sum_{i=1}^{n} \frac{\partial \dot{p}_i}{\partial p_i} \stackrel{(2.1)}{=} \sum_{i=1}^{n} \frac{\partial}{\partial q^i}\left(\frac{\partial H}{\partial p_i}\right) + \sum_{i=1}^{n} \frac{\partial}{\partial p_i}\left(-\frac{\partial H}{\partial q^i}\right) = 0.$$

And hence Hamiltonian dynamics preserves volumes in phase-space. ∎

**Definition C.0.30** (Regular complex Borel measure). *A complex Borel measure $\mu$ is called regular if the finite positive Borel measure given by the total variation $|\mu|$ is regular.*

**Lemma C.0.31.** *On any Hilbert space $\mathcal{H}$ that is either separable or of finite dimension $> 2$, for any non-negative frame function $f$ there exists a bounded, self-adjoint operator $T \in \mathfrak{B}(\mathcal{H})$ such that $f(x) = \langle x, Tx \rangle$, for every $x \in \mathbb{S}(\mathcal{H})$.*

*Proof:* To prove this lemma we will need other 10 lemmas, 2 propositions, 4 theorems and 4 definitions; we will break the proof in two parts, one for the finite dimensional case and the other for the infinite dimensional case.

We begin with the **finite dimensional case**. We first define a **completely real subspace** of a Hilbert space:

**Definition C.0.32** (Completely real subspace). *Let $\mathcal{H}$ be a Hilbert space, a closed real-linear subspace $S \subset \mathcal{H}$ is a **completely real subspace** if the inner product $\langle \cdot, \cdot \rangle : \mathcal{H} \times \mathcal{H} \to \mathbb{C}$ of the Hilbert space, takes only real values in $S \times S$.*

We then categorize frame functions as **bounded**, **semibounded** and **regular**.

**Definition C.0.33** (Frame function types). *A **frame function**, on a Hilbert space $\mathcal{H}$, $f : \mathbb{S}(\mathcal{H}) \to \mathbb{R}$ can be categorized in the following ways:*

(i) *A **bounded frame function** $f$ is such that $\sup\{\,|f(x)| \,|\, x \in \mathbb{S}(\mathcal{H})\} < \infty$.*

(ii) *A **semibounded frame function** $f$ is such that $\inf\{f(x) \,|\, x \in \mathbb{S}(\mathcal{H})\} > -\infty$.*

(iii) *A **regular frame function** $f$ is such that there exists a Hermitian operator $T$ on $\mathcal{H}$ such that:*
$$f(x) = \langle Tx, x \rangle, \quad x \in \mathbb{S}(\mathcal{H}). \tag{C.4}$$

**- Finite dimensional** Hilbert space case:

**Proposition C.0.34.** *Let $\dim(\mathcal{H}) = n$. A frame function $f$ on $\mathcal{H}$ is bounded if and only if $f$ is semibounded.*

*Proof:* Clearly a bound function is semibounded, otherwise there would be a sequence $\{x_n\}_{n \in \mathbb{N}}$ such that $f(x_n) \xrightarrow{n \to \infty} -\infty$ and therefore $|f(x_n)| \xrightarrow{n \to \infty} \infty$. To prove the other direction consider that $f$ is semibounded and that $\dim(\mathcal{H}) \geq 2$, since for dimension 1 it is immediate since $f(x) = W, \forall x \in \mathbb{S}(\mathcal{H})$. Being $K = \inf\{f(x) \,|\, x \in \mathbb{S}(\mathcal{H})\}$ and for a arbitrary $y \in \mathbb{S}(\mathcal{H})$ such that this unit vector together with $\{x_1, \ldots, x_{n-1}\} \subset \mathbb{S}(\mathcal{H})$ form a Hilbert basis. Then by the frame function property:

$$f(t) = W - \sum_{i=1}^{n-1} f(x_i)$$

Hence, $|f(x)| \leq W - (n-1)K, \forall x \in \mathbb{S}(\mathcal{H})$. ∎

We begin by proving that any semibounded frame function on $\mathbb{R}^3$ is regular.

**Proposition C.0.35.** *Let $f : \mathbb{S}(\mathbb{R}^3) \to \mathbb{R}$ be a semibounded frame function with $M = \sup\{f(x) \,|\, x \in \mathbb{S}(\mathbb{R}^3)\}$ and $m = \inf\{f(x) \,|\, x \in \mathbb{S}(\mathbb{R}^3)\}$. Let $\varepsilon > 0$ and $s \in \mathbb{S}(\mathbb{R}^3)$ with $f(s) > M - \varepsilon$ be given. Then there exists $t \in \mathbb{S}(\mathbb{R}^3)$ with $s \perp t$ and $f(t) < m + \varepsilon$.*

*Proof:* Consider a $\delta > 0$ such that $f(s) > M - \varepsilon + \delta$ and find a $t' \in \mathbb{S}(\mathbb{R}^3)$ such that $f(t') < m + \delta$. We then consider the great circle determined by $s$ and $t'$, which can be obtained by considering the component of $t'$ that is orthogonal to $s$ and considering a circle parametrization given by $C_{s,t'}(\tau) := \cos(\tau)s + \sin(\tau)\frac{(t' - (s \cdot t')s)}{\|t' - (s \cdot t')s\|}$, from this great circle we then choose $t, s' \in C_{s,t'} \equiv \{C_{s,t'}(\tau) \,|\, \tau \in [0, 2\pi)\}$ such that $t \perp s$ and $s' \perp t'$.

Since $f$ is a frame function on $\mathbb{R}^3$ defined on the sphere $\mathbb{S}(\mathbb{R}^3)$, then for any closed subspace $\mathcal{R}$ of $\mathbb{R}^3$ the function $f\big|_{\mathbb{S}(\mathcal{R})}$ is then, obviously, also a frame function for $\mathcal{R}$, this works in general for any Hilbert space $\mathscr{H}$. Since $\overline{\operatorname{span} C_{s,t'}}$ is a closed subspace of $\mathbb{R}^3$ then $f\big|_{C_{s,t'}}$ is also a frame function for this subspace and therefore by the properties of frame functions:

$$f(t) + f(s) = f(t') + f(s') \Rightarrow f(t) = f(t') + f(s') - f(s) < M + (m + \delta) - (M - \varepsilon + \delta) = m + \varepsilon.$$

∎

**Lemma C.0.36.** *Let $C$ be a finite or enumerable subset of $(0,1)$. Let $f : [0,1] \setminus C \to \mathbb{R}$ be a function such that:*

*(i) $f(0) = 0$;*

*(ii) If $a, b \in [0,1] \setminus C$ and $a < b$, then $f(a) < f(b)$;*

*(iii) If $a, b, c \in [0,1] \setminus C$ and $a + b + c = 1$, then $f(a) + f(b) + f(c) = 1$;*

*(iv) The limit $\lim_{\substack{t \to 0 \\ t \in [0,1] \setminus C}} f(t)$ exists and is equal to $f(0)$.*

*Then $f(a) = a$ for all $a \in [0,1] \setminus C$.*

*Proof:* The set $\hat{C} = \{rc \,|\, r \in \mathbb{Q}, c \in C\} \cup \{r(1-c) \,|\, r \in \mathbb{Q}, c \in C\}$ is clearly at most enumerable, since $\mathbb{Q}$ is enumerable and $C$ is at most enumerable, so we can find a $a_0 \in (0,1)$ that does not belong to $\hat{C}$, this implies that $ra_0 \notin C$ and $(1-r)a_0 \notin C$ for any rational $r \in \mathbb{Q}$. Considering in property *(iii)* of the function $f$ the values $a = ra_0$, $b = 1 - ra_0$ and $c = 0$ we get:

$$f(ra_0) = 1 - f(1 - ra_0), \tag{C.5}$$

If, instead of zero, we had set $a = ra_0, b = sa_0$ and $c = 1 - (ra_0 + sa_0)$ for $r, s \in \mathbb{Q}$ such that $ra_0, sa_0$ and $(r+s)a_0$ are in $[0,1]$, we would get that:

$$f(ra_0) + f(sa_0) = 1 - f(1 - (ra_0 + sa_0)) \stackrel{(C.5)}{=} f((r+s)a_0), \tag{C.6}$$

From this we can see that $\forall q \in \mathbb{Q}$ such that $qa_0 \in [0,1]$, and since $q = \frac{n}{d}$, $n, d \in \mathbb{N}$ by simple calculation, using the last equation recursively, we see that:

$$f(a_0) = f\left(d\frac{1}{d}a_0\right) = f\left(\sum_{i=1}^{d} \frac{1}{d}a_0\right) \stackrel{(C.6)}{=} d\, f\left(\frac{1}{d}a_0\right) \Rightarrow f\left(\frac{1}{d}a_0\right) = \frac{1}{d}f(a_0), \tag{C.7}$$

$$f(qa_0) = f\left(\frac{n}{d}a_0\right) = f\left(\sum_{i=1}^{n} \frac{1}{d}a_0\right) \stackrel{(C.6)}{=} nf\left(\frac{1}{d}a_0\right) \stackrel{(C.7)}{=} \frac{n}{d}f(a_0),$$

So that:
$$f(qa_0) = qf(a_0). \tag{C.8}$$

Since by *(iv)* we have that $\lim\limits_{\substack{t \to 0 \\ t \in [0,1] \setminus C}} f(t) = f(0) \stackrel{(i)}{=} 0$, then by trying to calculate $\lim\limits_{\substack{t \to \frac{1}{a_0} \\ t \in \mathbb{Q} \cap ([0,1] \setminus C)}} f(ta_0) \stackrel{(C.5)}{=}$
$\lim\limits_{\substack{t \to \frac{1}{a_0} \\ t \in \mathbb{Q} \cap ([0,1] \setminus C)}} (1 - f(1 - ta_0)) = 1 - \lim\limits_{\substack{k \to 0 \\ k \in [0,1] \setminus C}} f(k) = 1$. Therefore $1 = \lim\limits_{\substack{r \to \frac{1}{a_0} \\ r \in \mathbb{Q} \cap ([0,1] \setminus C)}} f(ra_0) = \lim\limits_{\substack{r \to \frac{1}{a_0} \\ r \in \mathbb{Q} \cap ([0,1] \setminus C)}} rf(a_0) =$
$= \frac{1}{a_0} f(a_0) \implies f(a_0) = a_0$ therefore $f(ra_0) = ra_0$ for any $r \in \mathbb{Q}$ such that $ra_0 \in [0,1]$. For a arbitrary $a \in [0,1] \setminus C$ and considering sequences of rational numbers $\{q_k\}_{k \in \mathbb{N}}$ and $\{p_k\}_{k \in \mathbb{N}}$, with $\{q_k a_0\}_{k \in \mathbb{N}} \subset [0,1] \supset \{p_k a_0\}_{k \in \mathbb{N}}$ such that $q_k a_0 \underset{k \to \infty}{\nearrow} a$ and $p_k a_0 \underset{k \to \infty}{\searrow} a$ therefore by *(ii)* and the squeeze theorem we get that $f(a) = a$, $\forall a \in [0,1] \setminus C$. ∎

**Lemma C.0.37.** *Let a frame function $f$ satisfy:*

(a) $f(p) = \sup\{f(x) \,|\, x \in \mathbb{S}(\mathbb{R}^3)\}$.

(b) $f$ *is constant on* $E_p \equiv \{y \in \mathbb{S}(\mathbb{R}^3) \,|\, y \perp p\}$ *(The equator with relation to $p$)*.

*Then for any $s \in N_p \equiv \{x \in \mathbb{S}(\mathbb{R}^3) \,|\, 0 < \arccos(x \cdot p) < \frac{\pi}{2}\}$ (The north pole with relation to $p$ excluding $p$ itself) and $s' \in D_s^p \equiv \{\cos(\alpha)s + \sin(\alpha)v \,|\, v \in E_p \cap E_s, \alpha \in [-\frac{\pi}{2}, \frac{\pi}{2}]\}$ (The descent through $s$ on $N_p$, or the great circle passing through $s$ such that the point closest to the north pole $p$ is $s$ itself). we have:*
$$f(s) \geq f(s').$$

*Proof:* From *(i)* $\exists \sup\{f(x) \,|\, x \in \mathbb{S}(\mathbb{R}^3)\}$ and since $\sup\{f(x) \,|\, x \in \mathbb{S}(\mathbb{R}^3)\} = -\inf\{-f(x) \,|\, x \in \mathbb{S}(\mathbb{R}^3)\}$ then the frame function $-f$ is semibounded and by proposition C.0.34 $-f$ is bounded, that is, $f$ is bounded. Calling $M = f(p)$ and $m = \inf\{f(x) \,|\, x \in \mathbb{S}(\mathbb{R}^3)\}$. If in proposition C.0.35 we put $s = p$, then we find $t \in E_p$ such that $f(t) < m + \varepsilon$. Letting instead $s \in N_p$ and $s' \in D_s^p$ and choosing $t, t' \in D_s^p$ in such a way that $s \perp t$, and $s' \perp t'$.

Since $\overline{\text{span}\, D_s^p}$ is a closed subspace of $\mathbb{R}^3$ then $f$ keeps being a frame function when constrained to $D_s^p$ and by the frame function property $f(s) + f(t) = f(s') + f(t')$, that $f(t') \geq m \Rightarrow f(t') - m \geq 0$ and the fact that $t \in E_p$ and the choice of $s \in N_p$ can make the choice of $\varepsilon$ that is dependent on $s$ arbitrarily small, whereas $f$ is constant on $E_p$ independently of all that, this makes $f(t) \leq m + \varepsilon$:
$$f(s) - f(s') = f(t') - f(t) \geq f(t') - (m + \varepsilon) \geq -\varepsilon,$$
since $\varepsilon > 0$ is arbitrarily small then $f(s) \geq f(s')$. ∎

**Lemma C.0.38.** *Let $f$ be a semibounded frame function on $\mathbb{R}^3$ and let $\varepsilon > 0$ be such that:*

(a) $f(p) > \sup\{f(x) \,|\, x \in \mathbb{S}(\mathbb{R}^3)\} - \varepsilon$.

(b) $f$ *is constant on $E_p$.*

*Then for $s \in E_p$ and $s' \in D_s^p$, $f(s) > f(s') - \varepsilon$.*

*Proof:* From proposition C.0.35 we conclude that $f(e) < m + \varepsilon$, $e \in E_p$, and:
$$f(s) - f(s') = f(t') - f(t) > f(t') - m - \varepsilon \geq -\varepsilon.$$
∎

**Lemma C.0.39.** *Let $s, t \in N_p$ and let the latitudes $l^p(w) := (w \cdot p)^2$ be such that $l(t) < l(s)$. Then there exists an integer $n \geq 1$ and a finite sequence $s_0, \ldots, s_n \in N_p$ such that $s_0 = s$, $s_n = t$ and $s_i \in D_{s_{i-1}}$ for $1 \leq i \leq n$.*

*Proof:* To facilitate the proof we visualize the azimuthal projection of the sphere $\mathbb{S}(\mathbb{R}^3)$, in which rays coming from the center of $\mathbb{S}(\mathbb{R}^3)$ project vectors on $\mathbb{S}(\mathbb{R}^3)$ to points on the tangent plane at point $p$. All the vectors with a given latitude in $\mathbb{S}(\mathbb{R}^3)$ are projected onto circles centered at $p$ in the tangent plane, and the descent $D_s^p$ becomes the straight line through a point $s$ that is tangent to the latitude circle that $s$ belongs.

If in the tangent plane, the projections of $s$ and $t$ lie on a ray beginning from the origin of the plane at point $p$, we may choose $n = 2$ and the point $s_1$ as one of the intersection points of the circle in the plane that passes through both the origin $p$ and $t$ with its center lying on the ray that joins $t$ and $s$, intersecting the line formed in the plane by $D_s^p$. To explain why that choice is justified, we first note that by the nature of the required sequence $s_1 \in D_s^p$, and choosing the previous circle makes any triangle formed with $p, t$ and a arbitrary point on the circle be a right triangle with the right angle on the vertex of the arbitrary point on the circle, by the inscribed angle theorem, this guarantees that the line joining the intersection point $s_1$ with $t$ is the descent $D_{s_1}^p$ since it is orthogonal to the straight line joining $p$ and $s_1$ on the plane.

For the general case, we first consider the following observation, fixing $s_0 = s = (x, 0)$ (in the plane coordinates) and $k \geq 1$, we choose finitely many points $s_1, \ldots, s_k$ successively and such that $s_i \in D_{s_{i-1}}^p$, choosing the angle between $s_i$ and $s_{i-1}$ to be $\frac{\pi}{k}$. Then the plane coordinates of $s_k$ will be $(-y_k, 0)$ with $y_k \geq x$, we shall then prove that $y_k - x \to 0$ as $k \to \infty$. Let $d_i$ be the distance from $s_i$ to the origin, then $d_0 = x$ and $d_k = y_k$, and for any $i$ we have $\frac{d_{i+1}}{d_i} = \frac{1}{\cos(\frac{\pi}{k})}$ which then gives:

$$1 \leq \frac{y_k}{x} = \prod_{i=1}^{k} \frac{d_i}{d_{i-1}} = \left(\cos\left(\frac{\pi}{k}\right)\right)^{-k} \leq \left(1 - \frac{1}{2}\frac{\pi^2}{k^2}\right)^{-k} \xrightarrow{k \to \infty} 1.$$

So that we can travel half a rotation of the plane while maintaining the radius of the points as close to that of $s$ as desired, in fact, this result shows that we can give even a full rotation on the plane with a ending radius arbitrarily close to that of $s$. So considering that the angle that is formed by $s, p$ and $t$ in the plane is $\theta$ then choosing $k \in \mathbb{N}$ big enough for:

$$x \left(\cos\left(\frac{\theta}{k}\right)\right)^{-k} < \tan\left(\arccos\left(t \cdot p\right)\right),$$

where $\tan(\arccos(t \cdot p))$ is the perceived size of $t$ in the plane in the same way that $x = \tan(\arccos(s \cdot p))$, then after arriving in this point we can use the construction with the inscribed angle, which we did for the case in which both $s$ and $t$ were aligned with a straight line passing through the origin $p$ in the plane, for the last point $s_{k+1}$, in such a way that $n = k + 2$. ∎

**Lemma C.0.40.** *Let $f$ be a frame function of $\mathbb{R}^3$ such that:*

(a) *$f(p) = M = \sup\{f(x) \mid x \in \mathbb{S}(\mathbb{R}^3)\}$.*

(b) *$f$ has a constant value $m$ on $E_p$. Then for any $s \in \mathbb{S}(\mathbb{R}^3)$ we have:*

$$f(s) = m + (M - m)(s \cdot p)^2.$$

*Proof:* From proposition C.0.35 we conclude that $m = \inf\{f(x) \mid x \in \mathbb{S}(\mathbb{R}^3)\}$, so that if $M = m$, then the lemma is true. If $M \neq m$, without loss of generality we can assume that $m = 0$ and $M = 1$ ( replacing $f(x)$ by $\left(\frac{f(x) - m\|x\|^2}{M-m}\right)$ if necessary). Let $s, t \in N_p$ with $l(s) > l(t)$ be given. Lemmas C.0.39 and C.0.37 yield $f(s) \geq f(t)$. For any $\ell \in [0,1]$ we put:

$$\overline{{}^p f}(\ell) = \sup\{f(s) \mid s \in N_p, \, l(s) = \ell\},$$

$$\underline{{}^p f}(\ell) = \inf\{f(s) \mid s \in N_p, \, l(s) = \ell\}.$$

Then $\overline{{}^p f}(1) = \underline{{}^p f}(1) = 1$, $\overline{{}^p f}(0) = \underline{{}^p f}(0) = 0$ and if $\ell_1, \ell_2 \in [0,1]$ with $\ell_1 < \ell_2$, we conclude $\overline{{}^p f}(\ell_1) \leq \underline{{}^p f}(\ell_2)$. Therefore the set $C = \{\ell \in [0,1] \mid \overline{{}^p f}(\ell) > \underline{{}^p f}(\ell)\}$ is at most enumerable due to $\sum_{\ell \in C} (\overline{{}^p f}(\ell) - \underline{{}^p f}(\ell)) \leq 1$. Define $\tilde{f}$ on $[0,1] \setminus C$ by $\tilde{f}(\ell) = \overline{{}^p f}(\ell) = \underline{{}^p f}(\ell)$.

Let $a, b, c \in [0,1]$ satisfy $a + b + c = 1$. We can find a frame $(x, y, z)$ with $l(x) = a, l(y) = b, l(z) = c$. Consequently, $\tilde{f}$ satisfies the conditions of lemma C.0.36, which means that $\tilde{f}(\ell) = \ell$ for $\ell \in [0,1] \setminus C$. From the definition of $\tilde{f}$ and from the property that $f(t) \leq f(s)$ for $l(s) < l(t)$, $s, t \in N_p$, we obtain that $C = \varnothing$, so that $f(s) = \tilde{f}(l(s)) = l(s) = (s \cdot p)^2$ holds for all $s \in N_p$. ∎

**Lemma C.0.41.** *Any semibounded frame function on $\mathbb{R}^3$ attains its extremal values.*

*Proof:* Denote by $M$ and $m$ the supremum and infimum of a semibounded frame function $f$ on $\mathbb{S}(\mathbb{R}^3)$. Choose a sequence $\{p_n\}_{n \in \mathbb{N}}$ in $\mathbb{S}(\mathbb{R}^3)$ such that $\lim_n f(p_n) = M$. Due to the compactness of the strong topology in the finite-dimensional $\mathbb{S}(\mathbb{R}^3)$, we may assume passing to a subsequence, if necessary, that $\{p_n\}_{n \in \mathbb{N}}$ converges, and we put $p = \lim_n p_n$. Without loss of generality we can assume that $p_n \in N_p$ for any $n \in \mathbb{N}$. Our aim is to show that $f(p) = M$ in four steps.

(a) Suppose that $e_0 \in E_p$, and denote by $C_0$ the great circle segment from $p$ to $e_0$. Let $R_n : \mathbb{S}(\mathbb{R}^3) \to \mathbb{S}(\mathbb{R}^3)$ be the rotation of $\mathbb{S}(\mathbb{R}^3)$ which takes $p$ to $p_n$ and some point $c_n \in C$ to $p$. Then $\lim_n c_n = p$. Now we define the sequence $\{g_n\}_{n \in \mathbb{N}}$ of functions on $\mathbb{S}(\mathbb{R}^3)$ via $g_n(s) = f(R_n(s))$, $s \in \mathbb{S}(\mathbb{R}^3)$. Then we have:

  a) $\lim_n g_n(p) = M$.

  b) $M = \{g_n(x) \mid x \in \mathbb{S}(\mathbb{R}^3)\}$ and $m = \inf\{g_n(x) \mid x \in \mathbb{S}(\mathbb{R}^3)\}$ for any $n \geq 1$.

  c) $g_n(c_n) = f(p)$ for any $n \geq 1$.

(b) For any $n \geq 1$ we define a frame function $h_n(s) = g_n(s) + g_n(p(s))$, $s \in \mathbb{S}(\mathbb{R}^3)$. Then we have:

  a) $2m \leq h_n(s) \leq 2M$, $s \in \mathbb{S}(\mathbb{R}^3)$, $n \geq 1$.

  b) $\lim_n h_n(p) = 2M$.

  c) Each $h_n$ is constant on $E_p$.

  d) $h_n(c_n) \leq M + f(p)$, $n \geq 1$.

(e) Any $h_n$ can be consider as the element of the compact space $[2m, 2M]^{\mathbb{S}(\mathbb{R}^3)}$ in the product topology, and $\{h_n\}_{n \in \mathbb{N}}$ has an accumulation point, say $h$. Then $h$ is a bounded frame function such that:

  a) $h(p) = 2M = \sup\{h(s) \,|\, s \in \mathbb{S}(\mathbb{R}^3)\}$.

  b) $h$ is constant on $E_p$.

  c) $h$ is continuous in view of lemma C.0.40.

(i) We shall show that $f(p) = M$. Choose $\varepsilon > 0$ and $c \in C_0$ such that $h(c) > 2M - \varepsilon$. By lemma C.0.39, $c$ can be reached from $c_n$ in two steps. Applying lemma C.0.38 to $h_n$ we see that $h_n(c_n) > h_n(c) - 2\delta_n$, where $\delta_n > 2M - h_n(p)$ and $\lim_n \delta_n = 0$. Now we choose a subsequence $\{h_{n_j}\}$ such that $\lim_j h_{n_j}(c) > 2M - \varepsilon$. From the above we have

$$M + f(p) \geq \liminf_j h_{n_j}(c_{n_j}) \geq \lim_j \left(h_{n_j}(c) - 2\delta_n\right) > 2M - \varepsilon$$

so that, $f(p) > M - \varepsilon$.

We note that passing to $-f$, $f$ obtains also its infimum, and the lemma is proved. ∎

**Theorem C.0.42.** *Let $f$ be a semibounded frame function on $\mathbb{R}^3$ with the weight $W$. Define $M = \sup\{f(x) \,|\, x \in \mathbb{S}(\mathbb{R}^3)\}$, $m = \inf\{f(x) \,|\, x \in \mathbb{S}(\mathbb{R}^3)\}$, $\alpha = W - M - m$. There is a frame $(p, q, r)$ such that if the frame coordinates with respect to $(p, q, r)$ of $s \in \mathbb{S}(\mathbb{R}^3)$ are $(x, y, z)$, then:*

$$f(s) = Mx^2 + \alpha y^2 + mz^2, \quad s \in \mathbb{S}(\mathbb{R}^3).$$

*Proof:* Choose $p \in \mathbb{S}(\mathbb{R}^3)$ such that $M = f(p)$ and $r \in E_p$ such that $f(r) = m$; this is possible in view of lemma C.0.41 and proposition C.0.35 Then $f(q) = \alpha$ and we can assume that $m < \alpha < M$, since otherwise, applying lemma C.0.40 to $f$ or to $-f$, we have (3.11). Let us denote the right-hand of (3.11) by $g(s)$ and put $h(s) = g(s) - f(s)$, $s \in \mathbb{S}(\mathbb{R}^3)$. We claim to show that $h = 0$, and we shall proceed in three steps.

(a) Let $\hat{p}, \hat{q}, \hat{r}$ denote the 90° right-hand rotations about the axes given by the vectors $p, q$ and $r$. The function $f(s) + f(\hat{p}(s))$ has the constant value $W - M = \alpha + m$ on $E_p$, and it attains its supremum $2M$ at $p$. From lemma C.0.40 we conclude that

$$f(s) + f(\hat{p}(s)) = 2Mx^2 + (m + \alpha)(1 - x^2) = g(s) + g(\hat{p}(s))$$

and similarly, if we take $-f$ and $-m = \sup\{-f(s) \,|\, s \in \mathbb{S}(\mathbb{R}^3)\}$, we have

$$f(s) + f(\hat{r}(s)) = g(s) + g(\hat{r}(s)).$$

(b) We show that *(i)* if $x = y$, or $x = z$, or $y = z$, then $f(s) = g(s)$; *(ii)* if $x = -y$, or $x = -z$, or $y = -z$, then $f(s) = g(s)$.

Let us identify a point $s \in \mathbb{S}(\mathbb{R}^3)$ with its frame coordinates $(x, y, z)$, then $\hat{r}(x, y, z) = (-y, x, z), \hat{p}(x, y, z) = (x, -z, y)$. More generally, we have:

$$\hat{p}\hat{p}\hat{r}(x, x, z) = (-x, -x, -z),$$
$$\hat{p}\hat{r}\hat{r}(x, z, z) = (-x, -z, -z),$$
$$\hat{r}\hat{p}\hat{p}\hat{p}\hat{r}(x, y, x) = (-x, -y, -x).$$

Suppose $x = y$. Since $f(s) = f(-s)$, $g(s) = g(-s)$, we conclude from:

$$f(s) + f(\hat{r}(s)) = g(s) + g(\hat{r}(s)),$$

$$f(\hat{r}(s)) + f(\hat{p}\hat{r}(s)) = g(\hat{r}(s)) + g(\hat{p}\hat{r}(s)),$$

$$f(\hat{p}\hat{r}(s)) + f(\hat{p}\hat{p}\hat{r}(s)) = g(\hat{p}\hat{r}(s)) + g(\hat{p}\hat{p}\hat{r}(s)).$$

Subtracting the second equation from the sum of the first and third, we obtain that $f(s) = g(s)$. Similarly, we obtain the remaining cases in *(i)* and *(ii)*.

(c) For the frame function $h = g - f$ we have $h(p) = h(q) = h(r) = 0$, so that the weight of $h$ is 0. Moreover, in view of (b), $h$ is zero on six great circles $x = \pm y, x = \pm z, y = \pm z$. We claim to show that $h$ is identical to zero. Suppose the converse and put $M' = \sup h = h(p')$; $m' = \inf h = h(r')$, $\alpha' = h(q')$, $q' \perp r'$, $q' \perp p'$. We have four possibilities:

(i) $M' = -m'$. If not, then $m' > -M'$ which gives $\alpha' < 0$, and $\alpha'$ is the maximal value of $h$ on the great circle orthogonal to $p'$. The great circle $x = y$ must intersect the former great circle in at least two points, where $h$ takes the value 0. Considering $-h$, we give a contradiction to the assumption $m' < -M'$.

(ii) $\alpha' = 0$. It follows from (i) and the fact that the weight of $h$ is 0.

(iii) $h(x', y', z') = M(x'^2 - z'^2)$, where $(x', y', z')$ denote the frame coordinates in the frame $(p', q', r')$. This follows from (b) when we change $f$ by $h$ and $g$ by $M(x'^2 - z'^2)$.

(iv) The function $h$ attains the value 0 on the great circle $x' = y'$ exactly at four points $(x'x'x')$, $(x'x' - x')$, $(-x' - x'x')$ and $(-x' - x' - x')$ This follows from (iii).

The great circles $x = y$, $x = z$, and $y = z$ intersect at two points, $(x, x, x)$ and $(-x, -x, -x)$. Since $h$ is zero on these great circles, the great circle $x' = y'$ must pass through the points $(x, x, x)$ and $(-x, -x, -x)$ since in the opposite case there would be six points on the great circle $x' = y'$ at which $h$ takes the value 0 (contradiction with (iv)).

The great circles $x = -y$ and $x = -z$ intersect at $(x, -x, -x)$ and $(-x, x, x)$. The great circle $x' = y'$ must also intersect at these points, since, otherwise, it has to intersect $x = -y$ and $x = -z$ at four points, making six points at which $h$ has to take the value 0 on $x' = y'$. On the other hand, there is only one great circle $y = z$ passing through the four points $(x, x, x)$, $(-x, -x, -x)$, $(x, -x, -x)$ and $(-x, x, x)$. It follows that $y = z$ and $x' = y'$ describe the same great circle, and, consequently, $h$, by (b), must take the value zero at all points $x' = y'$, which contradicts (iv).

The theorem is completely proved. ∎

**Lemma C.0.43.** *Let $f$ be a semibounded frame function on an $n$-dimensional Hilbert space $\mathscr{H}$ which is regular on each completely real subspace. Then $f$ is regular.*

*Proof:* Put $M = \sup\{J(x) \mid x \in \mathbb{S}(\mathscr{H}_n)\}$ and choose a sequence $\{x_n\}_{n \in \mathbb{N}}$ in $\mathbb{S}(\mathscr{H}_n)$ such that $\lim_n f(x_n) = M$. Since $\mathbb{S}(\mathscr{H}_n)$ is compact in the strong topology, passing to a subsequence, we can assume $\lim_n x_n = x$, $x \in \mathbb{S}(\mathscr{H}_n)$. First of all we show that $f(x) = M$. Let $\lambda_n = \frac{\langle x, x_n \rangle}{\|\langle x, x_n \rangle\|}$, $n \geq 1$.

Since $x_n \to x$, $\lambda_n \to 1$. The inner product $\langle \lambda_n x_n, x \rangle$ is real and, consequently, $\lambda_n x_n$ and $x$ generate over $\mathbb{R}$ a completely real subspace $\mathcal{V}$. Therefore, there is a Hermitian operator $T$ on $\mathcal{V}$ such that $f(x) = \langle Tx, x \rangle$, $x \in \mathbb{S}(\mathcal{V})$. Since $f(x_n) = f(\lambda_n x_n)$, we have:

$$|f(x) - M| \leq |(f(x) - f(\lambda_n x_n)| + |f(\lambda_n x_n) - M| =$$
$$= |\langle Tx, x \rangle - \langle Tx_n, x_n \rangle| + |f(x_n) - M| \leq$$
$$\leq 2\|T\| \|x - x_n\| + |f(x_n) - M| \to 0.$$

Let us now extend the function $f$ to a function $F$ defined on the whole $\mathcal{H}_n$ via:

$$F(x) = \begin{cases} 0 & \text{, if } x = 0, \\ \|x\|^2 f(\frac{x}{\|x\|}) & \text{, if } x \neq 0. \end{cases} \quad x \in \mathcal{H}_n.$$

It is clear that for $c \in \mathbb{K}$ we have:

$$F(cx) = |c| F(x), \, x \in \mathcal{H}_n.$$

We show that if $y$ is any vector of $\mathcal{H}_n$ orthogonal to $x$, then:

$$F(cx + y) = |c|^2 M + F(y), \quad c \in \mathbb{K}.$$

Let $L$ be the two-dimensional real linear manifold generated by $x$ and $y$. This is completely real. Since $f$ is regular, there is a Hermitian bilinear form $t_L$. But $x$ is the point at which the maximum of the bilinear form is reached, and $y$ is orthogonal to $x$ in $L$. Hence the matrix of the bilinear form $t_L$ is diagonal with respect to the orthonormal basis $\{x, y\}$ of $L$, and, consequently, for all real numbers $r, s$ we have:

$$F(rx + sy) = |r|^2 M + |s|^2 F(y).$$

Assume now that $v$ is any vector of $\mathcal{H}_n$ orthogonal to $y$, and $c$ is a non-zero element of $\mathbb{K}$. Calculating:

$$F(xc + v) = F\left(c(x + c^{-1}v)\right) = |c|^2 F(x + c^{-1}v) = |c|^2 F(x + |c|^{-1}|c|c^{-1}v),$$

and $y = |c|c^{-1}v$ is orthogonal to $x$. Then:

$$F(x + c^{-1}v) = M + |c|^{-2} F(|c|c^{-1}v) = M + |c|^{-1} F(v),$$

which gives $F(cx + v) = |c|^2 M + F(v)$.

In other words, we have proved that if the restriction of a semibounded frame function to any completely real subspace is regular, and so is $f$ on $\mathcal{H}_2$. The restriction of $f$ to $K_{n-1} = \text{span}(x)^\perp$. is a semibounded frame function on $K_{n-1}$ and it has the same properties as $f$. Using mathematical induction, we can assume that the present lemma is valid for any $k < n$. Then $F\Big|_{K_{n-1}}$ is regular by the induction hypothesis and, consequently, there is an orthonormal basis $\{e_1, \ldots, e_{n-1}\}$ for $K_{n-1}$ and real numbers $M_1, \ldots, M_{n-1}$ such that:

$$F(c_1 e_1 + \ldots + c_{n-1} e_{n-1}) = |c_1|^2 M_1 + \ldots + |c_{n-1}|^2 M_{n-1}, \quad \forall c_1, \ldots, c_{n-1} \in \mathbb{K}.$$

By the proof for $\mathcal{H}_2$, we see that:

$$F(cx + c_1 e_1 + \ldots + c_{n-1} e_{n-1}) = |c|^2 M^2 + |c_1|^2 M_1 + \ldots + |c_{n-1}|^2 M_{n-1}.$$

This means that $f(x) = \langle Ux, x \rangle$, $x \in \mathbb{S}(\mathcal{H}_n)$, where $U$ is a Hermitian operator on $\mathcal{H}_n$ such that $Ux = Mx$, $Ue_j = M_j e_j$, $j = 1, \ldots, n-1$. ∎

**Theorem C.0.44.** *Any semibounded frame function on $\mathscr{H}_n$, $\dim(\mathscr{H}_n) = n \geq 3$, is regular.*

*Proof:* Since $n \geq 3$, there is a completely real three-dimensional subspace, and the restriction of $f$ to any completely real three-dimensional subspace $\mathscr{V}$ of $\mathscr{H}_n$ induces, by theorem C.0.42, the regularity of $f\Big|_{\mathbb{S}(\mathscr{V})}$. On the other hand, every completely real two-dimensional subspace $\mathscr{W}$ of $\mathscr{H}_n$ can be embedded in a completely real three-dimensional subspace $\mathscr{V}$. Therefore, there is a bilinear form $t_\mathscr{V}$ on $\mathscr{V} \times \mathscr{V}$ such that $f(x) = t_\mathscr{V}(x, x)$, $x \in \mathbb{S}(\mathscr{V})$. It is easy to see that $t_\mathscr{V}\Big|_{\mathscr{W} \times \mathscr{W}}$ gives a bilinear form; consequently, $f\Big|_{\mathbb{S}(\mathscr{W})}$ is regular.

In view of lemma C.0.43, it means that $f$ is regular, that is, there is a unique Hermitian operator $T$ on $\mathscr{H}_n$ such that $f(x) = \langle Tx, x \rangle$, $x \in \mathbb{S}(\mathscr{H}_n)$. ■

- **Infinite dimensional** Hilbert space case:

**Definition C.0.45.** *Let $\mathscr{U}$ be a linear submanifold of $\mathscr{H}$, dense in $\mathscr{H}$. The mapping $f : \mathbb{S}(\mathscr{U}) \to \mathbb{R}$ is said to be a **frame type function** on $\mathscr{H}$ if:*

(a) *For any orthonormal set of vectors $\{x_i\}_{i \in I}$ in $\mathscr{U}$, $\{f(x_i)\}_{i \in I}$ is summable,*

(b) *For any finite-dimensional subspace $K$ of $\mathscr{U}$, $f\Big|_{\mathbb{S}(K)}$ is a frame function on $K$.*

It is clear that if $f$ is a frame type function on $\mathscr{H}$ with given $\mathscr{U}$, then if $K$ is a closed subspace of $\mathscr{H}$ such that $\mathscr{U} \cap K$ is dense in $K$, then $f_{K \cap \mathscr{U}} := f\Big|_{\mathbb{S}(K \cap \mathscr{U})}$ is a frame type function on $K$.

**Definition C.0.46** (Unbounded frame type functions). *Denote by $\mathfrak{R}(\mathscr{H})$ the set of all unbounded frame type function on $\mathscr{H}$.*

Our goal is to show that $\mathfrak{R}(\mathscr{H}) = \varnothing$ whenever $\dim(\mathscr{H}) = \infty$.

**Lemma C.0.47.** *Let $\mathfrak{R}(\mathscr{H}) \neq \varnothing$, $\dim(\mathscr{H}) = \infty$.*

(a) *There exist a submanifold $\mathscr{U}$ dense in $\mathscr{H}$ and an $f \in \mathfrak{R}(\mathscr{H})$, $f : \mathbb{S}(\mathscr{U}) \to \mathbb{R}$ such that $f_{\mathscr{U} \cap \{x\}^\perp} := f\Big|_{\mathscr{U} \cap \{x\}^\perp}$ is bounded on $\{x\}^\perp$ whenever $x \in \mathbb{S}(\mathscr{U})$ and $|f(x)| > 1$.*

(b) *There exist $f \in \mathfrak{R}(\mathscr{H})$ and a unit vector $x_0$ in $\mathscr{U}$ with:*

$$\max\left\{|f(x_0)|, \sup\{|f(y)| \,\big|\, y \in \mathbb{S}(\{x_0\}^\perp \cap \mathscr{U})\}\right\} = 1, \tag{C.9}$$

*where $f$ satisfies (i).*

*Proof:* (a) If $x \in \mathscr{U}$, then $\mathscr{U} \cap \{x\}^\perp$ is dense in $\{x\}^\perp$, so that $f_{\mathscr{U} \cap \{x\}^\perp}$ is a frame type function on $\{x\}^\perp$. Suppose that (i) does not hold. Then for any $f \in \mathfrak{R}(\mathscr{H})$ we find an $x_1 \in \mathbb{S}(\mathscr{U})$ such that $|f(x_1)| > 1$ and $f_{\mathscr{U} \cap \{x_1\}^\perp} \in \mathfrak{R}(\{x_1\}^\perp)$. Since $\mathscr{H}$ is infinite-dimensional, it is isomorphic with its subspace $\{x_1\}^\perp$. Consequently, any frame type function from a non-void set $\mathfrak{R}(\{x_1\}^\perp)$ also does not fulfill (i). In particular, for $f_{\mathscr{U} \cap \{x_1\}^\perp}$ we can find a unit vector $x_2$ in $\mathscr{U} \cap \{x_1\}^\perp$ such that $|f(x_2)| > 1$ and $f_{\mathscr{U} \cap \{x_1, x_2\}^\perp} \in \mathfrak{R}(\{x_l, x_2\}^\perp)$. Continuing this process, we find a sequence of orthonormal vectors $\{x_n\}_{n\in\mathbb{N}}$ from $\mathscr{U}$ such that $\sum\limits_{n=1}^{\infty}|f(x_n)|=\infty$, which contradicts the definition of a frame type function.

(b) Let $f$ and $\mathscr{U}$ satisfy (i). Multiplying $f$ by some non-zero constant, if necessary, we obtain C.9. ∎

**Lemma C.0.48.** *Let $f \in \mathfrak{R}(\mathscr{H})$, $\dim(\mathscr{H}) = \infty$, satisfy the condition (ii) of the previous lemma C.0.47. Then there exist orthonormal vectors $e_l, e_2, e_3 \in \mathscr{U}$ such that $|f(e_i)| > 1$, $i = 1, 2, 3$.*

*Proof:* Let $f$ and $x_0 \in \mathbb{S}(\mathscr{U})$ satisfying (ii) of lemma C.0.47 be given. From the unboundedness of $f$ we find an $e_3$ such that $|f(e_3)| > 6$. Put $K = \mathrm{span}\{e_3, x_0\}$ and choose $g, g_0 \in \mathbb{S}(K)$ such that $x_0 \perp g_0$ and $e_3 \perp g$. Since $f_{\mathscr{U}\cap K}$ is a frame function on $K$, we have $f(e_3) + f(g) = f(x_0) + f(g_0)$. Hence,

$$|f(g)| \geq |f(e_3)| - |f(g_0)| - |f(x_0)| > 4.$$

Let $h \in \mathbb{S}(K^\perp)$, so that, $h \in \mathbb{S}(\{x_0\}^\perp)$ and $|f(h)| \leq 1$. Since $|f(e_3)| > 1$, $f_{\mathscr{U}\cap\{e_3\}^\perp}$ is bounded on $\mathbb{S}(\{e_3\}^\perp)$ in view of (i) of lemma C.0.47. Put $E = \mathrm{span}\{g, h\}$ and embed it into a three-dimensional subspace $\mathscr{V}$ of $\mathscr{U} \cap \{e_3\}^\perp$. Then $f_{\mathscr{U}\cap\mathscr{V}}$ is a bounded frame function, and by Gleason's theorem for finite-dimensional Hilbert spaces, which follows from theorem C.0.44 applied in the proof of theorem 6.1.1, $f_{\mathscr{U}\cap\mathscr{V}}$ is continuous, so is $f_{\mathscr{U}\cap E}$.

Since $|f(g)| > 4$ and $|f(h)| \leq 1$, by a classical result of mathematical analysis, we can find a unit vector $e_1$ in $E$ such that $|f(e_1)| = 2$. Finally, let $e_2$ be any unit vector in $E$ orthogonal to $e_1$. Then $e_1, e_2, e_3$ are those in question. Indeed, since $f(e_2) + f(e_1) = f(g) + f(h)$, we obtain

$$|f(e_2)| \geq |f(g)| - |f(h)| - |f(e_1)| > 1.$$

∎

**Lemma C.0.49.** *Let $\mathscr{H}$ be a real four-dimensional Hilbert space. Let $e_l, e_2, e_3, e \in \mathbb{S}(\mathscr{H})$ such that $e_l, e_2, e_3$ are mutually orthogonal, and $e \notin \{e_1\}^\perp \cup \{e_2\}^\perp$, be given. Then there exist two non-zero vectors $x$ and $y$ in $\mathscr{H}$ such that:*

(a) $e = x + y$.

(b) $\langle x, e_1 \rangle = \langle y, e_2 \rangle = \langle x, y \rangle = \langle y - \|y\|^2 e, e_3 \rangle = 0$.

*Proof:* Let $e_4$ be any unit vector of $\mathscr{H}$ such that $\{e_1, e_2, e_3, e_4\}$ is an orthonormal basis of $\mathscr{H}$, and let $e^j = \langle e, e_j \rangle$, $j = 1, 2, 3, 4$, be the coordinates of the vector $e$ with respect to $\{e_1, e_2, e_3, e_4\}$. Regard the following real algebraic equations with unknown $u$ and $v$:

$$u(\alpha - u) + v(\beta - v) = 0, \tag{C.10}$$

$$u - (\gamma^2 + u^2 + v^2)\alpha = 0, \tag{C.11}$$

where $\alpha^2 + \beta^2 < 1$. This system is solvable since from (3) and (4), $u = \alpha\frac{\gamma^2+\beta v}{1-\alpha^2}$, and putting it into (3), we obtain a quadratic equation with respect to $v$, which in view of $\alpha^2 + \beta^2 < 1$, has real roots. Then the vectors $x = (0, e^2, e^3 - u, e^4 - v)$ and $y = (e^1, 0, u, v)$, with respect to our orthonormal basis, where $u$ and $v$ are the solutions to (3) and (4) with $\alpha = e^3$, $\beta = e^4$, $\gamma = e^1$, are vectors in question. In fact, since $e \notin \{e_1\}^\perp \cup \{e_2\}^\perp$, we have $e_1, e_2 \neq 0$, so that:

$$(e^3)^2 + (e^1)^2 < \sum_{j=1}^{4}(e^j)^2 = \|e\|^2 = 1.$$

∎

**Theorem C.0.50** (Dorofeev-Sherstnev). *Any frame type function on $\mathscr{H}$, with $\dim(\mathscr{H}) = \infty$, is bounded.*

*Proof:* Suppose the converse, i.e., let $\mathfrak{R}(\mathscr{H}) \neq \varnothing$, and let $f \in \mathfrak{R}(\mathscr{H})$ satisfy lemma C.0.47. Select orthonormal vectors $e_1, e_2, e_3$ from lemma C.0.48 with $|f(e_i)| > 1$, $i = 1, 2, 3$, and define the constant

$$C = \max_{1 \leq i \leq 3} \{ |f(e_i)|, \sup\{ |f(x)| \mid x \in \mathbb{S}(\{e_i\}^\perp \cap \mathscr{U}) \} \}.$$

From the unboundedness of $f$ it follows that there is a vector $h \in \mathbb{S}(\mathscr{U})$ such that $|f(h)| > 3C$. It is clear that $h \notin \bigcup_{i=1}^{3} \mathscr{U} \cap \{e_i\}^\perp$ and put $\lambda_i = \frac{\langle h, e_i \rangle}{|\langle h, e_i \rangle|}$, $i = 1, 2, 3$. Then $\sum_{i=1}^{3} \lambda_i = 1$ and $\langle h, \lambda_i e_i \rangle$ is real for $i = 1, 2, 3$. Let $\mathscr{V}$ be a completely real subspace of dimension 4 containing $h$ and all $\lambda_i e_i$'s. Applying lemma C.0.49 to vectors $\lambda_i e_i$'s and $h$, we find two non-zero vectors $x$ and $y$ in $\mathscr{V}$ such that:

$$\langle x, \lambda_2 e_2 \rangle = \langle y, \lambda_3 e_3 \rangle = \langle x, y \rangle = \langle z, \lambda_1 e_1 \rangle = 0, \quad h = x + y,$$

where $z = y - \|y\|^2 h$ is a non-zero vector. Since $\mathrm{span}\{z, h\} = \mathrm{span}\{x, y\} = \mathrm{span}\{y, h\}$, we have $f(h) + f\left(\frac{z}{\|z\|}\right) = f\left(\frac{x}{\|x\|}\right) + f\left(\frac{y}{\|y\|}\right)$. From the construction we conclude that $z \in \{e_1\}^\perp$, so that $\left|f\left(\frac{z}{\|z\|}\right)\right| \leq C$. Similarly, $\left|f\left(\frac{x}{\|x\|}\right)\right| \leq C$, $\left|f\left(\frac{y}{\|y\|}\right)\right| \leq C$.

Since $|f(h)| \leq \left|f(h) + f\left(\frac{z}{\|z\|}\right)\right| + \left|f\left(\frac{z}{\|z\|}\right)\right|$, then $\left|f(h) + f\left(\frac{z}{\|z\|}\right)\right| \geq |f(h)| - \left|f\left(\frac{z}{\|z\|}\right)\right| > 3C - C = 2C$, we finally obtain from the last equality:

$$2C \geq \left|f\left(\frac{x}{\|x\|}\right) + f\left(\frac{y}{\|y\|}\right)\right| = \left|f(h) + f\left(\frac{z}{\|z\|}\right)\right| > 2C,$$

which is a desired contradiction. ∎

**Theorem C.0.51.** *Let $\mathscr{U}$ be a dense submanifold of an infinite-dimensional Hilbert space $\mathscr{H}$. Let $f : \mathbb{S}(\mathscr{U}) \to \mathbb{R}$ be a frame type function. Then there is a unique Hermitian trace operator $T$ on $\mathscr{H}$ such that:*

$$f(x) = \langle Tx, x \rangle, x \in \mathbb{S}(\mathscr{U}).$$

*Proof:* In view of theorem C.0.50, $f$ is bounded. Let $\mathscr{N}$ be a finite-dimensional subspace of $\mathscr{J}$, so that, $f\big|_{\mathbb{S}(\mathscr{N})}$ is bounded on $\mathscr{N}$. There exists a bounded bilinear form $t_{\mathscr{N}}$ on $\mathscr{N} \times \mathscr{N}$ such that $f(x) = t_{\mathscr{N}}(x, x)$, $x \in \mathbb{S}(\mathscr{N})$.

We shall define a bilinear form $t$ on $\mathscr{U} \times \mathscr{U}$ as follows: Let $x$ and $y$ be two vectors of $\mathscr{U}$ and let $\mathscr{M}$ be any finite-dimensional subspace of $\mathscr{U}$ containing $x$ and $y$. Put $t(x, y) = t_{\mathscr{N}}(x, y)$. This $t$ is well-defined on $\mathscr{U} \times \mathscr{U}$ since if $\mathscr{M}$ is an another finite-dimensional subspace of $\mathscr{U}$ containing $x$ and $y$, then for $\mathscr{Q} = \mathscr{M} \vee \mathscr{N}$ we have, in view of the fact that for any real bilinear form $G$ and induced quadratic form $Q(x) := G(x, x)$ we have the identity $G(x, y) = \frac{1}{4}(Q(x+y) - Q(x-y))$ and for any complex bilinear form $G_{\mathbb{C}}$ and induced quadratic form $Q_{\mathbb{C}}(x) := G_{\mathbb{C}}(x, x)$ we have the identity $G_{\mathbb{C}}(x, y) = \frac{1}{4}(Q_{\mathbb{C}}(x+y) - Q_{\mathbb{C}}(x-y)) + \frac{i}{4}(Q_{\mathbb{C}}(x+iy) - Q_{\mathbb{C}}(x-iy))$, then $t_{\mathscr{N}}(x, y) = t_{\mathscr{Q}}(x, y) = t_{\mathscr{M}}(x, y)$. It is clear that $f(x) = t(x, x)$, $x \in \mathbb{S}(\mathscr{U})$.

The boundedness of $f$ entails that $t$ can be uniquely extended to a bounded symmetric bilinear form $\bar{t}$ on $\mathscr{H} \times \mathscr{H}$, consequently, by corollary B.0.14, there is a Hermitian operator $T$ on $\mathscr{H}$ such that $f(x) = \langle Tx, x \rangle$, $x \in \mathbb{S}(\mathscr{U})$.

We claim to show that $T \in \mathcal{J}_1(\mathcal{H})$. Suppose the converse. Then there is an orthonormal set of vectors $\{f_1, \ldots, f_{n_1}\}$ in $\mathcal{H}$ such that $\sum_{k=1}^{n_1} |\langle Tf_k, f_k \rangle| > 1$. Choose an $\varepsilon > 0$ such that $\sum_{k=1}^{n_1} |\langle Tf_k, f_k \rangle| > 1 + \varepsilon$. It is easy to see that for $\{f_1, \ldots, f_{n_1}\}$ we can find an orthonormal set of vectors $\{h_1, \ldots, h_{n_1}\}$ in $\mathcal{U}$ such that $\|h_k - f_k\| < \frac{\varepsilon}{2n_1\|T\|}$, $k = 1, \ldots, n_1$. Then:

$$|f(h_k) - \langle Tf_k, f_k \rangle| \leq |\langle T(h_k - f_k), f_k \rangle| + |\langle Th_k, h_k - f_k \rangle| \leq 2\|T\|\|h_k - f_k\| < \frac{\varepsilon}{n_1},$$

so that:

$$\sum_{k=1}^{n_1} |f(h_k)| \geq \sum_{k=1}^{n_1} |\langle Tf_k, f_k \rangle| - \sum_{k=1}^{n_1} |\langle Tf_k, f_k \rangle - f(h_k)| > 1.$$

Put $\mathcal{H}_1 = \{h_1, \ldots, h_{n_1}\}^\perp$, then $\mathcal{U}_1 = \mathcal{U} \cap \mathcal{H}_1$ is a dense submanifold in $\mathcal{H}_1$, so that, $f\big|_{\mathbb{S}(\mathcal{U}_1)}$ is a frame type function on $\mathcal{H}_1$. Therefore, as in the beginning of the present proof, there is a Hermitian operator $T_1 = P_{\mathcal{H}_1} T P_{\mathcal{H}_1}$ on $\mathcal{H}_1$ such that $f(x) = \langle T_1 x, x \rangle = \langle Tx, x \rangle$, $x \in \mathbb{S}(\mathcal{U}_1)$. Here $T_1$ is not any trace operator since $T \notin \mathcal{J}_1(\mathcal{H})$.

Repeating the same reasonings as above, we find an orthonormal set of vectors $\{f_{n_1+1}, \ldots, f_{n_2}\}$ in $\mathcal{H}_1$ such that $\sum_{k=n_1+1}^{n_2} |\langle Tf_k, f_k \rangle| > 1$, and for it we find and orthonormal basis $\{h_{n_1+1}, \ldots, h_{n_2}\}$ in $\mathcal{U}_1$ with $\sum_{k=n_1+1}^{n_2} |f(h_k)| > 1$. Continuing this process, we find a countable family of orthonormal set of vectors $\{h_n\}_{n \in \mathbb{N}} \subset \mathcal{U}$ such that $\sum_{k=1}^{\infty} |f(h_k)| = \infty$ which contradicts our assumption, so that, $T \in \mathcal{J}_1(\mathcal{H})$. ∎

**Corollary C.0.52.** *Any frame function $f$ on an infinite-dimensional Hilbert space $\mathcal{H}$ is regular. Moreover, there is a unique Hermitian trace operator $T$ on $\mathcal{H}$ such that C.4 holds.*

*Proof:* Let $f$ be a frame function on $\mathcal{H}$. Then it is a frame type function on $\mathcal{H}$ with $\mathcal{U} = \mathcal{H}$. Calling theorem C.0.51, we obtain the assertion of the corollary. ∎

∎

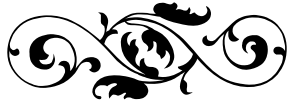

# Symbols List

$\omega$ State on a $C^*$-algebra, 52

$\mathfrak{S}$ Set of states on a $C^*$-algebra, 52

$\mathfrak{Im}$ Imaginary part 52

$\omega_\psi$ Vector state for the normalized vector $\psi$, 54

$C(X)$ The set of continuous functions from $X$ to $\mathbb{C}$, 55

$C_b(X)$ The set of bounded continuous functions from $X$ to $\mathbb{C}$, 55

$C_c(X)$ The set of continuous functions with compact support from $X$ to $\mathbb{C}$, 55

$C_0(\Omega)$ The set of continuous functions from $\Omega$ to $\mathbb{C}$ that vanish at infinity, 55

$\mathcal{J}_1(\mathscr{H})$ Set of operators of trace class in $\mathscr{H}$, 97

$\mathbb{S}$ Surface of sphere of radius 1, 98

$\mathcal{C}(\{A_n\}_{n\in I})$ Context generated by $\{A_n\}_{n\in I}$, 110

$\leftsquigarrow$ Trust relation, 121

**Obj**$_C$ Collection of objects of the category $C$, 124

**Mor**$_C$ Collection of morphisms of the category $C$, 124

$\mathfrak{C}(\mathscr{A})$ Context category of $\mathscr{A}$, 125

$\hookrightarrow$ Inclusion morphism, 125

$\Xi_\mathfrak{V}$ Gel'fand spectrum of the abelian von Neumann algebra $\mathfrak{V}$, 134

$\lambda$ Character, 134

$\hat{\tilde{A}}$ Gel'fand transform, 134

$\underline{\Xi}$ Gel'fand presheaf, 141

$\pi$ Usually means a representation 155

$\big|_\mathcal{W}$ Restriction, 156

$\mathscr{H}$ Hilbert space, 161

$\mathbb{N}_0$ Natural numbers including 0, 161

$\mathcal{V}^\#$ Continuous dual space of $\mathcal{V}$, 170

$\mathcal{V}'$ Algebraic dual space of $\mathcal{V}$, 175

$r(u)$ Spectral radius, 176

$B^{\mathscr{B}}(\sigma(A))$ Bounded Borel measurable functions in $\sigma(A)$, 182